\documentclass[11pt,oneside]{report}

\usepackage[letterpaper,margin=1in]{geometry}
\usepackage{setspace}
\usepackage{graphicx}
\usepackage{amsmath,amssymb,amsthm}
\usepackage[authoryear]{natbib}
\usepackage{bibentry}
\nobibliography*
\usepackage{booktabs} 
\usepackage[ruled]{algorithm2e} 
\usepackage{amsmath}
\usepackage{amsthm}
\usepackage{caption}
\usepackage{amssymb}
\usepackage{thmtools}
\usepackage{thm-restate}
\usepackage{comment}
\usepackage{xspace}
\usepackage[usenames,dvipsnames]{xcolor}
\usepackage[colorlinks=true,citecolor=BrickRed,linkcolor=BrickRed,urlcolor=BrickRed]{hyperref}
\usepackage{cleveref}
\usepackage{tikz}
\usetikzlibrary{arrows,positioning}

\newtheorem{theorem}{Theorem}[section]
\newtheorem{lemma}[theorem]{Lemma}
\newtheorem{proposition}[theorem]{Proposition}
\newtheorem{definition}[theorem]{Definition}
\newtheorem{corollary}[theorem]{Corollary}
\newtheorem{innercustomthm}{Theorem}
\newenvironment{customthm}[1]
  {\renewcommand\theinnercustomthm{#1}\innercustomthm}
  {\endinnercustomthm}
\newtheorem{example}{Example}[section]

\DeclareMathOperator*{\argmax}{arg\,max}
\DeclareMathOperator*{\argmin}{arg\,min}

\SetAlFnt{\small}
\SetAlCapFnt{\small}
\SetAlCapNameFnt{\small}
\SetAlCapHSkip{0pt}
\IncMargin{-\parindent}

\newcommand{\p}{\mathbf{p}}

\usepackage{footnote}
\makesavenoteenv{algorithm}

\newcommand{\alloc}{\mathbf{a}}
\newcommand{\x}{\mathbf{x}}
\newcommand{\eq}{\mathsf{EQ}}
\newcommand{\E}{\mathbb{E}}
\newcommand{\on}{\text{on}}
\newcommand{\off}{\text{off}}

\usepackage{multirow}
\usepackage{tabularray}
\usepackage{enumitem}
\usepackage{xspace}
\usepackage[usenames,dvipsnames]{xcolor}

\newtheorem{remark}{Remark}

\newcommand{\BR}{\text{BR}}
\newcommand{\ut}{\underline{t}}
\newcommand{\bt}{\mathbf{t}}
\newcommand{\w}{\mathbf{w}}
\newcommand{\z}{\mathbf{z}}
\usepackage{mathtools}
\definecolor{english}{rgb}{0.0, 0.5, 0.0}
\definecolor{lilac}{RGB}{230,215,255}
\definecolor{salmon}{RGB}{255,55,40}
\definecolor{customblue}{RGB}{53,140,219}
\usepackage{url}            
\usepackage{amsfonts}       
\usepackage{nicefrac}       
\usepackage{microtype}      
\usepackage{float}
\usepackage[noend]{algpseudocode}
\usepackage{subcaption}
\usepackage[labelformat=simple]{subcaption}

\newcommand{\B}{\mathcal{B}}
\renewcommand{\S}{\mathcal{S}}
\newcommand{\V}{\mathcal{V}}
\newcommand{\N}{\mathcal{N}}
\newcommand{\vr}{\mathbf{r}}
\newcommand{\I}{\mathcal{I}}
\newcommand{\Parens}[1]{\left(#1\right)}
\newcommand{\X}{\mathcal{X}}
\newcommand{\A}{\mathcal{A}}
\newcommand{\R}{\mathcal{R}}
\newcommand{\ind}{\mathbb{I}}
\newcommand{\vtheta}{\boldsymbol{\theta}}
\newcommand{\norm}[1]{\lVert#1\rVert}
\newcommand{\wo}{\backslash}
\usepackage[suppress]{color-edits}
\usepackage{bm}
\addauthor{gm}{cyan}
\addauthor{dcp}{blue}

\makeatletter
\def\@makechapterhead#1{%
  \vspace*{-20pt}%
  {\parindent \z@ \raggedright \normalfont
    \ifnum \c@secnumdepth >\m@ne
      \huge\bfseries \@chapapp\space \thechapter
      \par\nobreak
      \vskip 10pt
    \fi
    \LARGE\bfseries #1\par\nobreak
    \vskip 12pt
  }}
\def\@makeschapterhead#1{%
  \vspace*{-20pt}%
  {\parindent \z@ \raggedright \normalfont
    \LARGE\bfseries #1\par\nobreak
    \vskip 12pt
  }}
\makeatother
\begin{document}
\pagenumbering{roman}
\begin{titlepage}
\thispagestyle{empty}
\addcontentsline{toc}{chapter}{Title Page}
\begin{center}

{\LARGE Pricing, Matching, and Bundling: an Equilibrium Analysis of Online Platforms}

\vspace{1in}

a dissertation presented by

Gary Qiurui Ma

\vspace{0.35in}

to

John A. Paulson School of Engineering and Applied Sciences

\vspace{0.35in}

in partial fulfillment of the requirements

for the degree of

Doctor of Philosophy

in the subject of

Computer Science

\vfill

Harvard University

Cambridge, Massachusetts

April 7, 2026 

\end{center}
\end{titlepage}
\clearpage
\addtocounter{page}{1}
\thispagestyle{empty}
\addcontentsline{toc}{chapter}{Copyright}

\null
\vfill
\begin{center}
\copyright\ 2026 Gary Qiurui Ma\\
All rights reserved.
\end{center}
\vfill
\null
\clearpage

\thispagestyle{plain}
\addcontentsline{toc}{chapter}{Abstract}
\noindent Dissertation Advisor: David C. Parkes, Yannai A. Gonczarowski
\hfill Gary Qiurui Ma

\vspace{1.5cm}

\begin{center}
Pricing, Matching, and Bundling: An Equilibrium Analysis of Online Platforms

\vspace{0.4cm}

Abstract
\end{center}
Modern online platforms such as marketplaces, ride-hailing services, and food-delivery systems serve a dual role: they are both markets where participants interact and transact, and operators that design and govern how these markets function. 
These platforms connect multiple sides, for example buyers, sellers, and couriers, facilitating access that would otherwise be difficult to achieve. 
By setting the rules of the market, platforms determine who participates, how interactions take place, and how value is created and distributed. In response to these rules, participants may behave strategically, deciding whether to join the platform and which transactions to pursue.

This thesis studies how platform design affects market outcomes through three key levers: pricing that determines participants' gains when operating on a platform; matching that governs which interactions are feasible among participants; and bundling that shapes the structure of supply when the platform itself acts as a market participant. Across these levers, the goal in this thesis is to understand how platforms can be designed to balance platform profitability with overall market welfare. 

The first part of this thesis studies pricing, including both the commission fees that participants pay to a platform and the prices associated with each transaction. For commission fees, I analyze how they affect participants' decisions to join a market platform. For prices on transactions, I examine delivery fees and courier compensation on delivery platforms, showing that well-designed pricing can improve overall market welfare.

The second part of this thesis studies matching. By shaping recommendation systems and consumer search, platforms influence which transactions take place. I first demonstrate the computational challenges facing a platform when optimizing for matching decisions, as matching decisions for one participant impose externalities on other participants. Making use of a simulation approach, I then show that matching and pricing interact closely: when commissions fees are regulated, platforms can adjust matching to sustain revenue and support smaller sellers, helping maintain overall market health after an economic crisis.

The third part of this thesis analyzes bundling. As a marketplace operator, a platform may be able to source products from sellers and offer them as bundled packages to buyers. I analyze which sets of products a platform will choose to bundle, giving a characterization result about which items are included in bundles made available on the platform. 

Collectively, this thesis shows how pricing, matching, and bundling serve as complementary design levers through which platforms can shape market outcomes.
\clearpage

\singlespacing
\addcontentsline{toc}{chapter}{Table of Contents}
\tableofcontents
\clearpage
\doublespacing
\chapter*{Acknowledgments}\label{chap:acknowledgement}
\addcontentsline{toc}{chapter}{Acknowledgments}

I owe tremendously to my advisor, David C. Parkes and Yannai A. Gonczarowski. David took me in when I had limited prior experience and met my early ideas, however unformed or occasionally misguided, with patience and support. He gave me the freedom to explore while gently nudging me towards what matters and offering judgments when I needed it most. During times when I struggled with imposter syndrome, his quiet confidence in me became something I could lean on, giving me courage and calmness that I could not find on my own. Beyond research, his generosity shows in so many ways: in teaching where he goes the extra miles for students; in hosting me for lunch in Copenhagen and London; and forgiveness when I occasionally got into mischief around the SEC building. The attentiveness and care will stay with me long after this thesis.

I am equally grateful to Yannai. He taught me how to think like a theorist, if I may claim the title. In our meetings, he asked the sharpest questions, often cutting straight to the core of a problem and sending me back to first principles. Over the many hours we spent together in the fifth-floor conference rooms and on Zoom, going line by line and word by word through drafts, he imparted to me a wealth of wisdom, including lessons passed down from his own advisors, such as the "the introduction is a contract". These lessons, especially the philosophical ones, will continue to shape how I think and work in the years ahead.

My advisors have been remarkably kind and generous, telling me they are proud of me. If I am fortunate enough to enter academia, I hope to pass on the same love and aspire to be as good an advisor as they have been to me.

I am grateful to my committee member, Yiling Chen, in whose class I first developed my early ideas on matching in platforms. I also want to thank my undergraduate research advisors at Michigan, Michael Wellman and Tilman Borgers, who introduced me to the field of algorithmic game theory, and my undergraduate mentor Sirui Xie for leading me into research.

Over the course of my PhD, I have been incredibly fortunate to learn from and work with many mentors. Xintong Wang drew me into research on platforms; Alon Eden guided me through modeling a market theoretically; Shreyas Sekar and Auyon Siddiq taught me an operation research perspective, and Nicole Immorlica and Brendan Lucier taught me how to conduct research efficiently and craft a compelling narrative, including how to distill complex theories into simple, tractable frameworks while preserving a coherent economic story. I am also grateful to have collaborated with Luca D'Amico-Wong during his undergraduate years, whose trajectory offered a glimpse of how a top-tier researcher develops, and to have had the opportunity to work with Alex, Stephan, and Clara.

I am deeply grateful to my friends and labmates: Daniel, Eric, Fang-Yi, Francisco, Gianluca, Gili, He, Hugh, Jamie, Jeff, Jessie, Kehang, Manuel, Mark, Matheus, Sadie, Safwan, Sai, Sara, Shi, Shirley, Shuran, Siddarth, Tao, Tomer, Tonghan, and Zhou. Thank you for the companionship, support, and the many shared moments along the way. I am especially thankful to Junia, for being by my side and supporting me through these years, and to my parents, for their quiet and unwavering support from afar throughout this journey.
\clearpage

\chapter*{Citations to Previous Work}
\addcontentsline{toc}{chapter}{Citations to Previous Work}

This thesis is based on joint work with collaborators described below. 

{\noindent\large\bfseries \Cref{chap:platform_eq}\par}
\noindent\bibentry{platformEq}\par

{\noindent\large\bfseries \Cref{chap:pricing_with_tips}\par}
\noindent\bibentry{gonczarowski2025pricing}, Working paper\par

{\noindent\large\bfseries \Cref{chap:platform_disruption}\par}
\noindent\bibentry{platformDisruption}\par

{\noindent\large\bfseries \Cref{chap:simulations}\par}
\noindent\bibentry{platformSim}\par

{\noindent\large\bfseries \Cref{chap:bundling}\par}
\noindent\bibentry{OptimalSourcing2026}, Working paper\par

\listoffigures
\addcontentsline{toc}{chapter}{List of Figures}
\clearpage

\listoftables
\addcontentsline{toc}{chapter}{List of Tables}
\clearpage
\doublespacing
\pagenumbering{arabic}
\setcounter{page}{1}

\chapter{Introduction} \label{chap:intro}
Modern online platforms are central intermediaries in economic activities. Online marketplaces such as Amazon, ride-hailing services such as Uber, and food-delivery platforms such as Instacart facilitate interactions among large populations of participants.
At their core, these platforms are forms of markets. As defined by the economists Alfred Marshall \citep{marshall1961principles}, 

\begin{quote}{\em ``a market is the area within which buyers and sellers are in such free intercourse that the prices of the same goods tend to equality''.}
\end{quote}

In this view, online platforms enable participants to trade with one another, overcoming frictions that would otherwise hinder such interactions.

Modern online platforms also depart from this classical definition of a market in a few important ways. First, markets are multi-sided and participants on each side have their own incentive structures \citep{hagiu2015multi,evans2016matchmakers, weyl2010price}. Depending on the market context, platforms may connect buyers and sellers, riders and drivers, or eaters, shoppers, and stores. Each participant has their own preferences, constraints, and outside options, and decides whether to join a platform and how to transact once they do: buyers decide what to search for and purchase, sellers decide whether to list their products and at what prices, and couriers choose which tasks to accept and where to relocate next for better opportunities. Furthermore, these decisions are interdependent, where buyers respond to the availability of seller items as well as delivery conditions, sellers respond to expected buyer demand, and couriers respond to compensation and available orders. In this way, market outcomes are determined by the equilibrium behavior of participants.

Second, the notion of ``free intercourse'' is untenable in online markets due to their scale \citep{bakos1997reducing, ellison2009search}. In the classical sense, free intercourse implies that participants can readily observe and trade with all relevant counterparties. In modern online markets, hundreds of thousands if not millions of participants coexist. Even with digital catalogs provided by the platform, it remains costly for participants to do search over all potential trading partners \citep{diamond1982aggregate,einav2016peer, fradkin2017search}. Instead, they only see a subset of options presented to them, and must rely on the platform to mediate access to the market. Moreover, platforms are not neutral intermediaries, they are also operators that hold much power in actively designing and governing the environments where the economic interactions occur \citep{roth2007art, tirole2017economics,roth2018marketplaces}. As profit-maximizing operators, they design and govern the market in ways that shape access and interaction to their own advantage \citep{rochet2003platform, weyl2010price}. Platform-charged commission fees determine the cost of accessing the other side of the market, recommendation systems and matching algorithms determine which participants are visible to one another, and sourcing decisions shape which goods are available. In this sense, platforms both enable and restrict access: they reduce search and information frictions while introducing new ones through their design choices.

Taken together, modern online platforms operate by mediating access among participants. Market outcomes are jointly determined by the rules of interaction designed by the platform and the strategic behavior of participants under those rules.

This thesis studies how platforms use various levers to directly and indirectly regulate access between participants, and how these levers shape market outcomes. In particular, three levers will be studied: (1) indirect access control with pricing; (2) direct access control with matching; and (3) access control with bundling.

\section{Indirect Access Control with Pricing} 
The first lever through which platforms mediate access is pricing. In the classical view of markets, as described by Alfred Marshall, prices emerge from interactions among participants and are paid from one participant to another in exchange for goods or services. Under perfect competition, prices adjust so that supply equals demand and identical goods have equal price. In this sense, prices serves a coordinating role: they determine the terms of trade between participants but do not restrict access, as any participant who is willing to trade at the prevailing price can do so.

In online platforms, access to the market is mediated, and pricing takes on an additional role as a tool for controlling access \citep{rochet2003platform,Armstrong2006,weyl2010price}. Platforms charge fees for participation, typically in the form of commission or subscription fees. For participants, these fees represent the cost of accessing a larger and more connected market, and directly affect their decision as to whether to join. From the platform's perspective, setting these fees involves a tradeoff between expanding access, by attracting a larger pool of participants, and extracting revenue from those who participate. While many prior works have studied platform's fees \citep{banerjee2017segmenting,birge2021optimal}, Chapter 2 of this thesis focuses on how such fees affect access: even without the platform, participants may retain limited access to trade; the platform expands this access, but conditions it on the payment of fees.

Beyond setting fees for participation, platforms also shape the transaction-level prices exchanged between participants \citep{johnson2017agency}. While these prices arise from interactions among participants, the platform can influence how and when they are paid. For example, features such as ``Buy Now Pay Later'' on market platforms or  options in regard to tips on delivery platforms alter the timing and structure of payments. These design choices affect participants' incentives and thereby affect which trades take place once participants are on the platform. Chapter 3 studies one such design choice, examining how a delivery platform's decision over whether and when buyers can tip couriers influences market outcomes.

\paragraph{Chapter 2: Indirect Access Control with Commission Fee} This chapter studies how platform-imposed commission fees regulate access through participants' entry decisions. In a market platform with unit-demand buyers and unit-supply sellers, each seller chooses whether to join the platform or remain off-platform. Joining the platform grants a seller access to a larger pool of buyers, while remaining off-platform restricts trade to a limited, seller-specific set of buyers. The platform charges a commission fee to sellers who join the platform, in the form of a percentage of transaction price. The resulting equilibrium of this market is determined jointly by sellers' entry decisions and transaction prices: prices emerge from interactions among buyers and both on-platform and off-platform sellers, reflecting the access available to each seller.

Within this framework, this chapter studies the existence and structure of equilibrium outcomes and analyzes their welfare properties. A useful observation is that even when the platform sets fees to maximize its own revenue, the resulting equilibrium retains a bounded fraction of the optimal welfare benchmark in which all sellers have full access to all buyers. Moreover, even modest regulation of commission fees can further improve welfare. These results illustrate how pricing serves as an indirect lever for shaping access in platform-mediated markets.

\paragraph{Chapter 3: Influencing Market Outcomes through Tipping Design} Chapter 3 studies how platforms influence market outcomes through transaction-level prices exchanged between participants, in the context of tipping by buyers on delivery platforms. In a three-sided market with buyers, stores, and couriers, buyers pay for stores' goods and courier delivery through the platform, and may additionally tip couriers even before the delivery. A platform can control whether tipping is allowed and when the tipping amount is revealed to a courier. In particular, when the tipping amount is revealed before couriers decide whether to deliver an order, couriers prioritize high-tip orders, which in turn affects how buyers select orders and choose tipping strategies in equilibrium.

For this three-sided market, the chapter analyzes equilibrium outcomes where delivery prices, worker compensations, tips, and transactions jointly clear the market. I show that the presence of tipping fundamentally alters market structure: it expands the set of feasible outcomes and weakly improves optimal welfare. Furthermore, in markets where couriers have distance-based delivery costs, I show that revealing the tipping amount to couriers before they accept an order can achieve optimal welfare. In contrast, disallowing tipping or only revealing the tipping amount after delivery can lead to significantly lower welfare. These results illustrate how the design of transaction-level payments between participants can have a substantial effect on market outcomes.

\section{Direct Access Control with Matching}
The second lever through which platforms mediate access is matching, which operates as a direct form of access control. In the classical view of markets, participants can access and trade with all relevant counterparties in a decentralized manner. In online platforms, however, such access is mediated due to the sheer number of participants in the market. Rather than interacting with the full set of participants, each participant is matched only to a subset of participants as determined by the platform. This need for mediation gives platforms substantial control over which participants are visible to one another, this control typically effected through search and recommendation systems.

While these recommendation systems are often designed to surface options that best match participant's preferences \citep{aggarwal2016recommender, ricci2021recommender}, they also allow platforms to shape access in ways that serve their own objectives \citep{derakhshan2022product,chu2020position}. Platforms may promote products that generate higher margins than their substitutes, charge for more prominent positions in consumer search, or favor sellers who use the platform's fulfillment services \citep{chen2016empirical, ec2022amazon}. However, for the platform, promoting access for some sellers creates externalities on other sellers: \gmedit{By increasing the visibility of certain sellers, the platform can divert demand way from others, reducing their sales and expected profits. This may induce some sellers to exit the platform altogether, which shrinks the pools of paricipants from whom the platform can extract commission fee from. As a result,}\dcpcomment{\bf xx say more xx, xx As a result of this, and since ... ?xx}\gmdelete{Since all participants pay commission fees,}such changes in access may not generate a net positive effect on platform revenue. Thus, platforms must be careful when designing matching policies. Chapter 4 examines this problem in detail.

\paragraph{Chapter 4: Direct Access Control with Matching} This chapter models a market platform that introduces buyers to sellers in a setting where each seller initially has access only to a subset of buyers. By mediating which buyer-seller pairs are introduced, the platform directly determines the set of feasible interactions and thus the structure of the market. Prices and transactions arise from competitive forces under these access constraints, while the platform earns revenue from the trades it facilitates.

With this model, this chapter analyzes the platform's problem of selecting buyer-seller pairs to maximize revenue. I show that this problem is computationally intractable in general, reflecting the complex externalities introduced by direct access control: promoting access for some participants changes the prices and outcomes for others. I also show that the platform can extract a significant fraction of the additional welfare it creates through the access that it facilitates while still guaranteeing that the overall welfare remains within a logarithmic factor of optimal welfare. These results highlight how direct control over access through matching shapes both platform revenue and market outcomes.

\paragraph{Chapter 5: Matching under Regulatory Intervention during Economic Shocks} 
Direct access control through matching is not only a lever for a platform's own revenue maximization, but can also serve as a tool for maintaining overall economic health during periods of market volatility. Motivated by economic shocks like the COVID-19 lockdown, which led to surges in online demand, this chapter adopts a simulation-based approach to study a market platform in a dynamic, multi-period setting. Buyers and sellers are modeled as economically-motivated agents, choosing whether or not to pay corresponding fees to use a platform, while the platform sets the fees and \gmedit{determines which sellers each buyer can buy from}\gmdelete{matching} \dcpcomment{xx explain matching xx} per period. This chapter examines regulatory interventions aimed at preserving market resilience during such shocks. The results show that while many such interventions are ineffective when used together with a sophisticated platform, one particular kind of regulation--- fixing commission fees to optimal, pre-shock level while allowing the platform to retain flexibility in its approach to matching--- can improve efficiency, preserve seller diversity, and promote the resilience of the overall economic system.

\section{Access Control with Bundling}
While the previous chapters focus on platforms operating as matchmakers, controlling access between multiple sides of the market while not directly participating in transactions, the third part of my thesis considers platforms operating as middlemen \citep{hagiu2006merchant,hagiu2015marketplace,Anderson2024hybrid,abhishek2016agency}. Here, a platform may form contracts with sellers, sourcing products from them, and then reselling these products to buyers. In this setting, access control is not only about which participants can transact with each other, but also about which goods are carried by the platform and, consequently, which goods buyers can purchase through it. This role becomes especially salient in the agentic economy \citep{rothschild2026agentic, lucier2026agentic}, where personal AI assistants reduce search and transaction frictions and allow buyers to discover and contact sellers directly, \gmedit{thereby reducing the platform's role in providing and mediating access between buyers and sellers.}
\gmdelete{making connectivity far less scarce as the source of a platform' advantage.}\dcpcomment{xx `connectivity` the right word? `less scare`? xx} In this new kind of economy, participants facilitated by AI can again find the most suitable trading partners with almost zero search cost, and platforms cannot rely on charging for mediated access alone, and must instead monetize capabilities that individual buyers and sellers with AI agents cannot replicate.

The capability that I focus on is a platform’s power to contract with sellers. By signing exclusivity agreements with selected sellers, the platform can purchase a subset of products and effectively remove them from the open market, so that buyers cannot access these sourced sellers directly even with the help of AI assistants. Platform's access control therefore takes on a new meaning: instead of regulating whom can transact with whom on a platform, it governs which products remain available off-platform and how the platform sells the products that it has sourced. In Chapter 6 we examine this setting in the particular context of a platform that sells its sourced products as a whole bundle to buyers. 

\paragraph{Chapter 6: Access Control with Bundling}
This chapter studies access control when a platform operates as an intermediary rather than a matchmaker. The platform can form exclusive contracts with third-party sellers to acquire their items and then sell the acquired items as a single bundle to buyers with additive valuations, while sellers who are not contracted remain accessible to buyers directly off-platform. Sellers are assumed to be monopolists for their respective items, which differ in quality. The platform's profit equals the bundle's optimal posted-price revenue minus the payments required to obtain the sellers' items.

In this setting, the results show that when qualities are observable, the platform chooses to form contracts with sellers of contiguous qualities, and that the associated bundle may exclude both high-quality and low-quality sellers. When qualities are privately known to sellers, a posted price \gmedit{such that all sellers below some threshold enters the platform} is asymptotically profit-maximizing in large markets. However, to incentivize sellers to truthfully report their qualities, lower-quality sellers are contracted with the platform with higher probabilities. As a result of this incentive constraint, the platform must pay information rents to bring high-quality sellers on-platform. \gmedit{When the platform has in-house production capabilities, the information rents motivate the platform to produce its own high-quality products instead of sourcing from other sellers.} These results characterize the platform’s access control in regard to which sellers remain accessible off-platform versus which are included on-platform and made available only \gmedit{through the bundle that the platform offers.}

\part{Indirect Access Control with Pricing}
\chapter{Indirect Access Control with Commission Fee} \label{chap:platform_eq}
\section{Introduction}
During the COVID-19 crisis, entire populations were put under stay-at-home mandates and many businesses closed due to government restrictions. Nonetheless, some businesses weathered the situation well, even reporting a positive impact~\citep{bloom2021impact}. One of the factors that has been credited with alleviating the issue is internet economy. Specifically, online platforms such as Amazon and DoorDash were able to bridge the gap created by consumers' inability to directly transact with firms~\citep{raj2020covid}. At the same time, platforms used their increased market power to set high fees for merchants, leaving some  with very low or even negative profit margins~\citep{mckinsey2021}. As a response, states like New York and California, imposed caps on commissions, prompting some platforms to shift fees towards customers.\footnote{\url{https://www.protocol.com/delivery-commission-caps-uber-eats-grubhub}} We see that platforms have an increasing role in facilitating trade, but are also seen to act strategically in setting fees in order to maximize their gain. In this paper,  we introduce a new  model that  gives a  clean theoretical framework to  explain the interplay between the two, and provide insights into how platform regulation can play a role.

We consider a market with a set of \emph {unit-demand} buyers and \emph{unit-supply} sellers. Each buyer has a set of sellers they are able to directly transact with off-platform. Each seller can join a trading platform to transact with any buyer. Given the transaction constraints dependent on sellers' decisions, prices and transactions are induced by a competitive (Walrasian) equilibrium. For a seller, joining the platform weakly increases her price as a result of improved access to buyers, but she must also pay a {\em transaction fee} to the platform, which takes the form of a fraction of her revenue. The fraction is between 0 and 1, and chosen by the platform. Given a transaction fee fraction, a set of sellers each buyer can trade with off-platform, and buyers' valuations, a platform game is defined where each seller {simultaneously}
chooses whether or not to join the platform (or the probability with which they join). We name the Nash equilibrium that is formed in this game a {\em Platform Equilibrium}. We define the \emph{social welfare} as the sum of transacting buyers' valuations, and \emph{optimal welfare} as the social welfare when all sellers join the platform. 

We first study the existence of pure Nash equilibria in this game, and how to compute an equilibrium efficiently. We then study ratio of social welfare to optimal welfare, when the platform sets the transaction fee fraction to maximize its own revenue in the induced Platform Equilibrium. We continue to analyze the effect of regulation on social welfare, modelled by a cap on the transaction fee fraction. Finally, we extend our analysis to markets with multiple platforms, and beyond unit-demand buyers, and also consider sellers with production costs.

\subsection{Results and Techniques}
\label{sec:results}

\paragraph{Existence of pure Nash equilibrium and computation.}
In Section~\ref{sec:existence_of_pure}, we explore the existence of pure Platform Equilibria, in the platform game under a fixed transaction fees and market conditions. Throughout the paper, we assume the market clears at the maximum competitive prices. {This reflects a model in which sellers have  market power, so that it is sellers and not buyers that are, in effect, setting prices for trades (subject to the standard requirements of competitive equilibrium, which require balance of supply and demand).} 
As shown in Appendix~\ref{app:prices-min}, some of our results cease to hold with other choices for
competitive equilibrium prices, such as the minimum competitive equilibrium prices. We show when a buyer values all sellers' items equally, also known as the {\em homogeneous-goods} case, pure Platform Equilibrium always exist, and can be found by a simple procedure. The existence of a pure equilibrium is no longer guaranteed in general markets.

\begin{customthm}{1}[Informal Version of Proposition~\ref{prop:no_pure} and Theorems~\ref{thm:PE_pure}]\label{thm:informal_PE_pure}
    In any homogeneous-goods market, a pure Platform Equilibrium exists for any transaction fee fraction and can be found in polynomial time by Algorithm~\ref{alg:PE_pure}. In contrast, there exists general markets where pure Equilibrium does not exist for some values of transaction fee fraction.
\end{customthm}

The proof of this result is involved and draws upon the unique structure for homogeneous-goods market. \citet{kranton2000competition} defines the notion \emph{opportunity path}, which alternates between buyers and sellers to capture competitive prices for sellers' items. We use this combinatorial structure to decompose social welfare into optimal welfare achieved by sellers off-platform, and the remaining on-platform component. This in turn offers a simple expression for sellers' on-platform prices. Finally, we reason combinatorially to show as the transaction fee fraction lowers from 1 to 0, Algorithm~\ref{alg:PE_pure} can find pure equilibrium with any number of on-platform sellers.

The above theorem implies different ways to study social welfare in homogeneous-goods and general markets. In the former, we can use pure equilibria to bound platform revenue and social welfare. In the latter we need to reason about mixed equilibria when studying the effect of a platform on the market. 

\vspace{0.1in}

After studying the existence and computational aspects of Platform Equilibria, in Sections~\ref{sec:poa_rev_max} and \ref{sec:poa-regulated}, we analyze how the platform affects the social welfare in the market. We consider a platform setting the transaction fee fraction strategically to maximize its own revenue, and the sellers respond by playing a Platform Equilibrium. Since there can be multiple Platform Equilibria, we assume the platform can use its market power 
to choose a favorable equilibrium, given its choice of transaction fee fraction. We stress our results for Section~\ref{sec:poa_rev_max} and \ref{sec:poa-regulated} are proved for mixed as well as pure equilibrium.
\vspace{0.1in}

\noindent\textbf{Social welfare with an unregulated platform.}
As a first study, in Section~\ref{sec:poa_rev_max}, we consider the case where the platform is unregulated in setting fees. Let $n$ and $m$ be number of buyers and sellers, respectively. In a homogeneous-goods market, the obtained welfare can be as low as $O(1/\log (\min\{n,m\}))$-fraction of the optimal welfare, and this is tight. We show tightness by showing the platform is able to extract $\Omega(1/\log (\min\{n,m\}))$-fraction of optimal welfare as revenue. This is done by analyzing different possible market structures, and showing how the  platform extracts the desired revenue target in each one. As platform revenue provides a lower bound on the welfare, this yields the desired bound. In the proof, we utilize the pure equilibria finding algorithm to show how to choose a fee and a corresponding equilibrium to extract enough revenue (recall that we assume the platform can select a desirable equilibrium). The platform might choose a different fee and possibly mixed equilibrium to extract even more revenue, which implies the revenue guarantee of the platform, and as a result, the welfare guarantee of the buyers and sellers.

\begin{customthm}{2}[Informal Version of Theorem~\ref{thm:poa_upper_bound_homo} and Theorems~\ref{thm:poa_lower_bound_homo}]\label{thm:informal_poa_upper_bound_homo}
    When the  platform is unregulated, there exists a homogeneous-goods market in which the Platform Equilibrium is $O(1/\log(\min\{n,m\}))$-fraction of the optimal welfare. For every homogeneous-goods market, {the optimal fee for} a revenue-maximizing platform results in a Platform Equilibrium that guarantees at least $\Omega(1/\log(\min\{n,m\}))$-fraction of the optimal welfare.
\end{customthm}

The above result shows even though the platform is maximizing its own revenue, we still get welfare guarantees as a byproduct. However, this guarantee is modest, and only for a very restrictive market setting. For general markets, the prospect is worse: there exists a market where the Platform Equilibrium is only $O(1/\min\{n,m\})$-fraction of the optimal welfare. This motivates us to study the effect of imposing regulations on the fees that a platform can charge, something that is typical in practice \citep{FeeCapNews}. As we show, even light regulation can provide much stronger welfare guarantees, and even for general markets.

\paragraph{Social welfare with fee cap.} In Section~\ref{sec:poa-regulated}, we present the main positive result of the paper. Let $\alpha$ be the transaction fee fraction the platform posts and Price of Anarchy (PoA) of Platform Equilibrium be the ratio of optimal welfare to social welfare across all markets. We study PoA when regulators cap the highest possible $\alpha$ in general unit-demand unit-supply markets. Since social welfare is larger for lower transaction fees, we bound the PoA assuming the platform posts fee $\alpha$ exactly at this cap. If the platform posts a lower fee for higher revenue, this bound still holds. This analysis does not assume the platform uses market power to select a particular Platform Equilibrium, and thus work for all equilibria. Concretely, we show the following.

\begin{customthm}{3}[Informal Version of Theorem~\ref{thm:pure_poa}, \ref{thm:mixed_poa} and \ref{thm:poa_tight}]\label{thm:informal_poa}
    For any fixed fee $\alpha\in [0,1)$, the Price of Anarchy of Platform Equilibrium is at most $\frac{2-\alpha}{1-\alpha}$. Moreover, for every $\alpha$, there is a market for which the ratio of optimal welfare to social welfare is indeed $\frac{2-\alpha}{1-\alpha}$, and sellers use pure strategies.
\end{customthm}

The theorem implies if the platform's fee is capped at $30\%$, a quantity higher than most real world delivery platform fees as shown in Table~\ref{table:platforms and their coommission rate}, the resulting social welfare in every Platform Equilibrium is at least $41\%$ of the optimal welfare. Without the platform, the obtained welfare can be arbitrarily low.

As discussed above, a pure Platform Equilibrium might not exist for general markets. Thus, for this result we also analyze PoA when sellers employ mixed strategies.\footnote{As the strategy space is compact and the game is finite, a mixed Platform Equilibrium is guaranteed to exist.} Our proof takes a mixed Platform Equilibrium $\mathbf{x}=(x_1,\ldots,x_m)$, where $x_i$ is the probability seller $i$ joins the platform, and builds a
corresponding \emph{Bayesian game} for theoretical analysis. In this Bayesian game, each seller can always trade with its off-platform buyers in the original game, but with probability $x_i$ it can trade with all buyers. Each seller needs to decide whether to join the platform, and pay an $\alpha$ fee to the platform, or stay off the platform, \emph{before} it knows whether it can trade with all buyers. We show that if $\mathbf{x}$ is an equilibrium in the original game, then no sellers joining is an equilibrium of the new Bayesian game, and the expected welfare is at least $\frac{1-\alpha}{2-\alpha}$-fraction of the optimal welfare.

\begin{table}[t]
\centering
\begin{tabular}{|c|c c c c|} 
 \hline
 Platforms & Amazon & UberEats & DoorDash & Grubhub\\ [0.5ex] 
 \hline
 Commission Rate & 8\%-20\% & 15\%-30\% & 15\%-30\% & 15\%-25\%  \\  
 \hline
\end{tabular}
\caption[Platforms and their commission rate in the US from 2021-2022.]{Platforms and their commission rate in the US from 2021-2022.\protect\footnotemark}
\label{table:platforms and their coommission rate}
\end{table}
\footnotetext{This data is gathered from platforms' US websites in 2023. Amazon referral fees are charged according to item categories. 8\%-20\% for all categories except `Amazon device accessories'.
\url{https://sell.amazon.com/pricing}. For UberEats, DoorDash and Grubhub see \url{https://merchants.ubereats.com/us/en/pricing/},\url{https://get.doordash.com/en-us} 
and \url{https://get.grubhub.com/products/marketplace/}}

\paragraph{Generalizations.} Finally, we extend our model and results beyond the single platform unit-demand and zero cost setting. Section~\ref{sec:multi_platform} proves the Price of Anarchy results under regulation still hold in a market where multiple platforms compete for buyers and sellers. In Section~\ref{sec:seller_with_cost}, we show when sellers have production costs, the Price of Anarchy results extend naturally, with the results factoring out the cost. In Section~\ref{sec:beyond-ud}, we extend our results beyond unit-demand and consider additive-over-partition matroids valuation. That is, for each buyer, sellers are partitioned into a set of categories, and the buyer values at most a {\em capacity} number of sellers from each category. This generalizes both unit-demand and additive valuations, as well as $k$-demand valuations where a buyer is additive over their top $k$ items (e.g., in~\cite{BergerEF20,ZhangC20}).

\subsection{Related Work}
There are three works that use models similar to ours to examine how an online platform facilitates trade between buyers and sellers for revenue. \citet{birge2021optimal} study the optimal commission and subscription fee structure in a bipartite market. Also in a bipartite market, \citet{banerjee2017segmenting} consider a platform matching buyers to sellers for revenue. We share with these two works the idea that prices and trades are shaped by competitive forces endogeneous to the market, rather than being directly set by the platform. The main difference is that in these two works platform is the only venue where buyers and sellers interact, while we emphasize on the off-platform trading opportunities. This distinction allows us to model sellers' decision processes to join the platform. In fact, if off-platform options are not present in our model, sellers will always join the platform, resulting in arbitrarily high platform fees, while social welfare remains equal to optimal welfare. These two works are more general in other ways; for instance each node on the bipartite graph represents a continuum of buyers or sellers with varying valuations or costs. The third related work is \citep{platformDisruption}, which models off-platform options. However, their focus is on platform matching buyers to sellers for revenue, whereas ours centers on platform setting fees.

Besides the three closest papers, our work build upon previous results on how network structure influence competitive equilibrium prices. \citet{kranton2000competition} study a homogeneous-goods buyer-seller market and relate the highest and lowest competitive prices to \emph{ opportunity paths} of trading agents. We make use of this structural result in the analysis in the present paper. \citet{elliott2015inefficiencies} generalizes the opportunity path argument to markets with general unit-demand valuations. \citet{kakade2004economic} further study the way in which equilibrium prices depend on the statistical structure of the buyer-seller bipartite graph. These works focus on price and network structure, and do not consider an intermediary as we do.   

Another strand of work that is related to ours is the study on network formation games with competitive equilibrium models of trade, subject to edge-constrained trading relationships. \citet{kranton2001theory} model buyers purchasing links to sellers in a bipartite economy before their valuations are realized, and show the Nash equilibrium of the network-formation game is fully efficient. Under a similar setting, \citet{even2007network} characterize all possible, equilibrium  bipartite graph topologies, based on the edge cost and the number of sellers and buyers, but do not discuss welfare. Similar to the two works, we also build on top of foundational studies on competitive equilibria in bipartite economies \citep{kelso1982job, gul1999walrasian}, and analyze the effect of market participants' incentives on network formation. But the platform-based emphasis leads our model to depart from theirs in various ways:
1) in our work, there is a revenue-optimizing platform that mediates trade and chooses transaction fees. In their work, sellers and buyers interact without a selfish mediator;  
2) as discussed in the previous paragraph, buyers and sellers in our work start with existing trading opportunities;
3) in our work, sellers acquire edges to all buyers simultaneously by joining the platform, and pay a fraction of the final competitive price for joining platform. In theirs, each edge has to be purchased individually with a fixed price per edge. 

Our work also relates to the broader literature that studies platform economics, which tends to ask questions about how to build network effects through subsidy under various forms of platform competition~\citep{caillaud2003chicken,Armstrong2006,rochet2003platform}. Although our methods differ, we draw inspiration from this realm that a platform may subsidize one side of the market, while charging a fee to the other side (in our case, subsidizing buyers and charging sellers). For example, \citet{caillaud2003chicken}
study the  role  of platforms in matchmaking, with a platform that charges registration fees for entry and commission fees per transaction. Similarly, \citet{platformSim}, with registration and transaction fees, use simulation and reinforcement learning to investigate different regulatory approaches for platform prices to foster a more resilient economic system.

\section{Preliminaries}
\label{sec:prelims}

We adopt the buyer-seller network model of \citet{kranton2000competition}. There is a set of $n$ buyers,
 $B = \{1,\ldots, n\}$, and $m$ differentiated 
sellers, $S=\{1,\ldots, m\}$. Each seller has a single product to sell, and 
each buyer $i$ has a {\em unit-demand valuation}, $v_i$. Buyer $i$'s value for seller $j$'s product is $v_{ij}\geq 0$. A bipartite graph $G = \{g_{ij}\}_{\substack{i\in B\\ j\in S}}$ models which buyers and sellers can directly transact, without the use of the platform, where  
\begin{eqnarray}
    g_{ij} = \begin{cases} 1 \quad &\mbox{Buyer } i \mbox{ can directly transact with seller } j\\
                           0 \quad &\mbox{Otherwise}
            \end{cases}.
\end{eqnarray}

For buyer $i$,  $N(i) = \{j \ : \ g_{ij}=1\}$ denotes the list of sellers linked to the buyer in $G$. We define a competitive (Walrasian) equilibrium in the buyer-seller network model.
\begin{definition}[Competitive Equilibrium]
    A {\em competitive equilibrium} for the buyer-seller network $G$ is a tuple $(\p,\alloc)$. $\p=(p_1,\ldots, p_m)$ are none-negative item prices, $\alloc = \{a_{ij}\}_{\substack{i\in B\\ j\in S}}\in \{0,1\}^{n\times m}$ is an allocation of the goods to the buyers, and:
    \begin{itemize}
    \item Transactions must respect links: $a_{ij}\leq g_{ij}\quad \forall i\in B\ j\in S$.
    \item Buyers are allocated at most one good: $\sum_{j} a_{ij} \leq 1 \quad \forall i\in B$.
    \item Goods are sold at most once: $\sum_{i} a_{ij} \leq 1\quad \forall j\in S$.
    \item Buyers gets their most preferred outcome:  $u_i(\p,\alloc) \geq v_{ij} - p_j \quad \forall i\in B\ j\in S$. where 
    \begin{eqnarray}
    u_i(\p,\alloc) = \sum_{j}a_{ij}(v_{ij}-p_j),    \label{eq:buyer_util}
    \end{eqnarray}
    is buyer $i$'s utility for the allocation.
    \item Buyers have non-negative utility: $u_i(\p,\alloc)\geq 0\quad \forall i\in B$.
    \item Unassigned goods have price $0$.
    \end{itemize} 
\end{definition} \label{def:comp_eq}

A competitive equilibrium is a canonical model of the steady state in a market, capturing 
the notion of prices that are set such that supply meets demand. It follows from standard existence results~\citep{kelso1982job} that a competitive equilibrium always exists in a unit-demand buyer-seller network (a missing edge can be represented as $v_{ij}=0$). Moreover, competitive equilibria have a number of desirable properties. 
\begin{theorem}[First Welfare Theorem]
    In a competitive equilibrium, the social welfare is maximized with respect to 
the set of allocations that respect 
the transaction constraints posed by $G$.\label{thm:first_welfare} 
\end{theorem} 

\begin{theorem}[Second Welfare Theorem \citep{gul1999walrasian}]
    Let $(\p, \alloc)$ and $(\p',\alloc')$ be two competitive equilibria of a buyer-seller network $G$, then $(\p,\alloc')$ is also a competitive equilibrium (and so is $(\p',\alloc)$).
\end{theorem}

The Second Welfare Theorem implies that prices have the property of either forming a competitive equilibrium with any social-welfare optimal allocation, or forming a competitive equilibrium with none of them. We refer to prices $\p$ that are part of a competitive equilibrium as \emph{competitive prices}. It is also well known that 
 competitive prices   have a lattice structure.
\begin{theorem}[Lattice structure for competitive prices \citep{gul1999walrasian}]
    Let $\p_1$ and $\p_2$ be competitive prices, then $\p_1 \vee \p_2$ and $\p_1 \wedge \p_2$ are also  competitive prices, where $\vee$ is the coordinate-wise maximum and $\wedge$ is the coordinate-wise minimum.\label{thm: price_lattice}
\end{theorem}

As a result, there are {\em minimum} and {\em maximum} competitive prices, denoted $\underline{\p}$ and $\overline{\p}$ respectively,
with item-wise minimum and item-wide maximum prices
 denoted $\underline{p}_j$ and $\overline{p}_j$. 
For  $S$ sellers, $B$  buyers, network $G$,
 and buyer values $\mathbf{v}$, we use $W(S,B,\mathbf{v},G)$ to denote the 
{\em optimal welfare} from all feasible transactions between  sellers $S$ and  buyers $B$. When clear from context, we omit $\mathbf{v}$ from the notation. We will make use 
of the following characterization result.
\begin{theorem}[Characterization of competitive prices~\cite{gul1999walrasian}]
    The minimum and maximum competitive prices for an item $j$ have the following form:
\begin{align}
    \underline{p}_j &= W(S \cup \{j\}, B, G)-W(S, B, G), \label{eq:max_price}\\
    \overline{p}_j &= W(S, B, G) - W(S\setminus \{j\},B, G). \label{eq:min_price}
\end{align}

Here, $S \cup \{j\}$ denotes adding another copy of seller $j$ with all its edges to the market, and $S\setminus \{j\}$ is removing seller $j$ and all its connected edges from the market. The resulting graph changes correspondingly when adding or removing $j$ with all its edges, but for notation convenience we still use $G$ to denote the graph.
\end{theorem}

We next present our framework for analyzing the effect of the presence of a 
platform on a buyer-seller network.

\subsection{Platform Equilibrium} 
\label{sec:platform_eq}

Consider a buyer-seller network defined on graph $G$. Without a platform, a competitive equilibrium $(\p, \alloc)$ is formed subject to $G$ (see Definition~\ref{def:comp_eq}), where $\p$ are  competitive prices that support a welfare-optimal allocation, $\alloc$, subject to $G$. We do not assume anything regarding $\p$ except that it is competitive price vector. 

The utility of a buyer is given by Eq.~\eqref{eq:buyer_util}, while the utility of a seller $j$ is the revenue, $p_j$ (which is necessarily zero if the seller does not trade). By the First Welfare Theorem (Theorem~\ref{thm:first_welfare}), the obtained welfare in the market that forms in the absence of a platform is $W(S,B,G)$.  

Consider now a platform that declares a {\em transaction fee fraction}, $\alpha \in [0,1]$. Given that a set of sellers $P\subseteq S$ joins the platform, a new network $\hat{G}=G(P)= \{\hat{g}_{ij}\}_{\substack{i\in B\\ j\in S}}$ is formed with the following connections,
\begin{eqnarray}
    \hat{g}_{ij} = \begin{cases} 1 \quad & \mbox{Seller } j\in P\\
                           g_{ij} \quad & \mbox{Otherwise}
            \end{cases}.
\end{eqnarray}

That is, the constraints faced by a seller that does not join the platform are unchanged, 
whereas a seller that joins the platform can transact with any buyer.

Given the modified constraints on transactions, $\hat{G}$, a 
modified competitive equilibrium $(\hat{\p},\hat{\alloc})$ is formed, where $\hat{\alloc}$ is the optimal allocation constrained to $\hat{G}$, and $\hat{\p}$ are the supporting competitive prices. The utility of buyer $i$ in this competitive equilibrium 
 is $u_i(\hat{\p},\hat{\alloc})\geq 0$. The utility of a seller $j$ who does not join the platform is $\hat{p}_j$, while the utility of a seller $j$ who joins the platform is $(1-\alpha)\hat{p}_j$.\footnote{In reality, a seller joining the a  platform such as a food-delivery platform can still transact with a buyer off platform. 
But in the unit-supply unit-demand market, a seller has weakly no incentive to join the platform if it transacts directly with the buyer,  through an existing link. Therefore, we assume all sellers on platform pay transaction fee.} The platform's utility is $\alpha\cdot \sum_{j\in P}\hat{p}_j$. We assume the market clears with the maximum competitive prices, as it reflects a model where sellers have market power and in effect set prices for trades. That is, $\hat{p}_j=\overline{p}_j$. As we show in appendix~\ref{app:prices-min}, max price is meaningful to discuss welfare loss due to a selfish platform. 

For a seller $j$, and holding other sellers' strategy fixed, let $\p^{\on}$ and $\p^{\off}$ be the competitive prices when seller $j$ joins and not joins the platform, 
respectively. $j$ joins the platform if the utility from joining is more than that from not joining. That is, if $(1-\alpha)p^{\on}_j > p^{\off}_j$. If there is a tie in utility, the seller  may join or not. If the utility for joining is strictly smaller than not joining, the seller does not join. We can now define a Platform Equilibrium for
 pure strategies.
\begin{definition}[Pure Platform Equilibrium]
 Let $\p(P)$ denote the competitive prices formed when a set of sellers, $P$, join the platform, where for sellers who join, their prices are the maximum competitive price for the obtained network structure. The  set $P$ corresponds to  a \textit{Platform Equilibrium} if and only if,
    \begin{itemize}
        \item For each seller $j\in P$,  then $(1-\alpha)p^{\on}_j \geq p^\off_j$, for $\p^{\on} = \p(P)$ and $\p^{\off} = \p(P\setminus\{j\})$, and
        \item For each seller $j\notin P$, then $p^\off_j \geq (1-\alpha)p^\on_j$, for $\p^\off = \p(P)$ and $\p^\on = \p(P \cup \{j\})$.
    \end{itemize}
\end{definition}

In this way, a pure Platform Equilibrium is a Pure nash equilibrium of the game where the sellers can choose whether or not to join the platform. Similarly, we define the larger set of mixed Platform Equilibria.
\begin{definition}[Mixed Platform Equilibrium]
   Let $x_j\in [0,1]$ denote the probability
with    which seller $j$ joins the platform,
and assume that for sellers who join, their prices are the maximum 
competitive prices for the obtained network structure. The vector $\x=(x_1,\ldots, x_m)$ is a {\em Mixed Platform Equilibrium} if and only if, each seller $j$ maximizes their utility by choosing $x_j$ given $\x_{-j}$,
\begin{eqnarray}
    \E_{P\sim \x} \left[u_j(P,\alpha)\right] \geq \max\left(\E_{P\sim \x_{-j}} \left[u_j(P,\alpha)\right], \E_{P\sim \x_{-j}} \left[u_j(P\cup\{j\},\alpha)\right]\right),
\end{eqnarray}
where $u_j(P,\alpha)$ is the utility of seller $j$ when a set of sellers $P$ joins the platform, the platform charges transaction-fee $\alpha$, and the competitive prices are $\p(P)$.
\end{definition}

\subsection{Modeling the Effect of a Selfish Platform on social welfare} \label{sec:prelims-platform-effect}

We now present our model to analyze the effect of a selfish platform on social welfare. A Selfish platform sets a transaction fee $\alpha$ to maximize its revenue, assuming as before that the market clears with the maximum competitive prices. As there can be a plethora of Nash equilibria, let $\eq(S,B,\mathbf{v},G,\alpha)$ denote the set of equilibria defined by parameters $S,B,\mathbf{v}, G$, at transaction-fee $\alpha$. Further we assume the platform can use its market power to induce the equilibrium with highest revenue in $\eq(S,B,\mathbf{v},G,\alpha)$. The platform chooses $\alpha^{\star}$ such that
\begin{eqnarray}
    \alpha^{\star} \in \argmax_{\alpha}\max_{\x\, \in\, \eq(S,B,\mathbf{v},G,\alpha)}\E_{P\sim \x}\left[\alpha\cdot\sum_{j\in P} \p_j(P)\right]. \label{eq:platform_opt}
\end{eqnarray}

This is the transaction fee that maximizes the platform's revenue for the best possible Nash equilibrium in terms of revenue to the platform. If the platform is regulated in setting fees, and cannot set a fee higher than $\overline{\alpha}$, we simply add the constraint of $\alpha\le \overline{\alpha}$ to the $\argmax$ expression in the platform's optimization problem in Equation~\eqref{eq:platform_opt}.
In this way, the platform acts as a leader that optimizes for an optimistic Stackelberg Equilibrium in a Stackelberg game where both the platform and the sellers act strategically. We denote the set of Platform Equilibria that maximizes the platform's revenue as
\begin{eqnarray}
    \mathbf{X}^{\star} = \argmax_{\x\,\in\, \eq(S,B,\mathbf{v},G,\alpha^{\star})} \E_{P\sim\x}\left[\alpha^{\star}\cdot \sum_{j\in P}\p_j(P)\right].
\end{eqnarray}

We aim to quantify the efficiency of the resulting Platform Equilibrium. Let $k_{B,S}$ be the 
{\em complete bipartite graph between buyers and sellers},
 and let $W^{\star}(S,B,\mathbf{v})=W(S,B,\mathbf{v}, k_{B,S})$ be the \textit{optimal welfare}; i.e., the optimal social welfare in the case that there are no logistical constraints
and each buyer can  transact with any  seller. When clear from context, we denote the optimal welfare $W^{\star}(S,B,\mathbf{v})$ by $W^\star$. 
We use the  {\em price of anarchy}  to quantify the worst-case social welfare in a Platform Equilibrium. 
\begin{definition}[Price of Anarchy]\label{def:poa}
    The {\em price of anarchy} (PoA) of a Platform Equilibrium
is the worst case ratio between the optimal welfare to the lowest-welfare platform equilibrium that maximizes the platform's revenue for any market. Here the worst case is over all possible buyer-seller networks, and buyers' valuations.
\begin{eqnarray}
    PoA  = \max_{S,B,G,\mathbf{v}, \x\in \mathbf{X}^{\star}}\frac{W^\star(S,B,\mathbf{v},G)}{\E_{P\sim \x}\left[W(S,B,\mathbf{v}, G(P)\right]}.
\end{eqnarray}
\end{definition}
Though PoA is defined for all equilibria, in later sections we will slightly abuse notation and use "PoA of pure (mixed) Platform Equilibrium" to denote the the worst case ration between the ideal welfare to the lowest-welfare \textit{pure} (or \textit{mixed}) Platform Equilibrium respectively.

\section{Existence of a Pure Platform Equilibrium}\label{sec:existence_of_pure}
In this section, we show first show that a pure Platform Equilibrium does not always exist in general buyer-seller markets, and then give a polynomial-time algorithm that find a pure Platform Equilibrium in homogeneous-goods market regardless of transaction fee $\alpha$. Most proofs are deferred to Appendix~\ref{app:existence_of_pure}.

\subsection{Non-Existence of Pure Equilibria}

We first show that for some markets, pure Platform equilibria need not exist. 

\begin{restatable}{proposition}{propNoPure}
\label{prop:no_pure}
    There exists a general buyer-seller network and a transaction fee, $\alpha\in [0,1]$, for which there is no pure Platform Equilibrium. 
\end{restatable}

The proof follows the example in Figure~\ref{fig:no_pure_equilibrium}, and 
is given in Appendix~\ref{app:app_no_pure_eq}.
At a high level, a seller's competitive price is related to the externality they impose on the whole market, and is influenced by other sellers' connections. In the proof, we show that sellers' best-response dynamics can exhibit a cyclic behavior: some sellers joining the platform can cause other sellers already on platform to leave, or cause other sellers to join the platform. 

\begin{figure} 
    \centering
    \begin{tikzpicture}[scale=1.5]
        \foreach \i/\label in {1/A, 3/B, 5/C, 7/D}
            \node[draw, shape=rectangle, minimum size=0.6cm] (\label) at (\i, 2) {\label};
        
        \foreach \i/\label in {1/a, 3/b, 5/c}
            \node[draw, shape=circle, minimum size=0.6cm] (\label) at (\i, 0) {\label};
        
        \foreach \x/\y/\w in {A/a/1, B/b/1, C/c/1}
            \draw[line width=2pt] (\x) -- node[midway, right, font=\footnotesize] {\w} (\y);
        
        \foreach \x/\y/\w/\pos/\loc in {B/a/3.05/midway/below right, B/c/1.15/at start/right, C/b/1.1/near start/below right, D/c/0.05/midway/below right}
            \draw[line width=1.5pt, blue, dashed] (\x) -- node[\pos, \loc, font=\footnotesize] {\w} (\y);

        \node[left] at (0.5, 2)  {Buyers};

        \node[left] at (0.5, 0)  {Sellers};
    \end{tikzpicture}
    \caption{An example with no Platform Equilibrium in pure strategies at $\alpha=\frac{1}{2}$. Black solid lines are direct links, as captured by $N(i)$ for buyer $i$, and blue dotted lines indicate missing links. Buyer values are  annotated adjacent to each edge, and all any value that is omitted is zero.} \label{fig:no_pure_equilibrium}
\end{figure}
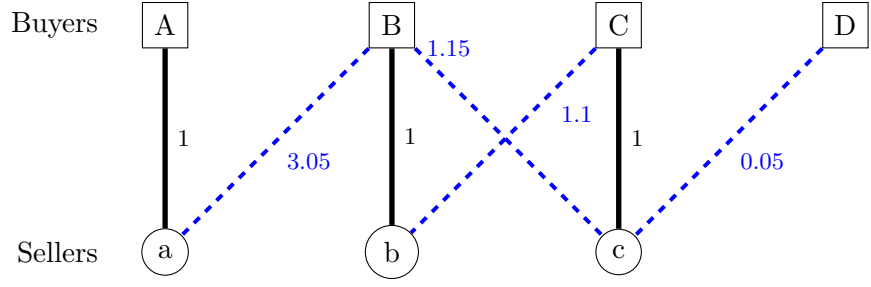 

\subsection{Computing a Pure Nash Equilibrium in a Homogeneous-Good Market}

We now study homogeneous-good markets, where sellers have identical items, and buyer valuation is captured by a single scalar $v_i\in \mathbb{R}^+$. For this case, a pure equilibrium can always be found in polynomial time. To show this, in Lemma~\ref{lem:optimal_welfare} we first decompose social welfare into optimal welfare achieved by sellers off-platform, and the remaining on platform component. Then in Lemma~\ref{lem:same_on_price} we show all on platform sellers have the same price, and is related to the number of sellers on platform. Finally, we give the algorithm that finds the pure equilibrium. As we show in Theorems~\ref{thm:PE_pure} and~\ref{thm:equilibrium_set_enlarges_1}, in homogeneous goods markets, there is an order of adding sellers to the platform with the following desirable property: when adding a new seller to the platform, (i) 
no seller that is previously added would want to leave, and (ii) a seller that had negative gain from joining before, would rather still not join. All lemmas and theorems in this subsection hold \textit{only} for homogeneous-goods markets. Missing proofs are deferred to Appendix~\ref{app:existence}.

\citet{kranton2000competition} proves when a new seller is added to a market, at most one new buyer transacts (Lemma~\ref{lem:add_one_link}). Building on this we state two useful lemmas on how a set of sellers joining the platform affects the welfare and what is their price. 

\begin{restatable}{lemma}{OptimalWelfare}
\label{lem:optimal_welfare}
    Consider a buyer-seller network $(S,B,G)$. Let $P\subseteq S$ be the sellers joining the platform. Let $B^G$  be buyers transacting in $W(S\setminus P, B,G)$ and $\bar{B}^{G}=B\setminus B^{G}$. Let $m'=|P|$ and let $\bar{v}(m')$ be the sum of the $m'$ largest buyers in $\bar{B}^G$, or the sum of all buyers in $\bar{B}^G$ if there are less then $m'$ buyers in $\bar{B}^G$. Then 
   \begin{eqnarray}\label{eq:optimal_welfare}
        W(S,B,G(P)) = W(S\setminus P, B^G, G) + \bar{v}(m').
    \end{eqnarray}
\end{restatable}

\begin{restatable}{lemma}{SameOnPrice}
\label{lem:same_on_price}
    All sellers joining the platform have the same price of $\bar{v}(m')-\bar{v}(m'-1).$
\end{restatable}

We now present the main result of this section. For any transaction fee $\alpha$ in a homogeneous-goods market, Algorithm~\ref{alg:PE_pure} finds a pure Platform Equilibrium in polynomial time. The algorithm adds sellers to $P$ iteratively, which eventually contains all sellers joining the platform in the pure Platform Equilibirum computed. At every round, the algorithm computes for all sellers who haven't joined $P$ yet, what is their utility gain in joining the platform, given the current $P$ (captured by $\phi_j^P$). The set of sellers with the largest utility gain is labeled as $\Phi_{\max}^P$ in step~\ref{alg:step}. If $|\Phi_{\max}^P|=1$, the algorithm adds the seller with the largest gain to $P$. If $|\Phi_{\max}^P|>1$, it chooses a seller with the lowest off-platform price among all in $\Phi_{\max}^P$. The following lemma shows the seller added to $P$ each round has the lowest off platform price out of all agents in $S\setminus P$.

\begin{algorithm}[h!]
    \begin{enumerate}
        \item Initialize $P\leftarrow \emptyset$.
        \item Let $G(P)$ denote the graph $G$ after connecting all sellers in $P$ to all buyers.  
        \item For each $j\in S$, let $p_j^\on(P)$ be $j$'s price on platform, $p_j^\off(P)$ be $j$'s price off platform given $P\setminus\{j\}$ is on platform.
        \item For each $j\in S$, let $\phi_j^P = (1-\alpha) p_j^\on(P) - p_j^\off(P)$. 
        \item Let $\Phi_{\max}^P=\argmax_{j\in S\setminus P}\{\phi_j^P\}$, 
        $\hat{j}\in\argmin_{j\in \Phi_{\max}^P}\{p_{j}^{\off}(P)\}$.  \label{alg:step}
        \item While $(P\neq S) \ \wedge \ (\phi_{\hat{j}}^P \geq 0)$:
         \begin{itemize}
            \item $P\gets P\cup \{\hat{j}\}$.
            \item Update $G(P)$.
            \item Recompute seller $\hat{j}$ according to the new prices $p_j^\on(P)$ and $p_j^\off(P)$ and the new $\phi_j^P$.
        \end{itemize}
        \item output $P$
    \end{enumerate}
\caption{Finding a pure Platform Equilibrium at fixed $\alpha$.}
\label{alg:PE_pure}
\end{algorithm}

\begin{restatable}{lemma}{AlgoSelect}
\label{lem:algo_select}
    At an iteration, if Algorithm~\ref{alg:PE_pure} adds seller $\hat j$ to platform, then for every seller $j \notin P$, $p_{\hat{j}}^{\off}(P)\leq p_{{j}}^{\off}(P)$. 
\end{restatable}

Using the above lemmas, we now provide a sketch of the proof of the main theorem of the section, in order to demonstrate the main ideas used to proof it. The full proof is quite technical and appears in Appendix~\ref{app:existence}. 

\begin{restatable}{theorem}{PEpure}
\label{thm:PE_pure}
    For any $\alpha\in [0,1]$, Algorithm~\ref{alg:PE_pure} outputs a pure strategy Platform Equilibrium.
\end{restatable}

\begin{proof}[Proof Sketch.]
    The gist of the proof is showing whenever a new seller $\hat{j}$ is added to $P$, the sellers $j$ which are already in  $P$ do not want to leave $P$ (that is, their utility is still higher on-platform). In particular, we want to prove $\forall j\in P$, their benefit of staying on platform is larger than $\hat{j}$'s benefit of joining. That is, given $P$ and $\hat{j}\in \argmax_{j\in S\setminus P}\{\phi_j^P\}$,
    \begin{eqnarray}
        \forall j\in P\quad \phi_j^{P\cup\{\hat{j}\}}\geq \phi_{\hat{j}}^{P}\geq 0.\label{eq:pure_eq_invariant_main_section}
    \end{eqnarray}
    Since $\hat{j}$ is the next seller joining, $\phi_{\hat{j}}^{P}\geq 0$. Expanding Eq.~\eqref{eq:pure_eq_invariant_main_section}, we need to prove the following,
    \begin{eqnarray}
        \phi_j^{P\cup\{\hat{j}\}} - \phi_{\hat{j}}^{P} &= (1-\alpha)[p_j^{\on}(P\cup\{\hat{j}\})-p_{\hat{j}}^{\on}(P)]+[p_{\hat{j}}^{\off}(P)-p_{j}^{\off}(P\cup\{\hat{j}\})]\ge 0.\label{eq:pure_eq_invariant_2_main_section}
    \end{eqnarray}
    Since $j\in P$, both $p_j^{\on}(P\cup\{\hat{j}\})$ and $p_{\hat{j}}^{\on}(P)$ consider the price of $j$ and $\hat{j}$ when the set of sellers on platform is $P\cup \{\hat{j}\}$. Therefore, according to Lemma~\ref{lem:same_on_price}, their price is the same, and the first term in Lemma~\ref{lem:same_on_price} is equal to 0.
    Thus, what's left to prove is 
    \begin{eqnarray}
        p_{\hat{j}}^{\off}(P) \geq p_{j}^{\off}(P\cup\{\hat{j}\}).\label{eq:second-term}    
    \end{eqnarray}
     
    The intuition for this is the following. By Lemma~\ref{lem:algo_select} $j$ has a lower off-platform price than $\hat{j}$ when it is selected by the algorithm. If, as opposed to Eq.~\eqref{eq:second-term}, we have  $p_{j}^{\off}(P\cup\{\hat{j}\}) >p_{\hat{j}}^{\off}(P)$, we show there must be a seller $j'$ joining after $j$, such that its joining the platform frees up a high value buyer for $j$. This implies $j'$ should have high off platform price when it joins, and should be added after $\hat{j}$, forming a contradiction. The full proof also uses a submodularity condition on max weight matching, and is deferred to the Appendix~\ref{app:existence}.
    
\end{proof}

\subsubsection{Useful Properties of Algorithm~\ref{alg:PE_pure}}

In this section, we state other useful properties of pure Equlibria in homogeneous goods markets that arise from Algorithm~\ref{alg:PE_pure}. In Section~\ref{sec:poa_homo_n_ub}, these properties will be useful in lower bounding revenue, and in turn welfare for markets with selfish platforms. 

Algorithm~\ref{alg:PE_pure} can be used to find a pure equilibrium at any $\alpha$. The following two lemmas allows one to continuously decrease transaction fee from $\alpha=1$ with algorithm~\ref{alg:PE_pure}, and gradually enlarge the set of on-platform sellers in equilibrium.

\begin{restatable}{lemma}{PEpureAlpha}\label{lem:pe_pure_alpha1}
    In a homogeneous-goods market $(S,B,G)$, let $S^G$ be the set of transacting sellers, $\bar{S}^G$ be the set of none-transacting sellers. Then any $P\subseteq \bar{S}^G$ joining is a pure equilibrium at $\alpha=1$.
\end{restatable}

\begin{restatable}{lemma}{PEpureContinuous}\label{lem:pe_pure_continuous}
    Let $P_1$ be the set of sellers that forms an equilibrium by Algorithm~\ref{alg:PE_pure} for $\alpha_1$. Then for any $P_2\supsetneq P_1$ there exists an $\alpha_2<\alpha_1$ for which Algorithm~\ref{alg:PE_pure} outputs $P_2$.    
\end{restatable}

Algorithm~\ref{alg:PE_pure}, Lemmata~\ref{lem:pe_pure_alpha1} and \ref{lem:pe_pure_continuous} offer a way to find sets of pure equilibrium at different transaction fees without cold starting the algorithm at $P=\emptyset$ for each $\alpha$. We can continuously decrease the transaction fee from $\alpha=1$ to $0$, adding a seller $j$ to platform when the on-platform gain $\phi_j(P)$ reaches zero. The following theorem further says as $\alpha$ decreases, the set of on-platform sellers in pure equilibrium enlarges by one seller each time. With this procedure, we can check and compare revenue each time a new seller joins. 

\begin{restatable}{theorem}{EnlargeEq}
\label{thm:equilibrium_set_enlarges_1}
    For any number $m_p= 1,...,m$, there exists a pure Platform Equilibrium $P$ such that $|P|=m_p$. Furthermore, all such equilibria can be found by lowering $\alpha$ from $1$ to $0$ with Algorithm~\ref{alg:PE_pure}.
\end{restatable}
\begin{proof}[Proof Sketch.]
    As $\alpha$ decreases continuously from $1$ to $0$, $\phi_j^P$ continuously increases for all $j$ when the equilibrium set $P$ stays the same. When some sellers $\Phi_{\max}^P$ reaches zero on-platform gain, Algorithm~\ref{alg:PE_pure} adds a seller $j\in \Phi_{\max}^P$ to $P$. It suffices to show after $j$ is added to $P$, $\forall \hat{j}\notin P$, either
    \begin{eqnarray*}
    \phi_{\hat{j}}(P\cup \{j\}) & \leq & \phi_{\hat{j}}(P) \text{\; or \;} \phi_{\hat{j}}(P\cup\{j\})<0.
    \end{eqnarray*}
    Therefore, the on platform gain is non-positive for sellers not on platform. Thus, $P\cup \{j\}$ is indeed a pure equilibrium, larger than $P$ by 1.
\end{proof}

Recall that in the example in Proposition~\ref{prop:no_pure}, one seller joining the platform can incentize off-platform sellers to join the platform, or on-platform sellers to leave the platform. This is not true for homogeneous-goods markets. Algorithm~\ref{alg:PE_pure} provides a way to build the set of pure equilibrium. Theorem~\ref{thm:PE_pure} shows that a seller joining would not incentivize on-platform sellers to leave, while Theorem~\ref{thm:equilibrium_set_enlarges_1} proves a seller joining would not incentivize off-platform sellers to join. In Section~\ref{sec:poa_homo_n_ub}, we will use these properties to prove the first results on the positive effect of a selfish platform on the social welfare of the market.

\section{Market Efficiency with an Unregulated Platform}\label{sec:poa_rev_max}

In this section, we study the social welfare with an unregulated platform. Recall our definition of the platform's optimization problem and the definition of the price of anarchy in measuring market efficiency, as in Section~\ref{sec:prelims-platform-effect}. In Section~\ref{sec:poa_homo_n_ub}, we present the first result in the positive effect of a selfish platform on market efficiency. We show that in any homogeneous goods market, the obtained Platform Equilibrium, be it pure or mixed, guarantees at least $1/\log(\min\{m,n\})$-fraction of the ideal social welfare. In Section~\ref{sec:poa_homo_n_lb}, we show that this is tight, up to constant factors. In Appendix~\ref{app:poa_mn_general} we show that in markets with general unit-demand valuations, the state of affairs is worse--the price of anarchy can be as bad as $\min\{n,m\}$. This motivates us in showing that even light regulation of the platform can lead to substantial welfare gain, as presented in Section~\ref{sec:poa-regulated}.

\subsection{$O(\log(\min(n,m)))$ PoA for Homogeneous-Goods Markets}\label{sec:poa_homo_n_ub}

We now show the upper bound of the price of anarchy. 
\begin{theorem}\label{thm:poa_upper_bound_homo}
    The PoA is $O(\log (\min\{n,m\}))$.
\end{theorem}
The proof follows immediately from Lemma~\ref{lem:sppoa_n_leq_m} and~\ref{lem:sspoa_n_geq_m}, which prove this claim for the range $n\leq m$ and for the range $n \geq m$. Interestingly, the proofs give a lower bound on the welfare by giving a lower bound on the platform's optimal revenue, which we denote by $Rev^\star$ in terms of the optimal welfare. As the platform's revenue is not more than the sellers' competitive prices which are no more than buyers values, this immediately gives a lower bound on the social welfare. As we lower bound welfare through revenue, the established bound works for mixed equilibrium.

For the proof of the two lemmas, we use the following notation. Let $B^G$ and $S^G$ be the set of buyers and sellers that transact when there's no platform $(S,B,G)$. Let $\bar{B}^G=B\setminus B^G$ and $\bar{S}^G = S\setminus S^G$. We normalize the ideal welfare to be $W^\star=1$. Denote the $k$-th harmonic number by $H_k=1+\frac{1}{2}+\frac{1}{3}+\ldots+\frac{1}{k}$.

\begin{lemma}
    When there are more sellers than buyers $n\leq m$, the PoA is $O(\log n)$. \label{lem:sppoa_n_leq_m}
\end{lemma}
\begin{proof}
    If $W(S^G,B^G,G) \ge 1/\log n$ we are done. Therefore, assume $W(S^G,B^G,G) < 1/\log n\leq 1/2$, thus $W^\star(S,\bar{B}^G,G) > 1/2$. Let $m^G=|S^G|=|B^G|$, $\bar{m}^G=|\bar{S}^G|$ and $\bar{n}^G= |\bar{B}^G|$ (thus $m=m^G+\bar{m}^G$). Notice that $\bar{m}^G \geq \bar{n}^G$. Sort and denote the values of the buyers in $\bar{B}^G$ by $v_1\geq v_2\geq ... \geq v_{\bar{n}^{G}}$. 

    By Lemma~\ref{lem:pe_pure_alpha1}, the platform can always post $\alpha=1$ and select any set of sellers in $\bar{S}^G$ to join the platform. By Lemma~\ref{lem:optimal_welfare}, if the platform chooses $k=1,\ldots ,\bar{n}^G$ sellers from $\bar{S}^G$ to join, then matching them to buyers $v_1,v_2,...,v_k$ maximizes the welfare. Lemma~\ref{lem:same_on_price} then says all on platform sellers have price $v_k$. Therefore $Rev^{\star}\geq k\cdot v_k$ for any $k\leq \bar{n}^{G}$. As $\bar{m}^G\geq \bar{n}^G, W^{\star}(\bar{S}^G,\bar{B}^G,G) =  W^{\star}(S,\bar{B}^G,G)>\frac{1}{2}$. We get that
    $$ \frac{1}{2}<W^{\star}(\bar{S}^G,\bar{B}^G,G)=\sum_{i=1}^{\bar{n}^G}v_i \leq \sum_{i=1}^{\bar{n}^G}Rev^{\star}/i \leq H_n \cdot Rev^{\star}$$
    which implies $Rev^{\star}=\Omega(1/\log n)$. As the platform's revenue is always smaller than the market's welfare, we get that the welfare is also $\Omega(1/\log n)$.
\end{proof}

\begin{lemma}
    When there are more buyers than sellers $n> m$, the PoA is $O(\log m)$. \label{lem:sspoa_n_geq_m}
\end{lemma}
\begin{proof}
    We follow the same notations as in the proof of Lemma~\ref{lem:sppoa_n_leq_m}. If $W(S^G,B^G,G) \ge 1/\log m$ we are done. Therefore, assume $W(S^G,B^G,G) < 1/\log m$, and thus $W^\star(S,\bar{B}^G,G) > 1-1/\log m$. Denote the values of the buyers in $B^G$ by $\hat{v}_1\geq \hat{v}_2\geq \ldots \geq \hat{v}_{m^G}$ and the values of buyers in $\bar{B}^G$ by $v_1\geq v_2 \geq \ldots \geq v_{n-m^G}$. We consider the following cases:

    \vspace{0.1cm}

    \noindent\textbf{Case 1:} $m^G < \bar{m}^G.$  The reasoning is same with proof in Lemma~\ref{lem:sppoa_n_leq_m}. The platform can always post $\alpha=1$ and select any set of sellers in $\bar{S}^G$ to join the platform. If the platform chooses $k=1,\ldots ,\bar{n}^G$ sellers from $\bar{S}^G$ to join, then on platform sellers have price $v_k$. Therefore $Rev^\star\geq k\cdot v_k$ for any $k\leq \bar{m}^G$. As $m^G < \bar{m}^G$, $W^\star(\bar{S}^G,\bar{B}^G,G)\geq  W^\star(S,\bar{B}^G,G)/2 > (1-1/\log m)/2\geq 1/3$. We get that $$1/3 < W^{\star}(\bar{S}^G,\bar{B}^G,G) =  \sum_{i=1}^{\bar{m}^G} v_i \leq \sum_{i=1}^{\bar{m}^G}Rev^\star/i \leq H_m\cdot Rev^\star,$$ which implies $Rev^\star=\Omega(1/\log m)$ and the welfare is also  $\Omega(1/\log m)$.

    \vspace{0.1cm}

    \noindent\textbf{Case 2:} $m^G \geq \bar{m}^G$ and $v_1\geq 1/12.$ If $\bar{m}^G>0$, then posting $\alpha=1$ guarantees a seller from $\bar{S}^G$ joining and paying $v_1\geq 1/12$ to the platform, which implies that the welfare is also at least $1/12$. Otherwise, by Theorem~\ref{thm:equilibrium_set_enlarges_1} decrease $\alpha$ until one seller $j\in S^G$ joins in pure equilibrium.
    
    As $j$'s off platform price is at most the value of a buyer in $B^G, p^{\off}_j\leq 1/\log m$ before joining and at least $p^{\on}_j\geq v_1-1/\log m\geq 1/24$ after joining. To have $j$ join, $\alpha$ is the highest value for $j$ to break even: $(1-\alpha)p^{on}_j = p^{\off}_j\leq  1/\log m\ \Rightarrow\ \alpha\geq 1-24/\log m \geq 1/2.$ Thus, $Rev^\star \geq \alpha/24 \geq 1/48$, which again implies the welfare at the equilibrium is a constant-fraction of the optimal welfare. 
    
    \vspace{0.1cm}

    \noindent\textbf{Case 3:} $m^G \geq \bar{m}^G$ and $v_1< 1/12.$ Since there are many sellers transacting off-platform, in optimal matching $S^G$ and $\bar{B}^G$ create a lot of welfare.
    \begin{eqnarray*}
        W^\star(S^G,\bar{B}^G,G) &>& W^\star(S^G,B,G)-W^\star(S^G,B^G,G)\\
        &>& \frac{1}{2}W^\star(S,B,G)-W(S^G,B^G,G) > \frac{1}{2}-1/\log m>\frac{1}{3}
    \end{eqnarray*}
    Now let $\ell = \argmax_{\hat{\ell}} \{\sum_{i=1}^{\hat{\ell}} v_i \mbox{ s.t. } \sum_{i=1}^{\hat{\ell}} v_i < 1/6.\}$ that is, the largest index of buyer in $\bar{B}^G$ such that the sum of the $\ell$ highest buyers values in $\bar{B}^G$ is smaller than $1/6$. As $v_1\leq 1/12$, we have $\sum_{i=1}^{\ell} v_i \geq 1/12$. Further since $l$ largest buyers have smaller than $1/6$ welfare but $\bar{B}^G$ provides at least $1/3$ welfare, there are more than $2\ell$ buyers in $\bar{B}^G$. This in turn requires
    \begin{eqnarray}
        \ell\leq m^G/2 \label{eq:ell_ub}
    \end{eqnarray} because otherwise $2\ell>m^G$, sellers in $S^G$ match to no more than $2\ell$ largest buyers in $\bar{B}^G$, creating welfare smaller than $\sum_{i=1}^{2\ell} v_i< 1/6 *2=1/3$. This contradicts with $W^{\star}(S^G,\bar{B}^{G},G)> 1/3$.
    
    As a next step, we lower bound platform's optimal revenue. Theorem~\ref{thm:equilibrium_set_enlarges_1} says as we lower $\alpha$ with Algorithm~\ref{alg:PE_pure}, for every $k\in \{1,2,...,\ell\}$, there is a pure equilibrium with $k$ sellers on platform. Denote the highest transaction fee when $k$ sellers join by $\alpha_k$. By Lemma~\ref{lem:same_on_price}, if $k$ sellers are on platform, they each has a price of the $k$-th largest buyer not transact off-platform, which is weakly larger than $v_k$. Then platform's optimal revenue satisfies $Rev^\star \ge \alpha_k \cdot k \cdot v_k \geq \alpha_l \cdot k \cdot v_k$ for every $k\in\{1,2,...,\ell\}$. We next show that $\alpha_l \geq 1/2$, which implies $$1/12\leq \sum_{i=1}^\ell v_i\leq \frac{1}{\alpha_l}Rev^\star \sum_{k=1}^\ell\frac{1}{k}\leq 2H_\ell Rev^\star \leq 2H_m Rev^\star,$$ and $Rev^\star = \Omega(1/\log m)$, meaning that we get our welfare guarantee.

    We now show $\alpha_{\ell}\geq 1/2$. By Theorem~\ref{thm:equilibrium_set_enlarges_1}, consider when the Algorithm decrease $\alpha$ to $\alpha_{\ell}$ and finally the $l$-th seller $j_{\ell}$ joins the platform in equilibrium. At $\alpha_{\ell}$, the on-platform gain of $j_{\ell}$ just reaches zero: $(1-\alpha_{\ell})p_{j_{\ell}}^{\on}(P) = p_{j_{\ell}}^{\off}(P)$. We lower bound $p_{j_{\ell}}^{\on}(P)$ and upper bound $p_{j_{\ell}}^{\off}(P)$. By Lemma~\ref{lem:same_on_price}, on platform price equals to the valuation of the $\ell$-th largest buyer not transacting off-platform, which is weakly larger than $v_\ell$. If $\ell\leq \bar{m}^G$, by Lemma~\ref{lem:pe_pure_alpha1} all $\ell$ on platform sellers belong to $\bar{m}^G$ and $\alpha_{\ell}=1$. When $j_{\ell}$ considers joining, there are in total $m^G-(\ell-1)$ sellers not joining, and one of them is matched to a buyer with valuation no larger than $\hat{v}_{m^G-\ell+1}$, as each seller from $S^G$ can unmatch at most one buyer in $B^G$. Since the algorithm adds seller $j_\ell$, by Lemma~\ref{lem:algo_select} $j_\ell$ has the lowest off platform price and is at most $p^{\off}_{j_{\ell}}\leq \hat{v}_{m^G-\ell+1}$. Then
    \begin{eqnarray*}
        (1-\alpha_k)v_\ell \leq (1-\alpha_k)p_{j_{\ell}}^\on(P) = p_{j_k}^\off(P) \leq \hat{v}_{m^G-\ell+1}. 
    \end{eqnarray*}
    Assume towards a contradiction that $\alpha_k< 1/2$, we get that 
    \begin{eqnarray}
        \hat{v}_{m^G/2}\geq \hat{v}_{m^G-\ell} \geq \hat{v}_{m^G-\ell+1}> v_\ell/2,
    \end{eqnarray}
    where the first inequality follows Eq.~\eqref{eq:ell_ub}. Thus, we have that
    \begin{eqnarray}
        W(S^G,B^G,G)> \frac{m^G}{2}\cdot\frac{v_\ell}{2} \geq \frac{m\cdot v_\ell}{8},  \label{eq:lb_off_platform_welfare}    
    \end{eqnarray}
    where the second inequality follows $m^G\geq m/2$. On the other hand, we know that 
    \begin{eqnarray}
        m\cdot v_\ell \geq \sum_{i=\ell+1}^{m^G} v_i = \sum_{1}^{m^G} v_i - \sum_{i=1}^{\ell} v_i = W^\star(S^G,\bar{B}^G,G)-\sum_{i=1}^{\ell} v_i \geq 1/3-1/6 = 1/6. \label{eq:up_on_platform_welfare}
    \end{eqnarray}

    Combining Equations~\eqref{eq:lb_off_platform_welfare} and~\eqref{eq:up_on_platform_welfare} yields that  $$W(S^G,B^G,G)\geq 1/48,$$ contradicting $W(S^G,B^G,G)<1/\log m$. Thus, $\alpha_k\geq 1/2$ which concludes the proof.    
\end{proof}

\subsection{$\Omega(\log(\min(n,m)))$ PoA for Homogeneous-Goods Markets}\label{sec:poa_homo_n_lb}

In this section, we give an example market to lower bound PoA $\Omega(\log(m,n))$, showing the tightness of our analysis in Section~\ref{sec:poa_homo_n_ub}.

\begin{figure} 
    \centering
    \begin{tikzpicture}[scale=1.5]
        \foreach \i/\label in {1/$b_1$, 2.5/$b_2$, 4/$b_3$, 7/$b_i$,10/$b_n$}
            \node[draw, shape=rectangle, minimum size=0.6cm] (\label) at (\i, 2) {\label};
        
        \foreach \i/\label in {1/$s_1$, 2.5/$s_2$, 4/$s_3$, 7/$s_i$,10/$s_n$}
            \node[draw, shape=circle, minimum size=0.6cm] (\label) at (\i, 0) {\label};
            
        \foreach \i/\j in {1/$n+\epsilon$, 2.5/$\frac{n}{2}$, 4/$\frac{n}{3}$, 7/$\frac{n}{i}$, 10/$1$}
            \node at (\i, 2.5) {\j};
        
        \foreach \i in {$b_1$,$b_2$,$b_3$,$b_i$,$b_n$}
            \foreach \j in {$s_1$,$s_2$,$s_3$,$s_i$,$s_n$}
                \draw[line width=1.1pt, blue, dashed] (\i) -- (\j);
     
        \node at (5.5, 2) {$\ldots$};
        \node at (8.5, 2) {$\ldots$};
        \node at (5.5, 0) {$\ldots$};
        \node at (8.5, 0) {$\ldots$};
        \node at (5.5, 2.5) {$\ldots$};
        \node at (8.5, 2.5) {$\ldots$};

        \node[left] at (0.5, 2)  {Buyers};
        \node[left] at (0.5, 0)  {Sellers};
        \node[left] at (0.5, 2.5)  {Values};
    
    \end{tikzpicture}
        \caption{A homogeneous goods market with price of anarchy $H_n$. There are no off-platform edges. Buyer $b_1$ has  value $n+\epsilon$ for any seller, and buyer $b_i$, $i=2,\ldots,n$, has value $n/i$ for any seller.} \label{fig:logn_poa}
\end{figure}
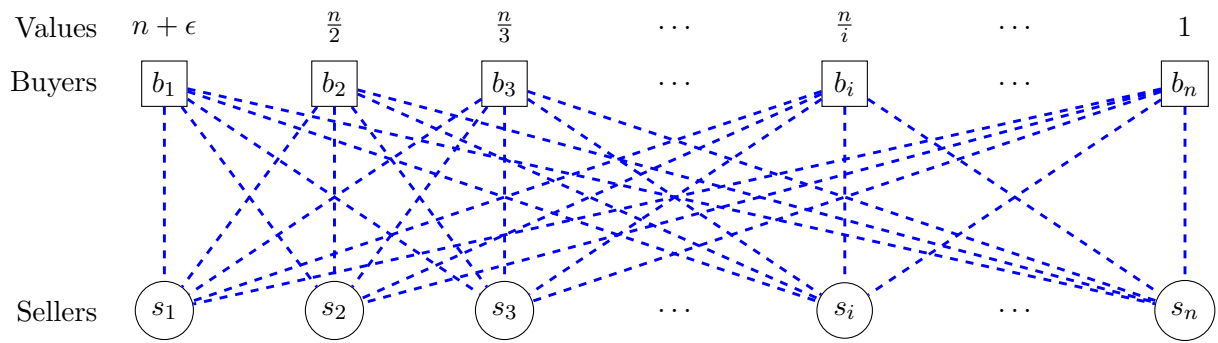

\begin{theorem}\label{thm:poa_lower_bound_homo}
    There exists a market for which the price of anarchy is $H_{\min\{n,m\}}=\Omega(\log(\min\{n,m\}))$.
\end{theorem}
\begin{proof}
        Figure~\ref{fig:logn_poa} depicts a $n$-buyer $n$-seller market where no transaction can take place without the platform. Buyer $b_1$ has value $n+\epsilon$ for any seller,
and buyer $b_i$, $i=2\ldots n$, has value $\frac{n}{i}$ for any seller.
 Since there are no possible  transactions without joining the platform, each seller weakly prefers to join the platform, so $\alpha^{\star}=1$ and there are multiple equilibria. 
As each buyer values the sellers equally, the platform is in effect choosing, through its choice of $\alpha$, how many sellers join the platform, and in turn the number of buyers that transact. 
    
    If only one seller joins the platform, the competitive price is $n+\epsilon$, and any other number $1<k\leq n$ of sellers joining the platform results in the  competitive price of $n/k$ for each seller
 and total platform revenue $n$. 
For this reason, the platform  prefers one seller to join the platform, in order to maximize its revenue. This results in PoA $\frac{(1+\frac{1}{2}+\frac{1}{3}+\dots+\frac{1}{n})n+\epsilon}{n+\epsilon} \approx H_n$ when $\epsilon$ is small.

If we increase the number of sellers $m$ to be larger than $n$, then still the revenue optimal equilibrium is the one where only one seller joins, whereas the welfare-optimal allocation matches $n$ sellers to $n$ buyers, obtaining a  the $H_n= H_{\min\{n,m\}}$ lower bound. If $m$ is smaller than $n$, we can have $m$ buyers with values $1+\epsilon$, $1/2$, $1/3$,\ldots,$1/m$ and then $n-m$ buyers with value 0, and get a $H_m =H_{\min\{n,m\}}$ lower bound.
\end{proof}

\section{Main Positive Result: Market Efficiency with a Regulated Platform} \label{sec:poa-regulated}

The previous section shows PoA can scale linearly in unregulated platforms. In this section, we show even light regulation can substantially increase welfare in markets with general unit-demand valuations. We prove this by providing a lower bound on the worst possible welfare at a given fee cap $\alpha$. Since the welfare guarantee is monotonically decreasing in $\alpha$, we assume the platform posts the cap $\alpha$ as its fee. It might be the case that in order to maximize its revenue, the revenue sets a lower transaction fee, but this only gives a higher welfare guarantee. Interestingly, the results in this section do not assume the platform can choose a specific Platform Equilibrium, as we prove the guarantee for every fee smaller than the cap, and for every equilibrium at a given fee.

This section is structured as follow. In Section~\ref{sec:poa_cap_pure}, we give a tight bound on the Price of Anarchy of pure Platform equilibria for a fixed transaction fee $\alpha$. As Proposition~\ref{prop:no_pure} shows a a pure equilibrium need not exist, in Section~\ref{sec:poa_mixed} we extend the analysis to the set of mixed Platform Equilibria. In Section~\ref{sec:poa_tight}, we show our analysis is tight for any fee $\alpha$.

\subsection{Pure Equilibrium Bound} \label{sec:poa_cap_pure}

Bounding the PoA of a Platform Equilibrium requires buyers' valuations for the allocation at the equilibrium while the strategic actors in the model are the sellers. In order to relate the two we use the characterization of maximum competitive prices given by Eq.~\eqref{eq:max_price}, as they directly relate to the sellers' utility as well as to buyers' welfare. 
\begin{theorem}
    The Price of Anarchy of pure Platform Equilibrium, when the platform sets a transaction-fee $\alpha\in [0,1)$, is at most $\frac{2-\alpha}{1-\alpha}$. \label{thm:pure_poa}
\end{theorem}
\begin{proof}
    Let $v_1,\ldots, v_n$ be buyers' valuations,
 and let $G$ be a buyer-seller network. We first show that when no seller chooses to join the platform, we already get the welfare guarantee: 
$$\frac{W^\star}{W(S,B,G)} \leq \frac{2-\alpha}{1-\alpha}.$$ 

    Consider this case of $P=\emptyset$, and denote $G(j)$ as the network when only seller $j$ joins the platform. Let 
    \begin{eqnarray}
        p_{j}^{\on} = W(S,B,G(j)) - W(S\setminus\{j\},B,G(j)) = W(S,B,G(j)) - W(S\setminus\{j\},B,G)\label{eq:p_on}
    \end{eqnarray} 
    denote $j$'s price when $j$ joins the platform. Let $p_{j}^{\off}$ denote seller $j$'s price when not joining the platform. We can bound this price by $j$'s maximum competitive price,
    \begin{eqnarray}
        p_{j}^{\off} \leq W(S,B,G) - W(S\setminus\{j\},B,G). \label{eq:p_off}
    \end{eqnarray} 

    As seller $j$ does not join the platform, we have  $$p_{j}^{\off} \geq (1-\alpha)p_{j}^{\on}.$$ By Equations~\eqref{eq:p_on} and~\eqref{eq:p_off}, we have
    \begin{eqnarray*}
        W(S,B,G(j)) - W(S,B,G) & \leq & p^{\on}_j-p^\off_j \\ 
        &\leq & \alpha\cdot p^\on_j \\ 
        & \leq & \frac{\alpha}{1-\alpha}\cdot p^\off_j \\
        & \leq & \frac{\alpha}{1-\alpha}\left(W(S,B,G) - W(S\setminus\{j\},B,G)\right).
    \end{eqnarray*}

    Rearranging gives 
    \begin{eqnarray}
    W(S,B,G(j)) \leq
        \frac{1}{1-\alpha}W(S,B,G) - \frac{\alpha}{1-\alpha}W(S\setminus\{j\},B,G). \label{eq:bound1}
    \end{eqnarray}

    Let $i^\star(j)$ be the buyer matched to $j$ in the ideal matching $W^\star$. Since this is also one of the options to match $j$ in $G(j)$ is to $i^\star(j)$, we have
    \begin{eqnarray}
        W(S,B, G(j)) & \geq& v_{i^\star(j)j} + W(S\setminus\{ j\}, B\setminus\{i^\star(j)\}, G). \label{eq:bound2}
    \end{eqnarray}

    Combining Equations~\eqref{eq:bound1} and~\eqref{eq:bound2}, we get
    \begin{eqnarray}
        v_{i^\star(j)j} \leq \frac{1}{1-\alpha}W(S,B,G)-\frac{\alpha}{1-\alpha}W(S\setminus\{j\},B,G)-W(S\setminus\{j\},B\setminus\{i^\star(j)\},G). \label{eq:bound3}
    \end{eqnarray}

    We wish to relate the right-hand side of Eq.~\eqref{eq:bound3} to terms that relate to $W(S,B,G)$ and $W^\star$.
 Let $i^G(j)$ be the buyer matched to $j$ in $W(S,B,G)$. First, consider $W(S\setminus\{j\},B,G)$. By 
definition, we have
    \begin{eqnarray*}
        W(S,B,G) = W(S\setminus\{j\}, B\setminus\{i^G(j)\}, G) + v_{i^G(j)j}. 
    \end{eqnarray*}
    Thus, we have,
    \begin{eqnarray}
        W(S\setminus\{j\},B,G) \geq W(S\setminus\{j\}, B\setminus\{i^G(j)\}, G) = W(S,B,G)-v_{i^G(j)j}.\label{eq:bound4}
    \end{eqnarray}

    As for $W(S\setminus\{j\},B\setminus\{i^\star(j)\},G)$, 
seller  $j$ is matched to $i^G(j)$ in $G$, while $i^\star(j)$ is matched to some potentially different vertex in $G$, which we call \textit{the twin of $j$}
 and denote by $t(j)$. We  have the following inequality,
    \begin{eqnarray*}
        W(S,B,G) &\leq & W(S\setminus\{j,t(j)\},B\setminus\{i^G(j),i^\star(j)\},G) + v_{i^G(j)j} + v_{i^\star(j)t(j)}\nonumber \\
        &\leq & W(S\setminus\{j\},B\setminus\{i^\star(j)\},G) + v_{i^G(j)j} + v_{i^\star(j)t(j)}, 
    \end{eqnarray*}
    where the first inequality is an equality if $j\neq t(j)$. Rearranging gives
    \begin{eqnarray}
        W(S\setminus\{j\},B\setminus\{i^\star(j)\},G) \geq W(S,B,G) - (v_{i^G(j)j} + v_{i^\star(j)t(j)}).\label{eq:bound5}
    \end{eqnarray}

    Combining Equations~\eqref{eq:bound3},~\eqref{eq:bound4}, and~\eqref{eq:bound5}, we get
    \begin{eqnarray}
        v_{{i^\star}(j)j} &\leq& \frac{1}{1-\alpha}W(S,B,G)-\frac{\alpha}{1-\alpha}\left(W(S,B,G)-v_{i^G(j)j}\right) - \left(W(S,B,G) - (v_{i^G(j)j} + v_{i^\star(j)t(j)})\right) \nonumber \\
        & = & \frac{1}{1-\alpha}\cdot v_{i^G(j)j} + v_{i^\star(j)t(j)}. \label{eq:combine_three_terms_right}
    \end{eqnarray}

    Summing over all sellers $j$, we have
    \begin{eqnarray*}
        W^\star = \sum_j v_{{i^\star}(j)j} & \leq & \sum_j \left(\frac{1}{1-\alpha}\cdot v_{i^G(j)j} + v_{i^\star(j)t(j)}\right)\\ & = &  \frac{1}{1-\alpha}\cdot\sum_j v_{i^G(j)j} + \sum_j v_{i^\star(j)t(j)}\\
        & \leq & \frac{1}{1-\alpha} W(S,B,G) + \sum_j v_{i^G(j)j}\\
        & = & \frac{2-\alpha}{1-\alpha} W(S,B,G),
    \end{eqnarray*}
    where the last  inequality follows because each seller $j$ has a distinct twin $t(j)$, so we sum over distinct edges of $G$.   

    To finish the proof, 
we consider a case where a non-empty
set $P$ of sellers join in a Nash equilibrium. Let $G(P)$ denote the network where sellers in $P$ are connected to all buyers,
 and sellers not in $P$ are connected only via links in the original network $G$.
 Now consider another market where the buyers have the same valuations as the original market, but the  initial network is $G'=G(P)$. For the same transaction fee $\alpha$, it is
 an equilibrium  for no seller to join the platform  in $(S,B,G')$.
 Indeed, sellers in $P$ are already linked to all buyers in this modified market,
 so their utility can only decrease in joining,
 and if there was a seller $j\notin P$ that would strictly prefer to join at $\alpha$, then
 $P$ is not an equilibrium in the original market.
    By the above, we have $\frac{W^\star(S,B,G')}{W(S,B,G')}\leq \frac{2-\alpha}{1-\alpha}$. Clearly, $W^\star(S,B,G')=W^\star(S,B,G)$ and  $W(S,B,G')=W(S,B,G(P))$, which concludes the proof.
\end{proof}

Theorem~\ref{thm:pure_poa} bounds the PoA of pure equilibrium at $\alpha$, should it exist. However,
we know from Proposition~\ref{prop:no_pure} that  there exists markets and transaction fees $\alpha$ for which there is no pure Platform Equilibrium.
 This motivates us to extend our analysis to mixed equilibria.

\subsection{Mixed Equilibrium Bound} \label{sec:poa_mixed}

In this section, we prove the same bound works for mixed equilibria. For mixed equilibria, sellers randomize between joining the platform and not joining (and thus only being able to transact with buyers who whom they have an existing connection). This possibly results in $2^{m}$ graph realizations and a competitive price for each of them. To reason about the probability of each realization through sellers' incentives, and taking expectation over the max welfare of all graphs is quite hard.

We thus reduce the argument for PoA of mixed equilibria to the case of pure equilibria. One way to do this is to reduce randomization of sellers' mixed strategy to randomization of nature's choice, in a \textit{Bayesian game}. Take a mixed Nash equilibrium, $\x$, and define a Bayesian game where sellers have pure strategies (joining vs.~not joining), but where in the event
that a seller $j$ does not join they are still connected to all buyers with 
 probability $x_j$.
We show that one of the \textit{Bayes-Nash Equilibrium} in this Bayesian game has no seller choosing to join, which corresponds to the smallest-welfare equilibrium. We then show this smallest-welfare equilibrium has the same expected social welfare as the mixed Nash equilibrium 
in the corresponding complete information game. The rest of the proof follows the same arguments as the first part of Theorem~\ref{thm:pure_poa}, but where the quantities for each seller $j$ are in expectation over the links formed. In the following, we give a sketch that gives the idea of the construction of the Bayesian game, and defer the full proof to  Appendix~\ref{app:mixed_proof}. 
\begin{restatable}{theorem}{mixedPoA}
\label{thm:mixed_poa}
    The Price of Anarchy of mixed Platform Equilibrium,  when the platform sets a transaction-fee $\alpha\in [0,1)$, is at most $\frac{2-\alpha}{1-\alpha}$.
\end{restatable}
\begin{proof}[Proof Sketch]
    Consider a mixed Platform Equilibrium, $\x = (x_1,\ldots, x_n)$, for a buyer-seller network $(S,B,G)$, where $x_j$ is the probability that seller $j$ joins the platform. We define the following Bayesian game:
    \begin{itemize}
        \item For each seller $j$, with probability $x_j$, $j$ can transact with all buyers ("type 1"), and with probability $1-x_j$, it can only transact with the buyers linked to $j$ in $G$ ("type 2").
        \item The platform posts a transaction fee $\alpha$.
        \item Before knowing the realization of its type,
 a seller can choose whether or not to join the platform, in which case the seller 
can with certainty transact with all buyers.
        \item Given the realized graph $G'$,which depends on the realized types as well as actions of each seller, a competitive equilibrium with the maximum price is formed. 
        \item A seller joining the platform pays an $\alpha$-fraction of their revenue to the platform, regardless of its type.
    \end{itemize}

    We  now show that there exists a smallest-welfare equilibrium in the Bayesian game where no seller joins. 
For this, consider a seller $j$ that adopts probability $x_j$ in the mixed Nash Platform Equilibrium
of the original, complete information platform game. There are three cases to consider:

\noindent\textbf{Case 1:} $x_j=0$ In this case, $j$'s expected utility from not joining the platform \textit{in the original, complete information game} is at least as much as $j$'s expected utility from joining given $\x_{-j}$. \textit{In the Bayesian game}, given that no other seller joins the platform, $j$'s utility from either joining or not joining the platform is exactly $j$'s utility for joining or not joining in the complete information game,
 as other links are formed according to $\x_{-j}$.

\noindent\textbf{Case 2:} $x_j=1$: In this case, in the Bayesian game, $j$ is linked to all buyers with probability $1$. Thus,  joining the platform in the Bayesian game does not increase $j$'s utility.

\noindent\textbf{Case 3:} $x_j\in (0,1)$: In this case, in the original, complete information game, $j$'s expected utility is the same for joining and not joining given $\x_{-j}$; otherwise, $j$ would deviate and $\x$ would not be an equilibrium.
 Let $u_j^\x$ denote $j$'s utility at $\x$ in the original game, 
which is also $j$'s utility from either joining or not joining the platform in the original game. 
        
\textit{In the Bayesian game}, consider the case where $j$ does not join the platform, and when no other seller joins the platform. If $j$'s links stay as they were in $G$, which happens with probability $1-x_j$, then $j$'s expected utility is exactly $u_j^\x$, as this is $j$'s utility from not joining the platform in the original, complete information game when all other sellers' links are formed according to $\x_{-j}$.
 If $j$'s links are formed, which happens with probability $x_j$,  then  $j$'s expected utility is at least as much as when it joins in the original game given all other sellers' links are formed according to $\x_{-j}$. This is because the  distribution on link formation is the same, while the seller does not have to pay the $\alpha$ fee to the platform. 
Therefore, $j$'s expected utility from not joining is at least $u_j^\x$. 

If $j$ does join the platform in the Bayesian game, then $j$'s utility is exactly the utility of $j$ for joining the platform in the original game, as the  distribution on link formation is the same as if they join in the original setup, and the fee $j$ pays is the same. Therefore, $j$'s utility is exactly $u_j^\x$. We get that $j$'s utility for not joining in the Bayesian game is at least as $j$'s utility for joining.

\vspace{0.3cm}

In all the above cases, for each seller $j$,  their expected utility for not joining the platform in the Bayesian game, given all other sellers do not join, is at least as much as  their 
expected utility for joining the platform in the Bayesian game.
Thus,  there is an equilibrium  
 in the Bayesian game where no seller joins.
The expected welfare of this equilibrium is exactly the same as the expected welfare of the original complete information game in mixed Platform Equilibrium $\x$, as we have the same distribution over the formation of links, and the competitive equilibrium formed 
always maximizes the welfare given  the links. 
    
    To conclude the proof, we
 show that if no sellers join in the Bayesian game,
 then the PoA 
with respect to pure strategies in the Bayesian game
is at most $\frac{2-\alpha}{1-\alpha}$. 
The proof follows the same arguments of the proof of this case in Theorem~\ref{thm:pure_poa}, but the quantities we reason about for seller $j$ are in expectation over $\x_{-j}$. For completeness, we give the full analysis in Appendix~\ref{app:mixed_proof}.      
\end{proof}

In markets without a platform, social welfare is defined as the sum of buyers' and sellers' utilities. We notice that the above result also implies welfare guarantees if we consider the benchmark used in settings without a platform. 

\begin{corollary}
    For any Platform Equilibria with $\alpha$ transaction fee, the sum of buyers and sellers' utility is at least $\frac{(1-\alpha)^2}{2-\alpha}$ fraction of the optimal welfare $W^\star$.
\end{corollary}
\begin{proof}
    When the set of sellers $P$ join the platform in equilibrium, platform's revenue is at most $\alpha W(S,B,G(P))$. The sum of buyers and sellers' utility is 
    $$ (1-\alpha)W(S,B,G(P))\geq (1-\alpha)\frac{1-\alpha}{2-\alpha}W^\star$$
\end{proof}

\subsection{Tight Instance} \label{sec:poa_tight}

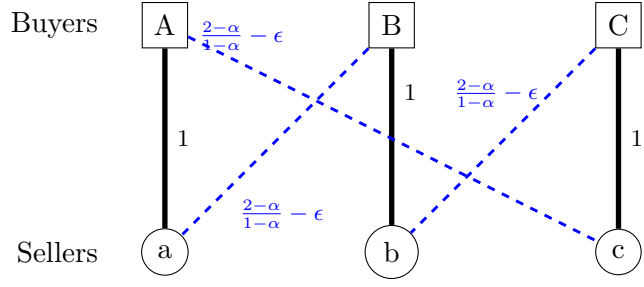
\begin{figure} 
    \centering
    \begin{tikzpicture}[scale=1.5]
        \foreach \i/\label in {1/A, 3/B, 5/C}
            \node[draw, shape=rectangle, minimum size=0.6cm] (\label) at (\i, 2) {\label};
        
        \foreach \i/\label in {1/a, 3/b, 5/c}
            \node[draw, shape=circle, minimum size=0.6cm] (\label) at (\i, 0) {\label};
        
        \foreach \x/\y/\w/\pos in {A/a/1/midway, B/b/1/near start, C/c/1/midway}
            \draw[line width=2pt] (\x) -- node[\pos, right, font=\footnotesize] {\w} (\y);
        
        \foreach \x/\y/\w/\pos/\loc in {B/a/$\frac{2-\alpha}{1-\alpha}-\epsilon$/near end/below right, A/c/$\frac{2-\alpha}{1-\alpha}-\epsilon$/at start/right, C/b/$\frac{2-\alpha}{1-\alpha}-\epsilon$/near start/left}
            \draw[line width=1.1pt, blue, dashed] (\x) -- node[\pos, \loc, font=\footnotesize] {\w} (\y);

        \node[left] at (0.5, 2)  {Buyers};

        \node[left] at (0.5, 0)  {Sellers};
    
    \end{tikzpicture}
    \caption{An example with a tight Bound for PoA for an $\alpha$ transaction fee, for any $\alpha \in [0,1)$. Black solid lines are direct links, as captured by $N(i)$ for buyer $i$, and blue dotted lines indicate missing links. Buyer values are annotated adjacent to each edge, and all any value that is omitted is zero.} \label{fig:tight_bound_poa_for_alpha_transaction_fee}
\end{figure}

We now show that our PoA analysis is tight: for any transaction fee $\alpha$, there exists an instance of the platform game in which the price of anarchy of Platform Equilibria with pure strategies is exactly $\frac{2-\alpha}{1-\alpha}$.
\begin{theorem}\label{thm:poa_tight}
    For every transaction fee $\alpha\in[0,1)$,  
    there exists a market for which the ratio of optimal welfare to social welfare of a Platform Equilibrium is indeed $\frac{2-\alpha}{1-\alpha}$, and in that Platform Equilibrium sellers use pure strategies.
\end{theorem}
\begin{proof}
For any given transaction fee $\alpha\in [0,1)$, 
Figure~\ref{fig:tight_bound_poa_for_alpha_transaction_fee} gives a three-buyer three-seller market with a tight PoA. At transaction fee $\alpha$, no seller joining the platform is a Nash equilibrium, 
since the off-platform price of $1$ is larger than the seller's on-platform utility,
which is $(1-\alpha)p^{on}=(1-\alpha)\left[\frac{2-\alpha}{1-\alpha}-\epsilon-1\right]=1-\epsilon+\alpha\epsilon$. As $\epsilon$ goes to zero in the parameterized values of buyers in the market,
 the price of anarchy goes to $\frac{2-\alpha}{1-\alpha}$. Note all sellers joining the platform is another Nash equilibrium. But the price of anarchy depends 
on the worst possible Nash equilibrium.
\end{proof}

\section{Matching with Costs}
In reality the platform incurs a fixed cost when bringing sellers on platform, that covers setting the seller in the system, providing the ordering tablet, etc. Because of the cost, regulators cannot simply impose the zero cap on $\alpha$. In this section, we use simulations to ask ``what is the minimum fee cap $\alpha$ that regulators can impose on a platform, such that platform incurring fixed cost of $c$ per on-platform seller still makes a net revenue and do not exit the market?"

\paragraph{Homogeneous good market}
\begin{lemma}
    As we lower $\alpha$ from 1 to 0 with \cref{alg:PE_pure}, $P$ the set of on-platform sellers enlarges, $B^G$ the set of buyers transacting in $W (S \setminus P, B, G)$ shrinks, $\bar{B}^G=B\setminus B^G$ enlarges, the $|P|$-th largest buyer valuation in $\bar{B}^G$ weakly decreases. All on-platform sellers have the same onplatform price, which weakly decreases.  
\end{lemma}
\begin{proof}
    For any homogeneous-goods markets, from \cref{lem:same_on_price} and \cref{lem:optimal_welfare}, all on-platform sellers have the same price $\bar{v}(m')-\bar{v}(m'-1)$, where $\bar{v}(m')$ is the sum of the $m'$ largest buyers valuation among all the buyers 

and $\bar{v}(m'-1)$ are tje  that is directly determined by the number of sellers joining the market. 

$\bar{v}(m')-\bar{v}(m'-1)$, where $\bar{v}(m')$ is the largest .
Thus, the regulators can easily run     
\end{proof}
This means that the regulator can simply decrease $\alpha$ from 1 to 0, until the on platform price reaches the fixed cost $C$, and select the minimum $\alpha$ such that on-platform prices is higher than fixed cost $C$. 

\paragraph{Hetergeneous good market} As we have shown before in \cref{prop:no_pure}, sellers' best response behavior might exhibit cyclic behavior, and a lower transaction fee $\alpha$ does not always lead to a weakly larger set of on-platform sellers, nor necessarily a weakly higher sum of all platform prices. Thus, it is much more complex to theoretically analyze the minimum fee cap that guarantees positive revenue for the platform. To do so, we turn to simulations. 

We randomly generate 100 markets with heterogeneous valuations of size $m=n=10$. For each of the 100 markets, we grid search through 100 values of transaction fee $\alpha\in\{0.01,0.02,0.03,...,0.99\}$. We plot the average number of sellers joining the market, the average platform revenue (without fixed cost), and the average of platform revenue divided by the number of on platform sellers. Hopefully it will show us some monotonicity.

\section{Generalizations}
In this section, we extend our previous results to more general settings. We analyze markets with multiple platforms in Section~\ref{sec:multi_platform}, sellers with production costs in Section~\ref{sec:seller_with_cost}, and show all of our results hold when buyers have additive-over partition valuation in Section~\ref{sec:beyond-ud}. Most proofs are delayed until Appendix~\ref{app:generalizations}.

\subsection{Market with Multiple Platforms}\label{sec:multi_platform}

For this section, we prove that our results in Section~\ref{sec:poa-regulated} hold in markets with multiple competing platforms. Consider the buyer-seller network on graph $G$, as defined in Section~\ref{sec:prelims}. There is a set of $R$ platforms $F=\{f_1,f_2,...,f_R\}$, where each platform $f_r$ has a subset of buyers $B_r\subset B$ and transaction fee $\alpha_i \leq \alpha$. A buyer can be on multiple platforms.
The regulator imposes an $\alpha$ cap on transaction-fee for all platforms. A seller $j$ can join multiple platforms $F_j\subseteq F$ and transact with all buyers on the platforms it joins, $\cup _{f_r\in F_j}B_r$, as well as the buyers it knows on $G$. The market clears according to competitive equilibrium. If $j$ transacts through platform edges with some buyer $i$, it pays $\alpha_r p_j^{\on}$ to the platform $f_r$ corresponding to buyer $i$. If $i$ is on multiple platforms that $j$ joins, then $j$ pays to the platform with the lowest fee. Everything else remains the same with Section~\ref{sec:poa-regulated}.

\begin{restatable}{theorem}{MultiPlatform}
    For a market with multiple platforms and transaction-fee cap $\alpha \in [0,1)$, the Price of Anarchy of Platform Equilibrium is at most $\frac{2-\alpha}{1-\alpha}$. \label{thm:pure_poa_multiple_platform}
\end{restatable}
The proof is similar to that of Theorem~\ref{thm:pure_poa} and \ref{thm:mixed_poa}, and is deferred to Appendix~\ref{app:multiple_platforms}.

\subsection{Sellers with Production Costs}\label{sec:seller_with_cost}
In this section, we extend our results to sellers $j\in S=\{1,2,...,m\}$ with production cost $c_j\geq 0$. We only present the main result and defer the readers to Appendix~\ref{app:seller_with_cost} for the full section.

\begin{restatable}{theorem}{purePoACost}
\label{thm:pure_poa_cost}
    In a market with cost, let $\beta=\sum_{j\in S}c_j / W^\star(S,B,\mathbf{v},\mathbf{c},G)$ denote the fraction of cost to optimal welfare. The Price of Anarchy of the Platform Equilibrium,  when the platform sets a transaction-fee $\alpha\in [0,1)$, is at most $\frac{2-\alpha}{1-\alpha-\alpha\beta}$.
\end{restatable}

The above theorem implies that at the same transaction fee, the social welfare guarantee deteriorates as costs get higher. This was specially relevant during the COVID-19 crisis, where demand and supply surged on digital platforms \citep{raj2020covid} and the cost of production grew \citep{felix2020us}.
At $30\%$ of transaction fee (Table~\ref{table:platforms and their coommission rate}) and $30\%$ of food cost \footnote{Most profitable restaurants aim for a food cost percentage between 28 and 35\%. This does not include other cost factors such as labor, rentals, etc. Figure is taken in 2023 from Doordash website \url{https://get.doordash.com/en-us/blog/food-cost-percentage}}, the resulting welfare at a  Platform Equilibrium  is at least $23.5\%$ of the ideal welfare. This is smaller than the $41\%$ without costs, further demonstration the need to regulate transaction fee during periods where production costs increase.

\subsection{Additive-over-Partition Buyer Valuation} \label{sec:beyond-ud}
Our results also extend beyond unit demand valuation. For example, consider the case where buyers have additive valuation. In this case, we can use $m$ vertices to represent each buyer in the bipartite graph, where each vertex only values one seller. We show how to generalize our results to more complex valuations, such as buyers valuing at most $c_{\ell}$ items  each group of sellers $S_\ell$ who sell similar items. Thus, the valuation of a buyer is additive constrained by the feasible sets defined by a partition matroid $\mathcal{M}$.

We formally define the class of buyers' valuations, which we term \textit{additive-over partition valuations}, and show results in Section~\ref{sec:poa-regulated} extend.

For each buyer $i\in\{1,2,...,n\}$, there is a partition of sellers $S_i = S_{i,1}\cup S_{i,2}\cup...\cup S_{i,T_i}$ and a  capacity $c_{i,\ell}$ associated with each seller group $S_{i,\ell}$.  Buyer $i$ values seller $j$'s item at $v_{ij}$, and the bundle of items $\alloc$ at
\begin{eqnarray}
    v_i(\alloc) = \sum_{\ell\in [T_i]} \max_{T\subset S_{i,\ell} \ : \ |T|\le c_{i,\ell}} \sum_{j\in T}v_{ij}. 
\end{eqnarray}

We now show how to encode such valuations into our  bipartite graph model. For each buyer $i$ and for each seller group $\ell$, we introduce $c_{i,\ell}$ unit-demand buyers. Each such buyer $i'\in [c_{i,\ell}]$ will have the following unit-demand valuation 
\begin{eqnarray*}
    v'_{i',j} = \begin{cases} v_{i,j} \quad & j\in S_{i,\ell}\\
                           0 \quad & j\notin S_{i,\ell}
            \end{cases},
\end{eqnarray*}
which ensures such buyer transacts with only sellers in $S_{i,\ell}$. For each such buyer $i'$, we also keep the same links to sellers such as the original buyer $i$. Via this reduction, our results in Section~\ref{sec:poa-regulated} trivially hold for additive-over-partition buyers as well.

\section{Conclusion and Future Directions}
In this paper, we initiate the rigorous study of the effect of platforms on the welfare of a constrained market, modeled by a buyer-seller network with missing links. We study the network structure of the market, and give an algorithm that always finds a pure equilibrium in homogeneous-goods markets. We show when the platform is unregulated, and can set fees to selfishly maximize its revenue, then the Price of Anarchy can be as bad as $\min\{m,n\}$ in general markets, and $\Theta(\log(\min\{n,m\} ) )$ even in homogeneous-goods markets. However, if regulation requires the platform to charge no more than $\alpha$-fraction of sellers' revenue, then we get a tight $\frac{2-\alpha}{1-\alpha}$-fraction of the optimal \textit{unconstrained} welfare in any equilibrium.
Moreover, even when discounting the platform's revenue, the surplus left for the buyers and sellers in any equilibrium is  at least $\frac{(1-\alpha)^2}{2-\alpha}$-fraction of the optimal unconstrained welfare. 
Finally, we extend our analysis to multiple platforms, buyer valuations beyond unit demand, and sellers with non-zero production cost.

While this work already provides strong insights into understanding a platform's effect on a market's welfare,  there are many exciting directions to take:
\begin{enumerate}
    \item Although the $\Omega(\log (\min\{n,m\}))$ lower bound on the Price of Anarchy is discouraging, this is a  contrived instance. We believe that for constrained and natural market structures, we can avoid the logarithmic dependence in the number of buyers/sellers. 
    \item While we extended the analysis to additive-over-partition valuations in Appendix~\ref{sec:beyond-ud}, there remains the need to incorporate more general classes of valuation functions, such as the commonly studied like gross-substitutes valuations. 
    \item Extending our model to accommodate multi-supply sellers is not straightforward. A seller with multiple items can limit supply to raise revenue. And her price becomes zero if some of her items are left unsold. This requires further extension of the model to explain how and why such sellers join the platform.
    \item While seller-side transaction-fees are common, other fees are used by platforms, such as buyer and seller monthly/annual registrations fees. 
    This changes buyers and sellers' strategic behaviors to join the platform, and will required further extensions of the model.
\end{enumerate}


\chapter{Influencing Market Outcomes through Tipping Design} \label{chap:pricing_with_tips}

\section{Introduction}
Online delivery platforms such as UberEats, DoorDash, Grubhub, and Instacart are an essential part of modern life. These platforms facilitate \emph{three-sided} transactions: In each transaction, a \emph{buyer} receives food from a \emph{store} via delivery by a \emph{courier}. Buyers pay for these transactions, while stores and couriers are compensated for them.

A distinctive feature of these platforms is that they include buyer-specified tips, which are 
a core component of courier compensation. A study \citep{jacobs2024gig} estimates that tips account for 36\% of the median hourly wage for delivery drivers in Los Angeles and San Francisco—two cities with binding minimum-wage regulations for couriers—and as much as 53\% in Boston, Chicago, and Seattle.

Since tips constitute such a substantial share of courier compensation, the timing at which buyers specify tips and couriers observe them becomes an important platform design choice. This choice differs across platforms. While on most ride-hailing platforms, tips are determined after the ride is completed,\footnote{See \citet{UberRideTipPolicy} and \citet{LyftTipPolicy}. While it is possible on Lyft to tip before business rides \cite{LyftBusinessTipPolicy}, for most trips tipping occurs when riders rate the driver.} on food-delivery platforms, in contrast, tips are typically chosen by buyers and observed by couriers before delivery—often even before couriers decide whether to accept an order.\footnote{On UberEats, DoorDash and Grubhub, couriers see the total fare of an order, which includes the tip, before deciding whether to accept an order \citep{UberEatTipPolicy, UberEatTipDriverView,DoorDashTipPolicy,GrubhubTipPolicy,GrubhubDriverView}. On Instacart, tips are listed alongside platform compensation before couriers decide whether to accept an order \citep{ InstacartDriverView}.}

While moving tipping to the pre-delivery stage may be a way to shift courier compensation onto buyers,\footnote{Following the introduction of minimum-wage regulations for couriers in New York City in 2023, which required platforms to meet minimum-wage obligations to couriers independently of buyer tips, UberEats and DoorDash removed the option for buyers to specify tips before delivery in New York; this change reignited public debate over whether platforms rely on tips to shift compensation costs to buyers \citep{NYCUber}.} this design choice has further implications. By allowing couriers to observe the tips before agreeing to a delivery, platforms let buyers pay to influence couriers' decisions on which orders to deliver.  
In practice, couriers often observe a range of nearby orders rather than a single order recommended by the platform, with each order displaying information including buyer's tip amount, platform payment, and delivery origin and destination.\footnote{UberEats matches couriers to orders in two ways: 1) Exclusive offers where a courier is assigned to one delivery order exclusively, and 2) TripRadar offers where a courier can choose between multiple orders simultaneously when they are driving low speed or not moving. Instacart couriers can see multiple orders simultaneously in their ``batch list.'' DoorDash and Grubhub show each courier a single order at a time. See \citet{UberTripRadar,InstacartAccessBatches,DoorDashOneMatch,GrubhubOneMatch}.}
Couriers then compare available orders and may strategically accept those with higher compensation, higher tips, or shorter distances---a practice commonly referred to as “cherry-picking” (see \citet{randomizedFifo} for a comparison of several mechanisms that allow cherry-picking).\footnote{For a concrete example, a Philadelphia delivery courier declined 75\% of orders and hung out only in wealthy areas to ensure good tips and claimed to receive as much as 45\$ per hour \citep{CherryPickingExample}.}

Since couriers may prioritize orders with higher tips, lower-tip orders are at times left unattended.\footnote{Some DoorDash couriers have adopted a ``no tip, no trip'' strategy, where they refuse to deliver orders with zero tips \citep{notipnotrip,NoTipNoTripNewsweek}.}
Even when delivery platforms implement various forms of dynamic pricing to encourage courier supply,\footnote{UberEats employs surge pricing to pay couriers during busy hours \citep{UberEatSurgePricing}. DoorDash couriers receive ``Peak Pay'' during busy hours \citep{DoorDashSurgePricing}. Grubhub offers fixed incentives like Missions and Special Offers to boost courier earnings through goal-based rewards, but unlike dynamic pricing, these are preset and not responsive to real-time market changes \citep{GrubhubSurgePricing, GrubhubSurgePricingSecond}. Instacart adds ``Pay Boost'' to an order if an order remains unaccepted for a while \citep{InstacartDriverView}.} not all orders are delivered in a timely manner. In extreme cases, untipped orders  go undelivered for hours \citep{notipnotrip,NoTipNoTripNewsweek}.
This has caused platforms to nudge buyers to increase tips in order to receive faster delivery, and buyers have increasingly become aware that higher tips can lead to faster service.\footnote{For instance, DoorDash warned customers that not tipping may lead to long wait times \citep{notipwaitlonger}. Admittedly, platforms might nudge buyers for tips to shift courier compensation onto buyers.} 
Pre-delivery tipping has thus become  a distinctive feature on major food-delivery platforms that affects both buyer and courier incentives. 

\begin{table}
\begin{tabular}{|c|ccc|ccc|}
\hline
 & \multicolumn{3}{|c|}{Delivery} & \multicolumn{3}{|c|}{Ride-Share}\\
 & without tips & with tips & tips amount & without tips & with tips & tips amount \\
\hline
CA & \$19.37 & \$31.35 & \$11.98(38\%) & \$32.22 & \$35.62 & \$3.4(9.5\%)\\
Outside CA & \$13.16 & \$28.23 & \$15.07(53\%) & \$35.8 & \$37.98 & \$2.18(5.7\%) \\
\hline
\end{tabular}

\caption{Median courier gross hourly earnings on ride-share and delivery platforms in 2022, taken from \citet{jacobs2024gig}. CA (California) refers to Los Angeles and San Francisco Bay, while Outside CA refers to  Boston, Chicago and Seattle. In California, gig workers are guaranteed to receive 120\% of local minimum wage.
\label{tab:larger_tip_table}}
\end{table}

A distinctive feature of food-delivery platforms is that transactions involve three sides---couriers, buyers, and sellers. 
Compared to traditional two-sided markets, this three-sided structure makes it more difficult for the platform to clear the market---that is, to balance buyer demand with courier and store supply---using prices and compensation alone. In principle, a delivery platform could balance supply and demand by price discrimination---tailoring buyer prices and courier compensation across transactions.\footnote{Consider a transaction-specific pricing---charging the buyer her willingness to pay, paying the courier his delivery cost, and the store its production cost. Buyers, couriers, and stores are all left with zero surplus and are therefore indifferent among the transactions the platform assigns them to.}
However, regulatory and societal considerations typically require buyer prices and courier compensation to be non-discriminatory: buyers placing orders that are identical face the same price, and couriers delivering the same order receive the same compensation.\footnote{Laws such as the Equal Pay Act require equal pay for equal work in the United States \citep{EEOCEqualPayAct}. Similar regulations exists in Europe \citep{EUPlatformWorkDirective2024}. For buyer prices, platforms such as UberEats explicitly state the components of the prices charged to buyers, implying that prices depend on order characteristics rather than buyer identity \citep{UberEatBuyerPricing,DoorDashBuyerPricing}. } While non-discriminatory prices and compensation are often sufficient for market clearing in two-sided ride-hailing platforms,\footnote{See \Cref{rem:two_sided} in \Cref{sec:role_tips} for a detailed discussion.}
we show that this is not the case in three-sided delivery platforms, and instead, pricing compensation together with pre-delivery tipping can guarantee market clearing.

In this paper, 
we incorporate pre-delivery tipping into a theoretical framework to study three-sided delivery platforms, where both prices and tips facilitate market-clearing. We seek to address the following two questions:
\begin{enumerate}
    \item Are there benefits to allowing buyers to specify tips before delivery, compared with after delivery or not at all? And, if so, what are these benefits?
    \item How should a platform that is interested in maximizing welfare set buyer prices and courier compensation, considering that pre-delivery tipping factors in buyers' and couriers' decision-making? How about a platform that is interested in maximizing profit? 
\end{enumerate}

To answer the first question, we formalize an equilibrium concept that incorporates pre-delivery tipping, which we term a \emph{with-tip equilibrium}, and contrast it with a \emph{without-tip equilibrium}, which captures both settings with no tips as well as those with post-delivery tips. We show the superiority of with-tip equilibrium to without-tip equilibrium in terms of existence, welfare, and profit. To address the second question, we first show broad impossibility results in general markets, and then contrast these with positive results for markets that exhibit natural structures.

\subsection{Our Results}

In \Cref{sec:model}, we introduce a platform economy with \emph{unit-demand} buyers, \emph{unit-capacity} couriers, and \emph{unit-supply} stores.\footnote{Our results extend to buyers with additive valuations and couriers with additive costs.} Each buyer has a \emph{valuation} for each store, which is the sum of the buyer's value for the item offered by the store and the buyer's value for having the item delivered. Each buyer views couriers interchangeably. Couriers incur store--buyer-specific \emph{delivery costs}. Each store sets a fixed \emph{product price}, and has a \emph{production cost}.

An \emph{allocation} is a three-sided, one-to-one-to-one, matching of buyers, stores, and couriers. It specifies a set of pairwise disjoint transactions, in which buyers purchase from stores and receive deliveries from couriers. The \emph{welfare} of an allocation is the sum of allocated buyers' valuations less the sum of both the allocated couriers' delivery costs and the allocated stores' production costs. The \emph{optimal welfare} is the maximum welfare of all allocations.

An equilibrium consists of market clearing purchase prices, delivery compensation and tips, and an (optimal, given the prices, compensation and tips) allocation. In more detail, in an equilibrium, in addition to the allocation:
\begin{itemize}
    \item For each store there is a store-specific \emph{purchase price} (for the item offered by this store), which is weakly greater than the store's fixed product price and is the same price for all buyers, i.e., there is no price discrimination for the same product. If a buyer buys an item from a store, then the buyer pays the purchase price and the store receives its fixed product price.
    \item For each buyer--store pair, there is a buyer--store-specific \emph{delivery compensation} (for delivering an order from the store to the buyer), which is the same for all couriers, i.e., there is no wage discrimination for the same delivery. A courier that delivers a buyer--store order receives its delivery compensation.
    \item For each buyer--store pair, there is also a (possibly zero) buyer--store-specific \emph{tip}. A courier that delivers a buyer--store order gets paid, by the buyer, the tip for this buyer--store pair \emph{in addition} to receiving its delivery compensation. While delivery compensations, like prices, are exogenous to the market participants (``set by the invisible hand'', or in our setting, set by the platform), tips are endogenous outcomes of buyer optimization as explained below.
\end{itemize}

Couriers are modeled as ``delivery compensation + tip'' takers. That is, in equilibrium, each courier delivers a buyer--store order that maximizes the courier's utility, defined as the delivery compensation plus the tip for the order, less the delivery cost for the order, or zero if not delivering any order. Buyers use tips to influence courier behavior, that is, having defined the courier's behavior, in equilibrium every buyer optimizes her tips to maximize her utility, defined as her valuation for the store from which she gets a delivery less both the purchase price for that store and the buyer's tip for that store. Finally, as is standard in the definition of Walrasian equilibrium in two sided-markets, at equilibrium for each store that does not sell, its purchase price and all the tips associated with delivery from the store are zero; and for each buyer--store pair that is not delivered, its delivery compensation is zero.

The platform collects purchase prices from buyers and pays compensation to couriers. We emphasize that rather than requiring budget balance at equilibrium, as has been considered in the multilateral matching-with-contracts literature \citep{hatfield2011multilateral,ostrovsky2008stability}, we model the platform as being able to subsidize some of the delivery cost, allowing the delivery compensation to be larger than the purchase price minus the fixed product price. This has support in practice; e.g., Instacart and DoorDash both use ``Pay Boost'' to guarantee delivery when an order is not accepted for a long time, where the boost is paid for by the delivery platforms and not charged to the buyer. The ``Pay Boost'' on Instacart can be as high as \$12 \citep{InstacartPeakBoost}, while the base-pay for a courier to complete an order is around \$4 \citep{InstacartBasePay}. Indeed, UberEats, DoorDash and Instacart only  turned  profitable as of 2023 and have, in effect, been subsidizing the economic activity on their platforms, while Grubhub is still reporting losses as of 2025.\footnote{Doordash generated positive net income for the first time as a public company in Q3, 2024 \citep{DoorDashPositiveEarnings}; Instacart reached positive net income in Q4, 2023 \citep{InstacartPositiveEarnings}; UberEats is a subsidiary of Uber, which itself turned profitable in Q2, 2023 \citep{UberPositiveEarnings}. Grubhub belongs to its parent company Just Eat Takeaway in 2024, which reported net losses for both 2023 and 2024 \citep{GrubhubNegativeEarnings,GrubhubNegativeEarnings2025}.}

We are interested in understanding whether the platform can implement an equilibrium 
that supports an optimal-welfare allocation. One potential source of inefficiency arises if a store's product price is greater than the product cost, as in this case buyer with valuation larger than product cost but lower than product cost won't be able to buyer, and optimal-welfare allocation may not be achieved. To turn off this source of inefficiency, we restrict attention in this paper to the setting in which the store's fixed product price equals its cost. (Alternatively, one may think of the store's fixed price as the cost.)

In \Cref{sec:motivate_use_tip}, we explore the benefits of allowing buyers to specify tips before delivery, as opposed to not allowing tips, or only allowing tips after delivery. An equilibrium in the first setting is a \emph{with-tip equilibrium} and  an equilibrium in either of the latter two settings is a \emph{without-tip equilibrium}. From the perspective of the incentives of risk-neutral couriers, not allowing tips is equivalent to allowing them only after delivery.\footnote{We expand on this in more detail in \Cref{sec:model}.}

Our first result is a dichotomy: without-tip equilibria are only guaranteed to exist when there are at least as many couriers as buyers or stores, while with-tip equilibria always exist. This is because in the without-tip case, courier shortages leave some buyers without delivery despite willingness to pay; with tips, the platform can set high compensation for some buyers--store pairs, making others require prohibitively high tips for delivery. 

We also show that every allocation supported in a without-tip equilibrium is also supported in a with-tip equilibrium, implying that the optimal with-tip equilibrium welfare is  weakly higher than optimal without-tip equilibrium welfare. We illustrate this with a market instance where all without-tip equilibria are highly inefficient, yet there exists a with-tip equilibrium that achieves the optimal welfare. 
Especially interesting is that the welfare gain from with-tip equilibria arises not from the actual (on-path) payment of tips, but rather from the off-path need for buyers to offer them to secure delivery. In fact, we show that every with-tip equilibrium allocation can be supported by a with-tip equilibrium in which all tips are zero---even on realized deliveries---highlighting that tips serve  to prevent off-path deviations rather than facilitate on-path transfers. The platform can subsidize delivery for buyer--store pairs in equilibrium, allowing buyers to pay zero tips for the equilibrium allocation, while requiring prohibitively high tips to attain delivery outside of the equilibrium allocation.
 
While tips guarantee the existence of equilibria and improve welfare, we show in \Cref{sec:general_markets} that even with-tip equilibria can, in general, lack desirable properties. First, there exist markets where every with-tip equilibrium is inefficient. 
Second, it is NP-hard to calculate the optimal equilibrium welfare. These limitations raise the question as to whether there are natural structural assumptions under which these economic and computational impossibilities can be circumvented. We give a positive answer to this question in \cref{sec:market_structures}. 

In \Cref{sec:market_structures_couriers}, we identify two natural structural assumptions on delivery costs that overcome these impossibilities: either delivery costs are decomposable into courier--store and store--buyer components, which reflects realistic distance-based costs that account for couriers' trips to stores and then trips to buyers; or delivery costs are  decomposable into store--buyer and buyer--courier components, which reflects distance-based costs that account for couriers' trips from stores to buyers, and couriers' return-home trips.
The courier--store component can also account for a courier's familiarity with a store, where grocery pickups can be time-consuming for unfamiliar locations.
In markets that satisfy one of these two delivery cost structures, 
there always exists an efficient with-tip equilibrium that can be found in polynomial time. In contrast, there exist markets with each of these structures in which {\em without-tip} equilibria do not exist, or exist and have low optimal without-tip equilibrium welfare. This underscores the importance of tips, even in these settings with naturally structured delivery costs.

In \Cref{sec:market_structures_buyers}, we instead consider structure on buyer valuations rather than on delivery costs. When buyers are {\em single-minded} (i.e., each buyer has positive valuation for just one store),
we show there exists an efficient with-tip equilibrium that can be found in polynomial time regardless of delivery costs. Again, 
just as in the case with structure on delivery costs, without-tip equilibria are not guaranteed to exist, and if exist can have low welfare. 

While we allow the platform to subsidize delivery costs and set courier compensation above buyer prices to maximize welfare, this need not be profit-maximizing for the platform. 
In \Cref{sec:profit}, we study a profit-maximizing platform, which selects an equilibrium while trading off higher buyer prices against lower courier compensation and smaller platform subsidies on delivery. As in the welfare case, the optimal with-tip equilibrium profit is always weakly higher than the optimal without-tip equilibrium profit. This profit gain comes at a computational cost: computing the optimal with-tip equilibrium profit is NP-hard via a reduction from the vertex cover problem, even in markets where each courier can deliver for only one store. In contrast, in these markets the profit-maximizing without-tip equilibrium can be found in polynomial time. Finally, in \Cref{sec:dis} we discuss some limitations of our model and suggest directions for future work. We present a summary of our major results in a list in \Cref{sec:summary_of_results}.

\subsection{Related Work}
Our work joins several strands of research, including the analysis of competitive equilibrium on trading networks \citep{ostrovsky2008stability,hatfield2013stability} and multilateral contracting \citep{hatfield2011multilateral}. These works extend results on matching and equilibria in two-sided markets \cite{shapley1971assignment,kelso1982job,gul1999walrasian} to richer environments, proving competitive or Walrasian equilibria are efficient and exist with suitable conditions on valuations. However, all  these works either require bilateral contracts, or permit agents to engage in fractional participation within a multilateral contract. In contrast, the matching on delivery platforms is discrete and between three sides.

For discrete, three-sided matching, competitive equilibrium may not exist \citep{alkan1988nonexistence}, and we further show in \Cref{app:ce_existence} that determining the existence of a competitive equilibrium is NP-hard in our setting.\footnote{See also \Cref{app:ce_existence} for a discussion as to whether, and to what extent, each of the first- and second-welfare theorems still hold for competitive equilibria in 3-sided markets.} We therefore move away from a three-sided competitive equilibrium model,  instead allowing the platform to subsidize delivery and abstracting away stores' incentives while still keeping the stores' capacity constraints.
With this notion of equilibrium---which, together with the availability of tips, restores the existence of equilibrium but loses the first welfare theorem---we ask how much welfare can be attained.

Our market structure assumptions in \Cref{sec:market_structures} satisfy local additivity, a sufficient condition for the existence of a core in multi-sided assignment games \citep{quint1991core, stuart1997supplier, atay2016generalized}. In the buyer–middlemen–seller setting, \citet{oishi2014middlemen} and \citet{atay2023matching} define competitive equilibria with price entries for each buyer–middleman–seller triplet and each seller to correspond with the core. By contrast, motivated by delivery platforms, our notion of competitive equilibrium does not allow price or wage discrimination, involving only seller prices and buyer–seller pairs. The main contribution of \Cref{sec:market_structures} is to show that, under our market structure assumptions and with tips, core allocations can indeed be supported by this equilibrium.

Other works in the envy-free pricing literature \citep{guruswami2005profit,chen2014envy} have studied how a seller in a two-sided market can set prices to maximize revenue, including under various supply constraints \citep{cheung2008approximation,im2012envy}. 
As with these works, we require the equilibrium allocation to be envy free for buyers given purchase prices. However, unlike supply constraints in a two-sided market, and hence unlike all of the above works, our three-sided problem also requires envy freeness for couriers, with buyers’ choices needing to align with couriers’ delivery decisions. Additionally, while these earlier works assume a single seller that sets prices and allow for nonzero prices for unsold items, our model differs in requiring unsold stores to have zero purchase prices and orders not delivered to have zero compensation.

We establish the computational hardness of finding an optimal-welfare allocation in a three-sided delivery platform by reducing from the 3-dimensional matching problem. While some works find a matching that achieves half of the maximum weight in weighted 3-dimensional matching in polynomial time ~\citep{arkin1998local,halldorsson1995approximating,chan2009linear}, they rely on local search algorithms that swap subsets of individual matches without considering envy-free requirements, making them unsuitable for our setting. The NP-hardness of finding an optimal-welfare allocation even without equilibrium requirements precludes the possibility of devising an algorithm for finding an efficient, i.e., welfare-optimal,
equilibrium in our setting. Therefore, we focus on identifying suitable structural assumptions on courier costs or on buyer valuations that limit couriers' and buyers' envy, and allow for the existence of equilibria that achieve the optimal welfare.

Previous theoretical works analyze tips paid after service completion and seek to rationalize tipping through repeated interactions \citep{ben1976tip}, social norms \citep{azar2005social,azar2007pay,debo2018tipping,snitkovsky2021modeling}, and altruism \citep{lynn2015service}. In contrast, works analyzing tips before service are scarce. \citet{lei2023two} suggest that delivery platforms may use upfront tips to reduce courier wages under competitive pressure. Our work provides a new perspective, showing that equilibria with tips before service weakly Pareto dominate those without tips or after service. Since buyers in our model account for both the utility impact of tipping and which store they are served by, we offer a novel rationale for why tipping—traditionally a post-service gesture—is specified upfront on delivery platforms. There is also a body of empirical literature on tipping, including studies on how default tip suggestions affect buyer satisfaction \citep{alexander2021effects, haggag2014default} and courier welfare \citep{shy2015tips,castillo2022designing}.

This study also contributes to the broader literature, on how online platforms set prices and facilitate matching in markets with strategic market participants, in three-sided markets \citep{wang2025recommending,liu2023operating,bahrami2023three}, as well as two-sided ones
\citep{platformSim, platformEq, platformDisruption,birge2021optimal,ma2020spatio,feldman2023managing,chen2022food,chen2024courier,randomizedFifo}.  While we share the common feature of a central platform using pricing and matching tools for market design, our work focuses on a different market dimension---the role of tips.

\section{Model}\label{sec:model}
We present a static model, which captures settings with time-sensitive deliveries---buyers who do not receive their food within a short period may no longer value the delivery. A delivery platform facilitates transactions between three sides of a market. There is a set of $m$ unit-demand buyers, $B=\{b_1,...,b_m\}$, $n$ unit-supply stores $S=\{s_1,...,s_n\}$, and $l$ unit-capacity couriers, $D=\{d_1,...,d_l\}$.\footnote{All our results generalize beyond unit-demand to buyers with additive valuations, and beyond unit-capacity to couriers with additive costs. For buyers with additive valuations, duplicate each buyer $n$ times, where the $k$-th duplicate of buyer $i$ (denoted as $v_{b_{i,k}}$) has valuation $v_{b_{i,k}}(s_k)=v_{b_i}(s_k)$ and $v_{b_{i,k}}(s'_k)=0$ for $s'_k\neq s_k$. For couriers with additive costs, duplicate each courier $mn$ times, each duplicate having unit capacity to serve one buyer--store pair, and infinite cost to serve any other buyer--store pairs.} An {\em order} (from a store, not a mathematical order) $o=(b,s)\in B\times S$ is a buyer--store pair, and the set of all orders is denoted as $O=B\times S$. 
Buyer $b$ has {\em valuation} $v_{b}(s)\geq 0$ for store $s\in S$, which includes both the buyer's valuation for the item of store $s$, and the buyer's valuation for the delivery. Courier $d$ incurs finite {\em delivery cost} $c_{d}(b,s)\geq 0$ for delivering from store $s$ to buyer $b$.
Each store $s$ charges the platform a predetermined product price when a buyer buys from it through the platform. We assume this price equals its cost.\footnote{We model product prices as being equal to stores' costs to 
remove the following source of inefficiency: If a buyer's valuation is smaller than the product price but larger than cost, a trade cannot occur despite the positive welfare generated. In this case, the optimal welfare cannot be obtained even without requiring trades occur in an equilibrium.} 
Without loss of generality, we further set each store's cost (and hence the product price) to zero, and show how to generalize all of our results to the case of nonzero store costs, with product prices still equaling store costs, in \Cref{app:zero_store_cost}. 

We now  introduce some of the components of an equilibrium, which include purchase prices, delivery compensation, and later we will also introduce tips.
For each store $s\in S$, there is a {\em purchase price}, $p_{s}\geq 0$, which includes both the predetermined product price (assumed equal to a store's cost, which is set to zero) and the platform delivery fee. We require the purchase price for a store being the same for all buyers, i.e., no price discrimination. We abstract away from discriminating purchase prices based on a buyer's distance from the store, as such distance-based pricing is often coarse; for example, Instacart applies a long-distance service fee only if the delivery route exceeds 30 minutes or includes a toll \citep{InstacartDistanceBasePrice}. That said, platforms do charge stores varying fees for fulfilling orders to buyers at different distances.\footnote{For example, DoorDash charges restaurants and partners who use the DoorDash DriveAPI distance-based fees for delivery, but these fees are not charged to buyers \citep{DoorDashDriverAPI,DoorDashDeliveryRadius}. On Grubhub and UberEats, individual stores can choose their delivery plans from Basic, Plus, or Premium, each of which comes with different delivery fees that all buyers ordering from the store has to pay. \citep{GrubhubDeliveryRadius,UberEatDeliveryRadius}.}

There is also a {\em delivery compensation}, $w_{b s}\geq 0$, for each order $(b,s)$, the same for every courier, i.e., there is no wage discrimination for the same delivery. Let $\mathbf{p}\in R^m$ and $\mathbf{w}\in R^{mn}$ denote the vector of purchase prices and delivery compensation. The platform charges purchase prices from buyers and pays delivery compensation to couriers. The platform can
subsidize some of the cost of delivery, allowing delivery compensation $w_{bs}$ to a courier to be larger than purchase price $p_s$ collected
from a buyer. This reflects the ``Pay Boost'' adopted by some delivery platforms to guarantee delivery when an order is not accepted for a long time. For example, pay boost on Instacart can be as high as \$12 when the base pay per order for a courier is only around \$4 \citep{InstacartPeakBoost,InstacartBasePay}. The source of subsidy $w_{bs}-p_s$ can come from membership fees or other fixed fees that are not charged per order.

\paragraph{Without-tip.} The utility associated with a trade $(b,s,d)$ is as follows: buyer $b$ has utility $u_b(s)=v_b(s)-p_s$; courier $d$ has utility $u_d(b, s)=w_{bs}-c_d(b,s)$; the platform has utility that equals the purchase price $p_s$ minus the courier compensation $p_{bs}$; a store has utility zero. The welfare created by a trade is the sum of the four utilities, which adds up to $v_b(s)-c_d(b, s)$.

An allocation, or a three-sided matching $\mathbf{x}$, is defined as  
\begin{equation*}
    x_{bsd} =
    \begin{cases}
      1 & \text{if $b$ buys from $s$, and $d$ serves the delivery,}\\
      0 & \text{otherwise.}
    \end{cases}       
\end{equation*}

An allocation  $\mathbf{x}$ is feasible if it satisfies unit-demand $\forall b, \sum_{sd}x_{bsd}\leq 1$, unit-supply $\forall s, \sum_{bd}x_{bsd}\leq 1$ and unit-capacity $\forall d, \sum_{bs}x_{bsd}\leq 1$. For an allocation $\mathbf{x}$, let $s_\mathbf{x}(b)$ be the store buyer $b$ buys from, $d_\mathbf{x}(b)$ be the courier that delivers for buyer $b$, and $o_\mathbf{x}(d)=(b_\mathbf{x}(d),s_\mathbf{x}(d))$ the buyer--store pair that courier $d$ delivers.
If $b$ does not purchase, let $s_\mathbf{x}(b)=\emptyset$ be a null store, and $d_\mathbf{x}(b)=\emptyset$ be a null courier. If $d$ does not deliver, then let $o_\mathbf{x}(d)=\emptyset$ be a null buyer--store pair. When the context is clear, we use $u_b(\mathbf{x})=u_b(s_\mathbf{x}(b)),u_d(\mathbf{x})=u_d(o_\mathbf{x}(d))$ as short hand for the utility of a buyer and courier, respectively, in allocation $\mathbf{x}$. A buyer who does not buy and a courier who does not deliver have zero utility. All values, costs and utilities associated with null are zero. 

The welfare of an allocation $\mathbf{x}$ is the sum of the welfare of all realized trades, and equals the sum of allocated buyers’ valuations minus the sum of allocated couriers' delivery costs: 
$$W(\mathbf{x})=\sum_{bsd}x_{bsd}(v_b(s)-c_d(b, s)).$$ 
The optimal welfare is the maximum welfare over all feasible allocations, $\mathit{OPT}=\max_{\mathbf{x}}W(\mathbf{x})$.

Given delivery compensation $\mathbf{w}$, denote the set of orders that maximize courier $d$'s utility as
\begin{equation*}
    \BR_d(\mathbf{w})=
    \begin{cases}
      \{(b,s)\;|\; \forall (b',s')\in O, u_d(b,s)\geq u_d(b',s')\} & \text{ if } \exists (b,s) \text{ such that } u_d(b,s)>0\\
      \{\emptyset\}\cup \{(b,s)\;|\; u_d(b,s)=0\} & \text{otherwise.}
    \end{cases}       
\end{equation*}
Given purchase price $\mathbf{p}$, denote the set of stores that maximize buyer $b$'s utility as 
\begin{equation*}
    \BR_b(\mathbf{p})=
    \begin{cases}
      \{s\;|\; \forall s'\in S, u_b(s)\geq u_b(s')\} & \text{ if } \exists s \text{ such that } u_b(s)>0\\
      \{\emptyset\}\cup \{s\;|\; u_b(s)=0\} & \text{otherwise.}
    \end{cases}       
\end{equation*}

\begin{definition}[Without-tip equilibrium]\label{def:withouttipequil}
    A without-tip equilibrium is a triple of price, compensation, and feasible allocation $(\mathbf{p},\mathbf{w},\mathbf{x})$ that satisfies \begin{itemize} 
    \item Every buyer buys from the favorite store $\forall b, \; s_\mathbf{x}(b)\in \BR_b(\mathbf{p})$.
    \item Every courier delivers the favorite order $\forall d, \; o_\mathbf{x}(d) \in \BR_d(\mathbf{w})$.
    \item Stores not bought from (i.e., $\sum_{bd}x_{bsd}=0$) have zero purchase price $p_s=0$.
    \item Order not delivered (i.e., $\sum_d x_{bsd}=0$) have zero delivery compensation $w_{bs}=0$.
\end{itemize}
\end{definition}
From the perspective of courier decision-making, the without-tip equilibrium also captures scenarios in which buyers specify tips after delivery, as such tips would have no or minimal influence on courier behavior.\footnote{One may consider the case where risk-neutral couriers hold a calibrated belief, that $\alpha$ percentage of purchase price is specified as tip after delivery. Couriers might favor stores with higher purchase price in the hope of a higher tip. This case can also be captured by a without-tip equilibrium, where purchase prices for buyers are increased by $\alpha$ percent, and courier compensation for any delivered order $(b,s)$ is also increased by $\alpha p_s$.}

The equilibrium definition assumes that couriers can observe all available orders, whereas real-world platforms typically use centralized matching systems that limit courier choices. We justify this assumption with two reasons. First, on platforms like UberEats and Instacart, couriers often see a range of nearby orders rather than just a single platform-recommended option.\footnote{See Footnote~4.} 
In most cases, the order that maximizes a courier's utility is one that begins close to the courier's current location. It is reasonable therefore, that we model couriers as choosing the choices among all choices that maximizes the utility. Second, modeling couriers as facing competitive prices across all orders guarantees that they are envy-free and always accept the platform-dispatched order, assuming the platform assigns each courier their most preferred available order.

It will also be useful to define a \emph{Walrasian equilibrium} for the two-sided market of buyers and stores $(B,S)$, which completely ignores couriers. We will use this concept to analyze couriers' and buyers' incentives in a three-sided allocation in \Cref{sec:eq_existence_sec} and \Cref{sec:market_structures}.
A \emph{buyer allocation} $\mathbf{z}$ is a two-sided matching in $(B,S)$ where 
\begin{equation*}
    z_{bs} =
    \begin{cases}
      1 & \text{if $b$ buys from store $s$,}\\
      0 & \text{otherwise.}
    \end{cases}       
\end{equation*}

A feasible buyer allocation $\mathbf{z}$ satisfies unit-demand buyer $\forall b, \sum_s z_{bs}\leq 1$ and unit-supply store $\forall s, \sum_b z_{bs}\leq 1$. Let $s_\mathbf{z}(b)$ be the store that buyer $b$ buys from, and $s_\mathbf{z}(b)=\emptyset$ if $b$ does not buy. A Walrasian equilibrium for $(B,S)$ is a pair of purchase prices and buyer allocation $(\mathbf{p},\mathbf{z})$, such that all buyers buy from the store that maximizes buyers' utility $s_\mathbf{z}(x)\in \BR_b(\mathbf{p})$, with purchase prices set to zero for stores that do not sell. A Walrasian equilibrium $(\mathbf{p},\mathbf{z})$ always exists in this unit-demand-supply setting \citep{gul1999walrasian}.\footnote{Although our definition of a without-tip equilibrium resembles a Walrasian equilibrium in two-sided markets, we do not use the term Walrasian since our model abstracts away the stores' incentives, and allows the platform to subsidize delivery. See \Cref{app:ce_existence} for a discussion of store incentives and \Cref{sec:dis} for a discussion of the assumption that the platform can subsidize delivery.}

\paragraph{With-tip.} When buyers are allowed to tip, let $t_{bs}\geq 0$ be the tip associated with an order $o=(b,s)$. Buyer $b$ pays $t_{bs}$ to the courier who delivers the order $(b,s)$, but does not pay out tips for orders that are not delivered. Let $\mathbf{t}\in R^{mn}$ be the vector of tips, $\mathbf{t}_b\in R^{n}$ be the vector of tips associated with buyer $b$, and $\mathbf{t}_{-b}\in R^{(m-1)n}$ the tips associated with all buyers except $b$. Buyer $b$ buying from store $s$ with tip $t_{bs}$ has utility $u_b(s)=v_b(s)-p_s-t_{bs}$.
courier $d$ delivering an order $(b,s)$ has utility $u_d(b,s)=p_{bs}-c_d(b,s)+t_{bs}$. The utility of the platform and stores, as well as the definition of an allocation and welfare remain the same as in the without-tip setting. 

Given the set of delivery compensation $\mathbf{w}$ and tips $\mathbf{t}$, the set of orders that maximize courier $d$'s utility is still defined as
\begin{equation*}
    \BR_d(\mathbf{w},\mathbf{t})=
    \begin{cases}
      \{(b,s)\;|\; \forall (b',s')\in O, u_d(b,s)\geq u_d(b',s')\} & \text{ if } \exists (b,s) \text{ such that } u_d(b,s)>0\\
      \{\emptyset\}\cup \{(b,s)\;|\; u_d(b,s)=0\} & \text{otherwise.}
    \end{cases}       
\end{equation*}

Since in online platforms buyers typically decide the tip amount, defining the set of stores that maximize a buyer's utility is more complex. Let a buyer $b$ consider buying from a store $s$. Given compensation $\mathbf{w}$ and all other buyers' tips $\mathbf{t}_{-b}$, 
there is a minimum tip required to have some courier deliver from the store $s$ to $b$, denoted as $$\underline{t}_{bs}(\mathbf{w},\mathbf{t}_{-b}) = \min\{t_{bs} \;|\; \exists d \text{ such that } (b,s)\in \BR_d(\mathbf{w},\mathbf{t}_b,\mathbf{t}_{-b}).\}$$

This minimum tip $\underline{t}_{bs}(\mathbf{w},\mathbf{t}_{-b})$ has a closed form solution, which we present in \Cref{app:min_tip}. For simplicity of notation, we omit the dependency on $\mathbf{w},\mathbf{t}_{-b}$ and write $\underline{t}_{bs}$ from now on. The buyer can calculate the minimum tip $\underline{t}_{bs'}$ for any store $s'\in S$. 
Given this, the set of stores that maximize buyer $b$'s utility is defined as
\begin{equation*}
    \BR_b(\mathbf{p},\mathbf{w},\mathbf{t}_{-b})=
    \begin{cases}
        \{s\;|\;\forall s'\in S, v_b(s)-p_s-\underline{t}_{bs}\geq v_b(s')-p_{s'}-\underline{t}_{bs'} \} & \text{ if } \exists s \text{ such that } v_b(s)-p_s-\underline{t}_{bs}> 0\\
        \{\emptyset\}\cup \{s\;|\; v_b(s)-p_s-\underline{t}_{bs}=0\} & \text{otherwise.}
    \end{cases}
\end{equation*}
By the definition, a buyer $b$ views $v_b(s)-p_s-\underline{t}_{bs}$ as the  utility of buying from a store $s$, and similarly $v_b(s')-p_{s'}-\underline{t}_{bs'}$ as the  utility of buying from store $s'$. The model of utility used in this definition implies every buyer is optimistic in believing that she can get the item even if there are other buyers who contend for the same store $s$.

\begin{definition}[With-tip equilibrium]\label{def:with_tip_eq}
A with-tip equilibrium is a quadruple of price, compensation, tips, and feasible allocation $(\mathbf{p},\mathbf{w},\mathbf{t},\mathbf{x})$ that satisfies
\begin{itemize}
    \item Every buyer buys from the favorite store $\forall b, \; s_\mathbf{x}(b)\in \BR_b(\mathbf{p},\mathbf{w},\mathbf{t}_{-b})$; and pays the minimum tip if buying from a store, $\forall b, s_\mathbf{x}(b)\neq \emptyset, t_{bs_\mathbf{x}(b)}=\underline{t}_{bs_\mathbf{x}(b)}$.
    \item Every courier delivers the favorite order $\forall d, \; o_\mathbf{x}(d) \in \BR_d(\mathbf{w},\mathbf{t})$.
    \item Stores not bought from (i.e., $\sum_{bd}x_{bsd}=0$) have zero purchase price $p_s=0$.
    \item Order not delivered (i.e., $\sum_d x_{bsd}=0$) have zero delivery compensation $w_{bs}=0$ and zero tips $t_{bs}$=0.
\end{itemize}
\end{definition}

This definition of a with-tip equilibrium allows buyers to reason in a counterfactual way. Given tip vector $\mathbf{t}_{-b}$, a buyer $b$ can infer the minimum tip required to secure delivery from stores, reflecting that buyers adjust tips in response to delivery conditions and courier supply. \footnote{This perspective aligns with the reported views from \citet{BuyerCompete}, which describes how ``demand for delivery pushed buyers to unknowingly compete against each other, with a willingness to pay more for delivery and more in tips to get their food delivered.''} Although the model permits buyers to specify different tips across stores, in equilibrium each buyer places at most one non-zero tip. Finally, we emphasize that tips are not the only mechanism through which the market clears. Instead, the tips work alongside the purchase prices and delivery compensation.

\begin{remark} 
An alternative, simpler definition of a with-tip equilibrium, which does not capture the ability of  buyers to set the tips is: let $u_b(s)=v_b(s)-p_s-t_{bs}$, and define
\begin{equation*}
    \BR_b(\mathbf{p},\mathbf{w},\mathbf{t})=
    \begin{cases}
        \{s\;|\;\forall s'\in S, u_b(s)\geq u_b(s')\} & \text{ if } \exists s \text{ such that } u_b(s)> 0\\
        \{\emptyset\}\cup \{s\;|\; u_b(s)=0\} & \text{otherwise.}
    \end{cases}
\end{equation*}

The definition would otherwise remain the same as \Cref{def:with_tip_eq}. In this alternative equilibrium definition, tips are a given amount that buyers must pay to buy from a store, rather than the minimum amount that a buyer could offer in incentivizing delivery. As such, this alternative is not economically sensible, removing buyer agency in setting tips. 
\end{remark}

We illustrate Definition~\ref{def:withouttipequil} and Definition~\ref{def:with_tip_eq} of with-tip and without-tip equilibria, respectively, with a market illustrated by \Cref{example:without_tip_bad} and \Cref{fig:without_tip_bad}. This is also a market where with-tip equilibria achieve optimal welfare but all without-tip equilibria have negative welfare.

\section{Motivating the Use of Tips}\label{sec:motivate_use_tip}
In this section, we highlight the benefits of incorporating tips into pricing. We show in \Cref{sec:eq_existence_sec} that while with-tip equilibria always exist, the existence of without-tip equilibria require a sufficient number of couriers.
In \Cref{sec:role_tips}, we show every allocation in a without-tip equilibrium is also in a with-tip equilibrium, meaning the existence of a with-tip equilibrium with weakly larger welfare than all without-tip equilibria.
Lastly, we demonstrate that when tipping is allowed, every equilibrium allocation is also in an equilibrium where all tips equal zero, pinpointing the role of tips as preventing deviations off path rather than for on-path transfers.

\subsection{Equilibrium Existence}\label{sec:eq_existence_sec}
In this section, we examine the existence of equilibrium in the without-tip and with-tip regime. We show there always exists a with-tip equilibrium. In contrast, a without-tip equilibrium is only guaranteed to exist when the number of couriers is weakly larger than the minimum number of buyers and stores $l\geq \min\{m,n\}$.

As a high-level intuition, a without-tip equilibrium requires that the two-sided market of buyers and stores forms a Walrasian equilibrium. When $l<\min\{m,n\}$, there can be insufficient number of couriers to deliver all orders in this walrasian equilibrium; e.g., the market in \Cref{fig:without_tip_not_exists} has $l=1<m=n=2$, and no without-tip equilibrium exists. To explain it in another way, courier shortage leads to some buyers unable to secure delivery despite their high willingness to pay. However, in the with-tip regime platform can heavily subsidize delivery, so that buyers will have to tip a very high amount should they buy from a store that does not sell.

To analyze equilibrium existence, we start from the courier side and look into a two-sided matching consisting of all orders and couriers. \Cref{lem:exist_courier_plan_serve,lem:exist_courier_plan_serve_equal_size} show that when there are enough couriers, there is always a delivery compensation that incentivizes couriers to deliver any subset of orders. Consequently, checking for equilibrium existence in the three-sided market reduces to checking for equilibrium existence in the two-sided buyer-store market.
\Cref{lem:exist_courier_plan_serve} also gives a tight upper bound on courier utility when delivering a subset of orders, which will be crucial in \Cref{sec:general_markets} to test whether an allocation is supported in some equilibrium.

We need more notations to describe the two-sided market of the orders and couriers. Let $G_D=(O,D)$ be a bipartite graph where each edge $(o,d)$ has cost $c_d(o)$. A \emph{courier allocation} $\mathbf{y}$ is a two-sided matching in $G_D$ where 
\begin{equation*}
    y_{od} =
    \begin{cases}
      1 & \text{if $d$ delivers the order $o$,}\\
      0 & \text{otherwise.}
    \end{cases}       
\end{equation*}

A feasible courier allocation $\mathbf{y}$ satisfies unit-capacity courier $\forall d, \sum_o y_{o,d}\leq 1$, unit-demand buyers and unit-supply store (i.e., $\forall o=(b,s),o'=(b',s') \text{ such that } y_{od}=y_{o'd}=1, \text{ it satisfies that } b\neq b, s\neq s'$). Denote $o_\mathbf{y}(d)$ as the order that courier $d$ delivers. If $d$ does not deliver, $o_\mathbf{y}(d)=\emptyset$. A \emph{courier plan} is a pair of delivery compensation and feasible courier allocation $(\mathbf{w},\mathbf{y})$ such that every courier delivers their most preferred order; i.e., $\forall d, o_{\mathbf{y}}(d)\in \BR_d(\mathbf{w})$.

Consider any subset of orders $\Omega\subset O$ where each buyer and store appears at most once.
A courier plan $(\mathbf{w},\mathbf{y})$ \emph{serves} $\Omega$ if all orders in $\Omega$ are delivered and only orders in $\Omega$ are delivered; i.e., $\forall o\in \Omega, \sum_{d}y_{od}=1 \text{ and } \forall o\notin \Omega, \sum_{d}y_{od}=0$. 

We now formalize the claim that, when there are sufficiently many couriers, one can always chose delivery compensation to induce couriers to deliver any subset of orders. We consider separately the cases in which the number of couriers equals the number of orders and in which it strictly exceeds it, since the couriers' maximal utility differs across these two regimes.
\begin{restatable}{lemma}{existscourierPlan}
\label{lem:exist_courier_plan_serve}
    Consider the case where $|\Omega|< l$. Any courier plan $(\mathbf{w},\mathbf{y})$ that serves $\Omega$ satisfies (i) $\mathbf{y}$ is the minimum-cost matching that covers $\Omega$ in $G_D$; and (ii) courier utility is upper bounded by $$u_d(o_\mathbf{y}(d))\leq \bar{u}_d:=C_{\Omega}(G_D\backslash d)-C_{\Omega}(G_D)$$ where $C_{\Omega}(G_D)$ is the cost of the minimum cost matching that covers $\Omega$ in $G_D$, and $C_{\Omega}(G_D\backslash d)$ is the cost of the minimum-cost matching that covers $\Omega$ without courier $d$. There always exists a courier plan $(\bar{\mathbf{w}},\mathbf{y})$ that serves $\Omega$ and achieves $\bar{u}_d$, where $\bar{w}_{bs}=0$ for $(b,s)\notin \Omega$. 
\end{restatable}

\begin{restatable}{lemma}{existscourierPlanEqualSize}
\label{lem:exist_courier_plan_serve_equal_size}
    Consider the case where $|\Omega|=l$. Any courier plan $(\mathbf{w},\mathbf{y})$ that serves $\Omega$ satisfies that $\mathbf{y}$ is the minimum-cost matching that covers $\Omega$ in $G_D$. There always exists a courier plan $(\bar{\mathbf{w}},\mathbf{y})$ that serves $\Omega$ where $\bar{w}_{bs}=0$ for $(b,s)\notin \Omega$, and each courier has utility $$\bar{u}_d=H+ C_{\Omega_{-1}}(G_D\backslash d)-C_{\Omega}(G_D)\geq \max_{b,s}\{v_b(s)\},$$ where $H=\sum_{bs}v_b(s)+\sum_{bsd}c_d(b,s)$, while $C_{\Omega_{-1}}(G_D\backslash d)$ is the cost of the minimum-cost matching that covers all but one order in $\Omega$, and $C_{\Omega}(G_D)$ is the cost of the minimum-cost matching that covers $\Omega$.
\end{restatable}

The proof of these two lemmas establishes a one-to-one relationship between courier plans that serve $\Omega$ and Walrasian equilibria in a two-sided order-courier market $M$ where one side is order $\Omega$ and another side is courier $D$.
As a Walrasian equilibrium always exists in a two-sided market $M$, a courier plan that serves $\Omega$ always exists. By further appealing to the largest Walrasian price in $M$, we are able to write down the exact form of the maximum courier utility $\bar{u}$ in a courier plan that serves $\Omega$, and the corresponding maximum courier compensation $\bar{\mathbf{w}}$. $\bar{u}$ and $\bar{\mathbf{w}}$ play a crucial role in testing if an allocation is in some with-tip equilibrium in \Cref{sec:general_markets}. Using these two lemmas, we show when there are sufficiently many couriers $l\geq\min\{m,n\}$, a without-tip equilibrium exists. 
\begin{restatable}{theorem}{WithoutTipEqExistence}
\label{thm:without_tip_eq_existence}
    In a market where the number of couriers is weakly larger than the minimum number of buyers and stores $l\geq \min\{m,n\}$, a without-tip equilibrium always exists.
\end{restatable}
The intuition is straightforward. Take any set of orders that arises in a Walrasian equilibrium of the two-sided buyer-store market. By definition, there exists buyer prices that make each buyer choose her preferred store consistent with that set of orders. Moreover, by the two lemmas above, there is a delivery compensation that incentives couriers to delivery exactly the set of orders. Now we discuss the existence of with-tip equilibrium.

\begin{restatable}{theorem}{WithTipEqExistence}
\label{thm:with_tip_eq_existence}
    There always exists a with-tip equilibrium.
\end{restatable}
When $l\geq\min\{m,n\}$, the argument for the existence of a with-tip equilibrium mirrors \Cref{thm:without_tip_eq_existence}. When $l<\min\{m,n\}$, the platform can restrict attention to a subset of orders whose size equals $l$. The platform can subsidize delivery and choose delivery compensation to attain the courier utility in \Cref{lem:exist_courier_plan_serve_equal_size}. No buyer has a profitable deviation from the prescribed set of orders because any deviation would require paying a tip exceeding the buyer’s own valuation.

Although equilibria always exists for with-tip equilibrium, they can be inefficient. We will discuss the welfare aspect of equilibrium in  \Cref{sec:general_markets} and \Cref{sec:market_structures}.

\subsection{Tips Weakly Improve Welfare}\label{sec:role_tips}
We now further illustrate the role of tips in pricing. Recall  \Cref{example:without_tip_bad} where the only without-tip equilibrium allocation is also supported in a with-tip equilibrium. We show this is not a coincidence.
\begin{restatable}{proposition}{paretoDominant}
\label{lem:pareto_dominant}
    For a market, if $(\mathbf{p},\mathbf{w},\mathbf{x})$ is a without-tip equilibrium, then $(\mathbf{p},\mathbf{w},\mathbf{t},\mathbf{x})$ is a with-tip equilibrium where $\mathbf{t}=0$. 
\end{restatable}
To prove the proposition, the set of orders that maximize courier utility remains the same in $(\mathbf{p},\mathbf{w},\mathbf{t},\mathbf{x})$ as that in $(\mathbf{p},\mathbf{w},\mathbf{x})$. A buyer $b$ who buys from store $s_\mathbf{x}(b)$ does not need to pay tips to get delivery from $s_\mathbf{x}(b)$, i.e., $\underline{t}_{bs_\mathbf{x}(b)}=0$. But $b$ does need to pay tips weakly larger than zero to get delivery for some other stores $s'\neq s_\mathbf{x}(b)$. So $s_\mathbf{x}(b)$ is still $b$'s favorite store.

This proposition implies that with-tip equilibria weakly Pareto dominate without-tip equilibria. From a welfare perspective, the benefit of including tips in pricing is made even more apparent when we do not require an equilibrium to clear the market; i.e., unsold stores can have positive prices and undelivered orders can have positive compensation and tips. In \Cref{app:tips_eq_not_clear_market}, we prove that when equilibria do not need to clear the market, the optimal with-tip equilibrium welfare is always lower bounded above zero while the optimal without-tip equilibrium welfare can be zero.

While it may seem intuitive that the added flexibility of tips leads to higher welfare, we show next that the welfare gain does not come from the actual (on-path) payment of tips, but rather from the off-path need of buyers to offer tips to secure delivery, pinpointing the role of tips as preventing deviations off path rather than for on-path transfers.
\begin{restatable}{proposition}{ZeroTipEquivalence}
\label{lem:zero_tip_equivalence}
    If there exists a with-tip equilibrium $(\mathbf{p},\mathbf{w},\mathbf{t},\mathbf{x})$, then there exists a with-tip equilibrium $(\mathbf{p}',\mathbf{w}',\mathbf{t}',\mathbf{x})$ with zero tips $\mathbf{t}'=0$.
\end{restatable}
To prove this, we construct $\mathbf{p}',\mathbf{w}'$ in a way that directly incorporates tips $\mathbf{t}$ into the purchase prices and delivery compensation. 
The crux of this proof is to show that the minimum tip for a buyer $b$ to get delivery from a store $s\neq s_\mathbf{x}(b)$ is the same between $(\mathbf{p},\mathbf{w},\mathbf{t},\mathbf{x})$ and $(\mathbf{p}',\mathbf{w}',\mathbf{t}',\mathbf{x})$. With the same minimum tip, we can prove that $s_\mathbf{x}(b)$ is still in the set of stores that maximize buyer utility.
The proposition will play a crucial role in \Cref{sec:general_markets} to determine whether an allocation is supported in some with-tip equilibrium.

Together, \Cref{lem:pareto_dominant,lem:zero_tip_equivalence} provide an interpretation of the role of tips on delivery platforms. In the with-tip equilibrium, the platform can set prices and compensation that require buyers to pay zero tips in equilibrium, but face very high tips if they deviate. In this sense, tips facilitate market-clearing by discouraging buyers from deviating from the equilibrium outcome.

\begin{remark}[Tips in Two-Sided Platforms]\label{rem:two_sided}
Ride-hailing platforms are two-sided, matching riders with drivers. Because Walrasian equilibria always exist in two-sided markets, there are non-discriminatory rider prices and driver compensations that balance rider demand with driver supply. Moreover, since Walrasian equilibria are welfare-maximizing, these prices and compensations implement the socially efficient outcome. This suggests that allowing tips before delivery would be of limited value for social welfare in such two-sided market. And indeed, tips are specified post-ride on ride-hailing platforms.

As three-sided food-delivery platforms like UberEats begin to adopt robot delivery alongside courier delivery\footnote{As of mid-2025, UberEats provides a robot-delivery option in Los Angeles, Miami, the Dallas–Fort Worth area, and Atlanta \citep{UberEatRobotDelivery}.}, the delivery platform becomes a mixture of a two-sided (robot-mediated) market and a three-sided (courier-mediated) market, with the relative importance of the latter shrinking as robots take over a larger share of deliveries. In such settings, tipping before delivery is no longer as valuable for improving welfare as it is in a three-sided market. And indeed, these delivery robots do not receive tips \citep{UberEatRobotDeliveryNoTip}.
\end{remark}

\section{Markets without any Structural Constraints}\label{sec:general_markets}
In this section, we focus on the equilibrium welfare properties. For simplicity, we use ``equilibrium'' to refer to a with-tip equilibrium and ``without-tip equilibrium'' to denote the alternative. 
We begin by presenting, in \Cref{prop:negative_opt_welfare} and \Cref{fig:market_clearing}, a market in which every equilibrium attains strictly lower welfare than the optimal welfare. As a second negative result, we show that computing the optimal equilibrium welfare, as well as the optimal welfare over all feasible allocations, is NP-hard. In light of these results, we turn in \Cref{sec:char_eq_alloc} to a characterization of equilibrium allocations, and show that for any given allocation, one can determine in polynomial time whether it can be supported by some equilibrium. This will pave the road for \Cref{sec:market_structures} to identify markets where optimal equilibrium welfare equals to optimal welfare.

The following theorem shows that the computation problem associated with finding the maximum welfare equilibrium is intractable.
\begin{restatable}{theorem}{OPTEqHard}
\label{lem:OPT_eq_hard}
    In a market where all buyers' valuations equal 1, and all couriers' costs lie in $\{0,1\}$, it is NP-hard to determine 
    \begin{itemize}
        \item whether the optimal with-tip equilibrium welfare exceeds $k$
        \item whether the optimal without-tip equilibrium welfare exceeds $k$
        \item whether the optimal welfare exceeds $k$
    \end{itemize}
\end{restatable}
We prove NP-hardness through a reduction from 3-dimensional matching (3DM). Given a 3-uniform hypergraph $G=(B,S,D)$ with $|B|=|S|=|D|=l$, we construct a market with $l$ buyers, $l$ stores, and $l$ couriers. A courier's cost for delivering for a buyer-store pair is zero if and only if there is a hyper-edge in $G$ that corresponds to the buyer-store-courier triple. We then show that $G$ admits a perfect matching if and only if the optimal welfare, optimal with-tip equilibrium welfare, and without-tip equilibrium welfare all equal to $l$.

Seeing the computational complexity, one natural direction is to extend the approximation results in the maximum-weight version of 3DM to our setting, and develop a polynomial-time algorithm that finds equilibria with good welfare. The obstacle is that most approximation results for maximum-weight 3DM rely on local-search algorithms that swap subsets of individual matches. These local-search algorithms do not consider buyers and couriers' incentives, and the swaps break the equilibrium requirements, making them unsuitable for our economic setting. Therefore, we turn our attention to finding markets where optimal equilibrium welfare equals to optimal welfare. To do that, we will first need to characterize the set of equilibrium allocations.

\subsection{Characterizing Equilibrium Allocations}\label{sec:char_eq_alloc}
Given a feasible allocation $\mathbf{x}$, we ask whether there is an efficient way to determine if there is an equilibrium $(\mathbf{p},\mathbf{w},\mathbf{t},\mathbf{x})$ for some $\mathbf{p},\mathbf{w},\mathbf{t}$. Through this section, we answer this question affirmatively, by first characterizing the courier compensation and tips that might support $\mathbf{x}$ in \Cref{lem:highest_courier_price_eq}, then checking whether $\mathbf{x}$ is in an equilibrium in a constructed two-sided market in \Cref{lem:test_alloc}.
\begin{restatable}{lemma}{highestcourierPriceEq}
\label{lem:highest_courier_price_eq}
    If a feasible allocation $\mathbf{x}$ is in some equilibrium $(\p,\w,\bt,\x)$ where $\bt=0$, then $(\p,\bar{\w},\bt,\x)$ is also an equilibrium where $\bar{\w}$ is the highest courier compensation defined in \Cref{lem:exist_courier_plan_serve} and \Cref{lem:exist_courier_plan_serve_equal_size}.
\end{restatable}
This lemmas shows that any with-tip equilibrium with zero tips remains an equilibrium if the platform subsidizes delivery and pays couriers the maximum compensation $\bar\w$ that incentivizes the courier to deliver the same set of orders. Courier incentives hold by definition of $\bar\w$. A buyer $b$ deviating to a store $s\neq s_\x(b)$ needs to pay at least tip $\ut_{bs}$ to have some courier willing to deliver $$\ut_{bs}=\min_d c_d(b,s)+\bar{u}_d$$ where $\bar{u}_d$ is the highest possible courier utility that is achieved by $\bar\w$ defined in \Cref{lem:exist_courier_plan_serve} and \Cref{lem:exist_courier_plan_serve_equal_size}. In contrast, if a buyer $b$ buys from $s_\x(b)$, she doesn't need to tip because $\bar\w$ guarantees some courier will deliver. Thus, in $(\p,\bar\w,\bt,\x)$, buyer incentives hold because $\ut_{bs}$ is higher than the minimum tip required for deviation in $(\p,\w,\bt,\x)$. If a buyer doesn't deviate with lower tips, she doesn't with higher tips.

Combining \Cref{lem:zero_tip_equivalence} and \Cref{lem:highest_courier_price_eq}, to check if an allocation $\mathbf{x}$ is in some equilibrium, it suffices to check for an equilibrium where $\bt=0$ and delivery compensation equals to $\bar{\w}$. We now check if $\x$ is in some equilibrium by ``folding'' the tip and compensation into a buyer-store two-sided market.

For an allocation $\x$, define a buyer allocation $\mathbf{z}$ that allocates stores to buyers to be $z_{bs}=\sum_d x_{bsd}$. Let $G_\x=(B,S)$ be the bipartite graph where one side are buyers, and the other side are stores. Each edge $(b,s)$ in $G_\x$ has weight 
\begin{equation*}
    v_{bs}^\x = \begin{cases}
        v_b(s) - \ut_{bs} & \text{ if } s \neq s_\x(b),\\
        v_b(s) & \text{otherwise.}
    \end{cases} 
\end{equation*}
Here, we  use $v^\x_{bs}$ to denote the edge weight in $G_\x$, because this edge weight corresponds to a buyer $b$'s realized value when buying from store $s$, after adjusting for the tips $\ut_{bs}$ she pays for the delivery.

\begin{restatable}{lemma}{testAlloc}
\label{lem:test_alloc}
    A feasible allocation $\mathbf{x}$ is in some equilibrium if and only if $\mathbf{z}$ is a maximum weight matching in $G_\x$. 
\end{restatable}
The proof for this lemma shows that $\mathbf{x}$ is an equilibrium if and only if $\mathbf{z}$ is a Walrasian equilibrium allocation in the two-sided market $G_x$, where buyer valuations for stores are adjusted by tips. An immediate corollary is that any feasible allocation $\x$ where all couriers make a delivery is supported in some equilibrium. This is because the platform can subsidize couriers, so that the minimum tip $\ut_{bs}$ is much higher than buyer $b$'s valuations for $s\neq s_\x(b)$. We state this \Cref{cor:all_courier_deliver} in \Cref{app:char_eq_alloc}. 
We are ready to state the main theorem for this section.
\begin{restatable}{theorem}{PolyCheck}
\label{thm:poly_check}
    For any allocation $\mathbf{x}$, there is a polynomial time procedure to determine if there exists an equilibrium that includes $\mathbf{x}$.
\end{restatable}
For any allocation $\x$, the procedure calculates $ \ut_{bs}$ for every pair of $(b,s)$ that $s\neq s_\x (b)$, and find the max weight matching in $G_\x$. As $\ut_{bs}$ depends on $\bar{u}_d$, which can all be computed in polynomial time according to \Cref{lem:exist_courier_plan_serve} and \Cref{lem:exist_courier_plan_serve_equal_size}, the entire procedure takes polynomial time.

\section{Markets with Structures}\label{sec:market_structures}
While the previous section demonstrates that all equilibria can be inefficient and it is computationally intractable to find the optimal equilibrium welfare, in this section we present our main positive results. We identify structures on couriers' costs or buyers' valuations that overcome both impossibilities. We continue to use ``equilibrium'' to refer to with-tip equilibrium and ``without-tip'' equilibrium to denote the alternative.

\subsection{Structures on Courier Costs}\label{sec:market_structures_couriers}
The structure on courier costs identifies components of cost. The three components of $c_d(b,s)$ that we identify are: (i) a buyer--store part $c(b,s)$, (ii) a courier--store part $c_d(s)$, and (iii) a courier--buyer part $c_d(b)$. These reflect distance-based costs that account for couriers' deliveries to buyers, trips to stores, and return-home trips. The courier--store cost can also reflect a courier's familiarity with a store, while the courier--buyer cost can account for familiarity with the buyer's neighborhood.

We first show that when for all couriers $c_d(b,s)=c(b,s)+c_d(s)$ or for all couriers $c_d(b,s)=c(b,s)+c_d(b)$, i.e., when one of the two possible structural assumptions holds, the optimal welfare across all allocations can be found in polynomial time.

\begin{figure}[hbtp!] 
    \centering
    \begin{tikzpicture}[scale=0.8]
        \node[draw, shape=circle] (Source) at (0, 2.25) {\scriptsize So};
        \node[draw, shape=circle] (Sink) at (11, 2.25) {\scriptsize Si};
        \foreach \i/\label in {0.9/$b_1$, 3.6/$b_2$}
            \node[draw, shape=circle,minimum size=0.1cm] (\label) at (1, \i){\scriptsize\label};
        \foreach \i/\label in {0/$b_1 s_1$, 1.5/$b_1 s_2$, 3/$b_2 s_1$, 4.5/$b_2 s_2$}
            \node[draw, shape=circle,minimum size=0.1cm] (\label) at (3, \i){\scriptsize\label};
        \foreach \i/\label in {0.9/$s_1$, 3.6/$s_2$}
            \node[draw, shape=circle,minimum size=0.1cm] (\label) at (5, \i){\scriptsize\label};
        \foreach \i/\label in {0.9/$s'_1$, 3.6/$s'_2$}
            \node[draw, shape=circle,minimum size=0.1cm] (\label) at (7, \i){\scriptsize\label};
        \foreach \i/\label in {0.6/$d_1$, 2.25/$d_2$, 3.9/$d_3$}
            \node[draw, shape=circle,minimum size=0.1cm] (\label) at (9, \i){\scriptsize\label};

        \foreach \x/\y in {Source/$b_1$, Source/$b_2$}
            \draw[->] (\x) -- node[midway, right]{} (\y);
        \foreach \x/\y/\w in {$b_1$/$b_1 s_2$/$-v_{b_1}(s_2)$, $b_2$/$b_2 s_2$/$-v_{b_2}(s_2)$}
            \draw[->] (\x) -- node[above,font=\tiny]{\w} (\y);
        \foreach \x/\y/\w in {$b_1$/$b_1 s_1$/$-v_{b_1}(s_1)$, $b_2$/$b_2 s_1$/$-v_{b_2}(s_1)$}
            \draw[->] (\x) -- node[below,font=\tiny]{\w} (\y);

        \foreach \x/\y/\w in {$b_2 s_2$/$s_2$/$c(b_2\text{, }s_2)$, $b_2 s_1$/$s_1$/,$b_1 s_2$/$s_2$/$c(b_1\text{, }s_2)$, $b_1 s_1$/$s_1$/$c(b_1\text{, }s_1)$}
            \draw[->] (\x) -- node[above,font=\tiny]{\w} (\y);
        \foreach \x/\y in {$s_2$/$s'_2$, $s_1$/$s'_1$}
            \draw[->] (\x) -- node[midway,font=\tiny]{} (\y);
        \foreach \x/\y/\w in {$s'_2$/$d_3$/$c_{d_3}(s_2)$, $s'_2$/$d_2$/$c_{d_2}(s_2)$,$s'_2$/$d_1$/, $s'_1$/$d_3$/,$s'_1$/$d_2$/, $s'_1$/$d_1$/$c_{d_1}(s_1)$}
            \draw[->] (\x) -- node[above,font=\tiny]{\w} (\y);
        \foreach \x in {$d_1$,$d_2$,$d_3$}
            \draw[->] (\x) -- node[]{}(Sink);
            
        \node[] at (1, 5.5)  {\footnotesize Buyer};
        \node[] at (3, 5.5)  {\footnotesize Order};
        \node[] at (5, 5.5)  {\footnotesize Store};
        \node[] at (7, 5.45)  {\footnotesize Dummy};
        \node[] at (9, 5.5)  {\footnotesize Courier};

    \end{tikzpicture}
    \caption{An example min-cost flow network for a market where $c_d(b,s)=c(b,s)+c_d(s)$. The market has two buyers, two stores, and three couriers. There is a dummy vertex $s'$ for each store $s$ to enforce the store's unit-capacity constraint. Each edge has capacity one. An edge between a buyer $b$ and an order $(b,s)$ has cost $-v_b(s)$. An edge between an order $(b,s)$ and a store $s$ have cost $c(b,s)$. An edge between a dummy vertex of a store $s$ and a courier $d$ has cost $c_d(s)$.
    \label{fig:OPTDivisibleCost}}
    \end{figure}
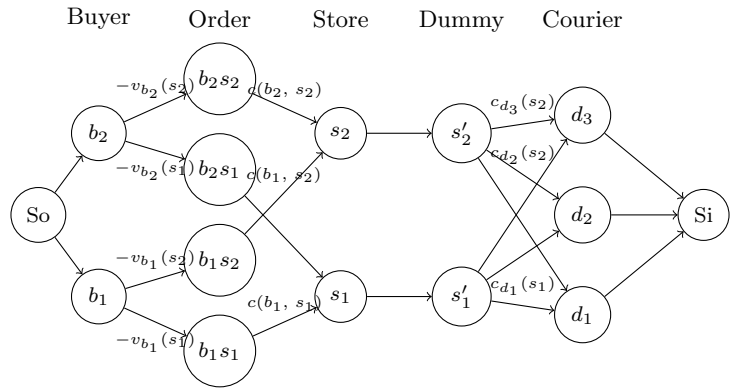

\begin{restatable}{theorem}{divisiblecourierCostOpt}
    \label{thm:divisible_courier_cost_opt}
    The optimal welfare can be computed in polynomial time when the couriers' costs are decomposable into a buyer--store and a courier--store part $c_d(b,s)=c(b,s)+c_d(s)$, or a buyer--store and a courier--buyer part $c_d(b,s)=c(b,s)+c_d(b)$.
\end{restatable}
The proof reduces the problem of finding the optimal welfare to a minimum cost-flow problem. \Cref{fig:OPTDivisibleCost} illustrates such a flow network for a market with two buyers, two stores, and three couriers when $c_d(b,s)=c(b,s)+c_d(s)$. As courier costs are decomposable, the edges on the flow network encode each component of courier costs. With unit capacity on all edges, every integral flow corresponds to a feasible allocation, and maximizing welfare is equivalent to finding a minimum-cost integral flow.

In the next theorem we show when $c_d(b,s)=c(b,s)+c_d(s)$, or $c_d(b,s)=c(b,s)+c_d(b)$, there always exists a with-tip equilibrium that achieves the optimal welfare, which is not true for without-tip equilibria.\footnote{Note the market in \Cref{fig:without_tip_not_exists} does not have a without-tip equilibrium, despite all courier costs being zero. And even when a without-tip equilibrium does exist, as in \Cref{example:without_tip_bad}, it can still yield low welfare, despite the structured courier costs.}

\begin{restatable}{theorem}{DivisiblecourierCost}
\label{thm:divisible_courier_cost}
    When the courier costs consist of a buyer--store and a courier--store part $c_d(b,s)=c(b,s)+c_d(s)$, or a buyer--store and a courier--buyer part $c_d(b,s)=c(b,s)+c_d(b)$, there always exists a with-tip equilibrium that achieves the optimal welfare. However, there exists a market with such courier costs, where every without-tip equilibrium has welfare lower than the optimal welfare.
\end{restatable}
The proof of this theorem is more technical than some of the other results. It verifies that the welfare-optimal allocation $\x$, found by \cref{thm:divisible_courier_cost_opt}, is a max-weight matching in the bipartite graph $G_\x$. It then applies \Cref{lem:test_alloc} to conclude that $\x$ is an equilibrium allocation. To establish $\x$ being a max-weight matching, the proof analyzes alternating paths, a graph-theoretic notion, separating the cases where an alternating path starts from an unmatched buyer or from a matched store in $\x$. In either case, welfare optimality in the three-sided market implies that $\x$ has no improving alternating path, and hence is a max-weight matching in $G_\x$.

\Cref{thm:divisible_courier_cost_opt,thm:divisible_courier_cost} imply that in markets where courier costs are distance-based, a with-tip equilibrium always achieve the optimal welfare, and can be computed in polynomial time. A without-tip equilibrium may not exist, and even if exists can have low welfare compared to the optimal welfare.

\subsection{Structures on buyer valuations}\label{sec:market_structures_buyers}

As an alternative to requiring some structure to courier costs, we introduce a structure on buyer valuations. The structure on buyer valuations is that of {\em single-mindedness}: each buyer only has positive valuation for one store. We first show that the optimal welfare can be found in polynomial time for a market with single-minded buyers, regardless of courier costs.
\begin{restatable}{theorem}{SingleMindedBuyer}\label{thm:buyer_value_opt}
    The optimal welfare can be computed in polynomial time when each buyer only has a positive valuation for one store.
\end{restatable}
Just as \Cref{thm:divisible_courier_cost_opt}, we reduce the problem of finding the optimal welfare to a minimum cost flow problem. Figure~\ref{fig:OneBuyerOneSeller} illustrates such a flow network. 
The full reduction is similar to that in \Cref{thm:divisible_courier_cost_opt} and omitted. We now examine the equilibrium welfare with single-minded buyers. 
\begin{restatable}{theorem}{OneBuyerOneStore}
\label{thm:one_buyer_one_store}
    When each buyer only has a positive valuation for one store, there always exists a with-tip equilibrium that achieves the optimal welfare. However, there exists a market with such buyer valuations, where every without-tip equilibrium has welfare lower than the optimal welfare.
\end{restatable}
The proof proceeds similarly to \Cref{thm:divisible_courier_cost}. When each buyer only values one store, any relevant alternating path has length at most three, and takes the form $(b_0,s_1,b_1)$, where both $b_0$ and $b_1$ only values $s_1$ positively. Note that without-tip equilibria need not exist even when buyers are single-minded when there are fewer couriers than buyers or stores. Moreover, even when they do exist, they can be inefficient, as illustrated in \Cref{example:without_tip_bad}.

\section{Platform Profit Maximization}\label{sec:profit}
While previous sections have assumed the platform can subsidize delivery costs and set courier compensation above buyer prices to maximize welfare, this need not be profit-maximizing for the platform. In this section we analyze platform profit---defined as sum of buyer prices minus courier compensation. 
We ask what is the maximum profit that a platform can extract, subject to equilibrium constraints. 
Throughout this section, we operate in markets where the number of couriers is weakly larger than the minimum number of buyers and stores, so that a without-tip equilibrium exists by \Cref{thm:without_tip_eq_existence}.

Following from \Cref{lem:pareto_dominant}, any without-tip equilibrium is also a with-tip equilibrium of zero tip with the same price and compensation. It naturally follows that the maximum platform profit in all with-tip equilibria is weakly higher than that of without-tip equilibria. We leave the formal statement \Cref{prop:profit_weakly_higher} in \Cref{app:profit}. 

We next turn to the computational problem of maximizing platform profit. Using the same reduction from 3-dimensional matching as in \Cref{lem:OPT_eq_hard}, we show that it is NP-hard to find the maximum profit of both with-tip and without-tip equilibrium. We leave the formal presentation of the theorem to \Cref{app:profit}. Although this hardness result is driven by the intrinsic difficulty of three-sided matching, we will show in \Cref{thm:profit_hard_with_tip} that for with-tip equilibria the hardness also stems from a buyer's ability to specify tips. 

To add some structures, consider markets where each courier can deliver orders from only one store. This captures settings where couriers wait near specific stores, for example by staying in store parking lots to obtain priority matching based on proximity \citep{InstacartCourierParkingLot}. In this restricted environment, the profit-maximizing without-tip equilibrium can be found in polynomial time.
\begin{restatable}{theorem}{ProfitPolyWithoutTipEqStoreLimitedCourier}
\label{thm:profit_polytime_without_tip}
    In markets where each courier can deliver orders from only one store and each store has at least one available courier, there exists a polynomial-time algorithm to find the profit-maximizing without-tip equilibrium.
\end{restatable}

We leave the formal proof in the appendix and provide the intuition behind the proof. When buyers cannot specify tips, any equilibrium allocation must coincide with a Walrasian equilibrium in the two-sided buyer-store market. By the lattice structure, all such equilibria share the same highest buyer prices \citep{gul1999walrasian}. When each courier can only deliver for one store, equilibrium assigns each store's order to its lowest-cost courier, and the minimum courier compensation equals that cost. The algorithm chooses, among Walrasian equilibria of the buyer-store market, the one that minimizes total courier costs.

In the above theorem, restricting each courier to a single store has the effect of reducing the analysis of the market to a two-sided one, yielding a polynomial-time algorithm for profit maximization without-tip. In contrast, the next theorem shows that computing a profit-maximizing with-tip equilibrium remains intractable. This shows that the hardness of the with-tip setting arises not only from the three-sidedness of the market, but also from buyers' ability to specify tips.
\begin{restatable}{theorem}{ProfitHardEqStoreLimitedCourier}
\label{thm:profit_hard_with_tip}
    In markets where each courier can only deliver for one store, it is NP-hard to determine whether there exists a with-tip equilibrium with profit at least $k$, even when 
    \begin{itemize}
        \item each store is served by one unique courier whose cost consists of courier-store and buyer-store part, or
        \item each courier can only deliver for one store-buyer pair.
    \end{itemize}
\end{restatable} 
Similar to \citet{platformDisruption} and \citet{guruswami2005profit}, we prove NP-hardness via a reduction from the vertex cover problem where vertices have maximum degree $K\geq 3$, which is NP-hard. For an instance of vertex cover problem with $V$ vertices and $E$ edges, we construct a market where each store has a unique dedicated courier, and where the maximum profit for with-tip equilibrium is equal to $|V|+M(|V|+|E|)-k$ where $k$ is the size of minimum vertex cover. Like \citet{platformDisruption}, our equilibrium requires stores not purchased to have zero price and delivery compensation, and we also use dummy buyers that generate zero profit to buy the edge stores not bought by edge buyers. Unlike \citet{platformDisruption} where the main lever is matching, the profit constraint in our problem arises from vertex buyers' ability to specify tips to deviate to edge stores. Our problem also differs in that both prices and compensations can be adjusted simultaneously, which introduce additional challenges to prove edge buyers must always pay their full valuation as profit. 

In summary, despite the NP-hardness of profit maximization for with-tip equilibrium, the platform can guarantee profit equal to the maximum without-tip profit; when restricting each courier to deliver for one (e.g., the closest) store, computing such profit takes polynomial time.

\section{Discussion}\label{sec:dis}
We have examined the role of tips in pricing and market clearing on a three-sided delivery platform. We define a with-tip equilibrium to model buyers specifying optional tips before delivery, and a without-tip equilibrium where buyers are not allowed to tip or only tip after delivery. We show the optimal with-tip equilibrium welfare and platform profit are respectively weakly higher than that without, highlighting the benefits of tips. However, even with-tip equilibria can suffer from inefficiencies, and computing the optimal equilibrium welfare is NP-hard. To address the welfare challenges, we identify market structures that ensure the existence and polynomial-time computability of efficient with-tip equilibria. Specifically, these are markets where delivery costs are decomposable into store--buyer and courier--store components, or store--buyer and buyer-courier components, as well as markets where buyers are single-minded. For profit, despite calculating the maximum-profit for both with-tip and without-tip equilibrium is intractable, the platform can find the maximum without-tip profit in polynomial time when restricting each courier to deliver for one (e.g., the closest) store.
Several interesting questions remain, offering avenues for future research:
\begin{itemize}
    \item To abstract away from the inefficiency arising because welfare-optimal trade cannot occur when store prices are higher than costs, we model stores set prices equal to their costs. 
    %
    However, in a free market, store prices are influenced by demand and supply, which in our model correspond to buyer orders and courier availability.
    It remains an open problem to define an equilibrium that accounts for store incentives, in setting store prices, while ensuring the  existence of an equilibrium in a three-sided market.
    \item The model presented here is static, which captures settings with time-sensitive deliveries---buyers who do not receive their food within a short period may no longer value the delivery. 
    It is worth considering a multi-period game, where buyers not attaining delivery this period can wait until the platform boosts delivery compensation associated with their orders in next periods.
    \item The structure in courier costs or buyer valuations we identify are sufficient conditions to ensure the existence of efficient with-tip equilibria. What are the necessary conditions for the existence of with-tip efficient equilibria?
    \item Can we find a polynomial-time algorithm that outputs with-tip equilibrium with good welfare in every market, not just the ones with specific structures on delivery costs and buyer valuations?
\end{itemize}


\part{Direct Access Control with Matching}
\chapter{Direct Access Control with Matching} \label{chap:platform_disruption}

\section{Introduction}
Online platforms, such as Amazon and Uber Eats, have an increasingly important role to play in the modern economy. As discussed in~\citet{platformSim}, such platforms bring together buyers and sellers, erase the physical and geographic barriers preventing trade, and generate value in the form of new transaction opportunities. The significance of these platforms was perhaps most evident during the COVID-19 pandemic. As entire populations were put under lockdown, consumers flocked to these platforms, and Uber Eats saw a surge in both supply and demand \citep{raj2020covid} while Amazon's stock price nearly doubled between March 13th and late August of 2020. \citep{icsik2021impact,tiananalysis}.

However, the growing market power of these platforms has also been associated with unwanted strategic practices, such as a platform favoring specific merchants over other, less profitable ones to maximize its own revenue. In 2019, the European Union Commission filed an antitrust lawsuit against Amazon, alleging that the latter had directed buyers to third-party sellers who paid it hefty delivery and storage fees, obscuring better deals elsewhere~\citep{veljanovski2022algorithmic}. While platforms play a crucial role in disrupting markets and increasing social welfare, a platform may also benefit by strategically matching buyers and sellers to further its own gain. In response to these concerns, in this paper we present a theoretical framework for studying the platform's revenue-maximization problem and the 
impact of the platform's strategic behavior on social welfare. 

We adapt a model from~\citet{platformEq} to our setting. We consider a market of \emph{unit-demand} buyers and \emph{unit-supply} sellers that is mediated by an online platform. Each buyer 
can transact with a set of sellers whom she already knows, and in addition with any 
sellers that might be introduced to her by the platform.
Given these constraints on trade, prices and transactions are induced by a competitive (Walrasian) equilibrium.
 The platform can choose to introduce a (possibly distinct) set of sellers to each buyer, and its revenue is proportional to the total prices of all trades between agents that it introduces.\footnote{For simplicity, we do not model this revenue as explicitly coming from the seller or the buyer.} The platform thus seeks to match buyers to sellers in a way that maximizes its revenue. We define the \emph{social welfare} as the sum of transacting buyers' valuations\footnote{We assume that sellers have no costs for trading.} and 
the \emph{optimal (social) welfare} as the social welfare when all sellers are introduced to each buyer. 

We study the computational complexity of the platform's revenue-maximization problem. While we show that the problem is generally intractable, we give structural results for revenue-optimal matchings, identifying specific market settings where the platform can maximize revenue in polynomial time. We also analyze the relationship between revenue optimization and social welfare: We lower bound the platform's revenue as a function of the maximum increase in social welfare that the platform can create, and we upper bound the loss in welfare that results from the platform's self-interest. Leveraging these results, we give a polynomial-time algorithm that guarantees a logarithmic approximation of the optimal revenue in markets where the platform can substantially increase social welfare. We also give tighter bounds for both revenue and social welfare in markets with \textit{homogeneous goods}, where each buyer values all sellers' items equally, but with different buyers having potentially different values.

\subsection{Results and Techniques}
\paragraph{Hardness of the Platform's Revenue-Maximization Problem.}

In Section~\ref{sec:hardness}, we study the computational complexity of maximizing the platform's revenue, to which we refer as the \emph{platform's problem}. 
We assume that the market clears according to {\em maximum} competitive equilibrium prices (i.e., the prices in the seller-optimal competitive equilibrium). In the event that multiple competitive allocations exist, we assume that the platform can break ties in its favor (selecting the equilibrium that maximizes its revenue). This reflects a setting
 in which the platform has sufficient market power to set prices and direct trades.
Intuitively, the platform's problem is challenging in that the platform must simultaneously consider the prices of potential trades, their feasibility in a competitive equilibrium, and the externalities they impose on the prices of other trades.

In a market with $n$ buyers and $m$ sellers, denote the set of sellers that buyer $b_i$ originally knows by $N(i)$, the size of this set as buyer $i$'s \emph{degree}, and buyer $i$'s value for seller $j$'s item by $v_{ij}$. For the platform's problem (with input size $\mathrm{poly}(n, m)$), we establish the following complexity result.

\begin{customthm}{1}[Informal Version of Theorems~\ref{thm:nphard} and \ref{thm:vc_apx_proof}]\label{thm:informal_hardness}
    The platform's revenue-maximization problem is NP-hard, even when every buyer has degree at most two ($|N(i)| \leq 2$ for all $i$) and values all \emph{desired} items equally (i.e., $v_{ij} \in \{0, v_i\}$ for all $j$). If we do not restrict to instances where the buyer has degree at most two, then the problem is APX-hard.
\end{customthm}

The proof of Theorem \ref{thm:informal_hardness} leverages techniques previously used in the profit-maximizing envy-free pricing literature, where we reduce from a version of $3$-SAT in \citet{chen2014envy} to prove NP-hardness and reduce from a version of minimum-vertex cover in \citet{guruswami2005profit} to prove APX-hardness. In contrast to the envy-free pricing problem, where there are no constraints on trade and unsold items can command high prices, our setting exhibits two main differences.
First, buyers have limited access to sellers, and access is mediated by the platform. When no buyer knows any seller, the platform can facilitate trades to achieve the welfare-optimal matching, extracting the optimal welfare as revenue. When buyers know some subsets of sellers, the existing trading possibilities both constrain the platform's revenue per trade and introduce externalities on the prices of trades that the platform facilitates. This is different from envy-free pricing where any buyer can envy any other buyers. When adapting the reductions from \citet{guruswami2005profit} and \citet{chen2014envy}, we introduce extra sellers to the market and carefully select the set of buyers and sellers who can trade outside the platform in order to mimic the envy structure present in the original reductions. Second, while competitive equilibrium prices are envy-free in this setting, competitive equilibria additionally require that unsold items have price $0$. In our proofs, we deal with this latter difference by introducing extra ``dummy'' buyers to transact with unsold items. 

\paragraph{Tractable Special Cases of the Platform's Problem.} 
Theorem~\ref{thm:informal_hardness} shows hardness, even when each buyer has degree at most two and values all desired items equally. We show that the restrictions in \cref{thm:informal_hardness} on the degree and values not being even stronger is not simply an artifact of our proof; rather, this
 corresponds to the frontier of intractability. In Section~\ref{sec:poly-special_case}, we introduce two market settings at the boundary of the
 hardness result of Theorem~\ref{thm:informal_hardness}  where the revenue-optimal matching can be found in polynomial time. 
The first market setting tightens the degree restriction to be at most $1$ and additionally assumes goods are homogeneous, that is, $v_{ij}=v_i$ for all $j$.
The second setting restricts the class of valuations to markets with homogeneous goods and buyers, that is, all buyers value all items equally at some constant positive amount (e.g., $v_{ij}=1$ for all $i,j$); the second setting also places a further restriction on the graph structure, which is already satisfied by the reduction of \cref{thm:informal_hardness}. These two cases demonstrate that starting from the setting of Theorem \ref{thm:informal_hardness}, imposing even slightly more structure on the underlying network and/or the valuations is enough to make the platform's problem tractable.

\begin{customthm}{2}[Informal Version of Theorems~\ref{thm: poly_AMOS} and \ref{thm:poly_identity}]\label{thm:informal_amos_iden}
    The platform's revenue-maximization problem can be solved in polynomial time (in $n$ and $m$) in the following two markets: (i) Homogeneous-goods markets with buyer degree at most $1$; (ii) Markets with homogeneous goods and buyers, in which each buyer has degree at most $2$ and there are sparse connections, so that for any pair of sellers, at most one buyer knows them both.
\end{customthm}

In the homogeneous-goods setting, we give a structural characterization result that shows that the revenue-optimal transactions introduced by the platform connect groups of buyers who know the same seller into \emph{chains} and \emph{cycles} of bounded length. Ranking all buyer groups by the largest buyer value in each group, we show that buyer groups that form a cycle are adjacent in rank. This characterization leads to a  dynamic programming algorithm to pair buyer groups into chains and cycles for maximum revenue, which is at the heart of the proof of Part (i) of \cref{thm:informal_amos_iden}. In markets with homogeneous goods and buyers, we show that finding the revenue-optimal matching is related to finding the maximum set of buyers with non-positive \textit{graph surplus}. In graph theory, the surplus of a vertex set is used in finding maximum matchings \citep{lovasz2009matching}, and is defined as the size of the neighborhood of the vertex set minus the size of the vertex set itself. We prove Part (ii) of \cref{thm:informal_amos_iden} by showing that this buyer set can be found in polynomial time under the structural assumptions of markets in the second special case. 

\paragraph{Bounds between Revenue Optimization and Social Welfare.} After studying the computational aspects of the platform's problem, we shift our focus in Sections~\ref{sec:ch4_general_markets} and~\ref{sec:hom_markets} to understanding the relationship between revenue optimization and welfare. On the one hand, when the platform has the potential to increase social welfare substantially, we ask whether such improvements come with a corresponding guarantee for the platform's revenue.
 Conversely, we investigate the impact on social welfare when the platform acts in its own self-interest and optimizes for revenue.
The theorems for each of the two directions are as follows:

\begin{customthm}{3}[Informal Version of Theorem~\ref{thm:gen_welfare_conversion}]
\label{thm:informal_gen_welfare_conversion}
    In any market, if there exists a set of transactions that the platform can add that increases social welfare by $\Delta W$, then the platform can add some subset of this set to generate $\Omega(\Delta W/\log(\min\{n,m\})$ in revenue. This subset can be found from the original set in polynomial time.\footnote{Note that the set of platform edges that maximizes the (increase in) social welfare can be found in polynomial time.} Furthermore, this bound is tight: 
There exists a market where the platform can increase welfare by $\Delta W$ and optimal revenue is $O(\Delta W/\log(\min\{n,m\})$. 
\end{customthm}

\begin{customthm}{4}[Informal Version of Theorem~\ref{thm:poa_upper_bound}]\label{thm:inform_poa}
   In any market, the social welfare is at least an $\Omega(1/\log(\min\{n,m\}))$-fraction of the optimal welfare when the platform maximizes its revenue.
\end{customthm}

Theorem~\ref{thm:informal_gen_welfare_conversion} illustrates a method for the platform to translate the potential welfare increase into revenue. In markets with a substantial welfare gap ($\Delta W$ is large), this
further suggests a polynomial-time approach to guaranteeing as revenue a logarithmic fraction of optimal welfare and,
 consequently, of
 optimal revenue. Combined with Theorem~\ref{thm:informal_hardness}, this result suggests that although the exact revenue maximization problem is hard in general markets, one can nevertheless extract a substantial fraction of optimal revenue in a computationally efficient manner. From another perspective, by giving a revenue guarantee in markets with high inefficiencies (i.e., large $\Delta W$), this theorem can be seen as establishing a motivation for revenue-interested platforms to disrupt such markets, providing a possible explanation for their real-life tendency to do so. In addition, and with an eye toward markets where efficiency is most lacking, we are less interested in settings where $\Delta W$ is small, even when
 optimal revenue can significantly exceed $\Delta W$. Theorem~\ref{thm:inform_poa} discusses the other side of the relationship between welfare and revenue: for all markets, revenue maximization comes with modest welfare guarantees --- namely, revenue maximization cannot harm overall welfare by more than a logarithmic factor.

\paragraph{High Revenue and Social Welfare in Homogeneous-Goods Markets.} 
We also explore the special case of homogeneous-goods markets; recall that these are markets where each buyer values all sellers' items equally but with different buyers potentially having different values. In these markets, optimal welfare is achieved when the buyers with the largest values transact. 
We show in Section~\ref{sec:hom_markets} 
that platform self-interest in this setting is perfectly
aligned with welfare maximization;
i.e., whenever the platform maximizes its own revenue, the optimal welfare is attained. 
Additionally, the platform can obtain revenue at least as large as the welfare gap $\Delta W$ in polynomial time. This implies a polynomial-time procedure offering a constant-factor approximation for revenue when $\Delta W$ is large.

\begin{customthm}{5}[Informal Version of Theorems~\ref{thm:hom_conversion} and \ref{thm:hom_poa}]
\label{custom_thm:hom}
   In homogeneous-goods markets, if there exists a set of transactions that the platform can add that increase welfare by $\Delta W$, there is a polynomial-time algorithm to guarantee platform's revenue of at least $\Delta W$. Furthermore, if the platform's revenue is optimal, then social welfare equals the optimal welfare. 
\end{customthm}

The proof of Theorem \ref{custom_thm:hom} utilizes the structural notion of \textit{opportunity paths}~\citep{kranton2000competition}
 to characterize the competitive price of a trade in homogeneous-goods markets. We show that in such markets, the buyers with the highest values must trade in the revenue-optimal matching, implying that the revenue-optimal matching attains the optimal welfare. We also show that the platform can match buyers to sellers according to the optimal welfare matching while obtaining the full increase in welfare ($\Delta W$)
 as revenue. Unlike in \cref{thm:informal_gen_welfare_conversion} for the general setting, this construction does not restrict itself to a subset of these transactions, offering both optimal welfare and good revenue guarantees.

Overall, our work intertwines computational complexity results with economic findings for the platform's revenue-maximization problem. Despite the computational hurdles that present a barrier to exact maximization in general settings, we develop ways to extract a substantial fraction of the optimal revenue in polynomial time in markets with a large welfare gap, and we further show that social welfare is reasonably well-aligned with revenue maximization, providing welfare guarantees for the revenue-optimal matching in fully general markets.

\subsection{Related Work}
Two works share particularly close models and research questions to our own. \citet{banerjee2017segmenting} model a platform controlling visibility between buyers and sellers of different types in a bipartite market. They study how the platform controls the visibility in order to maximize revenue and welfare. In a similar setting but with full visibility, \citet{birge2021optimal} ask how the platform optimally chooses to set transaction fees. The primary difference of both of these papers from our work is that buyers and sellers in these papers are assumed to have zero utility (equivalently, cannot trade) outside the platform. In contrast, we emphasize the existence of these outside trading opportunities, which play a crucial role both in our modeling and the complexity of the problem. These outside opportunities allow us to model markets that exist prior to platform disruption, rather than focusing only on markets that are initiated by the platform. Outside opportunities also mediate the complexity of the platform's revenue maximization. Without them, the platform's problem can solved in polynomial time in our setting; as more outside trading opportunities are introduced, the externalities they impose significantly complicate the question. We remark that \citet{banerjee2017segmenting} and \citet{birge2021optimal} are more general in other aspects; notably, each node there does not represents a single buyer or seller as in our paper but instead a continuum of buyers or sellers, each represented by a particular demand or supply curve.

Another closely related work is that of \citet{platformEq}, who similarly model a platform that joins a market with existing  trading opportunities. Their focus is on the platform strategically setting a transaction fee, whereas we analyze the platform's matching strategy.

Besides these three papers, our work aligns with the literature that studies the formation of economic markets modeled by networks. \citet{kranton2001theory,kranton2000competition} and \citet{elliott2015inefficiencies} examine a unit-demand, unit-supply market where buyers and sellers strategically invest in costly opportunities to trade. They analyze the effect of the strategic behavior on social welfare. For infinitely divisible goods, \citet{even2007network} analyze all possible outcomes of the market structure when agents strategize to invest in trading opportunities, and \citet{kakade2004economic} examine a Fisher market on a graph. Like these studies, we build on the foundational research on competitive equilibria \citep{kelso1982job, gul1999walrasian}. However, complementing these works where market participants form trading links in a decentralized manner, we focus on a centralized, revenue-maximizing platform that facilitates trades.

Our work relates to the literature at large studying a platform's role in a market. Platforms have been modeled as intermediaries stocking goods, setting prices, and reselling to downstream market participants on a fixed network \citep[e.g.][]{blume2007trading,condorelli2017bilateral,manea2018intermediation,kotowski2019trading}; as guiding search and discovery processes of different market segments \citep[e.g.][]{immorlica2021designing,hu2022dynamic,halaburda2018competing,huttenlocher2023matching}; and as strategic entities that maximize revenue through pricing or matching tools \citep{ke2022information,platformSim}. Our work differs crucially in the ways in which prices are set; in these works, all prices are set by the platform, whereas we assume that prices are beyond the platform's control, instead induced by market forces modeled by a competitive equilibrium.

For our computational complexity hardness results, we use similar methods to those used in the {\em envy-free pricing} literature \citep{guruswami2005profit,chen2014envy}. There, the goal is to find a revenue-maximizing set of prices (and corresponding allocation) such that each agent receives their most desired good. While we draw on similar techniques, our problem is distinct in that the platform does not directly set prices and is limited in its ability to influence the prices due to buyers' and sellers' existing connections without the platform. This makes it challenging to translate approximation algorithms from this literature to our setting.

\section{Model \& Preliminaries}

Following \citet{platformEq} and \citet{kranton2000competition}, we model a buyer-seller network as a bipartite graph $G = (B, S, E)$. Later, we significantly deviate from these models by introducing a centralized platform that directs trades between buyers and sellers. The network is composed of a set of $n$ unit-demand buyers $B = \{b_1, \dots, b_n\}$ and $m$ unit-supply sellers $S = \{s_1, \dots, s_m\}$. The set of edges $E$ represents the feasible transaction opportunities available to buyers and sellers – that is, buyer $b_i$ and seller $s_j$ can transact if and only if $(b_i, s_j) \in E$. Each buyer $b_i$ has a valuation $v_{ij} \geq 0$ for seller $s_j$'s good.

Given a graph $G$ and valuation profile $\textbf{v}$, the market clears according to a competitive equilibrium, defined by the tuple $(\textbf{x}, \textbf{p})$, where $\textbf{x} \in \{0, 1\}^{n \times m}$ represents the allocation and $\textbf{p} \in \mathbb{R}^m$ corresponds to the item prices (where $\textbf{p} \geq 0$). Formally, $x_{ij} = 1$ if and only if buyer $b_i$ receives seller $s_j$'s item, for which they pay price $p_j$.

\begin{definition}[Competitive Equilibrium on a Buyer-Seller Network]
    Given a graph $G = (B, S, E)$ and valuations $\textbf{v}$, a competitive equilibrium is defined by an allocation, price pair $(\textbf{x}, \textbf{p})$ satisfying:
    \begin{itemize}
        \item Transactions are feasible: $x_{ij} = 0$ for all $(b_i, s_j) \not \in E$.
        \item All items are allocated at most once: $\sum_i x_{ij} \leq 1$ for $1 \leq j \leq m$.
        \item Each buyer receives at most one good: $\sum_j x_{ij} \leq 1$ for $1 \leq i \leq n$.\footnote{While unit-demand buyers could receive more than one good in a Walrasian equilibrium, it is WLOG to assume that they receive only their favorite item in the bundle.}
        \item Each buyer receives their favorite item: $\sum_j x_{ij} \cdot (v_{ij} - p_j) = \max(0, \max_j v_{ij} - p_j)$.
        \item All unsold items have price zero: $\sum_i x_{ij} = 0 \implies p_j = 0$.
    \end{itemize}
\end{definition}

Let $\mathrm{Eq}(G,\mathbf{v})$ denote the set of all competitive equilibria for a graph $G$ and valuations $\mathbf{v}$. \citet{kelso1982job} proved that this set is non-empty when buyers have unit-demand valuations. Furthermore, the First Welfare Theorem states that any competitive equilibrium allocation $\textbf{x}$ maximizes social welfare; in other words, $\textbf{x}$ corresponds to a maximum weight matching on the graph $G$, where edge weights are given by valuations.

\begin{theorem}[First Welfare Theorem]
    If $(\textbf{x}, \textbf{p})$ is a competitive equilibrium, then $\textbf{x}$ maximizes welfare, respecting the constraints on transactions given by edges.
    \begin{align*}
        \sum_{i,j} x_{ij} \cdot v_{ij} = W(G) 
    \end{align*}
    where $W(G)$ denotes the value of the maximum weight matching on $G$.
\end{theorem}

For a market $G=(B,S,E)$ with valuations $\textbf{v}$, we use $W(G) \coloneqq W(B,S,\textbf{v},E)$ to denote the 
{\em social welfare} of this market. This corresponds to the value of the maximum weight matching on $G$. Letting $C_{B,S}$ denote the edges of the \textit{complete bipartite graph}, the \textit{optimal welfare} is denoted as $W^\star(G) \coloneqq W(B, S,\textbf{v},C_{B,S})$. This represents the social welfare when all buyers and sellers can freely transact. When it is clear from the context, we omit $G$ and directly use $W^\star$ to denote optimal welfare.

While the First Welfare Theorem characterizes the set of allocations belonging to a competitive equilibrium, it says nothing about the item prices. In general, there are many sets of prices that could form a competitive equilibrium, with these price vectors forming a lattice \citep{gul1999walrasian}. Throughout this paper, we use the maximum competitive prices, which have a natural interpretation as the seller's contribution to the social welfare.
\begin{theorem}[\citet{gul1999walrasian}]
    Seller $s_j$'s maximum competitive price is given by
    \begin{align*}
        p_j = W(G) - W(G \setminus \{s_j\})
    \end{align*}
    where $G \setminus \{s_j\}$ represents the graph $G$ after removing seller $s_j$ and all incident edges.
\end{theorem}

\subsection{Introducing the Platform}

Now, we model the platform's role in our buyer-seller network. Let $G_w = (B, S, E_w)$ denote the \textit{world graph} without the platform. The \textit{world edges} $E_w$ represent existing transaction opportunities between buyers and sellers; i.e., these buyers and sellers can already transact without the platform, perhaps in-person or through some other non-platform related channels. The platform possesses \textit{full information} about the world graph and the valuation profile $\textbf{v}$. It then chooses a set of \textit{platform edges} $E_p$ to add between buyers and sellers who are not already connected by world edges ($E_p \cap E_w = \emptyset$). We refer to the graph with platform edges as the \textit{platform graph}: $G_p = G_w\cup E_p = (B, S, E_w \cup E_p)$.

With these new edges in the platform graph, the market clears according to a competitive equilibrium with maximum competitive prices. The platform charges a fixed percentage, which we assume to be exogenous, of the total price of all trades on the platform edges.
 We will use $\mathrm{Rev}(E_p)$ to denote the sum of the prices associated with trades that are completed on
 platform edges $E_p$ and $\mathrm{Rev}^\star$ to denote the maximum possible revenue. In reality, the platform's revenue is a fixed percentage of this  quantity. For simplicity, we  refer to the sum of prices along transacting platform edges as platform ``revenue,'' noting that maximizing the two quantities is equivalent.
 
To reflect the platform's market power, we assume it can break ties between multiple equilibria, in addition to selecting for the maximum competitive equilibrium prices. The goal of the platform is to add a set of platform edges (and potentially break ties between competitive equilibria) to maximize its revenue.
\begin{definition}[The Platform's Problem]
    Given a world graph $G_w=(B,S,E_w)$ and valuation profile $\textbf{v}$, the platform seeks to introduce a set of platform edges $E_p$ that maximizes revenue. The optimal revenue is given by
    \begin{align*}
        \mathrm{Rev}^\star = \max_{E_p} \max_{(\textbf{x}, \textbf{p}) \in \mathrm{Eq}(G_p, \textbf{v})} \sum_{(b_i,s_j)\in E_p} p_j \cdot x_{ij}.
    \end{align*}
\end{definition}

\begin{figure}[t]
    \centering
    \begin{tikzpicture}[scale=1.3]
        \foreach \i/\label in {1/$b_1$, 2.5/$b_2$, 4/$b_3$, 6/$b_n$}
            \node[draw, shape=rectangle, minimum size=0.6cm] (\label) at (\i, 2) {\label};
            
        \foreach \i/\label in {1/$s_1$, 2.5/$s_2$, 4/$s_3$, 6/$s_n$}
            \node[draw, shape=circle, minimum size=0.6cm] (\label) at (\i, 0) {\label};

        \foreach \x/\y/\w/\pos/\loc in {$b_1$/$s_1$/1/midway/right, $b_2$/$s_1$/2/midway/below right,$b_2$/$s_2$/2/midway/right, $b_3$/$s_2$/3/midway/below right, $b_3$/$s_3$/3/midway/right, $b_n$/$s_n$/n/midway/right}
            \draw[line width=1.5pt, customblue, dashed] (\x) -- node[\pos, \loc, font=\footnotesize] {\w} (\y);
     
        \node at (5, 2) {$\ldots$};
        \node at (5, 0) {$\ldots$};

        \node[left] at (0.5, 2)  {Buyers};
        \node[left] at (0.5, 0)  {Sellers};
    
    \end{tikzpicture}
        \caption{A market where adding all platform edges is arbitrarily bad for revenue. There are no world edges. All nonzero valuations are indicated by dashed blue lines. Buyer $b_1$ has value $1$ for seller $s_1$, Buyer $b_i$ has value $i$ for sellers $s_{i-1}$ and $s_{i}$ for $i=2,\ldots,n$. All other valuations are zero.} \label{fig:all_platform_edge_bad}
\end{figure}
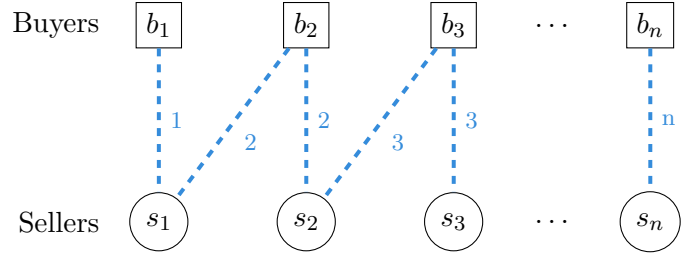

\begin{example}
  To illustrate the complexity of the platform's problem, consider the simple market in Figure~\ref{fig:all_platform_edge_bad} where there are no world edges and buyers have valuations $v_{1,1}=1,v_{i,i-1}= v_{i, i} = i$ for $i=2,\ldots n$, where $v_{ij}$ is buyer $b_i$'s valuation for seller $s_j$'s item. The naive strategy of adding all possible platform edges performs very poorly, obtaining revenue $n$ as every transaction occurs at a price of $W(G_p) - W(G_p\setminus\{s_i\}) = 1$. The revenue optimal matching would draw a single platform edge connecting each $b_i$ to $s_i$, obtaining revenue $n(n+1)/2$. The platform must strike a balance between facilitating as many transactions as possible and not adding too many edges and creating unwanted competition, driving transaction prices down.
\end{example}

\subsection{Competitive Prices in Homogeneous-Goods Markets}
To facilitate proofs and more easily identify competitive equilibrium prices in homogeneous-goods markets, we borrow insights from \citet{kranton2000competition}. They introduce the concept of \textit{opportunity paths} to delineate a buyer's direct and indirect competitors and to calculate a buyer's forgone trading opportunity.

\begin{definition}[Opportunity Path \citep{kranton2000competition}]\label{def:ch4_oppo_path}
    For a market $G=(B,S,E)$ and an allocation $\textbf{x}$, a buyer $b_i$ is connected to another buyer $b_j$ through opportunity path $$ b_i \mbox{ --- } s_1 \to b_1 \mbox{ --- } s_2 \ldots\ldots b_{j-1} \mbox{ --- } s_j \to b_j$$ where an undirected edge $b_p \mbox{ --- } s_q$ means buyer $b_p$ can trade with seller $s_q$ through existing edges but does not: $(b_p,s_q)\in E, x_{pq}=0$; and a directed edge $s_q \to b_p$ means seller $s_q$ sells to buyer $b_p$: $x_{pq}=1.$
\end{definition}
As an example, consider the market in Figure~\ref{fig:all_platform_edge_bad} when all platform edges are added. Buyer $b_i$ buys from seller $s_i$ and knows $s_{i-1}$. Thus, $b_n$ has opportunity path $b_n \mbox{ --- } s_{n-1} \to b_{n-1} \ldots b_2 \mbox{ --- } s_1 \to b_1$, alternating between transacting and non-transacting edges. Just like alternating or augmenting paths in graph theory, an opportunity path alternates between active edges where transactions occur, and inactive edges where transactions could take place but do not. The ``opportunity" refers to a buyer's outside trading options and uniquely determines the price of an transaction as follows.  
\begin{theorem}[Opportunity Path Theorem \citep{kranton2000competition}]\label{thm:oppo_path}
    For a maximum price competitive equilibrium $(\textbf{x}, \textbf{p})$ in a homogeneous-goods market $G=(B,S,E)$ where $x_{ij}=1$, seller $s_j$'s price is equal to the lowest valuation of any buyer connected to $b_i$ through an opportunity path. If $b_i$ is connected to a seller that does not sell through an opportunity path, $p_j = 0$.
\end{theorem}

\section{Hardness of the Platform's Problem}
\label{sec:hardness}

We begin by showing the platform's problem is computationally intractable, even under relatively significant restrictions on both the underlying structure of the world graph and the valuation structure of the buyers.

\begin{theorem}
\label{thm:nphard}
    The decision version of the platform's problem is NP-hard, even when buyers have degree at most two and value all desired goods equally ($v_{ij} \in \{0, v_i\}$).
\end{theorem}

Similar to the hardness proof for revenue-maximizing envy-free pricing in \citet{chen2014envy}, 
Theorem~\ref{thm:nphard} is the result of a reduction from a variant of 3-SAT and Lemma~\ref{lem:reduction_proof}. In contrast to their proof, where an item's price can be high despite the item being unsold, we design our reduction with world edges to provide buyers with outside trading opportunities and hence limit the price they pay. We also introduce ``dummy'' buyers to absorb any items that are not sold – ensuring that these unsold items do not introduce unwanted externalities in the form of low prices. 

Clauses and variables in the 3-CNF are associated with buyers, and literals are associated with sellers. Based
on the value of literals satisfying the assignment, prices that buyers with world edges to corresponding literal sellers pay change, reflected in the optimal revenue; the reduction is designed such that a valid assignment to the 3-CNF exists if and only if the optimal revenue exceeds a given threshold. 

\paragraph{Reduction from a Variant of 3-SAT}
Consider a modified version of 3-SAT, where each variable $x_i$ appears positively ($x_i$) and negatively $(\bar{x}_i)$ an equal number of times. Given a 3-CNF $\varphi$, we add clauses $d_i = (x_i \vee \bar{x}_i)$ for each variable $x_i$. Additionally, we pad $\varphi$ with these clauses such that each variable $x_i$ appears exactly $2t$ times for some $t > 0$.  Denote this modified 3-CNF by $\varphi' = (c_1 \wedge \dots \wedge c_k)$. Let $q$ be the number of unique variables, and let $Z = q \cdot k \cdot t$. 

Given this 3-CNF $\varphi'$, we construct a corresponding instance of the platform's problem as follows. There are $k + 2(t-1)q + (2tq - k) = (4t-2)q$ buyers in total and an equal number of items. We describe the construction below: 

\begin{itemize}
    \item For each variable $x_i$, we add $2t$ items $\alpha_{i,j}, \tau_{i,j}$ for $j \in \{0, \dots, t-1\}$. The $\alpha_{i,j}$ represent the positive instances of $x_i$ while the $\tau_{i,j}$ represent the negative instances. WLOG, let $\alpha_{i,0}, \tau_{i,0}$ be the items that correspond to clauses $d_i = (x_i \vee \bar{x}_i)$. 
    \item For each clause $c_i$ in $\varphi'$, we add a buyer $U_i$ that has value $Z$ for each item that represents the corresponding literals in clause $c_i$. 
    \item For each variable $x_i$, we add $2(t-1)$ buyers $A_{i,j}, T_{i,j}$, $j \in \{1, \dots, t-1\}$. We also add $2(t-1)$ items $\gamma_{i,j}, \delta_{i,j}$, $j \in \{1, \dots, t-1\}$. Each buyer $A_{i,j}$ has value $N+1$ for $\alpha_{i,0}, \alpha_{i,j}, \gamma_{i,j}$, and each buyer $T_{i,j}$ has value $Z+1$ for $\tau_{i,0}, \tau_{i,j}, \delta_{i,j}$. For each of these buyers, we add world edges to the $\alpha_{i,0},\alpha_{i,j},\tau_{i,0},\tau_{i,j}$ they have positive value for.
    \item Finally, we add $2tq - k$ dummy buyers who have value $H \geq Z+1$ for all $\alpha_{i,j}, \tau_{i,j}$.
\end{itemize}
All buyers have value $0$ for all sellers not mentioned above, and no other world edges are present besides those explicitly mentioned. We prove this is a valid reduction.
\begin{restatable}{lemma}{SATReduction}
\label{lem:reduction_proof}
    There is a valid assignment to the original CNF if and only if the optimal revenue for the platform is at least $D \coloneqq kZ + q(t-1)(2Z+1) + H(2tq - k)$. 
\end{restatable}
The proof argument shares similar logic with that of \citet{chen2014envy} and is given in Appendix~\ref{app:sat_np_proof}. The main difference with our proof arises in proving the ``if" direction. In our construction, buyers $A_{i,j}$ and $T_{i,j}$ always transact with $\gamma_{i,j}$ and $\delta_{i,j}$ through platform edges. At the same time, we use world edges to connect these buyers to $U_i$, forming opportunity paths. Buyer $U_i$ corresponds to the literal $x_i$. In this construction, $A_{i,j}$ and $T_{i,j}$ pay $Z$ or $Z+1$, depending on the true value of the literals in the CNF. This change is reflected in the optimal revenue and helps to establish the ``if" direction.

Theorem~\ref{thm:nphard} shows that even when introducing a small number of existing interdependencies between buyers and sellers through world edges, computing the revenue-optimal matching  becomes intractable. By further relaxing the constraint on the degree of the buyers, we prove the platform's problem is APX-hard. In other words, unless $P = NP$, the platform's problem does not admit a polynomial-time approximation scheme.
\begin{restatable}{theorem}{VertexCoverReduction}
\label{thm:vc_apx_proof}
    The decision version of the platform's problem is APX-hard, even when buyers value all desired goods equally ($v_{ij} \in \{0, v_i\}$). 
\end{restatable}
Like \citet{guruswami2005profit} for maximum revenue envy-free pricing, our proof reduces from minimum vertex cover on graphs with degree at most $K$. For an instance of the vertex cover problem, we construct a market with buyers that correspond to vertices and edges. The edge buyers always buy at price $1$, but a vertex buyer buys at price $1$ or $2$ depending on whether the vertex belongs to the minimum vertex cover. To achieve the different prices, we use world edges to build opportunity paths between the vertex buyers and edge buyers, so that the number of the vertex buyers paying price $1$ is at least the size of the minimum vertex cover. Finally we add extra ``dummy'' buyers to make sure no seller has price zero. The full reduction and proof is presented in Appendix \cref{app:vc_apx_proof}.

\paragraph{Differences from Envy-Free Pricing.} While there are connections between the platform's problem and profit-maximizing envy-free pricing \citep{guruswami2005profit, balcan2008item}, positive results there are less easily portable to our setting for two major reasons. First, the platform does not directly set prices; rather, prices are determined as a function of the edges that the platform chooses to add. Second, envy-free pricing allows unsold items to have high prices, effectively eliminating them from the market, where they do not affect prices of other trades. In contrast, a competitive equilibrium requires that unsold items clear at price zero. We bypass this difference in our proofs of hardness by adding dummy buyers so all sellers transact. However, this presents a major barrier when adapting approximation algorithms from the envy-free pricing literature. Since the platform cannot directly set prices, to remove the effect of a seller $s_j$ on other trades, there needs to be a buyer with high valuation for $s_j$, but if no such buyer exists, then there is nothing the platform can do. In contrast, in envy-free pricing, one could simply set $p_j$ to be high enough such that the item is irrelevant.

\section{Tractable Special Cases}\label{sec:poly-special_case}
While the previous section shows the platform's problem is hard in general, here we demonstrate there are structural properties of the markets that reduce the complexity of the problem. In Section~\ref{sec:structural_prop}, we characterize properties that hold for the platform's optimal matching in general unit-supply, unit-demand markets. In Sections~\ref{sec:special-case-homogeneous} and \ref{sec:special-case-identical}, we present two classes of markets where the platform's problem can be solved in polynomial time. These markets represent slight restrictions of the settings that are NP-hard (Theorem \ref{thm:nphard}), showing that our proof of NP-hardness is in this sense  ``tight.''

\subsection{Structural Properties of Revenue Maximum Matchings}\label{sec:structural_prop}

In spite of the hardness of the platform's problem, we are able to prove some structural properties that hold for the platform's optimal matching strategy more generally. 
Both lemmas below will play a role when finding revenue-optimal matchings in special markets in Section~\ref{sec:special-case-homogeneous} and \ref{sec:special-case-identical}, as well as the proof of hardness in Appendix~\ref{app:vc_apx_proof}.

First, we show that the platform has no incentive to add more than one edge incident on any one buyer/seller. Thus, we can think of the platform as recommending a single transaction to a given buyer or seller; any further recommendations are extraneous.
\begin{restatable}{lemma}{atmostone}
\label{lem:at_most_one_edge}
In any general market, there exists a revenue-optimal matching that adds at most one edge to each buyer and each seller. Furthermore, in this matching, all platform edges transact at positive prices in the competitive equilibrium.
\end{restatable}
\begin{proof}
    Let $E_t$ denote the set of edges where transactions take place in the competitive equilibrium. Assume that the platform adds two or more edges to a buyer or seller in $E_p$, or an edge that does not transact. Since buyers are unit-supply and sellers are unit-demand, there exists an edge $e_{ij}=(b_i, s_j) \in E_p$ such that $e_{ij} \notin E_t$. Removing $e_{ij}$ does not change seller $s_j$'s price, because 
    \begin{eqnarray*}
        W(G_p \setminus \{e_{ij}\}) & = &W(G_p)\\
        W(G_p\setminus\{s_j, e_{ij}\}) & = & W(G_p \setminus \{s_j\})
    \end{eqnarray*}
     However, removing $e_{ij}$ weakly increases other sellers' $s'_{j}$ prices, because 
     \begin{eqnarray*}
         W(G_p\setminus\{s'_j, e_{ij}\})\leq W(G_p \setminus \{s'_j\})
     \end{eqnarray*}
    Thus the platform weakly prefers to drop all edges $e_{ij}$ that are not transacting in the final competitive equilibrium. The same argument applies for transactions with price zero.
\end{proof}

Intuitively, any non-transacting edges that the platform adds only increase the level of competition between different sellers. By removing these edges, sellers are able to charge higher prices, and in turn, the platform's revenue from the transactions it facilitates increases. So when searching for the revenue-optimal matching, the platform has no strict incentives to add platform edges that do not transact.

had Next, we show that the platform has an incentive to make sure that as many buyers and sellers transact as possible. This lends further support to the hypothesis that the platform's incentive to maximize revenue can come hand-in-hand with social welfare maximization.
\begin{restatable}{lemma}{allTransact}
\label{lem:all_transact}
    In any general market, there exists a revenue-optimal matching where either all sellers sell, or all buyers buy, or buyers and sellers who do not trade have value zero for each other.
\end{restatable}
\begin{proof}
    We consider the case when there are weakly more buyers than sellers, though the opposite case follows similarly. Take any equilibrium where a seller $s_j$ does not transact. As there are at least as many buyers as sellers, there must be a buyer $b_i$ who also does not transact. For the sake of contradiction, assume that $v_{ij} > 0$. Note that there cannot exist an edge connecting $b_i$ to $s_j$, or else this would violate the First Welfare Theorem. Consider connecting the two via a new platform edge $e_{ij}=(b_i, s_j)$. 

     From this new transaction, we obtain revenue
    \begin{align*}
        W(G_p \cup \{e_{ij}\}) - W(G_p \cup \{e_{ij}\} \setminus \{s_j\}) = W(G_p \cup \{e_{ij}\}) - W(G_p) = v_{ij} > 0
    \end{align*}
    which is strictly positive.
    
    Now, consider any other seller $s'_j$, and let $G_p$ be the graph before adding edge $e_{ij}$. We have that
     \begin{align*}
         p_{s'_j}(G_p) &= W(G_p)-W(G_p \setminus \{s'_j\})\\
         p_{s'_j}(G_p\cup\{e_{ij}\}) &= W(G_p)+v_{ij}-W(G_p\cup\{e_{ij}\} \setminus\{s'_j\})
     \end{align*}
     We will show that $p_{s'_j}(G_p\cup\{e_{ij}\})\geq p_{s'_j}(G_p)$. It suffices to prove that
    \begin{align*}
        v_{ij} \geq W(G_p\cup\{e_{ij}\} \setminus\{s'_j\}) - W(G_p \setminus \{s'_j\})
    \end{align*}
    If the max weight matching in $G_p\cup\{e_{ij}\} \setminus\{s'_j\}$ does not use the new edge $e_{ij}$, then the right hand side is equal to zero and we are done. Thus, suppose that $e_{ij}$ is in the max weight matching. Then
    \begin{align*}
       &W(G_p\cup\{e_{ij}\} \setminus\{s'_j\})=W(G_p \setminus \{b_i, s_j, s'_j\})+v_{ij}\leq W(G_p \setminus \{s'_j\})+v_{ij}
    \end{align*}
    and this is precisely the inequality we want to prove. It follows that any other seller $s'_j$'s price weakly increases, so by matching $s_j$ to $b_i$, the platform's revenue strictly increases.
\end{proof}

\subsection{Single World Seller, Homogeneous-Goods Markets}\label{sec:special-case-homogeneous}

We begin by considering the class of homogeneous-goods markets ($v_{ij} = v_i$),
and in addition requiring that each buyer has at most one incident world edge. We call these markets SWSH (single world seller, homogeneous-goods) markets. This structure allows us to nicely decompose the world graph into what we will call {\em seller subgraphs}, {\em dangling buyers}, and {\em dangling sellers}.

\begin{definition}[Seller Subgraphs]
     In SWSH markets, take any seller $s_j$ who knows at least one buyer via a world edge. The {\em seller subgraph} $S_j$ is the collection of buyers and edges $S_j:=\{s_j\}\cup \{b_i|(b_i,s_j)\in E_w\} \cup \{(b_i, s_j) \in E_w\}$ adjacent to seller $s_j$ via world edges. We will use $v(S_j):=\max_{(b_i,s_j)\in E_w}v_i$ to denote the value of the largest buyer connected to $s_j$.
\end{definition}

There can be multiple buyers in a seller subgraph. All buyers and sellers are either part of a seller subgraph or have no incident world edges. The buyers and sellers with no incident world edges are termed \emph{dangling buyers} and \emph{dangling sellers}. By definition, platform edges cannot be part of any seller subgraphs. In a platform graph $G_p$, platform edges connect seller subgraphs into \emph{cycles} and \emph{chains}. We define these structures and show that they are closely related to the platform's revenue.

\begin{definition}[Cycles]
    For a SWSH market $G=(B,S,E)$ and an allocation $\textbf{x}$, a cycle of $k$ seller subgraphs is defined by 
    $$ S_1 \to S_2 \to S_3 \ldots S_k \to S_1,$$
    where $S_i \to S_j$ means that in allocation $\textbf{x}$, seller $s_i$ sells to a buyer $b_j\in S_j$ in seller subgraph $S_j$.
\end{definition}

\begin{definition}[Chains]
    For a SWSH market $G=(B,S,E)$ and an allocation $\textbf{x}$, a chain of $k$ seller subgraphs is defined by 
    $$ (s_0) \to S_1 \to S_2 \ldots S_k \to (b_k),$$
    where $S_i \to S_j$ means in allocation $\textbf{x}$, where $s_i$ buys from a buyer $b_j\in S_j$ in seller subgraph $S_j$. For the {\em starting subgraph} $S_1$, there can be a dangling seller $s_0$ selling to a buyer in $S_1$. For the {\em terminal subgraph} $S_k$, $s_k$ either does not sell, sells to a buyer $b_k$ who is dangling or $b_k\in S_k$, or belongs to a cycle that does not contain subgraphs $S_1,\ldots S_k$. A chain must include a terminal subgraph.
\end{definition}
We use {\em $k$-cycle} and {\em $k$-chain} to denote cycles and chains consisting of $k\geq 1$ seller subgraphs, or of length $k$. We use {\em $0$-chain} to denote a dangling seller $s_0$ selling to a buyer in a chain or cycle. A $1$-cycle is a seller subgraph $S_i$ where the seller $s_i$ transacts through world edges with a buyer $b_i\in S_i$. All sellers in a cycle transact within the cycle, while one seller $s_k\in S_k$ in the terminal subgraph of a chain can sell outside of the chain. This makes it possible for a chain to be attached to a cycle through $s_k$. 
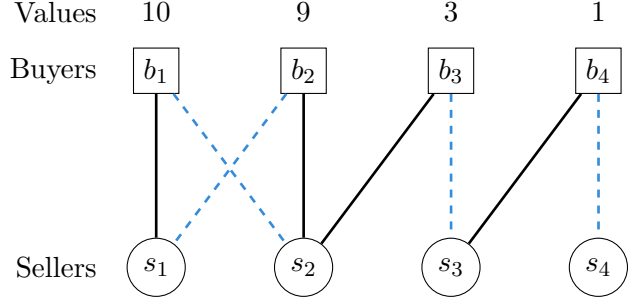
\begin{figure}[hbtp]
    \centering
    \begin{tikzpicture}[scale=1.3]
        \foreach \i/\label in {1/$b_1$, 2.5/$b_2$, 4/$b_3$, 5.5/$b_4$}
            \node[draw, shape=rectangle, minimum size=0.6cm] (\label) at (\i, 2) {\label};
            
        \foreach \i/\label in {1/$s_1$, 2.5/$s_2$, 4/$s_3$, 5.5/$s_4$}
            \node[draw, shape=circle, minimum size=0.6cm] (\label) at (\i, 0) {\label};

        \foreach \x/\y in {$b_1$/$s_2$, $b_2$/$s_1$, $b_3$/$s_3$, $b_4$/$s_4$}
            \draw[line width=1pt, customblue, dashed] (\x) -- (\y);

        \foreach \x/\y in {$b_1$/$s_1$, $b_2$/$s_2$, $b_3$/$s_2$, $b_4$/$s_3$}
            \draw[line width=1pt, black, solid] (\x) -- (\y);

        \node at (1, 2.6)  {10};
        \node at (2.5, 2.6)  {9};
        \node at (4, 2.6)  {3};
        \node at (5.5, 2.6)  {1};

        \node[left] at (0.5, 2.6)  {Values};
        \node[left] at (0.5, 2)  {Buyers};
        \node[left] at (0.5, 0)  {Sellers};
    
    \end{tikzpicture}
        \caption{An example of a SWSH market. World edges are depicted in black, and the optimal set of platform edges is in blue. In revenue optimal matching, $s_1$ sells to $b_2$, $s_2$ sells to $b_1$, $s_3$ sells to $b_3$ and $s_4$ sells to $b_4$. 
        } \label{fig:hom_char_example}
\end{figure}

\begin{example} \label{example:cycle_chain_illustrate}
To illustrate the definitions, consider the example market presented in Figure~\ref{fig:hom_char_example}. The set of optimal platform edges decomposes seller subgraphs into cycles and chains as follows:
    \begin{align*}
       \underbrace{s_4 \; {\to}}_{\text{0-Chain}}\;  \underbrace{S_3\; {\to} }_{\text{$1$-Chain}} \; \underbrace{S_2 \to S_1 \to S_2}_{\text{$2$-Cycle}}
    \end{align*}
\end{example}

The definitions above and the structure of SWSH markets directly yield the following proposition.
\begin{proposition}\label{prop:cycles_chains}
    In SWSH markets, any set of transacting platform edges uniquely connects seller subgraphs into chains and cycles.
\end{proposition}
Lemma~\ref{lem:at_most_one_edge} says there exists a revenue optimal matching where all platform edges transact. By Proposition~\ref{prop:cycles_chains}, these platform edges uniquely define a set of cycles and chains. Thus, finding the revenue-optimal matching is equivalent to finding the chains and cycles with the highest revenue. For ease of illustration, in this section we consider a market where the number of buyers $n$ equals the number of sellers $m$ \footnote{Our proof in Appendix~\ref{app:swsh_reduction} works for all cases of $n$ and $m$.}, and all buyers and sellers transact according to Lemma~\ref{lem:all_transact}. In addition, from here forward, we will suppress any mention of 0-chains and deal only with chains and cycles of positive length, leaving implicit the fact that any buyer who does not transact with sellers as part of a cycle or a chain will be matched to a dangling seller.

Recall that Theorem~\ref{thm:oppo_path} states that the price of a trade in a homogeneous-goods market is determined by the lowest buyer valuation on a buyer's opportunity path. This is precisely what the cycle and chain structure captures. Consider the $0$-chain and the $1$-chain in Example~\ref{example:cycle_chain_illustrate}. Buyer $b_4$'s opportunity path goes through the chain, reaching all buyers. As she has the lowest valuation on this chain, she pays price $1$. Buyer $b_3$ also has the lowest valuation on her own opportunity path, so she pays price $3$. Similar logic applies to the cycle. 
    
However, if $b_2$ and $b_3$ were to swap the sellers that they trade with, both $b_2$ and $b_1$ would have $v_3=3$ as the lowest valuation on their opportunity paths, and they would thus pay $3$ instead of $9$. As one can design the set of cycles and chains to go through only the buyer with the largest valuation in a given seller subgraph, these highest value buyers (i.e., $v(S_j)$ for $S_j$) are crucial for revenue maximization in SWSH markets.

We are now ready to present our results. Assume there are $\ell\leq m$ seller subgraphs in a SWSH market. We first sort all seller subgraphs by their valuations $v(S_1) \geq v(S_2) \geq \cdots \geq v(S_\ell)$. This induces a partial order over all seller subgraphs. We call a $k$-cycle \emph{contiguous} if it comprises of seller subgraph $S_i, S_{i+1}, \ldots S_{i+k-1}$ for some $i$ – that is, no intermediate seller subgraph is skipped. We call a chain \emph{contiguous} if there exists an index $i$ such that the chain is of form $S_\ell \to S_{\ell-1} \to \ldots \to S_{i-1} \to S_i$ and has terminal subgraph $S_i$. Our algorithm to compute a revenue-optimal matching relies on the following structural result, whose proof we give through a series of lemmas in Appendix~\ref{app:swsh_char}.
\begin{restatable}{theorem}{SWSHChar}
\label{thm:opt_char}
    In any SWSH market, there exists a set of revenue-optimal platform edges that, through the largest buyers in each seller subgraph,  connects seller subgraphs into contiguous cycles of length at most $3$, and at most one contiguous chain.
\end{restatable}
 
At a high level, the proof first shows that any cycle of length more than three can be split up into smaller cycles yielding weakly more revenue. Then if a chain or a cycle is not contiguous, we can further divide and/or join them to weakly increase revenue. We now use dynamic programming to leverage this characterization result and efficiently search over the possible optimal configurations as described in Theorem~\ref{thm:opt_char}.
\begin{theorem}\label{thm: poly_AMOS}
    For SWSH markets where $n=m$, there exists a polynomial-time algorithm to find the set of revenue-optimal platform edges.
\end{theorem}

\begin{proof} 
Fix a seller subgraph $S_k$ as the terminal subgraph for the potential chain. $S_k$ uniquely identifies a contiguous chain. There are $\ell\leq m$ such possible chains. Consider the induced graph $G'$ obtained by removing all the subgraphs in the chain from the original graph. 

Let $\mathrm{DP}[i]$ denote the maximum obtainable revenue from connecting the largest $i$ subgraphs in $G'$ via cycles. Let $\mathrm{Rev}(S_{i,j})$ denote the maximum obtainable revenue from the contiguous cycle formed by subgraphs $S_i, S_{i+1},\ldots, S_{j}$. When $j=i$, $\mathrm{Rev}(S_{i,i})=0$. As contiguous cycles have length at most three, $\mathrm{DP}[i]$ can be calculated through dynamic programming. Filling in the base cases for $DP[1], DP[2], DP[3]$, we have the following recurrence for $i=4,...,k-1$:
\begin{align*}
    &DP[i] = \max (DP[i-3] + \mathrm{Rev}(S_{i-2,i}),  DP[i-2] + \mathrm{Rev}(S_{i-1,i}), DP[i-1] + \mathrm{Rev}(S_{i,i}))
\end{align*}
Let $\mathrm{R}[k]$ be the maximum obtainable revenue from the potential chain that is identified by $S_k$. Since this chain is contiguous, $\mathrm{R}[k]$ can be calculated in $O(n^2)$ time, by comparing the revenue when connecting the chain to all possible buyers in $G'$ through $s_k$. It follows that the optimal revenue when there is a contiguous chain identified by $S_k$ is given by $DP[k-1] + R[k]$. Repeating this process for all $\ell$ possible chains, the optimal revenue is given by $\max_{k=1}^{\ell} DP[k-1] + R[k]$. Finding the overall maximum takes $O(n^3)$ time.

Note the above procedure exhaustively searches all configurations of contiguous cycles of length at most 3 and at most one contiguous chain. The procedure thus finds the optimal configuration in terms of revenue. By Theorem~\ref{thm:opt_char}, there exists an optimal solution that satisfies these properties.
\end{proof}

While Theorem~\ref{thm: poly_AMOS} gives a polynomial-time algorithm for the case in which $n = m$, we show that the more general case reduces to this case by carefully discarding some buyers and sellers from the graph. We present the proof of Theorem~\ref{thm:swsh_reduction} in Appendix~\ref{app:swsh_reduction}.

\begin{restatable}{theorem}{SWSHReduction}
\label{thm:swsh_reduction}
    There exists a polynomial-time algorithm to maximize platform's revenue in general SWSH markets.
\end{restatable}
\begin{proof}[Proof Sketch]
    When $n > m$, we show that we can discard all buyers besides those with the $m$ highest valuations without affecting optimality. Similarly, when $m > n$, we show that we can get rid of $m - n$ dangling sellers without affecting optimal revenue. Thus, we are left with the case in which $m = n$, for which Theorem \ref{thm: poly_AMOS} gives us the desired result.
\end{proof}

\subsection{Sparse Homogeneous-Goods-and-Buyers Markets}\label{sec:special-case-identical}
While Theorem~\ref{thm:nphard} only states hardness for the case where buyers have degree at most two, one can check that the reduction yields a world graph that satisfies an additional property:
    {\em For any pair of sellers, at most one buyer knows them both.}
We call the world graphs that satisfy this additional property ``sparse graphs''. By considering these sparse graphs with buyers of degree at most 2, and restricting the class of valuations from $v_{ij} \in \{0, v_i\}$ to $v_{ij} = c$ for some constant $c>0$, we show that the platform's problem becomes tractable. We call these markets SHGB (Sparse Homogeneous-Goods-and-Buyers) markets. We begin by showing that finding the optimal set of platform edges in SHGB markets reduces to a graph-theoretic problem involving sets of buyers with non-positive {\em surplus}. Proofs for this section are intricate and are all presented in Appendix~\ref{app:hv_identity}. 
\begin{definition}[Surplus]
    In a bipartite market $G=(B,S,E_w)$, consider a set of buyers $B_v$. Let $N(B_v)$ be the set of all sellers who are adjacent to at least one buyer in $B_v$ via a world edge. The surplus of $B_v$ is defined as $|N(B_v)| - |B_v|$\footnote{For a set $B_v$, we use $|B_v|$ to denote its cardinality.}.
\end{definition}
\begin{restatable}{lemma}{IdentityChar}
\label{lemma:identity_char}
    In SHGB markets, there exists a set of platform edges generating revenue $x$ if and only if there exists a set of buyers $B_v$ of non-positive surplus such that
    $$\min\{|B_v|, |S|\} - |N(B_v)| + k_v = x,$$
    where $k_v$ is the cardinality of the maximum matching between $B_v$ and $N(B_v)$ using platform edges.
\end{restatable}

For a set of buyers $B_v$ with non-positive surplus, the proof minimizes the number of buyers that connect to an unsold item through opportunity paths. Revenue $k_v$ is generated from buyers that trade with sellers in $N(B_v)$, and revenue $\min\{|B_v|,|S|\} - |N(B_v)|$ is attained from the remaining buyers who do not transact with $N(B_v)$.
We then prove that except for a few cases that can be checked in polynomial time, $k_v=|N(B_v)|$. This further simplifies the platform's problem. 

\begin{restatable}{lemma}{MaxCard}
\label{lemma:max_cardinality}
    In SHGB markets, the platform can find the set of revenue-optimal platform edges by finding the maximum set of buyers with non-positive surplus.
\end{restatable}

With this reduction in hand, we give a polynomial-time algorithm to find the maximum set of buyers with non-positive surplus. To do this, we rely on the graph structure in SHGB markets, identifying the buyers and sellers in these markets with edges and vertices of a general graph. Our problem then translates to finding the set of edges in this graph whose induced subgraph contains no more than a certain number of vertices. This is achievable using standard graph algorithms.
\begin{restatable}{theorem}{PolyIdentity}
\label{thm:poly_identity}
    There exists a polynomial-time algorithm to solve the platform's problem
 in SHGB markets.
\end{restatable}

\section{Guarantees for General Markets}\label{sec:ch4_general_markets}
Having established the hardness of the platform's problem in the general case and shown some special cases that can be solved in polynomial time, we  turn to analyzing the relationship between revenue optimization and social welfare.

In Section~\ref{sec:welfare_into_revenue}, we provide a lower bound for revenue based on the potential welfare increase the platform can generate. This shows when the platform can effectively disrupt the market and improve overall welfare, the platform's revenue is guaranteed to be large. Specifically, we show in Section~\ref{sec:log_rev_guarantee} that there is a polynomial-time algorithm to extract a logarithmic approximation of the optimal welfare as revenue in such markets. In Section~\ref{sec:PRM}, we study the other direction of this relationship. We prove that in markets where the platform optimizes for revenue, social welfare is at least an $\Omega(1/\log(\min\{n,m\}))$-fraction of the optimal welfare. These results explain the sense in which revenue maximization can go hand-in-hand with welfare considerations in a platform economy.

\subsection{Converting Potential Welfare Increase to Revenue}\label{sec:welfare_into_revenue}

There are two primary ways that the platform can make revenue. The first way is by ``monopolizing'' the existing welfare of the world graph. 

\begin{example}\label{exam:mono_welfare}
    To illustrate this ``monopolization,'' consider a graph with two buyers and a single seller. Buyer $b_1$ has value $1$ for the item and is connected to the seller via a world edge while buyer $b_2$ has value $1 + \epsilon$ and has no incident world edges. In this case, the platform can add a single platform edge from $b_2$ to the seller. Now, buyer $b_2$ transacts instead of buyer $b_1$, and the platform obtains revenue $1 + \epsilon$ while providing only $\epsilon$ in added welfare.
\end{example}

The second way is by facilitating new high-value transactions that were not possible without the assistance of the platform. In these cases, the platform has the potential to add a substantial amount of welfare to the world graph and claim some of the value for itself. The question is how much of this increase in welfare the platform is able to retain as revenue.
 We give a polynomial-time process to convert the potential increase in welfare into revenue below.
\begin{theorem}
\label{thm:gen_welfare_conversion}
    In a general market, suppose there exists a set of $k$ platform edges that increase social welfare by $\Delta W=W(G_p)-W(G_w)$. Then one can find a subset of these edges that yields revenue at least $\Delta W/H_k$ in polynomial time, where $H_k=\sum_{i=1}^k 1/i$ is the $k$th harmonic number.
\end{theorem}
\begin{proof}
    We prove this inductively. When $k = 1$, there is a single platform edge $(b_1, s_1)$ that can be added to increase welfare by $W$. As removing $s_1$ is weakly worse than removing edge $(b_1, s_1)$, the revenue/price of this edge is given by $W(G_p) - W(G_p \setminus \{s_1\})\geq W(G_p) - W(G_p \setminus \{(b_1,s_1)\})\geq \Delta W$.

    Now, suppose that it holds for $k = n - 1$. Consider a set of $k = n$ platform edges that adds welfare $W$. If every edge yields revenue at least $\frac{\Delta W}{n \cdot H_n}$, then it follows that we are guaranteed revenue at least $n \cdot \frac{\Delta W}{n \cdot H_n} = \frac{\Delta W}{H_n}$ and we are done. If not, then we can remove an edge $(b_i,s_j)$ that yields revenue strictly less than $\frac{\Delta W}{n \cdot H_n}$. By removing this edge, we can only decrease the added welfare by at most $W(G_p)-W(G_p\setminus\{(b_i,s_j)\})\leq W(G_p)-W(G_p\setminus\{s_j\}) \leq   \frac{\Delta W}{n \cdot H_n}$. Thus, we are left with a set of $n - 1$ platform edges that adds welfare at least
    $$
        \Delta W - \frac{\Delta W}{n \cdot H_n} = \frac{n H_n - 1}{n \cdot H_n} \cdot \Delta W = \frac{H_n - \frac{1}{n}}{H_n} \cdot \Delta W = \frac{H_{n-1}}{H_n} \cdot \Delta W.
    $$
    Applying our inductive hypothesis, we can guarantee revenue at least
    $$
        \frac{1}{H_{n-1}} \cdot \frac{H_{n-1}}{H_n} \cdot \Delta W = \frac{\Delta W}{H_n}
    $$
    from some subset of these $n-1$ platform edges, concluding the proof.
\end{proof}

The proof argument in Theorem~\ref{thm:gen_welfare_conversion} yields the following simple greedy algorithm.
 Given a set of $k$ platform edges that increase social welfare by $\Delta W$, iteratively remove the edge that offers the least revenue to the platform until left with a single edge. 
Keep track of the revenue at each step, and select the set of edges that yields the highest revenue overall. The proof argument
 shows that this set will guarantee revenue 
 at least $\Delta W/H_k$, and this process  takes
 polynomial time.

In order to achieve $\Delta W/H_k$ as revenue, the platform adds a subset of all edges that contribute to the $\Delta W$ welfare increase, not necessarily all of the edges. Proposition~\ref{prop:tight_welfare_conversion} further shows the $H_k$ conversion rate from the potential welfare increase to revenue is tight.
\begin{figure}[t]
    \centering
    \begin{tikzpicture}[scale=1.3]
        \foreach \i/\label in {1/$b_1^d$, 2/$b_2^d$, 3.5/$b_k^d$}
            \node[draw, shape=rectangle, minimum size=0.6cm] (\label) at (\i, 2) {\label};

        \foreach \i/\label in {5.5/$b_1$, 6.5/$b_2$, 8/$b_k$}
            \node[draw, shape=rectangle, minimum size=0.6cm] (\label) at (\i, 2) {\label};
            
        \foreach \i/\label in {1/$s_1^d$, 2/$s_2^d$, 3.5/$s_k^d$}
            \node[draw, shape=circle, minimum size=0.6cm] (\label) at (\i, 0) {\label};

        \foreach \i/\label in {5.5/$s_1$, 6.5/$s_2$, 8/$s_k$}
            \node[draw, shape=circle, minimum size=0.6cm] (\label) at (\i, 0) {\label};

        \foreach \x/\y in {$b_1^d$/$s_1^d$, $b_2^d$/$s_2^d$, $b_k^d$/$s_k^d$}
            \draw[line width=1pt, customblue, dashed] (\x) -- (\y);

       \foreach \x/\y in {$b_1^d$/$s_1$, $b_1^d$/$s_2$, $b_1^d$/$s_k$, $b_2^d$/$s_1$, $b_2^d$/$s_2$, $b_2^d$/$s_k$, $b_k^d$/$s_1$, $b_k^d$/$s_2$, $b_k^d$/$s_k$, $b_1$/$s_1$, $b_1$/$s_2$, $b_1$/$s_k$, $b_2$/$s_1$, $b_2$/$s_2$, $b_2$/$s_k$, $b_k$/$s_1$, $b_k$/$s_2$, $b_k$/$s_k$}
            \draw[line width=1pt, black, solid] (\x) -- (\y);
     
        \node at (2.75, 2) {$\ldots$};
        \node at (7.25, 2) {$\ldots$};
        \node at (2.75, 0) {$\ldots$};
        \node at (7.25, 0) {$\ldots$};

        \node[left] at (0.5, 2)  {Buyers};
        \node[left] at (0.5, 0)  {Sellers};
    
    \end{tikzpicture}
        \caption{The market used in the proof of Proposition~\ref{prop:tight_welfare_conversion}. Buyers $B_k=\{b_1,b_2,...,b_k\}$ and dummy buyers $B^d_k=\{b^d_1,b^d_2,...,b^d_k\}$ are fully connected to sellers $S_k=\{s_1, s_2,...,s_k\}$ through world edges, denoted by solid black edges. Dummy sellers $S^d_k=\{s^d_1, s^d_2,...,s^d_k\}$ are not connected to any buyers through world edges.
        Buyer $b_i\in B_k$ values all sellers in $S_k$ at $1/i$. Buyer $b^d_i\in B^d_k$ values all sellers at $1$. All other valuations are zero. The maximum welfare the platform can add through platform edges is $\Delta W = H_k$, indicated in dashed blue edges. The optimal revenue is $1$, obtained by adding any non-empty subset of the blue edges.} \label{fig:conv_tight}
\end{figure}
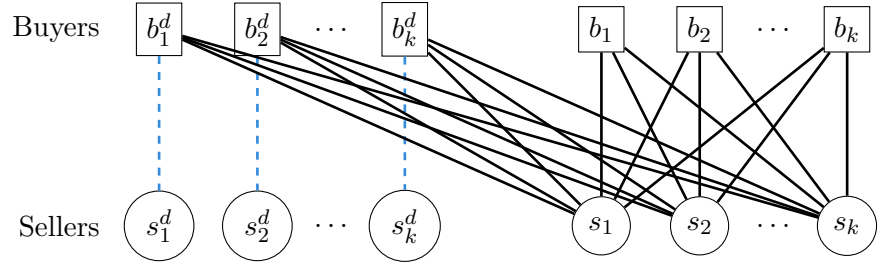

\begin{restatable}{proposition}{tightWelfareConversion}
\label{prop:tight_welfare_conversion}
     For all $k$, there exists a world graph $G_w$ and a set of $k$ edges adding welfare $\Delta W$ such that the optimal platform revenue is precisely $\Delta W/H_k$.
\end{restatable}
\begin{proof}
    For any $k$, Figure~\ref{fig:conv_tight} presents such a market.
 World edges are solid in black and platform edges are dashed in blue. There are buyers $B_k=\{b_1,b_2,...,b_k\}$, dummy buyers $B^d_k=\{b^d_1,b^d_2,...,b^d_k\}$, sellers $S_k=\{s_1, s_2,...,s_k\}$ and dummy sellers $S^d_k=\{s^d_1, s^d_2,...,s^d_k\}$. All buyers $B_k$ and $B^d_k$ are fully connected to sellers $S_k$ through world edges. Dummy sellers $S^d_k$ have no connections in the world graph. Buyer $b_i\in B_k$ values all sellers in $S_k$ at $1/i$. Buyer $b^d_i\in B^d_k$ values all sellers at $1$. All other valuations are zero. By adding platform edges to form a perfect matching between the dummy buyers and the dummy sellers, the platform can add welfare $\Delta W = \sum_{i=1}^k 1/i = H_k$ to the world graph. However,   the platform's optimal revenue is bounded above by $1$. The platform has no incentive to add platform edges to $B_k$ as they have value $0$ for all dummy sellers. Thus, the only thing the platform can do is match $\ell\in \{1,\ldots,k\}$ dummy buyers to the dummy sellers. If the platform matches $\ell$ of the dummy buyers to the dummy sellers, each transaction occurs at price $1/\ell$, and total revenue is $\ell \cdot \frac{1}{\ell} = 1$. Thus, we 
 have that the optimal revenue is given by $\Delta W/H_k = H_k/H_k = 1$.
\end{proof}

\subsection{Logarithmic Revenue Guarantees in Inefficient Markets}\label{sec:log_rev_guarantee}

Theorem~\ref{thm:gen_welfare_conversion} shows that the platform can convert a potential welfare increase into revenue in polynomial time. Consider now a minimal set of platform edges such that optimal welfare $W^\star$ is achieved in the resulting platform graph. If we apply Theorem~\ref{thm:gen_welfare_conversion} to this set of platform edges, then we can lower bound the platform's revenue as a function of the maximum increase in welfare.
\begin{corollary}
\label{corr:gen_welf_conv}
    In a general market where the platform's maximum contribution to welfare is $W^\star-W(G_w)$, the platform can obtain revenue $\frac{W^\star - W(G_w)}{H_{\min(n, m)}}$ in polynomial time.
\end{corollary}

If the optimal welfare $W^\star$ is large relative to welfare in the world graph $W(G_w)$, then the platform can add a substantial amount of welfare. In this case Corollary~\ref{corr:gen_welf_conv} implies that the platform can also obtain a substantial amount of revenue. Theorem \ref{thm:log_approx} makes this  precise.
\begin{theorem}
\label{thm:log_approx}
    In general markets where $W^\star - W(G_w) = \Omega(W(G_w))$, the platform can obtain $\frac{W^\star}{O(\log(\min(n, m)))}$ in revenue. This implies an $O(\log(\min(n, m)))$ polynomial-time approximation algorithm for revenue.
\end{theorem}
\begin{proof}
    We have $W^\star = (W^\star - W(G_w)) + W(G_w) = \Theta(W^\star - W(G_w))$. The rest of the proof follows from Corollary~\ref{corr:gen_welf_conv} and the fact that $\mathrm{Rev}^\star \leq W^\star$.
\end{proof}

\begin{remark}
  Theorem~\ref{thm:log_approx} does not just imply a logarithmic approximation to the optimal revenue in inefficient markets. It shows a stronger result: the platform can efficiently obtain a logarithmic fraction of the optimal welfare as revenue in this class of markets. 
\end{remark}

Though Theorem~\ref{thm:log_approx} only provides a revenue guarantee for markets with large potential welfare improvements, these are the markets that we care about and the ones platforms typically aim to enter and disrupt. This theorem can also be seen as a motivation for why revenue-interested platforms tend to disrupt markets with large inefficiencies. In markets that are already relatively efficient, optimal revenue can indeed be much larger than the maximum welfare increase, as shown in Example~\ref{exam:mono_welfare}, but sellers and buyers may have few incentives to use the platform in the first place.

\subsection{Bounding the Welfare Loss from Revenue Maximization}\label{sec:PRM}

Having shown how to lower bound revenue with optimal welfare, we now explore the converse question: how much welfare is guaranteed under the revenue-optimal matching? In other words, we want to understand the welfare loss, compared to the optimal welfare, due to the platform inefficiently matching buyers and sellers in order to maximize its own revenue.
 To formalize this loss in welfare, we define the {\em price of revenue maximization}.
\begin{definition}[Price of Revenue Maximization (PRM)]
    For any market $G_w=(B,S,E_w)$ and valuation profile $\textbf{v}$, let $\Pi_p^\star(G_w, \textbf{v})$ denote the set of all revenue-optimal platform edge configurations. Each element $E^\star_p\in \Pi_p^\star(G_w, \textbf{v})$ is a configuration of revenue-optimal platform edges in market $G_w$.
    The {\em price of revenue maximization} is the largest ratio of the optimal welfare to the welfare of revenue-maximizing platform graph, across any market and valuation profile:
    \begin{align*}
        \mathrm{PRM} = \max_{G_w, \textbf{v}, E^\star_p \in \Pi_p^\star(G_w, \textbf{v})} \frac{W^\star(G_w)}{W(B,S,E_w\cup E^\star_p)}.
    \end{align*}
\end{definition}

To better understand the extent to which the platform's incentives can negatively affect social welfare, we present the following bounds on the PRM for general markets.
\begin{figure} 
    \centering
    \begin{tikzpicture}[scale=1.3]
        \foreach \i/\label in {1/$b_1$, 3/$b_2$}
            \node[draw, shape=rectangle, minimum size=0.6cm] (\label) at (\i, 2) {\label};
            
        \foreach \i/\label in {1/$s_1$, 3/$s_2$}
            \node[draw, shape=circle, minimum size=0.6cm] (\label) at (\i, 0) {\label};

        \foreach \x/\y in {$b_2$/$s_1$}
            \draw[line width=1pt, blue, dashed] (\x) -- (\y);

        \foreach \x/\y/\w/\pos/\loc in {$b_1$/$s_1$/1/midway/left, $b_2$/$s_2$/$\epsilon$/midway/right}
            \draw[line width=1.5pt, customblue, dashed] (\x) -- node[\pos, \loc, font=\footnotesize] {\w} (\y);

        \foreach \x/\y/\w/\pos/\loc in {$b_1$/$s_2$/1/near end/below left}
            \draw[line width=1.5pt, violet, dashed] (\x) -- node[\pos, \loc, font=\footnotesize] {\w} (\y);

        \foreach \x/\y/\w/\pos/\loc in {$b_2$/$s_1$/1/near end/below right}
            \draw[line width=1.5pt, black, solid] (\x) -- node[\pos, \loc, font=\footnotesize] {\w} (\y);

        \node[left] at (0.5, 2)  {Buyers};
        \node[left] at (0.5, 0)  {Sellers};
    
    \end{tikzpicture}
        \caption{A two-buyer, two-seller market in which the $\mathrm{PRM}$ approaches $2$. The world edge is marked as a black solid edge. Buyer values are annotated next to edges. The welfare-optimal platform edge is in purple while the revenue-optimal set of platform edges is shown in blue. \label{fig:poa_lower_bound}}
\end{figure}
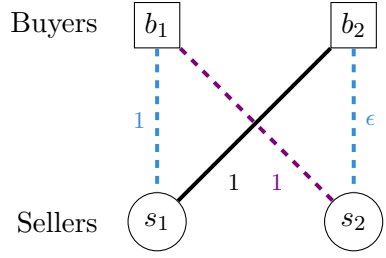 

\begin{restatable}{proposition}{poaLowerBound}
\label{prop:poa_lower_bound}
    The price of revenue maximization is bounded below by $2$.
\end{restatable}
\begin{proof}
    Consider the  two-seller, two-buyer market in Figure~\ref{fig:poa_lower_bound}. The welfare-optimal matching adds edge $(b_1, s_2)$ for a total welfare of $2$ and revenue of $1$. The revenue-optimal matching adds edges $(b_1, s_1), (b_2, s_2)$ for a total welfare and revenue of $1 + \epsilon$. This yields a ratio of $\frac{2}{1 + \epsilon} \to 2$ as $\epsilon \to 0$. Here, we again see the two primary ways in which the platform can obtain revenue. In the revenue-optimal strategy, the platform forgoes adding the socially optimal edge $(b_1, s_2)$ in favor of "monopolizing" the revenue from seller $s_1$ by adding edge $(b_1, s_1)$ instead. This allows it to facilitate another transaction, namely $(b_2, s_2)$, that would not be possibly under the welfare-optimal configuration.
\end{proof} 

While Proposition~\ref{prop:poa_lower_bound} demonstrates that the platform's strategic behavior may come at the cost of social welfare, the procedure of Theorem~\ref{thm:gen_welfare_conversion} allows us to bound this cost from above. Namely, the difference in optimal welfare and the welfare of revenue-optimal matching cannot exceed a logarithmic factor; otherwise, we could convert this difference in welfare into strictly more revenue than we obtain from the revenue-optimal matching.
\begin{theorem}
\label{thm:poa_upper_bound}
    In general markets, the PRM is bounded above by $H_{\min(n, m)} + 1 = O(\log(\min(n, m)))$.
\end{theorem}
\begin{proof}
    We show $PRM \leq H_{\min(n, m)} + 1$. Suppose otherwise, and there exists a market $G_w$ that satisfies $W^\star > (H_{\min(n, m)} + 1) \cdot W(G_w \cup E_p^\star)$, where $E_p^\star$ is a set of revenue-optimal platform edges. Then, there would exist a set of $k\leq \min(n, m)$ edges $E'_P$ such that, by Theorem \ref{thm:gen_welfare_conversion}, we have
    \begin{align*}
        \mathrm{Rev}(E'_P) &\geq \frac{W^\star - W(G_w)}{H_{k}} \geq \frac{W^\star - W(G_w \cup E_p^\star)}{H_{\min(n, m)}} > \frac{(H_{\min(n, m)} + 1) W(G_w \cup E_p^\star) - W(G_w \cup E_p^\star)}{H_{\min(n, m)}}\\
        &= W(G_w \cup E_p^\star) \geq \mathrm{Rev}(E_p^\star),
    \end{align*}
    and this is a contradiction as $E_p^\star$ is revenue-optimal.
\end{proof}

\section{Guarantees for Homogeneous-Goods Markets}\label{sec:hom_markets}

While the results of Section \ref{sec:ch4_general_markets} offer  guarantees on both welfare loss and approximately optimal revenue, it is natural to consider if other markets might admit stronger still
 guarantees. As seen in the proof of Proposition \ref{prop:tight_welfare_conversion}, even when we turn to the case in which each buyer  values all desired items equally $(v_{ij} \in \{v_i, 0\})$, the bound on welfare conversion into revenue is already tight.
 This motivates us to consider the next most natural class of valuations $(v_{ij} = v_i)$.
 In this section, we develop very strong welfare and revenue guarantees in homogeneous-goods markets.

\subsection{Converting Potential Welfare Increase to Revenue without Loss}

We first consider whether the platform can more efficiently convert the potential welfare increase in homogeneous-goods markets into revenue. Theorem \ref{thm:hom_conversion} shows that the platform can achieve this conversion 
perfectly, without any loss.
\begin{theorem}
\label{thm:hom_conversion}
    In homogeneous-goods markets, the platform can always extract
 revenue at least $W^\star - W(G_w)$ in polynomial time.
\end{theorem}
\begin{proof}
    As goods are homogeneous, the optimal welfare is obtained by always matching the buyers with the top $\min(m, n)$ highest valuations. Fix some arbitrary maximum weight matching $M^\star$ in the complete graph, and let $M$ be the max weight matching in $G_w$. Let $B_{old} = \{\tilde{b}_1, \dots, \tilde{b}_{\ell}\}$ represent the set of buyers matched in $M$ but unmatched in $M^\star$, and let $B_{new} = \{b^\star_1, \dots, b^\star_k\}$ represent the set of buyers that are matched in $M^\star$ but are unmatched in $M$. Weakly more buyers are matched in $M^\star$ than in $M$; otherwise the leftover sellers could always match with the buyers in $B_{old}$ that they were matched to in $M$. Thus, we have $k \geq \ell$. We  show that the platform can add edges and extract full welfare as revenue from each buyer in $B_{new}$, implying a lower bound of revenue $W^\star - W(G_w)$.

    Let $v_{(\min(n, m))}$ denote the $\min(n, m)$-th largest valuation among all buyers. Since in $M^\star$ the largest $\min(n, m)$ buyers transact, it must be the case that every buyer in $B_{new}$ has valuation weakly greater than $v_{(\min(n, m))}$ and every buyer in $B_{old}$ has valuation strictly less than $v_{(\min(n, m))}$. For each buyer $b^\star_i \in B_{new}$, add a platform edge to the seller that $\tilde{b}_i$ transacted with in $M$, or to a seller who does not sell in $M$ if $i > \ell$. Since $k-\ell$ more buyers transact in $M^\star$ compared to $M$, there are at least $k-\ell$ sellers who do not sell in $M$. Denote the set of platform edges added as $E_p$. Then all buyers in $B_{new}$ still transact in the maximum matching of $G\cup E_p$ through $E_p$. We now prove that every buyer $b_i^\star\in B_{new}$ pays the full price $v_{i}^\star$ by reasoning about the buyers' opportunity paths.

    Since $b^\star_i$ does not transact in $M$, there are no sellers with strictly smaller values on any of its opportunity paths. Otherwise, transactions would change along the opportunity path and $b^\star_i$ would transacted in $M$. As buyers in $B_{old}$ have valuation strictly less than $v_{\min(n,m)}$, they are not on $b^\star_i$'s opportunity path. Thus, in the max weight matching of $G_w\cup E_p$, where $B_{old}$ no longer transacts, $b^\star_i$ remains the buyer with smallest valuation on its own opportunity path and pays $v^\star_i$.
\end{proof}

We make two observations about the procedure in Theorem~\ref{thm:hom_conversion} to extract $W^\star-W(G_w)$ in revenue.
First, the procedure  also maximizes social welfare. Second, to extract revenue $W^\star-W(G_w)$, the platform needs to carefully add platform edges. There are platform edges that increase total welfare by $W^\star-W(G_w)$ 
that do not yield the desired revenue. As an example, consider the market  in Figure~\ref{fig:conv_tight}
and used in the proof of Proposition~\ref{prop:tight_welfare_conversion}. If we modify the market such that all buyers have homogeneous valuations, then adding the dashed blue edges, while yielding the optimal welfare, still only gives revenue $1$. To extract the total amount of added welfare, $H_k$, the platform must add edges between $b_i$ and $s_i^d$. 

As in the case of general markets, Theorem~\ref{thm:hom_conversion} naturally leads to an approximation algorithm for optimal welfare (and hence optimal revenue) when the platform has a way to add
 a large amount of welfare. In these market settings, we can guarantee a constant-factor approximation to optimal revenue, improving on the logarithmic guarantee given for general valuations.
\begin{theorem}
\label{thm:hom_const_approx}
    In homogeneous-goods markets where $W^\star - W(G_w) = \Omega(W(G_w))$, there exists a polynomial-time algorithm that yields a constant-factor approximation of the optimal welfare as revenue to the platform. This implies a constant-factor approximation to the optimal revenue.
\end{theorem}
\begin{proof}
    By Theorem \ref{thm:hom_conversion}, we can always guarantee $W^\star - W(G_w)$ in revenue in polynomial time. Since $W^\star - W(G_w) = \Omega(W(G_w))$, we have that $W^\star - W(G_w) = \Omega(W^\star)$. The final part of the theorem follows from the fact that $\mathrm{Rev}^\star \leq W^\star$.
\end{proof}

While we choose to focus on the case where $W^\star - W(G)$ is large, as we believe these are the markets platforms are most likely to enter, we show that we can more generally extract a $1/\min\{n,m\}$-fraction of the optimal revenue in polynomial time in any homogeneous-goods market. We defer the proof to Appendix~\ref{app:min_nm_approx_rev}.
\begin{restatable}{theorem}{HomoMinNM}
\label{thm:hom_min_nm_approx}
    In homogeneous-goods markets, there exists a polynomial-time algorithm that yields a $\min\{n,m\}$-approximation of the optimal revenue.
\end{restatable}
In order to guarantee a $\min\{n,m\}$-approximation of the optimal revenue, we show that the platform can efficiently maximize the revenue it obtains from a single transaction. As the platform can facilitate at most $\min\{n,m\}$ transactions in total, this yields the desired approximation ratio.

\subsection{Zero Welfare Loss from Revenue Maximization}
As with general markets, we now turn to bounding the welfare loss incurred as a result of the platform's self-interest. Adopting the same approach as  used in proving Theorem~\ref{thm:poa_upper_bound} for general markets, Theorem~\ref{thm:hom_conversion} implies that the price of revenue maximization is bounded above by $2$. While this is already a strong result, we can in fact  show that the price of revenue maximization in homogeneous-goods markets is  $1$. That is, revenue maximization  aligns exactly with welfare maximization in homogeneous-goods markets. 
\begin{restatable}{theorem}{PRMhomo}
\label{thm:hom_poa}
    In homogeneous-goods markets, the price of revenue maximization equals $1$.
\end{restatable}

\begin{proof}
    Let $\tilde{B}$ represent the set of buyers with the top $\min(n, m)$ highest valuations – i.e. the buyers who transact in the welfare-optimal matching. Suppose for the sake of contradiction that the revenue-optimal matching is not welfare optimal and thus some buyer $b_i \in \tilde{B}$ with $v_i \neq 0$ does not transact in the revenue-optimal matching. All opportunity paths from $b_i$ must lead to buyers with weakly higher values. If not, transactions can change along the opportunity path, resulting in a matching with strictly larger weight. This contradicts with the first welfare theorem of competitive equilibrium. 

    If there is a non-transacting seller $s_j$ in the revenue-optimal matching, then adding the edge $(b_i, s_j)$ strictly increases revenue. All buyers with opportunity paths previously ending at $s_j$ paid a price zero as seller $s_j$ did not transact; after the addition of this edge, they pay price $v_i > 0$, leading to a strict increase in revenue.

    If all sellers transact in the revenue-optimal matching, let $b_k$ be the transacting buyer with the smallest valuation and suppose that they currently transact with seller $s_\ell$. As the revenue-optimal matching is assumed not to be welfare optimal, we must have that $v_k < v_i$. By adding the edge $(b_i, s_\ell)$, revenue again strictly increases. All buyers with opportunity paths previously leading to $s_\ell$ must pay a weakly higher price now, since $v_i > v_k$ and all opportunity paths from $b_i$ lead to buyers with weakly higher values. In addition, we obtain revenue $v_i$ from the edge $(b_i, s_\ell)$ and lose revenue at most $v_k < v_i$ from the exclusion of edge $(b_k, s_{\ell})$, in the event that this was a platform edge. Thus, this leads to a strict increase in revenue.

    In either case, there exists a modification to the revenue-optimal matching that yields strictly higher revenue, concluding the proof.
\end{proof}

\section{Discussion}

We have studied the incentives facing platforms when  facilitating transactions between buyers and sellers, modeled by a bipartite buyer-seller network to which the platform can choose to add new links, introducing buyers to sellers.
The general problem of maximizing the platform's revenue is computationally hard, even under restrictive structural assumptions. By imposing additional structure, we give polynomial-time algorithms for  special classes of markets, establishing a ``frontier of tractability.'' We also examine the relationship between social welfare and platform's revenue, and show that  where the platform can substantially increase social welfare, it
 can also extract substantial revenue. This yields an $O(\log(\min\{n, m\}))$-approximation algorithm for revenue in these kinds of inefficient
markets, as well as an upper bound of $O(\log(\min\{n, m\}))$ on the impact of platform self interest on welfare, relative to the optimal welfare in general markets. These bounds can be substantially improved in homogeneous-goods markets, where we prove that revenue maximization is perfectly aligned with welfare maximization, and give a constant-factor approximation algorithm for optimal revenue in  inefficient  markets. There are many interesting directions for future work:

\begin{itemize}
    \item The current gap between the lower and upper bounds for the effect of platform self interest
    on welfare (the {\em price of revenue maximization} or PRM) is quite large: $2$ vs $H_{\min(n, m)}=\Theta(\log_{\min(n, m)})$. We conjecture that $\mathrm{PRM} = 2$ in general markets. Closing this gap has important consequences for the impact of platforms on social welfare.
    \item  Can we resolve the complexity of  intermediate classes of markets; e.g.,  is it possible to exactly maximize revenue in homogeneous-goods markets in polynomial time? 
    \item We have assumed that sellers are indifferent between transacting off- and on-platform, whereas in real life sellers may choose to leave and transact with buyers off the platform if transaction fees are too high. It is interesting to model these kinds of considerations and understand which of our results still hold.
    \item Our model assumes full information with both valuations and world edges known to the platform. In reality, platforms only have partial information. Extending the model to a Bayesian setting, in which the platform faces uncertainty about buyer values and world edges, is also an interesting direction for follow up. 
\end{itemize}
\chapter{Matching under Regulatory Intervention during Economic Shocks} \label{chap:simulations}

\section{Introduction}
Market-driven platforms, such as  Amazon, DoorDash, Uber, and TaskRabbit, play an increasingly important role in today’s economy, bringing together parties  to facilitate trade and presenting new ways to create value.
First, they reduce \emph{search cost} by introducing potential matches that were not known before, and  second, they reduce \emph{fulfillment cost} by  taking care of service or product delivery, thus reducing the effort made  to complete transactions.

The importance of the platform-based economy became even more apparent during the COVID-19 pandemic, especially for the restaurant industry.
Stay-at-home orders, together with the closure of dine-in channels and caution in regard to visiting brick-and-mortar businesses increased the fulfillment cost of consumers transacting in physical locations.
This led to an increasing number of users and restaurants to adopt food-delivery platforms. 
One study of Uber Eats from February through May 2020 showed a surge in both demand and supply after the shelter-in-place guidance was issued in the U.S. \citep{raj2021}.
%
At the same time,  this new prominence gave platforms  increased market power. 
As a demonstration, restaurant commission fees are set by some platforms to  30\% per order, leaving traditional restaurants, many of whom no longer have dine-in revenue, with low or even negative margins \citep{mckinsey2021}. 
In December 2020, the National Restaurant Association reported that more than 110,000 U.S.~restaurants---one in six---have permanently closed down since the start of pandemic \citep{NRA2020}.
To support restaurants,  states such as New York and California have imposed commission caps and yet platforms responded with countermeasures: the day after Jersey City enforced a 10\% cap on fees, Uber Eats added a \$3 delivery fee to customers and reduced the delivery radius for restaurants.%
\footnote{\url{https://www.restaurantbusinessonline.com/operations/another-city-caps-third-party-delivery-commissions} data accessed: April 03, 2026}
This speaks to the complexity of the ecosystem, and regulatory policies on delivery platforms have been continuously proposed and debated.\looseness=-1

In this work, we use the methods of AI and multi-agent simulation to study a platform-based economy under market shocks, with this as a first step towards reproducing phenomena observed in the real-world  economy and as a tool for conducting counterfactual analysis to understand platform behavior in response to different regulations. 
We develop a multi-agent Gym environment to capture major aspects of the economy in a multi-period setting, with key modeling choices (e.g., epoch-based decision making, user
behavior inertia, the discrete-logit choice model) based on the
 economic literature. 
Our model further captures a full cycle of market shock, designed to represent the pre-, during, and post-crisis periods.

We formulate the platform's problem as a \emph{partially observable Markov decision process} (POMDP), with both commonly observable components (e.g., shock intensity) and private elements (e.g., buyers' knowledge about sellers, off-platform transactions).
We model the platform as a rational agent that uses reinforcement learning (RL) to set fees and match buyer queries to sellers. 
Buyers and sellers decide whether to join the platform, considering both fees and search and fulfillment costs; on-platform buyers further decide whether to transact with a matched
 platform seller or off platform.

We conduct extensive simulations to explore a range of settings that differ in market structures (i.e.,  locations of buyers and sellers in product and preference space), knowledge levels of buyers about sellers, and the cost of off-platform fulfillment.
The goal is to use RL to model the optimal behavior of a platform under different regulations, and study the effect of a platform on the efficiency and resilience of the overall economic system. 
%
%
In the absence of any regulation, we find that a revenue-maximizing platform, even while helping to facilitate trades, tends to leverage its increased market power during the shock to raise fees and extract surplus from buyers and sellers. This leads to seller shutdowns and lower post-shock economic welfare, compared with that of the pre-shock period.\looseness=-1

%

As a first kind of regulation, we consider a class of {\em taxation policies} that impose different tax rates on different categories of profit (e.g., revenue made from user subscriptions, or revenue made from transaction referrals), and study the platform response. 
We further consider regulations that {\em cap platform fees} in a particular way.
Our results show that either taxation, or capping some subset of fees, simply leads the platform to transfer loss to users by adjusting other fees, demonstrating the power of making use of RL to model
the rational behavior of a platform.
On the other hand, capping {\em all} fees  and not just a subset
 has a moderately positive effect on protecting sellers from bankruptcy and promoting a resilient ecosystem. 
In practice, however, this  intervention requires a regulator to have detailed knowledge in setting
 these caps (and also assumes  the platform will follow a particular, myopic matching policy).
%
%
%

The second part of our study allows the platform to retain
full flexibility in choosing how to match buyers and sellers while restricting it to keep the same fee structure that it chooses in the absence of shocks.
Thus, it requires no special knowledge on the part of a regulator, and  
gives the platform full flexibility in regard to behavior (matching)
 that is  proprietary and not easy to regulate.
We show that under this intervention, a revenue-maximizing platform learns to use matching to retain a more diverse set of sellers on platform, so as to increase its long-term revenue from user registrations.
This also helps to promote the efficiency, resilience, and seller diversity in
 the overall economic system.

The present framework, which introduces a multi-agent Gym environment and uses RL to derive optimal platform responses, provides a tool for understanding regulations and platform economies. 
Our hope is that this can 
complement economic theory, which can become analytically intractable in complex, dynamic environments,
as well as pure data-driven approaches that cannot answer questions about changing market and agent behaviors.
We return to a discussion of the opportunities and outstanding challenges with this kind of AI-based approach to economic study at the end of the paper.\looseness=-1

\vspace{-2ex}
\subsection{Related Work}
%
\paragraph{\textbf{Platform Models.}} 
An extensive economic literature  focuses on how to establish network effects through fee-setting or subsidizing one side of the market under various forms of platform competition (e.g., single- vs. multi-homing) \citep{caillaud2003chicken, rochet2003platform, Armstrong2006}.
To facilitate equilibrium analysis, these models often require simplified assumptions for tractability, e.g., restricting to a single round of platform fee-setting and agent subscribing, adopting a fixed platform matching policy, and assuming homogeneous non-platform actors.
%
%
%
Besides related literature that characterizes platform behavior, there are also works on complexity results in regard to setting optimal fees (even for a market with one buyer and two periods) \citep{PapadimitriouPP16} and optimally matching buy queries to sellers (even for a single period)~\citep{mladenov2020optimizing}.

%

In contrast, the  literature on the role and behavior of economic platforms under shocks is fairly limited.
Empirical studies have been conducted to study the impact of COVID-19 on the demand of food-delivery platforms \citep{raj2021}, the effect of fee controls on on-demand services \citep{li2021,sullivan2022}, as well as the extent to which pandemic has persistently changed customers' purchasing behavior even after the shock calms down, due to habit formation around delivery \citep{oblander2022}.


\paragraph{\textbf{RL for Economic System Design.}}
Recent work has demonstrated the effectiveness of using RL for the design and understanding of complex economic systems, including dynamically setting reserve prices in auctions \citep{ShenPLZQHGDLT20}, selling user impressions to advertisers \citep{Tang17abc}, designing tax policies \citep{Zheng2021},  optimizing user satisfaction for recommender systems \citep{Chen2019,Zhan2021},  and designing sequential price mechanisms \citep{BreroEGPR21}.
Many of these works rely on agent-based simulation to model the economic system and conduct counterfactual analysis through the dynamic interactions of agents. For example,~\citet{Zheng2021} use RL to model a social planner who designs income taxes in multi-period, simulated spatial economies.
We are not aware of previous work on the modeling and study of market-based platforms.
Most similar to ours is the work of \citet{Zhan2021},  built on the {\em RecSim environment} \citep{Ie2019}, which  uses RL 
to optimize the long-term social welfare of  users and content providers in a dynamic, recommender system. 
Besides the presence of economic shocks, our setting differs in the use of platform fees, the existence of an alternate sales channel (i.e., off platform), and the implication that agents can choose to join or quit the platform.

\section{A Multi-Agent Platform Model}
\label{sec:gym_env}


\subsection{Market Environment and Agent Dynamics}

The market is populated with heterogeneous buyers $\B$,  heterogeneous sellers $\S$, and a single platform $p$.
Following embedding-based representations in recommender systems~\citep{Salakhutdinov2007}, we adopt a common {\em latent space} to represent each buyer or seller, $v_b, v_s \in \V \subseteq [0, 1]^2$. 
The first dimension of an agent's {\em location}, denoted $v^0$, describes product features (e.g., in the case of food, Italian, Japanese; spicy or not) and the second dimension, denoted $v^1$, the (normalized) price level (e.g., from the cheapest to the most expensive $\$\ldots\$\$\$\$$). 
A seller $s$ sells  food at a price $v^1_s$, 
of which an $\omega_s$ fraction is the production cost.
Each buyer $b$ knows a subset of sellers $\S_b \subseteq \S$, and may transact with $s \in \S_b$ without using a platform. 
Buyers who use the platform are also introduced to additional sellers on platform.

We formulate an {\em epoch-based} decision problem for agents, similar to~\citet{mladenov2020optimizing}. 
There are multiple epochs, indexed $k$, within an episode. Each epoch has a fixed length of $T$ time steps (e.g., a month of 30 days). 
%
We consider a {\em world transaction friction}, denoted $\mu_k>0$, which varies by epoch and represents the cost of buyers completing a transaction off-platform (see Eq.~\eqref{eq:potential_world_matching_surplus}).
We model {\em shocks} corresponding to changes in this friction; e.g., during a pandemic, transaction frictions for food-service industry were extremely high, due to fears of sharing indoor spaces and absence of dine-in options.

\paragraph{\textbf{At the start of an epoch $k$.}}
The platform sets fees, including the  buyer and seller {\em subscription fees}, denoted $P_{\B, k}\geq 0$ and $P_{\S, k}\geq 0$ respectively, 
and a per-transaction seller {\em referral rate} $P_{R, k}\in [0,1]$, the fraction of prices as transaction fee paid by the seller to the platform. 
We discuss the platform's fee-setting policy in Section~\ref{sec:pricing_policy}.
%

Buyers and sellers observe platform fees and the world transaction friction $\mu_k$,  and decide whether to pay the subscription fee to use the platform for epoch $k$. 
We denote the sets of subscribed buyers and sellers  in epoch $k$ as $\B_k$ and $\S_k$.
We assume that the platform knows the locations of buyer queries and on-platform sellers in the latent space. 
This reflects that platforms tend to have good data on the market-relevant properties of sellers and that the combination of a search interface and historical buyer information gives good information on the current demand context of a buyer.
Each buyer has a {\em per-epoch budget}, $\psi_b>0$, linearly proportional to the buyer's price preference  $v^1_b$. 
This controls the number of transactions that a buyer will undertake in a given epoch.
%
%
Beyond fees, the platform matches queries from platform buyers to platform sellers, and in selective treatments, with a matching policy learned through RL
 (see Section~\ref{sec:matching_policy}). 

\paragraph{\textbf{For each time step $t$ within an epoch $k$.}} 
We follow the sequence of ``query, match, and transact'':\\
~(1) \emph{Query}: a buyer $b \in \B$ is randomly selected to issue a query according to their taste  and price preferences (e.g., \$\$\$\$ sushi or \$\$ pizza), $q_{b,t} \sim \N(v_b, \sigma^2_b)$, where $\sigma^2_b$ specifies the query variance of $b$.\looseness=-1\\
%
~(2) \emph{Match}: only for an on-platform buyer $b$, the platform observes $q_{b,t}$ and matches it to an on-platform seller, denoted $s_{p,t} \in \S_k$.\\
~(3) \emph{Transact}: the buyer $b$ can pick a seller $s \in \{s_{p, t}\} \bigcup \S_b$ if $b$ is on-platform, and $s \in \S_b$ otherwise. A buyer may also choose not to transact if the matching surplus is negative (details in Section~\ref{sec:choice_transact}).

We refer to a transaction that is matched via the platform as \emph{a platform transaction}, and otherwise as a \emph{a world transaction}. 
Even an on-platform buyer can choose a world transaction if this is better than that recommended
 by the platform.
For each world transaction, the buyer suffers a fulfillment cost of $\mu_k$,  whereas for each platform transaction, the seller pays a {\em referral fee}, which is a fraction $P_{R, k}$ of the seller's price. 

\vspace{-1ex}
\paragraph{\textbf{At the end of an epoch $k$.}}
Buyers and sellers  evaluate their surplus from transactions and fees paid to guide future subscription decisions (see Section~\ref{sec:choice_subscribe}).
The platform evaluates revenue made through subscription and referral fees to inform adjustments to fees
or its matching policy (see Section~\ref{sec:pdp}).
Each seller has a {\em shutdown threshold}, $\lambda_s \in \mathbb{N}_{> 0}$, and will go bankrupt if they do not obtain positive surplus for a consecutive $\lambda_s$ epochs.  
Once shutdown, a seller is unable to engage in transactions for future epochs.


\vspace{-1ex}
\subsection{Transaction-Level Decisions}
\label{sec:choice_transact}

A buyer with query $q$ and transacting with seller $s$ receives a \emph{matching utility}, $u_{\B}(q,s)$, reflecting matching quality.
A buyer with a choice of sellers prefers the one that maximizes immediate \emph{matching surplus}, defined as matching utility minus any transaction friction. 

\vspace{-1ex}
\paragraph{\textbf{Choice in the world.}}
\label{sec:choice_world_only}
A buyer $b$ with query $q_{b, t}$ can choose from  known sellers  whose prices (denoted by $v_s^1$ for seller $s$) are within their epoch-budget left at $t$, denoted $\psi_{b,t}$, i.e.,
$\S_{b, t} := \{s \in \S_b \ :\  v^1_s \leq \psi_{b,t}\}$.
%
For $\S_{b, t} \neq \emptyset$, the best world choice is
$s^*_w := \argmax_{s \in \S_{b,t}} u_{\B}(q_{b,t}, s)$. 
Since the friction $\mu_k$ could be high enough to prevent a buyer from transacting, we let $s^w_{b, t}$ be $s^*_w$ if $u_{\B}(q_{b,t}, s^*_w) > \mu_k$, and $\phi$ otherwise, 
to denote no preferred world seller.
%
Writing $u_{\B}(q_{b,t}, \phi)=0$, the \emph{world surplus} to buyer $b$ at time $t$  is
\begin{equation}
	\label{eq:potential_world_matching_surplus}
	u^w_{b, t} = 
	\max(u_{\B}(q_{b,t}, s^w_{b, t}) - \mu_k,0).
\end{equation}

\vspace{-2ex}
\paragraph{\textbf{Choice on the platform.}}
\label{sec:choice_platform_only}
For  an on-platform buyer, the \emph{platform surplus}, $u^p_{b, t}$, at time $t$ is $u^p_{b, t}=u_{\B}(q_{b,t}, s_{p, t})$
in the case the platform-recommended seller, $s_{p, t}$, is within remaining budget  $\psi_{b,t}$, and zero otherwise.
We set $u^p_{b, t}=0$ if $b$ is off platform.

\vspace{-1ex}
\paragraph{\textbf{Overall choice.}}
%
If no seller provides a positive surplus, the buyer will choose not to transact. Otherwise, a
 buyer $b$ chooses $s^w_{b, t}$ if it is off-platform, and the most preferred of $s^w_{b, t}$ and $s_{p, t}$ when it is on-platform.\footnote{Sellers $s_{p, t}$ and $s^w_{b, t}$ may be the same seller, in which case the buyer will choose to transact via the platform since the world transaction friction is always positive.}
We write $s_{b,t}$ to denote the choice of the buyer at time $t$,
and denote a buyer's query, seller options, and transaction as 4-tuple $(q_{b,t}, s^w_{b, t}, s_{p, t}, s_{b,t})$, where $s_{p, t}$ is $\phi$ for an off-platform buyer.\looseness=-1

\vspace{-1ex}
\paragraph{\textbf{Surplus from a transaction.}}
The buyer surplus is $r_{b, t} = \max$ $\{u^w_{b, t}, u^p_{b, t}\}$, and zero if the buyer does not transact.
We define the \emph{world matching surplus} and \emph{platform matching surplus} respectively as $r^w_{b, t}=u^w_{b, t}\cdot \I^w_{b,t}$ and $r^p_{b, t}=u^p_{b, t}\cdot (1-\I^w_{b,t})$, where $\I^w_{b,t}$ is an indicator of whether buyer $b$ transacted in the world or not at time $t$. 
A seller when chosen by a buyer cannot decline a transaction.
The surplus of seller $s$  (i.e., net profit) is
\begin{equation}
	r_{s, t} = 
	\begin{cases}
		v^1_s(1 - \omega_s - P_{R, k}) & \text{for a  platform transaction,} \\
		v^1_s(1 - \omega_s) & \text{for a  world transaction,}\\
		0 & \text{otherwise.}
	\end{cases}
\end{equation}
We denote $n^p_{s, k}$ the number of transactions completed by seller $s$ via the platform during epoch $k$,  and  $n^w_{s, k}$   the number of transactions completed by the seller in the world.

\subsection{Epoch Surplus and Platform Revenue}
\label{sec:revenue_surplus}
Buyer $b$'s {\em epoch surplus}, $r_{b,k}$, is their total surplus from matching minus any subscription fee paid,
\begin{equation}
	r_{b, k} = \sum_{t \in k} r_{b, t} - \I^p_{b,k} P_{\B, k} = \sum_{t \in k} r^w_{b, t} + \sum_{t \in k} r^p_{b, t} - \I^p_{b,k} P_{\B, k},
\end{equation}
where $\I^p_{b,k} \in \{0, 1\}$ indicates whether the the buyer is off- or on-platform during epoch $k$. 
%
Similarly, a seller $s$'s {\em epoch surplus}, $r_{s,k}$, is their total net profit from transactions minus any fee paid,
\begin{equation}
	r_{s, k} = \sum_{t \in k} r_{s, t} - \I^p_{s,k} P_{\S, k} = n^w_{s, k} v^1_s(1 - \omega_s) + n^p_{s, k} v^1_s (1 - \omega_s - P_{R, k}) -  \I^p_{s,k} P_{\S, k},
\end{equation}
where $\I^p_{s,k} \in \{0, 1\}$ is an indicator to denote whether the the seller is off- or on-platform. 
%
%
The {\em total platform revenue} in epoch $k$ is simply the sum of the subscription and referral fee it charges,
\begin{equation}
	\label{eq:platform_epoch_profit}
	r_{p, k} = \sum_{b \in \B} \I^p_{b,k} P_{\B, k} + \sum_{s \in \S} \Parens{\I^p_{s,k} P_{\S, k} + n^p_{s, k} v^1_s P_{R, k}}.
\end{equation}
The {\em total welfare} of the economy in epoch $k$ is the sum of all buyer and seller surplus and platform revenue.

\subsection{Subscription-Level Decisions}
\label{sec:choice_subscribe}
At the start of each epoch, each buyer or seller ``wakes up'' with some probability, reevaluates their current state, and decides whether or not to subscribe to the platform.
We provide in the sequel a high-level description as to how buyers and sellers  make such decisions, and defer detailed analysis to Appendix~\ref{app:counterfactual_estimates}.


\vspace{-1ex}
\paragraph{\textbf{Estimating the effect of a subscription decision.}}
Each agent, whether buyer or seller, subscribes by comparing the estimated surplus on and off platform under the newly proposed platform fees and the observed world transaction friction. 
This is done assuming a unilateral change, i.e., they are the only one who wakes up and makes a different subscription decision, and queries from the agent itself as well as others remain the same as in the previous epoch.
For buyers or sellers who were on platform in the past epoch, this estimate means to re-evaluate their transaction decisions  under new fees and friction.
For agents who were off platform in the past epoch, we assume  the platform can provide information (honestly) to facilitate this estimation. This could occur, for example, through a trial period on the platform or by  providing an estimate of costs and benefits based on recent history.

\vspace{-1ex}
\paragraph{\textbf{Agent-specific decision inertia.}}
Our decision model incorporates  {\em behavior inertia}, capturing an agent's tendency to stick with their current decision (e.g., subscribing to a platform).
This kind of inertia has been empirically observed in platform adoption decisions post-pandemic~\citep{oblander2022}, along with other settings, including choosing consumer packaged goods~\citep{shum2004does,dube2010state} and health and automobile insurance~\citep{handel2013adverse,honka2014quantifying}. 
Following prior models \citep{dube2009switching,MacKay2021,farrell1988dynamic}, we incorporate inertia as an additive term to an agent's surplus from the current decision. 
Specifically, we model inertia logarithmically in the number of epochs for which an agent has committed to the same choice, and ``reset'' this inertia upon a change to the subscription decision. 
Based on this adjusted surplus, both buyer and seller agents decide whether to subscribe or not according to probabilities calculated from the standard {\em discrete-choice logit model}~\citep{dube2009switching,MacKay2021}.  See Appendix~\ref{app:counterfactual_estimates} for details. \looseness=-1

\section{The Platform's Decision Model}
\label{sec:pdp}

%

We formulate the platform's problem as a POMDP~\citep{KAELBLING1998}, with buyers' knowledge about sellers and transactions in the world as private information, and thus not observable to the platform. 
The platform learns a fee-setting and matching policy based on observations of the decisions of on-platform buyers and sellers.
In this work, for reasons of computational tractability, we study platforms that either follow a default, myopic matching policy and learn to set fees or follow fixed, regulated fees and learn how to match. 
%

\vspace{-1ex}
\subsection{Learning to Set Fees under Myopic Matching}
\label{sec:pricing_policy}

We study the sequential decision-making problem of
a platform that learns to set fees while using {\em myopic query matching}---recommending the closest on-platform seller to a query in the latent space, and thus the seller that yields the highest utility to the buyer. 
%
%
Here, the platform decides fees for epoch $k$ based on its experience from epoch $k-1$.
We describe the fee-setting POMDP model:
%
\begin{itemize}
	\item The {\em state} $x_k \in \X$ at the start of epoch $k$ is comprised of 
	\begin{enumerate}
		\item buyer attributes: the latent location, epoch budget, query distribution, and knowledge of world sellers,
		\item seller attributes: the latent location, cost fraction,  and shutdown threshold,
		\item agent subscription states:  either on- or off-platform for the past epoch, $\I^p_{\B, k-1}$ and $\I^p_{\S, k-1}$,
		\item the agent inertia levels: $\chi_{b,k-1}$ and $\chi_{s,k-1}$,
		\item a sequence of query, seller candidates, and buyer's choices of previous epoch: $Q_{k-1} = \{(q_{b,t}, s^p_{b,t}, s^w_{b,t}, s_{b,t})\}_{t \in k-1, b \in \B}$,
		\item the shutdown states for sellers: whether a seller has shut down at the end of epoch $k-1$, denoted $\I_{s, k-1}$,
		\item the platform fees for the past epoch: $P_{\B, k-1}, P_{\S, k-1}, P_{R, k-1}$,
		\item the world transaction friction for the current epoch: $\mu_k$.
	\end{enumerate}
	\item An {\em action} $a_k = (P_{\B, k}, P_{\S, k}, P_{R, k})$ defines the  fees for the upcoming epoch $k$. We model a discrete action space $\A$ where fees take discrete values at integer multiples of a tick (or percentage) size.
	\item For the {\em state transition} $\P: \X \times \A \rightarrow \Delta(\X)$, buyers and (viable) sellers follow their choice model to subscribe to the platform (Section~\ref{sec:choice_subscribe}), leading to new subscription states and inertia levels. 
	For each time step $t \in k$, we follow the ``query, match, transact'' dynamics, 
	which gives a full sequence  $Q_k$.
	Each viable seller may shut down based on the surplus in epoch $k$ and their shutdown threshold.
	Fees follow from the actions taken, and the world transaction friction evolves according to a lognormal process.
	Altogether, this gives a new state $x_{k+1} \sim \P(x_k, a_k)$.
	\item A {\em reward} $r_k \sim \R(x_k, a_k)$ is provided to the platform at the end of epoch $k$, when agent subscription and transaction outcomes are available. 
	The reward can be set to model different platform objectives, integrating considerations that come from regulation. 
	%
	\item The platform's {\em observation} $o_{k+1} \in \Omega$ consists
	of the sequence of queries generated by on-platform buyers, as well as their decisions on whether or not to transact via the platform, i.e., 
	$Q^p_k := \{(q_{b,t}, s^p_{b,t}, \ind \{s_{b,t} = s^p_{b,t}\})\}_{t \in k, b \in \B_k}$ (but not counterparties in off-platform transactions). 
\end{itemize}

\subsection{Learning to Match under Fixed Fees}
\label{sec:matching_policy}
In a second setting, we study the sequential decision-making problem of a platform that learns how to match  buyer queries to sellers when the fees are fixed, for example,
due to regulation.
Myopic query matching, as described above, favors the buyer side of the market, by directing a query to the buyer's utility-maximizing, on-platform seller.
To complement this, we model  matching strategies that can choose to benefit other parties in the economy, i.e., sellers or the platform itself.
For interpretability, we define a {\em matching strategy} by two parameters: (1)~a {\em matching utility threshold} $\eta \in [0, 1]$, that specifies the minimum utility that a recommended seller should provide to the buyer, as a fraction of utility from the myopically-optimal match, 
and (2) a {\em matching rule}, which  directs how to pick a seller amongst those that meet this utility threshold.
We  consider two rules: 
\begin{itemize}
	\item {\em The seller-aware matching rule}: Among sellers who meet the utility threshold,
	match a query to the seller who has the lowest surplus on the platform so far during the epoch,%
	\footnote{We assume  for simplicity that the platform knows a seller's production cost, and thus can calculate its surplus from transactions. In practice, this idealized seller-aware rule could be modified and defined in terms of the number of platform transactions completed by a seller, or other  inferred quantities about seller surplus.} 
	breaking ties in the buyer's favor.
	\item {\em The profit-driven matching rule}: Among sellers who meet the utility threshold,
	match a query to the seller who brings the largest revenue to the platform, breaking ties in the buyer's favor.  
\end{itemize}
%
%
As a special case,  myopic query matching corresponds to setting the matching threshold $\eta =1$.
Intuitively, the seller-aware rule may be useful in promoting a more diverse set of sellers, by increasing sales to those who have been benefiting less from the platform,  
whereas the profit-driven rule is at the other end of the spectrum, aiming to maximize the platform's myopic transaction revenue. 
The goal of the platform is to learn a  {\em matching policy} that chooses a matching strategy for an epoch---a utility threshold and a matching 
rule, based on an observation as to which buyers and sellers choose to subscribe for the upcoming epoch.
To model this, we make several adjustments to the fee-setting POMDP in defining a {\em matching POMDP}, and  
defer these details for space reasons to Appendix~\ref{app:pdp}.
%

\vspace{-1ex}
\subsection{Finding the Optimal Policy} 

Interactions between the platform and buyers and sellers can be considered as a {\em Stackelberg game}: the platform agent is the {\em leader}, choosing the fee or matching policies, and buyers and sellers are the {\em followers}, responding to platform policies. In our simulation, the buyer and seller strategies are  a fixed mapping from prior matching experience, platform fees, and  world transaction friction to decisions in regard to joining or exiting the platform,  and we can handle this Stackelberg structure by modeling the agents within the POMDP environment of the platform (as part of the transition model).
%
Depending on the set-up, the platform learns a fee-setting policy or a matching policy, denoted $\pi(a | o_k)$,  to maximize its discounted cumulative reward across different episodes: 
\begin{equation}
	\small
	\label{eq:rl_objective}
	\max_{\pi} \quad \E_{a \sim \pi, x \sim \P} \left[\sum_{k=0}^{K} \gamma^k r_k \right],
\end{equation}
where $\gamma\in (0,1)$ is the discount factor, $K = |\tau|$ is the total number of epochs in an episode, and $r_k$ 
is based on~Eq.\eqref{eq:platform_epoch_profit} whose precise value can further depend on a regulatory structure. 
Following the success of  deep learning to solve POMDPs~\citep{Heess2015,Wierstra2007,Hausknecht2015}, we use deep RL to learn $\pi(a | o_k; \vtheta)$, specifically parameters $\vtheta$ that extract sufficient statistics from the observation history and map to actions that maximize the objective (more details in Section~\ref{sec:experiments}).

\section{Platform Behavior under Shocks and Regulatory Interventions}
\label{sec:experiments}

We study platform behavior and its effect on the simulated economy under the following regulations: (1)~taxation policies that directly change the reward to the platform, (2)~fee caps that are enacted through restricting the platform's action space of fee-setting, and (3)~fee freezes that are studied along with a platform who can continue to change its matching policy.
We first provide specifics on simulation configurations, and then present experimental results.

\vspace{-1ex}
\subsection{Experiment Settings}

We follow Section~\ref{sec:gym_env} in specifying a set of different market environments, with 10 buyers and 10 sellers, an
 episode that consists of $K=12$ epochs, and with each epoch containing $T=100$ time steps. 

\vspace{-1ex}
\paragraph{\textbf{Market structure and dynamics.}}
We consider three types of \textit{market structures}, corresponding  to distinct latent locations of buyers and sellers (see Appendix~\ref{app:market_structures}, Figure~\ref{fig:three_structures}):

\begin{itemize}
	\item \emph{Uniform:} $v \sim U[0, 1]^2$, representing an economy with diverse buyer interests and seller attributes. 
	\item \emph{Core-and-Niche:} $v \sim \text{truncated Gaussian}(\mu, \sigma^2, 0, 1)$ with $\mu = [0.5, 0.4]$ and $\sigma^2 = [0.2, 0; 0, 0.2]$, representing a "core" of agents around $\mu$, as well as "niche" agents located away from the core. 
	%
	\item \emph{Two-Core:} 
	One group $v\sim \text{truncated Gaussian}(\mu_1, \sigma^2_1, 0, 1)$ with $\mu_1 = [0.7, 0.3]$ and $\sigma^2_1 = [0.17, 0; 0, 0.17]$, and another group $v\sim \text{truncated}$ $\text{Gaussian}(\mu_2, \sigma^2_2, 0, 1)$ with $\mu_2 = [0.3, 0.7]$ and $\sigma^2_2 = [0.17, 0; 0, 0.17]$. 
	The first  group is centered around relatively cheap options, and the second around more expensive ones.
\end{itemize}

We present results of the Core-and-Niche market, and defer the other two market structures to supplemental material that is published along with the paper.
For each structure, we also consider markets that vary in the buyer knowledge level, $\rho$. 
In particular, each buyer $b$ samples i.i.d. 
 $\mathit{Bern}(\rho)$ for each seller, to generate its set of known sellers $\S_b$.

\vspace{-1ex}
\paragraph{\textbf{Agent attributes.}}
Within an epoch, buyers arrive in a round robin fashion.
A buyer $b$ who arrives at $t$  submits a query around the buyer's 
 latent location, $q_{b,t} \sim \N(v_b, \sigma^2_b)$, with $\sigma^2_b = [0.02, 0; 0, 0.02]$. 
When query $q$ is fulfilled by seller $s$, the buyer receives a matching utility of $u_\B(q, s) = \exp(-c\norm{q - v_s}_2)$, where we choose $c=2$ to have matching utilities span $[0, 1]$. 
%
We set a buyer's \emph{per-epoch budget} $\psi_b$ to be $v_b^1$ (their price preferences) times the number of queries it submits within an epoch.
%
Each seller's fractional cost is drawn from $\omega_s \sim U[0.2, 0.4]$.
We set the shutdown threshold,  $\lambda=2$, for all sellers. 
Each agent has an initial preference of staying in the world or joining the platform, with an initial inertia level $\chi$ selected uniformly on $\{-2, -1, 0, 1, 2\}$ (intuitively, the more positive/negative the value reflects a stronger preference to stay on/off the platform).

\vspace{-1ex}
\paragraph{\textbf{Shocks.}}
We vary the world transaction friction, $\mu_k$,  across epochs to model {\em pre-,  during}, and {\em post-shock} stages, fixing
the  pre- and post-shock stages to each last for three epochs and with low world friction, $\mu_k = 0.1$.
The shock stage  is   controlled by a {\em shock intensity}, $I \sim U[I_{\min}, I_{\max}]$, which specifies the largest value that will be attained.  
%
We sample $\mu_k \sim$ Lognormal$(\mu=0, \sigma=0.5)$, and multiply the values by the intensity $I$. 
%
Figure~\ref{fig:welfare_decomposition} (red line) in Appendix~\ref{app:market_structures} shows the average shock schedule for $I \sim U[0.8, 1]$.

\vspace{-1ex}
\paragraph{\textbf{Platform action space.}}
Without fee caps, the registration fees, $P_{\B, k}$ and $P_{\S, k}$,  range from 0 to 10, with discrete levels at intervals of 0.2. 
The seller referral rate, $P_{R, k}$, ranges from 0 to 1, with discrete levels at intervals of 0.1.
For matching, the matching utility threshold ranges from 0 to 1, and is discretized at intervals of 0.1. 

\vspace{-1ex}
\paragraph{\textbf{Implementation details.}}
We sample initial states with a {\em warm-up epoch},   used in addition to each of the
 twelve  epochs, where the friction is 0.1, the platform charges no fees, and buyers and sellers join the platform based on their initial preferences.
The experience gained by buyers and sellers in this warm-up epoch provides a basis for the platform to choose actions, and for the buyers and sellers to form estimates to guide their subscription decisions. 
We use the {\em Advantage Actor-Critic (A2C)}~\citep{rlbook,Degris2012} to learn the optimal  platform policy, and 
%
present the average results from training for two different seeds and 100 test episodes. 
We defer descriptions of the neural network structure and hyperparameters to Appendix~\ref{app:nn}.

\begin{figure}[t]
	\centering
	\begin{subfigure}{0.48\columnwidth}	
		\centering
		\includegraphics[width=\columnwidth]{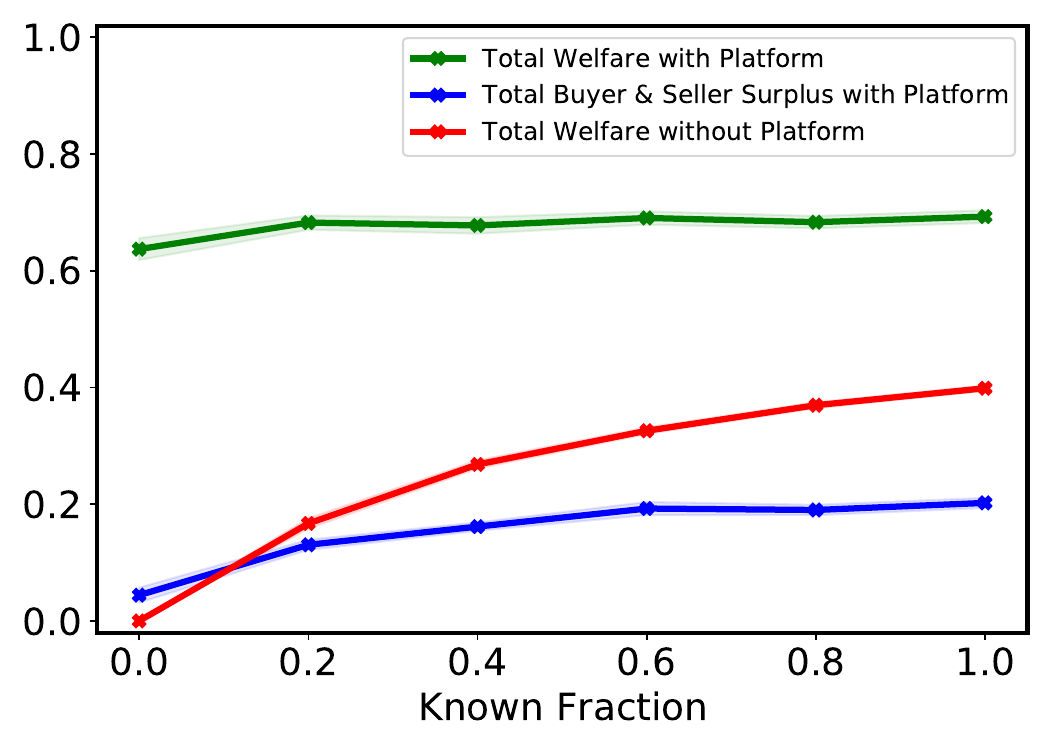}
  \vspace{-4ex}
		\caption{Varying buyers' knowledge level about sellers $\rho$, with $\mu = 0.6$.}
	\end{subfigure}
	\begin{subfigure}{0.48\columnwidth}	
		\centering
		\includegraphics[width=\columnwidth]{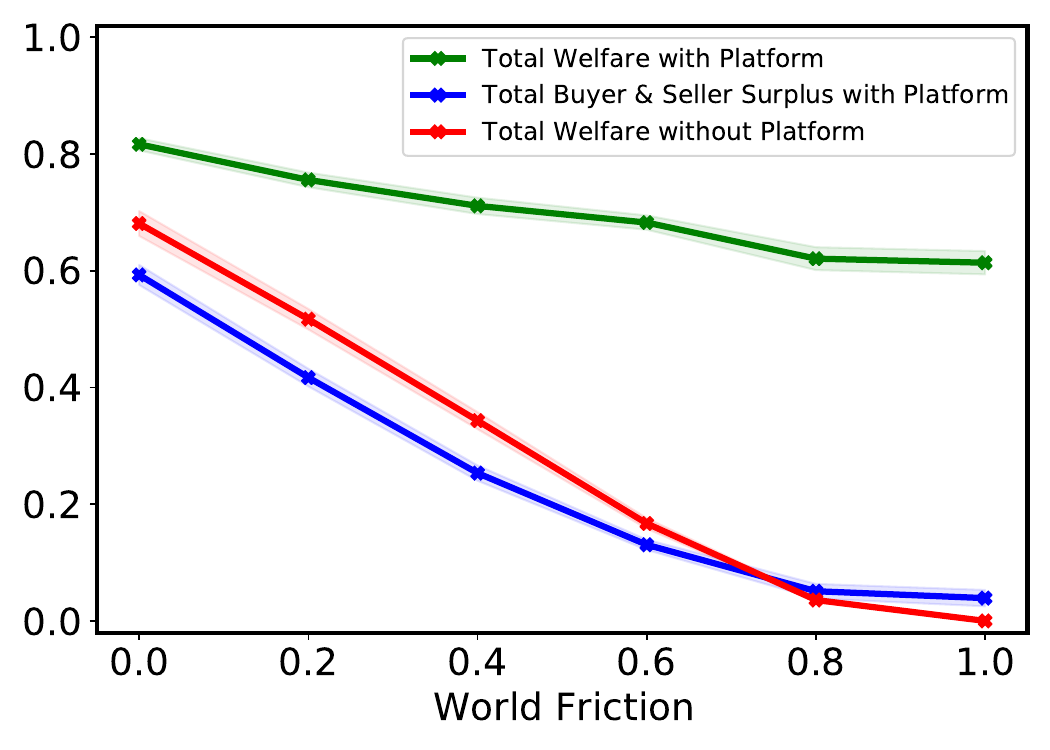}
    \vspace{-4ex}
		\caption{Varying the world transaction friction $\mu$ , with $\rho = 0.2$.\\}
	\end{subfigure}

	\caption{Total welfare and buyer and seller surplus achieved in environments that vary in $\rho$ and $\mu$, with and without a platform under the Core-and-Niche market structure. 
		\label{fig:platform_value_}}
  \vspace{-2ex}
\end{figure}

\begin{figure}[t]
    \centering
    \includegraphics[width=0.85\columnwidth]{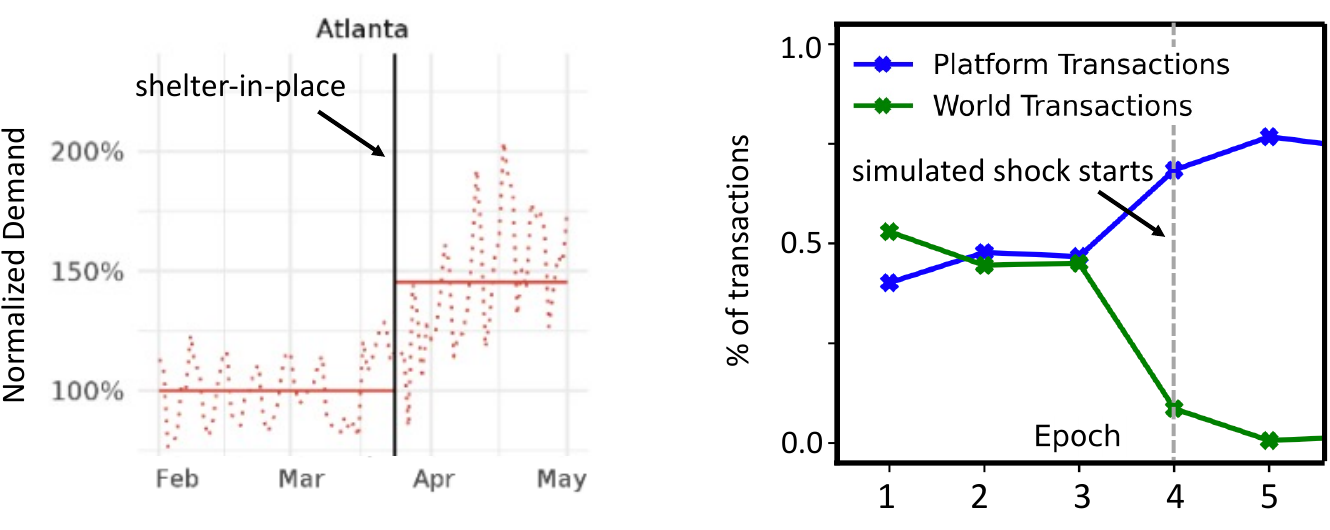}
    \caption{A comparison of empirically-observed demand surge after the shelter-in-place order as shown in~\citep{raj2021} (Left) to the increase in the number of on-platform transactions induced by our simulated economic shock (Right).
    \label{fig:my_label}}
\end{figure}

\vspace{-1ex}
\subsection{Baseline Economic Performance}
We first build intuition around our  model. 
We study the value generated by a revenue-maximizing platform across a range of \emph{single-epoch, no-shock} environments that vary in buyers' knowledge level about sellers, $\rho$, and the world transaction friction, $\mu$. 
%
For each market structure, we generate three samples of latent locations of buyers and sellers, and for each latent sample and knowledge level, $\rho$,  we sample ten knowledge structures, specifying which sellers are known by each buyer. 
%
We use {\em  Bayesian Optimization} (BO)~\citep{bo} to find platform fees that maximize the platform's revenue, and conduct controlled experiments with and without a platform. 

Figure~\ref{fig:platform_value_} shows the buyer and seller surplus, as well as total welfare (agents' surplus and platform revenue), achieved in Core-and-Niche markets.
We normalize surplus by the \textit{ideal welfare} achieved in a setting where buyers knows all sellers and there is no world transaction friction.
We validate the simulator by confirming that without a platform (Figure~\ref{fig:platform_value_}, red lines), the total welfare increases as buyers' knowledge about sellers increases and the world friction decreases.
Across all environments, 
a revenue-maximizing platform consistently increases total welfare relative to the absence of a platform, creating value by reducing search costs (i.e., matching buyers to unknown sellers) and facilitating transactions (i.e., reducing off-platform fulfilment costs). 
We  also see that the revenue a platform can extract (i.e., difference between green and blue lines) increases as buyers have less knowledge about sellers, and as the world transaction friction increases.

\vspace{-1ex}
\subsection{Platform Responses to Interventions}

\begin{figure*}[t]
	\centering
	\begin{subfigure}{\textwidth}	
		\centering
		\includegraphics[width=0.9\textwidth]{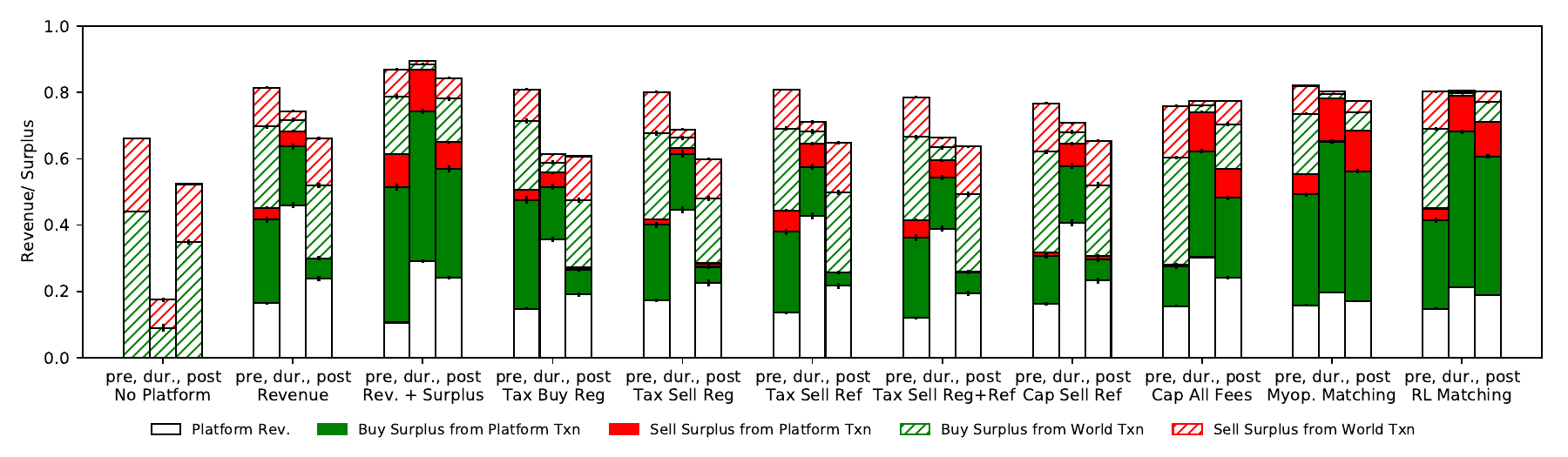}
		\caption{Welfare decomposition achieved by different learned fee-setting policies in each shock stage.}
		\label{fig:all_obj_results_profit_surplus}
	\end{subfigure}
	\begin{subfigure}{\textwidth}	
		\centering
		\includegraphics[width=0.9\textwidth]{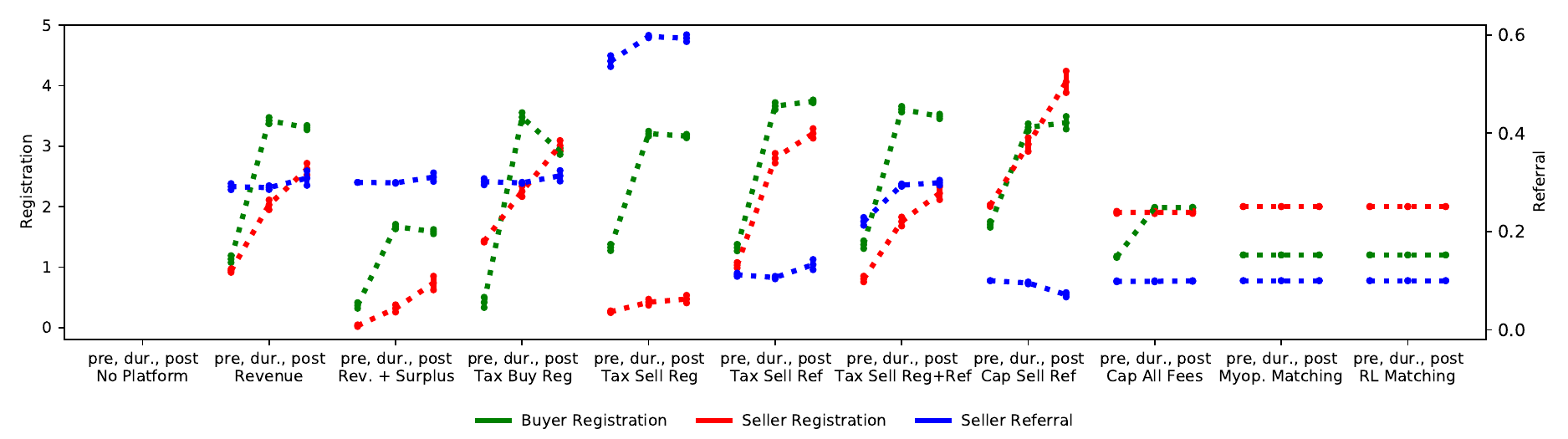}
		\caption{Platform fees set in each shock stage.}
		\label{fig:all_obj_results_price}
	\end{subfigure}
 	\begin{subfigure}{\textwidth}	
		\centering
		\includegraphics[width=0.9\textwidth]{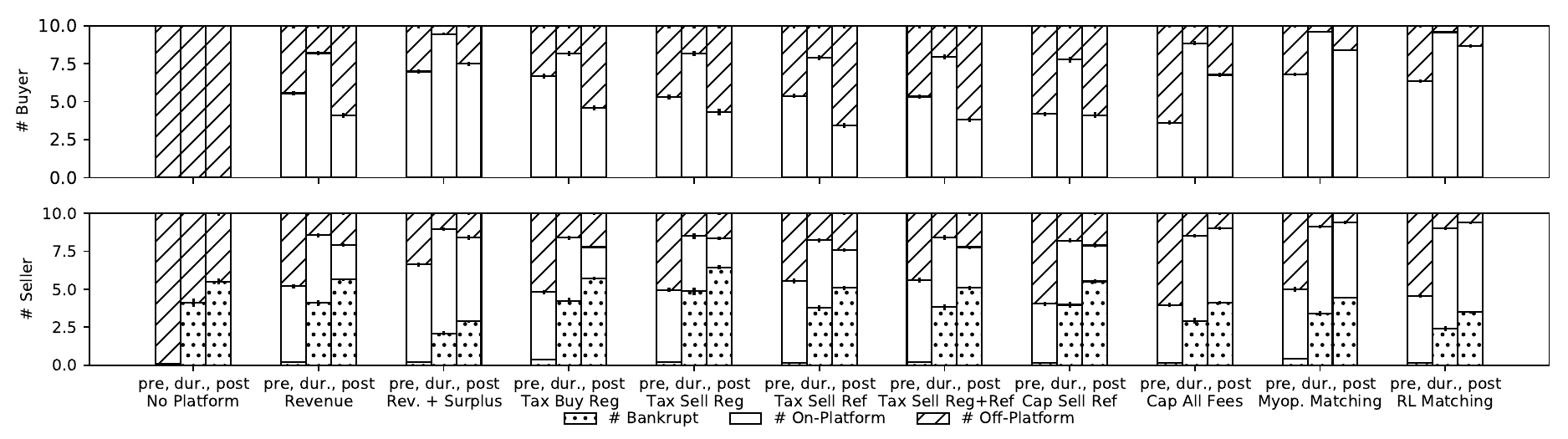}
		\caption{The number of on- or off-platform agents and bankrupt sellers in each shock stage.}
		\label{fig:all_obj_results_bankrupt_sellers}
	\end{subfigure}

	\caption{Welfare decomposition, platform fees, and agent states induced by different regulations (Core-and-Niche markets).}
	\vspace{-2ex}
	\label{fig:all_obj_results}
\end{figure*}

In this section, we present our main results, comparing no-intervention with three different kinds of interventions in the presence of market shocks. 
The headline result is that, whereas taxes and caps imposed on some subset of fees are largely ineffective, there are two forms of interventions under which a revenue-maximizing platform will promote  seller diversity and the efficiency and resilience of the  economy. These interventions either cap all fees (leaving flexibility to set fees subject to caps),
or fix fees to those that a self-interested platform chooses in an environment without shocks
but allow the platform to continue to optimize its matching policy.
%
%
%
The second intervention may be especially  relevant to practice: the knowledge of how to set fees comes from the platform's own behavior, and the platform can continue to flexibly make matching decisions.
%
%

\vspace{-1ex}
\paragraph{\textbf{Case 1: Platform fee setting in a laissez-faire system.}}
We first consider a platform that is free from any form of regulation
and learns to set fees to maximize revenue,  while
using myopic query matching. 
In Figure~\ref{fig:my_label}, we first illustrate the  similarity between the simulated increase in the number of platform transactions and the empirically-observed demand surge in Atlanta after the 2020 stay-in-place order, which is representative of other U.S.~cities 
 (i.e., on-platform transactions, or demand, increasing by around 50\% due to the shock). 
This gives a first, basic validation of the economic dynamics in our model. 

Figure~\ref{fig:all_obj_results}, columns 1-2, compare the outcomes achieved in markets with and without the platform.
%
%
%
As a first observation, from Figure~\ref{fig:all_obj_results}a,  the platform generally improves the overall economic welfare, considering the sum of the revenue to the platform and the surplus to buyers and sellers (i.e., the total heights of the bars).
This is especially salient during the shock stage (i.e., central bars), when the world friction is high and very few transactions can  generate surplus in the absence of a platform. 
At the same time, we see two less beneficial outcomes. 
First, by comparing pre- and post-shock periods (i.e., the left and right bars of the same column), we find that
the overall economic welfare falls after the shock, both with and without a platform.
This arises as a result of sellers going bankrupt, so that buyers can then only be matched with
 less-preferred sellers.
As further verified in Figure~\ref{fig:all_obj_results}c, we see that the number of bankrupt sellers in markets with a platform is almost the same as that in markets without a platform.
By comparing the surplus to buyers and sellers with and without the platform, we also 
find that the platform reduces the total surplus to buyers and sellers 
in the post-shock stage relative to without the platform. \looseness=-1
%

%

We further characterize the type of seller that is most likely to go bankrupt,
 classifying sellers into three groups: {\em core sellers}   within one standard deviation of the center and with at least two buyers nearby, 
{\em niche sellers} that are beyond two standard deviations from the center with at most one buyer nearby,
and {\em cheap sellers} with prices in the lower quartile. 
%
As shown in Table~\ref{table:seller_stats}, cheap sellers are much more likely to go bankrupt in the presence of a platform, whereas niche sellers are more likely to go bankrupt in the absence of a platform. 
A rational platform  learns to raise fees as much as possible during the shock (Figure~\ref{fig:all_obj_results}b), leaving sellers with lower margins unable to   afford the fees 
and facing bankruptcy.
\paragraph{\textbf{Idealized baseline.}}
As an exemplar on the possible effectiveness of regulations, we consider the effect of a \textit{surplus-aware platform} that sets fees to optimize some combination of its own revenue and the on-platform user (buyer and seller) surplus, i.e., $r_{p,k} + \alpha(r^p_{b, k} + r^p_{s, k})$,
 and here we choose $\alpha=0.4$.
As shown in Figure~\ref{fig:all_obj_results}a (column 3), this platform would restore a comparable level of overall economic welfare after the shock, and lead to fewer bankrupt sellers.

We next consider three possible regulatory 
interventions.

\vspace{-2ex}
\paragraph{\textbf{Case 2: platform fee setting under taxation policies.}}
We study the introduction of taxation on platform profits made from different categories of fees,  with tax schemes that charge a 20\% tax on profits made from buyer registration fees, seller registration fees, seller referral fees, and all fees charged on sellers, respectively.
%
In experiments, we tried a range of tax rates (i.e., 20\%, 40\%, 60\%). Despite differences in the absolute fee values, they lead to qualitatively similar trends.
As shown in Figure~\ref{fig:all_obj_results}, columns~4-7, 
taxation policies in general lead to similar outcomes as those observed in the laissez-faire system (column 2) and
 the overall economic welfare decreases after the shock due to bankrupt sellers. 
Specifically, charging taxes on one
 kind of fee leads a  platform to increase other fees, 
simply transferring the loss to another user group. 
For example, 
Figure~\ref{fig:all_obj_results}b compares fees when taxes are imposed on seller referral profits (column~6) to those
 in a laissez-faire system (column~2). We see that 
the platform learns to decrease referral rates while raising registration fees during and after the shock, with the effect that on-platform sellers achieve higher surplus (Figure~\ref{fig:all_obj_results}a, column~6, solid red) relative to laissez-faire but at the expense of lower on-platform buyer surplus (solid green).
%


\begin{table}[t]
	\centering
	\small
	\begin{tabular}{lllll}
		\textbf{Bankrupt freq.} & \textbf{No platform} & \textbf{Rev.-max.} & \textbf{Surplus-aware} \\
		\hline\hline
		All sellers & 0.55 (0.02) & 0.57 (0.01) & 0.29 (0.01)\\
		Core & 0.33 (0.03) & 0.46 (0.02) & 0.28 (0.02)\\
		Niche & 0.88 (0.08) & 0.45 (0.03) & 0.17 (0.02)\\
		Cheap & 0.52 (0.04) & 0.81 (0.03) & 0.43 (0.03)\\\hline
	\end{tabular}
	\vspace{1ex}
	\caption{Seller bankrupt frequencies in markets with no platform, a revenue-maximizing platform, and a surplus-aware platform. 
		\label{table:seller_stats}}
\end{table}

\paragraph{\textbf{Case 3: platform fee setting under fee caps.}}
As a third case, we first introduce a 10\% cap on just
the referral fee (a type of intervention that has been adopted in several U.S.~states).
Similar to   taxes imposed on referral profits, the platform   responds
 by raising the registration fees (Figure~\ref{fig:all_obj_results}b column 8). 
This leads more buyers and sellers to stay off platform, especially before the shock, and  reduces the on-platform surplus.
%
We also consider the case where all fees are capped, with $P_{\B, k} \leq 2.0$, $P_{\S, k} \leq 2.0$, and $P_{R, k} \leq 0.1$.
As shown in Figure~\ref{fig:all_obj_results}b column~9, under these  caps, the platform chooses to set the maximum possible fees except for a lower pre-shock buyer registration fee. 
These caps  are able to
 induce platform  policies that benefit the economic system,
where the overall welfare is not affected by the shock, with many
 viable sellers and good buyer matches remaining. 
At the same time, this intervention requires knowledge on the
part of the regulator. 

%
%
%
%
\vspace{-1ex}
\paragraph{\textbf{Case 4: platform matching under fixed fees.}}
In this case, 
we consider a regulatory policy in which
 the platform is required to keep the same fee structure as it picks, optimally, in a world without an economic shock. 
In addition, we allow the platform to adapt its matching policy in response to shocks. 
This case is of interest because the intervention
uses only knowledge available to a regulator, and does not assume a particular approach to matching by the platform. 
This is practically relevant, as matching 
 is proprietary and often hard to regulate due to lack of transparency.
%

\begin{figure}[t!]
    \centering
    \includegraphics[width=0.56\columnwidth]{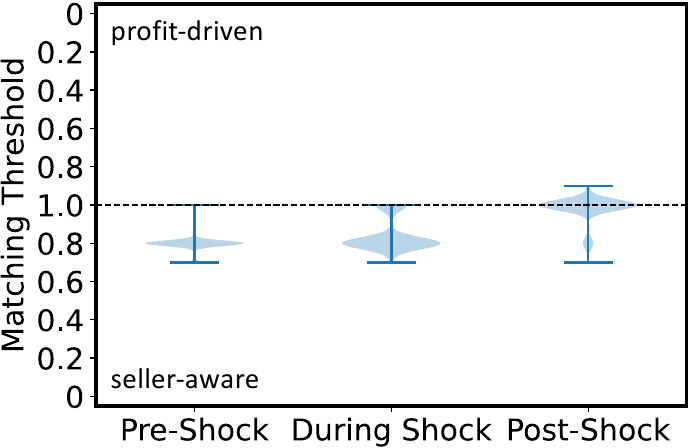}
    \caption{Probability density of learned matching strategy in the pre-, during, and post-shock stages. 
    \label{fig:matching_thresh}}
\end{figure}

We follow Section~\ref{sec:matching_policy}, and consider a platform that uses RL to learn a matching policy 
but with the fixed fees that are chosen by a revenue-optimizing platform based on the use of BO under the no-shock setting
($P_{\B, k} = 1.2$, $P_{\S, k}=2.0$, and $P_{R, k}=0.1$).
As shown in Figure~\ref{fig:all_obj_results}, right column, 
even together with flexibility in regard to choice of
 matching, the presence of the platform has the effect of 
helping the economy to 
 preserve a similar level of  welfare post-shock as pre-shock,
 and results in fewer bankrupt sellers.
We visualize the  matching policy  that is learned by the platform 
in Figure~\ref{fig:matching_thresh}.
%
To maximize revenue, the platform generally adopts the seller-aware rule, both before and during the shock,
which matches buyers to sellers with lower on-platform surplus. 
This can be explained as follows: (1)~the regulated fees motivate the platform to retain a larger number of sellers through matching in order to generate  revenue from registration fees, and (2)~the platform has more flexibility in matching as a result of the shock affecting world transactions, and thus can afford to compromise a bit on matching quality without losing buyer transactions. 
As the shock decays, the platform then
 learns to increase the matching utility threshold to provide buyers with better matches.

%
%
We further compare the use of RL matching with \textit{myopic query matching} under the same set of fixed fees.
As shown in Table~\ref{table:matching_stats}, the RL-matching platform achieves both higher revenue and total welfare compared to  myopic-matching (again, reflecting alignment with the interests of the regulator);
the RL-matching platform
 also learns to substantially reduce
 the probability of cheap and niche sellers going bankrupt. 
%

%

%
%

%

\begin{table}[t]
		\centering
		\small
		\begin{tabular}{llll}
			\textbf{} & \textbf{Myopic matching} & \textbf{RL matching} \\
			\hline\hline
			Welfare & 973.12 (5.29) & 989.44 (4.40)\\
			Revenue & 241.63 (3.06) & 261.28 (1.59)\\
			\hline
			Seller shutdown freq. & 0.44 (0.02) & 0.34 (0.01)\\
			Core  & 0.30 (0.01) & 0.21 (0.01)\\
			Niche  & 0.23 (0.02) & 0.10 (0.02)\\
			Cheap & 0.86 (0.02) & 0.78 (0.02)\\\hline
		\end{tabular}
        \vspace{1ex}
	\caption{Welfare, platform revenue, and seller bankrupt frequencies of markets mediated by a myopic-matching platform and a RL-matching platform.} 
	\label{table:matching_stats}
\end{table}

\section{Discussion}

We  have introduced a multi-agent simulation framework with which to study
 a platform-mediated economy in the presence of market shocks, making use of RL to model the decision making of a rational, self-interested platform.
Given the prominence of platforms in today's economic systems and complexity of these systems,
 this work illustrates  an important new application of AI methods. We 
 use the framework to study and interpret the effect of several regulations on the efficiency and resilience of the overall economic system under optimal platform responses, suggesting
 caution in regard to some interventions and giving support to a 
particular kind of intervention.\looseness=-1

Several extensions are of interest, including platform policies that combine fee-setting and matching, platforms with incomplete information about agent queries and locations, buyer preferences and seller offerings that can  adapt over time, and achieving greater scale.
%
There also remain several challenges to be addressed before these kinds of AI-based frameworks can guide economic system design in real-world settings. 
First and foremost 
is to collect data from existing markets  in order
 to calibrate  simulation frameworks, in regard to both environment settings and agent behavior.
Second, any learned AI policies should be interpretable to ensure transparency and fairness to all participants
in the digital economy. 

\part{Indirect Access Control with Pricing}
\chapter{Access Control with Bundling} \label{chap:bundling}
\section{Introduction}\label{sec:intro}

Online content platforms like Netflix and Spotify have traditionally provided value through advanced search and recommendation systems.  These make it possible for users to discover and navigate a vast array of product options while helping content providers reach new audiences. However, recent advances in personalized and AI-powered search technology make it increasingly easy for buyers to search for content through off-platform means. As search and recommendation increasingly move off-platform, will content follow, rendering centralized platforms obsolete?

In this paper we consider a scenario where content providers have the power to publish off-platform with no downsides in terms of searchability or buyer reach. 
The market also includes a platform that can act as an intermediary, signing exclusivity contracts with individual content providers and then charging users a subscription fee to access the resulting content. It might initially look like such a platform could not survive in the market, since in order to secure an exclusivity deal with a content provider it would need to pay them at least their outside option value of going off-platform. But it has long been understood that even when products are independent goods without complementarities nor substitution effects, a profit-maximizing seller can benefit by bundling many products together into a single package \citep{adams1976commodity,schmalensee1984gaussian}. This leaves space for the platform to make a profit by collecting an appropriate catalog of products. But which products? Should we expect the platform to target the highest-quality content in the market, or the lowest? The most niche, or the content with broadest appeal? And which content should a platform source externally, as opposed to producing in-house?

\paragraph{Our model.}
In our model, there are $N$ content providers (or \emph{sellers}). Each seller $i$ offers a single item with vertically differentiated quality. A population of buyers seek to purchase and consume content. For any seller, the buyer's valuation is drawn from a distribution whose mean equals that seller’s quality and whose variance is common across sellers. Buyer valuations are additive and independent across sellers.  There is also a platform with the power to sign and enforce exclusivity deals with sellers.  The platform can offer to each seller a (possibly personalized) payment to join the platform.  Each seller chooses independently whether to accept or decline this offer. Sellers who accept no longer sell their content off-platform, whereas those who reject set a price for their content to maximize their own revenue. The platform then sets a price for buyers that can depend on which content they were able to attract.  Buyers then choose whether or not to pay the platform's price to access the exclusive on-platform content, as well as which subset of off-platform content to purchase.  The platform seeks to maximize the net profit, that being the revenue gained from buyers less any payments to the sellers.


\paragraph{Our Results: Which Products Appear On-Platform.}
We begin our analysis in a complete information setting where the platform can observe the qualities of all items.  That is, the platform knows the buyer valuation distribution for each seller, but not the realized value of any given buyer.  In this case, the platform can attract any seller to join by paying them exactly their expected off-platform revenue, which is simply their monopolist revenue.  The platform's problem in the complete information setting therefore reduces to a procurement optimization task, and we study which set of items should be purchased.

We find that the platform's optimal strategy is always to purchase all items in a certain contiguous range of quality.  Notably, it can be strictly optimal for the platform to exclude all items with quality that is too low, and also exclude all items with quality that is too high.  The intuition behind this result is that acquiring high-quality items requires high payment, whereas bundling the low-quality items does not increase revenue as much. To balance the two effects of quality, the platform selects a bundle where the quantile of the revenue-maximizing price is similar for each good in the bundle.
If given the option to add another good whose monopolist target quantile is very different from the others, as is the case when the quality is much higher or much lower, the platform might find itself unable to effectively extract much more revenue from this addition. In the extreme, this makes the good worth less to the platform than the payment needed to offset its seller's outside option value of selling off-platform as a monopolist.

Having analyzed the complete information setting, we next consider a setting of incomplete information where the platform does not get to observe the quality of each seller's item in advance. The platform now faces a mechanism design problem with incentive constraints, where it can offer a menu of payments and a probability of joining the platform to each seller, who can then choose an option knowing their item's quality. To keep the analysis in the classical mechanism-design framework, we assume the buyer valuation distribution for each seller is strictly regular, which guarantees that the monopoly revenue function is differentiable with quality. This problem inherits all the classic challenges of multi-dimensional mechanism design: the valuation of the platform for different combinations of acquired items is inherently non-linear, and the sellers can extract information rents due to the platform's incentive compatibility constraints.

We address these challenges by first considering a surrogate proxy for platform profit, motivated by a large-market setting where the platform faces a continuum of sellers and no aggregate uncertainty about a buyer's valuation for a bundle of items. Under this surrogate profit, the platform's revenue from obtaining a bundle $S$ of items is simply the sum of seller's quality in the bundle. We emphasize that even in the case where the proxy revenue is always greater than the sum of monopolist revenues for all items, the platform does not necessarily want to contract with all sellers.  This is because of the incentive compatibility constraints and the sellers' ability to extract information rents: it can be beneficial to screen out some sellers in order to reduce the payments needed to attract others. What is true, however, is that the platform's optimal mechanism takes the form of a simple threshold rule, implementable by offering every seller a take-it-or-leave-it payment for joining the platform. The proof makes use of a virtual surplus representation of the platform's surrogate profit-maximization objective, which is non-monotone in general (even for regular buyer value distributions) but is amenable to a classic ironing approach. 

Having characterized that threshold mechanisms are optimal for surrogate profit, we then return to the original problem and establish approximate optimality for large but finite markets.  Under the additional assumption that buyer valuations are subexponential, we show that surrogate profit and true platform profit differ by a factor of $1 + O(N^{-1/3})$ where $N$ is the number of sellers. In particular, the approximation factor tends to $1$ polynomially quickly as the number of sellers $N$ grows large.  We conclude that, for large but finite markets, a threshold mechanism --- that is, the platform making a take-it-or-leave-it posted-price offer to all sellers --- is approximately optimal for reasonably large markets.  Under such mechanisms, all sellers with quality below some threshold will move on-platform, while the highest-quality sellers may reject the proposed price and remain off-platform.

\paragraph{Our Results: In-House Production.}
Having established that information asymmetry between the platform and individual sellers can harm the platform's ability to generate profit, we might expect that the platform has an incentive to produce content in-house.  Doing so allows the platform to directly observe and control quality, but we should expect this to be attractive only if such production increases platform revenue more than selling the resulting items off-platform or the outside option of contracting more items from third-party sellers. In contrast to signing exclusive contracts with individual sellers and buying their items---where the platform cannot control and can only estimate a seller’s quality---in-house production gives the platform full control over the quality it produces. Motivated in part by the observation that certain streaming platforms, such as Netflix, increasingly form direct partnerships with producers for in-house production, we ask: How valuable is this option to the platform, and is it more attractive to produce high-quality or low-quality goods in-house?

To model these choices, we focus our attention on a special case of our model in which there are only two levels of item quality in the market, high-quality and low-quality, and the buyer's value for a seller is drawn from a normal distribution. We remain in the incomplete information scenario: the platform knows the distribution of item quality in the market, but cannot ex ante distinguish high quality sellers from low quality sellers. The platform also has the option to contract directly with a producer to produce up to some maximum number of items in-house, at either desired quality level.  To secure this contract, the platform must pay the producer's outside option revenue from selling the resulting product off-platform; i.e., the off-platform monopolist price from the full-information scenario. Any items that the platform produces in-house will have known quality.  The platform can choose to produce any combination of high-quality and low-quality items, which can then be bundled together with any items sourced from external sellers.

We first show that, in every market, the platform will always want to produce either all high-quality items or all low-quality items.  Perhaps surprisingly, the platform sometimes prefers to produce low-quality items.  This can happen, for example, in small markets where the platform would like to bundle together many low-quality items to make a profit, but there are not enough such items on the market to make this strategy profitable.  In this case, producing extra low-quality items in-house may be preferable to producing high-quality items that do not bundle as well.  We show, however, that small market is the only scenario in which low-quality production is chosen: Whenever the platform attracts many on-platform sellers, bundling substantially reduces uncertainty in buyers’ valuation of the bundle, so the net profit from producing high-quality in-house and bundling it with acquired items (net production costs) always exceeds that from producing low-quality content.

We also fully characterize the platform’s optimal mechanism for contracting with external sellers to bundle with the items produced in-house, in markets where $N$ is large. Conditioned on being profitable, the platform purchases from all sellers when the ratio is low, as information rents are small; when the ratio is high, the platform purchases from only low-quality sellers.

Of course, the ability to produce in-house is always weakly valuable to the platform in our model: it removes a source of information asymmetry with no downside relative to purchasing items from external sellers. However, the platform’s in-house production capability is capped, so it cannot rely on in-house production alone. We show through examples that in-house production of high-quality items is additionally beneficial for extracting profit from low-quality items obtained from external sellers.  Indeed, in-house production and external sourcing are sometimes (but not always)  complementary to each other, with total profit being greater than the sum of what is achievable using either strategy on its own.  In other words, the ability to produce in-house can ``unlock'' potential profit from bundling externally-produced content.  In such cases, the ability to bundle with low-quality content allows the platform to extract more revenue from the production of high-quality content than a standalone seller, giving the platform a uniquely advantageous incentive to produce at a level of quality higher than what is otherwise found in the market. We interpret this as one potential reason for high-quality content to be produced and released directly in partnership with a subscription-based platform.

\subsection{Related Work}
Our results build on the classic bundling literature \citep{stigler1963united, adams1976commodity, schmalensee1984gaussian,mcafee1989multiproduct, salinger1995graphical}. Previous works show that among all possible bundling choices, a simple scheme that selects between pure bundling (selling all items as a single bundle) and component pricing (selling each item separately) can extract a large share of the optimal profit \citep{babaioff2020simple}, and give conditions for pure bundling being profit-optimal for a monopoly seller with costly items \citep{bankos1997aggregation,geng2005bundling,pavlov2010optimal,menicucci2015optimality,haghpanah2021pure,ghili2023characterization}. We do not study the optimality of pure bundling or component pricing. Instead in our model, the platform selects a subset of sellers to attract on-platform and sells their items as a single bundle, leaving the remaining sellers sell off-platform at monopoly prices. This corresponds, in the classical bundling framework, to a monopoly seller selecting a bundle that contains only a subset of items and selling the rest separately.

\Citet{sun2025partition} studies the same bundle-selection problem---choosing a single bundle while selling the remaining items separately---but focuses on the computation problem to select the best bundle. In contrast, we provide economic insights on which items are included on-platform and which are sold off-platform. In particular, we show that the optimal bundle is contiguous in item quality, which implies a polynomial-time algorithm to select the best bundle when buyer valuations are independent across sellers. The computational result in \citet{sun2025partition} is more general in other aspects, allowing for correlated buyer valuations. 

Most work that studies the optimality of pure bundling, including \citet{fang2006bundle,ibragimov2010optimal}, treats a monopolist's item costs as fixed primitives that are independent of buyer valuations. In our model, the platform does not own the items it bundles. Instead, it must source items from external sellers, at costs equal to their off-platform monopoly revenues. Including an item in the bundle therefore requires the platform to compare the acquisition cost to the increase in bundle revenue the item generates, both the cost and the revenue gain are dependent upon the item quality. The shared dependence on quality is what allows us to derive results on which sellers are included on-platform and which remain off-platform. 

In the incomplete information setting, the platform sources from external sellers with privately known qualities, resembling a Bayesian procurement mechanism design problem \citet{bei2012budget,deng2024procurement}. The difference is that in our setting, a seller's cost is not their private type, but a function of the private quality. Furthermore, as opposed to \citet{deng2024procurement} where the value function is assumed to monotonic and submodular, in our problem the platform's revenue from acquiring a set of items is a set function that is neither monotone, submodular, nor subadditive, rendering optimization techniques that rely on such structures inapplicable. We introduce a surrogate revenue function that is linear in seller quality and show that mechanisms maximizing this surrogate are approximately profit-maximizing mechanisms in large markets.

Our discussion of platform in-house production is motivated by streaming platforms' increasing production of proprietary original content \citep{adgate2024content}. \citet{leung2025dissecting} shows that Netflix's original content is substantially more popular than its licensed content, and argues the original content helps reduce Netflix's exposure to the risk of expiring licenses. \citet{du2025originality,wu2024optimal} analyze how optimal in-house production depends on consumer preferences, licensing fees, and the efficiency of a platform's production capabilities.
We offer another perspective by viewing in-house production as a way to avoid the information rents when sourcing from producers with privately known quality, further showing that in-house production can unlock additional profits from bundling licensed content.

Finally, our paper relates to the work at large on how online platforms shape the composition of on-platform sellers through a variety of design choices. Seller participation and the resulting distribution of seller qualities can be affected by pricing reasons \citep{light2024quality,platformEq}, by revenue-sharing rules \citep{alaei2022revenue}, by information disclosure policies \citep{guan2025platform}, and by the platform's choice of governance mode, operating as a marketplace versus a reseller \citep{hagiu2015marketplace}. In this work, seller selection in our model is driven by the economic force of bundling: both the platform's bundle revenue and the cost of acquisition depends on seller quality, offering results on which sellers are selected on-platform and which remain off-platform.

\section{Model}\label{sec:ch6_model}
There are $N$ sellers indexed by $i \in [N]$, each selling a single item. There is a unit mass of buyers with additive valuation over the items.  The buyer's value for item $i$ is a random variable $v_i = \mu_i + \sigma Z_i$
where $\{Z_i\}_{i=1}^N$ are i.i.d.\ random variables with zero mean and unit variance. We let $f$ and $F$ denote, respectively, the density function and the tail distribution function (one minus the cumulative distribution function) of the distribution from which each $Z_i$ is drawn, and we will tend to use $Z$ to denote a generic random variable drawn from this distribution. 
Parameter $\mu_i \in [\mu_L,\mu_H]$ captures seller $i$’s quality with $0<\mu_L<\mu_H$, 
while $\sigma > 0$ denotes the dispersion of buyer valuations for each item.
This specification places all value distributions 
within a common location-scale family.  

Since the buyer valuation is additive, each seller individually acts as a monopolist for their own item. 
We will write $Rev(\mu, \sigma)$ to denote the optimal revenue obtainable by a monopolist selling to buyers with value distributed as $\mu + \sigma Z$:
$$Rev(\mu,\sigma) 
= \sup_p p \cdot F\left[\frac{p-\mu}{\sigma}\right].$$
Since $Z$ has unit variance and a density function, the supremum exists and is attained at some \emph{optimal price} (See \Cref{app:maximizer_attained}). In particular, each seller $i$ has a monopolist revenue $Rev(\mu_i, \sigma)$ obtained at price $$p_i^* \in \arg\max_{p} \; p \cdot F\left[\frac{p-\mu_i}{\sigma}\right].$$ We refer to $F\left[\frac{p^*-\mu_i}{\sigma}\right]$ as the \emph{optimal demand}. More generally, given a random variable $V$, we will tend to write $Rev(V)$ for the optimal monopoly revenue when buyer value is distributed as $V$.

The platform may select a subset of sellers $S \subseteq [N]$, buy their items by compensating each seller $i \in S$ their optimal revenue $Rev(\mu_i,\sigma)$, and then bundle the items in $S$ into a single composite item. The platform then sells this bundle to buyers.
Buyers have additive valuations across sellers: For any subset $S \subseteq [N]$, buyer valuation for the bundle $S$ is $$v_S = \sum_{i \in S} v_i = \sum_{i \in S} \mu_i + \sum_{i \in S} \sigma Z_i.$$
Platform's optimal revenue from selling $S$ is expressed as 
$$Rev(v_S) = \sup_p p \cdot Pr[v_S \ge p].$$
Again the supremum exists and is attained at some optimal price (See \Cref{app:maximizer_attained_bundle}). The \emph{platform's profit} is defined as the optimal revenue from selling the bundle $S$ minus the compensation it pays to the sellers in the bundle, denoted as 
$$\Pi(S)= Rev(v_S)-\sum_{i\in S}Rev(\mu_i,\sigma).$$ The \emph{platform's problem} is to find the bundle of sellers $S$ that maximizes its profit. 


\subsection{Preliminaries}\label{sec:preliminaries}
\begin{figure}[htbp]
    \centering
    \begin{subfigure}[t]{0.48\textwidth}
        \centering
        \includegraphics[height=4cm]{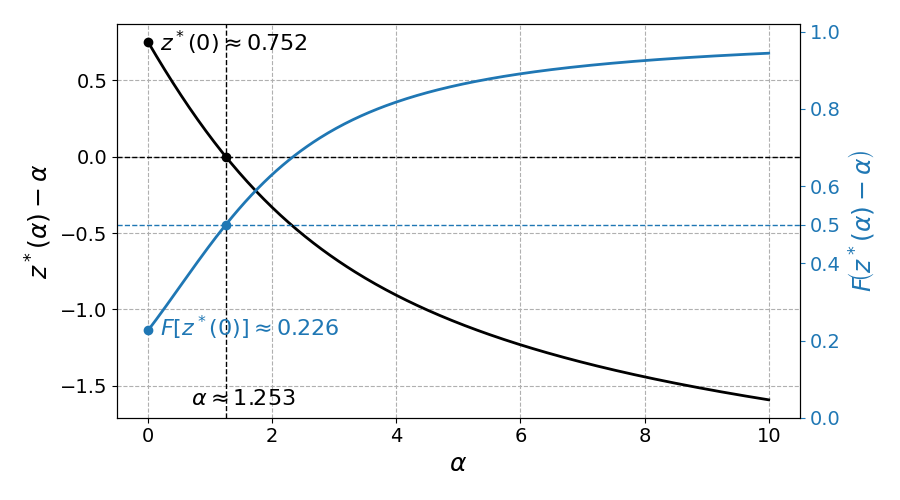}
        \caption{$z^*(\alpha)-\alpha$ and $F[z^*(\alpha)-\alpha]$ as functions of $\alpha=\mu/\sigma$. $z^*(\alpha)-\alpha=0$ occurs at $\alpha_0\approx1.253$.}
        \label{fig:core_niche}
    \end{subfigure}
    \hfill
    \begin{subfigure}[t]{0.48\textwidth}
        \centering
        \includegraphics[height=4cm]{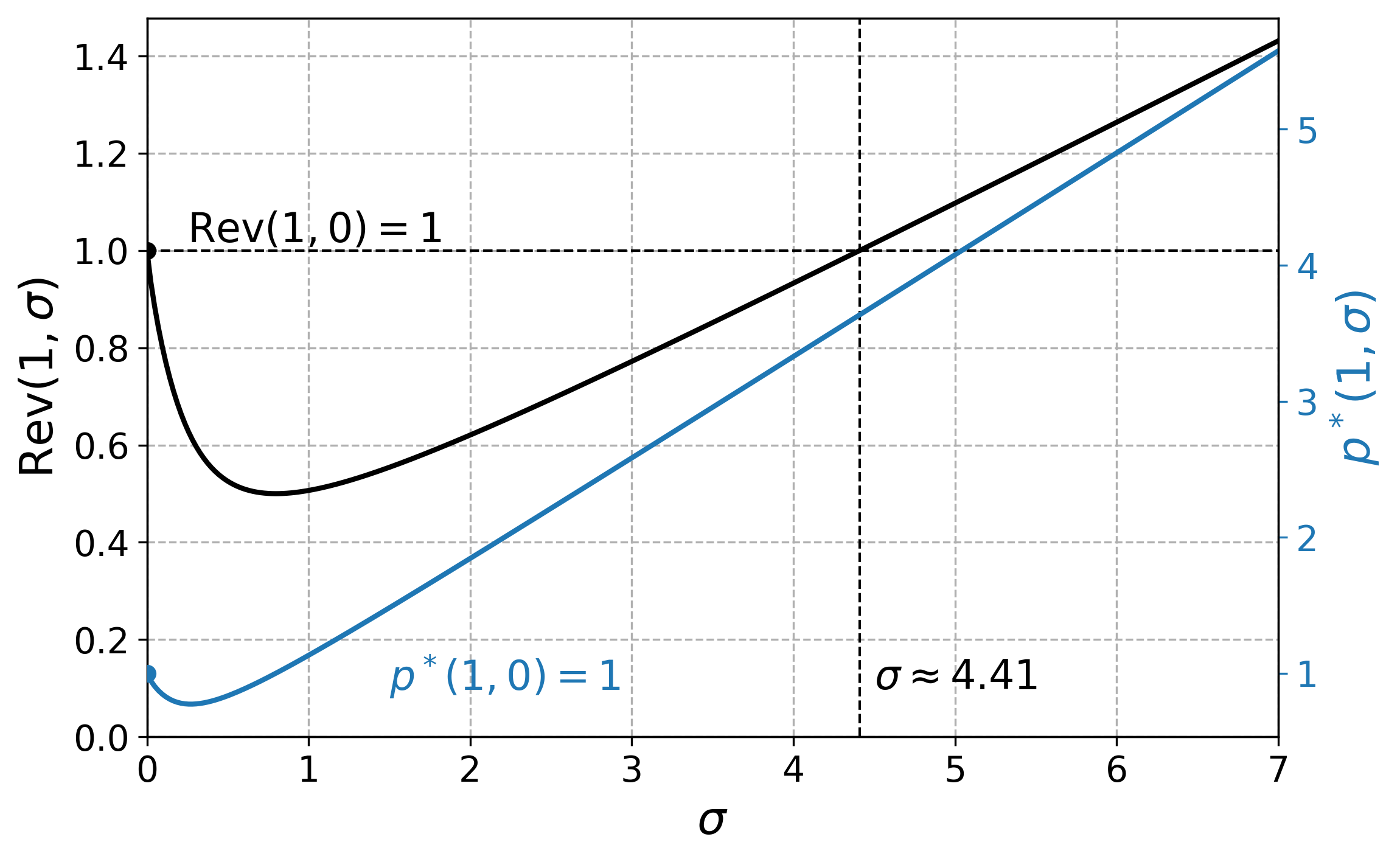}
        \caption{Optimal revenue $\mathrm{Rev}(1,\sigma)$ and optimal price $p^*(1,\sigma)$ as a function of $\sigma$.  $Rev(1,4.41)\approx1$;  $\argmin_\sigma Rev(1,\sigma)\approx 1/1.253$}
        \label{fig:rev_one_sigma}
    \end{subfigure}
    \caption{Numerical illustrations of demand, revenue, and price at optimality under normal buyer valuation. In \Cref{fig:core_niche}, as $\alpha\rightarrow\infty$, $z^*(\alpha)-\alpha$ goes to negative infinity, while the optimal demand $F[z^*(\alpha)-\alpha]$ approaches 1. (See \Cref{lem:decreasing_mills_ratio})}
\label{fig:gaussian_monopoly}
\end{figure}

To build intuition for the platform's optimal bundling strategy, we now analyze 
the optimal revenue function $Rev(\mu,\sigma)$ in the special case where 
$Z$ is a standard normal random variable. 
Define $\alpha=\mu/\sigma$ and $z^*=p^*/\sigma$, where $p^*$ is the optimal monopolist price for valuation $v$ distributed as $\mu+\sigma Z$. The optimal demand can be written as
$$F\left[\frac{p^*-\mu}{\sigma}\right]=F\left[z^*(\alpha)-\alpha\right]$$
where $z^*(\alpha)-\alpha$ is the optimal normalized price. \citet{schmalensee1984gaussian} show that for every $\alpha$ there exists a unique $z^*(\alpha)$, and that the optimal normalized price $z^*(\alpha)-\alpha$ monotonically decreases in $\alpha$, while the optimal demand $F[z^*(\alpha)-\alpha]$ monotonically increases in $\alpha$. We illustrate these properties in Figure~\ref{fig:core_niche}.

We interpret sellers with large $\alpha$ as \emph{core} sellers, who generate revenue by capturing a large fraction of demand at relatively low optimal prices. In contrast, sellers with small $\alpha$ are \emph{niche} sellers, who post relatively high optimal prices and sell only to a small fraction of buyers with high valuations. 

The effect of $\alpha$ on optimal revenue is mixed. While optimal revenue always increases with seller quality $\mu$ (See \Cref{lem:norm_rev_character}), it doesn't change monotonically with the dispersion of buyer valuation $\sigma$. \Cref{fig:rev_one_sigma} plots the optimal revenue $Rev(1,\sigma)$ fixing $\mu=1$. As $\sigma$ increases from zero, buyer valuations become more dispersed, and the seller becomes increasingly niche to capture the few buyers with high valuations. In this process, optimal demand decreases, and optimal prices slightly decreases then increases, resulting in optimal revenue first decreases than increases. Notably, when the dispersion parameter $\sigma$ is greater than $4.42$, 
$Rev(1,\sigma)>1=Rev(1,0)$.

This observation has important implications for how the platform selects sellers to bundle. Bundling effectively pools buyer valuations across items, thereby reducing dispersion and potentially creating value. However, for niche sellers whose profitability relies on high dispersion in buyer valuations, bundling reduces $\sigma$, which can lower their optimal revenue. We formalize this idea and use it explicitly in \Cref{sec:in_house_production}.

Throughout the paper, we rely heavily on convexity tools. As illustrated in \Cref{fig:gaussian_monopoly}, the function $Rev(\mu,\sigma)$ is convex in both $\mu$ and $\sigma$, and monotonically increases in $\mu$. 
See \Cref{sec:proofs.model} for the proof of \Cref{lem:rev_mu_sigma_convex_in_mu_sigma}, plus additional properties of the revenue function $Rev(\mu,\sigma)$ that will be helpful in our analysis.
\begin{restatable}{lemma}{RevMuSigmaConvex}
\label{lem:rev_mu_sigma_convex_in_mu_sigma}
    The optimal revenue function $Rev(\mu, \sigma)$ is convex in $(\mu,\sigma)$ and weakly increases in $\mu$.
\end{restatable}

\section{Complete Information and Heterogeneous Quality}
\label{sec:complete_info_heterogeneous_mu}
In this section, we study the platform's problem  
of procuring an optimal bundle
in a complete information environment where 
the platform can
observe each seller's quality $\mu_i$ perfectly. Under these assumptions, we will show that the optimal set of sellers to bundle is contiguous in their quality.

To illustrate the key ideas of our proof, we begin by considering the special case where $Z$ is drawn from a normal distribution, and hence $v_i \sim \mathcal{N}(\mu_i, \sigma^2)$ for each $i$.
For any set of sellers $S$, as $v_S\sim \mathcal{N}(\mu_S=\sum_{i\in S}\mu_i,\sigma_S^2=|S|\sigma^2)$ is still a normal random variable, we let $Rev(\mu_S,\sigma_S)=Rev(v_S)$ denote the optimal revenue for bundle $S$.
\begin{restatable}{theorem}{ContiguousMuNormal}
    \label{thm:contiguous_mu_normal}
    Consider $N$ sellers indexed so that $\mu_L \leq \mu_1 \leq \mu_2 \leq \ldots \leq \mu_N \leq \mu_H$. Buyer valuations satisfy $$v_i=\mu_i+\sigma Z_i, \qquad Z_i\sim \mathcal{N}(0,1)$$
    Then any profit-maximizing bundle is contiguous in quality: if seller $i$ and $j$ are selected with $i<j$, then every seller $k$ with $i<k<j$ is also selected.
    Moreover, define $$t_0=\min\biggl\{t\, \biggl|\, \frac{t}{\sqrt{t-1}+1}>\frac{\mu_H}{\mu_L}\biggr\}$$ For any profit-maximizing bundle of size $K\geq t_0-1$, the platform bundles the highest-quality sellers $N, N-1,...,N-K+1$.
\end{restatable}
\begin{proof}
    We first show if the platform selects a set of sellers $S$ with size $K$, out of a total $K+1$ sellers, it is optimal for the selected sellers to be contiguous in quality $\mu$. Let $Y=\sum_{j=1}^{K+1} \mu_j$. The platform bundling all but the $i$-th seller has profit $$Rev_i:= Rev(Y-\mu_i, \sqrt{K}\sigma)-\sum_{j\neq i}Rev(\mu_j,\sigma) = Rev(Y-\mu_i, \sqrt{K}\sigma)-\sum_{j=1}^{K+1} Rev(\mu_j,\sigma)+Rev(\mu_i,\sigma)$$
    Define a function $G(\mu)=Rev(Y-\mu,\sqrt{K}\sigma)+Rev(\mu,\sigma)$. It is clear that $G(\mu_i)>G(\mu_j)$ iff $Rev_i>Rev_j$. To select $K$ out of $K+1$ sellers is to leave out the seller $i$ with the largest $Rev_i$.
    Since $Rev(\mu,\sigma)$ is convex in $\mu$ by \Cref{lem:rev_mu_sigma_convex_in_mu_sigma}, $Y-\mu$ is affine in $\mu$, and the composition of an affine function with a convex function remains convex, it follows that $G(\mu)$ is convex. The convexity implies that $$\max\{G(\mu_1),G(\mu_2),...,G(\mu_K),G(\mu_{K+1})\}\in\{G(\mu_1),G(\mu_{K+1})\}$$ 
    This means $Rev_1$ or $Rev_{K+1}$ is the maximum among $\{Rev_1,Rev_2,...,Rev_K, Rev_{K+1}\}$. Thus, it is optimal to select sellers with contiguous quality $\mu$ when selecting $K$ out of $K+1$ sellers. Then it naturally follows that any profit-maximizing bundle of size $K$ out of $N$ 
    must be contiguous in $\mu$. Otherwise, it is not an optimal bundle of size $K$ out of a total of $K+1$ items. 

    Now, we prove the latter half of the theorem. Taking the derivative of $G(\mu)$ using \Cref{lem:norm_rev_character}, 
    \begin{align*}
        \frac{dG(\mu)}{d\mu} = -F\left[z^*(\frac{Y-\mu}{\sqrt{K}\sigma})-\frac{Y-\mu}{\sqrt{K}\sigma}\right]+F\left[z^*(\alpha)-\alpha\right]
    \end{align*}
    When $K+1\geq t_0$ it satisfies that for any $\mu\in [\mu_L,\mu_H]$ 
    \begin{align*}
        \frac{K+1}{\sqrt{K}+1}>\frac{\mu_H}{\mu_L} \Rightarrow (\sqrt{K}+1)\mu\leq (\sqrt{K}+1)\mu_H<(K+1)\mu_L\leq Y
        \Rightarrow \frac{\mu}{\sigma}<\frac{Y-\mu}{\sqrt{K}\sigma}
    \end{align*}
    Since \Cref{lem:norm_rev_character} shows $\frac{dF[z^*(\alpha)-\alpha]}{d\alpha}>0$, it holds that $$F[z^*(\alpha)-\alpha]<F\left[z^*(\frac{Y-\mu}{\sqrt{K}\sigma})-\frac{Y-\mu}{\sqrt{K}\sigma}\right] \text{ and } \frac{dG(\mu)}{d\mu}<0$$
    Thus, $G(\mu_1)$ is the largest so it is profit-maximizing to leave seller $1$ out when selecting $K$ out of $K+1$ sellers. This means that the profit-maximizing bundle of size $K\geq t_0-1$ out of $N$ sellers must not contain the lowest-quality sellers. Otherwise, there is a set of $K+1$ sellers that include the selected $K$ sellers and another unselected seller with larger $\mu$, forming a contradiction.
\end{proof}
The theorem says that any profit-maximizing bundle is composed of sellers with contiguous qualities. Fixing the bundle size, both the platform’s revenue from selling the bundle and the total payments required to acquire the bundle are convex in the seller quality that is excluded from the bundle. This convexity makes it optimal to exclude sellers with either low or high quality. From another perspective, high-quality sellers require high payments, whereas low-quality sellers do not increase bundle revenue as much. For sufficiently large bundles, bundling many items substantially reduces the dispersion in buyer valuations. If the reduction in dispersion is profitable, the increase in bundle revenue can offset the higher compensation needed to include higher-quality sellers, leading the platform to source from these sellers. Notably, the result is purely a statement about which sellers are included in the profit-maximizing bundle, conditional on bundling being profitable in the first place. \Cref{thm:in_house_production} in \Cref{sec:in_house_production} presents an example of a market where bundling yields negative profit and the platform's best action is to do nothing.

\begin{figure}[hbtp!]
    \centering
    \begin{subfigure}{0.48\textwidth}
        \centering
        \includegraphics[width=\textwidth]{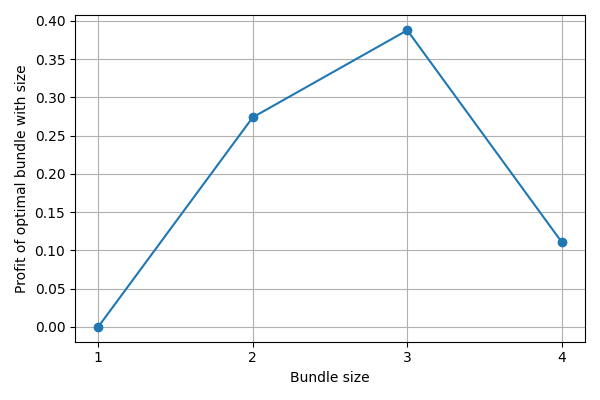}
        \caption{Optimal profit of bundle with different sizes. For the four-seller market, optimal bundle size is 3.}
        \label{fig:opt_bundle_size}
    \end{subfigure}
    \hfill
    \begin{subfigure}{0.48\textwidth}
        \centering
        \includegraphics[width=\textwidth]{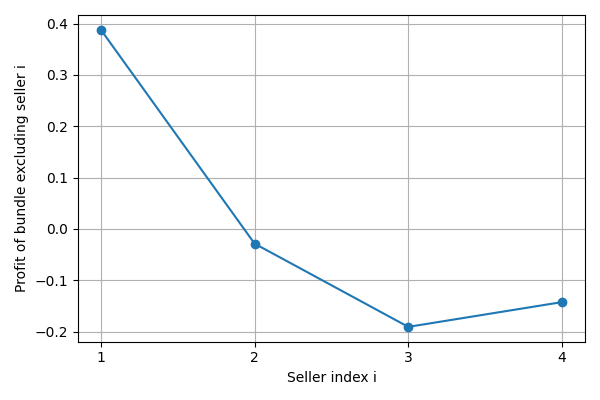}
        \caption{Profit of size-three bundle excluding seller i. Optimal bundles of size 3 excludes seller 1, where $\mu_1=0.1$}
        \label{fig:bundle_three_leave_out}
    \end{subfigure}

    \caption{Bundle profit in a four-seller market with $\sigma=1, (\mu_1,\mu_2,\mu_3,\mu_4)=(0.1,1.1,2.1,3.1)$. Buyer valuations are normally distributed.}
    \label{fig:four_seller_market}
\end{figure}

\Cref{fig:four_seller_market} further illustrates the proof idea in the above theorem with a four-seller market. 
\Cref{fig:opt_bundle_size} shows that the profit-maximizing bundle is of size three, while \Cref{fig:bundle_three_leave_out} plots the profit of a size-three bundle as a function of the quality $\mu_i$ of the excluded seller. Consistent with the proof, this function is convex in $\mu_i$, as illustrated in the figure.

There are two dimensions to consider when selecting the optimal seller bundle. One, the number of sellers in the bundle. Including one more seller doesn't necessarily increase revenue, especially with niche sellers who make revenue from the high dispersion in buyer valuation. Bundling very niche sellers decreases dispersion $\sigma$, hurting the revenue.\footnote{For a numerical example, consider two identical sellers with $\mu=1,\sigma=2$. Bundling the two shrinks dispersion by a $\sqrt{2}$ factor: $Rev(2\mu,\sqrt{2}\sigma)=2\mu Rev(1,\frac{\sigma}{\sqrt{2}\mu})\approx1.09$. The bundle revenue is lower than the revenue for selling the items of the two sellers separately $2Rev(\mu,\sigma)=2\mu Rev(1,\frac{\sigma}{\mu})\approx 1.24$.}

\begin{figure}[hbtp!]
    \centering
    \includegraphics[width=0.65\linewidth]{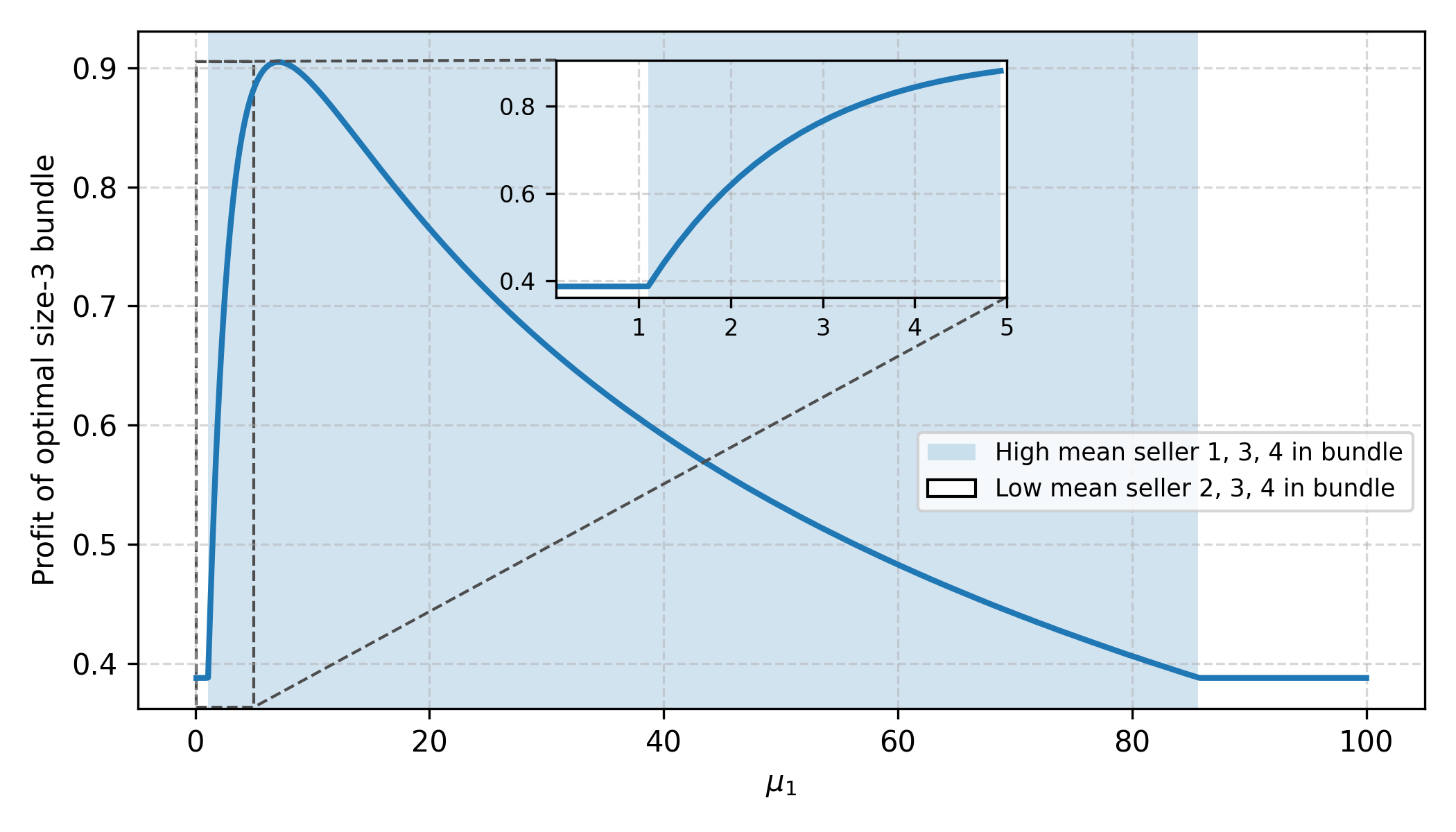}
    \caption{Profit of optimal size-3 bundle vs $\mu_1$ in a four-seller market with $\sigma=1, (\mu_2,\mu_3,\mu_4)=(1.1,2.1,3.1)$ and the first seller has changing $\mu_1$. The shaded area indicates optimal bundle includes seller $\mu_1, \mu_3,\mu_4$, the unshaded area indicates optimal bundle includes seller with $\mu_2, \mu_3, \mu_4$. As $\mu_1$ increases from 0, the optimal size-3 bundle starts to include $\mu_1$ when $\mu_1>\mu_2=1.1$. Bundle with larger mean benefits from reduces variance more. When $\mu_1>\bar{\mu}_1\approx 84$, the optimal size-3 bundle ceases to include $\mu_1$: Platform payment $Rev(\mu_1,1)$ to seller 1 becomes larger than the bundling gain.}
    \label{fig:bundle_or_not_vs_mu}
\end{figure}

A second dimension is which sellers to include in the bundle. Fixing the size of the bundle, the platform needs to balance between the gain of revenue from including high quality sellers, and the high compensation paid to high quality sellers. \Cref{fig:bundle_or_not_vs_mu} illustrates this balance. In a four-seller market with $(\mu_2,\mu_3,\mu_4)=(1.1,2.1,3.1)$, as $\mu_1$ increases from 0 to 100, the profit-maximizing size-three bundle includes seller 1 when $\mu_1 \in [1.1, \bar{\mu}]$ where $\bar{\mu}\approx 84$.

We now extend the result in \Cref{thm:contiguous_mu_normal} to buyer valuations beyond normal distribution, using this two-dimension perspective.
\begin{restatable}{theorem}{ContiguousMuGeneral}
    \label{thm:contiguous_mu_general}
    Consider $N$ sellers indexed so that $\mu_L \leq \mu_1 \leq \mu_2 \leq \ldots \leq \mu_N \leq \mu_H$. Buyer valuations satisfy $$v_i=\mu_i+\sigma Z_i$$
    Then any profit-maximizing bundle is contiguous in quality: if seller $i$ and $j$ are selected with $i<j$, then every seller $k$ with $i<k<j$ is also selected.
\end{restatable}
The proof decomposes the platform profit from any seller bundle $S$ into two components. The first component depends only on the size of $S$ and the price that the platform charges for $S$. The second aggregates seller-specific terms involving $\mu_i$ and $Rev(\mu_i,\sigma)$ for each seller $i\in S$. The proof then parallels that of \Cref{thm:contiguous_mu_normal}: Conditional on the size of $S$ and the bundle price, the platform's profit is shown to be convex in the qualities of the sellers excluded from $S$. Unlike \Cref{thm:contiguous_mu_normal}, \Cref{thm:contiguous_mu_general} does not show that large bundles favor highest-quality sellers, as that result require further assumptions on $Z$ that guarantee $Rev(v)$ being differentiable in $\mu$, and assumptions that guarantee the optimal demand function $F[\frac{p^*-\mu}{\sigma}]$ being monotonically increasing in $\mu$. 

A natural implication of \Cref{thm:contiguous_mu_general} is the existence of a polynomial-time algorithm to select the profit-maximizing bundle, as there are at most $N(N+1)/2$ bundles that are contiguous in seller quality.

\section{Incomplete Information and Heterogeneous Quality}\label{sec:incomplete_info_heterogeneous_mu}
In this section we study an incomplete-information environment. 
The departure from the complete-information setting is that each seller’s quality $\mu_i$ is now private information. The platform does not observe $\mu_i$, but knows that $\{\mu_i\}_{i=1}^N$ are independently and identically distributed according to a distribution with cdf $\Phi$, pdf $\phi$ and support $[\mu_L,\mu_H]$. We use $\mu$ to denote the quality of a representative seller.

Throughout this section, we assume $Z$ is strictly regular, meaning its density function $f$ and tail distribution $F$ are such that $\eta-\frac{F[\eta]}{f(\eta)}$ is strictly increasing in $\eta$. The strict regularity guarantees that $Rev(\mu,\sigma)$ is differentiable in $\mu$ (see \Cref{lem:rev_diff_mu}) and has a unique optimal price, which we will use to derive the platform's expected profit in \Cref{lem:expected_payment_virtual cost}. For our results in \Cref{sec:asymptotic_large_market} we will further assume that $Z$ is sub-exponential, i.e., 
there exist parameters $\gamma,\xi>0$ such that
$$E[e^{\lambda Z}]\leq e^{\frac{\lambda^2\gamma^2}{2}}, \quad \forall \lambda :|\lambda|<\frac{1}{\xi}.$$
For sub-exponential distributions, the difference between $Rev(\mu,\sigma)$ and $\mu$ is bounded. We will use this bound to show a simple mechanism being approximately profit-maximizing in \Cref{sec:asymptotic_large_market}. A range of common random variables are both strictly regular and sub-exponential, including normal, exponential, uniform, logistic, and gamma distributions with the scale parameter larger than one.\footnote{Note that $Z$ having mean zero and unit variance does not automatically guarantee it being sub-exponential. See \Cref{sec:app_incomplete_info} for a counterexample.}

To analyze how to procure the best bundle with incomplete information, we start by following the standard approach in Bayesian mechanism design and define the expected profit of a mechanism in \Cref{sec:myerson_pipeline}. The section proves a version of Myerson's lemma and the revenue equivalence theorem adapted to our setting.
One difference between our problem and standard Bayesian procurement mechanism design problems is that, in our setting, a seller's private type is not their cost; rather, each seller's opportunity cost is a function of their private quality. 
The resulting mechanism design problem is additionally complicated by the fact that the platform's value from obtaining a set of items $S$, $Rev(v_S)$, is a non-linear function of the qualities of the items procured.

\subsection{Defining expected profit-maximizing mechanism}\label{sec:myerson_pipeline}
The platform designs a symmetric mechanism $(x,\pi)$. The allocation rule $x:[\mu_L,\mu_H]\to [0,1]$ maps a seller’s reported quality to the probability that the seller is included in the platform’s bundle, while the payment rule $\pi:[\mu_L,\mu_H]\to\mathbb{R}^+$ specifies the transfer paid to a seller as a function of the reported quality. Although we restrict attention to symmetric mechanisms—reflecting the symmetry induced by quality drawn from identical distribution $\Phi$ and common $\sigma$—the problem remains non-trivial.

By the revelation principle, it is without loss of generality to focus on direct mechanisms that satisfy incentive compatibility (IC) constraints --  guaranteeing that sellers prefer to truthfully report their quality -- and individual rationality (IR) constraints that guarantee voluntary participation. For a seller with true quality $\mu$, its expected utility when reporting quality $\mu'$ is $\pi(\mu')-x(\mu')Rev(\mu,\sigma)$, where the latter term after the minus sign is a seller's expected opportunity cost for not being able to sell its own item when included into the platform's bundle. The IC and IR constraints therefore require that, $\forall \mu$,
\begin{align*}
    (\text{IR}) \qquad \pi(\mu) - x(\mu) Rev(\mu,\sigma) &\geq 0 \\
    (\text{IC}) \qquad \pi(\mu) - x(\mu) Rev(\mu,\sigma) &\geq \pi(\mu')-x(\mu') Rev(\mu,\sigma) \quad \forall \mu'
\end{align*}
As we restrict attention to symmetric mechanisms, a seller's report of its quality does not depend on other sellers' reports. As a consequence, the IC constraint listed above is just as strong as dominant strategy incentive compatibility.

Given an allocation rule $x$, seller inclusion in the bundle is stochastic. Conditional on reported types $\bm\mu =(\mu_1,\ldots,\mu_N)$, let $I_i(\mu_i)\sim\mathrm{Bernoulli}(x(\mu_i))$ be independent indicators of whether each seller $i$ is included. Let $S(\bm\mu, \bm I)$ denote the resulting random set of selected sellers. The platform sells the resulting bundle to buyers and earns revenue $Rev(v_{S(\bm{\mu},\bm I)})$. The \emph{platform’s problem} in the incomplete information setting is to design $(x,\pi)$ to maximize expected profit $\Pi(x,\pi)$ subject to the IC and IR constraints, where
$$\Pi(x,\pi)=E_{\bm\mu\sim \Phi}\left[E_{\bm I\sim \text{Bernoulli}(x(\bm\mu))} \left[Rev(v_{S(\bm\mu, \bm I)})\right]- \sum_{i\in [N]} \pi(\mu_i)\right].$$

The platform's problem is a Bayesian mechanism design problem \citep{hartline2013bayesian}. 
In our procurement setting, sellers have private quality, and hence private opportunity costs $Rev(\mu,\sigma)$ for their items, and have to get paid to part with the items and report the private quality truthfully. 
The platform runs a mechanism to pay to acquire sellers' items, and the value obtained by the platform is the revenue that can be obtained by selling those items as a bundle.

The platform's value function $Rev(v_{S(\bm\mu, \bm I)})$ is non-linear, but nevertheless each seller has a single-dimensional type $\mu_i$ that determines its opportunity cost $Rev(\mu_i, \sigma)$. Myerson's classic characterization of IC and IR allocation rules therefore applies. We restate it here in the notation of our model.

\begin{restatable}{lemma}{Myerson}
\label{lem:myerson_monotonic}
    Myerson's Lemma \citep{myerson1981optimal}
    \begin{itemize}
        \item For any $x$, there exists a payment rule $\pi$ such that the mechanism $(x,\pi)$ satisfies IC if and only if it is monotonically non-increasing in $\mu$: for $\mu\geq \mu'$, $x(\mu)\leq x(\mu')$.
        \item If $x$ is monotonically non-increasing, then there exists a unique payment rule $\pi$ such that $(x,\pi)$ satisfies the IC constraint, subject to a normalization term that sets the utility of a seller of quality $\mu_H$. 
        \item When the utility of a $\mu_H$-quality seller is set to zero, the above payment rule is given by  
        \begin{align*}
            \pi(\mu)=Rev(\mu,\sigma)x(\mu)+\int_{\theta=\mu}^{\mu_H} x(\theta)dRev(\theta,\sigma)
        \end{align*}
    \end{itemize}    
\end{restatable}
The difference from the standard Myerson's lemma is that the expected payment involves the derivative of $Rev(\mu,\sigma)$ on $\mu$.  This is because the seller IC constraint is with respect to the seller's private quality $\mu$, rather than their (opportunity) cost $Rev(\mu,\sigma)$.  This choice is crucial to our analysis of which seller qualities are attracted by the platform. It is also the reason we assume $Z$ is strictly regular: this regularity assumption on the buyer's values (not the seller types) is not driven by a desired monotonicity of virtual values, as is typical in mechanism design, but rather to ensure differentiability of $Rev(\mu,\sigma)$ with respect to $\mu$. We only write out Myerson's lemma when the allocation rule $x$ is differentiable. For any deterministic allocation rule where $x(\mu)\in\{0,1\}$, the first statement of the lemma means that $\forall \mu, x(\mu)=1 \text{ or } \forall \mu, x(\mu)=1$, or $x(\mu)$ is a step function: $\exists t$ such that $x(\mu)=1 \text{ if and only if } \mu\leq t$. All sellers included in the bundle gets paid equal to the revenue $Rev(t,\sigma)$ of threshold type $t$.

The above lemma suggests that, in order to induce sellers truthfully reporting the qualities, the platform must include lower-quality sellers in the bundle with higher probability and higher payments. This is a big contrast to the complete-information setting: in large markets, the profit-maximizing bundle includes high-quality sellers, as shown in \Cref{thm:contiguous_mu_normal}. From the platform's perspective, the loss in profit from incomplete information is costly, and this motivates our study of in-house production 
in \Cref{sec:in_house_production}.

In the mechanism-design framework, expected revenue equals expected virtual surplus. We now present the corresponding result in our setting.
\begin{restatable}{lemma}{VirtualCost}
\label{lem:expected_payment_virtual cost}
    A mechanism's expected payment to a seller is given by $$E_{\mu\sim\Phi}\left[\pi(\mu)\right]= E_{\mu\sim\Phi}\left[\varphi(\mu) x(\mu)\right]$$
    where $$\varphi(\mu)=Rev(\mu,\sigma)\left(1+\frac{\Phi(\mu)}{\phi(\mu)p^*(\mu)}\right)$$
    and $p^*(\mu) = \argmax_p\{p Pr[\mu+\sigma Z\geq p]\}$ is the unique optimal price for $Rev(\mu,\sigma)$.
\end{restatable}
The proof is an adaptation of Myerson's virtual surplus formulation of expected revenue, using the fact that $\frac{\partial Rev(\mu,\sigma)}{\partial \mu}= F[\frac{p^*-\mu}{\sigma}]$ (proved in \Cref{lem:rev_diff_mu} for strictly regular random variable $Z$) to expand the payment rule in \Cref{lem:myerson_monotonic}. Define the platform's expected profit when using a mechanism with allocation rule $x$ to be 
\begin{equation}\label{eq:expected_profit}
\Pi(x) = E_{\bm\mu\sim\Phi} \left[E_{\bm I} \left[Rev(v_{S(\bm\mu, \bm I)})\right]-\sum_{i\in [N]}\varphi(\mu_i)x(\mu_i)\right]
\end{equation}
We omit the dependency on $\pi$ as \Cref{lem:myerson_monotonic} uniquely determines $\pi$ with $x$. The main difficulty in directly characterizing the expected profit-maximizing mechanism is that the platform's revenue function $Rev(v_{S(\bm\mu,\bm I)})$ can be highly complex, as noted by the following lemma. 
\begin{restatable}{lemma}{NoneMonotonicNorSubmodular}
\label{lem:none_monotonic_nor_submodular}
    Define the set function $Rev(S):=Rev(v_S)$.
    Even when buyer valuations are normally distributed with identical variance, $Rev(S)$ is neither monotonic, submodular, supermodular, subadditive, nor superadditive.
\end{restatable}

To address this challenge, we introduce a surrogate profit function in \Cref{sec:deterministic_optimal} that approximates $Rev(v_{S(\bm\mu,\bm I)})$ and allows for profit-maximizing mechanisms in simple form. We then show in \Cref{sec:asymptotic_large_market} that, as the number of sellers grows large, mechanisms that maximize the surrogate profit are close to the true profit-maximizing mechanism.

\subsection{Optimal Mechanism for Surrogate Profit}\label{sec:deterministic_optimal}
The surrogate function that we employ is to use the mean of buyer valuation $\mu_{S(\bm\mu,\bm I)}=\sum_{i\in S(\bm\mu,\bm I)} \mu_i$ of the bundle to replace the optimal revenue $Rev(v_{S(\bm\mu,\bm I)})$ of the bundle. Intuitively, this corresponds to a very large market regime where, for any bundle $S$, the dispersion of buyer valuation towards this bundle drops to zero compared to the quality of this bundle. We will make this intuition clear in \Cref{sec:asymptotic_large_market}.
Define the expected 
surrogate profit to be 
\begin{align}
    \varpi(x) &= E_{\bm\mu\sim\Phi} \left[E_{\bm I}\left[\sum_{i:I_i(\mu_i)=1}\mu_i\right]-\sum_{i\in [N]}\varphi(\mu_i)x(\mu_i)\right]= N E_{\mu\sim \Phi} \left[x(\mu)(\mu-\varphi(\mu))\right] \label{eq:expected_surrogate_profit}
\end{align}
By replacing the complex revenue term $Rev(v_S)$ with the surrogate $\sum_{i\in S}\mu_i$, we linearized the objective inside the expectation. Because of symmetry, the optimal expected surrogate profit mechanism turns out to be a threshold mechanism. The proof follows the same structure as that for Myerson's revenue optimal mechanism with ironing.

\begin{figure}[hbtp!]
    \centering
    \includegraphics[width=0.5\linewidth]{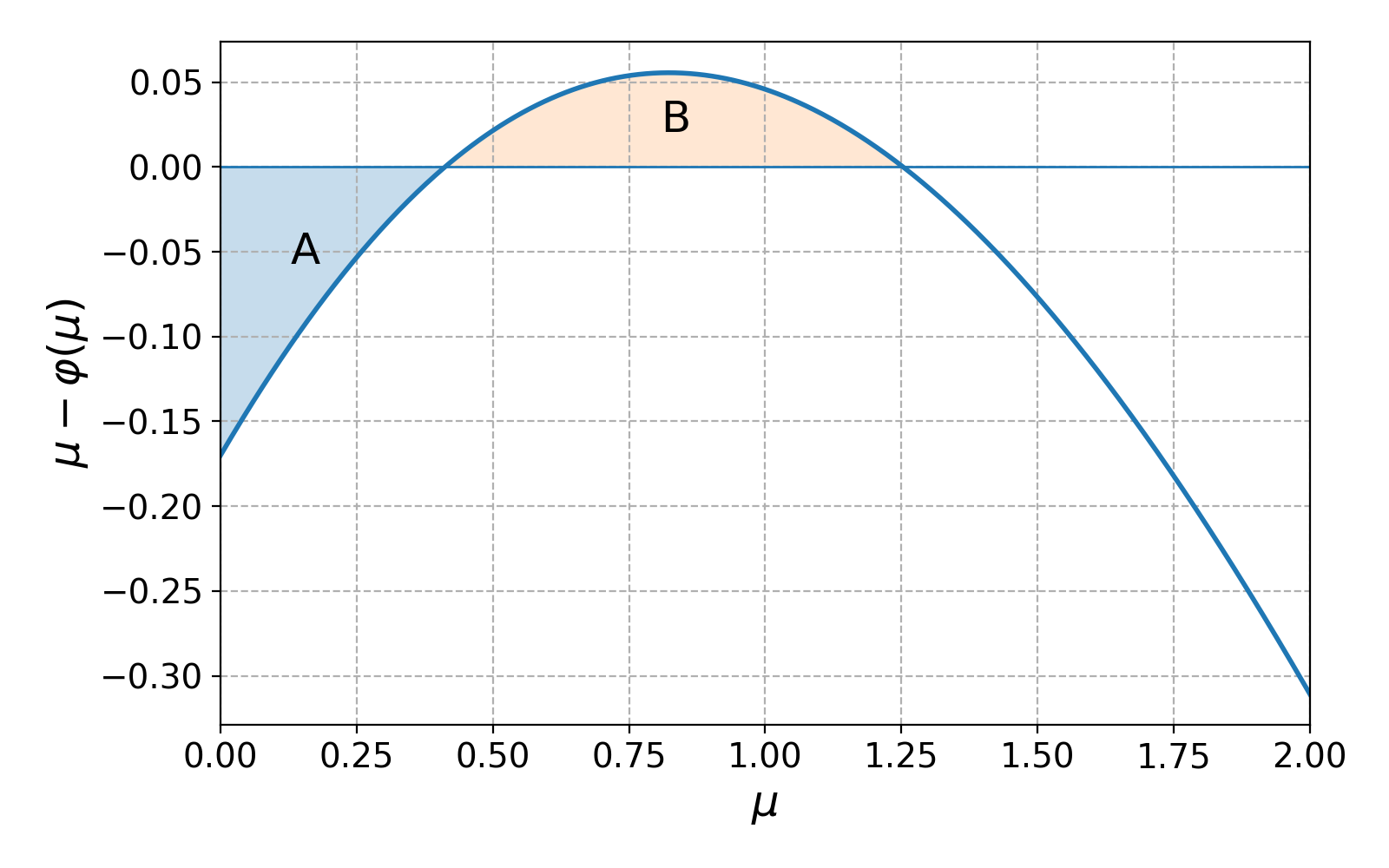}
    \caption{$\mu-\varphi(\mu)$ as function of $\mu$. illustrating expected surrogate profit $\varpi$ for buyer valuations drawn from a normal distribution with parameters $(\mu,\sigma=1)$, and seller qualities $\mu$ drawn from a uniform distribution $U[0,2]$.  The area under the curve are approximately equal $A\approx B\approx 0.031$.}
    \label{fig:virtual_surplus}
\end{figure}

\begin{restatable}{theorem}{ProfitMaxSurrogate}
\label{thm: profit_max_surrogate}
    There exists an IC and IR mechanism that maximizes expected surrogate profit, characterized by a threshold $t\in \mathbb{R}$, where
    \begin{align*}
    x(\mu)=
    \begin{cases}
       1 &\qquad {if }\; \mu\leq t \\
       0 &\qquad \text{else}
    \end{cases};
    \quad
    \pi(\mu)=\begin{cases}Rev(t,\sigma) &\qquad {if }\; \mu\leq t 
    \\0 &\qquad \text{else}\end{cases}
\end{align*}
\end{restatable}
Having shown that the optimal mechanism has a relatively simple form, let us briefly discuss the difference between this mechanism and the first-best expected surrogate profit.  That is, how much expected surrogate profit is lost due to the incomplete information on $\mu$? \Cref{fig:virtual_surplus} presents such an illustration. In a market where buyer valuation is drawn from normal distribution with $\sigma=1$, and seller quality $\mu$ drawn from uniform distribution $U[0,2]$, the figure plots $\mu-\varphi(\mu)$ as function of $\mu$. Under complete information, if the platform observes sellers' private types $\mu$, it can extract the entire area under the curve above zero, denoted by region B, as surrogate profit. Under incomplete information, however, any IC mechanism must employ an allocation rule that is monotone non-increasing in $\mu$. As a result, the platform can extract at most the surplus corresponding to region B net of the loss in region A. Consequently, the attainable surrogate profit in this example is equal to $\max\{0, B-A\}\approx 0$.

\subsection{Asymptotic Optimality in Large Markets}\label{sec:asymptotic_large_market}
This section provides a justification for using surrogate profit as the objective function. We show that in large markets, optimizing for the surrogate profit leads to a mechanism that closely approximate the true profit-maximizing mechanisms. 

Let $x^* \in \argmax_x \Pi(x)$ be an IC mechanism allocation rule that maximizes expected profit, and $\tilde{x}^*\in \argmax_x \varpi(x)$ be the IC mechanism allocation rule that maximizes expected surrogate profit, specified in \Cref{thm: profit_max_surrogate}. We show in \Cref{thm:asymptotic_profit} that $\tilde{x}^*$ is almost as good as $x^*$ in large markets.

As the expected surrogate profit function $\varpi$ uses the mean valuation $\mu_S=\sum_{i\in S}\mu_i$ to replace $Rev(v_S)$, we need to bound the difference between $Rev(v_S)$ and $\sum_{i\in S}\mu_i$ resulting from the dispersion in buyer valuation. This is where we make use of $Z$ being a sub-exponential random variable.
\begin{restatable}{lemma}{SigmaRevBound}
\label{lem:sigma_rev_bound}
    Assume $Z$ is a subexponential random variable $Z\in SE(\gamma^2,\xi)$ with parameter $\gamma,\xi>0$ : $$E[e^{\lambda Z}]\leq e^{\frac{\lambda^2\gamma^2}{2}}, \quad \forall \lambda :|\lambda|<\frac{1}{\xi}$$ 
    For any buyer valuation $v=C+ \sigma \sum_{i\in [N]}a_iZ_i$ where $C, \sigma$ are constants and $a_i\in [0,1]$, let $A=\sum_{i\in [N]}a_i^2$. Then 
    $$|Rev(C+\sigma\sum_{i\in [N]} a_i Z_i)-C|\leq \max\{\sigma \gamma e^{-\frac{1}{2}}A^{\frac{1}{2}},\frac{8\sigma\xi}{3 e},\sqrt{2}\sigma A^{\frac{1}{2}}, 2\sigma^{\frac{2}{3}}(CA)^{\frac{1}{3}}\}$$
\end{restatable}
To upper bound $Rev(v)-C$, we use the assumption that $Z$ is sub-exponential, which ensures that demand at high prices decays at least exponentially. To upper bound $C-Rev(v)$, we select a price slightly below $C$ and lower bound its demand and the revenue at this price. This lemma enables us to prove that the mechanism that maximizes expected surrogate profit is approximately optimal in large markets.
\begin{restatable}{theorem}{AsymptoticProfit}
\label{thm:asymptotic_profit}
    Consider buyer have valuation $\mu_i+\sigma Z$ towards seller $i$, where $Z$ is any subexponential random variable $Z\in SE(\gamma^2,\xi)$ with mean zero and standard deviation 1.
    If $\varpi(\tilde{x}^*)>0$, then $$\frac{\Pi(x^*)}{\Pi(\tilde{x}^*)}\leq 1+O(N^{-\frac{1}{3}})$$
\end{restatable}
The proof decomposes the difference between $\Pi(x^*)$ and $\Pi(\tilde{x}^*)$ into two parts. The first part arises from the random realization of $\bm I\sim \text{Bernoulli}(x(\bm \mu))$, and further splits to (1) the fluctuation in mean quality $\sum_{i\in [N]}(I_i(\mu_i)-x(\mu_i))\mu_i$, whose expectation is $O(N^{\frac{1}{2}})$ by standard concentration for Bernoulli sums, and (2) the difference resulting from dispersion of buyer valuations, which is controlled by \Cref{lem:sigma_rev_bound}. The second source arises from $\tilde{x}^*$ optimizing for the mean valuation instead of the real revenue $Rev(v_S)$ of a bundle $S$; the difference is $O(N^{\frac{2}{3}})$, also bounded by \Cref{lem:sigma_rev_bound}.

The results in this section show that, under incomplete information, bundling sellers with contiguous quality remains approximately profit-maximizing. The contiguity in quality is a result of seller's incentive compatibility constraints, rather than the trade-off between bundle revenue and acquisition costs as in \Cref{sec:complete_info_heterogeneous_mu}. Moreover, any incentive compatible mechanism selects lower-quality sellers with higher probability---opposite to the complete-information result that favors high-quality sellers in \Cref{thm:contiguous_mu_normal}. As a result, including high-quality sellers in platform's bundle necessitates paying high information rents, which can be detrimental to platform profit. This motivates our analysis in the next section on platform in-house production.

\section{Platform in House Production}\label{sec:in_house_production}
Having established that information asymmetry between the platform and sellers can harm profit, we next study how in-house production may help increase profit. Beyond selecting which sellers to contract with, the platform can produce its own items and bundle them with externally sourced content before selling to buyers. By adjusting both the quantity and the quality of in-house production, the platform can substitute for high-quality sellers and reduce the information rents it pay to low-quality sellers. We characterize which sellers the platform sources from and what the platform produces in-house to constitute the optimal bundle.
 
To make the analysis tractable, we focus on a stylized environment. We will assume that $Z$ is a standard normal random variable, where recall that the buyer's value for a seller $i$'s item is expressed as $v_i=\mu_i+\sigma Z$. Sellers come in only two possible quality levels $\mu\in \{\mu_L,\mu_H\}$, with $0<\mu_L<\mu_H$. There are $N$ sellers in the market. A fraction $\tau \in (0,1)$ are low-quality, so $N_L=\lceil \tau N\rceil$, and the remaining $N_H= N- N_L$ are high-quality sellers. The platform knows the total number of sellers $N$ and the composition $(N_L, N_H)$, but does not observe individual seller's quality. The platform has the ability to produce up to $M$ units of items in-house. Each in-house item can be produced at either quality $\mu_L$ or $\mu_H$, and the platform perfectly observes the quality of its own production.
The cost to the platform of producing an item of quality $\mu \in \{\mu_L, \mu_H\}$ is taken to be $Rev(\mu,\sigma)$. The reason for this choice is that, regardless of the actual physical cost of creating an item, there is a natural outside option of selling it individually at its monopoly price. We can therefore interpret this cost as either the opportunity cost of not selling the produced item individually off-platform, or as a contract payment needed to incentivize a producer to create the item for the platform directly rather than for individual sale.

Again for tractability, we restrict attention to deterministic mechanisms.
Since there are only two seller qualities, without loss of profit, the platform adopts a posted-price mechanism: it posts price $\pi_L:=Rev(\mu_L,\sigma)$ in which case all $N_L$ low-quality sellers are on-platform and are bundled with the in-house items; $\pi_H:=Rev(\mu_H,\sigma)$ in which case all $N$ sellers are bundled with in-house produced items, or $\pi_0=0$ in which case no sellers are on-platform and only in-house items are bundled. Let $S(\pi)$ be the set of sellers selected by posted price mechanisms, and $S_M$ denote the set of in-house produced items. The platform sells the resulting bundle $S(\pi)\cup S_M$ to buyers and earn revenue $Rev(v_{S(\pi)\cup S_M})$, with a profit of 
$$\Pi(\pi,S_M) = Rev(v_{(S(\pi)\cup S_M)})- |S(\pi)|\pi-\sum_{i\in S_M}Rev(\mu_i, \sigma)$$
where recall $Rev(\mu_i,\sigma)$ corresponds to platform's opportunity cost for selling its in-house produced item $\mu_i$ individually.
The platform's problem is to jointly choose a posted price and the set of in-house produced items $S_M$ to maximize profit. We first show that, however many units are produced in-house, the platform either always produces all of them at quality $\mu_L$ or all of them at quality $\mu_H$.
\begin{restatable}{lemma}{NoMix}
\label{lem:in_house_no_mix}
    Suppose the platform produces $m$ items in-house where $0<m\leq M$. Then the profit-optimal choice of quality is to either always produce low-quality items $\mu_L$, or always produce high-quality items $\mu_H$. Conditional on including some sellers in the bundle, a sufficient condition for the platform always producing high-quality items is $$\min\{F[\sqrt{N_L+m}\frac{\mu_L}{\sigma}],F[\frac{(N_L+m)\mu_L+N_H \mu_H}{\sqrt{N+m}\sigma}]\}>\frac{Rev(\mu_H,\sigma)-Rev(\mu_L,\sigma)}{\mu_H-\mu_L}.$$
    Conditional on including some sellers in the bundle, a sufficient condition for the platform always producing low-quality items is $$\max\{F[\frac{N_L\mu_L+m \mu_H}{\sqrt{N_L+m}\sigma}],F[\frac{N_L \mu_L+(N_H+m)\mu_H}{\sqrt{N+m}\sigma}]\}<\frac{Rev(\mu_H,\sigma)-Rev(\mu_L,\sigma)}{\mu_H-\mu_L}.$$ 
\end{restatable}
The proof shows that the profit function is convex in the number of low-quality items. The two sufficient conditions reflect the relationship between the size of the final bundle ($N+m$ or $N_L+m$) and the quality of platform's in-house production. When the bundle is large, the reduction in dispersion of buyer valuation is large, meaning there is large gain in bundling. This gain justifies the large cost of producing $\mu_H$. When the size of the bundle is small, the gain of revenue from is not enough to cover the high costs of producing $\mu_H$.

With this intuition, we fully characterize the platform's optimal strategy in a sufficiently large market where the platform's production capacity is small compared to the market size. 
\begin{restatable}{theorem}{BundleSeller}
\label{thm:in_house_production}
    There exists a threshold $N_0$ on the market size $N$, such that when the number of sellers is larger than this threshold $N>N_0$, platform's profit-maximizing posted price is equal to  
    \begin{itemize}
        \item If $\mu_L > Rev(\mu_L,\sigma)$ 
        $$
    \begin{cases}
        \pi_H & \text{ if } \tau \leq \frac{\mu_H-Rev(\mu_H,\sigma)}{\mu_H-Rev(\mu_L,\sigma)}\\
        \pi_L & \text{otherwise}
    \end{cases}
    $$
        \item If $\mu_L-Rev(\mu_L,\sigma)\leq 0$ but $\mu_H-Rev(\mu_H,\sigma)>0$
        $$
    \begin{cases}
        \pi_H & \text{ if } \tau \leq \frac{\mu_H-Rev(\mu_H,\sigma)}{\mu_H-\mu_L}\\
        \pi_0 & \text{otherwise}
    \end{cases}
    $$
        \item If $\mu_H-Rev(\mu_H,\sigma)\leq 0$, then profit-maximizing posted price is equal to $\pi_0=0$.
    \end{itemize}
    In these large markets, the platform never in-house produces low-quality items. If the profit-maximizing posted price is $\pi_L$ or $\pi_H$, the platform in-house produces high-quality item if and only if $\mu_H>Rev(\mu_H,\sigma)$. If the profit-maximizing posted price is $\pi_0$, the platform in-house produces high-quality item if and only if 
    $$\exists\, 1\leq m\leq M,\quad  Rev(m\mu_H,\sqrt{m}\sigma)-mRev(\mu_H,\sigma)>0$$
\end{restatable}
The intuition for this theorem is that in these large markets where buyer valuations are normally distributed, bundling shrinks the dispersion in buyers valuation at a rate of $1/\sqrt{N}$. So the revenue from selling a bundle approaches buyer's mean valuation towards the bundle, and platform's profit for including a seller of quality $\mu$ is roughly equal to $\mu - Rev(\mu,\sigma)$. This is why we present the optimal posted price by discussing three cases of $\mu-Rev(\mu,\sigma)$.
Platform's in-house production strategy follows from similar logic: When selling the in-house produced item as a part of the bundle, the profit derived approaches $\mu-Rev(\mu,\sigma)$ when the bundle size is large. By \Cref{lem:norm_rev_character}, $\mu-Rev(\mu,\sigma)$ monotonically increases in $\mu$, so the platform never bundle low-quality items.

\begin{figure}[hbtp!]
    \centering
    \begin{subfigure}{0.47\textwidth}
        \centering
        \includegraphics[width=\textwidth]{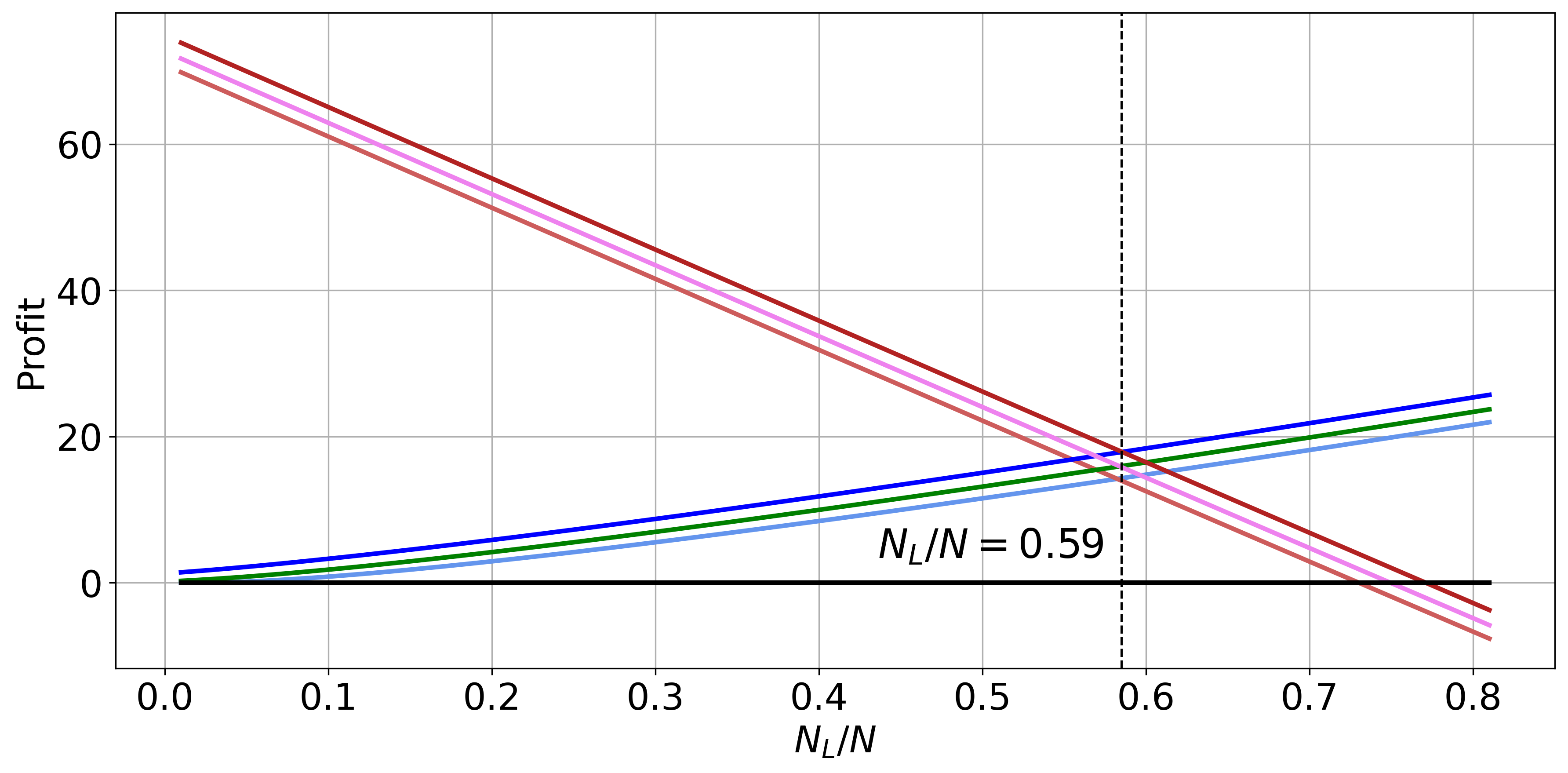}
        \caption{$\mu_L>Rev(\mu_L,\sigma)$, $\mu_L=1, \mu_H=2$. Profit for Just Produce Low: 0.159; Just Produce High: 1.278. In infinitely-large market, the optimal strategy changes at $\tau=\frac{\mu_H-Rev(\mu_H,\sigma)}{\mu_H-Rev(\mu_L,\sigma)}\approx 0.635$.\\\\}
        \label{fig:in_house_produce_case1}
    \end{subfigure}
    \hfill
    \begin{subfigure}{0.47\textwidth}
        \centering
        \includegraphics[width=\textwidth]{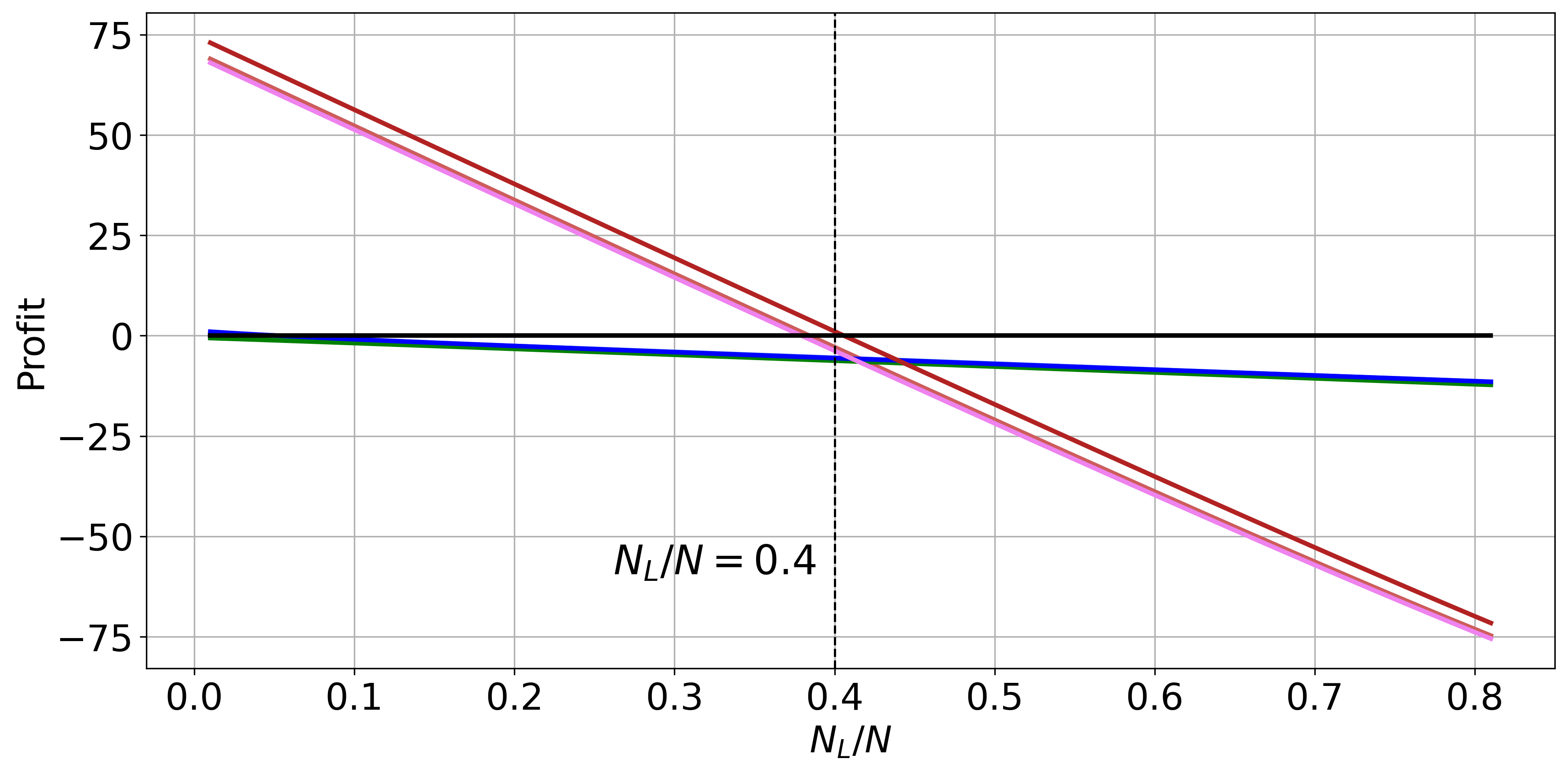}
        \caption{$\mu_L\leq Rev(\mu_L,\sigma),\mu_H> Rev(\mu_H,\sigma)$, $\mu_L=0.1, \mu_H=2$. Profit for Just Produce Low: -0.463; Just Produce High: 1.278. The profit-maximizing strategy when $N_L/N>0.4$ is to in-house produce high-quality items. In infinitely-large market, the optimal strategy changes at $\tau=\frac{\mu_H-Rev(\mu_H,\sigma)}{\mu_H-\mu_L}\approx 0.5$.}
        \label{fig:in_house_produce_case2}
    \end{subfigure}\\
    \begin{subfigure}{0.7\textwidth}
        \centering
        \includegraphics[width=\textwidth]{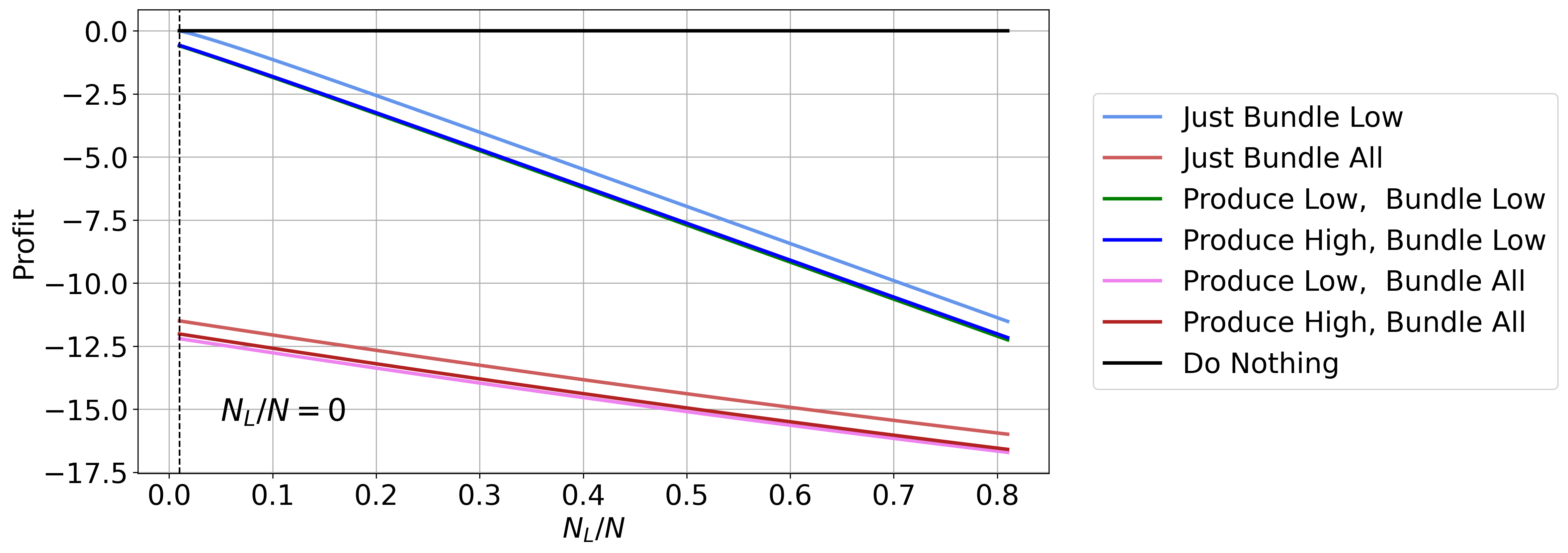}
        \caption{$\mu_H<Rev(\mu_H,\sigma)$, $\mu_L=0.1,\mu_H=0.2$. Profit for Just Produce Low: -0.463, Just Produce High: -0.443.}
        \label{fig:in_house_produce_case3}
    \end{subfigure}
    \caption{Profit for different strategies vs the ratio of low quality-sellers $N_L/N$ in three markets. All three markets have $N=100, \sigma=1$, and whenever the platform produces in house, it produces $m=5$ items. The three markets differ only in $\mu_L, \mu_H$. By \Cref{thm:in_house_production}, in large markets in-house production of low-quality items is never optimal, so strategies that produce low-quality items are omitted. The \emph{Just Produce} strategies (Just Produce Low/ Just Produce High) involve no sellers and therefore yield profits that do not vary with $N_L/N$; their values are reported in each panel caption. A vertical dotted line marks the value of $N_L/N$ at which the profit-maximizing strategy switches. The panel caption also notes out the value where profit-maximizing strategy switches when the market is infinitely large, as noted in \Cref{thm:in_house_production}. This value is approached by the value indicated in the dotted line, when the market size grows.}\label{fig:three_cases_in_house_production}
\end{figure}

We further illustrate this theorem with \Cref{fig:three_cases_in_house_production}, where all three markets have $N=100, \sigma=1$, and we assume whenever the platform produces in house, it produces $m=M=5$ items.

\Cref{fig:in_house_produce_case1} illustrates a market where sourcing from low-quality sellers is profitable (i.e., $\mu_L>Rev(\mu_L,\sigma)$), as well as from high-quality sellers (i.e., $\mu_H>Rev(\mu_H,\sigma)$).
As the fraction of low-quality sellers $\tau$ increases, the information rent when sourcing from all sellers rises, leading the platform to switch from sourcing from all sellers to only low-quality sellers. \Cref{thm:in_house_production} suggests that when $\mu_H>Rev(\mu_H,\sigma)$, the platform in-house produces high-quality items in large markets. And indeed the profit-maximizing strategy in \Cref{fig:in_house_produce_case1} in-house produces high-quality items.

\Cref{fig:in_house_produce_case2} illustrates a market where sourcing from high-quality sellers is profitable but not low-quality sellers: $\mu_L\leq Rev(\mu_L,\sigma) \text{ and } \mu_H>Rev(\mu_H,\sigma)$. Since the platform cannot tell low from high quality sellers, it needs to source from all sellers. As $\tau$ grows, the information rent eventually causes the platform to not source from any sellers. Since $\mu_H>Rev(\mu_H,\sigma)$, it is always profitable to in-house produce high-quality items, even when sourcing from external sellers is not profitable. This is also an example where, absent the platform's in-house production capability, the platform would earn non-positive profit. Finally, when $\mu_L<Rev(\mu_L,\sigma)$ and $\mu_H<Rev(\mu_H,\sigma)$, bundling sellers is not profitable at all and neither is in-house production. As a result, the optimal strategy is to do nothing, as illustrated in \Cref{fig:in_house_produce_case3}.

\paragraph{We next comment on when it can be profit-maximizing for the platform to produce low-quality items in-house.} \Cref{lem:no_low_in_house_production} proves that it is never optimal to only produce low-quality items in-house without sourcing from external sellers, since it is dominated by only producing high-quality items. When the quality gap is sufficiently large, specifically when $\frac{\mu_H}{\mu_L}> \frac{N_L}{\sqrt{m+N_L}-m}$ (which in particular requires $\sqrt{m+N_L}>m$), there are markets in which producing low-quality items in-house and sourcing from external sellers is optimal. Intuitively, in this regime the marginal revenue gain from high-quality in-house production is too small relative to its higher costs, so low-quality production can be preferable. We present a representative example in \Cref{exam:low_quality_in_house_optimal}: the platform would like to profitably bundle many low-quality items, but there are too few low-quality sellers available, so it fills the gap through in-house production. Such cases arise only in relatively small markets with the platform producing only a limited number of items; in contrast, in the large market regime of \Cref{thm:in_house_production}, low-quality in-house production is never optimal.
\begin{restatable}{lemma}{NoLowInhouseProduction}
\label{lem:no_low_in_house_production}
    It can never be profit-maximizing to only in-house produce low-quality items without sourcing from external sellers. A necessary condition for producing low-quality items to be profit-maximizing is that
    $$\frac{\mu_H}{\mu_L}> \frac{N_L}{\sqrt{m+N_L}-m}\quad \text{and} \quad N_L>m(m-1)$$
    where $m$ is the number of items the platform in-house produce.
\end{restatable}

\begin{figure}
    \centering
    \includegraphics[width=0.5\linewidth]{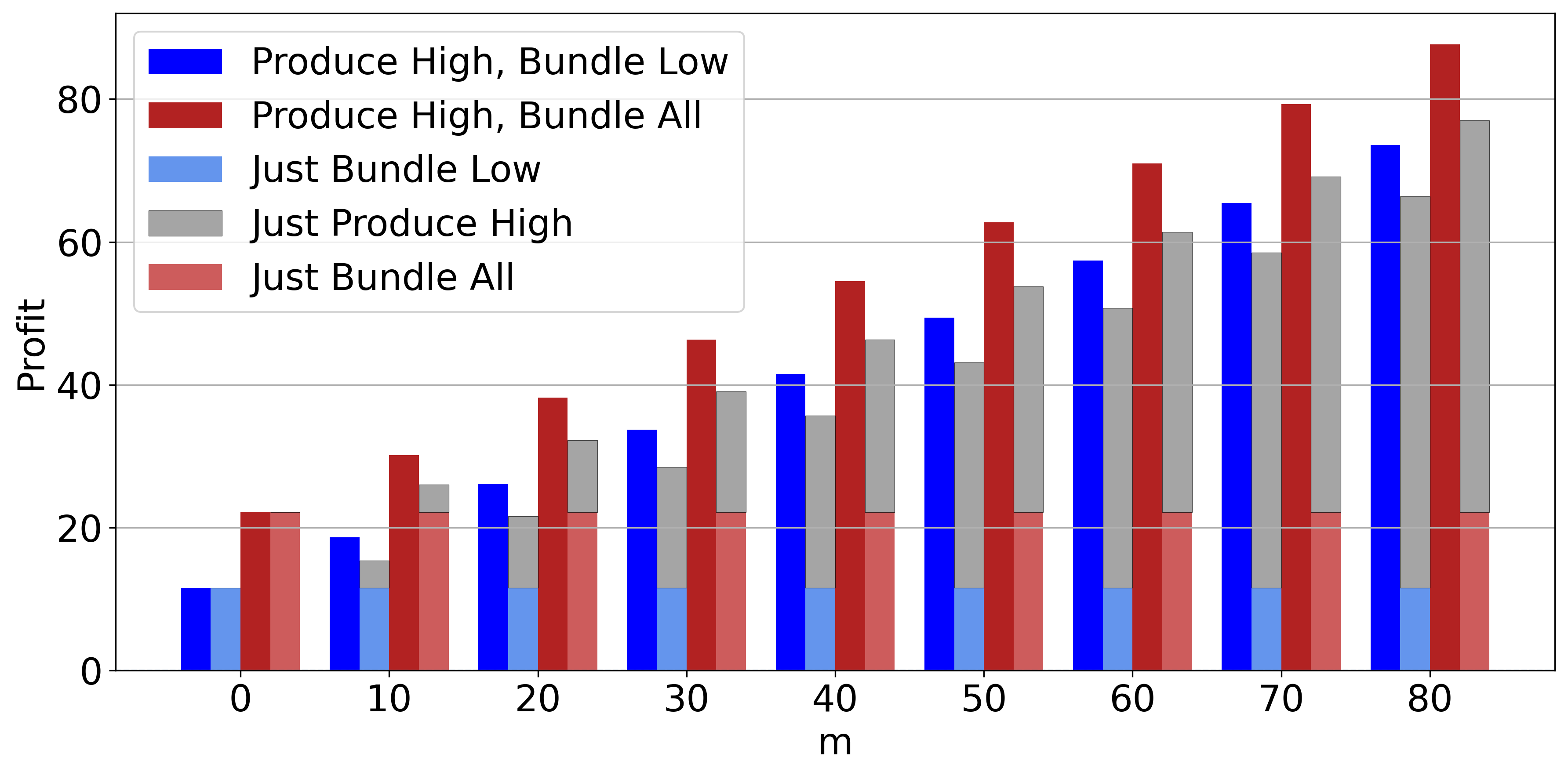}
    \caption{Profit for different strategies vs $m$, the number of items the platform in-house produces in market with $N=100, N_L=50, \mu_L=1, \mu_H=2$. For each $m$, the two stacked bars add the profit of just producing high-quality items, respectively with just bundle low-quality sellers, and just bundle all sellers.}
    \label{fig:in_house_production_changing_capacity}
\end{figure}

\paragraph{Complementarity between production and bundling external sellers.} Next, we present a market in \Cref{fig:in_house_production_changing_capacity} where joint in-house production and bundling yields higher profit than the sum of only producing in-house or only sourcing from external sellers. This complementarity implies that a platform with bundling capability can extract greater value from in-house production than a standalone seller.

For every in-house production level $m$, \Cref{fig:in_house_production_changing_capacity} uses a bar plot to compare the platform's profit from \emph{jointly} producing $m$ high-quality items and bundling the $N_L$ low-quality sellers, expressed as
$$Rev(m\mu_H+N_L\mu_L,\sqrt{m+N_L}\sigma)-mRev(\mu_H,\sigma)-N_LRev(\mu_L,\sigma),$$
against the profit obtained by treating producing $m$ high-quality items alone and bundling the $N_L$ low-quality sellers alone,
$$ Rev(m\mu_H,\sqrt{m}
\sigma)-mRev(\mu_H,\sigma) + Rev(N_L\mu_L,\sqrt{N_L}\sigma)-N_L Rev(\mu_L,\sqrt{N_L}\sigma).$$
The gap between these two profits reflects a \emph{complementarity} effect in profit.
In this example, a platform with the ability to bundle is in a uniquely advantageous position: it can extract more value from in-house production of high-quality content than a standalone seller. We interpret this as one potential reason for high-quality content to be produced and released directly in partnership with a subscription-based platform.

While the complementarity is market specific (See \Cref{exam:sub-modularity_bundling_producing} for a market where the profit in jointly bundling and producing is smaller than that of bundling alone plus bundling alone), it is a phenomenon that occurs in small markets.  Indeed, as the market size increases, the revenue of a set of items approaches the mean valuation, so the profit from jointly producing and bundling becomes approximately additive.

\section{Discussion}
We examine a platform that acquires items from third-party sellers with different qualities and sells them as a single bundle to buyers with additive valuations. When item quality is observable, the profit-maximizing bundle is contiguous in quality but may exclude both high-quality and low-quality items. When quality is privately known to each seller, we show that a simple posted-price mechanism is asymptotically profit-maximizing in large markets. Because information asymmetry can reduce platform profit, we study platform in-house production as a way to control quality while sourcing from third-party sellers with uncertain quality. When in-house production is profitable, the platform always produces high-quality items in large markets. In-house production can be complementary to external sourcing: total profit may exceed the sum achievable by only producing in-house and only sourcing externally alone. Many interesting questions remain, offering avenues for future research. 
\begin{itemize}
    \item In the incomplete information case, we derive an asymptotically profit-maximizing mechanism for sub-exponential buyer valuation distribution. Characterizing profit-optimal mechanisms under heavy-tailed valuation distribution remains an open question.
    \item While our analysis of in-house production assumes two quality levels and a fixed production capacity, it would be natural to generalize the model to a continuum of qualities. 
    \item We study a single platform that acquires items from third-party sellers. In real life, streaming platforms compete when licensing content from producers, raising question of how inter-platform competition would affect the quality composition of on-platform content.  
\end{itemize}

\chapter{Closing Remarks}
Markets have traditionally been viewed as systems in which participants freely trade with one another. In this classical view, participants can interact with all relevant counterparties. In modern online platforms, however, the scale of the market requires mediation. Because participants cannot realistically interact with all potential counterparties, platforms mediate these interactions through recommendation systems, often directing participants to options that best match their needs. At the same time, this intermediation gives platforms significant power to influence access: they determine which participants can view each another and how transactions are structured through design choices such as pricing, matching, and sourcing decisions. As profit-maximizing entities, platforms' objectives may not fully align with those of participants. As a result, market outcomes are shaped jointly by the platform's design choices and how participants strategically respond to them. 

This thesis explores how platforms use three design levers to shape access and, in turn, influence market outcomes. Part I studies \emph{Indirect Access Control with Pricing}, where commission fees charged by platforms affect sellers' decision to join the platform, and the timing of revealing the tip amount paid by buyers to couriers affects couriers decisions over which orders to accept. This illustrates how prices, which emerge endogenously in markets where participants freely trade with one another, in platform-mediated markets can be used as an instrument to shape market outcomes.

Part II of the thesis studies \emph{Direct Access Control with Matching}, where platforms determine which buyer-seller pairs are allowed to interact. From a regulatory perspective, because platforms maximize revenue which does not always align with overall market welfare, and because the set of feasible buyer-seller interactions shapes market outcomes, regulators should carefully examine platform's matching decisions. At the same time, even when platforms optimize for revenue, allowing platforms to determine matching can guarantee social welfare to be a fraction of optimal welfare, and can facilitate market recovery following economic shocks.

Part III of the thesis studies \emph{Access Control with Bundling}, in which platforms operate as middlemen who sign exclusivity contracts with selected sellers, and sell the items to downstream buyers as a bundle. Through exclusivity contracts, sellers sourced by the platform are no longer accessible by buyers outside of the platform. This part offers structural predictions on which sellers remain accessible by buyers outside of the platform.

Taken together, these results highlight the role of access control in shaping market outcomes in platform-mediated environments. Looking ahead, the questions studied in this thesis can be extended to more dynamic settings, enriched with data, and applied to contexts where emerging technologies reshape how access is created and mediated.

While this thesis primarily model platform-mediated online markets as static environments, real-world problems involving complex incentives are inherently dynamic, with participants optimizing current decisions in anticipation of future outcomes. In many platform settings, such as delivery services, orders arrive stochastically and must be matched quickly to couriers who can accept or reject proposed orders. Couriers may reject current matches in anticipation of better future opportunities, leading to delays and unserved orders. Understanding how platforms should design dynamic matching policies in such environments requires accounting for these forward-looking incentives. A key challenge is that the platform's matching decisions and couriers' acceptance behavior are jointly determined in equilibrium, giving rise to a fixed-point problem. Studying such systems can provide insights into how platforms can shape couriers' beliefs on future opportunities  to reduce rejections and improve reliability in real-time markets.

A second direction is to complement the theoretical modeling in this thesis with data-driven research that directly addresses real-world platform design questions. Many important market design questions, such as how to price and how to match participants, are shaped by operational considerations that are difficult to capture without empirical input. Engaging with industry partners to allow these considerations to be incorporated into the analysis enables the evaluation of competing design choices in realistic environments. 

Finally, the rise of AI agents introduces a new frontier for platform design. As agents increasingly act on behalf of users, fundamental questions arise regarding how to design mechanisms and market rules that interact with and coordinate these agents, how to elicit and aggregate underlying human preferences when agents may represent them imperfectly or strategically, and how pricing should be structured when decisions are made algorithmically rather than by humans. More broadly, the presence of AI agents may reshape workflows and the organization of production, altering both the cost structure of providing goods and services and the nature of labor itself. Understanding how these changes affect market outcomes, and how platforms can design access and incentives in such environments, is an important direction for future research.

\appendix

\chapter{Appendix to \Cref{chap:platform_eq}}\label{app:ch2}
\section{Using Maximum Competitive Price}\label{app:prices-min}
Theorem~\ref{thm: price_lattice} shows competitive prices form a lattice. Among all competitive equilibrium prices, the maximum price $$\overline{p}_j = W(S, B, G) - W(S\setminus \{j\},B, G) $$
is the most natural when considering sellers' decision to join the platform: when a benevolent social planner charges $\alpha=0$, all sellers joining is a Platform Equilibrium. This is because for any other set of sellers $P$ joining the platform, seller $j$'s gain from joining is $p^{\on}_j-p^{\off}_j=W(S,B,G(P)\cup\{i\})-W(S,B,G(P))\geq 0$. This is not true for any other competitive prices, making it inappropriate to analyze the welfare loss due to the platform. We formalize it into the following

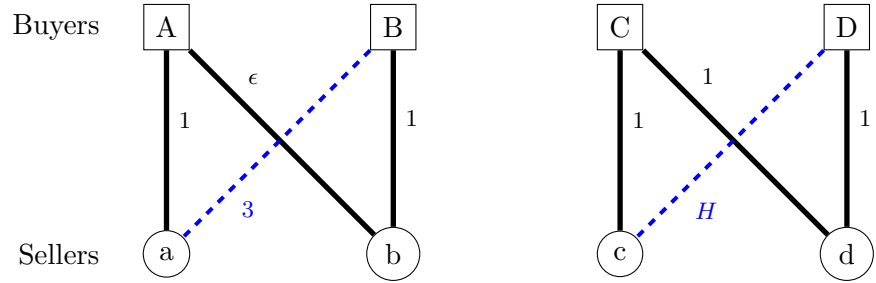
\begin{figure}[h!]
    \centering
    \begin{tikzpicture}[scale=1.5]
        \foreach \i/\label in {1/A, 3/B, 5/C, 7/D}
            \node[draw, shape=rectangle, minimum size=0.6cm] (\label) at (\i, 2) {\label};
        
        \foreach \i/\label in {1/a, 3/b, 5/c, 7/d}
            \node[draw, shape=circle, minimum size=0.6cm] (\label) at (\i, 0) {\label};
        
        \foreach \x/\y/\w/\pos in {A/a/1/midway, B/b/1/midway, C/c/1/midway, D/d/1/midway, A/b/$\epsilon$/near start, C/d/1/near start}
            \draw[line width=2pt] (\x) -- node[\pos, above right, font=\footnotesize] {\w} (\y);
        
        \foreach \x/\y/\w/\pos in {B/a/3/near end,  D/c/$H$/near end}
            \draw[line width=1.5pt, blue, dashed] (\x) -- node[\pos, below right, font=\footnotesize] {\w} (\y);

        \node[left] at (0.5, 2)  {Buyers};

        \node[left] at (0.5, 0)  {Sellers};
    
    \end{tikzpicture}
    \caption{A market with infinite price of anarchy for any $\alpha\in[0,1]$.  Black solid lines are direct links, as captured by $N(i)$ for buyer $i$, and blue dotted lines indicate missing links. Buyer values are annotated adjacent to each edge, and all any value that is omitted is zero.}
    \label{fig:min_price_poa}
\end{figure} 

\begin{theorem}
    If the market clears with the minimum competitive prices, the price of anarchy of the Platform Equilibrium with pure strategies, when the platform sets a zero transaction-fee,
    can be arbitrarily large.
\end{theorem}
\begin{proof}
    Figure~\ref{fig:min_price_poa} gives an infinite-PoA market. When no sellers join, the allocation of goods is seller $A$ to buyer $a$, seller $B$ to buyer $b$, seller $C$ to buyer $c$, seller $D$ to buyer $d$.
    Sellers' minimum competitive prices are all zero. If seller $a$ and seller $c$ join the platform, $a$ sells to $B$, $b$  to $A$, $c$ to $D$ and $d$ to $C$. The minimum competitive prices are now $\underline{p}_a=1-\epsilon$ and  $\underline{p}_b=\underline{p}_c=\underline{p}_d=1$. So for any $\alpha\in[0,1]$ there is an equilibrium where seller $c$ does not join the platform. When $H\rightarrow\infty$, the price of anarchy approaches infinity.
\end{proof}

\section{Missing Proofs in Section~\ref{sec:existence_of_pure}}\label{app:existence_of_pure}

\subsection{None Existence of Pure Equilibrium} \label{app:non-existence}

\label{app:app_no_pure_eq}
\propNoPure*
\begin{proof}
    A three-seller-four-buyer counterexample is given in Figure~\ref{fig:no_pure_equilibrium}. At transaction fee $\alpha=\frac{1}{2}$, there exists no pure strategy equilibrium. On a high level, this is because a seller's price considers the externality she imposes on the whole market and is easily influenced by other sellers' connections. So sellers' best response dynamic can exhibit a cyclic nature: an on-platform seller can drop off after some other sellers join.

    To explicitly show there is no pure equilibrium, go through all eight pure strategy profiles: 
    \begin{enumerate}
        \item When no one joins, seller $a$ joins the platform and sell to B because $(1-\alpha)p_{a}^{on}=\frac{1}{2}(3.05-1)>1=p_{a}^{off}$.
        \item When seller $a$ is on platform, seller $b$ will have to join the platform to sell to $C$ because $B$ is taken and staying off platform brings zero price.
        \item When both seller $a,b$ are on platform, seller $c$ does not have any demand when off-platform and joins the platform in order to sell to buyer $D$.
        \item When all sellers $a,b,c$ join, seller $a$'s price decreases to $p^{on}_{a}=(3.05+1.1+0.05)-(1.1+1.15)=1.95$. Seller $a$ drops off from platform because after paying platform fees $(1-\alpha)p^{on}_{a}<1=p^{off}_{a}$ the revenue is lower than staying off platform.
        \item When $a$ drops off and $b,c$ are on platform, seller $b$'s on platform utility is $\frac{1}{2}\times1.1$ smaller than off platform utility $1$ and hence $b$ will drop.
        \item When $a,b$ are off platform, $c$ will drop as well.
        \item When $a,c$ are on platform and $b$ not, $b$ will want to join the platform.
        \item When $a,c$ are off platform and $b$ on, $b$ will drop.
    \end{enumerate} 

\end{proof}

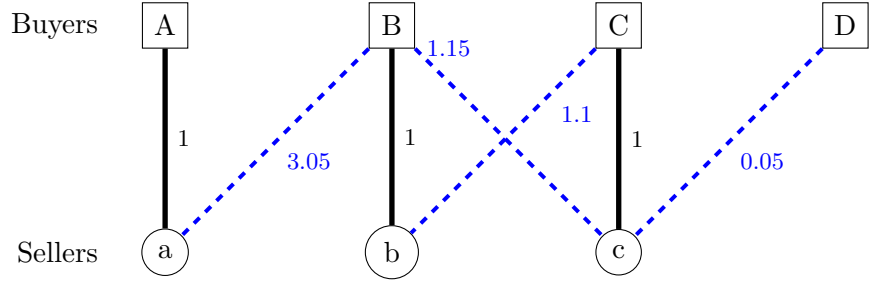
\begin{figure}[h!] 
    \centering
    \begin{tikzpicture}[scale=1.5]
        \foreach \i/\label in {1/A, 3/B, 5/C, 7/D}
            \node[draw, shape=rectangle, minimum size=0.6cm] (\label) at (\i, 2) {\label};
        
        \foreach \i/\label in {1/a, 3/b, 5/c}
            \node[draw, shape=circle, minimum size=0.6cm] (\label) at (\i, 0) {\label};
        
        \foreach \x/\y/\w in {A/a/1, B/b/1, C/c/1}
            \draw[line width=2pt] (\x) -- node[midway, right, font=\footnotesize] {\w} (\y);
        
        \foreach \x/\y/\w/\pos/\loc in {B/a/3.05/midway/below right, B/c/1.15/at start/right, C/b/1.1/near start/below right, D/c/0.05/midway/below right}
            \draw[line width=1.5pt, blue, dashed] (\x) -- node[\pos, \loc, font=\footnotesize] {\w} (\y);

        \node[left] at (0.5, 2)  {Buyers};

        \node[left] at (0.5, 0)  {Sellers};
    
    \end{tikzpicture}
    \caption*{Fig. 1. An example with no Platform Equilibrium in pure strategies at $\alpha=\frac{1}{2}$. Black solid lines are direct links, as captured by $N(i)$ for buyer $i$, and blue dotted lines indicate missing links. Buyer values are  annotated adjacent to each edge, and all any value that is omitted is zero.}
\end{figure}

\subsection{Computing Pure Equilibrium in Homogeneous-Goods Markets} \label{app:existence}
For homogeneous-goods markets, competitive prices are directly related to a buyer's next best forgone trade opportunity, or more specifically its direct and indirect competitors. This notion is characterized by \textit{opportunity path} in \citet{kranton2000competition} as follows.
\begin{definition}[Opportunity Path \cite{kranton2000competition}]\label{def:oppo_path}
    For an allocation $\mathbf{a}$ of goods on a buyer-seller network $G$, buyer $i_1$'s 
    opportunity path linking to another buyer $i_t$ is a path $$(i_1,j_1,i_2,j_2,\ldots, j_{t-1},i_t),$$ where for every $\ell\in \{1,\ldots, t-1\}$, $$g_{i_\ell,j_\ell}=1\mbox{ and }g_{i_{\ell+1},j_\ell}=0,$$ and  $$a_{i_\ell,j_\ell}=0\mbox{ and } a_{i_{\ell+1},j_\ell}=1.$$ 
\end{definition}
\citet{kranton2000competition} show the following connection between buyers' opportunity path and sellers' maximum competitive prices, which we make use of.
\begin{theorem}[Opportunity Path Theorem \cite{kranton2000competition}]\label{thm:ch2_oppo_path}
    Consider a competitive equilibrium with maximum competitive prices $(\bar{\p},\alloc)$ where $a_{ij}=1$. Seller $j$'s price $\bar{p}_j$ is equal to the lowest valuation of any buyer linked by an opportunity path from buyer $i$. $\bar{p}_j=0$ iff buyer $i$ has an opportunity path linking to a seller who does not sell.
\end{theorem}

\citet{kranton2000competition} then analyze how adding a link between a buyer $i$ and a seller $j$ to graph $G$ affects the optimal welfare. For our purpose, we consider the case where seller $j$ has no prior links or does not transact before the new link.
\begin{lemma}\cite{kranton2000competition}\label{lem:add_one_link}
    Consider a buyer-seller network where a seller $j_0$, that either does not transact or have no prior links, is added a single buyer link to buyer $i_0$. Let $\alloc$ and $\alloc'$ be the optimal allocations before and after adding the link. In $\alloc'$ at most one previously unmatched buyer is matched. Moreover, if in $\alloc$, $a_{i_0,j_1}=1$,  $a_{i_1,j_2}=1$, \ldots, $a_{i_{t-1},j_t}=1$ (and therefore, $a_{i_\ell,j_\ell}=0$ for $\ell \in \{0,\ldots, t-1\}$), and in $\alloc'$, $a_{i_\ell,j_\ell}=1$ for $\ell \in \{0,\ldots, t-1\}$, then we have the following. Before adding the link between $j_0$ and $i_0$, $(i_t,j_t,\ldots, i_1,j_1,i_0)$ is an opportunity path. After adding the link, $(i_0, j_1,i_1,\ldots,  j_t,i_t)$ is an opportunity path. Buyer $i_t$ obtains the good in $\alloc'$ but not in $\alloc$. Furthermore, $v_{i_t}\leq v_{i_\ell}$ for every $\ell< t$.
\end{lemma}

If we introduce a set of new sellers and their edges to a market and apply Lemma~\ref{lem:add_one_link} repeatedly on each new seller, every step at most one new buyer is introduced, and thus buyers transacting in the original competitive equilibrium allocation can be preserved. This is formalized as the following corollary.
\begin{corollary}\label{cor:adding_sellers_buyers_set}
Let $(S,B,G)$ be a buyer-seller network with a corresponding optimal allocation $\alloc$, and let $B^G$ be the set of transacting buyers. Now add a set of sellers $P$ with their edges who previously know no buyers and let $G(P)$ be resulting network. 
For the new market $(S\cup P,B,G(P))$, there exists an optimal allocation $\alloc'$ where all buyers in $B^G$ transact.
\end{corollary}

Using this, we can relate the optimal welfare in a network to the welfare of sellers not joining the platform.

\OptimalWelfare*
\begin{proof}
    Market $(S,B,G(P))$ can be viewed as adding sellers in $P$ one-by-one with all edges to $(S\setminus P, B,G)$. When adding each new seller, Corollary~\ref{cor:adding_sellers_buyers_set} reads the previously transacting buyer remains the same, while the largest remaining unmatched buyer is matched to the new seller.
\end{proof}

\SameOnPrice*
\begin{proof}
    By Lemma~\ref{lem:optimal_welfare} and Eq.~\eqref{eq:max_price}, expand the on platform price for each seller $j\in P$ equals to $\bar{v}(m')-\bar{v}(m'-1)$, where $|P|=m'$.
\end{proof}

\vspace{0.5cm}

Algorithm~\ref{alg:PE_pure} defines a seller's on platform gain. We show it is closely related with the buyer valuation that $j$ is matched to in some markets. We will use this quantity as an intermediate step to prove Lemma~\ref{lem:algo_select} and Theorem~\ref{alg:PE_pure}. 

\begin{lemma}\label{lem:on_platform_gain}
    Consider a market $(S,B,G(P))$ where sellers $j,\hat{j}\in S, j,\hat{j}\notin P$. Let $i_P(j),i_P(\hat{j})$ be the new transacting buyer when $j$ and $\hat{j}$ are added with their off-platform edges to market $(S\setminus P \setminus \{j,\hat{j}\},B,G)$ respectively. Then $\phi_j^P > \phi_{\hat{j}}^P \Rightarrow v_{i_P(j)} < v_{i_P(\hat{j})}$, and $v_{i_P(j)} < v_{i_P(\hat{j})} \Rightarrow \phi_j^P \geq \phi_{\hat{j}}^P$
    where $\phi_j^P=(1-\alpha)p_j^{\on}(P)-p_j^{\off}(P)$ is the on-platform gain.
\end{lemma}
\begin{proof}
    Expanding the definition of prices
    \begin{eqnarray}
        \phi_j^P-\phi_{\hat{j}}^P &=& (1-\alpha)[p_j^{\on}(P)-p_{\hat{j}}^{\on}(P)]+ [p^{\off}_{\hat{j}}(P)-p^{\off}_{j}(P)]\label{eq:on_platform_gain}\\
        p_j^{\on}(P)-p_{\hat{j}}^{\on}(P) &=& W(S,B,G(P\cup \{j\}))-W(S,B,G(P\cup \{\hat{j}\}))\nonumber\\
        & + & W(S\setminus\{\hat{j}\},B,G(P)) - W(S\setminus\{j\},B,G(P)) \nonumber\\
        p^{\off}_{\hat{j}}(P)-p^{\off}_{j}(P)   &=& W(S\setminus\{j\},B,G(P)) - W(S\setminus\{\hat{j}\},B,G(P))\nonumber
    \end{eqnarray}
    View the welfare terms as adding sellers to a base market $(S\setminus P\setminus\{j,\hat{j}\}, B, G)$. For example, $W(S\setminus\{j\},B,G(P))$ is adding $\hat{j}$ to market then $P$ to platform. Let $\bar{v}_{S\setminus P\setminus \{j\}}(|P|)$ be the sum of the $|P|$ largest buyers who do not transact in $W(S\setminus P \setminus\{j\},B,G)$. (There are more than $|P|$ none-transacting buyers otherwise the algorithm would not have chosen $P$ to join.) We can thus express the welfare term as follows using Lemma~\ref{lem:optimal_welfare}
    \begin{eqnarray*}
        W(S\setminus\{j\},B,G(P)) &=& W(S\setminus P\setminus \{j,\hat{j}\},B,G)+v_{i_P(\hat{j})}+\bar{v}_{S\setminus P\setminus \{j\}}(|P|)\\
        W(S\setminus\{\hat{j}\},B,G(P)) &=& W(S\setminus P\setminus \{j,\hat{j}\},B,G)+v_{i_P(j)}+\bar{v}_{S\setminus P\setminus \{\hat{j}\}}(|P|)\\
        W(S,B,G(P\cup\{j\})) &=& W(S\setminus P\setminus \{j,\hat{j}\},B,G)+v_{i_P(\hat{j})}+\bar{v}_{S\setminus P\setminus \{j\}}(|P|+1)\\
        W(S,B,G(P\cup\{\hat{j}\})) &=& W(S\setminus P\setminus \{j,\hat{j}\},B,G)+v_{i_P(j)}+\bar{v}_{S\setminus P\setminus \{\hat{j}\}}(|P|+1)
    \end{eqnarray*}
    Let $\bar{v}_{S\setminus P\setminus\{j\}}^{|P|+1}$ be the $|P|+1$-th largest buyer who does not transact in $W(S\setminus P\setminus \{j\},B,G)$, and similarly for $\bar{v}_{S\setminus P\setminus\{\hat{j}\}}^{|P|+1}$. 
    Bring these back to Equation~\eqref{eq:on_platform_gain}
    \begin{eqnarray*}
        \phi_j^P-\phi_{\hat{j}}^P = [v_{i_P(\hat{j})}+\bar{v}_{S\setminus P\setminus \{j\}}(|P|)+(1-\alpha)\bar{v}_{S\setminus P\setminus\{j\}}^{|P|+1}]-
        [v_{i_P(j)}+\bar{v}_{S\setminus P\setminus \{\hat{j}\}}(|P|)+(1-\alpha)\bar{v}_{S\setminus P\setminus\{\hat{j}\}}^{|P|+1}]
    \end{eqnarray*}
    Starting from market $W(S\setminus P\setminus \{j,\hat{j}\},B,G)$,
    the first bracket is the value of buyer that $\hat{j}$ introduces, plus the $|P|$ largest remaining buyers, plus $(1-\alpha)$ of the $|P|+1$-th remaining largest buyer. The second bracket is similar, after $j$ introduces a buyer. To guarantee $\phi_j^P>\phi_{\hat{j}}^P$, a necessary condition is $v_{i_P(\hat{j})}>v_{i_P(j)}$. Note both inequalities are strict: if $\phi_j^P=\phi_{\hat{j}}^P$, it can be that $v_{i_P(\hat{j})}\leq v_{i_P(j)}$. On the other direction, if $v_{i_P(\hat{j})}>v_{i_P(j)}$, we can only guarantee $\phi_j^P\geq \phi_{\hat{j}}^P$.
\end{proof}

\vspace{0.1cm}

\AlgoSelect*
\begin{proof}
    Lemma~\ref{lem:on_platform_gain} shows that the necessary condition for $\phi_{\hat{j}}^P>\phi_j^P$ is $v_{i_P(j)}>v_{i_P({\hat{j}})}$. Thus for all sellers $ {\hat{j}}\in \Phi_{\max}^P$ and all sellers  $j\in S\setminus P, j\notin \Phi_{\max}^P$
    \begin{eqnarray*}
        W(S\setminus P\setminus \{{\hat{j}},j\},B,G) + v_{i_P({\hat{j}})} & < & W(S\setminus P\setminus \{{\hat{j}},j\},B,G) + v_{i_P(j)}\\
        W(S\setminus P\setminus \{j\},B,G) &< & W(S\setminus P\setminus \{{\hat{j}}\},B,G).
    \end{eqnarray*}
    Let $\bar{v}_{S\setminus P\setminus\{j\}}(|P|)$ be the sum of the $|P|$ largest buyers who do not transact in $W(S\setminus P\setminus\{j\},B,G)$, and $\bar{v}_{S\setminus P\setminus\{{\hat{j}}\}}(|P|)$ be the sum of the $|P|$ largest buyers who do not transact in $W(S\setminus P\setminus\{\hat{j}\},B,G)$.
    \begin{eqnarray*}
        W(S\setminus P\setminus \{j\},B,G) + \bar{v}_{S\setminus P\setminus\{j\}}(|P|) &\leq & W(S\setminus P\setminus \{{\hat{j}}\},B,G) + \bar{v}_{S\setminus P\setminus\{{\hat{j}}\}}(|P|) \\
        W(S\setminus \{j\},B,G(P)) &\leq & W(S\setminus \{{\hat{j}}\},B,G(P)).
    \end{eqnarray*}
    By expanding the definition of prices, we have
    \begin{eqnarray*}
        p_{\hat{j}}^{\off}(P) &=& W(S,B,G(P))-W(S\setminus\{{\hat{j}}\},B,G(P))\\
        &\leq& W(S,B,G(P))-W(S\setminus\{j\},B,G(P))=p_{j}^{\off}(P)
    \end{eqnarray*}
    For sellers ${\hat{j}},j$ within set $\Phi_{\max}^P$, $\phi_{\hat{j}}^P=\phi_{j}^P$. The algorithm in Step~\ref{alg:step} breaks ties towards ${\hat{j}}$ with the lowest off platform price.
    
    For the other direction, one can verify if $p_{\hat{j}}^{\off}(P)<p_{j}^{\off}(P)$, then $v_{i_P}(j)>v_{i_P}({\hat{j}})$. Lemma~\ref{lem:on_platform_gain} then guarantees $\phi_{\hat{j}}^P\geq \phi_{j}^P$. So ${\hat{j}}$ is in $\Phi_{\max}^P$ at Step~\ref{alg:step} of the Algorithm and selected because it has lower off platform price.
\end{proof}

\vspace{0.1in}

To prove theorem~\ref{thm:PE_pure}, we will need one more conditions. \citet{maxweightsubmod} proves max weight matching on bipartite graphs satisfies the submodularity condition: with more vertices on one side, adding one new edge brings marginally lower increase in maximum weight. We rephrase it here and apply it to the buyer-seller markets.
\begin{lemma}[Max weight matching submodularity \cite{maxweightsubmod}]\label{lem:submodularity}
    For a set of sellers $S'\subseteq S$, a single seller $j\notin S$ and buyers $B$ connection $G$   
    \begin{eqnarray}\label{eq:submodularity}
        W(S\cup\{j\},B,G)-W(S,B,G) \leq W(S'\cup\{j\},B,G)-W(S',B,G)
    \end{eqnarray}
\end{lemma}
\begin{proof}
    we prove the above through an equivalent definition of submodularity. For set of sellers $S_1,S_2\subseteq S$, denote $W(S_1):=W(S_1,B,G),W(S_2):=W(S_2,B,G)$. Prove $$W(S_1)+W(S_2)\geq W(S_1\cup S_2)+W(S_1\cap S_2)$$ 
    Using $M_{\cup},M_{\cap}$ to denote the edges in max matching with sellers $S_1\cup S_2, S_1\cap S_2$, it suffices to prove that $M_{\cup}$ and $M_{\cap}$ can be partitioned into two disjoint matchings $M_1, M_2$ for $W(S_1)$ and $W(S_2)$ respectively. Then by max weight matching $W(S_1)$ is weakly greater than the weight of matching $M_1$, $W(S_2)$ greater than that of $M_2$.
    
    Let $C$ denote the collection of edges in $M_\cap$ and $M_\cup$. Edges in $C$ form alternating paths and cycles. A path alternate between edges in $M_\cap$ and $M_\cup$, while cycles are restricted to vertices within $S_1\cap S_2$: for vertices in $S_1\setminus S_2$ and $S_2\setminus S_1$, there is at most one edge in $C$ connecting them because they are not in $M_\cap$.

    Let $L_1$ be the set of paths with at least one vertex in $S_1\setminus S_2$, $L_2$ be the set of paths with at least one vertex in $S_2\setminus S_1$. Then $L_1\cap L_2 = \emptyset$. Suppose a path $l$ contains vertex $j_1\in S_1\setminus S_2, j_2\in S_2\setminus S_1$, then $j_1,j_2$ must be two end points of $l$ because they have one edge. For the same reason, there are no other vertex of $S_1\setminus S_2, S_2\setminus S_1$ in $l$. Since $j_1, j_2$ are both in $S$, there are even number of edges in this path. This cannot be because $l$ alternates between edges in $M_\cap$ and $M_\cup$: if $l_1$ is connected through edges in $M_\cup$, by even number of edges $l_2$ must be connected through edges in $M_\cap$. Vice versa. Either $j_1$ or $j_2$ must be in $S_1\cap S_2$, forming a contradiction.

    Now we are ready to partition $M_\cap$ and $M_\cup$. Let 
    \begin{eqnarray}
        M_1  &= (L_1\cap M_\cup) \cup (L_2\cap M_\cap) \cup (C\setminus \{L_1\cup L_2\}\cap M_\cap)\\
        M_2  &= (L_2\cap M_\cup) \cup (L_1\cap M_\cap) \cup (C\setminus \{L_1\cup L_2\}\cap M_\cup)
    \end{eqnarray}
    We verify $M_1\cup M_2 = M_\cup \cup M_\cap$, $M_1\cap M_2 = M_\cup \cap M_\cap$, confirming $M_1$ and $M_2$ are indeed two partitions of $M_\cap, M_\cup$. Now prove $M_1$ is a matching for $W(S_1)$. First $M_1$ uses no vertex from $S_2\setminus S_1$: $L_1$, $M_\cap$ and $C\setminus\{L_1,L_2\}$ contains no vertex from $S_2\setminus S_1$. Second the three terms in $M_1$ being independent, each vertex in $S_1$ is matched at most once in $M_1$. Otherwise, a vertex would be matched twice in either $M_\cup$ or $M_\cap$, contradiction. Similarly, prove $M_2$ is a matching for $W(S_2)$, concluding the proof.
\end{proof}

Now we are ready to prove Theorem~\ref{thm:PE_pure}.
\PEpure*
\begin{proof}

The algorithm always terminates because there are finite sellers. To prove it terminates at a pure strategy platform equilibrium, it suffices to show seller $j\in P$ already on the platform don't have incentive to drop off when a new seller $\hat{j}$ is added to $P$. In particular, we want to prove $\forall j\in P$, its benefit of staying on platform 
    is larger than $\hat{j}$'s benefit of joining.
    \begin{eqnarray}
        \forall j\in P, \phi_j^{P\cup\{\hat{j}\}}\geq \phi_{\hat{j}}^{P}\geq 0 \text{ for } \hat{j}\in \argmax_{j\in S\setminus P}\{\phi_j^P\};\label{eq:pure_eq_invariant}
    \end{eqnarray}
    Then sellers in $P$ do not drop off from platform when the algorithm terminates. 
    Since $\hat{j}$ is the next seller joining, $\phi_{\hat{j}}^{P}\geq 0$. Lemma~\ref{lem:on_platform_gain} does not apply because $j\in P$. So we expand Eq.~\eqref{eq:pure_eq_invariant} to prove the following is non-negative
    \begin{eqnarray}
        \phi_j^{P\cup\{\hat{j}\}} - \phi_{\hat{j}}^{P} = (1-\alpha)[p_j^{\on}(P\cup\{\hat{j}\})-p_{\hat{j}}^{\on}(P)]+[p_{\hat{j}}^{\off}(P)-p_{j}^{\off}(P\cup\{\hat{j}\})]\label{eq:pure_eq_invariant_2}
    \end{eqnarray}
    The first term $p_j^{\on}(P\cup\{\hat{j}\})-p_{\hat{j}}^{\on}(P)$ in Eq.~\eqref{eq:pure_eq_invariant_2} is zero because according to Corollary~\ref{lem:same_on_price}, all on-platform sellers in a same market have the same price. Now prove $p_{\hat{j}}^{\off}(P) \geq p_{j}^{\off}(P\cup\{\hat{j}\})$. 
    Intuitively this is true because $j$ is selected by the algorithm before $\hat{j}$ and had a weakly lower off-platform price when it was added. For the rest of the proof, we will prove this intuition.
    
    Let $P_{-j}$ be $P\setminus \{j\}$. Expand the two prices $p_{\hat{j}}^{\off}(P)=W(S,B,G(P))-W(S\setminus \{\hat{j}\},B,G(P))$ and $p_{j}^{\off}(P\cup\{\hat{j}\}) = W(S,B,G(P_{-j}\cup \{\hat{j}\}))-W(S\setminus\{j\},B,G(P_{-j}\cup \{\hat{j}\}))$. The two minus terms are equal $$W(S\setminus \{\hat{j}\},B,G(P))=W(S\setminus\{j\},B,G(P_{-j}\cup \{\hat{j}\}))$$ because the two terms have exactly the same off platform sellers and the same number of on-platform sellers. By Lemma~\ref{lem:optimal_welfare}, they have the same welfare. Rearranging, prove the following is non-negative.
    \begin{eqnarray}
        p_{\hat{j}}^{\off}(P) - p_{j}^{\off}(P\cup\{\hat{j}\}) = W(S,B,G(P))-W(S,B,G(P_{-j}\cup \{\hat{j}\}))\label{eq:pure_eq_target_price}
    \end{eqnarray}

    Just as in the proof for Lemma~\ref{lem:on_platform_gain}, view $W(S,B,G(P))$ as adding seller $\hat{j}$ to the market, then $P$ to platform in a base market $(S\setminus P\setminus\{\hat{j}\},B,G)$. Let $i_P(\hat{j})$ be the new buyer introduced by $\hat{j}$, $\bar{v}_{S\setminus P}(|P|)$ be sum of the $|P|$ largest buyers who do not transact in $W(S\setminus P,B,G)$. Similarly, let $i_P(j)$ be the new buyer introduced by adding $j$ to the base market, $\bar{v}_{S\setminus P_{-j}\setminus\{\hat{j}\}}(|P|)$ be sum of the $|P|$ largest buyers who do not transact in $W(S\setminus P_{-j}\setminus\{\hat{j}\},B,G)$. There are more than $|P|$ buyers not transacting in $W(S\setminus P,B,G)$ or $W(S\setminus P_{-j}\setminus\{\hat{j}\},B,G)$, otherwise all buyers transact in $(S,B,G(P))$, contradicting with the algorithm picking $\hat{j}$ next. Assume for contradiction Equation~\eqref{eq:pure_eq_invariant_2} and \ref{eq:pure_eq_target_price} is negative, expanding 
    \begin{eqnarray*}
        \phi_j^{P\cup\{\hat{j}\}} - \phi_{\hat{j}}^{P} = v_{i_P(\hat{j})}+\bar{v}_{S\setminus P}(|P|)-v_{i_P(j)}-\bar{v}_{S\setminus P_{-j}\setminus\{\hat{j}\}}(|P|) < 0
    \end{eqnarray*}
    A necessary condition for it being negative is 
    \begin{eqnarray}\label{eq:pure_eq_target_price_decompose}
        v_{i_P(\hat{j})} & < v_{i_P(j)} 
    \end{eqnarray}
    Now consider right before Algorithm~\ref{alg:PE_pure} adds $j$ to the platform. Denote the sellers joining the platform then by $P^{j}\subset P$. The Algorithm chooses $j$ at Step~\ref{alg:step}. Let $i_{P^j}(j), i_{P^j}(\hat{j})$ be the new transacting buyer when $j$ and $\hat{j}$ is added with their off-platform edges to market $(S\setminus P^j\setminus \{j,\hat{j}\},B,G)$ respectively.
    
    If $\phi_j^{P^j}=\phi_{\hat{j}}^{P^j}$ and $p^{\off}_j(P^j)=p^{\off}_{\hat{j}}(P^j)$. There are two cases.
    
    \textbf{Case 1:} $j$ and $\hat{j}$'s off platform prices are determined by two different buyers of the same valuation on two different opportunity paths. Then $j$ joining the platform won't change $\hat{j}$'s off platform price. So $\hat{j}$ is the next to join. Similarly, $\hat{j}$ joining won't change $j$'s off platform price and equation~\eqref{eq:pure_eq_target_price} equals to zero.
    
    \textbf{Case 2:} $j$ and $\hat{j}$'s off platform prices equal to the the valuation of the smallest buyer on the same opportunity path. This valuation equals to $p_j^{\off}(P^j)$. By Lemma~\ref{lem:optimal_welfare},
    $W(S,B,G(P^j))=W(S\setminus P^j,B^G,G)+\bar{v}(|P^j|)$. Sellers in $P^j$ are matched to largest buyers in $B\setminus B^G$. If the smallest buyer in $j$ and $\hat{j}$'s opportunity path is matched to an on platform seller in $P^j$, the next seller joining the platform will have a price weakly lower than $p_j^{\off}(P^j)$. This contradicts with $j$ joining the platform next. If an on platform seller is on $j$ and $\hat{j}$'s opportunity path but not matched to the smallest buyer on the path, one can remove this on platform seller, shift  transactions along the path and strictly increase $W(S\setminus P^j, B^G,G)$. This contradicts with off-platform sellers being matched optimally. So no seller in $P^j$ is on $j$ and $\hat{j}$'s opportunity path. 
    %
    Now, if the opportunity path ends at a seller who does not transact, then $v_{i_{P^j}(\hat{j})} = v_{i_{P^j}(j)}$ equals to the lowest buyer valuation on the path. If the opportunity path ends on the buyer with the lowest valuation, then $v_{i_{P^j}(\hat{j})} = v_{i_{P^j}(j)}$ equals to the second smallest buyer valuation on the path. This is discussed together with the next case in inequality~\eqref{eq:pure_eq_cond_price_decompose}.

    Otherwise $\phi_j^{P^j}>\phi_{\hat{j}}^{P^j}$ or $p^{\off}_j(P^j)<p^{\off}_{\hat{j}}(P^j)$. Lemma~\ref{lem:on_platform_gain} reads \begin{eqnarray}\label{eq:pure_eq_cond_price_decompose}
        v_{i_{P^j}(j)} \leq v_{i_{P^j}(\hat{j})} 
    \end{eqnarray}
    Let $P^{\text{mid}}=P_{-j}\setminus P^j$. The two base markets $S\setminus P\setminus \{\hat{j}\}$ and $S\setminus P^j\setminus\{j,\hat{j}\}$ differ by $P^{\text{mid}}$. When $P^{\text{mid}}=\emptyset$, the two base markets are the same, $v_{i_P(\hat{j})}=v_{i_{P^j}(\hat{j})}, v_{i_P(j)}=v_{i_{P^j}(j)}$. Inequality~\eqref{eq:pure_eq_target_price_decompose} directly contradicts with Inequality~\eqref{eq:pure_eq_cond_price_decompose}. So the difference in on-platform gain in equation~\eqref{eq:pure_eq_invariant_2} cannot be negative.
    
    We now analyze when $P^{\text{mid}}\neq \emptyset$. The intuition is according to inequality~\eqref{eq:pure_eq_target_price_decompose}, $i_P(j)$ as a buyer of high value is matched to $j$ in $(S\setminus P\setminus\{\hat{j}\},B,G)$. However, it is matched to another seller $j'\in P^{\text{mid}}$ in market $(S\setminus P^j\setminus\{j,\hat{j}\},B,G)$ because of inequality~\eqref{eq:pure_eq_invariant_2}. Then when adding $j'$ on platform, the algorithm should have picked $\hat{j}$ instead because $\hat{j}$ introduces a low valuation buyer. We now formalize this intuition.
    
    Comparing the two base markets, $S\setminus P\setminus \{\hat{j}\} \subset S\setminus P^j\setminus\{j,\hat{j}\}$, applying the submodularity condition in Lemma~\ref{lem:submodularity}
    \begin{eqnarray}
        v_{i_P(\hat{j})} & \geq v_{i_{P^j}(\hat{j})} \label{eq:apply_submodularity_to_target}\\
        v_{i_P(j)} & > v_{i_{P^j}(j)}\label{eq:apply_strict_submodularity_to_condition}
    \end{eqnarray}
    Inequality~\eqref{eq:apply_strict_submodularity_to_condition} is strict because otherwise combining inequality~\eqref{eq:pure_eq_target_price_decompose} and \eqref{eq:pure_eq_cond_price_decompose}
    we have $v_{i_{P^j}(\hat{j})} \geq v_{i_{P^j}(j)} = v_{i_P(j)} > v_{i_P(\hat{j})}$, which contradicts with \eqref{eq:apply_submodularity_to_target}. By definition, $j$ introduces $i_{P}(j)$ in the base market $(S\setminus P\setminus\{\hat{j}\},B,G)$. Thus by Corollary~\ref{cor:adding_sellers_buyers_set}, $i_P(j)$ still transacts in the second base market containing more sellers $(S\setminus P^j\setminus\{\hat{j}\},B,G)$. But the strict inequality ~\eqref{eq:apply_strict_submodularity_to_condition} says in this market, $j$ does not introduce $i_p(j)$ anymore. So $i_P(j)$ still transacts in the market $(S\setminus P^j\setminus\{j\},B,G)$. Note by definition $i_P(j)$ does not transact in market $(S\setminus P,B,G)$.
    
    Now examine when the algorithm adds seller $j$ to platform, and before it adds seller $\hat{j}$ to platform. The former market is $(S,B,G(P^j\cup \{j\}))$, where sellers in $S\setminus P^j\setminus\{j\}$ is off-platform. The latter market is $(S,B,G(P))$, where $S\setminus P$ is off-platform. By Corollary~\ref{cor:adding_sellers_buyers_set} and our discussion in the previous paragraph, $i_P(j)$ transacts with off-platform sellers in the former market but not in the latter. Then there $\exists j'\in P\setminus P^j$ such that it joining the platform results in $i_P(j)$ not transacting with off-platform sellers. Equivalently, adding $j'$ to off-platform sellers will introduce $i_P(j)$.
    
    Consider when the algorithm adds $j'$ to platform. By Lemma~\ref{lem:add_one_link}, $i_P(j)$ is the lowest-value buyer on $j'$'s opportunity path so $p^{\off}(j')=v_{i_P(j)}$. 
    Again because of submodularity, seller $\hat{j}$ introduces a buyer with valuation smaller than $v_{i_P(\hat{j})}$ so its off-platform price is $p^{\off}_{\hat{j}}\leq v_{i_P(\hat{j})}<p^{\off}_{j'}$. By Lemma~\ref{lem:algo_select} the algorithm should not have added seller $j'$, which forms a contradiction.
\end{proof}

\vspace{0.5cm}
The algorithm can terminate any time when $\forall j\notin P, \phi_j^P\leq 0$. This gives flexibility to find multiple pure equilibrium, particularly of interest is when $\alpha=1$. 
\PEpureAlpha*
\begin{proof}
    At $\alpha=1$ sellers in $S^G$ weakly prefer not to join. After adding a first $j_1\in \bar{S}^G$ seller to $P$, other sellers in $\bar{S}^G\setminus \{j_1\}$ still does not transact because Lemma~\ref{lem:add_one_link} reads adding a new link involves no new sellers. Thus, every seller in $\bar{S}^G$ is indifferent to join at $\alpha=1$.
\end{proof}

This following lemma allows the platform to continuously decrease $\alpha$ with algorithm~\ref{alg:PE_pure}, and gradually increase the set of on-platform sellers in equilibrium to calculate revenue.
\PEpureContinuous*
\begin{proof}
At $\alpha_2$, Step~\ref{alg:step} can first add all sellers in $P_1$ before any other sellers: a lower $\alpha$ does not change the seller with largest on-platform gain. 
\begin{eqnarray*}
    (1-\alpha_1)p_j^\on(P)-p_j^\off(P)\geq (1-\alpha_1)p_{j'}^\on(P)-p_{j'}^\off(P)\\
    \Rightarrow (1-\alpha_2)p_j^\on(P)-p_j^\off(P)\geq (1-\alpha_2)p_{j'}^\on(P)-p_{j'}^\off(P)
\end{eqnarray*}
\end{proof}

\EnlargeEq*
\begin{proof}
    Lemma~\ref{lem:pe_pure_alpha1} proves a pure platform equilibrium always exists for $m_p=1,...,\bar{S}^G$ at $\alpha=1$. At $P=\bar{S}^G$, for all sellers $j\notin \bar{S}^G, \phi_j^{\bar{S}^G}=-p_j^{\off}(\bar{S}^G)\leq 0$. As $\alpha$ decreases continuously from $1$ to $0$, $\phi_j^P$ continuously increases for all $j$ as long as the equilibrium set $P$ stays the same. When a first set of sellers $\Phi_{\max}^P$ have zero on-platform gain, Algorithm~\ref{alg:PE_pure} adds a seller $j\in \Phi_{\max}^P$ to platform. Then it suffices to show after $j$ is added to $P$, $\forall \hat{j}\notin P$, either
    \begin{eqnarray}
    \phi_{\hat{j}}(P\cup \{j\}) & \leq & \phi_{\hat{j}}(P) \nonumber\\
    \Leftrightarrow	 (1-\alpha)[p_{\hat{j}}^{\on}(P\cup\{j\})-p_{\hat{j}}^{\on}(P)] &\leq 
    &  p_{\hat{j}}^{\off}(P\cup\{j\})-p_{\hat{j}}^{\off}(P)\label{eq:equilibrium_set_enlarges_1}
    \end{eqnarray}
    or $\phi_{\hat{j}}(P\cup\{j\})<0$.
    Then for $\hat{j}\notin P\cup\{j\}$, on platform gain is non-positive. So $P\cup \{j\}$ is indeed a pure equilibrium of size one larger than $P$. And the algorithm can further add sellers whose on platform gain is zero, and after that continue to decrease $\alpha$.

    We first look at the left hand side of equation~\eqref{eq:equilibrium_set_enlarges_1}, the change in on platform prices. By Lemma~\ref{lem:same_on_price}, $p_{\hat{j}}^{\on}(P\cup\{j\})$ equals to the value of the $(|P|+2)$-st largest buyer who does not transact in market $(S\setminus P\setminus \{\hat{j},j\},B,G)$; $p_{\hat{j}}^{\on}(P)$ equals to the value of the $(|P|+1)$-st largest buyer who does not transact in market $(S\setminus P\setminus \{\hat{j}\},B,G)$. There is possibly one more none-transacting buyer in the former market than the latter, so $p_{\hat{j}}^{\on}(P\cup\{j\}) \leq p_{\hat{j}}^{\on}(P)$.

    Now we examine the right hand side of equation~\eqref{eq:equilibrium_set_enlarges_1}, the change in $\hat{j}$'s off platform price after $j$ joins.
    If $p_{\hat{j}}^{\off}(P\cup\{j\}) \geq p_{\hat{j}}^{\off}(P)$, naturally equation~\eqref{eq:equilibrium_set_enlarges_1} holds. We first find and rule out the sellers with weakly larger off platform prices. We then prove sellers $\hat{j}$ with smaller off platform price will be matched to the same buyer even joining the platform, thus $\phi_{\hat{j}}(P\cup\{j\})<0$.
    
    View seller $j$ joining the platform as first leaving the market $(S,B,G(P))$, and then transacting with the largest remaining buyer. By Lemma~\ref{lem:add_one_link} at most one buyer $i^{G(P)}(j)$ is unmatched because of $j$ leaving, and it is connected through opportunity path $$l_{G(P)}=(j_0=j,i_0,j_1,i_1,...,j_t,i_t=i^{G(P)}(j))$$ in $(S,B,G(P))$ where $a_{j_q,i_q}=1 \text{ for } q\in\{0,...,t\}$.  Furthermore, $i^{G(P)}$ is the smallest buyer on the opportunity path $p_j^{\off}(P)=v_{i^{G(P)}(j)}$.
    When $j$ joins the platform, it is connected to all buyers. Again by Corollary~\ref{lem:optimal_welfare}, it sells to the largest none-transacting buyer $i^{G(P\cup\{j\})}(j)$. By Lemma~\ref{lem:same_on_price}, $p_j^{\on}(P)=v_{i^{G(P\cup\{j\})}(j)}$. This in turn means $i^{G(P\cup\{j\})}(j)$ is the smallest value buyer in any opportunity path that extends from $i^{G(P\cup\{j\})}(j)$. We also have $v_{i^{G(P\cup\{j\})}(j)}>v_{i^{G(P)}(j)}$ otherwise seller $j$ would not have joined the platform.
    When $j$ joins the platform, opportunity path $l_{G(P)}$ changes and reverses into $$l_{G(P\cup\{j\})}=(j_t,i_{t-1},...,i_1,j_1,i_0,j_0=j,i^{G(P\cup\{j\})}(j))$$
    where $a_{j_0,i^{G(P\cup\{j\})}(j))}=1$ and $a_{j_q,i_{q-1}}=1$ for $q\in\{t,t-1,...,1\}$.
    Since price equals the lowest buyer valuation in opportunity path, sellers $j_1,j_2,...,j_t$'s off platform price in $(S,B,G(P))$ is $v_{i^{G(P)}(j)}$ and that in $(S,B,G(P\cup\{j\}))$ is weakly larger because $v_{i^{G(P\cup\{j\})}(j)}>v_{i^{G(P)}(j)}$, and $i^{G(P)}(j)$ is the smallest buyer in $l_{G(P)}$. So we have proved for $\hat{j}\in \{j_1,...,j_t\}$, off platform price weakly increases.

    We now show if off platform prices decreases for $\hat{j}\notin \{j_1,...,j_t\}$, they will be matched to the same buyer even joining the platform. Seller $\hat{j}$ transacts in $(S,B,G(P\cup\{j\}))$ with the same buyer as $(S,B,G(P))$ off-platform. Denote the buyer as $i(\hat{j})$. There are two possible changes to opportunity paths that begins at $i(\hat{j})$. First, if $i(\hat{j})$ is connected to $l_{G(P)}$ by an opportunity path, then this path will instead connect to $l_{G(P\cup\{j\})}$. The minimum buyer valuation on this opportunity path increases. Second, now that $j$ joins the platform, $i(\hat{j})$ knows $j$ and have an opportunity path to $i^{G(P\cup\{j\})}(j)$. If $i(\hat{j})$'s off platform price decreases, the smallest buyer connected through opportunity path must be $i^{G(P\cup\{j\})}(j)$, meaning  $$v_{i(\hat{j})}>p_{\hat{j}}^{\off}(P\cup\{j\})= v_{i^{G(P\cup\{j\})}(j)}$$ By Lemma~\ref{lem:same_on_price}, if $\hat{j}$ joins the platform, it is matched to the largest none-transacting buyer, which is of weakly smaller value than $i^{G(P\cup\{j\})}(j)$, and smaller than $i(\hat{j})$. So optimal matching requires $\hat{j}$ be matched to $i(\hat{j})$ even on platform. Then $\phi_{\hat{j}}(P\cup\{j\})<0$ and $\hat{j}$ doesn't want to join the platform after $j$ joins.
\end{proof}

\section{$\min(m,n)$ Lower Bound for the Price of Anarchy in General Markets} \label{app:poa_mn_general}

In this section, we show that without regulation, in general unit-demand markets, the price of anarchy can be as bad as $\min\{m,n\}$. This is in stark contrast to the positive results we have in Section~\ref{sec:poa-regulated} for the case the platform is regulated in the fees it can set.

\begin{figure} 
    \centering
    \begin{tikzpicture}[scale=1.5]
        \foreach \i/\label in {1/$b_1$, 3/$b_2$, 7/$b_{n-1}$, 9/$b_n$}
            \node[draw, shape=rectangle, minimum size=0.6cm] (\label) at (\i, 2) {\label};
        
        \foreach \i/\label in {1/$s_1$, 3/$s_2$, 7/$s_{n-1}$, 9/$s_n$}
            \node[draw, shape=circle, minimum size=0.6cm] (\label) at (\i, 0) {\label};
            
        \foreach \i\j in {$b_1$/$s_1$,$b_2$/$s_2$,$b_{n-1}$/$s_{n-1}$}
            \draw[line width=1.2pt] (\i) -- (\j);

        \node at (5, 2) {$\ldots$};
        \node at (5, 0) {$\ldots$};        

        \foreach \x/\y/\w/\pos/\loc in {$b_1$/$s_n$/1/at start/below right,  $b_2$/$s_n$/$x$/at start/below right, $b_{n-1}$/$s_n$/$x$/near start/below left, $b_n$/$s_n$/$x$/near start/below right}
            \draw[line width=1.1pt, blue, dashed] (\x) -- node[\pos, \loc, font=\footnotesize] {\w} (\y);

        \draw[line width=1.1pt, blue, dashed] (8.8,0.2) -- node[near end,above,font=\footnotesize] {$x$} (5.2,1.8);

        \foreach \x/\y/\w/\pos/\loc in {$s_1$/$b_2$/$\frac{n}{n-1}x$/near start/below right, $s_{n-1}$/$b_n$/$\frac{n}{n-1}x$/at start/below right}
            \draw[line width=1.1pt, blue, dashed] (\x) -- node[\pos, \loc, font=\footnotesize] {\w} (\y);

        \draw[line width=1.1pt, blue, dashed] (3.2,0.2) -- node[near start,below right,font=\footnotesize] {$\frac{n}{n-1}x$} (4.8,1.8);

        \node[left] at (0.5, 2)  {Buyers};
        \node[left] at (0.5, 0)  {Sellers};
    \end{tikzpicture}
    \caption{An $n$-buyer-$n$-seller general valuation market where $PoA=n$. Black solid lines are direct links, as captured by $N(i)$ for buyer $i$, and blue dotted lines indicate missing links. Buyer values are annotated adjacent to each edge. $x$ is a large constant. For the first seller $v_{11}=\frac{nx}{(n-1)(x+n)}+\frac{n}{x+n}, v_{21}=\frac{nx}{n-1}$. For sellers $j=2,...,n-1$, $v_{j+1,j}=\frac{n}{n-1}x, v_{jj}=\frac{n^2x}{(n-1)(x+n)}+(i-1)\epsilon$. For the last seller $v_{1n}=1,v_{jn}=x$ for $j=2,3,...,n$   All other values are zero.}
    \label{fig:poa_mn_general}
\end{figure}
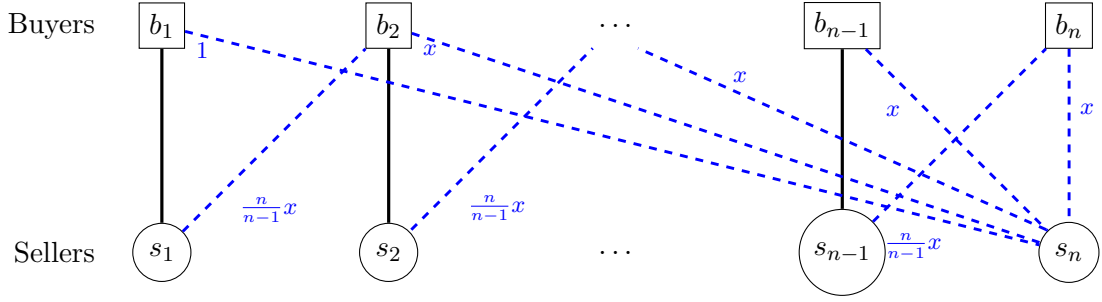

We show this through an example market in Figure~\ref{fig:poa_mn_general}. There are two main reasons for large PoA in this general valuation market. First, the unique market structure allows for only two pure equilibria. The platform either posts a low transaction fee where all sellers join, or a high fee where only one joins. Second, even though the equilibrium where all join has high social welfare, each seller has low price. Mixed equilibrium requires too low a transaction fee. The platform thus selects the pure equilibrium where only one seller joins.   

\begin{restatable}{proposition}{propPoAMNGeneral} \label{prop:poa_mn_general}
    There exists a $n$-buyer-$m$-seller  market for which the price of anarchy is $\min(\{n,m\})$.
\end{restatable}
\begin{proof}
Figure~\ref{fig:poa_mn_general} gives a $n$-seller-$n$-buyer market where PoA approaches $n$. Black solid lines are direct links, as captured by $N(i)$ for buyer $i$, and blue dotted lines indicate missing links. Seller $n$ has no off-platform links and always join the platform, and seller $i=1,2,...,n-1$ has a link to buyer $i$. Let $x$ be a large constant.  $V_{1n}=1, V_{11}=\frac{\frac{n}{n-1}x+n}{x+n}-\epsilon, V_{in}=x, V_{ii}=\frac{n^2x}{(x+n)(n-1)}+(i-1)\epsilon, V_{i,i-1}=\frac{n}{n-1}x ,\forall i\in [2,n]$. $\epsilon$ is a very small quantity to break ties. All other valuations are zero.
    
The optimal social welfare when all sellers join platform is $W^{\star}=nx+1$. We will show platform's revenue optimal transaction fee is $\alpha^{\star}=1$ and only seller $n$ joins the platform. Price of anarchy is then $\frac{nx+1}{x+\sum_{i=1}^{n-1}v_{ii}}\rightarrow n$ as $x\rightarrow \infty$. There are two unique properties to this market:
    
\begin{enumerate}
    \item whenever seller $j$ joins, seller $j+1$ joins because its off-platform edge is taken.
    \begin{eqnarray*}
        V_{j+1,j}=\frac{n}{n-1}x\geq V_{jj}+V_{j+1,j+1}, & \forall j=1,2,\ldots n-1
    \end{eqnarray*}
    This reads if $j+1$ does not join the platform, max weight matching matches seller $j$ to buyer $i=j+1$, and seller $j+1$ won't sell. 
    \item whenever all sellers $j'\geq j+1$ joins, seller $j$ joins. To see this, when only seller $n$ is on platform, seller $n-1$ joins when transaction fee $\alpha_{n-1}$ satisfies 
    \begin{eqnarray*}
        \alpha \leq \alpha_{n-1} = 1- \frac{p_{n-1}^{\off}(P=\{n\})}{p_{n-1}^{\on}(P=\{n\})} = 1-V_{n-1,n-1}/\frac{n}{n-1}x
    \end{eqnarray*}
    Now seeing seller $n-1$ joins, seller $n-2$ reasons about threshold of joining $\alpha_{n-2}$
    \begin{eqnarray*}
        \alpha \leq \alpha_{n-2} = 1- \frac{p_{n-2}^{\off}(P=\{n-1,n\})}{p_{n-1}^{\on}(P=\{n-1,n\})} = 1- v_{n-2,n-2}/\frac{n}{n-1}x
    \end{eqnarray*}
    So when $V_{n-2,n-2}< V_{n-1,n-1}$ seller $n-2$ joins whenever $n-1$ joins. This continues with $V_{2,2}\leq V_{3,3}\leq ...\leq V_{n-1}$ until the first seller. When seller $2,3,...,n$ joins, the first seller joins when
    \begin{eqnarray*}
        \alpha \leq \alpha_1 = 1-\frac{p_1^{\off}(P=S\setminus \{1\})}{p_1^\on(P=S\setminus \{1\})} = 1- v_{11}/(\frac{x}{n-1}+1) 
    \end{eqnarray*}
    The claim is true if $\alpha_1\geq \alpha_2 \geq ...\geq \alpha_{n-1}$. Verifying the buyers' valuations $v_{jj}$ for it $j=1,2,...,n-1$ we see \begin{eqnarray}\label{eqn:general_valuation_n_buyer_n_seller}
    V_{11}/(\frac{x}{n-1}+1) < V_{22}/\frac{n}{n-1}x <\ldots < V_{n-1,n-1}/\frac{n}{n-1}x    
    \end{eqnarray}
    This guarantees whenever $j+1$ joins, $j$ joins.
\end{enumerate}

Because of these two properties, only two pure equilibria exists. The platform can either post $\alpha^{\star}=1$ and charge $x$ from seller $n$, or posting $\alpha_1$ and have all sellers join. In the latter case, $p_j^\on=\frac{x}{n-1}+1, \forall j=1,2,...,n-1$
\begin{eqnarray*}
    Rev(\alpha_1) = \alpha_1[p_n^\on(S)+\sum_{j=1}^{n-1}p_j^\on(S)] = [1- v_{11}/(\frac{x}{n-1}+1)](x+n)
\end{eqnarray*}
We can verify $Rev(\alpha =1)=x> Rev(\alpha_1)$. This proves among all pure equilibrium, only seller $n$ joining is the revenue optimal one. We now prove no mixed equilibrium have larger revenue than $x$. If seller $j=1$ mixes, her revenue from not joining equals to that of joining. Off platform price is $v_{11}$. On platform price is at least as large as $\frac{n}{n-1}x-x=\frac{x}{n-1}$ because $x$ is the max externality seller $1$ imposes on others when transacting to buyer 2. This requires
\begin{eqnarray*}
    v_{11} = \alpha p_1^{\on} \geq \alpha\frac{x}{n-1}  & \text{or } & \alpha \leq (n-1)v_{11}/x
\end{eqnarray*}
Then the platform's revenue is no larger than $\alpha n x = n(n-1)v_{11} << x$ when $x\rightarrow \infty$. The same argument is true when seller 2 to n-1 mixes: referral fee $\alpha$ is too small for platform to make any revenue. 

The above verifies the $n$ PoA for a market with $n=m$ sellers and buyers. For a market $n>m$, we can have $m$ buyers with the above valuations and the $n-m$ buyers valuing no sellers, obtaining a $m=\min\{m,n\}$ lower bound. For a market $m>n$, we can have $n$ sellers with the above valuations and the $m-n$ sellers being valued by no buyers, obtaining a $n=\min\{m,n\}$ lower bound.

\end{proof}

\section{Missing Proofs for Theorem~\ref{thm:mixed_poa}}\label{app:mixed_proof}

Section~\ref{thm:mixed_poa} explains the intuition for the poa for mixed equilibrium. Here we give a complete and formal proof.  
\mixedPoA*

\begin{proof}
Consider a mixed Platform Equilibrium $\x = (x_1,\ldots, x_m)$ for a buyer-seller network $(S,B,G)$, where $x_j$ is the probability the seller $j$ joins the platform. We define the following Bayesian game:
    \begin{itemize}
        \item For each seller $j$, with probability $x_j$, $j$ can transact with all buyers (``type 1", $t_j=t_1$), and with probability $1-x_j$, it can only transact with the buyers linked to $j$ in $G$ (``type 2", $t_j=t_2$).
        \item The platform posts a transaction-fee $\alpha$.
        \item Before knowing the realization of its type, seller $j$ chooses a pure strategy $a_j\in\{\on,\off\}$ for joining the platform and being able to transact with all buyers, or staying off platform. Let $\mathbf{a}$ denote the joint strategy and $\mathbf{a}_{-j}$ the joint strategy except $j$'s.
        \item Given the realized graph $G'$, a competitive equilibrium is formed. Market clears according to the maximum competitive prices $p$.
        \item A seller has off platform utility $u'_j(\off,\mathbf{a}_{-j};\mathbf{t})=E_{G'\sim \mathbf{a},\mathbf{t}}[p_j]$ and on platform utility $u'_j(\on,\mathbf{a}_{-j};\mathbf{t})=(1-\alpha)E_{G'\sim \mathbf{a},\mathbf{t}}[p_j]$. A seller joining the platform pays an $\alpha$-fraction of their revenue to the platform.
    \end{itemize}

    We now show no seller joining $\mathbf{a}^{\off}=\{\off,\dots,\off\}$ is a pure-strategy Bayes-Nash equilibrium: $$\forall j, E_{\mathbf{t}\sim \x}[u'_j(\off,\mathbf{a}^{\off}_{-j};\mathbf{t})]\geq E_{\mathbf{t}\sim \x}[u'_j(\on,\mathbf{a}^{\off}_{-j};\mathbf{t})]$$ For this, consider a seller j that adopts probability $x_j$ in the mixed Nash Platform Equilibrium of the original, complete information platform game. There are three cases to consider
   
    \noindent\textbf{Case 1:} $x_j=0$ In this case, $j$'s expected utility from not joining the platform \textit{in the original, complete information game} is at least as much as $j$'s expected utility from joining given $\x_{-j}$. \textit{In the Bayesian game}, given that no other seller joins the platform, $j$'s utility from either joining or not joining the platform is exactly $j$'s utility for joining or not joining in the complete information game,
as other links are formed according to $\x_{-j}$.

    \noindent\textbf{Case 2:} $x_j=1$: In this case, in the Bayesian game, $j$ is linked to all buyers with probability $1$. Thus,  joining the platform in the Bayesian game does not increase $j$'s utility.

    \noindent\textbf{Case 3:} $x_j\in (0,1)$: In this case, in the \textit{complete information game}, $j$'s expected utility is the same for joining and not joining given $\x_{-j}$; otherwise, $j$ would deviate and $\x$ would not be an equilibrium. Let $u_j^\x$ denote this quantity. \textit{In the Bayesian game}, first consider $j$'s expected utility where it does not join
    \begin{eqnarray*}
        E_{\mathbf{t}\sim \x}[u'_j(\off,a^{\off}_{-j};\mathbf{t})] &=& (1-x_j)E_{\mathbf{t}_{-j}\sim\x_{-j}}[u'_j(\off,a^{\off}_{-j};t_2,\mathbf{t}_{-j})] + x_j E_{\mathbf{t}_{-j}\sim\x_{-j}}[u'_j(\off,a^{\off}_{-j};t_1,\mathbf{t}_{-j})]\\
        & \geq & (1-x_j)u^{\x}_j + x_j u^{\x}_j = u^{\x}_j
    \end{eqnarray*}
    The first term on the right reads with probability $1-x_j$, $t_j=t_2$ and $j$'s links stay as they were in $G$. Then $j$'s expected utility is exactly $u_j^\x$, as this is $j$'s utility from not joining the platform in the original game when all other sellers' links are formed according to $\x_{-j}$. 
    The second term on the right reads with probability $x_j$, $t_j=t_1$ and all $j$'s links are formed. Then $j$'s expected utility is at least as much as when it joins in the original game given all other sellers' links are formed according to $\x_{-j}$. This is because link formation distribution is the same, while the seller does not have to pay the $\alpha$ fee to the platform in the Bayesian game. Therefore, $j$'s expected utility from not joining is at least $u_j^\x$. 

    Then consider $j$'s expected utility if it joins the platform in the Bayesian game.
$$E_{\mathbf{t}\sim \x}[u'_j(\on,a^{\off}_{-j};\mathbf{t})] = u^{\x}_j$$ It exactly equals to the utility of $j$ joining the platform in the original setup, as the link formation distribution is the same as if $j$ joins in the original setup, and the fee $j$ pays is the same. So again $j$'s expected utility for not joining in the Bayesian game is at least as that for joining.

    In all above cases, each sellers' expected utility of not joining is at least as much as joining when no others join. Thus, $\mathbf{a}^{\off}$ is indeed a pure strategy Bayes-Nash equilibrium. The expected welfare of this equilibrium is exactly the same as the mixed Platform Equilibrium $\x$ in the original complete information game, as we have the same distribution over the formation of links, and the competitive equilibrium formed always maximizes the welfare given the links.
    
    For the last step of the proof, show if no sellers join is an equilibrium in the Bayesian game, then PoA with respect to pure strategies is at most $\frac{2-\alpha}{1-\alpha}$. The proof follows the same arguments of the proof of this case in Theorem~\ref{thm:pure_poa}, but the quantities we reason about for seller $j$ are in expectation over type probability $\x_{-j}$, which is no longer fixed to the mixed equilibrium. We prove $$\frac{W^\star}{E_{G'\sim \mathbf{a}^{\off},\mathbf{t}}[W(S,B,G')]} \leq \frac{2-\alpha}{1-\alpha}$$ where $G'$ is the realized network graph where no seller joins the market in the Bayesiain game. 

    Consider a single seller $j$ joining the platform, denote $G(j)$ as the realized graph where only $j$ joins. The on and off platform utility of seller $j$ can be expressed as 
    \begin{eqnarray*}
        E_{\mathbf{t}\sim\x}[u'_j(\on,\mathbf{a}^{\off}_{-j};\mathbf{t})] & = & (1-\alpha)E_{\mathbf{t}\sim\x}E_{G(j)\sim \mathbf{a}^{\off}_{-j},\mathbf{t}} [p_j] \\
        & = & (1-\alpha) E_{G(j)\sim a^{\off}_{-j},\x}[W(S,B,G(j))-W(S\setminus\{j\},B,G(j))]\\
        E_{\mathbf{t}\sim\x}[u'_j(\off,\mathbf{a}^{\off}_{-j};\mathbf{t})] & = & E_{\mathbf{t}\sim\x}E_{G'\sim a^{\off},\mathbf{t}} [\hat{p}_j] \\
        & = & E_{G'\sim \mathbf{a}^{\off},\x}[W(S,B,G')-W(S\setminus\{j\},B,G')]
    \end{eqnarray*}
    As $j$ does not join the platform, the off platform utility is larger than on utility. Rearranging
    \begin{eqnarray}
        E_{G(j)\sim \mathbf{a}^{\off}_{-j},\x}[W(S,B,G(j))] & \leq & E_{G(j)\sim \mathbf{a}^{\off}_{-j},\x}[W(S\setminus\{j\},B,G(j))]+\frac{1}{1-\alpha}E_{G'\sim \mathbf{a}^{\off},\x}[W(S,B,G')]\nonumber\\
        & - & \frac{1}{1-\alpha}E_{G'\sim \mathbf{a}^{\off},\x}[W(S\setminus\{j\},B,G')]\nonumber\\
        & = & \frac{1}{1-\alpha}E_{G'\sim \mathbf{a}^{\off},\x}[W(S,B,G')] -  \frac{\alpha}{1-\alpha}E_{G'\sim \mathbf{a}^{\off},\x}[W(S\setminus\{j\},B,G')] \label{eq:bound1_mixed}
    \end{eqnarray}

    Let $i^\star(j)$ be the buyer matched to $j$ in the ideal matching $W^\star$. Since $j$ can be matched to $i^\star(j)$ in $G(j)$,
    \begin{eqnarray}
        W(S,B, G(j)) & \geq& v_{i^\star(j)j} + W(S\setminus\{ j\}, B\setminus\{i^\star(j)\}, G(j)) \label{eq:bound2_mixed}
    \end{eqnarray}
    But since $G(j)$ and $G'$ only differ by seller $j$ always joining the platform 
    \begin{eqnarray}
        E_{G(j)\sim \mathbf{a}^{\off}_{-j},\x}[W(S\setminus\{j\},B\setminus\{i^\star(j)\},G(j))]=E_{G'\sim \mathbf{a}^{\off},\x}[W(S\setminus\{j\},B\setminus\{i^\star(j)\},G')] \label{eq:taking_out_j_same_mixed}
    \end{eqnarray}
    
    Combining Equations~\eqref{eq:bound1_mixed}, ~\eqref{eq:bound2_mixed} and ~\eqref{eq:taking_out_j_same_mixed}
    \begin{eqnarray}
        v_{i^\star(j)j} & \leq & \frac{1}{1-\alpha}E_{G'\sim \mathbf{a}^{\off},\x}[W(S,B,G')] -  \frac{\alpha}{1-\alpha}E_{G'\sim \mathbf{a}^{\off},\x}[W(S\setminus\{j\},B,G')] \nonumber\\
        & - & E_{G'\sim \mathbf{a}^{\off},\x}[W(S\setminus\{j\},B\setminus\{i^\star(j)\},G')]\nonumber \\
        &= & E_{G'\sim \mathbf{a}^{\off},\x}[\frac{1}{1-\alpha}W(S,B,G')-\frac{\alpha}{1-\alpha}W(S\setminus\{j\},B,G')-W(S\setminus\{j\},B\setminus\{i^{\star}(j)\},G')]
        \label{eq:bound3_mixed}
    \end{eqnarray}

    For a fixed realized graph $G'$, the three terms inside the expectation in Eq.~\eqref{eq:bound3_mixed} have been analyzed in Eq.~\eqref{eq:bound3}. Directly taking the results there we have a similar form to Eq. ~\eqref{eq:combine_three_terms_right}
    \begin{eqnarray*}
        v_{{i^\star}(j)j} &\leq &  E_{G'\sim \mathbf{a}^{\off},\x}[\frac{1}{1-\alpha}\cdot v_{i^{G'}(j)j} + v_{i^\star(j)t(j)}]
    \end{eqnarray*}
    Again $t(j)$ is the seller who transacts with $i^{\star}(j)$ in $G'$, also named as \textit{ the twin of $j$ }. 

    Summing over all sellers $j$, we get
    \begin{eqnarray*}
        W^\star =  \sum_j v_{{i^\star}(j)j} & \leq & \sum_j E_{G'\sim \mathbf{a}^{\off},\x}\left(\frac{1}{1-\alpha}\cdot v_{i^{G'}(j)j} + v_{i^\star(j)t(j)}\right)\\ 
        & = &  E_{G'\sim \mathbf{a}^{\off},\x} [\frac{1}{1-\alpha}\cdot\sum_j v_{i^{G'}(j)j} + \sum_j v_{i^\star(j)t(j)}]\\
        & \leq & E_{G'\sim \mathbf{a}^{\off},\x} [\frac{1}{1-\alpha} W(S,B,G') + \sum_j v_{i^{G'}(j)j}]\\
        & = & \frac{2-\alpha}{1-\alpha} E_{G'\sim \mathbf{a}^{\off},\x}[W(S,B,G')],
    \end{eqnarray*}
    where the last inequality follows because each seller $j$ has a distinct twin $t(j)$, so we sum over distinct edges of $G'$.
\end{proof}

\section{Generalizations}\label{app:generalizations}
\subsection{Market with Multiple Platforms}\label{app:multiple_platforms}
\MultiPlatform*
\begin{proof}[proof sketch]
    The proof follows from the same argument as that of Theorem~\ref{thm:pure_poa} and Theorem~\ref{thm:mixed_poa}. For the Price of Anarchy of pure Platform Equilibrium, we again first prove if no seller chooses to join any platform, welfare is guaranteed by, $$ \frac{W^\star}{W(S,B,G)}\leq\frac{2-\alpha}{1-\alpha}.$$
   
    Let $i^{\star}(j)$ be the buyer that $j$ transacts with in ideal matching $W^\star$, $f^\star(j)$ be the platform that $j$ pays in ideal matching, and $G^{\star}(j)$ be the network when only seller $j$ joins platform $f^\star(j)$. Then inequalities~\ref{eq:p_on} to \ref{eq:bound2} still hold after changing $W(S,B,G(j))$ for $(W,S,B,G^\star(j))$, and inequalities~\ref{eq:bound3} to \ref{eq:combine_three_terms_right} remain the same. Now consider any pure equilibrium where a set of sellers $P$ join some platforms. Denote the resulting graph by $G(P)$.
    Again we can construct another market where the buyers have the same valuations as the original market but the initial network is $G'=G(P)$. No seller joining is a pure Platform Equilibrium because otherwise some sellers would have deviated in the pure equilibrium in the original market and joined some other platforms. So $\frac{W^\star}{W(S,B,G(P))}=\frac{W^\star}{W(S,B,G')}\leq \frac{2-\alpha}{1-\alpha}$. The proof for the Price of Anarchy of mixed Platform Equilibrium follows similarly.
\end{proof}

\subsection{Sellers with Production Costs} \label{app:seller_with_cost}
In this section, we extend our results to sellers $j\in S=\{1,2,...,m\}$ with production cost $c_j\geq 0$. Sellers with a cost too high barely trade, and do not contribute to social welfare. Therefore, for the discussion we assume all sellers transact. This can be modeled as each seller $j$ being connected off-platform to a dummy buyer $i_j$ that only values seller $j$ and dislikes other sellers: 
$$v_{i_j,j'}= 
\begin{cases}
    c_j,     & \text{if } j'=j\\
    -\infty,        & \text{otherwise}
\end{cases}.$$

For any market $M=(S,B,\mathbf{v},\mathbf{c},G)$ with cost $\mathbf{c}$, define a many-to-one mapping of markets $T(M)=M'$, where $M'$ is a new market with zero seller cost and buyer valuations 
$$v'_{ij}= 
\begin{cases}
    v_{ij}-c_j,     & \text{if } v_{ij}\geq c_j\\
    -\infty,        & \text{otherwise}
\end{cases}$$

\begin{lemma}
\label{lem:ce_with_cost_exist}
    If $(\mathbf{p},\mathbf{a})$ is a competitive equilibrium in market $M$, define $p_j'=\max\{p_j-c_j,0\}$, then $(\mathbf{p}',\mathbf{a})$ is a competitive equilibrium in market $M'$. If $(\mathbf{p}',\mathbf{a})$ is a competitive equilibrium in market $M'$, define $p_j=p'_j+c_j$, then $(\mathbf{p},\mathbf{a})$ is a competitive equilibrium in market $M$.
\end{lemma}

\begin{proof}
We first prove if $(\mathbf{p}',\mathbf{a})$ is a competitive equilibrium in $M'$, then $(\mathbf{p},\mathbf{a})$ is a competitive equilibrium in $M$. 
The first three conditions for competitive equilibrium in Definition~\ref{def:comp_eq} are easy to check. Without loss of generality each seller $j$ in $M'$ transacts. Because otherwise $p_j=0$ equals to the dummy buyers' valuation in $M'$: $v'_{i_j,j}=0$. So the last condition is satisfied. We now show every buyer gets their most preferred outcome with $(\mathbf{p},\mathbf{a})$. When buyer $i$ does not obtain an item, $\forall j, p'_j\geq v'_{ij}$. If $v_{ij}\geq c_j$ expanding we have $p_j\geq v_{ij}$; If $v_{ij}< c_j$ with prices being non-negative $p_j=p'_j+c_j\geq c_j > v_{ij}$. In either case buyer $i$'s best choice in $M$ is not obtaining an item. When buyer $i$ does obtain an item $j^{\star}$ in $M'$, $v_{ij^\star}\geq c_{j^\star}$, and $\forall j, v'_{ij^\star}-p'_{j^\star}=v_{ij^\star}-p_{j^\star}\geq v'_{ij}-p'_j$. If $v_{ij}\geq c_j$ then $v_{ij^\star}-p_j^\star\geq v_{ij}-p_j$ and we are done; otherwise $v_{ij}-p_j<c_j-p_j \leq 0 \leq v_{ij^\star}-p_j^\star$. So indeed buyer $i$'s favorite item in $M$ is $j^\star$.

Now prove the other direction: if $(\mathbf{p},\mathbf{a})$ is a competitive equilibrium in $M$, $(\mathbf{p}',\mathbf{a})$ is a competitive equilibrium in $M'$. Again the first three criteria for Definition~\ref{def:comp_eq} are easy to verify. With dummy buyers in $M$, all items are sold in $\mathbf{a}$ so the last criteria is satisfied.
We now verify each buyer gets their most preferred outcome in $M'$. When buyer $i$ does not obtain an item, $\forall j, v_{ij}\leq p_j$. If $v_{ij}< c_j$ then $v'_{ij}=-\infty <0 \leq p'_j$; if $v_{ij}\geq c_j$ then expanding $v'_{ij}\leq p'_j$. In either case buyer $i$'s best choice in $M'$ is not obtaining an item.
If a dummy buyer $i_j$ acquires an item $j^\star=j$ in $M$. Then $v_{i_j,j^{\star}}=c_{j^\star}\geq p_{j^\star}$. With the same allocation $v'_{ij^\star}-p'_{j^\star}=0-\max\{0,p_{j^\star}-c_{j^\star}\}=0$. But for any other seller item $j'$, $v'_{i_j,j'}=-\infty$ so the dummy buyer strictly does not want other items. If buyer $i$ that is not a dummy buyer obtains an item $j^\star$ in $M$, then $p_{j^\star}\geq c_{j^\star}$ because otherwise dummy buyer $i_{j^\star}$ does not satisfy the demand. 
By non-negative of utility and price $v_{ij^\star}\geq p_{j^\star} \geq c_{j^\star}$ so $\forall j, v'_{ij^\star}-p'_{j^\star}=v_{ij^\star}-p_{j^\star}\geq v_{ij}-p_{j}$. 
If $v_{ij}\geq c_j$ and $p_j\geq c_j$ then we have $v'_{ij} - p'_j =v_{ij}-c_j-\max\{0,p_j-c_j\}=v_{ij}-p_j$ and we are done, if $v_{ij}\geq c_j$ and $p_j< c_j$ then we have  $v'_{ij} - p'_j =v_{ij}-c_j< v_{ij}-p_j$ and we are again done; 
if $v_{ij}< c_j$ then $v'_{ij}=-\infty$ and surely in $M'$ buyer $i$ does not demand $j$.
\end{proof}

The above lemma gives a one-to-one correspondence between competitive equilibrium in $M$ and $M'$. This allows us to analyze social welfare in $M$. The social welfare of an allocation $\mathbf{a}$ in $M$ is equal to valuation minus cost in $M'$, that is, $$\sum_{ij}a_{ij}(v_{ij}-c_{j}).$$  By applying the first welfare theorem for the market without costs, we get the following. 
\begin{corollary}
\label{cor:first_welfare_cost}
    In a competitive equilibrium for a market with production costs, the social welfare is maximized with respect to the set of allocations that respect the transaction constraints posed by $G$.
\end{corollary}
\begin{proof}
    By First Welfare Theorem~\ref{thm:first_welfare} in market without cost, the competitive allocation $\mathbf{a}$ in market $M'$ satisfies $\forall \mathbf{a}', \sum_{ij}a_{ij}v'_{ij}\geq \sum_{ij}a'_{ij}v'_{ij}$. If $a_{ij}=1$, by non-negative utility $v_{ij}\geq c_j$. So $\forall \mathbf{a'}, \sum_{ij}a_{ij}(v_{ij}-c_j)\geq \sum_{ij}a'_{ij}v'_{ij}$. For allocations $\mathbf{a'}$ that let buyers purchase despite $v_{ij}<c_j$, the social welfare is smaller than simply letting this transaction opportunity go. So we focus on $\mathbf{a'}$ that only transacts when $v_{ij}\geq c_j$. For these $\mathbf{a'}$,  $\sum_{ij}a_{ij}(v_{ij}-c_j)\geq \sum_{ij}a'_{ij}(v_{ij}-c_j)$. As social welfare incorporates cost in $M$, $\mathbf{a}$ maximizes social welfare.
\end{proof}

Let $W(S,B,\mathbf{v},\mathbf{c},G)$ denote optimal welfare in $M$. For any competitive equilibrium allocation $\mathbf{a}$, it maximizes social welfare in both $M$ and $M'$. By the construction of buyer valuation in $M'$, we observe the following.

\begin{corollary}
\label{cor:equal_welfare}
    Optimal welfare in $M$ and $M'$ are equal. $W(S,B,\mathbf{v},\mathbf{c},G)=W(S,B,\mathbf{v'},G)$.\label{thm:equal_welfare}
\end{corollary}
\begin{proof}
    Consider a competitive equilibrium allocation $\mathbf{a}$ for both market $M$ and $M'$. Optimal social welfare in $M'$ equals to $W(S,B,\mathbf{v}',G)=\sum_{ij}a_{ij}v'_{ij}=\sum_{ij}a_{ij}(v_{ij}-c_j)=W(S,B,\mathbf{v},\mathbf{c},G)$. The second inequality is because the First Welfare Theorem~\ref{thm:first_welfare} requires $\mathbf{a}$ to maximize welfare: whenever $a_{ij}=1, v'_{ij}\geq 0$.
\end{proof}

The two corollaries above are implications of Lemma~\ref{lem:ce_with_cost_exist} on the allocation side of competitive equilibrium. On the price side, though a competitive price $p_j$ in $M$ associates with $p'_j=\max\{p_j-c,0\}$, the maximum competitive price correspondence is simpler.

\begin{corollary}
\label{cor:max_price_cost}
    Given a of price $\mathbf{p}$, define $p'_j=p_j-c_j$. $\mathbf{p}$ is the max competitive price for $M$ if and only if $\mathbf{p}'$ is the max competitive price for $M'$.
\end{corollary}
\begin{proof}
    If $\mathbf{p}$ is the maximum competitive price in $M$, $p_j\geq c_j$: if $j$ is matched to a dummy buyer $i_j$ in equilibrium, the price can always increase until $p_j=c_j$; if $j$ is matched to other buyers, then $p_j>c_j$ because otherwise $i_j$ envies. By Lemma~\ref{lem:ce_with_cost_exist}, $p'_j=p_j-c_j$ is a competitive price in $M'$. If in $M'$ there is a higher competitive price $\mathbf{p}''$, by Lemma~\ref{lem:ce_with_cost_exist} in market $M$ there must be a competitive price $\mathbf{p}''+\mathbf{c}$ higher than $\mathbf{p}$, contradict. So $\mathbf{p}'$ is indeed the maximum competitive price. The other direction works similarly.
\end{proof}

We are now ready to inspect the effect of cost on sellers' decisions to join platform with maximum competitive price. If platform would charge a transaction fee as a percentage of sellers' utility, $p_j-c_j$, all our previous results without cost immediately hold. As the platform cannot directly observer sellers' costs, the transaction fee is a percentage sellers' price, which distorts sellers' incentives to join. At fee $\alpha$, seller $j$ in $M$ joins the platform if $(1-\alpha)p_j^{\on}-c_j \geq p_j^{\off}-c_j\Leftrightarrow (1-\alpha)(p_j^{'\on}+c_j) \geq p_j^{'\off}+c_j \Leftrightarrow (1-\alpha)p_j^{'\on}\geq p_j^{'\off}+\alpha c_j$. The higher the cost, the less likely seller $j$ joins the platform. We generalize our result in Section~\ref{sec:poa-regulated} to settings with costs.

\purePoACost*
\begin{proof}
    The proof logic is similar to that in section~\ref{sec:poa-regulated}, where we first prove for pure equilibrium then extend it to mixed equilibrium. If no seller joins the market at $\alpha$, then $\forall j\in P$
    $$p_{j}^{'\off} +\alpha c_j \geq (1-\alpha)p_{j}^{'\on}$$ Expanding the price 
    \begin{eqnarray*}
        W(S,B,\mathbf{v}',G(j)) - W(S,B,\mathbf{v}',G) & \leq & p^{'\on}_j-p^{'\off}_j \leq  \alpha[p^{'\on}_j+c_j] \\
        & \leq & \frac{\alpha}{1-\alpha}\cdot p^{'\off}_j + \frac{\alpha}{1-\alpha}c_j\\
        & \leq & \frac{\alpha}{1-\alpha}\left(W(S,B,\mathbf{v}',G) - W(S\setminus\{j\},B,\mathbf{v}',G)\right)+\frac{\alpha}{1-\alpha}c_j
    \end{eqnarray*}

    Rearranging gives 
    \begin{eqnarray*}
    W(S,B,\mathbf{v}',G(j)) \leq
        \frac{1}{1-\alpha}W(S,B,\mathbf{v}',G) - \frac{\alpha}{1-\alpha}W(S\setminus\{j\},B,\mathbf{v}',G)+\frac{\alpha}{1-\alpha}c_j
    \end{eqnarray*}

    Let $i^\star(j)$ be the buyer matched to $j$ in the ideal matching $W^\star(S,B,\mathbf{v},\mathbf{c},G)$. By Lemma~\ref{lem:ce_with_cost_exist}, $i^\star(j)$ is the buyer matched to $j$ in $W^\star(S,B,\mathbf{v}',G)$. Since matching $i^\star(j)$ to $j$ is also one of the options in $(S,B,\mathbf{v}',G(j))$
    \begin{eqnarray*}
        W(S,B,\mathbf{v}',G(j)) & \geq& v'_{i^\star(j)j} + W(S\setminus\{ j\}, B\setminus\{i^\star(j)\}, \mathbf{v}',G)
    \end{eqnarray*}

    Combining the two above inequalities
    \begin{eqnarray}
        v'_{i^\star(j)j} \leq \frac{1}{1-\alpha}W(S,B,\mathbf{v}',G)-\frac{\alpha}{1-\alpha}W(S\setminus\{j\},B,\mathbf{v}',G)-W(S\setminus\{j\},B\setminus\{i^\star(j)\},\mathbf{v}',G)+\frac{\alpha}{1-\alpha}c_j \label{eq:bound3_cost}
    \end{eqnarray}

    We wish to relate the right-hand side of Eq.~\eqref{eq:bound3_cost} to terms that relate to $W(S,B, \mathbf{v}',G)$ and $W^\star$.
 Let $i^G(j)$ be the buyer matched to $j$ in $W(S,B, \mathbf{v}',G)$ and $W(S,B,\mathbf{v},\mathbf{c},G)$. First, consider $W(S\setminus\{j\},B,\mathbf{v}',G)$. By 
definition,
    \begin{eqnarray*}
        W(S,B,
        \mathbf{v}',G) = W(S\setminus\{j\}, B\setminus\{i^G(j)\},\mathbf{v}', G) + v'_{i^G(j)j}. 
    \end{eqnarray*}
    Thus, we have,
    \begin{eqnarray}
        W(S\setminus\{j\},B, \mathbf{v}',G) \geq W(S\setminus\{j\}, B\setminus\{i^G(j)\}, \mathbf{v}', G) = W(S,B,\mathbf{v}',G)-v'_{i^G(j)j}.\label{eq:bound4_cost}
    \end{eqnarray}

    As for $W(S\setminus\{j\},B\setminus\{i^\star(j)\}, \mathbf{v}',G)$, seller  $j$ is matched to $i^G(j)$ in $G$, while $i^\star(j)$ is matched to some potentially different vertex in $G$, which we call \textit{the twin of $j$}
 and denote by $t(j)$. We  have the following inequality,
    \begin{eqnarray*}
        W(S,B,\mathbf{v}',G) &\leq & W(S\setminus\{j,t(j)\},B\setminus\{i^G(j),i^\star(j)\},\mathbf{v}',G) + v'_{i^G(j)j} + v'_{i^\star(j)t(j)}\nonumber \\
        &\leq & W(S\setminus\{j\},B\setminus\{i^\star(j)\},\mathbf{v}',G) + v'_{i^G(j)j} + v'_{i^\star(j)t(j)}, 
    \end{eqnarray*}
    where the first inequality is an equality if $j\neq t(j)$. Rearranging gives
    \begin{eqnarray}
        W(S\setminus\{j\},B\setminus\{i^\star(j)\},\mathbf{v}',G) \geq W(S,B,\mathbf{v}',G) - (v'_{i^G(j)j} + v'_{i^\star(j)t(j)}).\label{eq:bound5_cost}
    \end{eqnarray}

    Combining Equations~\eqref{eq:bound3_cost},~\eqref{eq:bound4_cost}, and~\eqref{eq:bound5_cost}, we get
    \begin{eqnarray}
        v'_{{i^\star}(j)j} &\leq& \frac{1}{1-\alpha}W(S,B,\mathbf{v}',G)-\frac{\alpha}{1-\alpha}\left(W(S,B,\mathbf{v}',G)-v'_{i^G(j)j}\right)\nonumber\\  &-& \left(W(S,B,\mathbf{v}',G) - (v'_{i^G(j)j} + v'_{i^\star(j)t(j)})\right)+\frac{\alpha}{1-\alpha}c_j \nonumber \\
        & = & \frac{1}{1-\alpha}\cdot v'_{i^G(j)j} + v'_{i^\star(j)t(j)} + \frac{\alpha}{1-\alpha}c_j \nonumber
    \end{eqnarray}

    Summing over all sellers $j$, we have
    \begin{eqnarray}
        W^\star(S,B,\mathbf{v},\mathbf{c},G)=W^\star(S,B,\mathbf{v}',G)   
        = \sum_j v'_{{i^\star}(j)j} & \leq & \sum_j \left(\frac{1}{1-\alpha}\cdot v'_{i^G(j)j} + v'_{i^\star(j)t(j)}+\frac{\alpha}{1-\alpha}c_j \right)\nonumber\\ & = &  \frac{1}{1-\alpha}\cdot\sum_j v'_{i^G(j)j} + \sum_j v'_{i^\star(j)t(j)}+\frac{\alpha}{1-\alpha}\sum_j c_j\nonumber\\
        & \leq & \frac{2-\alpha}{1-\alpha} W(S,B,\mathbf{v}',G)+\frac{\alpha}{1-\alpha}\sum_j c_j\nonumber\\
        & = & \frac{2-\alpha}{1-\alpha} W(S,B,\mathbf{v},\mathbf{c},G)+\frac{\alpha}{1-\alpha}\sum_j c_j \label{eq:poa_cost}
    \end{eqnarray}
    where the first and the last equality follows from Corollary~\ref{cor:equal_welfare}, and the last inequality is because each seller $j$ has a distinct twin $t(j)$, so we sum over distinct edges of $G$. Bring in $\sum_j c_j =\beta W^\star(S,B,\mathbf{v},\mathbf{c},G)$ 
    \begin{eqnarray}
        W(S,B,\mathbf{v},\mathbf{c},G) \geq \frac{1-\alpha-\alpha\beta}{2-\alpha} W^\star(S,B,\mathbf{v},\mathbf{c},G)
    \end{eqnarray}

    Now if a set $P$ sellers join the platform in a pure equilibrium, just like that in the proof for Theorem~\ref{thm:pure_poa}, we can construct another market where the initial network is $G'=G(P)$. For the same transaction fee, no seller joining the platform in $(S,B,\mathbf{v},\mathbf{c},G')$ is an equilibrium. And we can apply inequality~\ref{eq:poa_cost} to $G'$ and achieve the same result.
    
    The extension to mixed equilibrium works similarly to that in the proof for Theorem~\ref{thm:mixed_poa} and is omitted.
\end{proof}

The above theorem implies that at the same transaction fee, the social welfare guarantee (naturally) deteriorates as costs get higher. This was specially relevant during the COVID-19 crisis, where demand and supply surged on digital platforms \citep{raj2020covid} and the cost of production grew \citep{felix2020us}.
At $30\%$ of transaction fee (Table~\ref{table:platforms and their coommission rate}) and $30\%$ of food cost \footnote{Most profitable restaurants aim for a food cost percentage between 28 and 35\%. This does not include other cost factors such as labor, rentals, etc. Figure is taken in 2023 from Doordash website \url{https://get.doordash.com/en-us/blog/food-cost-percentage}}, the resulting welfare at a  Platform Equilibrium  is at least $23.5\%$ of the ideal welfare. This is smaller than the $41\%$ without costs, further demonstration the need to regulate transaction fee during periods where production costs increase.
\chapter{Appendix to \Cref{chap:pricing_with_tips}}\label{app:ch3}

\crefalias{section}{appendix}
\crefalias{subsection}{appendix}

\section{Summary of results}\label{sec:summary_of_results}
\Cref{sec:motivate_use_tip} --- Motivating the use of tips
\begin{itemize}[labelwidth=2.5cm, leftmargin=!, labelsep=0.6em, align=left]
    \item[$\bullet$ \Cref{thm:with_tip_eq_existence}] With-tips equilibrium always exists
    \item[$\bullet$ \Cref{thm:without_tip_eq_existence}] Without-tips equilibrium exists with sufficiently many couriers
    \item[$\bullet$ \Cref{lem:pareto_dominant}] Highest with-tips equilibrium welfare weakly larger than highest without-tips equilibrium welfare
\end{itemize}
\Cref{sec:general_markets} --- Markets without any structural constraints
\begin{itemize}[labelwidth=2.5cm, leftmargin=!, labelsep=0.6em, align=left]
    \item[$\bullet$ \cref{lem:OPT_eq_hard}] NP-hardness through reduction from 3DM: finding optimal welfare, maximum with-tip equilibrium welfare, and maximum without-tip equilibrium welfare
    \item[$\bullet$ \cref{thm:poly_check}] Polynomial-time algorithm to check if any allocation is supported by any with-tip equilibrium
\end{itemize}
\Cref{sec:market_structures} --- Markets with structures
\begin{itemize}[labelwidth=2.5cm, leftmargin=!, labelsep=0.6em, align=left]
    \item[$\bullet$ \cref{thm:divisible_courier_cost_opt}] With structured courier costs, polynomial-time algorithm to find optimal welfare
    \item[$\bullet$ \cref{thm:divisible_courier_cost}] With structured courier costs, optimal with-tip equilibrium equals to optimal welfare
    \item[$\bullet$ \cref{thm:buyer_value_opt}] With structured buyer valuations, polynomial-time algorithm to find optimal welfare
    \item[$\bullet$ \cref{thm:one_buyer_one_store}] With structured buyer valuations, optimal with-tip equilibrium equals to optimal welfare
\end{itemize}
\Cref{sec:profit} --- Platform profit maximization\\
Focusing on markets where one courier can only deliver for one store
\begin{itemize}[labelwidth=2.5cm, leftmargin=!, labelsep=0.6em, align=left]
    \item[$\bullet$ \cref{thm:profit_polytime_without_tip}] Polynomial-time algorithm for finding profit-maximizing without-tip equilibrium
    \item[$\bullet$ \cref{thm:profit_hard_with_tip}] NP-hardness through reduction from Vertex Cover: finding the maximum profit of with-tip equilibrium
\end{itemize}

\section{Closed-Form Solution for Minimum Tip}\label{app:min_tip}
We give the closed form solution for $\underline{t}_{bs}$.
\begin{lemma}
    Given delivery compensation $\mathbf{w}$ and other buyers' tips $\mathbf{t}_{-b}$, buyer $b$'s minimum tip required to have some courier deliver from a store $s$ to her is given by 
$$\underline{t}_{bs}=\min_d\{\max\{0,c_d(b,s)-w_{bs},\max_{(b',s')\neq (b,s)}\{w_{b's'}+t_{b's'}-c_d(b',s')-w_{bs}+c_d(b,s)\}\}\}$$
where $t_{b's'}=0$ for $b'=b, s'\neq s$.
\end{lemma}
\begin{proof}
    We first prove that when buyer $b$ finds the tip profile $\mathbf{t}_{b}$ with the lowest $t_{bs}$. W.l.o.g., $\mathbf{t}_{bs'}=0$ for $s'\neq s$. Suppose $\exists s'\neq s, t_{bs'}>0$ and $(b,s)\in \BR_d(\mathbf{w},\mathbf{t}_b,\mathbf{t}_{-b})$ for a courier $d$. Then $w_{bs}-c_d(b,s)+t_{bs}\geq w_{bs'}-c_d(b,s')+t_{bs'}$. By changing $t_{bs'}$ to $0$, it remains that $(b,s)\in \BR_d(\mathbf{w},\mathbf{t}_b,\mathbf{t}_{-b})$.
    
    \noindent The condition for a courier $d$ satisfying $(b,s)\in \BR_d(\mathbf{w},\mathbf{t}_b,\mathbf{t}_{-b})$ is
    \begin{align*}
        w_{bs}+t_{bs}-c_{d}(b,s) &\geq w_{b's'}+t_{b',s'}-c_{d}(b',s')  \text{\; for all \;} (b',s')\neq (b,s) \text{\; and}\\
        w_{bs}+t_{bs}-c_{d}(b,s) &\geq 0 \text{\; and \;} t_{bs}\geq 0
    \end{align*}
    For $b'\neq b$, the value of $\mathbf{t}_{b'}$ is already given by $\mathbf{t}_{-b}$. And we have already proved that w.l.o.g, $t_{bs'}=0$ for $s'\neq s$. So the minimum tip required for courier $d$ to delivery from $s$ to $b$ is \begin{equation}\label{eq:min_tip_for_courier_d}
    \max\{0,c_d(b,s)-w_{bs},\max_{(b',s')\neq (b,s)}\{w_{b's'}+t_{b's'}-c_d(b',s')-w_{bs}+c_d(b,s)\}\}
    \end{equation}
\end{proof}

\section{Example Market}\label{sec:example market}
\begin{figure}[t]
    \centering
    \begin{tblr}{}
    \centering
\begin{tikzpicture}[baseline=(current bounding box.center),scale=0.9]
    \node[draw, shape=circle, minimum size=0.1cm] (1) at (-1, 2) {\scriptsize $b_1$};
    \node[draw, shape=circle, minimum size=0.1cm] (2) at (-1, 0) {\scriptsize $b_2$};
    \node[draw, shape=circle, minimum size=0.1cm] (3) at (2, 1) {\scriptsize $s_1$};
    \node[draw, shape=circle, minimum size=0.1cm] (4) at (3.5, 2) {\scriptsize $d_1$};
    \node[draw, shape=circle, minimum size=0.1cm] (5) at (3.5, 0) {\scriptsize $d_2$};
    \draw (1) to (3);
    \draw (2) to (3);
    \node[] at (0.5, 1.2){\scriptsize 3};
    \node[] at (0.5, 0.15){\scriptsize 10};
\end{tikzpicture}
& 
\begin{tabular}{c|cc}
Cost & $d_1$  & $d_2$\\
\hline
$b_1 s_1$ & $0$  & $1$\\
$b_2 s_1$ & $11$  & $12$\\
\end{tabular}
\end{tblr}
    \caption{A market where the optimal without-tip equilibrium welfare is -1, but a with-tip equilibrium achieves the optimal welfare. Buyers $b_1$ and $b_2$ value the only store, $s_1$, at $3$ and $10$, respectively. Courier $d_1$ has costs of delivery $0$ and $11$ to $b_1$ and $b_2$, respectively; courier $d_2$ has costs of delivery $1$ and $12$ to $b_1$ and $b_2$, respectively. In \Cref{sec:market_structures}, we will use this market to illustrate \Cref{thm:divisible_courier_cost}. Note that courier costs can be decomposed either as $c(b_1,s_1)=0, c(b_2,s_1)=11, c_{d_1}(s_1)=0, c_{d_2}(s_1)=1$, or as $c(b_1,s_1)=0, c(b_2,s_1)=11, c_{d_1}(b_1)=c_{d_1}(b_2)=0, c_{d_2}(b_1)=c_{d_2}(b_2)=1$.
    \label{fig:without_tip_bad}}
\end{figure}
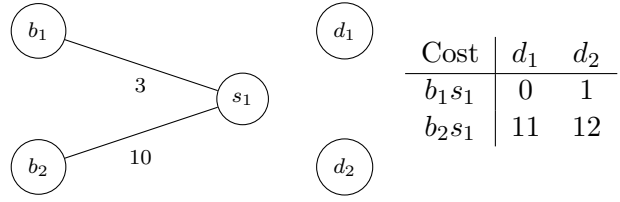

The following example markets, illustrated also in \Cref{fig:without_tip_bad} shows that optimal with-tip equilibrium achieves the optimal welfare, while the without-tip equilibrium is highly inefficient.
\begin{example}\label{example:without_tip_bad}
    Consider a simple market in Figure~\ref{fig:without_tip_bad}. Buyers $b_1$ and $b_2$ value the store $s_1$ at $3$ and $10$, respectively. Couriers $d_1$ and $d_2$ have delivery costs $c_{d_1}(b_1,s_1)=0,c_{d_1}(b_2,s_1)=11$ and $c_{d_2}(b_1,s_1)=1,c_{d_2}(b_2,s_1)=12$. The optimal welfare is $3$ with the allocation in which $d_1$ delivers from $s_1$ to $b_1$. 
    
    In the without-tip regime, the optimal without-tip equilibrium welfare is $-1$. The equilibrium is defined by $p_{s_1}\in[3,10], w_{b_2s_1}=11,w_{b_1s_1}=0$, and $d_1$ delivers to $b_2$. To see this, there is no equilibrium where $p_{s_1}<3$ because both $b_1$ and $b_2$ want to buy from $s_1$. If $p_{s_1}>10$, the market does not clear.

    In the with-tip equilibrium, the optimal with-tip equilibrium welfare is $3$. One such equilibrium is given by $p_{s_1}=1, w_{b_1s_1}=1, w_{b_2s_1}=0, t_{b_1s_1}=t_{b_2,s_1}=0$ and allocation $d_1$ delivers from $s_1$ to $b_1$. To verify, couriers best responses are $\BR_{d_1}(\mathbf{w},\mathbf{t})=\{(b_1,s_1)\}, \BR_{d_2}(\mathbf{w},\mathbf{t})=\{\emptyset,(b_1,s_1)\}$. Buyer $b_1$ has positive utility $u_{b_1}(s_1)=2>0$, and offers zero tip. Buyer $b_2$ has to tip at least 12 for courier $d_2$ and 12 for courier $d_1$ to deliver from $s_1$ to her. Observe that $\underline{t}_{b_2s_1}=12$, so if deviating, buyer $b_2$ has to pay $p_{s_1}+\underline{t}_{b_2s_1}=13>v_{b_2s_1}$.

    There is another with-tip equilibrium that shares the same purchase price, delivery compensation, and allocations as the without-tip equilibrium, but with zero tips. To check, couriers incentives remain unchanged with zero tips, $\BR_{d_1}(\mathbf{w},0)=\{\emptyset, (b_2,s_1), (b_1,s_1)\}$ and $\BR_{d_2}(\mathbf{w},0)=\{\emptyset\}$. For $b_1$ the minimum tip to have courier $d_1$ deliver for her is $\underline{t}_{b_1s_1}=0$. But $v_{b_1}(s_1)-p_{s_1}-\underline{t}_{b_1s_1}\leq 0$. For $b_2$ it holds that $\underline{t}_{b_2s_1}=0$.
\end{example}

\section{Missing Proofs and Lemmas in \Cref{sec:motivate_use_tip}}\label{app:eq_existence}
\begin{figure}[t]
    \centering
    \begin{tblr}{}
    \centering
\begin{tikzpicture}[baseline=(current bounding box.center),scale=0.9]
    \node[draw, shape=circle, minimum size=0.1cm] (1) at (-1, 2) {\scriptsize $b_1$};
    \node[draw, shape=circle, minimum size=0.1cm] (2) at (-1, 0) {\scriptsize $b_2$};
    \node[draw, shape=circle, minimum size=0.1cm] (3) at (2, 2) {\scriptsize $s_1$};
    \node[draw, shape=circle, minimum size=0.1cm] (4) at (2, 0) {\scriptsize $s_2$};
    \node[draw, shape=circle, minimum size=0.1cm] (5) at (3.5, 1) {\scriptsize $d_1$};
    \draw (1) to (3);
    \draw (1) to (4);
    \draw (2) to (3);
    \draw (2) to (4);
    \node[] at (0.5, 2.2){\scriptsize 4};
    \node[] at (-0.5, 1.4){\scriptsize 2};
    \node[] at (-0.5, 0.6){\scriptsize 1};
    \node[] at (0.5, 0.15){\scriptsize 3};
\end{tikzpicture}
& 
\begin{tabular}{c|c}
Cost & $d_1$ \\
\hline
$b_1 s_1$ & $0$ \\
$b_1 s_2$ & $0$ \\
$b_2 s_1$ & $0$ \\
$b_2 s_2$ & $0$
\end{tabular}
\end{tblr}
    \caption{A market where there is no without-tip equilibrium. As there is only one courier, one of the two stores are not bought in any feasible allocation. At the purchase price $0$, the buyer who does not buy will want to buy from a store that does not sell in the without-tip regime. A with-tip equilibrium is given by $p_{s_1}=1,p_{s_2}=0,\mathbf{t}=0,w_{b_1s_1}=4,w_{b_1s_2}=w_{b_2s_1}=w_{b_2s_2}=0$ and $d_1$ deliver $s_1$ to $b_1$.
    \label{fig:without_tip_not_exists}}
\end{figure}
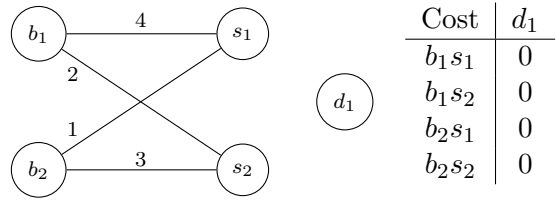

\subsection{Missing Proofs in \Cref{sec:eq_existence_sec}}
We prove \Cref{lem:exist_courier_plan_serve} and \Cref{lem:exist_courier_plan_serve_equal_size} together. 
\existscourierPlan*
\existscourierPlanEqualSize*
\begin{proof}
    Given $\Omega$, define a two-sided market $M=(\Omega, D)$ where $\Omega$ are consumers and $D$ the items. In $M$, a consumer $o=(b,s)$ values item $d\in D$ at $v_{o}(d)=H-c_d(o)=H-c_d(b,s)$, where $H=\sum_{bs}v_b(s)+\sum_{bs}w_{bs}+\sum_{bsd}c_d(b,s)$. We build a one-to-one correspondence between courier plans that serves $\Omega$ to walrasian equilibria in $M$.

    Given $(\mathbf{w},\mathbf{y})$, let $\mathbf{y}'$ be the allocation of items $D$ to consumers $\Omega$ that follow $\mathbf{y}$ (i.e., $\forall o\in\Omega, d\in D, y'_{od}=y_{od}$). Let $\mathbf{u}'$ be the vector of courier utility (i.e., for $o_\mathbf{y}(d)=(b,s), u'_d=u_d(o_\mathbf{y}(d))=w_{bs}-c_d(o)$). We prove if $(\mathbf{w},\mathbf{y})$ serves $\Omega$, then $(\mathbf{y}',\mathbf{u}')$ is a walrasian equilibrium in $M$, where $\mathbf{y}'$ is the walrasian allocation and $\mathbf{u}'$ the walrasian price.

    First, an item $d$ in $M$ that is not allocated corresponds to a courier $d$ who do not deliver, and have $u'_d=0$. Second, each consumer $o$ in $M$ receives the favorite item. Consider a consumer $o=(b,s)$ receiving item $d\neq \emptyset$ with utility $H-c_d(b,s)-u'_d=H-w_{bs}$. If $o$ buys from any other item $d'$ which is not allocated $u'_{d'}=0$, she has utility $H-c_{d'}(b,s)-u'_{d'}=H-c_{d'}(bs)$. But being a courier plan, $(\mathbf{w},\mathbf{y})$ satisfies for courier $d': w_{bs}-c_{d'}(bs)\leq 0$. This means consumer $o$ has no incentive to deviate to some item $d'$ unallocated. Alternatively, if $o$ buys from any other item $d'$ who is allocated to a consumer $o'=(b',s')\neq (b,s)$, she has deviation utility $H-c_{d'}(b,s)-u'_{d'}=H-c_{d'}(b,s)-w_{b',s'}+c_{d'}(b',s')$. Being a courier plan, $(\mathbf{w},\mathbf{y})$ satisfies for courier $d': w_{b's'}-c_{d'}(b',s')\geq w_{bs}-c_{d'}(b,s)$, and  there is no incentive for consumer $o$ to deviate to buy item $d'$. Finally customer $o$ has weakly positive utility $H-w_{bs}>0$.

    We now look at the other direction. Given a walrasian equilibrium $(\mathbf{y}',\mathbf{u}')$ in $M$, we construct a courier plan that serves $\Omega$. As $|\Omega|\leq l$ and each customer $o$ has valuation $v_o(d)=H-c_d(o)>0$ towards any item $d$, by the first welfare theorem, all customers $o\in \Omega$ are allocated to some courier. To construct the courier plan $(\mathbf{w},\mathbf{y})$, for $o=(b,s)\notin \Omega$, set $w_{b,s}=0, y_{od}=0, \forall d$. For $o=(b,s)\in\Omega$ where in the walrasian equilibrium $o$ buys an item $d$, set $\forall d', y_{od'}=y'_{od'}$ and $w_{b,s}=u'_d+c_d(b,s)$. This construction guarantees that $(\mathbf{w},\mathbf{y})$ serves $\Omega$ if it is a courier plan. We now prove that if $(\mathbf{y}',\mathbf{u}')$ is a walrasian equilibrium in $M$, then $(\mathbf{w},\mathbf{y})$ is a courier plan.
    
    For any courier $d$ that delivers $o_\mathbf{y}(d)=(b,s)\neq \emptyset$, her utility is $w_{b,s}-c_d(b,s)=u'_d\geq 0$ because walrasian equilibrium $(\mathbf{y}',\mathbf{u}')$ satisfies all prices are weakly larger than zero. Deviating to a store $o'\notin \Omega$ yields weakly less than zero utility. Deviating to a store $o'=(b',s')\in\Omega$, yields utility  $w_{b',s'}-c_d(b',s')$. But since every customer $o'\in \Omega$ are allocated an item $d'$ in the Walrasian equilibrium, $w_{b',s'}-c_d(b',s')=u'_{d'}+c_{d'}(b',s')-c_d(b',s')$. Since walrasian equilibrium $(\mathbf{y}',\mathbf{u}')$ guarantees customer $o'$ incentives, $H-c_{d'}(b',s')-u'_{d'}\geq H-c_{d}(b',s')-u'_d \Leftrightarrow u'_d\geq u'_{d'}+c_{d'}(b',s')-c_{d}(b',s')$. This means courier $d$ does not want to deviate to the store $o'$. For $d, o_\mathbf{y}(d)\in \BR_d(\mathbf{w})$.

    For any courier $d$ that does not deliver in $\mathbf{y}$, delivering an order $o'\notin \Omega$ yields utility weakly less than zero. Delivering an order $o'=(b',s')\in \Omega$, which is delivered by another courier $d'$ yields utility $w_{b',s'}-c_d(b',s')=u'_{d'}+c_{d'}(b',s')-c_d(b',s')$. But examining the incentive of customer $(b',s')$ in the walrasian equilibrium, $H-c_{d'}(b',s')-u'_{d'}\geq H - c_d(b',s')-0\Leftrightarrow u'_{d'}+c_{d'}(b',s')-c_d(b',s')\leq 0$. So $d$ does not deviate to $o'$. For $d, o_\mathbf{y}(d)\in \BR_d(\mathbf{w})$.

    We have established an one-to-one relationship between courier plans that serves $\Omega$ and walrasian equilibria in $M$. Since a walrasian equilibrium always exists for unit-demand supply market, there is always a courier plan that serves $\Omega$. By the first welfare theorem, walrasian equilibrium in $M$ achieves the optimal welfare, so every $o\in \Omega$ must be allocated in $\mathbf{y}$. Then in a courier plan, $\mathbf{y}$ must correspond to a minimum cost matching that covers all $\Omega$. This is because (i) edge weight in $M$ is defined as $H$ minus cost, (ii) $\mathbf{y}$ serves $\Omega$.
    
    Use $\mathbf{y}^{'*}$ to denote the max weight matching of the market $M$ with $SW(M)$ being its weight. Use $\mathbf{y}^{'*}_{-d}$ to denote the max weight matching of the market $M$ without item $d$, with $SW(M\backslash d)$ being its weight. If a courier in has $o_\mathbf{y}(d)=\emptyset$, then $u_d(o_\mathbf{y}(d))=0$. If $d$ is matched to $(b,s)\in\Omega$, its utility satisfies
    \begin{align*}
        u_d(o_\mathbf{y}(d))=w_{bs}-c_d(b,s)=u'_d\leq SW(M)-SW(M\backslash d)
    \end{align*}
    The inequality is the highest achievable walrasian equilibrium price, proved in \cite{gul1999walrasian}. 
    
    We now upper bound the utility of couriers in the courier plan that serves $\Omega$ when $l>|\Omega|$. We can define the highest possible courier compensation $\mathbf{w}$ with this highest walrasian equilibrium price $\mathbf{u}'$. 
    When $l>|\Omega|$, it satisfies that the size of $\mathbf{y}^{'*}$ is equal to the size of $\mathbf{y}^{'*}_{-d}$, and all customers in $\Omega$ in market $M$ are matched for $\mathbf{y}^{'*}_{-d}$.
    \begin{align*}
         SW(M)-SW(M\backslash d) &= \sum_{(o,d)\in \mathbf{y}^{'*}}[H-c_d(o)] - \sum_{(o,d)\in \mathbf{y}^{'*}_{-d}}[H-c_d(o)]\\
         &=\sum_{(o,d)\in \mathbf{y}^{'*}_{-d}}c_d(o) - \sum_{(o,d)\in \mathbf{y}^{'*}}c_d(o)\\
         &= C_{\Omega}(G_D\backslash d)-C_{\Omega}(G_D)
    \end{align*}
    where $C_{\Omega}(G_D)$ is the cost of the minimum cost matching that covers $\Omega$ in $G_D$, and $C_{\Omega}(G_D\backslash d)$ is the cost of the minimum-cost matching that covers $\Omega$ without courier $d$. 
    
    When $l=|\Omega|$, there is generally no upper bound to the courier utility. So we supply a courier plan that guarantees courier utility larger than $\max_{bs}v_b(s)$. The proof for constructing a courier plan $(\mathbf{y}, \mathbf{w})$ out of a walrasian equilibrium $(\mathbf{y}',\mathbf{u}')$ in $M$ does not depend on $H>w_{b,s}$. So we can instead set a market $M$ with $H=\sum_{bs}v_b(s)+\sum_{bsd}c_d(b,s)$. Consider the maximum walrasian equilibrium price given by $u_d'=SW(M)-SW(M\backslash d)$. The size of $\mathbf{y}^{'*}$ is larger than the size of $\mathbf{y}^{'*}_{-d}$ by 1. So $\mathbf{y}^{'*}_{-d}$ is the maximum weight matching that covers all but one customer in $\Omega$ in $M$. 
    \begin{align*}
         SW(M)-SW(M\backslash d) &= \sum_{(o,d)\in \mathbf{y}^{'*}}[H-c_d(o)] - \sum_{(o,d)\in \mathbf{y}^{'*}_{-d}}[H-c_d(o)]\\
         &=H+\sum_{(o,d)\in \mathbf{y}^{'*}_{-d}}c_d(o) - \sum_{(o,d)\in \mathbf{y}^{'*}}c_d(o)
         \geq \max_{b,s}v_b(s)
    \end{align*}
    which gives one possible courier compensation. 
\end{proof}

\WithoutTipEqExistence*
\begin{proof}
    Consider a Walrasian equilibrium $(\mathbf{p},\mathbf{z})$ in the two-sided market $(B,S)$. $\mathbf{z}$ defines a subset of orders $\Omega =\{(b,s) | z_{bs}=1\}$ satisfying $|\Omega|\leq l$. By \Cref{lem:exist_courier_plan_serve} and \Cref{lem:exist_courier_plan_serve_equal_size}, there is always a courier plan $(\mathbf{w},\mathbf{y})$ that serves $\Omega$ and satisfies $\forall (b,s)\notin \Omega, \; w_{bs}=0$. Define a feasible allocation $\mathbf{x}$, where for an order $(b,s)=o$, let $x_{bsd}= z_{bs}y_{od}$. $\mathbf{x}$ does not change what buyers buy in $(\mathbf{p},\mathbf{z})$ and does not change what couriers deliver in $(\mathbf{w},\mathbf{y})$. Then $(\mathbf{p},\mathbf{w},\mathbf{x})$ is a without-tip equilibrium. This is because the Walrasian equilirbium guarantees buyers incentives, the courier plan guarantees couriers incentives, and unsold store prices and undelivered orders delivery compensation are zero. 
\end{proof}

\WithTipEqExistence*
\begin{proof}
    \Cref{lem:pareto_dominant} shows that any allocation in a without-tip equilibrium is in a with-tip equilibrium of zero tips. When $l\geq \min\{m,n\}$, \Cref{thm:without_tip_eq_existence} says without-tip equilibria exist. So a with-tip equilibria also exists. When $l<\min\{m,n\}$, consider a subset of orders $\Omega$ where each buyer and store appears at most once and $|\Omega|=l$. \Cref{lem:exist_courier_plan_serve_equal_size} says there exists a courier plan $(\bar{\mathbf{w}},\mathbf{y})$ that serves $\Omega$ where $\bar{w}_{bs}=0$ for $(b,s)\notin \Omega$, and satisfies all courier utility being large $u_d=\bar{u}_d>\max_{bs}\{v_b(s)\}$. Define an allocation $\mathbf{x}$
    \begin{equation*}
        x_{bsd}=
        \begin{cases}
            1 & \text{ if } (b,s)=o\in\Omega \text{ and } y_{od}=1\\
            0 & \text{ otherwise}
        \end{cases}
    \end{equation*} 
    Define a purchase price
    \begin{equation*}
        p_{s}=
        \begin{cases}
            v_b(s) & \text{ if } \exists (b,s)\in\Omega\\
            0 & \text{ otherwise}
        \end{cases}
    \end{equation*} 
    Set tips to be zero $\mathbf{t}=0$. Then $(\mathbf{p},\bar{\mathbf{w}},\mathbf{t},\mathbf{x})$ is a with-tip equilibrium. To see this, first it satisfies that unsold store has purchase price zero, and undelivered orders have compensation and tips zero. Couriers incentives are satisfied from the definition of courier plan. For any buyer $b$ that buys from store $s_\mathbf{x}(b)$, the minimum tip for store $s'\neq s_\mathbf{x}(b)$ is larger than its valuation, i.e., $\underline{t}_{bs'}>v_b(s')$. This is because to have any courier $d$ delivers from $s'$ to $b$, the tip must satisfy $t+w_{bs'}=t\geq \bar{u}_d>\max_{bs}\{v_b(s)\}$. However, buyers $b$ only needs to pay zero tips for $s_\mathbf{x}(b)$: $\underline{t}_{bs_\mathbf{x}(b)}=t_{bs_\mathbf{x}(b)}=0$. Buyers incentives are also satisfied.
\end{proof}

\subsection{Missing Proofs in \Cref{sec:role_tips}}
\paretoDominant*
\begin{proof}
    As $\mathbf{t}=0$ and $(\mathbf{p},\mathbf{w},\mathbf{x})$ is a without-tip equilibrium, it satisfies that stores not bought from have zero purchase price, and orders not delivered have zero delivery compensation and zero tips $t_{bs}$=0. For couriers, as tips are zero, the set of orders that maximize utility with tips is the same as that without tips.
    
    Consider a buyer $b$ buying from a store $s\neq \emptyset$ in $\mathbf{x}$. As $\mathbf{t}_{-b}=0$ does not change couriers incentives, the minimum tip to have some courier delivers from store $s$ to her is zero $\underline{t}_{bs}=0$. So buying from $s$ remains the best option for $b$, i.e., $\forall s',\; v_b(s)-p_s-\underline{t}_{bs}=v_b(s)-p_s\geq v_b(s')-p_{s'}\geq v_b(s')-p_{s'}-\underline{t}_{bs'}$ where the first inequality is because $s\in \BR_d(\mathbf{p})$. A buyer $b$ not buying in $\mathbf{x}$ still does not buy now as purchase prices do not change.
\end{proof}

\ZeroTipEquivalence*
\begin{proof}
    For a store $s$ that is allocated in $\mathbf{x}$ to a buyer $b$, set $p'_{s}=p_s+t_{bs}$. For a store $s$ that is not allocated, set $p'_{s}=0$. For an order $(b,s)$ delivered by a courier $d$ in $\mathbf{x}$, set $w'_{bs}=w_{bs}+t_{bs}$. For an order $(b,s)$ not delivered by any courier in $\mathbf{x}$, set $w'_{bs}=0$. From the couriers' perspective, the delivery compensation plus tips associated with each order is the same: $w_{bs}+t_{bs}=w'_{bs}$, so the set of orders that maximizes each courier utility remains unchanged. We only need to prove that $s_\mathbf{x}(b)$ is still in the set of stores that maximizes each buyer utility in $(\mathbf{p}',\mathbf{w}',\mathbf{t}',\mathbf{x})$. 
    
    Consider a buyer $b$ buying from a store $s_\mathbf{x}(b)$. Let $\ut_{bs}(d)$ and $\ut'_{bs}(d)$ be the minimum tip required for a courier $d$ to deliver from store $s\neq s_\mathbf{x}(b)$ to $b$, under $(\mathbf{p},\mathbf{w},\mathbf{t},\mathbf{x})$ and $(\mathbf{p}',\mathbf{w}',\mathbf{t}',\mathbf{x})$ respectively. We first show that $\ut_{bs}(d)=\ut'_{bs}(d)$. According to \Cref{eq:min_tip_for_courier_d}, their values are given by 
    \begin{align*}
        \ut_{bs}(d) &= \max\{0,c_d(b,s)-w_{bs},\max_{(b',s')\neq (b,s)}\{w_{b's'}+t_{b's'}-c_d(b',s')-w_{bs}+c_d(b,s)\}\}\\
        &= \max\{0,c_d(b,s),\max_{(b',s')\neq (b,s)}\{w_{b's'}+t_{b's'}-c_d(b',s')+c_d(b,s)\}\}\\
        \ut'_{bs}(d) &= \max\{0,c_d(b,s)-w'_{bs},\max_{(b',s')\neq (b,s)}\{w'_{b's'}+t'_{b's'}-c_d(b',s')-w'_{bs}+c_d(b,s)\}\}\\
        &= \max\{0,c_d(b,s),\max_{(b',s')\neq (b,s)}\{w'_{b's'}-c_d(b',s')+c_d(b,s)\}\}
    \end{align*}
    Here we have used the condition that $w_{bs}=w'_{bs}=0$ for $s\neq s_\mathbf{x}(b)$ in equilibrium, and $\mathbf{t}'=0$. For an order $(b',s')\neq(b,s)$, by construction it always satisfies that $w'_{b's'}=w_{b's'}+t_{b's'}$. So $\ut_{bs}(d)=\ut'_{bs}(d)$ and the minimum tip to get a delivery from store $s$ is the same $$\ut_{bs}=\min_d \ut_{bs}(d)= \min_d \ut'_{bs}(d)=\ut'_{bs}$$

    Consider a buyer $b$ which buys from store $s=s_\mathbf{x}(b)$ in $\mathbf{x}$. Now we are ready to prove that $s$ is still in the set of stores that maximizes buyer $b$'s utility in $(\mathbf{p}',\mathbf{w}',\mathbf{t}',\mathbf{x})$.
    
    When $s\neq \emptyset$, let $u_b(s')=v_b(s')-p_s-\ut_{bs'}$ and $u'_b(s')=v_b(s')-p'_{s'}-\ut'_{bs'}$ denote the utility of $b$ when buying from a store $s'$ in $(\mathbf{p},\mathbf{w},\mathbf{t},\mathbf{x})$ and $(\mathbf{p}',\mathbf{w}',\mathbf{t}',\mathbf{x})$ respectively. 
     As $(\mathbf{p},\mathbf{w},\mathbf{t},\mathbf{x})$ is a with-tip equilibrium, buyer $b$ pays the minimum tip $t_{bs}=\ut_{bs}$. It satisfies for buyer $b$ 
    and any store $s'$ that
    \begin{align*}
        u_b(s)=v_b(s) - p_s - t_{bs}=v_b(s) - p_s - \ut_{bs}\geq v_b(s') - p_{s'} - \ut_{bs'}=u_b(s')
    \end{align*}

    When $s=\emptyset$, its utility satisfies $\forall s, 0 \geq u_{b}(s)=v_b(s)-p_s-\ut_{bs}$. Now in $(\mathbf{p}',\mathbf{w}',\mathbf{t}',\mathbf{x})$, buyer $b$ when buying from a store $s$ has utility
    \begin{align*}
        u'_b(s) &= v_b(s)-p'_s-\ut'_{bs}=v_b(s)-p'_s-\ut_{bs}\\
        &\leq v_b(s)-p_s-\ut_{bs}\leq u_b(s)\leq 0
    \end{align*} 
\end{proof}

\subsection{Pricing with tips when equilibrium do not clear the market}\label{app:tips_eq_not_clear_market}
In this section, we demonstrate the benefit of tips when equilibrium are not required to clear the market. That is, unsold stores can have positive prices and undelivered orders can have positive compensation and tips. Take the market in \Cref{example:without_tip_bad}, the best without-tip equilibrium has welfare 0, when no transactions take place. In the contrary, the best with-tip equilibrium achieves the optimal welfare of 3. 
\begin{proposition}
    When equilibrium do not need to clear the market, there always exists a with-tip equilibrium whose welfare is at least $\mathit{OPT}/\min\{m,n,l\}$.
\end{proposition}
\begin{proof}
    Let $\mathbf{x}^\star$ be an allocation that achieves the optimal welfare $\mathit{OPT}$. In $\mathbf{x}^\star$ there are at most $\min\{m,n,l\}$ transacting buyer--store--courier triples, or at most $\min\{m,n,l\}$ trades. So there the trade that generates the largest welfare is guaranteed to generate welfare at least $\frac{1}{\min\{m,n,l\}}\mathit{OPT}$. Denote this trade by $(b^\star, s^\star, d^\star)$. We show the allocation where only $(b^\star, s^\star, d^\star)$ trades is always in a with-tip equilibrium with zero tips $\mathbf{t}=0$.
    
    Given the order $(b^\star, s^\star)$, $d^\star$ is the courier with the lowest cost for delivering $(b^\star, s^\star)$. Hence, set courier compensation $\mathbf{w}$ to be  $w_{bs}=0, \forall (b,s)\neq (b^\star,s^\star)$ and $w_{b^\star,s^\star}=c_{d^\star}(b^\star, s^\star)$. This way, no other couriers have incentives to serve $(b^\star,s^\star)$.

    For purchase prices, set $p_s=\max_{b,s}v_b(s)$ for all $s\neq s^\star$ and $p_{s^\star}=v_{b^\star} (s^\star)$. Buyer $b^\star$ do not need to pay any tips for the delivery $\underline{t}_{b^\star s^\star}=0$.  This way $b^\star$ is indifferent to buy from $s^\star$, while all other buyers $b$ have to pay $p_{s^\star}+\underline{t}_{b s^\star}$ to buy from $s$. Furthermore, for any $b\neq b^\star$ the minimum tip for some couriers to deliver $s^\star$ satisfies $\underline{t}_{b s^\star}+w_{bs^\star}=\underline{t}_{b s^\star}\geq \min_d{c_d(b,s^\star)}$. So for any other buyer $b$, purchasing from $s^\star$ results in utility $v_b(s^\star)-\min_d{c^d_{b s^\star}}-v_b^\star(s^\star)$. But since $(b^\star, v^\star, s^\star)$ is the trade with the largest welfare, $v_{b^\star} (s^\star)\geq v_{b^\star}(s^\star)-c_{d^\star}(b^\star,s^\star)\geq v_{b}(s^\star)-c_d(b,s^\star) \; \forall b,d$. So any other buyer $b$ will have a weakly smaller than zero of utility of purchasing and paying tips for $s^\star$. So buyers incentives are satisfied.
\end{proof}

\section{Missing Proofs and Lemmas in \Cref{sec:general_markets}}\label{app:general_markets}

\begin{restatable}{proposition}{NegativeOptWelfare}
\label{prop:negative_opt_welfare}
    There exists a market where the only with-tip equilibria has welfare strictly less than the optimal welfare.
\end{restatable}
\begin{proof}
    The market in Figure~\ref{fig:market_clearing} is an example where all equilibria have low welfare compared with the optimal welfare. Let $\kappa>2$ be a constant.  Both buyers value both stores at $1$, and the two couriers have costs $c_{d_1}(b_1s_1)=0,c_{d_1}(b_1s_2)=c_{d_1}(b_2s_1)=c_{d_2}(b_1s_1)=c_{d_2}(b_2s_2)=\kappa,c_{d_1}(b_2s_2)=c_{d_2}(b_2s_1)=0.5,c_{d_2}(b_1s_2)=0.49$. The optimal welfare in this market is $1$, realized by $b_1$ buys from $s_1$ delivered by $d_1$. We show the optimal equilibrium welfare is $2-\kappa$ and can be very inefficient when $\kappa$ is large.

    Consider allocations where both buyers buy. These allocations have a maximum welfare of $2-\kappa<0$. One such equilibrium is specified by $x_{b_2s_2d_2}=x_{b_1s_1d_1}=1,p_{s_1}=p_{s_2}=1,w_{b_1s_2}=w_{b_2s_1}=0,w_{b_1s_1}=w_{b_2s_2}=\kappa, \mathbf{t}=0$.

    We now show there is no equilibrium where only one buyer buys. If such an equilibrium exists, the store not bought from has zero purchase price. And the courier with the lower cost between $d_1$ and $d_2$ delivers. Otherwise, the idle courier with a lower cost will want to deliver as well.
    \begin{itemize}
        \item If only $b_1$ buys from $s_1$. $d_1$ delivers, $d_2$ has zero utility. Buyer $b_1$ can buy from $s_2$ with the minimum tip that makes $d_2$ indifferent to deliver $\underline{t}_{b_1s_2}=0.49$, resulting in a buyer utility of $0.51$. To make $b_1$ buy from $s_1$, $p_1\leq 0.49$. But then $b_2$ can buy $s_1$ with tip $0.5$, resulting in a positive utility $1-0.5-0.49>0$.
        \item If only $b_1$ buys from $s_2$. $d_2$ delivers, $d_1$ has zero utility. $b_1$ can buy $s_1$ with zero tip and utility $1$, so $p_2=0$. But then $b_2$ can tip $d_1$ with $\underline{t}_{b_2s_2}=0.5$ to buy from $s_2$ with positive utility $1-0.5>0$.
    \end{itemize}
    With the same logic, one can show the other two allocations ``Only $b_2$ buys from $s_1$'' and ``Only $b_2$ buys from $s_2$'' are not in any equilibrium.
\end{proof}
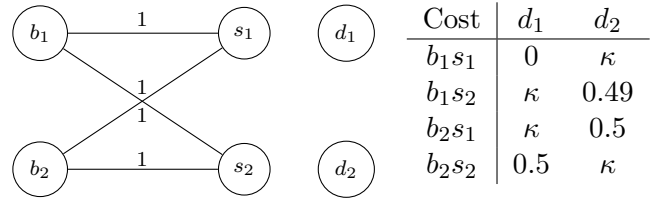
\begin{figure}[t]
    \centering
    \begin{tblr}{}
    \centering
\begin{tikzpicture}[baseline=(current bounding box.center),scale=0.9]
    \node[draw, shape=circle, minimum size=0.1cm] (1) at (-1, 2) {\scriptsize $b_1$};
    \node[draw, shape=circle, minimum size=0.1cm] (2) at (-1, 0) {\scriptsize $b_2$};
    \node[draw, shape=circle, minimum size=0.1cm] (3) at (2, 2) {\scriptsize $s_1$};
    \node[draw, shape=circle, minimum size=0.1cm] (4) at (2, 0) {\scriptsize $s_2$};
    \node[draw, shape=circle, minimum size=0.1cm] (5) at (3.5, 2) {\scriptsize $d_1$};
    \node[draw, shape=circle, minimum size=0.1cm] (6) at (3.5, 0) {\scriptsize $d_2$};
    \draw (1) to (3);
    \draw (1) to (4);
    \draw (2) to (3);
    \draw (2) to (4);
    \node[] at (0.5, 2.2){\scriptsize 1};
    \node[] at (0.5, 1.2){\scriptsize 1};
    \node[] at (0.5, 0.8){\scriptsize 1};
    \node[] at (0.5, 0.15){\scriptsize 1};
\end{tikzpicture}
& 
\begin{tabular}{c|cc}
Cost & $d_1$  & $d_2$\\
\hline
$b_1 s_1$ & $0$  & $\kappa$\\
$b_1 s_2$ & $\kappa$  & $0.49$\\
$b_2 s_1$ & $\kappa$  & $0.5$\\
$b_2 s_2$ & $0.5$  & $\kappa$
\end{tabular}
\end{tblr}
    \caption{A market where the only equilibrium has welfare strictly less than the optimal welfare. Both buyers value both stores at $1$. The two courier have different cost of delivery.}
    \label{fig:market_clearing}
\end{figure}

\OPTEqHard*
\begin{proof}
    We prove the three statements simultaneously by reducing from the 3-dimensional matching (3DM), one of the first 21 NP-complete problems proved \citep{karp1975computational}. 3DM asks the following question: Given a 3-uniform hypergraph $G=(B,S,D)$ with hyperedges $T\subset B \times S \times D$ where $|B|=|S|=|D|=l$, is there a perfect matching? 
    
    We construct a following market $M$ with $m$ buyers, $n$ stores and $l$ couriers where $m=n=l$. Buyers' valuations are always 1. Couriers' costs are set as 
    \begin{equation*}
    c_d(b,s) =
    \begin{cases}
      0 & \text{if $ (b,s,d)\in T$,}\\
      1 & \text{otherwise.}
    \end{cases}       
    \end{equation*}
  
    If $G$ has a perfect matching of size $l$, this perfect matching defines a subset of orders $$\Omega=\{(b,s) |\; \exists d \text{ such that } (b,s,d) \text{ is in the perfect matching}\}.$$ It holds that $|\Omega|=l$. By \Cref{lem:exist_courier_plan_serve_equal_size}, there exists a courier plan $(\bar{\mathbf{w}},\mathbf{y})$ that serves $\Omega$ where all orders $(b,s)\notin \Omega$ have delivery compensation $\bar{w}_{bs}=0$. We  define a purchase price $\mathbf{p}=1$, a tip $\mathbf{t}=0$, an allocation $\x$ where $x_{bsd}=1$ iff and only if $(b,s)=o\in\Omega \text{ and } y_{od}=1$.
    
    Then $(\mathbf{p},\bar{\mathbf{w}},\mathbf{t},\mathbf{x})$ is an equilibrium. This is because all buyers buy with minimum tip 0 and utility 0. If a buyer $b$ buy from a store other than $s_\mathbf{x}(b)$, she needs to pay nonzero tips and a purchase price of 1, resulting in a deviation utility weakly less than 0. The construction of courier plan guarantees courier incentives. By our construction, this equilibrium has welfare $l$. Similarly $(\mathbf{p},\bar{\mathbf{w}},\mathbf{x})$ is a without-tip equilibrium with welfare $l$. And naturally, the optimal welfare is $l$ as well.

   For the other direction, If $G$ does not have a perfect matching of size $l$, then by the construction of the market $M$, there does not exists any allocation $\mathbf{x}$ with welfare $l$. So the optimal equilibrium welfare is smaller than $l$. We have shown that $G$ has a perfect matching if and only if there exists an equilibrium with welfare $l$. Similarly, if $G$ does not have a perfect matching, then there doesn't exist an allocation with welfare $l$. So $G$ has a perfect matching if and only if optimal welfare is equal to $l$.

   Since valuations and costs are integral, as deciding whether the optimal equilibrium welfare (optimal welfare) equals $l$ is NP-hard, it follows from standard reduction that deciding whether there exists an equilibrium with welfare at least $k$ is NP-hard as well.
\end{proof}

\subsection{Missing Proofs and Lemmas in \Cref{sec:char_eq_alloc}}\label{app:char_eq_alloc}
\highestcourierPriceEq*
\begin{proof}
    Let $\Omega_\mathbf{x}=\{(b,s)|\sum_d x_{bsd}=1\}$ and $y$ a courier allocation defined by 
    $$ y_{od} =\begin{cases}
    1 & o=(b,s)\in \Omega \text{ and } x_{bsd}=1,\\
    0 & \text{otherwise.}
    \end{cases} $$
    
    For $(\p,\w,\bt,\x)$ to be an equilibrium, $(\w,\mathbf{y})$ must be a courier plan that serves $\Omega$. \Cref{lem:exist_courier_plan_serve} and \Cref{lem:exist_courier_plan_serve_equal_size} state the existence of a courier plan $(\bar{\w},\mathbf{y}')$. These two lemmas also require that both $\mathbf{y}$ and $\mathbf{y}'$ are a minimum-cost matching that covers $\Omega$ in the bipartite graph $G_D=(O,D)$. Then $(\bar{\w}, \mathbf{y})$ is a courier plan that serves $\Omega$.\footnote{The proof for \Cref{lem:exist_courier_plan_serve} and \Cref{lem:exist_courier_plan_serve_equal_size} build a one-to-one correspondence between courier plans and Walrasian equilibrium in a two-sided market $(\Omega,D)$. $\mathbf{y}$ and $\mathbf{y}'$ corresponds to the Walrasian allocation, and $\w,\bar{\w}$ corresponds to the Walrasian price. By the second welfare theorem, $(\mathbf{y},\bar{\w})$ corresponds to a Walrasian equilibrium, so is a courier plan as well.} This shows that courier incentives are satisfied.  

    We now check buyer incentives. As $(\bar{\w}, \mathbf{y})$ is a courier plan that serves $\Omega$ and $\bar{\w}$ is the highest courier compensation, the couriers achieve the highest possible utility $\bar{u}_d$, defined in \Cref{lem:exist_courier_plan_serve} and \Cref{lem:exist_courier_plan_serve_equal_size}. For $(b,s)\in\Omega_\x$, the courier plan $(\bar{\w}, \mathbf{y})$ serving $\Omega_\x$ already guarantees that some couriers are willing to deliver $(b,s)$ so $\ut_{bs}=0$. When $(b,s)\notin\Omega_\x$, to have courier $d$ deliver for store $s$, the tip must be large enough to match couriers current utility in the courier plan $$\ut_{bs}+w_{bs}-c_d(b,s)=\ut_{bs}+0-c_d(b,s) \geq \bar{u}_d$$
    Combining the two cases, the minimum tip buyer $b$ needs to offer for some courier to deliver from a store $s$ is \begin{equation}\label{eq:highest_minimum_tip}
        \ut_{bs} = \begin{cases}
            0 & \text{if } (b,s)\in\Omega_x, \\
            \min_d c_d(b,s)+\bar{u}_d & \text{otherwise.}
        \end{cases}     
    \end{equation}
    We now proceed to show that since a buyer $b$ does not offer a lower tip to buy from any other store than $s_\x(b)$ in $(\p,\w,\bt,\x)$, she does not offer a higher tip to buy from other stores in $(\p,\bar{\w},\bt,\x)$.
    There are two cases when the buyer $b$ deviates to buy from any other store $s\neq s_\x(b)$
    \begin{itemize}
        \item When $|\Omega_\x|=l$, it satisfies that for all couriers $\bar{u}_d>\max_{bs}\{v_b(s)\}$. So buyers utility buying from any other $s$ is negative because $\underline{t}_{bs}\geq\bar{u}_d$ is high.
        \item When $|\Omega_\x|<l$, $\bar{u}_d$ is the highest possible courier utility over all courier plans. So the minimum tip $\ut_{bs}$ required for buyer $b$ to get delivery from store $s\neq s_\x(b)$ is the highest possible tip. As buyers do not deviate in $(\p,\w,\bt,\x)$, she does not deviate in $(\p,\bar{\w},\bt,\x)$ by paying an even higher tip.
    \end{itemize}
    And finally, by the definition of $\Omega_\x$, it cannot be larger than $l$. This completes the proof.
    
\end{proof}

\testAlloc*
\begin{proof}
    The only if direction. When $\x$ is in some equilibrium, we have proved that it is always in an equilibrium of the form $(\p,\bar{\w},\bt,\x)$ where $\bt=0$. The minimum tip for buyer $b$ to buy from store $s$ it $\ut_{bs}$. From a buyer $b$'s perspective, her realized valuation for store $s$ is exactly $v^x_{bs}$, as she also has to pay tips to get the delivery. In equilibrium $(\p,\w,\bt,\x)$ all buyers buy from their favorite store, and stores unsold has price zero. These two conditions correspond to a Walrasian equilibrium in the market defined by $G_\x$. By the first welfare theorem, $\mathbf{z}$ must be a maximum weight matching in $G_\x$. 

    The if direction. $\mathbf{z}$ being a maximum weight matching defines a Walrasian equilibrium $(\mathbf{z},\p)$ in the two-sided market $G_x$, where $p_s=0$ for unallocated stores. Then $(\p,\bar{\w},\bt,\x)$ is an equilibrium, where $\bt=0$. This is because we have already shown that $\bar{\w}$, zero tips, and $\x$ satisfy courier incentives. Buyer incentives are satisfied by the Walrasian equilibrium in $G_x$. 
\end{proof}

\begin{restatable}{corollary}{allcouriersDeliver}
\label{cor:all_courier_deliver}
    If a feasible allocation $\x$ satisfies that $|\Omega_\x|=l$, then $\x$ is in some equilibrium. 
\end{restatable}
\begin{proof}
\Cref{lem:exist_courier_plan_serve_equal_size} shows with courier compensation $\w$, any courier utility is at least as large as the maximum buyer valuations $\bar{u}_d >\max_{bs}v_b(s)$. This means the minimum tip $\ut_{bs}$ in  \Cref{eq:highest_minimum_tip} for $(b,s)\notin\Omega_\x$ is larger than buyer valuation. Thus any edge not in $\mathbf{z}$ has negative weights and any edge in $\mathbf{z}$ has weakly positive weights. By \Cref{lem:test_alloc}, $\x$ is in some equilibrium.
\end{proof}

\PolyCheck*
\begin{proof}
    For any feasible $\mathbf{x}$, we can calculate the 
    weight of edges in graph $G_\x$ to invoke \Cref{lem:test_alloc}. Following their expressions in \Cref{lem:exist_courier_plan_serve} and \Cref{lem:exist_courier_plan_serve_equal_size}, calculating $\bar{u}_d$ for all couriers takes polynomial time, as each $\bar{u}_d$ requires two computations for minimum cost matching that covers $\Omega_\x$ or all but one element in $\Omega_\x$. Calculating the minimum tip $\ut_{bs}$ for all buyers $b$ and stores $s$, and finding the value of max weight matching in $G_\x$ takes polynomial time as well.
\end{proof}

\section{Missing Proofs in \Cref{sec:market_structures}}\label{app:ch3_market_structures}
\divisiblecourierCostOpt*
\begin{proof}
    We reduce the problem of finding the optimal welfare to a standard \emph{minimum cost flow} problem, solvable in polynomial time using existing algorithms. We demonstrate this for $c_d(b,s)=c(b,s)+c_d(s)$, while the similar case $c_d(b,s)=c(b,s)+c_d(b)$ is omitted. 
    
    Given a market, construct a flow network. Figure~\ref{fig:OPTDivisibleCost} illustrates such a network for a market with two buyers, two stores and three couriers. There is a source node $S_o$, a sink node $S_i$, a node for each buyer $b\in B$, order $o\in B\times S$, store $s\in S$, courier $d\in D$, and an additional dummy node for each store. There exists a directed edge from the source to each buyer node, from each buyer to each order that involves the buyer, from each order to the store that is involved in the order, from each store to the dummy node of the same store, from each dummy store to each courier, and from each courier to the sink. All edges have capacity one. An edge between a buyer $b$ and an order $(b,s)$ has cost $-v_b(s)$, an edge between an order $(b,s)$ and a store $s$ has cost $c(b,s)$, an edge between a dummy store $s'$ and a courier have $c_d(s)$ as cost. The net supply or demand are zero for all nodes except the source and the sink node.

    Each integer flow corresponds to a feasible allocation. To see this, an integer flow ensures unit-demand, unit-capacity, and unit-supply constraints, as each buyer node receives at most one unit of flow from the source, and each store node send out one unit of flow to its dummy node, and each courier node send out one unit of flow to the sink. The path each unit of flow goes through defines a trade $x_{bsd}=1$.
    
    To solve for the optimal welfare, set net supply of the source node equal to the net demand of the sink node to be $f=1,2,...,\min\{m,n,l\}$. Solve the minimum cost flow problem for each $f$ and take the flow with the minimum cost over all $f$. By the integrality theorem, any minimum cost network flow problem instance whose demands data are all integers has an optimal solution with integer flow on each edge. 
    Each flow corresponds to a feasible allocation $\mathbf{x}$, and the cost of a flow equals to the negation of the welfare of $\mathbf{x}$. Hence finding the minimum cost flow among all $f=1,2,...,\min\{m,n,l\}$ is finding the allocation with the optimal welfare.
\end{proof}

\DivisiblecourierCost*
\begin{proof}
    Let $\x$ be a feasible allocation that achieves the optimal welfare $W(\x)=\mathit{OPT}$. If $|\Omega_\x|=l$, \Cref{cor:all_courier_deliver} shows $\x$ is in some equilibrium. We focus on the case where $|\Omega_\x|<l$. 
    
    \textbf{We first present the proofs here for $c_d(b,s)=c(b,s)+c_d(b)$ and all stores are bought from $|\Omega_x|=n$ where $n$ is the number of stores in the market. We present the case for $|\Omega_x|<n$ and $c_d(b,s)=c(b,s)+c_d(s)$ later.}
    
    Let $\z$ be the buyer allocation induced by $
    \x$ in $G_\x$, and $\z'$ be any other feasible buyer allocation in $G_\x$. We will show that $\z$ is of weakly larger weight than $\z'$. Then \Cref{lem:test_alloc} shows $\x$ is in an equilibrium. In $G_\x$, the two matching $\z$ and $\z'$ define some alternating paths and cycles $$\pi=((b_0),s_1,b_1,s_2,b_2,\ldots, b_{t-1},s_t, b_t),$$ where for every $q\in \{1,\ldots, t\}$, $$z_{b_q,s_q}=1\mbox{ and }z_{b_{q-1},s_q}=0 \mbox{ and }z'_{b_q,s_q}=0\mbox{ and } z'_{b_{q-1},s_q}=1.$$ 
    
    An alternating path can either start from 1) $b_0$, a buyer allocated by $\z'$ but not by $\z$; 2) $s_1$, a store allocated by $\z$ but not by $\z'$. In an alternating path buyer $b_t$ is allocated by $\z$ but not by $\z'$. An alternating cycle is defined by $b_t=b_0$, and all buyers and stores on the cycle is allocated by both $\z$ and $\z'$. We use $\pi$ to denote both alternating paths and cycles. As $|\Omega_\x|=n$, all stores in $\pi$ are allocated in $\z$. Each alternating path and cycle captures buyers comparing the store they buy from in $\x$, against stores they can buy from by paying tips.
    Figure~\ref{fig:alternative_path_cycle} shows some examples of the alternating paths and cycles.

    \begin{figure}[t]
    \centering
    \begin{subfigure}{.3\textwidth}
    \centering
    \begin{tikzpicture}[scale=0.7]
        \foreach \i/\label in {1/$b_1$, 2.5/$b_2$, 4/$b_3$}
            \node[draw, shape=rectangle,scale=0.8] (\label) at (\i, 2) {\tiny\label};
            
        \foreach \i/\label in {1/$s_1$, 2.5/$s_2$, 4/$s_3$}
            \node[draw, shape=circle, scale=0.8] (\label) at (\i, 0) {\tiny\label};

        \foreach \x/\y in {$b_1$/$s_2$, $b_2$/$s_3$}
            \draw[line width=0.5pt, dashed] (\x) -- (\y);

        \foreach \x/\y in {$b_1$/$s_1$, $b_2$/$s_2$, $b_3$/$s_3$}
            \draw[line width=0.5pt, black, solid] (\x) -- (\y);    
    \end{tikzpicture}
    \end{subfigure}%
    \begin{subfigure}{.3\textwidth}
    \centering
    \begin{tikzpicture}[scale=0.7]
        \foreach \i/\label in {-0.5/$b_0$,1/$b_1$, 2.5/$b_2$, 4/$b_3$}
            \node[draw, shape=rectangle,scale=0.8] (\label) at (\i, 2) {\tiny\label};
            
        \foreach \i/\label in {1/$s_1$, 2.5/$s_2$, 4/$s_3$}
            \node[draw, shape=circle, scale=0.8] (\label) at (\i, 0) {\tiny\label};

        \foreach \x/\y in {$b_0$/$s_1$,$b_1$/$s_2$, $b_2$/$s_3$}
            \draw[line width=0.5pt, dashed] (\x) -- (\y);

        \foreach \x/\y in {$b_1$/$s_1$, $b_2$/$s_2$, $b_3$/$s_3$}
            \draw[line width=0.5pt, black, solid] (\x) -- (\y);    
    \end{tikzpicture}
    \end{subfigure}
    \begin{subfigure}{.3\textwidth}
    \centering
    \begin{tikzpicture}[scale=0.7]
        \foreach \i/\label in {1/$b_1$, 2.5/$b_2$, 4/$b_3$}
            \node[draw, shape=rectangle,scale=0.8] (\label) at (\i, 2) {\tiny\label};
            
        \foreach \i/\label in {1/$s_1$, 2.5/$s_2$, 4/$s_3$}
            \node[draw, shape=circle, scale=0.8] (\label) at (\i, 0) {\tiny\label};

        \foreach \x/\y in {$b_1$/$s_2$, $b_2$/$s_3$, $b_3$/$s_1$}
            \draw[line width=0.5pt, dashed] (\x) -- (\y);

        \foreach \x/\y in {$b_1$/$s_1$, $b_2$/$s_2$, $b_3$/$s_3$}
            \draw[line width=0.5pt, black, solid] (\x) -- (\y);    
    \end{tikzpicture}
    \end{subfigure}
    \caption{In $G_x$, $\z$ and $\z'$ define some alternating paths and cycles that do no intersect. Solid edges means $z_{bs}=1$, dotted edges means $z'_{bs}=1$. From left to right: alternating path with no unmatched buyer in $\z$, alternating path with $b_0$ unmatched by $\z$, alternating cycle.\label{fig:alternative_path_cycle}}
    \end{figure}
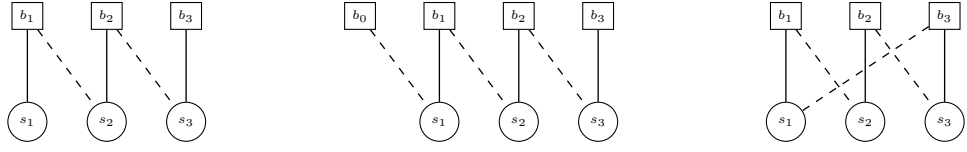

    Each $\pi$ defines a subgraph $G^\pi_\x$ that consists of buyers and stores in $\pi$. Since $\z$ and $\z'$ are matchings in $G_\x$, no two alternating paths or cycles intersect.
    We will prove $\z$ is a maximum weight matching when restricted to each such subgraph $G^\pi_\x$. This implies that $\z$ is a max weight matching in $G_x$. We define some notations that allow us to operate on each subgraph $G^\pi_\x$. Let $\z^\pi$ denote $\z$ restricted to $\pi$ and $\x^\pi$ denote $\x$ restricted to $\pi$: $$z^\pi_{bs}=\begin{cases}
        z_{bs} & \text{ if } b\in \pi, s\in \pi,\\
        0 & \text{otherwise.}
    \end{cases},\;
    x^\pi_{bsd}=\begin{cases}
        x_{bsd} & \text{ if } b\in \pi, s\in \pi,\\
        0 & \text{otherwise.}
    \end{cases}
    $$ 
    
    Similarly, let $\z^{'\pi}$ be $\z'$ restricted to $\pi$.
    Let $\z^{\backslash \pi}=\z-\z^\pi$ be the part of $\z$ not involving buyers and stores in $\pi$, and similarly $\x^{\backslash \pi}=\x-\x^\pi$ be the part of $\x$ not involving buyers and stores in $\pi$.
    
    \paragraph{Case I. Alternating path starting at $s_1$ or alternating cycle.} 
    All buyers matched in $\z^{'\pi}$ is matched in $\z^\pi$ as well.
    This allows us to define a feasible allocation $\mathbf{x'^\pi}$ on three-sided market for the buyer allocation $\z'^{\pi}$ in the following way: A buyer matched in $\z'^{\pi}$ receives delivery from the same courier as in $\x^\pi$
    \begin{equation*}
    x'^\pi_{bsd} = \begin{cases}
        1 & \text{if } z'^{\pi}_{bs}=1 \text{ and }  \sum_s x^\pi_{bsd}=1, \\
        0 & \text{otherwise.}
    \end{cases}     
    \end{equation*}
    As $\x'^\pi$ only allocates buyers stores and couriers allocated in $\x^\pi$, $\x'^\pi+\x^{\backslash \pi}$ is another matching. Since $\mathbf{x}$ is the allocation with the optimal welfare, we have $W(x)\geq W(\x'^\pi)+W(\x^{\backslash \pi})$.
    \begin{align*}
        \sum_{bs: x^{\pi}_{bs}=1}[v_b(s)-c(b,s)]-\sum_{bd: \sum_s x^{\pi}_{bsd}=1}c_d(b) 
        &= W(\x^\pi) =W(\x)-W(\x^{\backslash \pi})
        \geq W(\x'^\pi)\\
        &= \sum_{bs: x'^{\pi}_{bs}=1}v_b(s) - \sum_{bs: x'^{\pi}_{bs}=1}c(b,s) - \sum_{bd: \sum_s x'^{\pi}_{bsd}=1}c_d(b)\\
        &\geq \sum_{bs: x'^{\pi}_{bs}=1}[v_b(s)-\ut_{bs}] - \sum_{bd: \sum_s x^\pi_{bsd}=1}c_d(b).
    \end{align*}
    The last inequality comes from tip $\ut_{bs}\geq c(b,s)$ for $s\neq s_\x(b)$, and $\mathbf{x'^\pi}$ using the same couriers to deliver for buyers in $\mathbf{x^\pi}$. Simplifying the inequality we have $$ \sum_{bs: z^{\pi}_{bs}=1}v_b(s) \geq \sum_{bs: \z'^\pi_{bs}=1}[v_b(s)-\ut_{bs}]=\sum_{bs: z'^{\pi}_{bs}=1}v^\x_b(s),$$
    which means $\z$ is a max weight matching when restricted to $G^\pi_\x$.

    \paragraph{Case II. Alternating path starting at $b_0$.} The buyer $b_t$ at the end of the alternating path is matched in $\z$ but not $\z'$. 
    To denote the dependency on $\x$, let $\bar{u}_d(\x):=\bar{u}_d$ be the highest courier utility in \Cref{lem:exist_courier_plan_serve}. 
    For an alternating path starting at $b_0$, let $d_0$ be the courier who is willing to deliver order $(b_0,s_1)$ with the lowest tip $d_0=\argmin_{d}\{c_d(b_0,s_1)+\bar{u}_d(\x)\}$. As $|\Omega_\x|<l$ \Cref{lem:exist_courier_plan_serve} gives an expression for $d_0$'s highest utility $\bar{u}_{d_0}(\x)=C_{\Omega_\x}(G_D\backslash d_0)-C_{\Omega_\x}(G_D)$. 
    Here $C_{\Omega_\x}(G_D)$ is the minimum cost of delivering all orders in $\Omega_\x$, also the total courier cost for $\x$ because $\x$ is optimal.
    $C_{\Omega_\x}(G_D\backslash d_0)$ is the cost of the minimum cost matching that covers $\Omega_\x$ without courier $d_0$ in the bipartite graph $G_D=(O,D)$. Let $\mathbf{y}^{\backslash d_0}$ be the courier allocation in $G_D$ that covers $\Omega_\x$ with the minimum cost without courier $d_0$. The buyer-courier part of the cost for $\mathbf{y}^{\backslash d_0}$ is equal to total cost minus the buyer--store part of the cost $C_{\Omega_\x}(G_D\backslash d_0)-\sum_{bs:z_{bs}=1}c(b,s)$.
    
    Now define a feasible allocation $\x'$ for the buyer allocation $\z'^{\pi}+z^{\backslash \pi}$ where 
    $$x'_{bsd}=\begin{cases}
        1 & \text{ if } (b,s,d)=(b_0,s_1,d_0),\\
        1 & \text{ if } (b,s,d)\neq(b_0,s_1,d_0) \mbox{ and } z'^\pi_{bs}+z^{\backslash \pi}_{bs}=1 \mbox{ and } \sum_{s'}y^{\backslash d_0}_{o,d}=1\mbox{ for } o=(b,s'),\\
        0 & \text{otherwise.}
    \end{cases}$$
    
    This means, $\x'$ fulfills all orders in $z'^\pi$ and $z^{\backslash \pi}$, by having $d_0$ delivering for $b_0$, and have the buyers receive delivery from the same couriers that they receive deliver in $\mathbf{y}^{\backslash d_0}$. The buyer-courier part of the delivery cost for $\x'$ is weakly smaller than $c_{d_0}(b_0)+C_{\Omega_\x}(G_D\backslash d_0)-\sum_{bs:z_{bs}=1}c(b,s)$. This is because an order $(b_t,s_t)$ is matched in $\mathbf{y}^{\backslash d_0}$, but $b_t$ is no longer matched in $\x'$. The welfare of $\mathbf{x'}$ can be expressed as
    \begin{align*}
        W(\mathbf{x'})&\geq \sum_{bs: z'^\pi_{bs}=1} [v_b(s)-c(b,s)] +\sum_{bs: z^{\backslash \pi}_{bs}=1} [v_b(s)-c(b,s)] - c_{d_0}(b_0)- C_{\Omega_\x}(G_D\backslash d_0)+\sum_{bs:z_{bs}=1}c(b,s)
    \end{align*}
    We can also write out the welfare of $\mathbf{x^{\backslash \pi}}$ by expressing the couriers cost indirectly through the courier cost for $\mathbf{x^\pi}$.
    \begin{align*}
        W(\mathbf{x^{\backslash \pi}})=\sum_{bs: x^{\backslash \pi}_{bs}=1} v_b(s)-[C_{\Omega_\x}(G_D)-\sum_{bs: z^{\pi}_{bs}=1}c(b,s)-\sum_{bd: \sum_s x^{\pi}_{bsd}=1}c_d(b)].
    \end{align*}
    Putting the two together we have,
    \begin{align*}
        W(\mathbf{x'})-W(\mathbf{x^{\backslash \pi}})&\geq \sum_{bs: z'^\pi_{bs}=1} [v_b(s)-c(b,s)] - [c_{d_0}(b_0)+ C_{\Omega_\x}(G_D\backslash d_0)-C_{\Omega_\x}(G_D)]\\
        & + (\sum_{bs:z_{bs}=1}c(b,s)-\sum_{bs: z^{\pi}_{bs}=1}c(b,s)- \sum_{bs: z^{\backslash \pi}_{bs}=1} c(b,s))- \sum_{bd: \sum_s x^{\pi}_{bsd}=1}c_d(b)\\
        &\geq \sum_{bs: z'^\pi_{bs}=1} [v_b(s)-c(b,s)] - (c_{d_0}(b_0)+\bar{u}_{d_0}(\x))-\sum_{bd: \sum_s x^{\pi}_{bsd}=1}c_d(b).
    \end{align*}
    
    By $\mathbf{x}$ having larger welfare than $\mathbf{x'}$,
    \begin{align*}
        \sum_{bs: z^{\pi}_{bs}=1}[v_b(s)-c(b,s)]-\sum_{bd: \sum_s x^{\pi}_{bsd}=1}c_d(b) 
        &= W(\mathbf{x}^\pi) =W(\mathbf{x})-W(\mathbf{x}^{\backslash \pi})
        \geq W(\mathbf{x}')-W(\mathbf{x}^{\backslash \pi}).
    \end{align*}
    
    Combining the latter two inequalities and simplifying,
    \begin{align*}
        \sum_{bs: z^{\pi}_{bs}=1}v_b(s) &\geq \sum_{bs: z'^\pi_{bs}=1} [v_b(s)-c(b,s)] - (c_{d_0}(b_0)+\bar{u}_{d_0}(\x)) \\
        &\geq v_{b_0}(s_1)-\ut_{b_0s_1}+ \sum_{bs: z'^\pi_{bs}=1, b\neq b_0} [v_b(s)-\ut_{bs}] = \sum_{bs: z'^{\pi}_{bs}=1}v^\x_b(s).
    \end{align*}
 
    In the last inequality we have used that for an order $(b,s)$ matched in $z'^\pi_{bs}=1$ but $z^\pi_{bs}=0$, $\ut_{bs}=\min_d (c_d(b,s)+\bar{u}_d(\x))\geq \min_d c_d(b,s)\geq c(b,s)$.
  We have proved that $\z$ is a max weight matching when restricted to $G^\pi_\x$.

    \textbf{We now continue the proof for $c_d(b,s)=c(b,s)+c_d(b)$ when not all stores are bought from $|\Omega_\x|<n$.} The main difference from the previous case $|\Omega_\x|=n$ is that alternative paths can include stores not bought from in $\x$ now. But the analysis are similar.
    
    The two matching $\z$ and $\z'$ define some alternating paths and cycles. The proof for the case $|\Omega_\x|=n$ already works for alternative paths and cycles that do not contain unmatched stores in $\z$. So we can focus on alternating paths that involve stores that are not matched in $\z$. Stores unmatched in $\z$ cannot exist in any alternating cycles. An alternating path $p$ is denoted as 
    $$((b_0),s_1,b_1,s_2,b_2,\ldots, b_{t-1},s_t),$$ where for every $q\in \{1,\ldots, t-1\}$, $$z_{b_q,s_q}=1\mbox{ and }z'_{b_q,s_q}=0,$$ and for every $q\in \{1,\ldots, t\}$,  $$z_{b_{q-1},s_q}=0\mbox{ and } z'_{b_{q-1},s_q}=1.$$ Per discussion of the previous paragraph, an alternating path always end at some store $s_t$ that is not matched by $\z$. An alternating path can either start from 1) $b_0$, a buyer allocated by $\z'$ but not by $\z$; 2) or $s_1$, a store allocated by $\z$ but not by $\z'$. 

    Similar as the proof for $\Omega_n$, define $\z^\pi$ to be the buyer--store allocation of $\z$ when restricted to the alternating path $\pi$, and $\z^{\backslash \pi}=\z-\z^{\pi}$ the buyer--store allocation of $\z$ not involving buyers or stores in $\pi$. Let $\mathbf{x}^\pi$ and $\mathbf{x}^{\backslash \pi}$ be the part of $\mathbf{x}$ that does or does not involve $\pi$ respectively. Similarly, define $\mathbf{z}^{'\pi}$ to be $\z'$ restricted to the alternating path $\pi$. 

    \paragraph{Case I. alternating path starting at $s_1$.} This case is entirely the same as Case I. of $|\Omega_\x|=n$.
    
    \paragraph{Case II. alternating path starting at $b_0$.} If the alternative path starts from $b_0$, let $d_0$ be the courier who are willing to deliver the order $(b_0,s_1)$ with the lowest tip $d_0=\argmin_d{c_d(b_{0},s_1)+\bar{u}_d(\mathbf{x})}$. Use $C_{\Omega_\x}(G_D)$ and $C_{\Omega_\x}(G_D\backslash d_0)$ to denote the cost of minimum cost matching in $G_D$ that covers $\Omega_\x$ with and without $d_0$ respectively. As $\x$ is the welfare-optimal allocation, $C_{\Omega_\x}(G_D)$ is also the total courier cost for $\x$. Let $\mathbf{y}^{\backslash d_0}$ be the minimum cost matching in $G_D$ that covers $\Omega_\x$ without $d_0$. 
    
    Let $\mathbf{x'}$ be the minimum-cost allocation that satisfies $\sum_d x'_{bs}= z^{'\pi}_{bs}+z^{\backslash \pi}_{bs}$. To deliver the orders in $ \z^{'\pi}+\z^{\backslash \pi}$, one possible way is to have courier $d_0$ deliver $(b_{0},s_1)$ and all other buyers' orders being delivered by the same couriers as in $\mathbf{y}^{\backslash d_0}$, that is, a buyer $b$ receives delivery from a courier $d$ that satisfies $y_{od}^{\backslash d_0}=1$ for $o=(b,s)\in \Omega_x$.
    The courier--buyer part of the cost of this matching is $c_{d_0}(b_0)+[C_{\Omega_\x}(G_D\backslash d_0)-\sum_{bs:z_{bs}=1}c(b,s)]$. So total courier cost in $\mathbf{x'}$ is weakly smaller than the above stated way of matching couriers.
    \begin{align*}
        W(\mathbf{x'}) \geq \sum_{bs:z_{bs}^{'\pi}=1}[v_b(s)-c(b,s)]+\sum_{bs:z_{bs}^{\backslash \pi}=1}[v_b(s)-c(b,s)] - c_{d_0}(b_0) -C_{\Omega_\x}(G_D\backslash d_0)+\sum_{bs:z_{bs}=1}c(b,s)
    \end{align*} 
    We also write out the total welfare of $W(\mathbf{x}^{\backslash \pi})$ by expressing the courier cost indirectly through the courier cost of $\mathbf{x}^\pi$.
    \begin{align*}
        W(\mathbf{x}^{\backslash \pi}) = \sum_{bs:z_{bs}^{\backslash \pi}}v_{b}(s) - [C_{\Omega_\x}(G_D)-\sum_{bs:z^\pi_{bs}=1}c(b,s)-\sum_{bd:\sum_s x^p_{bsd}=1}c_d(b)]
    \end{align*}
    Putting the two inequalities together
    \begin{align*}
        W(\mathbf{x}') - W(\mathbf{x}^{\backslash \pi}) & \geq \sum_{bs:z_{bs}^{'\pi}=1}[v_b(s)-c(b,s)] - c_{d_0}(s_1) - [C_{\Omega_\x}(G_D\backslash d_0)-C_{\Omega_\x}(G_D)]-\sum_{bd:\sum_d x^\pi_{bsd}=1}c_d(b)
    \end{align*}
    By $\mathbf{x}$ having larger welfare than $\mathbf{x'}$
    \begin{align*}
        \sum_{bs:z^\pi_{bs}=1}[v_b(s)-c(b,s)]-\sum_{bd:\sum_s x^\pi_{bsd}=1}c_d(b)=W(\mathbf{x}^\pi)=W(\mathbf{x})- W(\mathbf{x}^{\backslash \pi}) \geq W(\mathbf{x}') - W(\mathbf{x}^{\backslash \pi})
    \end{align*}
    Combining the last two inequalities and simplifying
    \begin{align*}
        \sum_{bs:z^\pi_{bs}=1}v_b(s) &\geq \sum_{bs:z^\pi_{bs}=1}[v_b(s)-c(b,s)] \geq \sum_{bs:z_{bs}^{'\pi}=1}[v_b(s)-c(b,s)] - c_{d_0}(s_1)  - \bar{u}_{d_0}(\x)\\
        &\geq \sum_{bs:z_{bs}^{'\pi}=1}[v_b(s)-\ut_{bs}] = \sum_{bs:z_{bs}^{'\pi}=1} v^\x_b(s)
    \end{align*}
    In the last inequality, we used $\ut_{bs}\geq c(b,s)$ for $(b,s)\notin \Omega_\x$. So for any alternating path $p$ that starts from $b_0$, $\z^\pi$ have weakly larger welfare than any other matching $\z'^{\pi}$ on $p$.

    The proof for courier costs being divisible to buyer--store and store--courier part $c^d(b,s)=c_d(s)+c(b,s)$ is quite similar to the proof above for $c^d(b,s)=c_d(b)+c(b,s)$. One just goes through all four types and alternating paths and cycles, and swapping out the buyer-courier cost to store-courier cost in the proof. So we omit that proof. 
\end{proof}

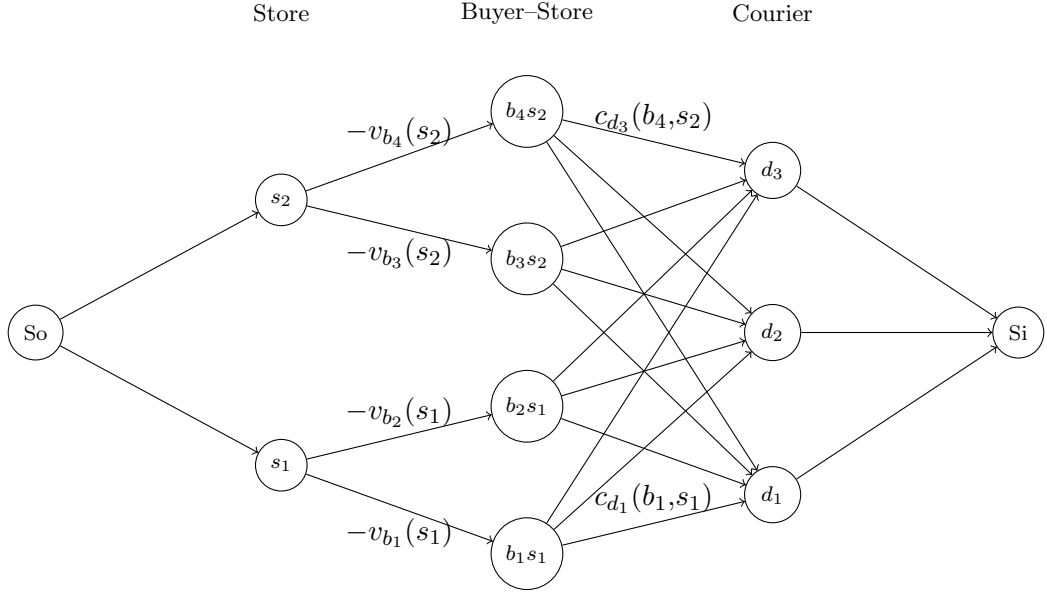
\begin{figure}[t] 
    \centering
    \begin{tikzpicture}[scale=1.3]
        \node[draw, shape=circle] (Source) at (0, 2.25) {\scriptsize So};
        \node[draw, shape=circle] (Sink) at (10, 2.25) {\scriptsize Si};
        \foreach \i/\label in {0.9/$s_1$, 3.6/$s_2$}
            \node[draw, shape=circle,minimum size=0.1cm] (\label) at (2.5, \i){\scriptsize\label};
        \foreach \i/\label in {0/$b_1 s_1$, 1.5/$b_2 s_1$, 3/$b_3 s_2$, 4.5/$b_4 s_2$}
            \node[draw, shape=circle,minimum size=0.1cm] (\label) at (5, \i){\scriptsize\label};
        \foreach \i/\label in {0.6/$d_1$, 2.25/$d_2$, 3.9/$d_3$}
            \node[draw, shape=circle,minimum size=0.1cm] (\label) at (7.5, \i){\scriptsize\label};

        \foreach \x/\y in {Source/$s_1$, Source/$s_2$}
            \draw[->] (\x) -- node[midway, right]{} (\y);
        \foreach \x/\y/\w in {$s_1$/$b_1 s_1$/$-v_{b_1}(s_1)$, $s_2$/$b_3 s_2$/$-v_{b_3}(s_2)$}
            \draw[->] (\x) -- node[below]{\w} (\y);
        \foreach \x/\y/\w in {$s_1$/$b_2 s_1$/$-v_{b_2}(s_1)$, $s_2$/$b_4 s_2$/$-v_{b_4}(s_2)$}
            \draw[->] (\x) -- node[above]{\w} (\y);

        \foreach \x/\y/\w in {$b_4 s_2$/$d_2$/$c_{d_2}^{b_4 s_2}$, $b_4 s_2$/$d_1$/$c_{d_1}^{b_4 s_2}$,$b_3 s_2$/$d_3$/$c_{d_3}^{b_3 s_2}$, $b_3 s_2$/$d_2$/$c_{d_2}^{b_3 s_2}$, $b_3 s_2$/$d_1$/$c_{d_1}^{b_3 s_2}$,
        $b_2 s_1$/$d_3$/$c_{d_3}^{b_2 s_1}$, $b_2 s_1$/$d_2$/$c_{d_2}^{b_2 s_1}$, $b_2 s_1$/$d_1$/$c_{d_1}^{b_2 s_1}$,$b_1 s_1$/$d_3$/$c_{d_3}^{b_1 s_1}$, $b_1 s_1$/$d_2$/$c_{d_2}^{b_1 s_1}$}
            \draw[->] (\x) -- node[]{}(\y);
        \foreach \x/\y/\w in {$b_4 s_2$/$d_3$/$c_{d_3}(b_4\text{,}s_2)$,$b_1 s_1$/$d_1$/$c_{d_1}(b_1\text{,}s_1)$}
            \draw[->] (\x) -- node[above]{\w} (\y);
        \foreach \x in {$d_1$,$d_2$,$d_3$}
            \draw[->] (\x) -- node[]{}(Sink);
            
        \node[] at (2.5, 5.5)  {\footnotesize Store};
        \node[] at (5, 5.5)  {\footnotesize Buyer--Store};
        \node[] at (7.5, 5.5)  {\footnotesize Courier};
    \end{tikzpicture}
    \caption{An example min-cost flow network for a market where each buyer only values one store. There is a vertex $bs$ for each buyer $b$ that values store $s$.
    \label{fig:OneBuyerOneSeller}}
    \end{figure}

\OneBuyerOneStore*
\begin{proof}
    Let $\x$ be a feasible allocation that achieves the optimal welfare $W(\x)=\mathit{OPT}$. If $|\Omega_\x|=l$, \Cref{cor:all_courier_deliver} shows $\x$ is in some equilibrium. We focus on the case where $|\Omega_\x|<l$. Let $\z$ be the buyer allocation induced by $
    \x$ in $G_\x$. We prove that $\z$ is a maximum weight matching in $G_\x$, and by \Cref{lem:test_alloc}, $\mathbf{x}$ is indeed in an equilibrium.

    For a store $s_0$ that is valued by some buyer $b_0$ but not matched in $\x$, let $d_0$ be the courier who is willing to deliver order $(b_0,s_0)$ with the lowest tip $d_0=\argmin_{d}\{c_{d'}(b_0,s_0)+\bar{u}_{d}(\x)\}$. By \Cref{lem:exist_courier_plan_serve}, courier $d_0$'s utility is written as $\bar{u}_{d_0}(\x)=C_{\Omega_\x}(G_D\backslash d_0)-C_{\Omega_\x}(G_D)$. Here $C_{\Omega_\x}(G_D)$ is the minimum cost of delivering all orders in $\Omega_\x$, also the total courier cost for $\x$ because $\x$ is optimal.
    $C_{\Omega_\x}(G_D\backslash d_0)$ is the cost of the minimum cost matching that covers $\Omega_\x$ without courier $d_0$ in the bipartite graph $G_D=(O,D)$. Let $\mathbf{y}^{\backslash d_0}$ be the courier allocation in $G_D$ that covers $\Omega_\x$ with the minimum cost without courier $d_0$.
    
    Consider an allocation where $d_0$ delivers the order $(b_0,s_0)$, and all buyer--store pairs in $(b,s)\in\Omega_x, b\neq b_0, s\neq s_0$ are delivered by the same couriers as in $\mathbf{y}^{\backslash d_0}$. As $\x$ is welfare-optimal, 
    \begin{align*}
        v_{b_0}(s_0) - c_{d_0}(b_0,s_0) + \sum_{bs:z_{bs}=1}v_b(s) - C_{\Omega_\x}(G_D\backslash d_0) &\leq W(\mathbf{x})=\sum_{bs:z_{bs}=1}v_b(s) - C_{\Omega_\x}(G_D)\\
        v_{b_0}(s_0) - c_{d_0}(b_0,s_0) &\leq C_{\Omega_\x}(G_D\backslash d_0) - C_{\Omega_\x}(G_D) = \bar{u}_{d_0}(\x)\\
        v_{b_0}(s_0) &\leq \ut_{b_0s_0}
    \end{align*}
    This shows all edges connecting to an unmatched store $s_0$ has weight weakly smaller than zero in $G_\x$.
    
    For a store $s_0$ that is matched to a buyer $b_0$ and delivered by a courier $d_0$ in $\x$, but also valued by some buyer $b_1$, we now show that $v_{b_1}(s_0)-\ut_{b_1s_0}\leq v_{b_0}(s_0)$. Similar as the above case, one feasible allocation $\mathbf{x}'$ is to replace the buyer--store pair $(b_0,s_0)$ by $(b_1,s_0)$, and have courier $d_0$ deliver for $(b_1,s_0)$. All other orders in $\Omega_\x$ are delivered as the same way as $\mathbf{y}^\backslash d_0$, which have a total cost weakly smaller than $C_{\Omega_\x}(G_D\backslash d_0)$ because the order $(b_0, s_0)$ no longer needs to be delivered.
    \begin{align*}
        W(\mathbf{x}) &= v_{b_0}(s_0) +\sum_{bs:z_{bs}=1,b\neq b_0} v_b(s) - C_{\Omega_\x}(G_D)\\
        &\geq W(\mathbf{x}') \geq v_{b_1}(s_0) - c_{d_0}(b_1,s_0) + \sum_{bs:z_{bs}=1,b\neq b_0} v_b(s) - C_{\Omega_\x}(G_D\backslash d_0)\\
        v_{b_0}(s_0) &\geq v_{b_1}(s_0) - c_{d_0}(b_1,s_0)-\bar{u}_{d_0}(\x)=v_{b_1}(s_0)-\ut_{b_1s_0}
    \end{align*}
    This shows for any store $s_0$ already matched in $\x$, its matched edge has weakly larger weight than any edges not matched in $G_\x$. This completes the proof that $\mathbf{x}$ is the maximum weight matching in $G_\x$.
\end{proof}

\section{Missing Lemmas and Proofs in \Cref{sec:profit}}\label{app:profit}
We first write the formal statement that the platform profit is weakly higher in with-tip equilibrium compared to without-tip equilibrium. The proposition automatically follows from \Cref{lem:pareto_dominant} so we omit the proof.
\begin{proposition}\label{prop:profit_weakly_higher}
    For any without-tip equilibrium profit, there exists a with-tip equilibrium that achieves the same profit.
\end{proposition}

\begin{restatable}{theorem}{ProfitGeneralHard}
\label{thm:profit_hard}
    It is NP-hard to determine whether the maximum profit of with-tip equilibrium is at least $k$. The same hardness result holds for without-tip equilibrium in markets where the number of couriers is weakly larger than the minimum number of buyers and stores.
\end{restatable}
\begin{proof}
    We use the same construction as in the proof for \Cref{lem:OPT_eq_hard}.
    
    If there exists a perfect matching of size $l$, then consider an allocation $\mathbf{x}$ that corresponds to the perfect matching, purchase price $\mathbf{p}=1$, courier compensation $\mathbf{w}=0$. This constitutes a without-tip equilibrium with profit $l$. For couriers, delivering or not delivering both provides utility zero. For buyers, purchasing means paying price one and having zero utility, while deviating also means zero utility. Furthering specifying $\bt=0$ gives a with-tip equilibrium with profit $l$ as well, because buyers deviating need to pay positive tips as well as a price of one. The equilibrium brings profit equal to $l$.
    
    On the other direction, if there exists some equilibrium with profit $k$, then this equilibrium defines a matching. For each trade in the matching, the maximum profit the platform can extract is one. So the profit of the equilibrium is no larger than the size of the matching it facilitates.

    We have proved that there exists a perfect matching in the 3DM instance if and only if the maximum profit is $l$. Now given the 3DM instance, we can set $k=l$ and find the maximum profit. If the max-profit is at least $l$, then a perfect matching exists, otherwise not. This shows determining whether the maximum profit is at least $k$ is also NP-hard.
\end{proof}

\ProfitPolyWithoutTipEqStoreLimitedCourier*
\begin{proof}
We first present the algorithm. Let $\epsilon$ be a small value such that $\epsilon \sum_{b,s,d} c_d{(b,s)}\leq \min_{bs}\{v_{bs}|v_{bs}>0\}$. Define a new buyer-seller bipartite market where buyer valuation is modified by the cost $v'_{bs}=v_{bs}+\epsilon\max_{b,s,d}c_d(b,s) -\epsilon \min_d c_d{(b,s)} \geq 0$. The walrasian equilibrium allocation $z$ in the new buyer-seller market is the profit-maximal buyer-store orders in the three-sided market, where couriers with the minimum cost delivers for orders. Formally, the profit maximum without-tip allocation $x$ and courier compensation is given by $$x_{bsd} =\begin{cases} 1 & \text{ if } z_{bs}=1 \text{ and } \forall d', c_d(b,s)\leq c_{d'}(b,s)\\ 0& \text{otherwise} \end{cases}, \quad w_{bs}=\begin{cases}\min_d c_d(b,s) & \text{ if } z_{bs}=1\\ 0 & \text{otherwise}\end{cases}$$ Buyer prices is the maximum walrasian price in the buyer-seller market with unaltered valuation. As calculating max-weight matching in bipartite graph, or a walrasian equilibrium takes polynomial time, the above procedure takes polynomial time.

We then explain why the algorithm produces the max-profit without-tip equilibrium. When buyers cannot specify tips, the set of allocations that satisfy buyer incentives must be the same set of allocation in the walrasian equilibrium of the buyer-store two-sided market. Since all walrasian equilibrium have the same highest price by the lattice structure \citep{gul1999walrasian}, the maximum prices the platform can charge to buyers is fixed in the three-sided market. Since each courier only deliver for one store, equilibrium requires that the courier with minimum costs deliver the order, and minimum compensation equal to costs. The $\epsilon$ modification to values serves to select the walrasian equilibrium in the two-sided market in the new buyer-store market with the minimum courier costs.
\end{proof}

\subsection{Proof of \Cref{thm:profit_hard_with_tip}}\label{sec:proof_vertex_cover}
Similar to \cite{platformDisruption} and \cite{guruswami2005profit}, we prove NP-hard by reducing from a version of the vertex cover problem. Given a graph $(V,E)$ where vertices have maximum degree $K\geq 3$, the problem to find the minimum set of vertices to cover all edges is known to be NP-hard. Given a graph $(V,E)$, we construct the following instance of profit maximization problem for the case when each store is served by one unique courier. (parenthesis points out the difference in proofs where each courier can only deliver for one store-buyer pair). 
\begin{itemize}
    \item For each vertex $v\in V$, create a vertex store $s_v$ and a vertex buyer $b_v$.
    \item For each edge $(u,v)\in E$, create an edge buyer $b_{(u,v)}$, two edge stores $s_{u,(u,v)}$ and $s_{v,(u,v)}$, and two dummy buyers $b_{u,(u,v)}$ and $b_{v,(u,v)}$.
    \item Each vertex buyer $b_v$ values $s_v$ and $s_{v,(v,u)}$ at $M+1$ for all $u$ satisfying $(v,u)\in E$; each edge buyer $b_{(u,v)}$ values $s_{v,(u,v)}$ and $s_{u,(u,v)}$ at $M$; each dummy buyer $b_{u,(u,v)}$ only values $s_{u,(u,v)}$ at $M+2$. Here $M>|V|$ is a constant and all other valuations are zero.
    \item For each store, create a courier that only delivers for this store: $d_{v}$ for $s_v$ and $d_{u,(u,v)}$ for $s_{u,(u,v)}$. A vertex courier $d_v$ has cost zero for delivering for store $s_v$. An edge courier $d_{v,(u,v)}$ has cost $M+2$ for delivering from $s_{v,(u,v)}$ to dummy buyer $b_{v,(u,v)}$, and cost zero for delivering to $b_v$ and $b_{(u,v)}$. A courier cannot deliver for any other store than its designated one, or equivalently all other costs are positive infinity. Note this cost can be decomposed into a courier-store part which is 0 or positive infinity, and a store-buyer part which is 0 or $M+2$.
    (For each buyer-store pair with positive valuation, create a courier that only serves this buyer-store pair. The cost for delivering to dummy buyer is $M+2$. The cost for delivering to other buyers from the designated store is zero. All other costs are positive infinity.)
\end{itemize}

\begin{lemma}\label{lem:vertex_cover_np}
    For a vertex cover of size $k$, there exists a with-tip equilibrium with profit $|V|+M(|V|+|E|)-k$.
\end{lemma} 
\begin{proof}
    For a vertex cover $Q$ of size $k$, the following constitutes a with-tip equilibrium. Set tip to be zero $\bt = 0$, vertex stores prices and compensation to be 
    $$p_{s_v} = \begin{cases}
        M & \text{if } v\in Q\\
        M+1 & \text{otherwise}
    \end{cases}; \quad w_{b_v,s_v}=0;\quad$$
     For edge stores for every edge $(u,v)$, if $u\in Q, v\notin Q$ then 
     $$p_{s_{u,(u,v)}} = M; \quad p_{s_{v,(u,v)}} = M+2;\quad w_{\cdot,s_{u,(u,v)}}=0;\quad w_{b,s_{v,(u,v)}}=\begin{cases}M+2&\text{if } b=b_{v,(u,v)}\\ 0&\text{otherwise}\end{cases}$$
     If $u\in Q, v\in Q$, then randomly select one of the two vertex, say $v$ to set a high buyer price and courier compensation
     $$p_{s_{u,(u,v)}} = M; \quad p_{s_{v,(u,v)}} = M+2;\quad w_{\cdot,s_{u,(u,v)}}=0;\quad w_{b,s_{v,(u,v)}}=\begin{cases}M+2&\text{if } b=b_{v,(u,v)}\\ 0&\text{otherwise}\end{cases}$$
     For the allocation, a vertex buyer $b_v$ always purchase from the vertex store $s_v$; an edge buyer $s_{(u,v)}$ always buys from the edge store $s_{u,(u,v)}$ where $u\in Q$ and $p_{s_{u,(u,v)}}=M$; an dummy buyer $b_{u,(u,v)}$ buys from $s_{u,(u,v)}$ if and only if $s_{(u,v)}$ doesn't buy from $s_{u,(u,v)}$. The only courier associated with each store delivers for the store. (Couriers associated with the buyer-store order delivers for that order.)

     Now we show the above constitutes a with-tip equilibrium. First, the allocation is feasible and all stores are allocated, with compensation and tips in undelivered orders being zero, and compensation being none-zero only for orders delivered to dummy buyers. Next we analyze the incentives. All couriers have zero utility in delivering and does not have beneficial deviations. For vertex buyer $b_v$, if $v\in Q$, then it faces price $M$ at $s_v$, and price $M$ or $M+2$ at at the edge stores $s_{v,(u,v)}$ so it is happy to purchase from $s_v$ with utility of 1; if $v\notin Q$, then it faces price $M+1$ at $s_v$ and any edge stores $s_{v,(u,v)}$ that it has valuation for so it is indifferent to purchase from $s_v$ with utility of 0. For edge buyer $b_{(u,v)}$, as it always buy from the edge store priced at $M$, while the other edge store it values have price $M+2$, it is indifferent to have utility zero. For dummy buyer $b_{u,(u,v)}$, if $u\in Q$ and $p_{s_{u,(u,v)}}=M$ then by the construction $w_{b_{u,(u,v)},s_{u,(u,v)}}=0$ so it needs to tip $M+2$ to secure delivery. Thus, this dummy buyer does not deviate to buy from $s_{u,(u,v)}$, instead $s_{u,(u,v)}$ is sold to the edge buyer $b_{(u,v)}$. For dummy buyer $b_{u,(u,v)}$, if $u\notin Q$ or $p_{s_{u,(u,v)}}=M+2$, then by construction the platform covers courier compensation $w_{b_{u,(u,v)},s_{u,(u,v)}}=M+2$ so the dummy buyer is indifferent to purchase at utility zero.

     In the above with-tip equilibrium, dummy buyer generates zero profit for the platform, each edge buyer generates profit $M$, each vertex buyer in the vertex cover generates profit $M$, each vertex buyer not in the vertex cover generates profit $M+1$. Total profit is $|E|M+ kM+(|V|-k)(M+1)=-k+|V|+M(|V|+|E|)$.
\end{proof}

\begin{lemma}\label{lem:converse_vertex_cover_np}
    If the maximum with-tip equilibrium profit is $|V|+M(|V|+|E|)-k$, then there exists a vertex cover of size no larger than $k$.
\end{lemma} 
\begin{proof}
     As dummy buyers have delivery cost equal to its valuation, the platform cannot extract any profit from dummy buyers. Vertex buyers can generate at most profit $|V|(M+1)$. Each edge buyer generate profit at most $M$. If there exists even one edge buyer who is not matched in the equilibrium, then the maximum profit is bounded by $M(|E|-1)+|V|(M+1)<-k+|V|+M(|E|+|V|)$ because $M>|V|\geq k$. So every edge buyer must be matched.

     (When each courier can only deliver for one store-buyer pair: As every edge buyer is matched to an edge store and each buyer-store pair can only be served by one courier, the maximum profit requires that compensation of delivering from the edge store to be zero because a higher compensation doesn't affect the minimum tip other buyers needed to attain delivery from the store. The max profit also requires that buyer prices for these edge stores to be $M$. Otherwise, by increasing the buyer prices, not only platform profit is increased, but also further dampens the incentives for the vertex buyers to deviate to these edge stores.)

     When each store is served by one unique courier, we also show that the profit extracted from each edge buyer is $M$ by showing that without loss of profit 1) the price the edge buyers pay are $M$; 2) the courier compensation associated with edge buyer orders are zero.

     For 1) suppose that an edge buyer $b_{(u,v)}$ pays price $M-\epsilon, 0<\epsilon\leq M$ for store $s_{u,(u,v)}$. We discuss the three cases for $s_{v,(u,v)}$ 
     \begin{itemize}
         \item unmatched at price $0$.  The courier $d_{v,(u,v)}$ does not deliver and has utility zero. Then buyer $b_{(u,v)}$ can deviate to buying from $s_{v,(u,v)}$ and have a utility of $M$, which is strictly larger than her current utility. So this case cannot occur.
         \item matched to $b_v$ at price $0$ because $b_v$ could have buy from $s_v$ at price $0$. To prevent $b_{(u,v)}$ from deviating to $s_{v,(u,v)}$, the courier utility has to be at least $M-\epsilon$, paid by the platform. This means in this case the platform loses profit by $M-\epsilon$ at least.
         \item matched to $b_{v,(u,v)}$. The platform can post a price of $M+2$ for $s_{v,(u,v)}$ and compensate the courier the same amount to have profit zero. For this case, the platform can also set a price $p_{s_{v,(u,v)}}<M$, but that will only weakly increase the courier compensation required to prevent $b_{(u,v)}$ and $s_v$ from deviating. So without loss of profit $p_{s_{v,(u,v)}}=M+2=w_{b_{v,(u,v)},s_{v,(u,v)}}$.
     \end{itemize}
    From these three cases, decreasing $\epsilon$ keeps buyer $b_{(u,v)}$ utility weakly above zero, while reducing the incentives for $b_u$ to deviate. So without loss $p_{s_{u,(u,v)}}=M$.

     For 2) suppose that an edge buyer $b_{(u,v)}$ pays price $M$ for store $s_{u,(u,v)}$ with courier compensation $w>0$. As the courier cost for delivery is zero, the positive compensation exists only to decrease the incentive for other buyer to deviate to $s_{u,(u,v)}$. As dummy buyer has valuation equal to cost equal to $M+2$, we only consider deviation by buyer $b_u$. The minimum tip for $b_u$ to attain delivery from $s_{u,(u,v)}$ is $w$, so the deviation utility for $b_u$ is $1-w$. If $p_{s_{u}}\leq M$, then $b_u$ has utility $1$ buying from $s_u$ and no incentives to deviate so $w>0$ is not profit maximal. If $M< p_{s_{u}}\leq M+1$, then buyer $b_u$ having no incentive to deviate in the equilibrium means that $M+1-p_{s_u}\geq 1-w \Leftrightarrow w\geq p_{s_u}-M$. Now consider setting $w=0$ and $p'_{s_u}=p_{s_u}-w$. With the new compensation and price buyer $b_u$ still doesn't deviate because $M+1-p_{s_u}+w \geq 1 \Leftrightarrow w\geq p_{s_u}-M$. Note total profit remains unchanged as well.

     From the above, without loss of profit each edge buyer contributes profit $M$. For each edge $(u,v)$, one of the two store $s_{v,(u,v)}$ and $s_{u,(u,v)}$ has price $M$ and 0 courier compensation. To prevent vertex buyer deviation, at least one of the two stores between $s_u$ and $s_v$ must have price no larger than $M$, otherwise one of the two buyers $b_u$ and $b_v$ would deviate to buy from $s_{u,(u,v)}$ or $s_{v,(u,v)}$ at price $M$ and zero tip. Let $Q=\{v| p_{s_v}\leq M\}$. Then the maximum profit the equilibrium can attain is $|Q|M+(|V|-|Q|)(M+1)+|E|M$, which must be larger than $|V|+M(|V|+|E|)-k$, so $|Q|\leq k$. Also $Q$ is a vertex cover because every edge is covered by it.
\end{proof}

Now a reduction from the vertex problem follows. If the maximum with-tip equilibrium profit is weakly larger than $|V|+M(|V|+|E|-k)$, then by \Cref{lem:converse_vertex_cover_np} there exists a vertex cover of size no larger than $k$. If the maximum with-tip equilibrium profit is smaller than $|V|+M(|V|+|E|)-k$, by \Cref{lem:vertex_cover_np} there cannot exists a vertex cover of size no larger than $k$. This gives the reduction and the NP-hardness follows.


\section{Zero Store Costs and Store Incentives}\label{sec:store_incentives}
Throughout the paper we made two statements on stores. First, we assume stores charge the platform its product price, which is equal to its cost. Second, we normalized the store costs to zero, and in doing so normalized the product prices to zero as well. 
In \Cref{app:zero_store_cost}, we show that it is without loss to assume zero store cost and product price. In \Cref{app:ce_existence}, we formulate a competitive equilibrium following the literature on multilateral contracting \citep{hatfield2011multilateral} and on competitive equilibrium in trading networks \citep{ostrovsky2008stability,hatfield2013stability}. This competitive equilibrium definition addresses store incentives, satisfies the first welfare theorem. But it is NP-hard to determine its existence. 

\subsection{Generalizing to Nonzero Store Costs}\label{app:zero_store_cost} 
All our results generalize to nonzero costs, under the first assumption that every store charges the platform its product price equaling its cost. We will illustrate the case for with-tip equilibrium. The without-tip equilibrium follows suit.

Consider a market $M'$ with store cost, where each store $s$ has cost $c_s>0$, a buyer has valuation $v'_b(s)$, and a courier has cost $c_d(b,s)$ delivering the order $(b,s)$. Welfare associated with a trade $(b,s,d)$ is $v'_b(s)-c'_s-c_d(b,s)$. The purchase price $\p'$ includes the stores' product prices and the platform's delivery fee, where the product price is set to store cost. Hence, $\forall s, p'_s\geq c_s$. 
Now define the \emph{store cost version of with-tip equilibrium} the same way as the with-tip equilibrium, with the only difference that stores not bought from (i.e., $\sum_{bd}x_{bsd}=0$) have purchase price equaling its cost $p'_s=c_s$.

Given $M'$, we can construct another market $M$ without store cost. A buyer $b$ has valuation $v_b(s)=v'_b(s)-c_s$ for store $s$, and a courier has cost $c_d(b,s)$ delivering the order $(b,s)$. Welfare associated with a trade $(b,s,d)$ in $M'$ is equal  to $v_b(s)-c_d(b,s)$.

Similarly, given a market $M$ without cost, we can construct a market $M'$ with cost by setting $v'_b(s)=v_b(s)+c_s$ for some store cost $c_s$.
\begin{proposition}
    $(\p,\w,\bt,\x)$ is a with-tip equilibrium in $M$ if and only if $(\p',\w,\bt,\x)$ is a store cost version of with tip equilibrium in $M'$, where $p'_s=p_s+c_s$. 
\end{proposition}
This is very easy to check from the definition of equilibrium. Couriers incentives do not change because delivery compensation and tips remains the same. For a buyer, we can show the following two equivalence
$$\begin{cases}
    v_b(s)-p_s-\ut_{bs}\geq v_b(s')-p_{s'}-\ut_{bs'} &\Leftrightarrow v'_b(s)-p'_s-\ut_{bs}\geq v'_b(s')-p'_{s'}-\ut_{bs'}\\
    v_b(s)-p_s-\ut_{bs}\geq 0 &\Leftrightarrow v'_b(s)-p'_s-\ut_{bs}\geq 0
\end{cases}$$
So given a market $M'$ with tip, we can first solve for the with-tip equilibrium in $M$, then transfer it back to $M'$.

\subsection{Store Incentives through Competitive Equilibrium}\label{app:ce_existence}
In this section, we follow the standard definition of competitive equilibrium in the multilateral contracting literature. This competitive equilibrium definition includes stores' incentives, but does not permit tips. We write out the first and second welfare theorem, and show that a competitive equilibrium does not always exist. In fact, it is NP-hard to determine the existence of a competitive equilibrium. Since this is a separate part from the main body of paper, it is easier to redefine notations. 

Let $B,S,D$ denote the set of unit demand buyers, unit supply stores, and unit capacity couriers where $|B|=|S|=|D|=n$, denoted by $i,j,k$ respectively. The sets of all trades is denoted by $\Omega = B\times S\times D$, where $w_{ijk}$ means the trade where buyer $i$ purchase from store $j$, delivered by courier $k$. An agent $a\in B\cup S\cup D$ has valuation $v^a(w_{ijk})\in R$ for trade $w_{ijk}$ concerning this agent $a\in\{i,j,k\}$ and 0 for trades that do not concern her. $v^a(w_{ijk})$ can be positive or negative, representing a buyer's valuation, or a store or courier's cost. Each trade is associated with prices $p_\omega=(p^i_\omega,p^j_\omega,p^k_\omega)$ denoting the payments agents make, where $p^i_\omega\geq 0$ is the amount the buyer pays, $p^j_\omega\leq 0$ is the amount the store pays, or the negation of what the store receives, and $p^k_\omega\leq 0$ being the negation of what the courier receives. A pricing scheme assigns a price for all trades in $\Omega$ and is \emph{budget balanced} if $p^i_\omega+p^j_\omega+p^k_\omega=0$. An agent $a$ participating in trade $\omega$ has utility $u^a(\omega)=v^a(\omega)-p^a(\omega)$. A \textit{competitive equilibrium} $(\psi^\star,p^\star)$ is a set of trades that occur $\psi^\star\subseteq \Omega$, and a pricing scheme $p^\star$, such that (i) no agent participates in more than one trade in $\psi^\star$, (ii) the pricing scheme is budget balanced, and (iii) agents participate in their favorite trades in $\psi^\star$ at $p^\star$, with an outside option of utility zero.
As is common in the multilateral matching literature, the vital difference from competitive equilirbium(CE) in two sided is that we do not require trades that do not take place have price zero. We require budget balance instead.

\subsection{Existence of Competitive Equilibrium}
Just like CE in two sided markets, we can write out the \textit{primal integer program (PIP)}  and the \textit{dual of the linear program relaxation (DLPR)} of the central planner's welfare maximization problem.
\begin{align*}
    \text{max \;} & \sum_{ijk}x_{ijk} (v^i(w_{ijk})+v^j(w_{ijk})+v^k(w_{ijk}))\\
     & x_{ijk}\in\{0,1\} \qquad \forall i,j,k\\
     & \sum_{jk}x_{ijk}\leq 1 \quad \forall i; \quad \sum_{ik}x_{ijk}\leq 1 \quad \forall j; \quad \sum_{ij}x_{ijk}\leq 1 \quad \forall k
\end{align*}

\begin{align*}
    \text{min \;} & \sum_{i}u_i +\sum_{j}u_j+\sum_{k}u_k\\
     & u_{a}\geq 0 \qquad \forall a\in B\cup S\cup D\\
     & v^i(w_{ijk})+v^j(w_{ijk})+v^k(w_{ijk})\leq u_i+u_j+u_k \quad \forall i,j,k
\end{align*}

\begin{proposition}[First Welfare Theorem]
    The competitive equilibrium trades $\psi^\star$ maximizes welfare, even over fractional solution of the primal linear program relaxation.
\end{proposition}
\begin{proof}
    The proof is almost the same with that in two-sided markets. Let $\psi^\star_a$ be the trade that agent $a$ takes part in $\psi^\star$, and $x$ be any feasible solution to the \textit{primal linear program relaxation (PLPR)}.
    \begin{align*}
        v^i(\psi_i^\star) - p^i(\psi^\star_i) \geq \sum_{jk}x_{ijk}(v^i(\omega_{ijk}) - p^i(\omega_{ijk}))
    \end{align*}
    The results follow by summing the left and right hand side of the inequality, and noting that all prices sum to zero $\sum_{a}p^a(\psi^\star_a)=0$, and agents not trading has $v^i(\psi^\star)=0$.
\end{proof}

\begin{proposition}[Second Welfare Theorem]
    Given an integral solution $x^\star$ to PLPR, and a solution $u^\star$ to DLPR, there exists a competitive equilibrium with trades $\psi^\star$ as directed by $x^\star$, and a pricing scheme $p^\star$, where $p_a^\star(w_{ijk}) = -u^\star_a+v_a(w_{ijk})+q_{ijk}/3$ where $q_{ijk}=(u_i^\star+u_j^\star+u_k^\star-v_i(w_{ijk})-v_j(w_{ijk})-v_k(w_{ijk}))$
\end{proposition}
\begin{proof}
    It is easy to verify that prices sum to zero $p^\star_i(w_{ijk})+p^\star_j(w_{ijk})+p^\star_k(w_{ijk})=-u^\star_i-u^\star_j-u^\star_k+v_i(w_{ijk})+v_j(w_{ijk})+v_k(w_{ijk})+q_{ijk}=0$. Now by complementary slackness, if $x_{ijk}=1>0$, $q_{ijk}=0$. So $u_i^\star=v_i(w_{ijk})-p_i^\star(w_{ijk})$. For any other trades $x_{ij'k'}=0$, $u_i^\star=v_i(w_{ij'k'})-p^\star_i(w_{ij'k'})+q_{ij'k'}/3\geq v_i(w_{ij'k'})-p^\star_i(w_{ij'k'})$ so indeed the trade $(i,j,k)$ is preferred.
\end{proof}
Combining the first and the second welfare theorem, we have
 \begin{corollary}
     A competitive equilibrium exists if and only if the PLPR admits an integral optimal solution. 
 \end{corollary}    
\begin{proposition}
    It is NP-hard to determine if a competitive equilibrium exists in a three-sided markets.    
\end{proposition}
\begin{proof}
    It is well known that three dimensional matching problem is NP-hard. Given a subset of hyperedges $T\subseteq V_1\times V_2\times V_3$ where $|V_1|=|V_2|=|V_3|$ and each vertex in $V_1\cup V_2\cup V_3$ has degree three, it is NP-complete to determine whether there is a perfect matching that each vertex is covered exactly once. Given an instance of such a three dimensional matching problem, construct a market where $|B|=|S|=|D|=n$ and $v_a(w_{ijk})=1$ if $a\in\{i,j,k\}$ for $w_{ijk}\in T$. If in polynomial time one can determine a CE exists, then an integer solution to PLPR exists. One can check if the value of PLPR equals to $3n$ and tell if a perfect matching exists. If in polynomial time one determines no CE exists, then PLPR has a fractional solution. This fractional solution cannot have value larger than $3n$ because of the PLPR constraints. So any solution to the PIP has value strictly less than $3n$. So no perfect matching exists. Thus determining if a CE exists is harder than 3DM matching.
\end{proof}

There is another version of second welfare theorem, which we append it here.
\begin{proposition}[Another Second Welfare Theorem]
    If $(\psi^\star,p^\star)$ is a competitive equilibrium, and $\phi^\star$ indicates another set of trades that is welfare efficient, then $(\phi^\star,p^\star)$ is a competitive equilibrium.
\end{proposition}
\begin{proof}
    \begin{align*}
        \sum_a u_a(\phi^\star_a,p^\star) &=\sum_{a} v_a(\phi^\star_a)-p^\star_a(\phi_a^\star)=\sum_{a} v_a(\phi^\star_a)\\
        &\geq \sum_{a} v_a(\psi^\star_a) = \sum_{a} v_a(\psi^\star_a)-p^\star_a(\psi_a) = \sum_a u_a(\psi^\star_a,p^\star)
    \end{align*}
    where the inequality is because $\phi^\star$ is welfare maximizing. But by $(\psi^\star,p^\star)$ being a CE, we know $u_a(\psi_a^\star,p^\star)\geq u_a(\phi_a^\star,p^\star),\forall a$. This with the inequalities above implies $u_a(\phi_a^\star,p^\star)= u_a(\psi_a^\star,p^\star),\forall a$.
\end{proof}
As even determining the existence of a competitive equilibrium is NP-hard, we move away from this definition. In \Cref{sec:model}, we instead define an equilibrium that allows the platform to subsidize delivery, breaking the budget balance requirement. This guarantees equilibrium to exist but sacrifices the first and the second welfare theorem. 

We abstract away stores' incentives, while still keeping the stores' capacity constraints. We further require prices to be nondiscriminatory, instead of having a unique price for each trade in the definition for competitive equilibrium. With this altered notion of equilibrium, we ask how much welfare can be attained.

\chapter{Appendix to \Cref{chap:platform_disruption}}\label{app:ch4}

\section{Missing Proofs in Section~\ref{sec:hardness}}
\subsection{Proof of NP-Hardness (Lemma~\ref{lem:reduction_proof})}\label{app:sat_np_proof}
\SATReduction*
\begin{proof}
Suppose there is a valid assignment to the original CNF $\varphi$. Then there exists a matching between buyers $U_i$ and items $\alpha_{i,j}, \tau_{i,j}$ such that no buyer $U_i$ is matched to an item $\alpha_{i,j}$ where $x_i$ is false or an item $\tau_{i,j}$ when $x_i$ is true. Connect these as platform edges. For each buyer $A_{i,j}, T_{i,j}$, add platform edges to items $\gamma_{i,j}$ and $\delta_{i,j}$ respectively. Finally, add platform edges between each dummy item and a single unsold item so that all items are sold on platform.

The maximum weight matching assigns buyers $U_i$ their corresponding items via the platform edges. Each $A_{i,j}$ gets item $\gamma_{i,j}$ and each $T_{i,j}$ gets item $\delta_{i,j}$. Finally, the dummy buyers are matched to the remaining $(2tq - k)$ items that have not already been sold.
    
The total revenue of this matching is $D$. Each buyer $U_i$ pays price $Z$, resulting in revenue $kZ$. All dummy buyers pay price $H$, resulting in revenue $H(2tq-k)$. If $x_i$ is true, $\alpha_{i,0}$ is sold to a buyer $U_i$, so buyer $A_{i,j}$ pays price $Z$ as it has an opportunity path $A_{i,j} \mbox{ --- } \alpha_{i,0} \to U_i$ to this buyer. However, for $T_{i,j}$, both $\tau_{i,0}$ and $\tau_{i,j}$ are sold to dummy buyers – thus, $T_{i,j}$ only has opportunity paths to buyers with weakly higher values and must pay $Z + 1$. Thus, $\forall j, A_{i,j}$ has price $Z$ and $T_{i,j}$ has price $Z+1$. The converse holds when $x_i$ is False. In conclusion, from each $A_{i,j}$ and $B_{i,j}$, we get revenue $(t-1)Z + (t-1)(Z+1) = (t-1)(2Z+1)$. Thus, in total we get revenue $kZ + q(t-1)(2Z+1) + H(2tq - k) = D$ as desired. \\

Now, suppose that there exists a set of platform edges and a matching that generates revenue at least $D$. Then it must be the case that each buyer $U_i$ gets an item as otherwise we could have maximal revenue $(k-1)Z + 2q(t-1)(Z+1) + H(2tq - k) < D$. For each $U_i$ corresponding to clause $d_i = (x_i \vee \bar{x}_i)$, they must receive either $\alpha_{i,0}$ or $\tau_{i,0}$. We construct our satisfying assignment as follows:
\begin{itemize}
    \item If $U_i$ receives $\alpha_{i,0}$, set $x_i$ to True.
    \item If $U_i$ receives $\tau_{i,0}$, set $x_i$ to False.
\end{itemize}
It suffices to show that this is a satisfying assignment. To do this, we show that no buyer $U_k$ receives an item $\alpha_{i,j}$ when $x_i$ is False or an item $\tau_{i,j}$ when $x_i$ is True.

If $x_i$ is True, each $A_{i,j}$ has an opportunity path to buyer $U_i$, so we can get at most revenue $Z$ from them. Similarly, if $x_i$ is False, each $T_{i,j}$ has an opportunity path to buyer $U_i$, so we can get at most revenue $Z$ from them. Thus, the platform makes revenue at most $Z$ from each buyer $A_{i,j}$ or $B_{i,j}$ that corresponds to the $q$ true literal. For the $q$ false literals, note that the platform has to get maximal revenue $Z+1$ from the corresponding buyer because $D-kZ-H(2tq-k)-q(t-1)Z=q(t-1)(Z+1)$. If $U_k$ receives an item corresponding to the false literal, then the buyer $A_{i,j}$ or $T_{i,j}$ that corresponds to the false literal will have an opportunity path to this $U_k$, meaning they will only pay price $Z$, contradicting the above. It follows that our assignment is indeed a satisfying assignment as desired, concluding the proof.
\end{proof}

\subsection{Proof of APX-Hardness (Theorem \ref{thm:vc_apx_proof})}\label{app:vc_apx_proof}
The APX-hardness proof is similar to that in \citet{guruswami2005profit} for revenue maximizing envy-free pricing, and reduces from a version of the vertex cover problem. Given a connected graph $G=(V,E)$ where vertices have maximum degree $K$, the problem asks to find the minimum set of vertices to cover all edges. This problem is known to be APX-hard for $K \geq 3$. Given $G$, we construct the following instance of the platform's problem:
\begin{itemize}
    \item For each vertex $v \in V$, we create a vertex buyer $B_v$ and item $S_v$. We also create items $\alpha_{v, i}$ for $i \in \{1, \dots, \mathrm{deg}(v)\}$.
    \item For each edge $(u, v) \in E$, we create an edge buyer $B_{(u, v)}$.
    \item Finally, we add $|E|$ dummy buyers $D_1, \dots, D_{|E|}$.
    \item Each buyer $B_v$ has value $2$ for items $S_v, \alpha_{v, \cdot}$. Each buyer $B_{(u, v)}$ has value $1$ for items $\alpha_{u, \cdot}$ and $\alpha_{v, \cdot}$. Each dummy buyer $D_i$ has value $H\geq 2$ for all $\alpha$ items.
    \item We add world edges from $B_v$ to all $\alpha_{v, \cdot}$.
\end{itemize}
We complete the reduction by showing the following.
\begin{lemma}
\label{lem:vertice_cover_reduction}
    The optimal revenue is given by $2|V| + (H+1)|E| - q$ where $q$ is the size of the smallest vertex cover.
\end{lemma}
\begin{proof}
Suppose there is a vertex cover $Q$ of size $q$. We add platform edges from $B_v$ to $S_v$ for each $v \in V$. Additionally, for each $B_{(u, v)}$, if $u \in Q$, we add a platform edge to some $\alpha_{u, i}$. Otherwise – note that this implies that $v \in Q$ – we add a platform edge to $\alpha_{v, i}$. We do so such that each $B_{(u, v)}$ is matched via a platform edge to a different item. Finally, we add a platform edge from each $D_i$ to an $\alpha$ item that is not incident on any already added platform edges. Thus, the platform edges act as a perfect matching between buyers and items.

The platform then picks all the platform edges as the max weight matching and get revenue $2|V|+(H+1)|E|-q$. All dummy buyers pay full price $H$ and all edge buyers $B_{(u,v)}$ pays full price $1$ because they are not connected to any other buyers through opportunity path. Buyers $B_v$ where $v\in Q$ has an opportunity path $B_v \mbox{ --- } \alpha_{v,i} \to B_{v,u}$ to some edge buyer $B_{(u,v)}$ and pays price 1. Buyers $B_v$ where $v\notin Q$ pays full price $2$ because they are connected through opportunity path to dummy buyers. So total revenue from vertex buyers is $q+2(|V|-q)=2|V|-q$. \\

Now, we show the optimal revenue is bounded above by $2|V| + (H+1)|E| - q$ when the minimum vertex cover has size $q$. By Lemma~\ref{lem:at_most_one_edge}, if an edge buyer $B_{(u,v)}$ is matched, it must be matched to some item $\alpha_{u,i}$ or $\alpha_{v, i}$ that it has positive value for. If $B_{(u,v)}$ is not matched, Lemma~\ref{lem:all_transact} says all items $\alpha_{u, i}$ and $\alpha_{v, i}$ are matched with vertex buyer $B_u, B_v$ or some dummy buyers. No matter which buyer $\alpha_{u, i}$ and $\alpha_{v, i}$ are matched with, the platform can always match $B_v$ to $S_v$, $B_u$ to $S_u$, and the dummy buyers to some other $\alpha$ buyers because the number of buyers and sellers are equal. In doing so, the platform gain revenue 1 from $B_{u,v}$, and loses at most revenue 1 from $B_u$ or $B_v$. So we can assume there is a revenue-optimal matching where all edge buyers $B_{(u,v)}$ are matched with the corresponding $\alpha$ seller. 

As the minimum vertex cover is of size $q$, edge buyers are matched to some items $\alpha_{u, i}$ with at least $q$ different vertex $u$. Thus, from each corresponding vertex buyer $B_u$, we can attain at most revenue $1$ because they are connected through opportunity path to $B_{(u,v)}$. It follows that the maximum revenue we can achieve is given by $2|V| + (H+1)|E| - q$, concluding the proof.
\end{proof}

\VertexCoverReduction*
\begin{proof}
    We have assumed the graph that we reduce from is connected. So $|E| \geq |V| - 1$. Further because of maximum degree $K$, $|E|\leq |V|K/2$. So $|E|=\Theta(|V|)$. Again because of maximum degree $K$, the minimum vertex cover has size $q$ at least $|E|/K = \Omega(|V|)$. 
    
    Let $\mathrm{Rev}^\star$ be optimal revenue and $q^\star$ be the minimum vertex cover. By Lemma~\ref{lem:vertice_cover_reduction}, $\mathrm{Rev}^\star=2|V|+(H+1)|E|-q^\star$. Then a constant factor approximation for $\mathrm{Rev}^\star$ translates into a constant factor approximation for $q^\star$, yielding a PTAS reduction and concluding the proof of APX-hardness.
\end{proof}

\section{Missing Proofs in Section~\ref{sec:special-case-homogeneous}}

\subsection{Proof of Theorem \ref{thm:opt_char}}
\label{app:swsh_char}

Here, we provide the proof of Theorem~\ref{thm:opt_char}, characterizing the optimal platform matching in SWSH markets.

\SWSHChar*

We start with the following lemma, which places a constraint of the length of any cycles in the optimal solution.
\begin{lemma}
    There exists an optimal set of transactions that can be decomposed into chains and/or cycles. Furthermore, there exists an optimal solution where all such cycles are of length at most $3$.
\end{lemma}
\begin{proof}
    Suppose that we had a cycle of length at least $4$. Then we could always split this cycle into smaller cycles of length at most $3$ (note that we can represent every integer larger than $3$ as a sum of multiples of $2$ and $3$). Furthermore, these cycles can only result in weakly higher revenue among sellers included in the original cycle as we are decreasing the number of opportunity paths. Finally, any chain connected to the original cycle can now be attached to the same subgraph in this new split set of cycles – again, there are fewer opportunity paths, so each seller in the chain also obtains weakly higher revenue. 

     It follows that there always exists an optimal solution where all cycles have length at most $3$.
\end{proof}

 We make the following simple observation about the optimal configuration of such cycles and chains.

\begin{remark}
    The cycle on $k$ subgraphs that yields optimal revenue connects the highest buyer in each subgraph together. The optimal-revenue chain connects subgraphs from smallest to largest in terms of their max-value bidder. The chain and cycle that yields optimal revenue is constructed by attaching the optimal chain to the largest second-highest bidder in any subgraph in the optimal cycle.
\end{remark}

 In fact, we can also show that there exists an optimal solution with only a single chain.
\begin{lemma}
    There exists an optimal solution with only a single chain. Additionally, this optimal solution can still restrict the cycle length to at most $3$.
\end{lemma}
\begin{proof}
    Suppose we had two chains, each attached to a different cycle, or potentially attached to the same cycle. We could always consolidate these two chains into a single chain and attach it to the cycle that yields the largest minimum opportunity path – all sellers in the chain now receive weakly more revenue than they were before. This is doable because each seller subgraph has at least one seller and one buyer to be connected into a chain. Note that this doesn't affect the restriction on cycle length. 
\end{proof}
To conclude the discussion of the structure of chains in the optimal solution, we show that these chains must be contiguous and moreover, chains ``fill out" the rest of the subgraphs once started.
\begin{lemma}
    If $S_i$ is part of a chain in the optimal solution, then $S_j$ is also part of this chain, for $j > i$, provided that $S_j$ is not part of the cycle that the chain attaches to.
\end{lemma}
\begin{proof}
    Suppose otherwise. Consider the optimal solution with at most one chain. Since $S_j$ is not part of a chain, it must be part of a cycle. As this solution is optimal, it must be the case that adding this whole cycle to the chain decreases the overall revenue. That is, the minimum opportunity path in $S_j$'s cycle is larger than the minimum opportunity path in the cycle that $S_i$'s chain is attached to. 

     However, this implies that we could add $S_i$ to $S_j$'s cycle, increasing the revenue generated from the seller in $S_i$ without affecting the revenue from any of the sellers in $S_j$ (as $S_i$ has a larger max-value buyer than $S_j$), which is a contradiction.
\end{proof}
\begin{lemma}
    Suppose that there is a chain in the optimal solution that connects to a cycle $\mathcal{C}$, and let $S_{min, \mathcal{C}}$ be the smallest subgraph belonging to $\mathcal{C}$. Then there exists an optimal solution where no subgraph larger than $S_{min, \mathcal{C}}$ is part of the chain.
\end{lemma}
\begin{proof}
    Suppose otherwise. If there are two or more such subgraphs, then we could always connect them in a cycle, and their minimum opportunity path would be weakly larger than that obtained by placing them in the chain. If there is a single such subgraph, we could add it to $\mathcal{C}$ – note that this does not affect the minimum opportunity path of the chain or of the sellers in $\mathcal{C}$ and weakly increases the revenue obtained from this subgraph. If at any point during this process, we obtain a cycle with more than three subgraphs, we can simply split it up into smaller cycles, preserving our desired property.
\end{proof}

Note that this implies that once a subgraph is part of a chain, all smaller subgraphs (in terms of max bidder value) must also be part of this chain. We make one final observation regarding the structure of the optimal solution – namely, all cycles in the optimal solution can be made to be contiguous.
\begin{lemma}
    There exists an optimal solution where all cycles are contiguous.
\end{lemma}
\begin{proof}
    Suppose we have an optimal solution satisfying all the above properties where at least one cycle is non-contiguous. Let $S_m$ be the smallest subgraph that belongs to a non-contiguous cycle (clearly $S_m \neq S_1$).

     Firstly, if $S_m$ is a part of a three-cycle $S_a - S_b -S_m$. If $a<b-1$, then we can connect $S_b - S_m$ in a two cycle. By Lemma 4.5, $S_{b-1}$ cannot be in the chain otherwise $S_b$ and $S_m$ would be in the chain. Thus, $S_{b-1}$ is in a cycle. We attach $S_a$ to the cycle that $S_{b-1}$ is in. This weakly increases $S_a$'s revenue because $S_{b-1}$ only connects to larger seller subgraphs.    

     Now we can focus on non-contiguous two-cycles $S_a - S_m$ and three-cycles $S_a - S_{a+1} - S_m$.
    Again, by Lemma 4.5, $S_{m-1}$ cannot be a part of the chain otherwise $S_m$ would have been in the chain as well.
    There are two cases. Suppose that the second highest bidder in $S_{m-1}$ has a weakly higher valuation than the highest bidder in $S_m$. Consider the following alternative configuration. Connect $S_m$ to $S_{m-1}$ as a chain. 
    (If a chain is already connected to $S_{m-1}$, add $S_m$ to this chain.)
    Combine the two cycles that $S_{m-1}, S_m$ belonged to (save for $S_m$, works also if $S_{m-1}$ is in a 1-cycle.) Additionally, keep any existing chains that were connected to any of the subgraphs involved in the two cycles. 
    
     Note that all subgraphs that were previously connected to $S_m$ via a cycle now have weakly increased revenue – the minimum opportunity path in the new cycle is now $S_{m-1}$ rather than $S_m$. All subgraphs that were previously connected to $S_{m-1}$ have the same revenue; all such subgraphs still have $S_{m-1}$ as their smallest opportunity path within the cycle. It is easy to verify that all chains also have weakly higher revenue. 

     Now, consider the second case where the second highest buyer in $S_{m-1}$ has a smaller valuation than the highest bidder in $S_m$, or $S_{m-1}$ does not have a second world buyer. There are now a couple of possible cases:
    \begin{itemize}
        \item Two or more subgraphs were previously connected to $S_m, S_{m-1}$. Connect $S_m, S_{m-1}$ in a two-cycle. In this case, connect all the subgraphs that were previously connected to $S_m, S_{m-1}$ together into a single cycle. Note that we lose $v(S_{m-1}) - v(S_m)$ in revenue from connecting $S_m$ and $S_{m-1}$ together, but we gain at least $v(S_{m-2}) - v(S_m)$ from the other subgraphs.
        \item $S_{m}$ belonged to a 2-cycle and $S_{m-1}$ belonged to a 1-cycle. If the other subgraph in the $S_m$ cycle was $S_{m-2}$, then connect $S_{m-2}, S_{m-1}$ together and leave $S_m$ as a 1-cycle. Note that we get weakly more revenue from $S_{m-2}$ in this case, and a 1-cycle with $S_m$ is less costly than a 1-cycle with $S_{m-1}$.
        \item $S_{m}$ belonged to a 2-cycle and $S_{m-1}$ belonged to a 1-cycle. Connect $S_m, S_{m-1}$ in a two-cycle. If the other subgraph in the $S_m$ cycle was not $S_{m-2}$, connect the other subgraph in $S_m$'s cycle to the cycle that $S_{m-2}$ is currently a part of. Note that we gain $v(S_m)$ from connecting $S_m$ and $S_{m-1}$ together, and at least $v(S_{m-2}) - v(S_m)$ from the other subgraphs.
    \end{itemize}
     In all the above cases, we keep all existing chains as they are – they are weakly better off under this new configuration because the second highest buyer in $S_{m-1}$ is smaller than $v(S_m)$. We continue this process until there are no non-contiguous cycles, at each step weakly increasing our revenue. If at step $t$, the smallest subgraph belonging to a non-contiguous cycle is $S_m$, then at step $t+1$, the process ensures that the smallest possible subgraph belonging to a non-contiguous cycle is $S_{m-1}$, meaning that this process must terminate.
\end{proof}

Combining the above lemmas, we have proved Theorem \ref{thm:opt_char}, which we restate below for completeness.

\SWSHChar*

\subsection{Proof of Theorem \ref{thm:swsh_reduction}}
\label{app:swsh_reduction}

Here, we complete the proof of Theorem~\ref{thm:swsh_reduction}, which states that there exists a polynomial-time algorithm to maximize the platform's revenue in general SWSH markets.

\SWSHReduction*

We prove the above theorem through two lemmas, addressing the case where $n > m$ and $n < m$ respectively. Note that the case in which $n = m$ is addressed by Theorem~\ref{thm: poly_AMOS}.

\begin{lemma}\label{lem:hom_sellers_transact_with_largest_buyers}
    When $n=|B|>|S|=m$, sort buyers by their valuation, and denote the set of the $m$ largest buyers by $B_m=\{b_1,b_2,...,b_m\}$. Then perform the revenue-optimal matching on $S, B_m$ while ignoring $B\setminus B_m$. This is revenue optimal.
\end{lemma}

\begin{proof}
    By Lemma \ref{lem:all_transact}, there exists a revenue-optimal transaction where all sellers trade. Otherwise, all sellers connecting to this non-trading seller through an opportunity path have a price of zero. By adding an edge between this seller and a non-trading buyer (guaranteed to exist as $n > m$), other sellers' prices weakly increase.
    
     Further, no buyer in $B\setminus B_m$ trades. Suppose otherwise and take any configuration where a buyer in $B \setminus B_m$ trades. Let $b_{min}$ be the smallest trading buyer, and let $b_{max}$ be the largest non-trading buyer, noting that $v(b_{min}) < v(b_{max})$. Suppose that $b_{min}$ currently transacts with seller $s$. 

     As the market clears according to the maximum weight matching, it must be that $s$ is not connected to $b_{max}$ – otherwise, matching them would yield a strictly higher welfare matching. Thus, we can add a platform edge from $s$ to $b_{max}$, ensuring that they transact in the new matching. Note that $b_{min}$ no longer transacts as it was the minimum value transacting buyer, and thus the maximum welfare matching can no longer include it. 
    
     We claim that all sellers' prices weakly increase. The only modified opportunity paths that previously existed are those which previously went through $b_{min}$ – as previously mentioned, there are no opportunity paths that terminate at a non-trading seller, so this was the minimum possible opportunity path. Now, no opportunity paths terminate at $b_{min}$ (as they no longer trade), meaning that the minimum opportunity path weakly increased (in the event of a tie with $b_{min}$). 

     We can continue this process until no buyers in $B \setminus B_m$ trade. Buyers not trading will not be in any opportunity path hence not affect the characterization results nor optimality of the procedure in Theorem~\ref{thm: poly_AMOS}.
\end{proof}

\begin{lemma}
\label{lemma:discard_dangling_sellers}
    When $n = |B| < |S| = m$, discard $m - n$ of the sellers who do not have world edges, and perform the revenue-optimal matching on the induced graph. This is revenue optimal.
\end{lemma}
\begin{proof}
    By Lemma \ref{lem:all_transact}, note that there exists a revenue optimal matching in which all buyers transact. Suppose that there is a seller $s$ who has at least one adjacent world edge who does not transact. Consider any buyer $b$ who is in seller $s$'s subgraph. 

     Clearly, buyer $b$ must transact with some seller $s'$ via a platform edge (as they do not transact with the seller in their subgraph). Consider eliminating this platform edge $(b, s')$. No revenue is lost from $b$ as they had an opportunity path to $s$, who did not transact, so they paid price $0$. Secondly, any sellers who previously had opportunity paths to $b$ also received price $0$, so removing this edge does not decrease their price either. 

     It follows that there exists a revenue optimal matching where all sellers with an adjacent platform edge transact, which implies that we can eliminate $m-n$ sellers with no adjacent world edges, as they would not transact in this revenue optimal matching regardless.
\end{proof}

Together, Lemmas~\ref{lem:hom_sellers_transact_with_largest_buyers} and \ref{lemma:discard_dangling_sellers} show that general SWSH markets can be reduced to the case in which $n = m$. Combining this with Theorem~\ref{thm: poly_AMOS} concludes the proof of Theorem~\ref{thm:swsh_reduction}.

\section{Missing Proofs in Section~\ref{sec:special-case-identical}}
\subsection{Proof of Lemma \ref{lemma:identity_char}}
\label{app:hv_identity}

Here, we prove Lemma~\ref{lemma:identity_char}, connecting the maximum revenue attainable in an identity-good market to the problem of finding buyer sets of non-positive surplus.

\IdentityChar*

We begin by first proving the following lemma, which characterizes the revenue attainable between a particular set of buyers and a set of sellers.

\begin{lemma}\label{cor:price_k_hall_violator}
    Given $k$ buyers and sellers $B_k=\{b_1,b_2,...,b_k\}, S_k=\{s_1,s_2,...,s_k\}$ where $B_k$ and $S_k$ have a perfect matching through only platform edges, total revenue $k$ can be attained if and only if $B_k$ belongs to a Hall violator $B_{vk}$ for graph $G^{-k}=(S\setminus S_k \cup B, E)$ of deficiency at least $k$.
\end{lemma}
\begin{proof}
Assume that $B_k$ belongs to a Hall violator $B_{vk}$ for graph $G^{-k}=(S\setminus S_k \cup B, E)$ of deficiency at least $k$. Since $B_{vk}$ has deficiency at least $k$, there exists a set of platform edges such that in the final maximum weight matching, $S_k$ transacts with $B_k$ and all sellers in $N(B_{vk})$ transact with buyers in $B_{vk}$. In this case, all of the buyers in $B_k$ only have opportunity paths to those sellers in $N(B_{vk})$, who all transact, so the platform obtains revenue $1$ from all these buyers, yielding revenue $k$ in total.

Now, suppose that total revenue $k$ can be attained from $B_k$ and $S_k$. Let $G_P^{-k}$ denote the corresponding platform graph, with the set of sellers $S_k$ removed. Let $B_{vk}$ be the set of all buyers on an opportunity path from any of the buyers in $B_k$. Note that all the sellers in $N_{G_P^{-k}}(B_{vk})$ must transact as otherwise at least one buyer in $B_k$ would transact at price zero, thus yielding less than $k$ revenue in total. Additionally, all these sellers must transact with a buyer in $B_{vk}$ as they are reachable via an opportunity path from $B_k$. Furthermore, all the sellers in $B_k$ transact with $S_k$. Thus, we have that $|B_{vk}| \geq |N_{G_P^{-k}}(B_{vk})| + k$. By Lemma \ref{lem:all_transact}, all platform edges transact, so it is also the case that $|B_{vk}| \geq |N_{G^{-k}}(B_{vk})| + k$, concluding the proof.
\end{proof}

Given $B_k, S_k$, finding such a Hall violator is equivalent to the max difference hall violator problem defined in Definition~\ref{def:max_dif_hall_violator}. The question for a revenue-optimizing platform is then to pick the sets $B_k, S_k$. The platform can first identify the buyers as follow.

\begin{corollary}\label{cor:price_k_hall_violator_buyer}
    The platform can extract revenue $k$ from a set of buyers $B_k$ if and only if there exists a buyer group $B_v$ where $B_k\subseteq B_v$ that satisfies the following two conditions. Let $E^{\neg}$ be the complement edges of the world graph, let $k_v\leq k$ be the value of maximum matching for $(B_k, N(B_v), E^{\neg})$.
    \begin{enumerate}
        \item Enough sellers to match with $B_k$ buyers $|S|-|N(B_v)|\geq k-k_v$
        \item $B_v$ has deficiency at least $k-k_v$
    \end{enumerate}
\end{corollary}
\begin{proof}
    If direction. Match the $k_v$ sellers in $N(B_v)$ through platform edges to $k_v$ buyers in $B_v$. For the rest of the $k-k_v$ buyers, match them with arbitrary sellers outside of $N(B_v)$. Denote the union of these two sets of sellers by $S_k$. Then $B_v$ is a Hall violator for graph $G^{-k}=(S\setminus S_k\cap B,E)$ of deficiency at least $k$. This satisfies Lemma~\ref{cor:price_k_hall_violator}. 
    
    Only if direction. Given a seller set $S_k$ that sells to $B_k$, let $B_v$ be the set of buyers reachable through opportunity paths from $B_k$, unioned with $B_k$ itself. $B_v$ satisfies the first criterion: at most $k_v$ buyers from $B_k$ can be matched with sellers from $N(B_v)$, the rest of the sellers are from $S\setminus N(B_v)$. All sellers in $N(B_v)$ transact with a unique buyer in $B_v$. If only $k'$ such sellers transact with $k'$ buyers in $B_k$, the rest of the $k-k'$ buyers must transact with sellers in $S\setminus N(B_v)$. Thus, $|B_{v}|=|N(B_{v})|+(k-k')$ and $B_v$ has deficiency $k-k'$. As $k'\leq k_v$, $B_v$ has deficiency weakly larger than $k-k_v$ in the platform graph. This does not change in the original world graph.
\end{proof}

\begin{corollary}\label{cor:price_k_hall_violator_vertex_set_revenue}
    Given a buyer set $B_v$ with non-positive surplus in $G$, 
    let $E^{\neg}$ be the complement edges of the world graph, and let $k_v$ be the value of maximum matching for $(B_v, N(B_v), E^{\neg})$. The platform can extract $$k = k_v + \min\{|B_v|, |S|\}-|N(B_v)|$$
    revenue from the buyers in $B_v$, and this is the maximum possible revenue achievable from the buyers in $B_v$. 
\end{corollary} 
\begin{proof}
    Firstly, note that the stated revenue is weakly smaller than the number of buyers in $B_v$: $k\leq k_v+|B_v|-|N(B_v)|\leq |B_v|$ since $k_v\leq |N(B_v)|$. We pick the buyer set $B_k$ as follows: 1) Take $k_v$ buyers in the maximum matching in $(B_v, N(B_v), E^{\neg})$, 2) Additionally, take $\min\{|B_v|, |S|\}-|N(B_v)|\geq 0$ buyers in $B_v$ besides the $k_v$ ones. We show that $B_k$ satisfies the two conditions in Corollary ~\ref{cor:price_k_hall_violator_buyer}.
    \begin{eqnarray*}
        k-k_v = \min\{|B_v|, |S|\}-|N(B_v)| =  \min\{|B_v|-N(B_v), |S|-N(B_v)\}
    \end{eqnarray*}
    The first and second terms in the min correspond to the first and second conditions of Corollary~\ref{cor:price_k_hall_violator_buyer} respectively. Additionally, a larger $k$ will violate at least one of the conditions so $k$ is the maximum attainable revenue.
\end{proof}

Combining the above results, we state and prove Lemma~\ref{lemma:identity_char}.

\IdentityChar*
\begin{proof}
    Suppose that the optimal revenue is $k$ and the platform matches $S_k$ to $B_k$. Corollary~\ref{cor:price_k_hall_violator_buyer} states that there is a set $B_k\subseteq B_v$ with non-positive surplus. Corollary~\ref{cor:price_k_hall_violator_vertex_set_revenue} gives the max revenue from $B_v$. As $k$ is the optimal revenue in the entire graph, it is also the maximum revenue from $B_v$. Thus, $k=k_v + \min\{|B_v|, |S|\}-|N(B_v)|$. If $k\geq x$, then $k_v + \min\{|B_v|, |S|\}-|N(B_v)|\geq x$. Conversely, if for all buyer sets $B_v$ with non-positive surplus, $k_v + \min\{|B_v|, |S|\}-|N(B_v)|< x$, then the optimal revenue is given by $k<x$.
\end{proof}

\subsection{Proof of Lemma \ref{lemma:max_cardinality}}
\label{app:max_cardinality}

In this section, we give the proof of Lemma~\ref{lemma:max_cardinality}, simplifying the platform's problem in SHGB markets to the problem of finding the maximum cardinality set of buyers with non-positive surplus. To do so, we first show in sparse graph with buyer degree at most 2, any set of buyers $B_v$ with non-positive surplus satisfies $k_v = |N(B_v)|$, except for certain special cases.

\begin{lemma}
\label{lemma:full_matching}
    Consider any set of buyers $B_v$ with non-positive surplus in SHGB markets. if $|B_v| \geq 3$ and there are more than 2 degree-two buyers in $B_v$, then $k_v = |N(B_v)|$.
\end{lemma}
\begin{proof}
    Construct the flow network $G^{\neg}_{\infty}$  from the complement graph  $G^{\neg}=(B_v, N(B_v),E^{\neg})$ in such a way that there are edges of capacity 1 from the source $p$ to every vertex in $N(B_v)$ and from every vertex in $B_v$ to the sink $q$, and of capacity $+\infty$ from $s$ to $b$ for any $(s,b)\in E^{\neg}$.
    \begin{figure}[!h]
        \centering
        \includegraphics[width=0.5\textwidth]{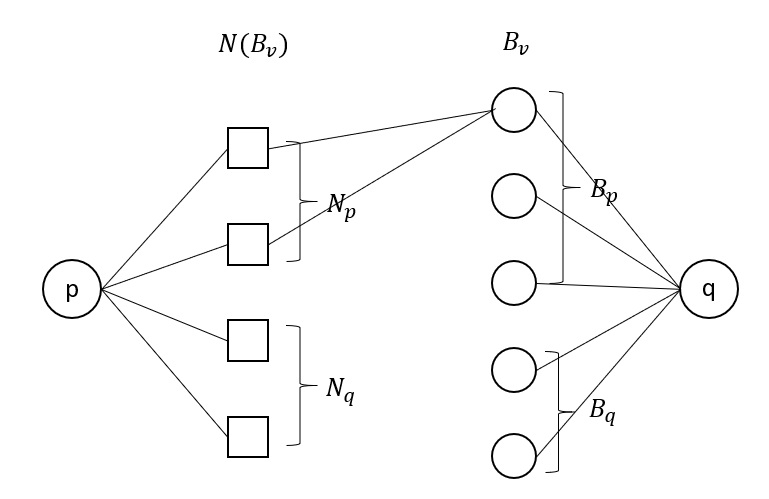}
    \end{figure}
    The size of the maximum matching in $G^{\neg}$ equals to that of the max flow/min cut in $G^{\neg}_{\infty}$. Let $(P,Q)$ be a min cut and $C\leq |N(B_v)|$ be the cut value. Let $N(B_v)=N_{p}\cup N_{q}, B_v=B_{p}\cup B_{q}$, such that $N_{p},B_{p}\subset P, N_{q}, B_{q}\subset Q$. The minimum cut is composed of edges going from $p$ to $N_{q}$ and from $B_{p}$ to $q$, as any edges between $N_q, B_p$ and $N_p, B_q$ would make the size of the cut infinite.

    We proceed by casework:
    \begin{itemize}
        \item Firstly, note that if any degree-zero buyer is in $B_q$, then it must be that $N_p = \emptyset$ as degree-zero buyers are adjacent to all sellers in $N(B_v)$ in the complement graph. Thus, $C \geq |N_q| = |N(B_v)|$.
        \item If $|B_q| = 0$, then we have that $C \geq |B_v| \geq |N(B_v)|$.
        \item Suppose that $|B_q| = 1$. As all buyers have degree at most $2$, it must be that $|N_p| \leq 2$. thus, we have that $C \geq |N_q| + |B_p| \geq (|N(B_v)| - 2) + (|B_v| - 1) \geq |N(B_v)|$ since $|B_v| \geq 3$.
        \item Suppose that $|B_q| = 2$. In any case, it must be that $|N_p| \leq 1$ since either we have a degree-one buyer present in $B_q$ or two degree-two buyers (in which case they cannot share the same two sellers). Thus, $C = |N_q| + |B_p| \geq  (|N(B_v)| - 1) + 1 = |N(B_v)|$.
        \item Suppose that $|B_q| \geq 3$. If there are at least $3$ degree-two buyers in $B_q$, then since a pair of sellers knows at most one common buyer, it follows that $|N_p| = 0$. Thus, we have that $C \geq |N(B_v)|$.

        Otherwise, there must be at least one degree-one buyer in $B_q$ so $|N_p| \leq 1$. Additionally, at least one degree-two buyer must be excluded from $B_q$. Thus, we have that $C = |N_q| + |B_p| \geq  (|N(B_v)| - 1) + 1 = |N(B_v)|$.
    \end{itemize}

    Thus, if $|B_v| \geq 3$ and there are more than $2$ degree-two buyers in $B_v$, it follows that $k_v = |N(B_v)|$, concluding the proof.
\end{proof}

We need one more lemma before proving Lemma~\ref{lemma:max_cardinality}. Lemma~\ref{lem:all_degree} shows that there exists an optimal set of buyers $B_v$, in the sense of Lemma~\ref{lemma:identity_char}, in which all buyers of degree zero and degree one are included.

\begin{lemma}
\label{lem:all_degree}
    There exists a set of buyers $B_v$ maximizing
    \begin{align*}
        \min(|B_v|, |S|) - |N(B_v)| + k_v
    \end{align*}
    such that all buyers of degree zero and degree one are included in $B_v$.
\end{lemma}
\begin{proof}
    Clearly, any degree-zero buyers can only contribute positively to the above expression. Thus, all buyers of degree zero are WLOG in the optimal set $B_v$. 
    
    Turning to the case of buyers of degree one, consider any alternative set of buyers $B'_v$. There are two cases for the change in the above expression when adding a buyer of degree one. Let $s_j$ be the seller adjacent to this buyer. If $s_j \in N(B'_v)$, note that the above expression can only increase.

    If $s_j \not \in N(B'_v)$, then there are again two cases. If $|B'_v| < |S|$, then the above expression can again only increase – there is an increase of $1$ from the first term and a decrease of $1$ from the second term. If $|B'_v| \geq |S|$, then since $s_j \not \in N(B'_v)$, we claim that the $k_v$ term must strictly increase. Consider any previous maximum matching between $B'_v$ and $N(B'_v)$. Take an arbitrary buyer $b'$ in $B'_v$ (note that one must exist since $|B'_v| \geq |S|$) and match them to $s_j$ while matching the new buyer to the seller $b'$ was matched to in the maximum matching (if one exists). This increases the cardinality of the maximum matching by $1$, so it follows that the value of above expression does not decrease. This concludes the proof.
\end{proof}

Given the above lemma, we now prove Lemma~\ref{lemma:max_cardinality}.

\MaxCard*
\begin{proof}
    Restricting our search to sets of buyers $B_v$ that satisfy the conditions of Lemma~\ref{lemma:full_matching}, we have that $k_v = |N(B_v)|$. This corresponds to a set of $B_v$ such that $|B_v| \geq 3$ and there are more than 2 degree-two buyers in $B_v$. By Lemma~\ref{lemma:identity_char}, the revenue from this given set of buyers $B_v$ is given by
    \begin{align*}
        \min(|B_v|, |S|) - |N(B_v)| + k_v = \min(|B_v|, |S|) - |N(B_v)| + |N(B_v)| = \min(|B_v|, |S|)
    \end{align*}
    so maximizing the platform's revenue over these sets is equivalent to finding the maximum cardinality such set of buyers.

    Now, consider the sets of buyers $B_v$ that do not satisfy the conditions of Lemma~\ref{lemma:full_matching}. These are sets of buyers with non-positive surplus $B_v$ such that either $|B_v| < 3$ or there are no more than $2$ degree-two buyers in $B_v$. Clearly, we can enumerate all $B_v$ such that $|B_v| < 3$ and check the revenue attainable from each of these sets. For those $B_v$ containing no more than $2$ degree-two buyers in $B_v$, Lemma~\ref{lem:all_degree} proves that it is WLOG to begin with all degree zero and degree one buyers in $B_v$. Thus, we can simply ennumerate over all $O\left(\binom{n}{2}\right)$ choices of two degree-two buyers to add, check the revenue attainable from each of these sets. This brute force check takes polynomial time, concluding the proof.
\end{proof}

\subsection{Proof of Theorem~\ref{thm:poly_identity}}
\label{app:poly_identity}

Here, we provide the proof of Theorem~\ref{thm:poly_identity}, showing that we can maximize the platform's revenue in SHGB markets in polynomial time. We begin with the following lemma, which will form the crux of our proof.

\begin{lemma}
\label{lemma:induced_edges}
    Consider a world graph where all buyers are of degree two and for every pair of sellers, at most one buyer knows them both. The maximum cardinality set $B_v$ of buyers with surplus at most $k$ – that is, $|N(B_v)| - |B_v| \leq k$ – can be identified in polynomial time.
\end{lemma}
\begin{proof}
    Note that such world graphs are in direct correspondence with general graphs $G = (V, E)$ where each buyer corresponds to an edge and each seller corresponds to a vertex. Specifically, $b_i$ corresponds to an edge $(s_i, s_j)$ between the vertices representing sellers $s_i, s_j$ whom $b_i$ is connected to in the original world graph.

    Restated in these terms, finding the maximum cardinality set $B_v$ of buyers with surplus at most $k$ is equivalent to finding the maximum cardinality set of edges $\tilde{E} \subseteq E$ such that the induced subgraph of $G$ corresponding to $\tilde{E}$ has at most $|\tilde{E}| + k$ vertices.

    We claim that $\tilde{E}$ must correspond to a union of connected components of $G$. Suppose that $\tilde{E}$ contained only a subset of edges within a given connected component $C$ of $G$. By adding all the edges in $C$, the number of vertices in the induced subgraph can increase by at most the number of additional edges, not harming the surplus of the induced subgraph in the process. It follows that $\tilde{E}$ must correspond to a union of connected components of $G$ as desired.

    Thus, our algorithm to find the maximum cardinality set of edges $\tilde{E} \subseteq E$ such that the induced subgraph of $G$ corresponding to $\tilde{E}$ has at most $|\tilde{E}| + k$ vertices proceeds as follows:
    \begin{itemize}
        \item Find all the connected components of $G$. If a connected component is not a tree, it contributes more edges than vertices, so we can add it to $E_{cur}$.
        \item Having added all non-tree connected components, compute the difference between the number of edges and number of vertices in the induced subgraph of $E_{cur}$. Call this difference $\ell$.
        \item Take the $k + \ell$ largest (in terms of the number of edges) connected components of $G$ that are trees and add their edges to $E_{cur}$. Return the final set as $\tilde{E}$.
    \end{itemize}
    We now prove the correctness of our algorithm below. Clearly, $\tilde{E}$ must contain all connected components of $G$ that are not trees. Given this fact, the remaining edges in $\tilde{E}$ must correspond to connected components of $G$ that are trees. Our algorithm takes precisely those trees with the largest number of edges, completing the proof.
\end{proof}

\PolyIdentity*
\begin{proof}
    It suffices to show that we can find the maximum cardinality set $B_v$ of buyers of non-positive surplus in polynomial time in SHGB markets, which together with Lemma~\ref{lemma:max_cardinality} implies the desired result.

    All buyers of degree zero and degree one contribute at most one to $|N(B_v)|$ while contributing at least one to $|B_v|$. Thus, they must all be included. Upon including all these buyers, consider the induced subgraph obtained by removing all sellers adjacent to a buyer of degree one or zero. This induced subgraph may have more buyers of degree one or zero, which by the same argument as above, must be included in $B_v$. We can repeat this process until we are left with a current set of buyers $B'_v$ which must be included in $B_v$ as well as a graph in which all buyers are of degree two and for every pair of sellers, at most one buyer knows them both.

    Thus, our problem reduces to finding the maximum cardinality set of these buyers with surplus at most $|B'_v| - |N(B'_v)|$. Lemma~\ref{lemma:induced_edges} shows that this is solvable in polynomial time, concluding the proof.
\end{proof}

\section{Missing Proofs in Section~\ref{sec:hom_markets}}
\subsection{Proof of Theorem~\ref{thm:hom_min_nm_approx}}\label{app:min_nm_approx_rev}
Here, we provide the proof of the following theorem, giving a $\min\{n, m\}$ approximation algorithm for revenue in general homogeneous-goods markets.
\HomoMinNM*

In order to guarantee this approximation ratio, we consider a buyer-seller pair $(b_i, s_j)$ and aim to find the maximum possible revenue the platform could extract from the single transaction between these two agents. Asking this same question for all possible pairs, we are guaranteed to obtain the maximum possible revenue the platform could attain from any single transaction. As there at most $\min\{n, m\}$ transactions conducted via the platform, being able to achieve this maximum transaction in polynomial time would prove Theorem~\ref{thm:hom_min_nm_approx}.

We begin by noting that we can restrict our attention to transactions involving buyers among those with the top $\min\{n, m\}$ highest valuations.

\begin{lemma}
    Up to ties, the buyer-seller pair $(b_i, s_j)$ from which the platform can extract the maximum revenue always involves a buyer among those with the top $\min\{n, m\}$ highest valuations. Additionally, the maximum revenue is either zero or at least $v_{\min\{n, m\}}$, where this is the $\min\{n, m\}^{th}$ largest valuation.
\end{lemma}
\begin{proof}
    The statement is trivial if $m \geq n$. Otherwise, if $n > m$, then either the top $m$ buyers already have all possible world edges, in which case we can extract $0$ revenue from the market or there exists a buyer among the top $m$ buyers that is missing at least one world edge. In this case, we can add this as a platform edge and match off the remaining top $m - 1$ buyers, guaranteeing us a revenue of at least $v_{(m)}$, where this is the $m^{th}$ largest valuation. As we can extract at most $v_i$ from buyer $b_i$, it follows that the revenue-optimal pair $(b_i, s_j)$ must include one of the top $m$ buyers.
\end{proof}

We will show that the maximum price for a $(b_i, s_j)$ pair is closely connected to the notion of Hall violators, most notable for their use in matching theory \citep{lovasz2009matching}.

\begin{definition}[Hall violator]
    A set $B_v$ of buyers is a Hall violator if $|B_v| > |N(B_v)|$, where $N(B_v)$ is the set of sellers connected to any buyer in $B_v$ via a world edge.
\end{definition}

Next, we state the relationship between Hall violators and the maximum price attainable from a buyer-seller pair. Let $\tilde{B}$ denote the set of buyers who have one of the top $\min\{n, m\}$ valuations. Let $v_{(\min\{n, m\})}$ denote the $\min\{n, m\}^{th}$ largest valuation. By Lemma 

\begin{lemma}
\label{lemma:hall_violator_prices}
    Consider a buyer-seller pair $(b_i, s_j)$, where $b_i \in \tilde{B}$ is not connected to $s_j$ via a world edge. There exists a set of platform edges the platform can add such that $s_j$ sells to $b_i$ at a positive price $k \geq v_{(\min\{n, m\})}$ if and only if $b_i$ belongs to a Hall violator $B_{vi}\subseteq \tilde{B}$ for $G^{-j} = (S \setminus \{s_j\} \cup \tilde{B}, E)$ where $k \leq \min_{b_i \in B_{vi}} v_i$.
\end{lemma}
\begin{proof}
    Suppose there exists a set of platform edges such that $b_i$ transacts with $s_j$ at a positive price $k \geq v_{(\min\{n, m\})}$. Consider the set of all buyers reachable via an alternating/opportunity path from $b_i$. As $k > 0$, this set of buyers $B_{vi}$ forms a Hall violator in $G^{-j}$; otherwise, there would be an opportunity path from $b_i$ to an unsold item. Additionally, $s_j$ transacts at price $\min_{b \in B_{vi}} v(b)$ by Theorem~\ref{thm:oppo_path}. Since $k \geq v_{(\min\{n, m\})}$, $B_{vi}$ cannot contain any buyers outside of $\tilde{B}$. 

    Now, suppose we have some Hall violator $B_{vi} \subseteq \tilde{B}$ of $G^{-j}$ that includes $b_i$, where $k \leq \min_{b_i \in B_{vi}} v_i$. We claim that the platform can add edges $E_P$ and induce platform graph $G_P=(S\cup B,E\cup E_P)$ such that in the max weight matching:
    \begin{itemize}
        \item Sellers in $N_{G_P}(B_{vi})$ all transact.
        \item The sellers in $N_{G_P}(B_{vi})$ only transact with the buyers in $B_{vi} \setminus b_i$.
        \item $s_j$ transacts with $b_i$.
    \end{itemize}

    Consider the graph $G^{-j}$. As we have a Hall violator $B_{vi}\subseteq \tilde{B}$ of this graph, it follows that $|N_{G^{-j}}(B_{vi})| < |B_{vi}| \implies |N_{G^{-j}}(B_{vi})| \leq |B_{vi} \setminus \{b_i\}|$. Note that any matching that perfectly matches the top $\min\{m, n\}$ buyers to the top $\min\{m, n\}$ sellers is a maximum weight matching.

    Thus, we add the following set of platform edges: connect $b_i$ to $s_j$, connect the largest $|N_{G^{-j}}(B_{vi})|$ buyers in $B_{vi} \setminus \{b_i\}$ to $N_{G^{-j}}(B_{vi})$, and connect any extra buyers in $\tilde{B}$ arbitrarily to the remaining sellers not in $N_{G^{-j}}(B_{vi})$. We can then arbitrarily break ties in favor of the max weight matching that matches $b_i$ with $s_j$, that matches $N_{G^{-j}}(B_{vi})$ with the buyers in $B_{vi} \setminus \{b_i\}$, and that matches the rest of the buyers in $\tilde{B}$ arbitrarily to sellers such that everyone in $\tilde{B}$ transacts. Note that this matching does indeed match the top $|S|$ buyers to the top $|S|$ sellers, so it must be a maximum weight matching.

    By construction, since $b_i \in B_{vi}$ and $N_{G}(B_{vi})$ only transacts with the largest buyers in $B_{vi}$, it follows that the minimum opportunity path must point to a buyer in $B_{vi}$ and the price is thus bounded below by $\min_{b_i \in B_{vi}} v_i \geq k$.
\end{proof}

As a subroutine that will be used in our final algorithm, we show that we can efficiently decide, given a graph $G$, whether there exists a Hall violator containing a buyer $b$. To begin, we define the following graph-theoretic problems.

\begin{definition}[vertex Hall violator problem]
    Given a bipartite graph $G=(B\cup S,E)$ and a vertex $b\in B$, is there a Hall violator that contains $b$?
    \label{def:hall_violator_vertex}
\end{definition}
\begin{definition}[max-difference Hall violator problem]
    Given a bipartite graph $G=(B\cup S, E)$ and a positive integer $k$, is there a Hall violator $B_v\subset B$ such that $|B_v|-|N_G(B_v)|\geq k$?
    \label{def:max_dif_hall_violator}
\end{definition}

\begin{lemma}
\label{lemma:hv_reduction}
    There is a polynomial-time reduction from the vertex Hall violator problem to the max-difference Hall violator problem.
\end{lemma}
\begin{proof}
    We show that the vertex Hall violator problem can be reduced in polynomial time to the max-difference Hall violator problem. Given an instance $(G, b)$ of the vertex Hall violator problem, let $G'=(B\setminus\{b\}\cup S\setminus\{N_G(b)\},E)$ and $k=|N_G(b)|$. Then, solve the max-difference Hall violator problem on the instance $(G', k)$. If there is a yes-certificate $B_{v}\subseteq B\setminus \{b\}$ for the max-difference Hall violator problem, then $B_v\cup\{b\}$ is a Hall violator that includes $b$ in $G$. This is because by the yes-certificate
    \begin{eqnarray*}
        |B_v| - |N_{G'}(B_v)| & \geq & k = |N_G(b)|\\
        |B_v\cup\{b\}| - 1 & \geq & |N_{G'}(B_v)|+|N_G(b)| = |N_G(B_v\cup \{b\})|
    \end{eqnarray*}
    If $(G', k)$ is a no-instance of the max-difference Hall violator problem, then there does not exist a Hall violator that contains $b$ in $G$.  Otherwise, if $B_v\cup\{b\}, b\notin B_v$ were such a Hall violator, then
    \begin{eqnarray*}
        |B_v\cup\{b\}| -1 & \geq & |N_G(B_v\cup\{b\})| = |N_{G'}(B_v)|+|N_G(b)|\\
        |B_v| - |N_{G'}(B_v)| & \geq & |N_G(b)| = k
    \end{eqnarray*}
    which contradicts with $(G',k)$ being a no-instance.
\end{proof}

Finding the maximum-difference Hall violator is closely related to the notion of the deficiency of a graph, defined below.

\begin{definition}
\label{def:deficiency}
    Define the deficiency of a subset $B_v$ as $\mathrm{def}(B_v) = \max \left(|B_v| - |N_G(B_v)|, 0 \right)$. Similarly, define the deficiency of $B$ as $\mathrm{def}(G; B) = \max_{B_v \subseteq B} \mathrm{def}(B_v)$.
\end{definition}

For bipartite graphs, we have the following well-known fact about the deficiency of a graph $G$.

\begin{lemma}
\label{lem:deficiency}
    Let $m$ be the size of the maximum-matching on $G$. Then
    \begin{align*}
    \mathrm{def}(G; B) = |B| - m
    \end{align*}
\end{lemma}

Moreover, one can find the subset of maximum deficiency in polynomial time.

\begin{theorem}
    There is a polynomial-time algorithm to find the max-difference Hall violator (or equivalently, the subset of $B$ with maximum deficiency).
\end{theorem}\label{thm:poly_max_diff_violator}
\begin{proof}
    Note that finding the max-difference Hall violator is precisely equivalent to finding the subset of maximum deficiency. Start with a maximum matching $M$ on $G$, and let $B_{unmatched}$ be the subset of vertices in $B$ that are not saturated by $M$. If $B_{unmatched}=\emptyset$ then Hall theorem says there are no Hall violator. So we only consider $B_{unmatched}\neq \emptyset$. Let $B_{max}$ be the set of all vertices that are reachable via an alternating path from $B_{unmatched}$.

    We claim that the set $N_G(B_{max})$ is fully saturated by $M$. Consider any $s \in N_G(B_{max})$. By construction, $s$ is reachable from some unmatched vertex via an alternating path. If $s$ were not saturated by $M$, this would represent an augmenting path, which is a contradiction since $M$ is maximal.

    Additionally, we have that every $s \in N_G(B_{max})$ is matched to a vertex in $B_{max}$, again by construction of $B_{max}$ by the opportunity path. Thus, there is a bijection between matched vertices in $B_{max}$ and $N_G(B_{max})$, so by Lemma \ref{lem:deficiency}, it follows that $|B_{max}| - |N_G(B_{max})| = |B_{unmatched}| = \mathrm{def}(G; B)$, and $B_{max}$ is the max-difference Hall violator.
\end{proof}

As a corollary of Lemma~\ref{lemma:hv_reduction} and Lemma~\ref{thm:poly_max_diff_violator}, we can solve the vertex Hall violator problem in polynomial time.

\begin{corollary}
    The vertex Hall violator problem can be solved in polynomial time.
\end{corollary}
\begin{proof}
    By Lemma~\ref{lemma:hv_reduction}, the vertex Hall violator problem can be reduced to the max-difference hall violator problem, which can be solved in polynomial time using the algorithm described in Lemma~\ref{thm:poly_max_diff_violator}.
\end{proof}

Finally, we have all the tools necessary to conclude the proof of Theorem~\ref{thm:hom_min_nm_approx}. We restate the theorem, along with the proof below.

\begin{theorem}
    Given a buyer-seller pair $(b_i, s_j)$ not connected via a world edge, with $b_i \in \tilde{B}$, we can find the set of platform edges that maximizes the revenue generated from this pair in polynomial time.
\end{theorem}
\begin{proof}
    By Lemma~\ref{lemma:hall_violator_prices}, we simply need to find the Hall violator $B_{vi} \subseteq \tilde{B}$ containing $b_i$ of $G^{-j} = (S \setminus \{s_j\} \cup \tilde{B}, E)$ that maximizes the minimum valuation among all buyers in $B_{vi}$.

    Note that $s_j$ sells to $b_i$ at one of $|\tilde{B}|$ prices. To find the max revenue, we can simply check these candidate prices by iteratively checking if there is a Hall violator $B_{vi}$ for $G^{-j} = (S \setminus \{s_j\} \cup \tilde{B}, E)$ such that $b_i \in B_{vi}$, removing the lowest-value bidders from the graph at each round.

    That is, we look for a Hall violator containing $b_i$, allowing all buyers in $\tilde{B}$ to be used. We repeat this process, allowing all buyers except for $b_{(m)}$ to be used. Next, we allow all buyers except for $b_{(m)}, b_{(m - 1)}$ to be used, continually removing the buyers with the smallest values until no such Hall violator exists. Note that this will indeed find us the Hall violator that includes $b_i$ with optimal maximin value.

    After finding the Hall violator with the optimal maximin value, the second statement of Lemma~\ref{lemma:hall_violator_prices} gives a way to construct platform edges to sell $s_j$ to $b_i$ at such price, concluding the proof.
\end{proof}

\HomoMinNM*
\begin{proof}
    The revenue optimal platform matching has at most $\min\{n,m\}$ edges that transact. One of the edges will generate more than $\frac{1}{\min\{n,m\}}$. By Lemma 13.8, the platform can iterate through all buyer-seller pairs $(b_i, s_j)$ and check for the maximum revenue attainable from this pair. The best pair guarantees a $\frac{1}{\min\{n, m\}}$-fraction of the optimal revenue.
\end{proof}
\chapter{Appendix to \Cref{chap:simulations}}\label{app:ch5}

\section{Subscription-Level Decision}
\label{app:model}

\label{app:counterfactual_estimates}



\paragraph{Calculations for Subscription Effect Estimation}
We first describe the subscription effect estimates for each type of the agents. 
To recap, these estimates, denoted $\xi_{k+1}$, are conducted based on the new platform fees and world transaction friction, under the same set of queries submitted or received in the last epoch (i.e., epoch $k$) and the same subscription decisions made by other agents.
%
%
\begin{itemize}
	\item \textbf{A buyer $b$, on-platform in epoch $k$}. 
	We denote the estimate of epoch surplus if buyer $b$ does not subscribe to the platform as $\xi^w_{b, k+1}$.
	For this, given each query of $b$  in epoch $k$, the choice in the world under new friction $\mu_{k+1}$ (Section~\ref{sec:choice_world_only}) is reevaluated
	\[\small \xi^w_{b, k+1} = \sum_{t \in k} u^w_{b, t} (\mu_{k+1}) = \sum_{t \in k} \max\{u_{\B}(q_{b,t}, s^*_w) - \mu_{k+1}, 0\},\]
	where we denote $u^w_{b, t} (\mu)$ the world matching surplus for buyer $b$ at $t$ under the world friction $\mu$, i.e., $u^w_{b, t} (\mu) = \max\{u_{\B}(q_{b,t}, s^*_w) - \mu, 0\}$.
	Similarly, we denote $\xi^p_{b, k+1}$ the estimate of epoch surplus if buyer $b$ remains on platform.
	The new friction may affect the choice between a platform seller and a world seller. 
	This surplus is estimated based on updated decisions under $\mu_{k+1}$, with 
	\[\small \xi^p_{b, k+1} = - P_{\B, k+1} + \sum_{t \in k} \max \{u^w_{b, t} (\mu_{k+1}), u^p_{b, t}\}.\]
	
	\item \textbf{A buyer $b$,  off-platform in epoch $k$}. 
	The surplus estimate of remaining off-platform depends on the new friction, i.e., 
	$\xi^w_{b, k+1} = \sum_{t \in k} u^w_{b, t} (\mu_{k+1})$.
	For $\xi^p_{b, k+1} $, we need to estimate the epoch surplus if $b$ subscribes to the platform.  We assume that buyer $b$ can observe platform-recommended sellers, e.g., this can be from trial periods to gain platform experience or a platform's estimate based on past orders.
	For a query sequence $\{q_{b,t}\}_{t \in k}$,  denote the corresponding best platform-recommended sellers as $\{s^{p*}_{b,t}\}$.  The estimated surplus if subscribing to the platform is
	\[\small \xi^p_{b, k+1} = - P_{\B, k+1} + \sum_{t \in k} \max \{u^w_{b, t} (\mu_{k+1}), u_{\B}(q_{b,t}, s^{p*}_{b,t})\}.\]

	\item \textbf{A  seller $s$, on-platform in epoch $k$}. 
	For $\xi^w_{s, k+1}$, we need to estimate the epoch surplus if $s$ does not subscribe to the platform, by reasoning about 
	(1)~how many more world transactions would happen if $s$ is not on the platform, and
	(2)~how buyers' transaction decisions may be affected by the epoch $k+1$ world friction. 
	To facilitate a precise estimate, we assume that seller $s$ can observe the sequence of query and seller candidate tuples $\{(q_{b, t}, s, s^p_{b,t})\}_{t\in k}$ in which seller $s$ is chosen as the best world option.
	Given $q_{b,t}$, $\S_k \wo s$, and a fixed platform matching strategy used in epoch $k$, we denote the updated, best-platform seller as $s^{p*}_{b,t}$.
	For this modified sequence $\{(q_{b, t}, s, s^{p*}_{b,t})\}$, we consider the choices buyers will make under the new friction $\mu_{k+1}$, and estimate the number of transactions  seller $s$ will receive without being on the platform, denoted $n^{w'}_{s, k}$.
	Thus, the estimated surplus if seller $s$  is off platform is 
	$\xi^w_{s, k+1} = n^{w'}_{s, k} v^1_s (1 - \omega_s)$.
	
	For $\xi^p_{s, k+1}$, we reason about how the new friction affects the number of transactions.
	Given the sequence $\{(q_{b, t}, s^w_{b,t}, s)\}_{t \in k}$ where $s$ is picked as the best platform seller, we consider  each buyer's choice and estimate the number of platform visits under $\mu_{k+1}$, denoted $n^{p^*}_{s, k}$.
	Given the sequence $\{(q_{b, t}, s, s^p_{b,t})\}$ where $s$ is picked as the best world seller, we estimate the number of world transactions under $\mu_{k+1}$ (even if seller $s$ chooses to stay on the platform), denoted $n^{w^*}_{s, k}$.
	Thus, the estimated  surplus that seller $s$ would receive  by subscribing to the platform is \[\small \xi^p_{s, k+1} = - P_{\S, k+1} + n^{p^*}_{s, k} v^1_s (1 - \omega_s - P_{R, k+1}) + n^{w^*}_{s, k} v^1_s (1 - \omega_s).\]
	
	\item \textbf{A seller $s$, off-platform in epoch $k$}. 
	For $\xi^p_{s, k+1}$, we reason about the surplus from subscribing to the platform, e.g., by asking the platform for an estimate on the number of platform transactions. 
	Given the sequence of queries on the platform, i.e., $\{q_{b, t}\}_{b \in \B_k, t\in k}$, the platform can update the matches it would suggest $s$ also on-platform by following its matching strategy. 
	Denote the sequence of query and updated seller matches with tuples $\{(q_{b, t}, s^w_{b, t}, s^{p^*}_{b, t})\}_{b \in \B_k, t\in k}$. 
	With $\mu_{k+1}$, we estimate the  numbers of platform transactions $n^{p^*}_{s, k}$ and world transactions $n^{w^*}_{s, k}$.
	The estimated surplus if seller $s$  subscribes to the platform is 
	\[\small \xi^p_{s, k+1} = - P_{\S, k+1} + n^{p^*}_{s, k} v^1_s (1 - \omega_s - P_{R, k+1}) + n^{w^*}_{s, k} v^1_s (1 - \omega_s).\]
	
	Seller $s$ also adjusts the surplus that they would get by remaining off platform, reasoning about how buyers' transaction decisions will be affected by the new world friction.
	Given the sequence of query and matched seller  tuples $\{(q_{b, t}, s, s^p_{b,t})\}_{b \in \B, t \in k}$ in which seller $s$ was chosen as the best world seller, we re-evaluate each buyer's choice to get $n^{w'}_{s, k}$.
	The estimated off-platform surplus is 
	\[\small \xi^w_{s, k+1} = n^{w'}_{s, k} v^1_s (1 - \omega_s).\]
\end{itemize}

\paragraph{Calculations for Agent-Specific Decision Inertia} 
\label{sec:apx-inertia}

There is a rich body of literature that establishes \emph{decision inertia}, modeling the presence of such inertia across different markets (see Section~\ref{sec:choice_subscribe}). 
In our setting an agent, either a buyer or seller, in  subscription state $\I^p_k \in \{0, 1\}$, is prone to stay in the same state in epoch $k+1$, due to habit formation, loyalty, or inattention. 
We treat buyers and sellers in the same way and illustrate the concept with a buyer $b$  for simplicity.
%
%
Each buyer starts with an initial preference in regard to adopting the platform or not, denoted by an integer $\chi_{b,0} \sim U[-\chi, \chi]$ for some integer $\chi$. 
%
If the initial $\chi_{b,0}$ is positive, the buyer subscribes at the warm-up epoch, otherwise they do not. 
The  inertia $\chi_{b,k}$ maps into an additive bonus to either the surplus for joining the platform or remaining in the world through the functional form:  
\begin{equation}
	\sigma^p_{b, k+1} := \I^p_{b, k} \log \Parens{\chi_{b,k}}, 
	\qquad
	\sigma^w_{b, k+1} := (1-\I^p_{b, k}) \log \Parens{-\chi_{b,k}}.%
\end{equation}
That is, the longer an agent sticks to their decision, the larger the bias term gets, which increases in a concave way (logarithmically) over time.
Such interpretation of decision inertia as an additive bonus is common in the literature~\citep{MacKay2021,dube2009switching,farrell1988dynamic}. 
Based on this adjusted utility, agents  decide whether to subscribe or not according to probabilities inferred by the standard {\em discrete-choice logit} model~\citep{MacKay2021,dube2009switching}, where the probability of subscribing to the platform is
$$\delta^p_{b, k+1} := \frac{\exp \Parens{\xi^p_{b,k+1} + \sigma^p_{b, k+1}}}{\exp \Parens{\xi^p_{b,k+1} + \sigma^p_{b, k+1}} + \exp \Parens{\xi^w_{b,k+1} + \sigma^w_{b, k+1}}}$$

The inertia is updated after each decision in the following way. 
If $\chi_{b,k}>0$ and the buyer subscribes, then $\chi_{b,k+1}:=\chi_{b,k}+1$, and otherwise $\chi_{b,k+1}:=-1$ (i.e., it resets for off-platform).
Similarly if $\chi_{b,k}<0$ and the buyer decides to stay off-platform, then $\chi_{b,k+1}:=\chi_{b,k}-1$, and otherwise reset $\chi_{b,k+1}:=1$. 

\section{The Platform Matching POMDP}
\label{app:pdp}
Based on the fee-setting POMDP (Section~\ref{sec:pricing_policy}), we make the following adjustments to define the matching POMDP:
\begin{itemize}
	\item We define $x_k \in \X$ for epoch $k$ as the state of the market after the platform has set fees and agents have chosen to subscribe, but still before the first query is submitted.  
	That is, the state includes agent-subscription states $\I^p_{\B, k}$ and $\I^p_{\S, k}$ and  platform fees, i.e., $P_{\B, k}, P_{\S, k}, P_{R, k}$, for the upcoming epoch $k$.
	We also include the matching utility threshold adopted in the previous epoch, $\eta_{k-1}$, to state $x_k$.  
	All other elements of the state remain the same.
	\item Here, the platform's action $a_k$ chooses (i) the matching utility threshold for epoch $k$, and (ii) the matching rule for the epoch, whether seller-aware or profit-driven.
	For the threshold, we consider $\eta_k$ that takes discrete values in $[0,1]$.
	\item Different from the fee-setting POMDP, the state transition for platform matching starts with buyer queries: (1)~a buy agent generates a query, (2)~if the buyer is on platform, the platform recommends a seller based on $\eta_k$ and its matching rule, and (3)~the buyer selects a seller with whom to transact (Section~\ref{sec:choice_transact}).
	This gives the full sequence of queries, $Q_k$.
	At the end of epoch $k$, buyers and sellers observe the new fees $P_{\B, k+1}$, $P_{\S, k+1}$ and $P_{R, k+1}$, as given by a fixed fee schedule, and decide whether or not to subscribe to the platform for the next epoch.\looseness=-1 
	%
	\item The reward $r_k \sim \R(x_k, a_k)$ of the platform for epoch $k$ includes both the referral fees from epoch $k$ and the subscription fees collected from buyers and sellers, reflecting the decision in regard to whether or not to join the platform for epoch $k+1$.
	%
	This avoids delayed reward for the  platform, as the registration decisions, and thus registration fees for the next epoch, are influenced by the platform's matching strategy (i.e., action $a_k$) during  epoch $k$. 
	%
	%
	\item The platform's observability of information in the state follows the same as for the fee-setting POMDP. 
\end{itemize}

\section{Deferred Materials for Section~\ref{sec:experiments}}
\label{app:experiments}
\subsection{Market Environments}
\label{app:market_structures}
%
\begin{figure}[t]
	\centering
	\begin{subfigure}{0.32\columnwidth}	
		\centering
		\includegraphics[width=\columnwidth]{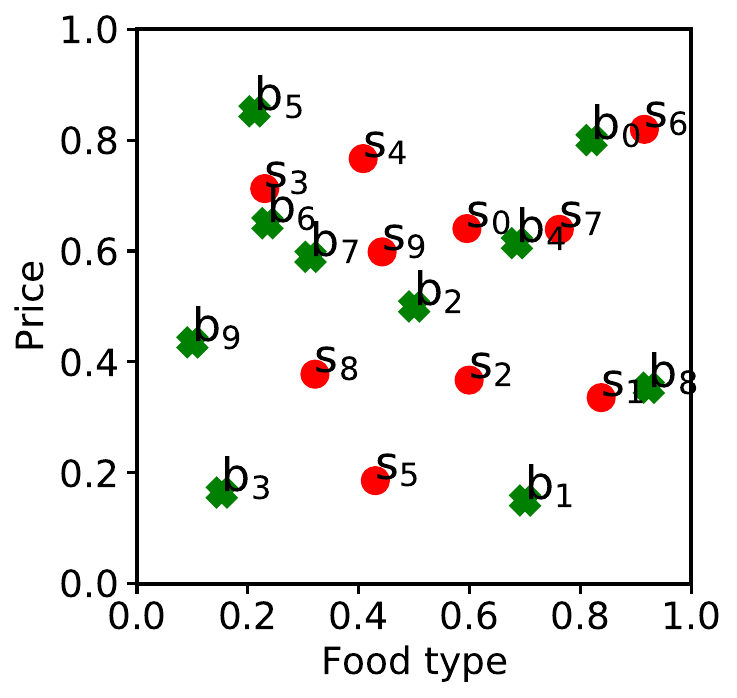}
		\caption{Uniform.}
	\end{subfigure}
	\begin{subfigure}{0.32\columnwidth}	
		\centering
		\includegraphics[width=\columnwidth]{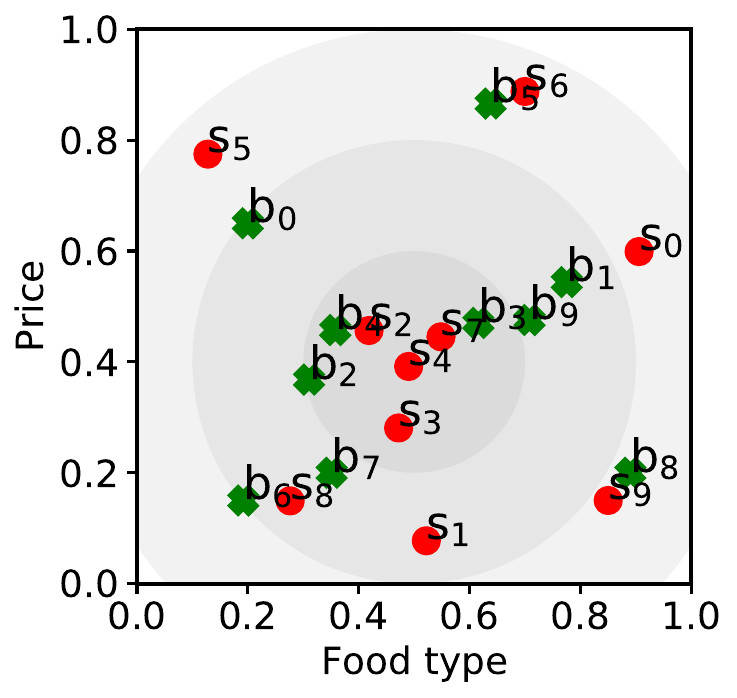}
		\caption{Core-and-Niche.}
	\end{subfigure}
	\begin{subfigure}{0.32\columnwidth}	
		\centering
		\includegraphics[width=\columnwidth]{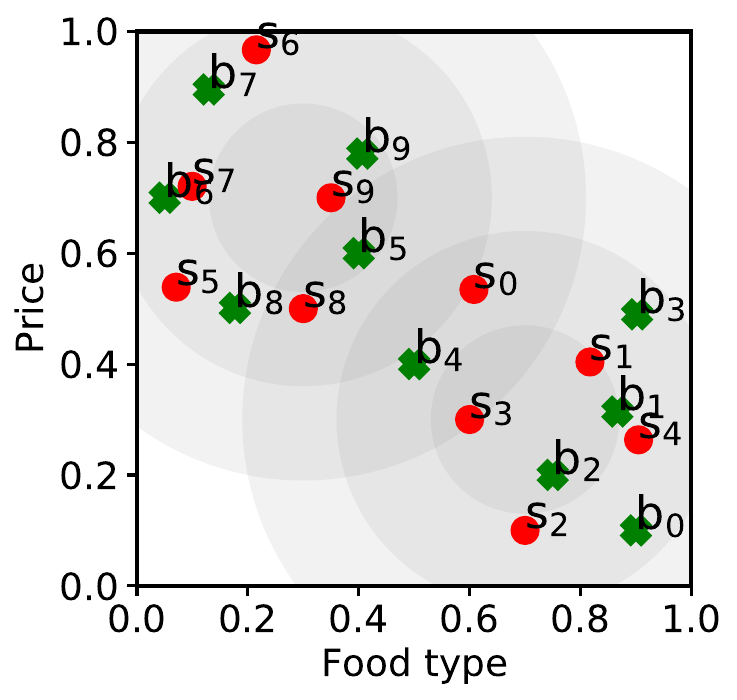}
		\caption{Two-Core.}
	\end{subfigure}
	\caption{Three types of latent locations of buyers and sellers.
	}
	\label{fig:three_structures}
\end{figure}
\begin{figure}[t]
	\centering
	\includegraphics[width=0.55\columnwidth]{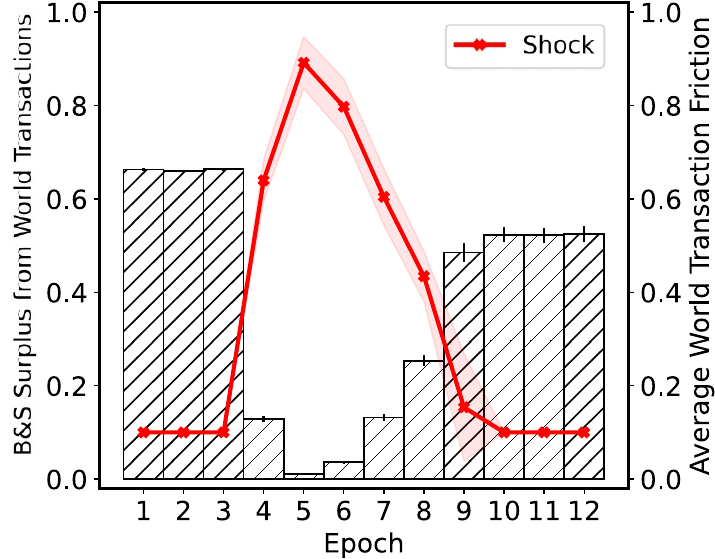}
	\caption{Buyer and seller surplus in markets without a platform across the 12 epochs with a full cycle of economic shock.
		\label{fig:welfare_decomposition}}
\end{figure}

%
%
Figure~\ref{fig:welfare_decomposition} red line plots the average shock of two-hundred simulation runs.
Each run includes a pre-shock stage (epoch 1-3) with $\mu=0.1$, a shock stage (epoch 4-9) where we sample the shock intensity $I \sim U[0.8, 1]$ and the world transaction frictions according to Section 4.1, and a post-shock stage with $\mu=0.1$ (epoch 9-12). 

\vspace{-1ex}
\subsection{Implementation Details}
\label{app:nn}
\paragraph{Observation features.}
Following our POMDP formulation, we make the following observations available to the platform:
\begin{itemize}
	\item On-platform buyers and sellers, represented by two binary vectors, and their latent locations, 
	\item Summary statistics of on-platform agents, including the number of platform transactions and platform surplus accumulated so far within an epoch, 
	\item The platform matching and transaction matrices between {\em on-platform} buyers and sellers for the past epoch, 
	\item The platform fees, the matching rule and utility threshold (if learn matching), and the current epoch's world friction. 
\end{itemize}

\begin{figure}[t]
	\centering
	\includegraphics[width=0.95\columnwidth]{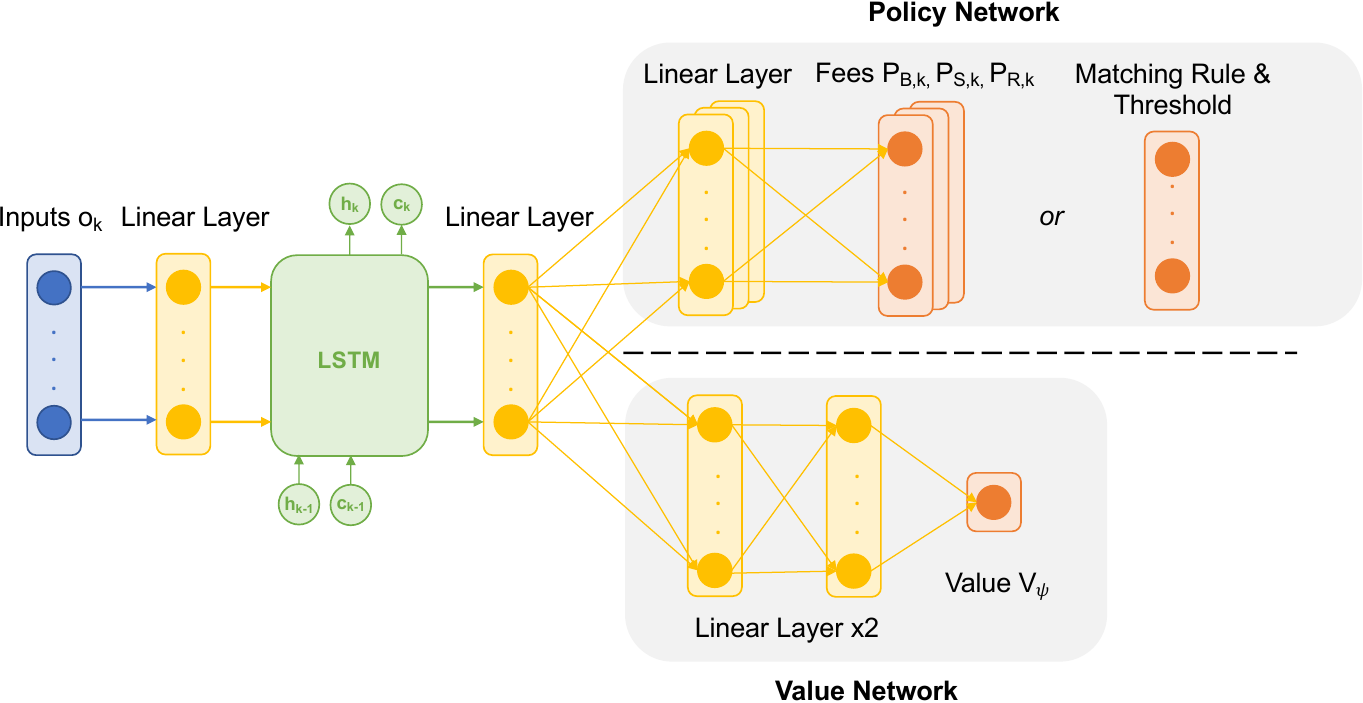}
   \vspace{-1ex}
	\caption{Neural network structure for the platform policies.
		\label{fig:NN_diagram}}
    \vspace{-3ex}
\end{figure}

\paragraph{Neural network structure and training parameters.}
Based on preliminary explorations, we design the actor and critic to share a fully-connected layer, LSTM cells of size 128, and again a fully-connected layer to recover sufficient statistics of the history, using this to in effect infer the knowledge structure of buyers and the demand elasticity of agents to platform fees.
Each network also has its own two fully-connected layers. 
The critic outputs the value $V_\psi(o)$ of an observation $o$, and the actor gives policy $\pi_\theta$ for an observation $o$. 
For the fee-setting actor, this includes three separate output layers, with each returning a vector of probabilities for one type of platform fee.
For the matching actor, this is a vector of probabilities over the matching utility thresholds that are applicable to both matching rules. 
Figure~\ref{fig:NN_diagram} illustrates the neural network structure we implement.
Besides the  policy gradient loss, we apply {\em entropy regularization} to the policy network to encourage exploration.  
The respective losses for the policy network and the value network are,
$\mathcal{L}_{\pi} = -\log \pi(a_k | o_k; \vtheta)(R_k - V_\psi(o_k)) - \beta \mathcal{H}(\pi(A_k | o_k ; \vtheta)) \text{ and } \mathcal{L}_{V} = (R_k - V_\psi(o_k))^2,$
where $\mathcal{H}$ denotes the entropy over learned action probabilities.
%
%
%

We tune the platform agent with various combinations of learning rates $\{0.0001, 0.0005, 0.001\}$, batch sizes $\{4, 16, 32, 64, 128\}$, and entropy weights $\{0.001, 0.01, 0.05\}$, and select hyperparameters that maximize the objective function. 
%
We report detailed training parameters in the supplemental material.

\chapter{Appendix to \Cref{chap:bundling}}\label{app:ch6}

\section{Missing lemmas and proofs in \Cref{sec:ch6_model}}
\label{sec:proofs.model}
We show it here with \Cref{app:maximizer_attained} and \Cref{app:maximizer_attained_bundle} that an optimal price exists for $Z$ with unit variance and a density function.
\begin{lemma}\label{app:maximizer_attained}
    For any zero mean, unit variance random variable $Z$ with a density function $f$ and tail distribution $F$, and any $\mu,\sigma>0$, there exists $p^*\in [0,\infty)$ such that $$p^*F[\frac{p^*-\mu}{\sigma}]=\sup_{p} p F[\frac{p-\mu}{\sigma}]$$
\end{lemma}
\begin{proof}
    Since $Z$ has a density $f$, the tail function $F[\cdot]$ and $pF[\frac{p-\mu}{\sigma}]$ is continuous. Let $\eta=\frac{p-\mu}{\sigma}$. By $Z$ having unit variance and Chebyshev's inequality $$\forall \eta>0, \quad Pr[Z\geq \eta] \leq Pr[|Z|\geq \eta] \leq \frac{1}{\eta^2}$$
    Thus, for any $\eta>0$ 
    $$0< (\mu+\sigma\eta)F[\eta] \leq \frac{\mu+\sigma\eta}{\eta^2}$$
    This means that $$\lim_{p\rightarrow\infty} p F[\frac{p-\mu}{\sigma}]=\lim_{\eta\rightarrow\infty}\frac{\mu+\sigma\eta}{\eta}=0$$
    Since $pF[\frac{p-\mu}{\sigma}]$ is contiguous, and tends to $0$ at infinity, there exists $\bar{P}>0$ such that $$\forall p > \bar{P},\quad pF[\frac{p-\mu}{\sigma}]< 1$$
    On the compact interval $[0,\bar{P}]$, continuity implies that $pF[\frac{p-\mu}{\sigma}]$ attains a maximum: there exists $p^*\in [0,P]$ such that $$p^*F[\frac{p^*-\mu}{\sigma}]=\max_{p\in [0,P]}pF[\frac{p-\mu}{\sigma}]=\sup_{p\geq 0}pF[\frac{p-\mu}{\sigma}]$$
    where the latter equality follows from $\forall p>\bar{P},\quad pF[\frac{p-\mu}{\sigma}]<1$ and $\lim_{p\rightarrow\infty} p F[\frac{p-\mu}{\sigma}]=0$.
\end{proof}

\begin{lemma}\label{app:maximizer_attained_bundle}
    For any zero mean, unit variance, i.i.d. random variables $\{Z\}_{i=1}^N$ with a common density function $f$ and tail distribution $F$, and any positive $\{\mu\}_{i=1}^N,\sigma$, there exists $p^*\in [0,\infty)$ such that $$p^*Pr\left[\sum_{i\in [N]}\mu_i+\sigma Z_i\geq p\right]=\sup_{p} p Pr\left[\sum_{i\in [N]}\mu_i+\sigma Z_i\geq p\right]$$
\end{lemma}
\begin{proof}
    Define $$Z'=\frac{\sum_{i\in [N]}\sigma Z_i}{\sqrt{\sum_{i\in [N]}\sigma^2}},\quad \mu'=\sum_{i\in [N]}\mu_i,\quad \sigma'=\sqrt{\sum_{i\in[N]}\sigma^2}$$
    By $Z_i$ being i.i.d. with zero mean, unit variance with a density function, it follows that $Z'$ is of zero mean and unit variance, and with some density function. Then $$p Pr\left[\sum_{i\in [N]}\mu_i+\sigma Z_i\geq p\right] = p Pr\left[Z'\geq \frac{p-\mu'}{\sigma'}\right]$$
    As the proof for \Cref{app:maximizer_attained} did not impose any assumption on the tail distribution of $F$, the same proof as \Cref{app:maximizer_attained} follows.
\end{proof}

\begin{restatable}{lemma}{RevMuSigma}
\label{lem:rev_mu_sigma}
    When $Z$ is a standard normal random variable, for any $\mu,\sigma>0$, $Rev(\mu,\sigma) = \mu Rev(1,\frac{\sigma}{\mu})=\sigma Rev(\frac{\mu}{\sigma},1)$.
\end{restatable}
\begin{proof}
    For any buyer valuation $v=\mu+\sigma Z$, let $v'=v/\mu$. When $Z$ is a standard normal random variable, $v\sim \mathcal{N}(\mu,\sigma^2)$ and $v'= X/\mu\sim \mathcal{N}(1,(\frac{\sigma}{\mu})^2)$. For any price $p$, let $p'= p/\mu$ and $Pr[v\geq p]=Pr[\mu v' \geq \mu p']=Pr[v'\geq p']$. The optimal revenue can be written as $$Rev(\mu,\sigma)=\max_{p\geq 0} p Pr[X\geq p]=\max_{p'\geq 0} \mu p' Pr[v'\geq p'] = \mu \max_{p'\geq 0} p' Pr[v'\geq p']$$
    But then as $Rev(1,\frac{\sigma}{\mu})=\max_{p'\geq 0} p' Pr[v'\geq p']$, this means that $Rev(\mu,\sigma)=\mu Rev(1,\frac{\sigma}{\mu})$.

    Similarly we can set $v'' = v/\sigma\sim \mathcal{N}(\frac{\mu}{\sigma},1)$ and for any price $p$, let $p''=p/\sigma$. Then $Pr[v\geq p]=Pr[v''\sigma \geq p''\sigma]= Pr[v''\geq p'']$. So the revenue can be written as 
    $$Rev(\mu,\sigma)=\max_{p\geq 0} p Pr[v\geq p]=\max_{p''\geq 0} p''/\sigma Pr[v''\geq p'']=\frac{1}{\sigma}Rev(\frac{\mu}{\sigma},1)$$
\end{proof}

\RevMuSigmaConvex*
\begin{proof} 
    Any price $p$ can be parameterized as $p=\mu+\sigma \eta$ where $\eta$ denotes the normalized price. As \Cref{app:maximizer_attained} shows the supremum exists and is attained at some optimal price,
    \begin{align*}
        Rev(\mu,\sigma) &= \max_p p Pr[\mu+\sigma Z\geq p] = \begin{cases}
            \max_{\eta} (\mu+\sigma\eta) F[\eta] \quad &\text{ if } \sigma>0\\
            \max_{\eta} (\mu+\sigma\eta)\quad &\text{ if } \sigma=0
        \end{cases}\\
        &= \max_{\eta} (\mu+\sigma\eta)F[\eta]
    \end{align*}
    Fixing $\eta$, $(\mu+\sigma\eta)F[\eta]$ is convex in $(\mu,\sigma)$. As maximum of convex function is convex, hence $Rev(\mu,\sigma)$ is convex in $(\mu,\sigma)$.
    
    Take two means $\mu\geq \mu'$. For the same price $p$, $Pr[\mu+\sigma Z\geq p]\geq Pr[\mu'+\sigma Z\geq p]$. Therefore for all $p\geq 0$, $p\cdot Pr[\mu+\sigma Z\geq p] \geq p\cdot  Pr[\mu'+\sigma Z\geq p]$. Taking the maximum over $p\geq 0$ on both sides yields $Rev(\mu,\sigma)$ being weakly increasing.
\end{proof}

In the paper, we also make use of the following lemma on $Rev(\mu,\sigma)$ when deriving proofs for normal buyer valuations.
\begin{restatable}[\cite{schmalensee1984gaussian}]{lemma}{NormalRevCharacter}
\label{lem:norm_rev_character}
    With buyer valuations normally distributed with mean $\mu$ and standard deviation $\sigma$, it holds that $$F[z^*(\alpha)-\alpha] = z^*(\alpha)f(z^*(\alpha)-\alpha) ; \quad 0<z^*_\alpha(\alpha)=\frac{d z^*(\alpha)}{d\alpha}=\frac{z^*(z^*-\alpha)-1}{z^*(z^*-\alpha)-2}<1 ; \quad \frac{d F[z^*(\alpha)-\alpha]}{d\alpha}>0$$
    The first and the second partial derivatives of $Rev(\mu,\sigma)$ satisfy:
    \begin{align*}
        \frac{\partial Rev(\mu,\sigma)}{\partial \mu} &= F[z^*(\alpha)-\alpha] > 0; & \quad \frac{\partial^2 Rev(\mu,\sigma)}{\partial\mu^2}& >0 \\
        \frac{\partial Rev(\mu,\sigma)}{\partial \sigma} &= (z^*(\alpha)-\alpha)F[z^*(\alpha)-\alpha]; & \quad \frac{\partial^2 Rev(\mu,\sigma)}{\partial \sigma^2}&>0
    \end{align*}
    There exists a threshold $\alpha_0\approx 1.253$ such that $$z^*(\alpha)-\alpha \geq 0 \text{ if and only if } \alpha \leq \alpha_0$$ 
\end{restatable}
\begin{proof}
    \cite{schmalensee1984gaussian} explicitly writes out the proof except the second order derivatives. For completeness we provide the proof.  
    Expanding the optimal revenue $Rev(\mu,\sigma)=p^*F[z^*(\alpha)-\alpha]=\sigma z^*(\alpha) F[z^*(\alpha)-\alpha]$. As $z^*$ being the optimal solution to revenue maximization, the first order condition requires that $$\sigma F\left[z^*(\alpha)-\alpha)\right]-\sigma z^*(\alpha) f(z^*(\alpha)-\alpha)=0$$
    Differentiating the above with regards to $\alpha$ and using the derivative of standard normal distribution $f'(x) =-xf(x)$ yields the expression for $z^*_\alpha(\alpha)$. Using the second order condition for revenue maximization determines the sign of $z_\alpha^*(\alpha)$. For the first and second order derivative, 
    \begin{align*}
        \frac{\partial Rev(\mu,\sigma)}{\partial\mu} &= \frac{\partial \sigma z^*(\alpha)F[z^*(\alpha)-\alpha]}{\partial\alpha}\frac{d\alpha}{d\mu} = z^*_\alpha(\alpha)F[z^*(\alpha)-\alpha]-z^*(\alpha)f(z^*(\alpha)-\alpha)(z_\alpha^*(\alpha)-1)\\
        &= z_\alpha^*(\alpha)[F[z^*(\alpha)-\alpha]-z^*(\alpha)f(z^*(\alpha)-\alpha)]+z^*(\alpha)f(z^*(\alpha)-\alpha)= F[z^*-\alpha]>0\\
    \frac{\partial^2 Rev(\mu,\sigma)}{\partial\mu^2}&=\frac{\partial F[z^*-\alpha]}{\partial \alpha}\frac{d\alpha}{d\mu}=-f(z^*-\alpha)[z^*_\alpha(\alpha)-1]/\sigma>0\\
    \frac{\partial Rev(\mu,\sigma)}{\partial \sigma} &= z^*(\alpha)F[z^*(\alpha)-\alpha]+\sigma \frac{d z^*(\alpha)F[z^*(\alpha)-\alpha]}{d\alpha}\frac{d\alpha}{d\sigma}=
    (z^*(\alpha)-\alpha)F[z^*(\alpha)-\alpha]\\
    \text{Letting } y(\alpha) &:=z^*(\alpha)-\alpha\\
    \frac{\partial^2 Rev(\mu,\sigma)}{\partial \sigma^2} &= \frac{d y(\alpha)F[y(\alpha)]}{d y}\frac{dy}{d\alpha}\frac{d\alpha}{d\sigma}= (F[y]-yf(y))(z^*_\alpha(\alpha)-1)(-\frac{\mu}{\sigma^2})\\
    &= [z^*(\alpha)f(z^*(\alpha)-\alpha)-(z^*(\alpha)-\alpha)f(z^*(\alpha)-\alpha)](z^*_\alpha(\alpha)-1)(-\frac{\mu}{\sigma^2})\\
    &= \frac{\mu}{\sigma^2}(1-z^*_\alpha(\alpha))\alpha f(z^*(\alpha)-\alpha)>0
    \end{align*}
\end{proof}

\section{Missing Proofs in \Cref{sec:complete_info_heterogeneous_mu}}
\label{sec:app_complete_info_heterogeneous_mu}
\ContiguousMuGeneral*
\begin{proof}
    For any seller bundle $S$, buyer valuation for the bundle is $\sum_{i\in S}v_i = \sum_{i\in S}\mu_i + \sigma \sum_{i\in S}Z_i$. Denote the tail distribution of $\sum_{i\in S}Z_i$ as $\Psi$, and $\mu_S =  \sum_{i\in S}\mu_i$. Any price can be expressed as $p=\mu_s + \sigma \eta$ where $\eta$ is the normalized price. The demand at normalized price $\eta$ is $$Pr\left[\sum_{i\in S}Z_i\geq \eta\right]=\Psi(\eta)$$
    And the revenue at normalized price $\eta$ is $(\mu_S+\sigma\eta)\Psi(\eta)$. Let $\Pi(S,\mu_S+\sigma\eta)$ be the platform's profit for selling $S$ at normalized price $\eta$, and it is written as $$\Pi(S,\mu_S+\sigma\eta)=\Psi(\eta)\eta \sigma + \sum_{i\in S}[\mu_i\Psi(\eta)-Rev(\mu_i,\sigma)]$$
    Fixing $\eta$ and size of the bundle $S$, $\Psi(\eta)\eta\sigma$ is fixed because $\Psi(\cdot)$ only depends on the size of the bundle. We show for any fixed $\eta$ and fixed size of the bundle, the profit-maximizing set of sellers is contiguous in quality, with the same technique as in the proof for \Cref{thm:contiguous_mu_normal}.

    Define a function $G(\mu) = Rev(\mu, \sigma) - \mu \Psi(\eta)$. By \Cref{lem:rev_mu_sigma_convex_in_mu_sigma}, $Rev(\mu,\sigma)$ is convex in $\mu$, so is $G(\mu)$. Assume the profit-maximizing bundle has size $|S|=N-1$ and define $S_{-k}=\{1,2,...,k-1,k+1,...,N\}$. Then 
    \begin{align*}
        \Pi(S_{-k},\mu_{S_{-k}}+\eta\sigma)= \Psi(\eta)\eta\sigma + \sum_{i\in [N]} [\mu_i\Psi(\eta)-Rev(\mu_i,\sigma)] + G(\mu_k)
    \end{align*}
    $\Pi(S_{-k},\mu_{S_{-k}}+\eta\sigma)$ is the profit of size-$N-1$ bundle excluding seller $k$, at normalized price $\eta$. This function only depends on the parameter $k$. For any $\eta$, as $G(\mu)$ is convex, $\Pi(S_{-k},\mu_{S_{-k}}+\eta\sigma)$ takes maximum value when $k\in\{1,N\}$, which means it is optimal to select all but sellers $i\in\{1,N\}$. As this analysis is general for any fixed size of $S$ (not only limited to $|S|=N-1$) and $\eta$, the optimal bundle is contiguous in $\mu$.
\end{proof}

\section{Missing Lemmas and Proofs in \Cref{sec:incomplete_info_heterogeneous_mu}}
\label{sec:app_incomplete_info}
\begin{example}
\label{exam:mean_zero_unit_variance_non_subexponential}
Let $Z = \frac{Z'}{\sqrt{3}}$ where $Z'$ is a drawn from a student's t distribution with parameter $\nu=3$. Since $E[Z']=0$ and $Var[Z']=\frac{\nu}{\nu-2}=3$, it holds that $E[Z]=0, Var[Z]=1$. It is well known that a student's t distribution is heavy-tailed, with a polynomial decay rate, and it does not have a moment generating function. This means that there does not exists any $h>0$, such that for all $\lambda$ satisfying $-h <\lambda <h$, $E[e^{\lambda Z'}]$ exists, or 
$$\forall\lambda\neq 0,\quad E[e^{\lambda Z'}]=\infty$$
So $E[e^{\lambda Z}]$ also diverges for any $\lambda$, meaning $Z$ is not sub-exponential.   
\end{example}

\begin{restatable}{lemma}{RegularZ}
\label{lem:rev_diff_mu}
    For any zero mean, unit variance random variable $Z$ with a density function $f$ and tail distribution $F$ that satisfies $$\eta - \frac{F[\eta]}{f(\eta)}$$
    being strictly increasing in $\eta$, then the optimal price 
    $p^*=\argmax_p p F[\frac{p-\mu}{\sigma}]$ is unique, and the optimal revenue $Rev(\mu,\sigma)$ is differentiable in $\mu$
    $$\frac{\partial Rev(\mu,\sigma)}{\partial \mu}=F\left[\frac{p^*-\mu}{\sigma}\right]$$
\end{restatable}
\begin{proof}
    \Cref{app:maximizer_attained} shows that some optimal price $p^*$ exists. We now show that it is unique. Denote an optimal price $p^*=\mu+\sigma\eta^*$. As $p^*>0$, it is an interior optimizer, the first order condition for $p^*$ maximizing $Rev(\mu,\sigma)$ requires that $$F\left[\frac{p^*-\mu}{\sigma}\right]=\frac{p^*}{\sigma}f\left(\frac{p^*-\mu}{\sigma}\right)  \Leftrightarrow \frac{p^*}{\sigma}=\frac{F\left[\frac{p^*-\mu}{\sigma}\right]}{f\left(\frac{p^*-\mu}{\sigma}\right)} \Leftrightarrow \eta^* - \frac{F[\eta^*]}{f(\eta^*)} = -\frac{\mu}{\sigma}$$
    As $\eta - \frac{F[\eta]}{f(\eta)}$ being strictly increasing, there is at most one solution to $\eta - \frac{F[\eta]}{f(\eta)}=-\frac{\mu}{\sigma}$. This shows uniqueness.

    For differentiability, we invoke Theorem 2 and Corollary 4 in \citep{milgrom2002envelope}, which presents a version of the envelop theorem. The following conditions are satisfied, listed in the order of conditions in the Theorem 2 of \citep{milgrom2002envelope}.
    \begin{itemize}
        \item $pF\left[\frac{p-\mu}{\sigma}\right]$ is absolute continuous in $p$ and $\mu$ because there exists a density function.
        \item The optimal price $p^*$ is taken in a compact space: In the proof of \Cref{app:maximizer_attained}, we showed that for unit variance $Z$, the optimal price $p^*$ for $Rev(\mu,\sigma)$ is attained at some compact space $[0,\bar{P}]$ for any $\mu, \sigma>0$. The compactness means the "integrable bound" condition of Theorem 2 in \cite{milgrom2002envelope} is satisfied. 
        \item $pF\left[\frac{p-\mu}{\sigma}\right]$ is differentiable in $\mu$ for all $p$, and is equal to $$\frac{d pF\left[\frac{p-\mu}{\sigma}\right]}{d\mu}=\frac{p}{\sigma}f\left(\frac{p-\mu}{\sigma}\right)$$
        \item There is a unique optimal price $p^*$, so the following set is a singleton $$\Bigl\{\frac{d p^*F\left[\frac{p^*-\mu}{\sigma}\right]}{d\mu} \,\vert \,p^*\in \argmax_p p F\left[\frac{p-\mu}{\sigma}\right]\Bigr\}$$ 
    \end{itemize} 
    Given all these conditions are satisfied, Corollary 4 in \citep{milgrom2002envelope} reads $Rev(\mu,\sigma)$ is differentiable and satisfies 
    $$\frac{\partial Rev(\mu,\sigma)}{\partial \mu} = F\left[ \frac{p^*-\mu}{\sigma} \right]$$
\end{proof}

\Myerson*
\begin{proof}
    First prove monotonicity. For any seller of quality $\mu$, IC requires that 
    \begin{align}
    &\begin{cases}
        \pi(\mu)-x(\mu)Rev(\mu,\sigma) &\geq \pi(\mu')-x(\mu')Rev(\mu,\sigma)\\
        \pi(\mu')-x(\mu')Rev(\mu',\sigma)&\geq \pi(\mu)-x(\mu)Rev(\mu',\sigma)
    \end{cases}\nonumber\\
    &\Rightarrow  Rev(\mu,\sigma)\left[x(\mu)-x(\mu')\right]\leq \pi(\mu)-\pi(\mu') \leq Rev(\mu',\sigma)\left[x(\mu)-x(\mu')\right]\label{eq:myerson_envelop}
    \end{align}
    $\forall \mu\geq \mu'$, \Cref{lem:rev_mu_sigma_convex_in_mu_sigma} shows that $Rev(\mu,\sigma)\geq Rev(\mu',\sigma)$, the above inequality requires that $x(\mu')\geq x(\mu)$, proving the monotonicity in allocation. The above inequality also requires that $\pi(\mu)\leq \pi(\mu')$, meaning the sellers with higher quality gets paid weakly less. 

    To prove the second statement and for ease of exposition, we assume the allocation rule $x$ and payment rule $\pi$ are both differentiable, although the proof idea can be applied to non-differentiable ones.

    By dividing each side of \Cref{eq:myerson_envelop} by $\mu-\mu'$ where $\mu\geq \mu'$ there is 
    $$ Rev(\mu,\sigma) \frac{x(\mu)-x(\mu')}{\mu-\mu'} \leq \frac{\pi(\mu)-\pi(\mu')}{\mu-\mu'} \leq Rev(\mu',\sigma) \frac{x(\mu)-x(\mu')}{\mu-\mu'}$$
    Taking the limit as $\mu'$ goes to $\mu$ yields $$Rev(\mu,\sigma)\frac{dx(\mu)}{d\mu}= \frac{d \pi(\mu)}{d\mu}\leq 0$$
    So 
    \begin{align*}
        \pi(\mu_H)-\pi(\mu)&=\int_{\theta=\mu}^{\mu_H} Rev(\theta,\sigma) \frac{dx(\theta)}{d\theta} d\theta = \left[Rev(\theta,\sigma)x(\theta)\right]|_{\mu}^{\mu_H}-\int_{\theta=\mu}^{\mu_H}x(\theta)d Rev(\theta,\sigma)\\
        \pi(\mu) &= \pi(\mu_H)-Rev(\mu_H,\sigma)x(\mu_H)+Rev(\mu,\sigma)x(\mu)+\int_{\theta=\mu}^{\mu_H}x(\theta)d Rev(\theta,\sigma)
    \end{align*}
    For payment minimization, since $\pi(\mu_H)$ is the lower bound on payments made to any seller, it is best to set $\pi(\mu_H)=Rev(\mu_H,\sigma)x(\mu_H)$. So we can write 
    $$ \pi(\mu) = Rev(\mu,\sigma)x(\mu)+\int_{\theta=\mu}^{\mu_H}x(\theta)d Rev(\theta,\sigma)$$
    Now checking if the $(x,\pi)$ indeed satisfy the IC condition. A seller of type $\mu$ reporting $\mu'$ have utility  $$u_{\mu}(\mu')=\pi(\mu')-Rev(\mu,\sigma)x(\mu')=  (Rev(\mu',\sigma)-Rev(\mu,\sigma))x(\mu')+\int_{\theta=\mu'}^{\mu_H}x(\theta)d Rev(\theta,\sigma)$$
    For $\mu'>\mu$, the difference of seller's utility is $$u_{\mu}(\mu')-u_{\mu}(\mu)=(Rev(\mu',\sigma)-Rev(\mu,\sigma))x(\mu') -\int_{\theta=\mu}^{\mu'}x(\theta)d Rev(\theta,\sigma)\leq 0$$
    because the monotonicity of allocation rule $\forall \theta \in [\mu,\mu'],\, x(\mu')\leq x(\theta)$. Similarly, for $\mu'<\mu$, the difference of seller's utility is $$u_{\mu}(\mu')-u_{\mu}(\mu)=(Rev(\mu',\sigma)-Rev(\mu,\sigma))x(\mu') +\int_{\theta=\mu'}^{\mu}x(\theta)d Rev(\theta,\sigma)\leq 0$$
    also because of the monotonicity of allocation rule $\forall \theta \in [\mu',\mu],\, x(\mu')\geq x(\theta)$.
\end{proof}

\VirtualCost*
\begin{proof}
    \begin{align*}
        E_{\mu\sim \Phi}[\pi(\mu)] &= \int_{\theta=\mu_L}^{\mu_H} \pi(\theta) \phi(\theta)d\theta = \int_{\theta=\mu_L}^{\mu_H}[ Rev(\theta,\sigma)x(\theta)+\int_{\mu=\theta}^{\mu_H}x(\mu)dRev(\mu,\sigma)]\phi(\theta)d\theta\\
        &= \int_{\theta=\mu_L}^{\mu_H}Rev(\theta,\sigma)x(\theta)\phi(\theta)d\theta + \int_{\mu=\mu_L}^{\mu_H}\left[\int_{\theta=\mu_L}^{\mu}\phi(\theta)d\theta \right]x(\mu) dRev(\mu,\sigma)\\
        &= \int_{\theta=\mu_L}^{\mu_H}Rev(\theta,\sigma)x(\theta)\phi(\theta)d\theta + \int_{\theta=\mu_L}^{\mu_H}\Phi(\theta)x(\theta) dRev(\theta,\sigma)\\
        &= \int_{\theta=\mu_L}^{\mu_H} Rev(\theta,\sigma)x(\theta)\phi(\theta) \left[ 1+\frac{\partial Rev(\theta,\sigma)}{\partial\theta}\frac{\Phi(\theta)}{Rev(\theta,\sigma)\phi(\theta)}\right]d\theta\\
        &= \int_{\theta=\mu_L}^{\mu_H}x(\theta)\phi(\theta)Rev(\theta,\sigma)\left(1+\frac{\Phi(\theta)}{p^*(\theta) \phi(\theta)}\right)d\theta
    \end{align*}
    Where the last equality follows from \Cref{lem:rev_diff_mu} that $\frac{\partial Rev(\theta,\sigma)}{\partial\theta}=F\left[\frac{p^*-\mu}{\sigma}\right]$.
\end{proof}

\NoneMonotonicNorSubmodular*
\begin{proof}
    For none-monotonicity, consider two sellers where $\mu_1=5,\mu_2=0.1,\sigma=1$. Their revenue can be written as $Rev(\mu_1,\sigma)\approx 3.37, Rev(\mu_2,\sigma)\approx 0.19$. Then $Rev(\mu_1+\mu_2,\sqrt{2})\approx 3.13 < Rev(\mu_1,\sigma)$. The reduction in revenue arises because bundling with seller $\mu_2$ increases the combined mean valuation only marginally, while substantially increasing the dispersion of buyer valuations. Since higher dispersion does not necessarily raise optimal revenue (see \Cref{fig:rev_one_sigma}), the net effect of bundling can be negative.

    For none-submodularity, consider two sellers with the same $\mu=1, \sigma=0.5$. Then 
    \begin{align*}
        Rev(\{s_1\}) &=Rev(\mu,\sigma)\approx 0.525\\
        Rev(\{s_1,s_2\})-Rev(\{s_1\}) &= Rev(2\mu,\sqrt{2}\sigma)-Rev(\mu,\sigma) \approx 0.618
    \end{align*}
    So $Rev(\{s_1\}) < Rev(\{s_1,s_2\})-Rev(\{s_1\})$ and that $Rev(S)$ does not satisfy submodularity. For none-supermodularity, consider two sellers with the same $\mu=1, \sigma=1$. Then 
    \begin{align*}
        Rev(\{s_1\}) &=Rev(\mu,\sigma)\approx 0.507\\
        Rev(\{s_1,s_2\})-Rev(\{s_1\}) &= Rev(2\mu,\sqrt{2}\sigma)-Rev(\mu,\sigma) \approx 0.497
    \end{align*}
    So $Rev(\{s_1\}) > Rev(\{s_1,s_2\})-Rev(\{s_1\})$ and that $Rev(S)$ does not satisfy supermodularity. For none-subadditive, again for $\mu=1,\sigma=0.5$, it holds that $Rev(2\mu,\sqrt{2}\sigma)\approx 1.143 > 1.05\approx 2Rev(\mu,\sigma)$. For none-superadditive, again for $\mu=1,\sigma=1$, it holds that $Rev(2\mu,\sqrt{2}\sigma)\approx 1.004 < 1.014\approx 2Rev(\mu,\sigma)$.
\end{proof}

\subsection{Missing Proofs in \Cref{sec:deterministic_optimal}}
We formally write out the proof for \Cref{thm:asymptotic_profit} using similar steps as the proof for profit-maximizing mechanisms.
\ProfitMaxSurrogate*

The proof follows from the similar structure as \cite{fu2017notes} on myerson's revenue-optimal mechanism. 
\paragraph{Step 1: Decomposition of $x$ into weighted-combination of posted-price mechanism} Consider a posted price mechanism of price $Rev(\mu,\sigma)$. The allocation rule of this posted price mechanism is a decreasing step function where the step occurs at $\mu$. We use $\mu$ to index this posted-price mechanism with a slight abuse of notation use $\varpi(\mu)$ to represent the expected surrogate profit of this posted price mechanism, which is given by $$\varpi(\mu)=-Rev(\mu,\sigma)\Phi(\mu)+\int_{\theta=\mu_L}^{\mu}\theta g(\theta) d\theta$$
    
Given any monotone non-increasing allocation rule $x$, we use these posted-price mechanism allocation rules of different $\mu$ as basis to construct the allocation rule $x$. For our proof we will focus on allocation rule $x$ that is differentiable, and the proof for none-differentiable $x$ shares the same idea. Construct the following weighted-sum mechanism. For each posted-price mechanism $\mu$, give it weight $-dx(\mu)$. Further give weight $1$ to another another constant mechanism that always allocate with probability $x(\mu_H)$ and pays $x(\mu_H)Rev(\mu_H,\sigma)$. For any quality $\mu$, the weighted total allocation probability is $$-\int_{\theta=\mu_L}^{\mu_H}\mathbf{I}[\mu\leq \theta]dx(\mu)+x(\mu_H)=-\int_{\theta=\mu}^{\mu_H}dx(\mu)+x(\mu_H)=x(\mu)$$ which is the same as the allocation probability of rule $x$. Then by \Cref{lem:myerson_monotonic}, the expected surrogate profit of $x$ is the same as the expected surrogate profit of the weighted-sum-mechanism, which can be expressed as 
\begin{align*}
    \varpi(x) &= -\int_{\mu=\mu_L}^{\mu_H}\varpi(\mu)dx(\mu)+x(\mu_H)E_\mu[\mu]-x(\mu_H)Rev(\mu_H,\sigma) \\
    &= [x(\mu)Rev(\mu,\sigma)\Phi(\mu)|_{\mu_L}^{\mu_H}]-\int_{\mu=\mu_L}^{\mu_H}(-x(\mu))(-\frac{dRev(\mu,\sigma)}{d\mu}\Phi(\mu)-Rev(\mu,\sigma)\phi(\mu))d\mu\\
    &- \int_{\mu=\mu_L}^{\mu_H}\int_{\theta=\mu_L}^\mu \theta \phi(\theta)d\theta dx(\mu)+x(\mu_H)E_\mu[\mu]-x(\mu_H)Rev(\mu_H,\sigma) \\
    &=E[x(\mu)\mu-x(\mu)Rev(\mu,\sigma)(1+\frac{\Phi(\mu)}{\phi(\mu)Rev(\mu,\sigma)}\frac{d Rev(\mu,\sigma)}{d\mu})]=E[x(\mu)(\mu-\varphi(\mu))]
\end{align*}
The second equality is through change of integration order. Note how exactly this equals to our previously derived revenue of $x$. 

\paragraph{Step 2: Quantile space} We transfer $\mu$ into quantile space. Define $q(\mu)=\Phi(\mu)$ and $y(q)=x(\Phi^{-1}(q))$. Clearly $y$ is monotone non-increasing as well. Give weight $-dy(q)= - dx(\mu)$ to each post-price mechanism indexed by $q$. Let $\varpi'(q)$ denote the expected surrogate profit of the post-price mechanism of $q$, where the $\prime$ sign denotes operating in the quantile space. The constant mechanism which allocates with probability $x(\mu_H)$ and always pay $x(\mu_H)Rev(\mu_H,\sigma)$ has expected surrogate welfare $x(\mu_H)[E_{\mu}[\mu]-Rev(\mu_H,\sigma)]$, which is equal to $y(1)\varpi'(1)$ in quantile space. This is because $\varpi'(1)=\varpi(\mu_H)=E_{\mu}[\mu]-Rev(\mu_H,\sigma)$ and $y(1)=x(\mu_H)$. Further note the expected surrogate profit for posted-price mechanism which posts $Rev(\mu_L,\sigma)$ is zero, or $\varpi'(0)=0$.

Combining weight $1$ of the constant mechanism and weight $-dy(q)$ of the post-price mechanisms yields expected surrogate profit of $x$ equaling $w$, expressed in quantile space as
\begin{equation}\label{eq:quantile_space_profit}-\int_{q=0}^{1}\varpi'(q)dy(q) + \varpi'(1)y(1)=[-\varpi'(q)y(q)|_0^1]+\int_{q=0}^{1}y(q)d\varpi'(q)+y(1)\varpi'(1)=\int_{q=0}^1 y(q)\frac{d\varpi'(q)}{dq}dq\end{equation}
From the left-hand side of the equality, fixing allocation rule $y$, the mechanism that operates on a higher point-wise $\varpi'$ will attain higher expected surrogate profit. Examining the right-hand side of the equality
\begin{align*}
    \varpi'(q) &= -qRev(\mu^{-1}(q),\sigma)+\int_{\theta=0}^q \mu^{-1}(\theta)d\theta\\
    \frac{d\varpi'(q)}{dq} &= \mu^{-1}(q)-Rev(\mu^{-1}(q),\sigma)-q\frac{dRev(\mu^{-1}(q),\sigma)}{d \mu^{-1}(q)}\frac{d\mu^{-1}(q)}{dq}\\
    &= \mu-Rev(\mu,\sigma)(1+\frac{\Phi(\mu)}{\phi(\mu)Rev(\mu,\sigma)}\frac{dRev(\mu,\sigma)}{d\mu}) = \mu^{-1}(q)-\varphi(\mu^{-1}(q))
\end{align*}
So we can express in quantile space the expected surrogate profit as 
$$\varpi'(y)=\int_{q=0}^1 y(q) ( \mu^{-1}(q)-\varphi(\mu^{-1}(q)))dq$$

\paragraph{Step 3: Ironing} If $\mu^{-1}(q)-\varphi(\mu^{-1}(q))$ is monotone non-increasing in $q$, then we can specify a monotone non-increasing allocation rule $y(q)$ that maximizes $\varpi(y)$. Such a $y$ will take a threshold form, such that $y(q)=1$ iff $\mu^{-1}(q)-\varphi(\mu^{-1}(q))\geq 0$. Note that $\mu^{-1}(q)-\varphi(\mu^{-1}(q))$ being monotone non-increasing means that $\frac{d\varpi'(q)}{dq}$ is monotone non-decreasing, or $\varpi'(q)$ is concave. Now define the ironing procedure
\begin{itemize}
    \item Define $\overline{\varpi}'(q)$ as the concave hull of $\varpi'(q)$
    \item Define the ironed virtual surplus function as $\varrho(q)=d \overline{\varpi}'(q)/dq$
\end{itemize}
The $\varrho(q)$ is the ironed value of $\mu^{-1}(q)-\varphi(\mu^{-1}(q))$, which is monotone non-increasing.
\begin{lemma}\label{thm:iron}
    For any monotone non-increasing allocation rule $y(q)$ and any function $\mu^{-1}(q)-\varphi(\mu^{-1}(q))$, the expected surrogate profit is upper bounded by the expected ironed virtual surplus 
    $$\varpi'(y)=E\left[y(q)(\mu^{-1}(q)-\varphi(\mu^{-1}(q)))\right]\leq E\left[y(q) \varrho(q)\right]$$ Furthermore, this inequality holds with equality if the allocation rule $y$ satisfies $\frac{dy(q)}{dq}=0$ for all $q$ where $\overline{\varpi}'(q)>\varpi'(q)$.
\end{lemma}
\begin{proof}
    Since $\overline{\varpi}'$ is the concave hull of $\varpi'$, it holds that $\overline{\varpi}'\geq \varpi'$ and $\overline{\varpi}'(1)=\varpi'(1)$. Using \Cref{eq:quantile_space_profit} to expand the terms
    \begin{align*}
        E\left[y(q)(\varrho(q)-  \mu^{-1}(q)+\varphi(\mu^{-1}(q)))\right]=-\int_{q=0}^1 (\overline{\varpi}'(q)-\varpi'(q))dy(q)\geq 0
    \end{align*}
    The second statement in the theorem follows naturally.
\end{proof}
Now we are ready to finally prove \Cref{thm: profit_max_surrogate}. Define the threshold $t$ to be $$t=\begin{cases}
    \mu_H & \text{ if } \varrho(1)>0\\
    \text{any value smaller than } \mu_L & \text{ if } \varrho(1)\leq 0 \text{ and } \varrho(0)\leq 0\\
    \max\{\mu|\varrho(q(\mu))>0\}& \text{ otherwise}
\end{cases}$$
For BIC: Since $\overline{\varpi}'(q)$ is concave, $\varrho(q)$ is monotone non-increasing. Thus $x$ defined as in \Cref{thm: profit_max_surrogate} is a reverse step function and monotone non-increasing. The $\pi$ defined satisfies the Myerson's lemma in \Cref{lem:myerson_monotonic} so $(x,\pi)$ is BIC. For profit-maximizing: \Cref{thm:iron} says any IC mechanism has expected surrogate profit upper bounded by the expected ironed virtual surplus. As $x$ achieves the maximum ironed virtual surplus, $(x,\pi)$ achieves the maximum expected surrogate profit.

\subsection{Missing Proofs in \Cref{sec:asymptotic_large_market}}
\SigmaRevBound*
\begin{proof}
    Denote $A = \sum_{i\in [N]}a_i^2$ and $J= \sum_{i\in [N]}\frac{a_i Z_i}{\sqrt{A}}$. Then buyer's valuation can be written in the form $$v = C+\sigma \sum_{i\in [N]}a_i Z_i = C+\sigma \sqrt{A} J$$
    Since $Z_i$ are independent random variables, and for any $\frac{|\lambda|}{\sqrt{A}}< \frac{1}{\xi} \Leftrightarrow |\lambda| <\frac{\sqrt{A}}{\xi}$, then $\frac{|\lambda| a_i}{\sqrt{A}}< \frac{1}{\xi}$ so
    \begin{align*}
        E[e^{\lambda J}] &= E[e^{\frac{\lambda}{\sqrt{A}} a_1Z_1}\times \ldots \times e^{\frac{\lambda}{\sqrt{A}} a_NZ_N}] = E[e^{\frac{\lambda}{\sqrt{A}} a_1Z_1}]\times \ldots \times E[e^{\frac{\lambda}{\sqrt{A}} a_NZ_N}]\\
        &\leq e^{\frac{\lambda^2\gamma^2 a_1^2}{2A}}\times \ldots \times e^{\frac{\lambda^2\gamma^2 a_N^2}{2A}} = e^{\frac{\lambda^2\gamma^2}{2}}
    \end{align*}
    Applying Chernoff bound yields
    \begin{align*}
        Pr[J\geq \eta] \leq \inf_{\lambda> 0} E[e^{\lambda J}]e^{-\lambda \eta} \leq  \inf_{0<\lambda<\frac{\sqrt{A}}{\xi}} e^{-\lambda \eta + \frac{\lambda^2\gamma^2}{2}}
    \end{align*}
    Let $\bar{\eta} = \frac{\gamma^2\sqrt{A}}{\xi}$. When $0<\eta< \bar{\eta}$, we can set $\lambda = \frac{\eta}{\gamma^2}$ which satisfies that $0<\lambda<\frac{\sqrt{A}}{\xi}$, and $$\forall 0<\eta<\bar{\eta},\quad Pr[J\geq \eta] \leq e^{-\frac{\eta^2}{2\gamma^2}}$$
    This is the tail region where subexponential distributions share similar behavior with sub-gaussian distributions. When $\eta\geq \bar{\eta}$, we can set $\lambda= \frac{\sqrt{A}}{2\xi}$. Further note that for any $\eta\geq \bar{\eta}$, it satisfies that $\frac{1}{4}\frac{\sqrt{A}}{2\xi}\eta \geq \frac{1}{4}\frac{\sqrt{A}}{2\xi} \frac{\gamma^2\sqrt{A}}{\xi}=\frac{A\gamma^2}{8\xi^2}$, so there exists the following bound 
    $$\forall \eta\geq \bar{\eta}, \quad Pr[J\geq\eta]\leq e^{-\frac{\sqrt{A}}{2\xi}\eta+\frac{A\gamma^2}{8\xi^2}} \leq e^{-\frac{3}{4}\frac{\sqrt{A}}{2\xi}\eta}$$
    
    
    \paragraph{First prove an upper bound on $Rev(v) - C$.} Any price can be written as $p= C+ \sigma\sqrt{A} \eta$
    \begin{align*}
        Rev(v) -C = \sup_\eta (C+\sigma\sqrt{A}\eta) Pr\left[C+\sigma\sqrt{A} J \geq C+\sigma \sqrt{A}\eta\right]-C
        \leq \sigma\sqrt{A}\sup_\eta \eta Pr\left[J\geq \eta\right]
    \end{align*}
    Without loss, we can focus on $\eta>0$. From the above analysis on $Pr[J\geq \eta]$, 
    \begin{align*}
        \sup_{\eta}\eta Pr[J\geq \eta] \leq \begin{cases}
            \eta e^{-\frac{\eta^2}{2\gamma^2}} & \text{ if } 0<\eta<\bar{\eta}\\
            \eta e^{-\frac{3\sqrt{A}}{8\xi}\eta} & \text{ if } \eta\geq \bar{\eta}
        \end{cases}\leq \begin{cases}
            \gamma e^{-\frac{1}{2}} & \text{ if } 0<\eta<\bar{\eta}\\
            \frac{8\xi}{3\sqrt{A}}e^{-1} & \text{ if } \eta\geq \bar{\eta}
        \end{cases}
    \end{align*}
    the maximizer of the two cases are respectively given by $\eta=\gamma$ when $0<\eta<\bar{\eta}$ and $\eta=\frac{8\xi}{3\sqrt{A}}$ when $\eta\geq \bar{\eta}$. Note the maximizer might not be attained in the two feasible regions, which only makes the upper bound more valid. 
    This gives an upper bound on $Rev(v)-C$ as follow
    $$Rev(v)-C \leq \sigma\sqrt{A}\max\{\gamma e^{-\frac{1}{2}}, \frac{8\xi}{3\sqrt{A}}e^{-1}\}$$
    \paragraph{Then prove an upper bound on $C-Rev(v)$ through proving a lower bound on $Rev(v)$.} As it is sufficient to find one price that lowers bound $Rev(v)$, we use a different technique than the above. Denote any price $p=C-\sigma \sqrt{A}\eta$. For any $\eta$, it satisfies that \begin{align*}
        Rev(v)\geq (C-\sigma \sqrt{A}\eta) Pr[C+\sigma\sqrt{A}J\geq C - \sigma\sqrt{A}\eta]&=(C-\sigma \sqrt{A}\eta)  (1-Pr[J<-\eta])\\
        C- Rev(v) \leq  \sigma\sqrt{A}\eta +(C-\sigma\sqrt{A}\eta) Pr[J<-\eta] &\leq \sigma\sqrt{A}\eta +C Pr[J<-\eta]
    \end{align*}
    As the bound holds for any $\eta$, we can restrict attention to the value of $\eta$ such that $0<\eta\leq \frac{C}{\sigma\sqrt{A}}$ so that $0\leq p<C$. Since $Var(J)=1, E[J]=0$, by Cantelli inequality, $$\forall \eta>0,\quad Pr[J<-\eta]\leq\frac{1}{1+\eta^2}<\frac{1}{\eta^2}$$
    Directly applying this to the above for $\eta\in (0,\frac{C}{\sigma\sqrt{A}}]$ means that
    $C-Rev(v)\leq \sigma \sqrt{A}\eta+\frac{C}{\eta^2}$. Let $\eta^*=(\frac{2C}{\sigma\sqrt{A}})^{\frac{1}{3}}$. If $\eta^*>\frac{C}{\sigma\sqrt{A}}$, then $$C-Rev(v)\leq C< \sigma\sqrt{2A}$$
    If $\eta^*\leq \frac{C}{\sigma\sqrt{A}}$, then $$C-Rev(v)\leq \sigma\sqrt{A}\eta^*+C/(\eta^*)^2=C^{\frac{1}{3}}(\sigma\sqrt{A})^\frac{2}{3}(2^{\frac{1}{3}}+2^{-\frac{2}{3}})\leq 1.89 \sigma^{\frac{2}{3}}(CA)^{\frac{1}{3}}$$
    Note that this is a crude bound that doesn't rely on the properties of subexponential distribution.
\end{proof}

\AsymptoticProfit*
\begin{proof}
    We decompose the difference between $\Pi(x^*)$ and $\Pi(\tilde{x}^*)$ into two parts. The first part captures the expected difference in profit due to the random realization of $\bm I\sim \text{Bernoulli}(x(\bm \mu))$, and in the second part bounds the expected difference in profit due to replacing the $Rev(v_S)$ by its mean valuation.
    Besides $\Pi(x)$ defined in \Cref{eq:expected_profit} and $\varpi(x)$ in \Cref{eq:expected_surrogate_profit}, define an intermediate surrogate profit $\varpi'(x)$ that doesn't depend on the random realization of $\bm I$. 
    $$\varpi'(x) = E_{\bm\mu\sim \bm\Phi}\left[Rev\left(\sum_{i\in[N]}x(\mu_i) (\mu_i+\sigma Z_i)\right)-\sum_{i\in [N]}x(\mu_i)\varphi(\mu_i)\right]$$
    For any mechanism allocation rule $x$, we first provide a bound on $|\Pi(x)-\varpi'(x)|$.
    For any $x$ and realization $\bm I(\bm\mu)$, denote 
    $$\mathcal{Y}(\bm I,\bm\mu)=\sum_{i\in [N]} I_i(\mu_i)\mu_i,\; \Upsilon(x,\bm\mu)=\sum_{i\in [N]} x(\mu_i)\mu_i$$
    Since $I_i(\mu_i)\sim \text{Bernoulli}(x(\mu_i))$, it satisfies that
    \begin{align*}
        E_{I}[\mathcal{Y}(\bm I,\bm\mu)|\bm\mu]& =\Upsilon(x,\bm\mu)\quad Var(\mathcal{Y}(\bm I,\bm\mu)|\bm\mu) \leq \mu_H^2\sum_{i\in [N]} x(\mu_i)(1-x(\mu_i))\leq \mu_H^2 N\\
         E_{\bm\mu,I}[(\mathcal{Y}(\bm I,\bm\mu)-\Upsilon(x,\bm\mu))^2] &= E_{\bm\mu}[E_{\bm I}[(\mathcal{Y}(\bm I,\bm\mu)-\Upsilon(x,\bm\mu))^2|\bm\mu]]= E_{\bm\mu}[Var(\mathcal{Y}(\bm I,\bm\mu)|\bm\mu)]\leq \mu_H^2 N
    \end{align*}
    Then by the definition of variance,
    \begin{align}\label{eq:expected_mu_difference}
        E_{I,\bm\mu}[|\mathcal{Y}(\bm I,\bm\mu)-\Upsilon(x,\bm\mu)|]=\sqrt{E_{I,\bm\mu}[(\mathcal{Y}(\bm I,\bm\mu)-\Upsilon(x,\bm\mu))^2]-Var(|\mathcal{Y}(\bm I,\bm\mu)-\Upsilon(x,\bm\mu)|)}\leq \mu_H\sqrt{N}
    \end{align}
    The above corresponds to the difference in $\Pi(x)$ and $\varpi'(x)$ due to the different mean quality of the set $S(\bm\mu,\bm I)$. We now bound the difference due to the dispersion of buyer valuation. Apply \Cref{lem:sigma_rev_bound} with $C=\sum_{i\in [N]}I_i(\mu_i)\mu_i$, $a_i=I_i(\mu_i), A = \sum_{i\in [N]}I_i(\mu_i)^2$, then
    \begin{align*}
        \left|Rev\left(\sum_{i\in [N]}I_i(\mu_i)(\mu_i+\sigma Z_i)\right)-\sum_{i\in [N]}I_i(\mu_i)\mu_i\right| \leq \max\{\sigma \gamma e^{-\frac{1}{2}}A^{\frac{1}{2}},\frac{8\sigma\xi}{3 e},\sqrt{2}\sigma A^{\frac{1}{2}}, 2\sigma^{\frac{2}{3}}(CA)^{\frac{1}{3}}\}\} = O(N^{\frac{2}{3}})
    \end{align*}
    Similarly $\left|Rev\left(\sum_{i\in [N]}x(\mu_i)(\mu_i+\sigma Z_i)\right)-\sum_{i\in [N]}x(\mu_i)\mu_i\right|=O(N^{\frac{2}{3}})$. For any realization of $\bm I(\bm\mu)$
    \begin{align*}
        &\left|Rev(v_{S(\bm\mu,\bm I)}) - Rev\left(\sum_{i\in[N]}x(\mu_i)( \mu_i+\sigma Z_i)\right)\right| = \left|Rev\left(\sum_{i\in [N]}I_i(\mu_i)(\mu_i+\sigma Z_i)\right) - Rev\left(\sum_{i\in[N]}x(\mu_i) (\mu_i+\sigma Z_i)\right)\right|\\
        \leq & \Bigg|Rev\left(\sum_{i\in [N]}I_i(\mu_i)(\mu_i+\sigma Z_i)\right)-Rev\left(\sum_{i\in [N]}I_i(\mu_i)\mu_i\right)+Rev\left(\sum_{i\in [N]}I_i(\mu_i)\mu_i\right)-Rev\left(\sum_{i\in [N]}x(\mu_i)\mu_i\right)\\
        &+Rev\left(\sum_{i\in [N]}x(\mu_i)\mu_i\right)-Rev\left(\sum_{i\in[N]}x(\mu_i) (\mu_i+\sigma Z_i)\right)\Bigg|
        \leq \left|\mathcal{Y}(\bm I,\bm\mu)-\Upsilon(x,\bm\mu)\right| + O(N^{\frac{2}{3}})
    \end{align*}
    We are now ready to analyze $|\Pi(x)-\varpi'(x)|$.
    \begin{align*}
        \left|\Pi(x)-\varpi'(x)\right|&= \left|E_{\bm\mu, \bm I}\left[Rev(S(\bm\mu,\bm I))-Rev\left(\sum_{i\in[N]}x(\mu_i) (\mu_i+\sigma Z_i)\right)\right] \right|\\
        &\leq E_{\bm\mu, \bm I}\left[\left|Rev(S(\bm\mu,\bm I))-Rev\left(\sum_{i\in[N]}x(\mu_i) (\mu_i+\sigma Z_i)\right)\right|\right]\\
        &\leq E_{\bm\mu, \bm I}\left[\left|\mathcal{Y}(\bm I,\bm \mu)-\Upsilon(x,\bm \mu)\right|\right] + O(N^{\frac{2}{3}})
    \end{align*}
    Referring to \Cref{eq:expected_mu_difference} there is $|\Pi(x)-\varpi'(x)|=O(N^{\frac{2}{3}})$. We can now proceed to bound $|\varpi'(x)-\varpi(x)|$. 
    \begin{align*}
        |\varpi(x)-\varpi'(x)| &= \left|E_{\bm\mu}\left[Rev\left(\Upsilon(x,\bm\mu)+\sigma\sum_{i\in [N]}x(\mu_i)Z_i\right)- \Upsilon(x,\bm\mu)\right]\right|\\
        &\leq E_{\bm\mu}\left[\left|Rev\left(\Upsilon(x,\bm\mu)+\sigma\sum_{i\in [N]}x(\mu_i)Z_i\right)- \Upsilon(x,\bm\mu)\right|\right]
    \end{align*}
    Directly applying \Cref{lem:sigma_rev_bound} with $C=\Upsilon(x,\mu), a_i=x(\mu_i)$ yields $|\varpi'(x)-\varpi(x)|=O(N^{\frac{2}{3}})$. Combined with $|\varpi'(x)-\Pi(x)|=O(N^{\frac{2}{3}})$ means that $$\Pi(x^*)\leq \varpi(x^*)+O(N^{\frac{2}{3}})\leq \varpi(\tilde{x}^*)+O(N^{\frac{2}{3}})\leq \Pi(\tilde{x}^{*})+O(N^{\frac{2}{3}})$$
    Finally, by the definition of $\varpi(x)=NE_{\bm\mu}[x(\mu)(\mu-\varphi(\mu))]$ in \Cref{eq:expected_surrogate_profit}, if $\varpi(\tilde{x}^*)>0$, it holds that $\varpi(\tilde{x}^*)=\Theta(N)$. And from our previous analysis, $|\Pi(\tilde{x}^*)- \varpi(\tilde{x}^*)|= O(N^{\frac{2}{3}})$ so $\Pi(\tilde{x}^*)=\Theta(N)$. Then 
    $$\frac{\Pi(x^*)}{\Pi(\tilde{x}^*)}\leq 1+O(N^{-\frac{1}{3}})$$
\end{proof}

\section{Missing Lemmas and Proofs in \Cref{sec:in_house_production}}
Denote platform's decision is to produce $m_L$ units of $\mu_L$ and $m_H=m-m_L$ units of $\mu_H$. As $m_L$ and $m$ uniquely determines $S_M$, we use $\Pi(\pi,m_L,m):=\Pi(x,S_M)$ to represent the platform's profit.
\NoMix*
\begin{proof}
    We examine $\Pi(\pi_L,m_L,m)$ first 
    \begin{align*}
        \Pi(\pi_L,m_L,m) &= Rev((N_L+m_L)\mu_L+(m-m_L)\mu_H,\sqrt{N_L+m}\sigma) \\&- (N_L+m_L)Rev(\mu_L,\sigma) - (m-m_L)Rev(\mu_H,\sigma)\\
        \frac{\partial \Pi(\pi_L,m_L,m)}{\partial m_L} &= (\mu_L-\mu_H) F[z^*(\alpha_L)-\alpha_L] - Rev(\mu_L,\sigma)+Rev(\mu_H,\sigma)\\
        \frac{\partial^2 Rev(p_L)}{\partial m_L^2} &= -(\mu_L-\mu_H) f[z^*(\alpha_L)-\alpha_L][z^*_{\alpha_L}(\alpha_L)-1]\frac{\mu_L-\mu_H}{\sqrt{N_L+m}\sigma}>0\\
        \text{where} \quad \alpha_L &= \frac{(N_L+m_L)\mu_L+(m-m_L)\mu_H}{\sqrt{N_L+m}\sigma}
    \end{align*}
    Now examine $\Pi(\pi_H,m_L,m)$
    \begin{align*}
        \Pi(\pi_H,m_L,m) &= Rev((N_L+m_L)\mu_L+(N_H+m-m_L)\mu_H,\sqrt{N+m}\sigma) \\
        &- m_L Rev(\mu_L,\sigma) - (N+m-m_L)Rev(\mu_H,\sigma)\\
        \frac{\partial \Pi(\pi_H,m_L,m)}{\partial m_L}&= 
        (\mu_L-\mu_H)F[z^*(\alpha_H)-\alpha_H] - Rev(\mu_L,\sigma)+Rev(\mu_H,\sigma)\\
        \frac{\partial^2 \Pi(\pi_H,m_L,m)}{\partial m_L^2} &= -(\mu_L-\mu_H) f[z^*(\alpha_H)-\alpha_H](z_{\alpha_H}^*(\alpha_H)-1)\frac{\mu_L-\mu_H}{\sqrt{N+m}\sigma}>0\\
        \text{where }\quad \alpha_H &= \frac{(N_L+m_L)\mu_L+(N_H+m-m_L)\mu_H}{\sqrt{N+m}\sigma} 
    \end{align*}
    and finally at posted price 0
    \begin{align*}
        \Pi(\pi_0,m_L,m) &= Rev(m_L\mu_L+(m-m_L)\mu_H,\sqrt{m}\sigma)-m_L Rev(\mu_L,\sigma) - (m-m_L)Rev(\mu_H,\sigma)\\
        \frac{\partial \Pi(\pi_0,m_L,m)}{\partial m_L} &= (\mu_L-\mu_H) F[z^*(\alpha_0)-\alpha_0]-Rev(\mu_L,\sigma)+Rev(\mu_H,\sigma)\\
        \frac{\partial^2 \Pi(\pi_0,m_L,m)}{\partial m_L^2} &=-(\mu_L-\mu_H)f(z^*(\alpha_0)-\alpha_0)(z^*_{\alpha_0}(\alpha_0)-1)\frac{-\mu_H}{\sqrt{m}\sigma}>0\\
        \text{ where } \alpha_0 &= \frac{m_L\mu_L+(m-m_L)\mu_H}{\sqrt{m}\sigma}
    \end{align*}
    Thus for either prices, $\Pi(\pi,m_L,m)$ is convex in $m_L$. This means the optimal quantity $m_L\in \{0,m\}$, and the platform either always produce low-quality or high-quality items. 
    
    We now give sufficient conditions on when $m_L=0$ is preferred over $m_L=m$. When the platform posts $\pi_L$, a sufficient condition for setting $m_L=0$ is that $\frac{\partial \Pi(\pi_L,m_L,m)}{\partial m_L}<0$ for $m_L\in [0,m]$. As $\frac{\partial \Pi(\pi_L,m_L,m)}{\partial m_L}$ increases with $m_L$, the sufficient condition further becomes
    \begin{equation}\label{eq:pL_mL_0}
        \frac{\partial \Pi(\pi_L,m_L,m)}{\partial m_L}|_{m_L=m}<0 \;\Leftrightarrow \; F[z^*(\frac{(N_L+m)\mu_L}{\sqrt{N_L+m}\sigma})-\frac{(N_L+m)\mu_L}{\sqrt{N_L+m}\sigma}]>\frac{Rev(\mu_H,\sigma)-Rev(\mu_L,\sigma)}{\mu_H-\mu_L}
    \end{equation}
    Similarly when the platform posts $\pi_L$, a sufficient condition for setting $m_L=m$ is that $\frac{\partial \Pi(\pi_L,m_L,m)}{\partial m_L}>0$ for $m_L\in [0,m]$. As $\frac{\partial \Pi(\pi_L,m_L,m)}{\partial m_L}$ increases with $m_L$, the sufficient condition further becomes
    \begin{equation}\label{eq:pL_mL_m}
        \frac{\partial \Pi(\pi_L,m_L,m)}{\partial m_L}|_{m_L=0}>0 \;\Leftrightarrow \; F[z^*(\frac{N_L\mu_L+m\mu_H}{\sqrt{N_L+m}\sigma})-\frac{N_L\mu_L+m\mu_H}{\sqrt{N_L+m}\sigma}]<\frac{Rev(\mu_H,\sigma)-Rev(\mu_L,\sigma)}{\mu_H-\mu_L}
    \end{equation}
    Clearly these two conditions cannot hold simultaneously, as $\frac{(N_L+m)\mu_L}{\sqrt{N_L+m}\sigma}<\frac{N_L\mu_L+m\mu_H}{\sqrt{N_L+m}\sigma}$. In the same way, we can write out the sufficient conditions when the platform posts $\pi_H$. A sufficient condition for setting $m_L=0$ is that $\frac{\partial\Pi(\pi_H,m_L,m)}{\partial m_L}<0$ for $m_L\in[0,m]$. As $\frac{\partial \Pi(\pi_H,m_L,m)}{\partial m_L}$ increases with $m_L$, the sufficient condition becomes
    \begin{equation}\label{eq:pH_mL_0}
        \frac{\partial \Pi(\pi_H,m_L,m)}{\partial m_L}|_{m_L=m}<0 \Leftrightarrow F[z^*(\frac{(N_L+m)\mu_L+N_H\mu_H}{\sqrt{N+m}\sigma})-\frac{(N_L+m)\mu_L+N_H\mu_H}{\sqrt{N+m}\sigma}]>\frac{Rev(\mu_H,\sigma)-Rev(\mu_L,\sigma)}{\mu_H-\mu_L}
    \end{equation}
    A sufficient condition for setting $m_L=m$ is that $\frac{\partial \Pi(\pi_H,m_L,m)}{\partial m_L}>0$ for $m_L\in[0,m]$. As $\frac{\partial \Pi(\pi_H,m_L,m)}{\partial m_L}$ increases with $m_L$, the sufficient condition becomes
    \begin{equation}\label{eq:pH_mL_m}
        \frac{\partial \Pi(\pi_H,m_L,m)}{\partial m_L}|_{m_L=0}>0 \Leftrightarrow F[z^*(\frac{N_L\mu_L+(N_H+m)\mu_H}{\sqrt{N+m}\sigma})-\frac{N_L\mu_L+(N_H+m)\mu_H}{\sqrt{N+m}\sigma}]<\frac{Rev(\mu_H,\sigma)-Rev(\mu_L,\sigma)}{\mu_H-\mu_L}
    \end{equation}    
    Then when the platform must in-house produce $m$ units, sufficient condition for always setting $m_L=0$ (produces $\mu_H$) is that both \Cref{eq:pL_mL_0,eq:pH_mL_0} satisfies; a sufficient condition for always setting $m_L=m$ (produces $\mu_L$) is that both \Cref{eq:pL_mL_m,eq:pH_mL_m} satisfies.
\end{proof}

\begin{lemma}\label{lem:decreasing_mills_ratio}
    With buyer valuations normally distributed, for any $\epsilon>0$, there exists a threshold $\alpha_\epsilon$ such that whenever $\alpha>\alpha_\epsilon$, the optimal demand satisfies $F[z^*(\alpha)-\alpha]>1-\epsilon$. 
\end{lemma}
\begin{proof}
    Let $t(\alpha)=z^*(\alpha)-\alpha$. By \Cref{lem:norm_rev_character} define $r(t(\alpha))=\frac{F[t(\alpha)]}{f(t(\alpha))} = \frac{F[z^*(\alpha)-\alpha]}{f(z^*(\alpha)-\alpha)}=z^*(\alpha)=\alpha+t(\alpha)$ to be the inverse hazard rate function. It is known that the normal distribution has strictly increasing hazard rate function or strictly decreasing $r(t)=\frac{F(t)}{f(t)}$ \citep{baricz2010mills, sampford1953some}. 

    For any $\epsilon\in (0,1)$, pick $t_\epsilon<0$ such that $F(t_\epsilon)>1-\epsilon$. Let $\alpha_\epsilon = r(t_\epsilon)-t_\epsilon$. Then $\forall \alpha >\alpha_\epsilon$, if $t(\alpha)\geq t_\epsilon$, then by $r(t)-t$ being strictly decreasing in $t$, $r(t(\alpha))-t(\alpha)<r(t_\epsilon)-t_\epsilon=\alpha_\epsilon<\alpha$, which forms a contradiction. So $t(\alpha)<t_\epsilon$. Then $F(t(\alpha))>F(t_\epsilon)>1-\epsilon$.
\end{proof}

\BundleSeller*
\begin{proof}
    First prove the optimal posted price. 
    We write down the platform's profit when in-house producing $m$ unit of item with quality $\mu$, when bundling every seller, only low-quality sellers, and no seller respectively. Since we are fixing the in-house production, let $\Pi(\pi)$ denote the platform's profit.
    \begin{align*}
        \lim_{N\rightarrow\infty}\frac{1}{N}\Pi(\pi_H) &=\lim_{N\rightarrow \infty}\frac{1}{N}\left[Rev(N_L\mu_L+(N-N_L)\mu_H+m\mu,\sqrt{N+m}\sigma) - NRev(\mu_H,\sigma) - mRev(\mu,\sigma)\right] \\
        &= \lim_{N\rightarrow\infty} Rev\left(\tau \mu_L+(1-\tau)\mu_H, \frac{\sigma}{\sqrt{N}}\right)- Rev(\mu_H,\sigma)\\
        &= -\tau (\mu_H-\mu_L)  + \mu_H-Rev(\mu_H,\sigma)\\
        \lim_{N\rightarrow\infty}\frac{1}{N}\Pi(\pi_L) &= \lim_{N\rightarrow \infty}\frac{1}{N}\left[Rev(N_L\mu_L+m\mu,\sqrt{N_L+m}\sigma) - N_LRev(\mu_L,\sigma) - mRev(\mu,\sigma)\right]\\
        &= \lim_{N\rightarrow \infty} Rev\left(\tau\mu_L,\frac{\sqrt{\tau}\sigma}{\sqrt{N}}\right)-\tau Rev(\mu_L,\sigma) = \tau (\mu_L -  Rev(\mu_L,\sigma))\\
        \lim_{N\rightarrow\infty}\frac{1}{N}\Pi(\pi_0) &= 0
    \end{align*}
    Define 
    \begin{align*}
        \Delta_L(\tau)&=\tau (\mu_L-Rev(\mu_L,\sigma))\\
        \Delta_H(\tau)&=-\tau(\mu_H-\mu_L)+\mu_H-Rev(\mu_H,\sigma)
    \end{align*}
    The platform chooses among $\{\Delta_L(\tau),\Delta_H(\tau),0\}$. We can thus discuss the three cases 
    
    \textbf{Case 1} If $\mu_L
    -Rev(\mu_L,\sigma) >0$. By \Cref{lem:norm_rev_character}, $\mu_H-Rev(\mu_H,\sigma)>\mu_L
    -Rev(\mu_L,\sigma) >0$. So $\Delta_L(\tau)$ monotonically increases from $\Delta_L(\tau=0)=0$ to $\Delta_L(\tau=1)=\mu_L-Rev(\mu_L,\sigma)$. And $\Delta_H(\tau)$ monotonically decreases from $\Delta_H(\tau=0)=\mu_H-Rev(\mu_H,\sigma)$ to $\Delta_H(\tau=1)=\mu_L-Rev(\mu_H,\sigma)< \Delta_L(\tau=1)$.
    So in this case, optimal posted price is $$
    \begin{cases}
        \pi_H & \text{ if } \tau \leq \frac{\mu_H-Rev(\mu_H,\sigma)}{\mu_H-Rev(\mu_L,\sigma)}\\
        \pi_L & \text{otherwise}
    \end{cases}
    $$
    \textbf{Case 2} If $\mu_L-Rev(\mu_L,\sigma)\leq 0$ but $\mu_H-Rev(\mu_H,\sigma)>0$, then $\Delta_L(\tau)$ monotonically decreases from 0, and $\Delta_H(\tau)$ decreases from $\Delta_H(\tau=0)=\mu_H-Rev(\mu_H,\sigma)>0$ to $\Delta_H(\tau=1)=\mu_L-Rev(\mu_H,\sigma)$. The optimal posted price is 
    $$
    \begin{cases}
        \pi_H & \text{ if } \tau \leq \frac{\mu_H-Rev(\mu_H,\sigma)}{\mu_H-\mu_L}\\
        \pi_0 & \text{otherwise}
    \end{cases}
    $$
    \textbf{Case 3} if $\mu_H-Rev(\mu_H,\sigma)<0$, then both $\Delta_L(\tau)$ and $\Delta_H(\tau)<0$. So optimal posted price is to post $\pi_0$.

    Now, for the in-house production part of the theorem. We first show that when in-house producing, the profit-maximizing strategy always produces high-quality items. 

    Whenever $\pi_0$ being the profit-maximizing posted price, it has to be in the above Case 2 or 3, which means that $\mu_L-Rev(\mu_L,\sigma)\leq 0 \Leftrightarrow 1\leq Rev(1,\frac{\sigma}{\mu_L})$. Then the platform producing in-house item of quality $\mu_L$ has profit $$Rev(m\mu_L,\sqrt{m}\sigma)-m Rev(\mu_L,\sigma)=m\mu_L (Rev(1,\frac{\sigma}{\sqrt{m}\mu_L})-Rev(1,\frac{\sigma}{\mu_L}))\leq 0$$
    The last inequality is because from \Cref{fig:rev_one_sigma} and \Cref{lem:norm_rev_character}, $Rev(1,\sigma)$, when $Rev(1,\frac{\sigma}{\mu_L})\geq 1$, any further reduction in buyer valuation harms optimal revenue. This means the platform never in-house produces low-quality items at $\pi_0=0$. The condition for in-house producing $m$ units of high-quality items is given by $$Rev(m\mu_H,\sqrt{m}\sigma)-mRev(\mu_H,\sigma)$$
    the sign of which depends on $m,\mu_H$, and $\sigma$.
    
    When profit-maximizing posted price is $\pi_L$ or $\pi_H$, let $\epsilon:=1-\frac{Rev(\mu_H,\sigma)-Rev(\mu_L,\sigma)}{\mu_H-\mu_L}$. By \Cref{lem:norm_rev_character},
    $$Rev(\mu_H,\sigma)-Rev(\mu_L,\sigma)= \int_{\mu=\mu_L}^{\mu_H}\frac{\partial Rev(\mu,\sigma)}{\partial\mu}d\mu < \mu_H-\mu_L$$
    so 
    $\epsilon\in (0,1)$. Let $\alpha_\epsilon$ be the threshold value satisfying \Cref{lem:decreasing_mills_ratio}, that is $\forall \alpha>\alpha_\epsilon, F[z^*(\alpha)-\alpha]>1-\epsilon$. Now choose $N(\tau) = \lceil(\frac{\sigma}{\tau \mu_L}\alpha_\epsilon)^2\rceil$. Then for $N\geq N(\tau)$ and any unit $m$ of items that platform in-house produces,
    \begin{align*}
        \sqrt{N_L+m}\frac{\mu_L}{\sigma}\geq \sqrt{\tau N}\frac{\mu_L}{\sigma}> \alpha_\epsilon\quad \text{ and }\quad
        \frac{(N_L+m)\mu_L+N_H\mu_H}{\sqrt{N+m}\sigma}> \tau \sqrt{N+m}\frac{\mu_L}{\sigma}>\alpha_\epsilon
    \end{align*}
    Invoking \Cref{lem:decreasing_mills_ratio}, 
    \begin{align*}
        F[z^*(\sqrt{N_L+m}\frac{\mu_L}{\sigma})-\sqrt{N_L+m}\frac{\mu_L}{\sigma}] &> \frac{Rev(\mu_H\sigma)-Rev(\mu_L,\sigma)}{\mu_H-\mu_L}\\
        F[z^*(\frac{(N_L+m)\mu_L+N_H \mu_H}{\sqrt{N+m}\sigma})-\frac{(N_L+m)\mu_L+N_H \mu_H}{\sqrt{N+m}\sigma}]&>\frac{Rev(\mu_H,\sigma)-Rev(\mu_L,\sigma)}{\mu_H-\mu_L}
    \end{align*}
    These are the sufficient condition for producing high-quality items when the platform bundles some sellers, given in \Cref{lem:in_house_no_mix}. So in large markets when the platform posts price $\pi_L$ or $\pi_H$, the platform always in-house produces high quality items if it in-house produces. As the market is large and the platform bundles some sellers, producing $m$ units of high-quality items increase platform revenue by $m\mu_H$ and costs the platform $m Rev(\mu_H,\sigma)$. The platform produces high-quality items in house if and only if $\mu_H> Rev(\mu_H,\sigma)$.
\end{proof}

\begin{example}\label{exam:low_quality_in_house_optimal}
    The following is a market where it is profit-maximizing to produce low-quality item in house. $N=4, N_L=3, M=1,\mu_L=1,\mu_H=20,\sigma=0.5$. In this market, profits are 
    \begin{itemize}
        \item 0.242 when only bundling low-quality seller with no in-house production
        \item -54.066 when bundling all sellers with no in-house production
        \item 0 for just in-house produce high or low-quality seller because the max capacity is 1
        \item 0.428 for bundling low-quality sellers and in-house producing 1 unit of low-quality item
        \item 0.303 for bundling low-quality sellers and in-house producing 1 unit of high-quality item
        \item -53.849 for bundling all sellers and in-house producing 1 unit of low-quality item
        \item -54.247 for bundling all sellers and in-house producing 1 unit of high-quality item
    \end{itemize}
    So the profit-maximizing strategy is to bundle low-quality sellers and in-house produce 1 unit of low-quality sellers.
\end{example}


\NoLowInhouseProduction*
\begin{proof}
    For any platform's bundling decision to include a set of sellers $S$, let $\mu_S$ be the mean of the bundle.
    We show that regardless of the bundling decision, producing low-quality items is dominated by producing high-quality items. By switching from producing $m$ units of low-quality to $m$ units of high-quality item, the revenue gain in the bundle is equal to 
    $$Rev(\mu_S+m\mu_H, \sqrt{m+|S|}\sigma)-Rev(\mu_S+m\mu_L, \sqrt{m+|S|}\sigma)=\int_{\mu=\mu_L}^{\mu_H} mF\left[ z^*\left(\frac{\mu_S+m\mu}{\sqrt{m+|S|}\sigma}\right)-\frac{\mu_S+m\mu}{\sqrt{m+|S|}\sigma}\right]d\mu$$
    where we used \Cref{lem:norm_rev_character} to write out the derivative of optimal revenue to mean for normal valuations. The lost due to opportunity cost is given by 
    $$mRev(\mu_H,\sigma)-mRev(\mu_L,\sigma)=m\int_{\mu=\mu_L}^{\mu_H} F\left[z^*\left(\frac{\mu}{\sigma}\right)-\frac{\mu}{\sigma}\right]d\mu$$
    So the total change in profit switching from producing low-quality to high-quality seller becomes
    $$ m\int_{\mu=\mu_L}^{\mu_H}F\left[ z^*\left(\frac{\mu_S+m\mu}{\sqrt{m+|S|}\sigma}\right)-\frac{\mu_S+m\mu}{\sqrt{m+|S|}\sigma}\right]-F\left[z^*\left(\frac{\mu}{\sigma}\right)-\frac{\mu}{\sigma}\right]d\mu$$
    If producing low-quality item is profit-maximizing, it cannot be that for all $\mu$
    $$ \frac{\mu_S+m\mu}{\sqrt{m+|S|}\sigma} > \frac{\mu}{\sigma}$$
    across all three cases of $S$ being an empty set, $S$ being all low-quality sellers, and $S$ being all sellers. Otherwise, by \Cref{lem:norm_rev_character}, $\frac{d F[z^*(\alpha)-\alpha]}{d\alpha}>0$ and the change of profit is always positive by switching to high-quality in-house production, so producing low-quality item is no longer profit-maximizing.
  
    When $S=\emptyset$, $\mu_S=0$. The above inequality indeed holds $$\frac{\sqrt{m}\mu}{\sigma}>\frac{\mu}{\sigma}$$
    This means that if the platform doesn't source from external sellers $S=\emptyset$, then producing high-quality items in-house have higher profit than low-quality items. Now we continue with the proof of the necessary condition. If producing low-quality item in-house is beneficial, then it is required that for some value of $\mu$ 
    \begin{align*}
        \frac{N_L\mu_L+m\mu}{\sqrt{m+N_L}\sigma} <\frac{\mu}{\sigma} \quad&\text{ or }\quad \frac{N_L\mu_L+(N-N_L)\mu_H+m\mu}{\sqrt{m+N}\sigma}<\frac{\mu}{\sigma} \\
        \Leftrightarrow N_L\mu_L<(\sqrt{m+N_L}-m)\mu \quad&\text{ or }\quad N_L\mu_L+(N-N_L)\mu_H < (\sqrt{m+N}-m)\mu
    \end{align*}   
    A necessary condition for the above is that $$\begin{cases}\sqrt{m+N_L}>(m+N_L)-N_L\\ \frac{\mu_H}{\mu_L}> \frac{N_L}{\sqrt{m+N_L}-(m+N_L)+N_L}\end{cases}\quad \text{ or }\quad
    \begin{cases}\sqrt{m+N}> (m+N)-N_L\\\frac{\mu_H}{\mu_L}>\frac{N_L}{\sqrt{m+N}-(m+N)+N_L}\end{cases}$$
    as $\sqrt{x}-x$ is decreasing when $x\geq 1$, the above necessary condition is equivalent to 
    $$ \begin{cases}\sqrt{m+N_L}>m\\ \frac{\mu_H}{\mu_L}> \frac{N_L}{\sqrt{m+N_L}-m}\end{cases} $$
\end{proof}

\begin{example}\label{exam:sub-modularity_bundling_producing}
    The following is a market where the profit in jointly bundling and producing is smaller than that of bundling alone plus producing alone. $N_L=3, m=2,\mu_L=1,\mu_H=20,\sigma=1$. In this market, profits respectively are
    \begin{itemize}
        \item Producing high-quality items in-house and bundling low-quality sellers $$Rev(m\mu_H+N_L\mu_L,\sqrt{m+N_L}\sigma)-mRev(\mu_H,\sigma)-N_LRev(\mu_L,\sigma)\approx 0.952$$
        \item Producing high-quality items
        $$ Rev(m\mu_H,\sqrt{m}
\sigma)-mRev(\mu_H,\sigma)\approx 0.018$$
\item Bundling low-quality items $$Rev(N_L\mu_L,\sqrt{N_L}\sigma)-N_L Rev(\mu_L,\sqrt{N_L}\sigma)\approx 1.196$$
    \end{itemize}
    So it satisfies the ``sub-modularity'' effect where the profit in jointly bundling and producing is smaller than that of bundling alone plus bundling alone.
\end{example}

\clearpage
\singlespacing
\bibliographystyle{plainnat}
\bibliography{ref}

@article{hagiu2015multi,
  title={Multi-sided platforms},
  author={Hagiu, Andrei and Wright, Julian},
  journal={International journal of industrial organization},
  volume={43},
  pages={162--174},
  year={2015},
  publisher={Elsevier}
}

@book{evans2016matchmakers,
  title={Matchmakers: The new economics of multisided platforms},
  author={Evans, David S and Schmalensee, Richard},
  year={2016},
  publisher={Harvard Business Review Press}
}

@article{tirole2017economics,
  title={Economics for the common good},
  author={Tirole, Jean},
  year={2017},
  publisher={Princeton University Press}
}

@article{roth2007art,
  title={The art of designing markets},
  author={Roth, Alvin E},
  journal={Harvard business review},
  volume={85},
  number={10},
  pages={118},
  year={2007}
}

@article{einav2016peer,
  title={Peer-to-peer markets},
  author={Einav, Liran and Farronato, Chiara and Levin, Jonathan},
  journal={Annual Review of Economics},
  volume={8},
  pages={615--635},
  year={2016},
  publisher={Annual Reviews}
}

@article{fradkin2017search,
  title={Search, matching, and the role of digital marketplace design in enabling trade: Evidence from Airbnb},
  author={Fradkin, Andrey},
  journal={Matching, and the Role of Digital Marketplace Design in Enabling Trade: Evidence from Airbnb (March 21, 2017)},
  year={2017}
}

@article{rochet2003platform,
  title={Platform competition in two-sided markets},
  author={Rochet, Jean-Charles and Tirole, Jean},
  journal={Journal of the european economic association},
  volume={1},
  number={4},
  pages={990--1029},
  year={2003},
  publisher={Oxford University Press}
}

@article{weyl2010price,
  title={A price theory of multi-sided platforms},
  author={Weyl, E Glen},
  journal={American economic review},
  volume={100},
  number={4},
  pages={1642--1672},
  year={2010},
  publisher={American Economic Association}
}

@article{birge2021optimal,
  title={Optimal commissions and subscriptions in networked markets},
  author={Birge, John and Candogan, Ozan and Chen, Hongfan and Saban, Daniela},
  journal={Manufacturing \& Service Operations Management},
  volume={23},
  number={3},
  pages={569--588},
  year={2021},
  publisher={INFORMS}
}

@inproceedings{banerjee2017segmenting,
  title={Segmenting two-sided markets},
  author={Banerjee, Siddhartha and Gollapudi, Sreenivas and Kollias, Kostas and Munagala, Kamesh},
  booktitle={Proceedings of the 26th International Conference on World Wide Web},
  pages={63--72},
  year={2017}
}

@article{ricci2021recommender,
  title={Recommender systems: Techniques, applications, and challenges},
  author={Ricci, Francesco and Rokach, Lior and Shapira, Bracha},
  journal={Recommender systems handbook},
  pages={1--35},
  year={2021},
  publisher={Springer}
}

@book{aggarwal2016recommender,
  title={Recommender systems},
  author={Aggarwal, Charu C and others},
  volume={1},
  year={2016},
  publisher={Springer}
}

@inproceedings{chen2016empirical,
  title={An empirical analysis of algorithmic pricing on amazon marketplace},
  author={Chen, Le and Mislove, Alan and Wilson, Christo},
  booktitle={Proceedings of the 25th international conference on World Wide Web},
  pages={1339--1349},
  year={2016}
}

@misc{ec2022amazon,
  author = {{European Commission}},
  title = {Antitrust: Commission accepts commitments by Amazon on e-commerce practices},
  year = {2022},
  howpublished = {\url{https://ec.europa.eu/commission/presscorner/detail/en/ip_22_7777}},
  note = {Press release, December 20, 2022}
}

@String{JACM = "J. ACM" }

@String{Computing = "Computing" }

@String{Computer = "{IEEE} Computer" }

@String{Springer = "Springer-Verlag" }

@article{elliott2015inefficiencies,
  title={Inefficiencies in networked markets},
  author={Elliott, Matthew},
  journal={American Economic Journal: Microeconomics},
  volume={7},
  number={4},
  pages={43--82},
  year={2015},
  publisher={American Economic Association 2014 Broadway, Suite 305, Nashville, TN 37203-2425}
}

@article{gul1999walrasian,
  title={Walrasian equilibrium with gross substitutes},
  author={Gul, Faruk and Stacchetti, Ennio},
  journal={Journal of Economic theory},
  volume={87},
  number={1},
  pages={95--124},
  year={1999},
  publisher={Elsevier}
}

@article{kelso1982job,
  title={Job matching, coalition formation, and gross substitutes},
  author={Kelso, Alexander S and Crawford, Vincent P},
  journal={Econometrica: Journal of the Econometric Society},
  pages={1483--1504},
  year={1982},
  publisher={JSTOR}
}

@article{caillaud2003chicken,
  title={Chicken \& egg: Competition among intermediation service providers},
  author={Caillaud, Bernard and Jullien, Bruno},
  journal={RAND journal of Economics},
  pages={309--328},
  year={2003},
  publisher={JSTOR}
}

@article{raj2020covid,
  title={COVID-19 and digital resilience: Evidence from Uber Eats},
  author={Raj, Manav and Sundararajan, Arun and You, Calum},
  journal={arXiv preprint arXiv:2006.07204},
  year={2020}
}

@techreport{bloom2021impact,
  title={The impact of COVID-19 on US firms},
  author={Bloom, Nicholas and Fletcher, Robert S and Yeh, Ethan},
  year={2021},
  institution={National Bureau of Economic Research}
}

@article{kotowski2019trading,
  title={Trading networks and equilibrium intermediation},
  author={Kotowski, Maciej H and Leister, C Matthew},
  year={2019},
  publisher={HKS Working Paper No. RWP18-001}
}

@article{condorelli2017bilateral,
  title={Bilateral trading in networks},
  author={Condorelli, Daniele and Galeotti, Andrea and Renou, Ludovic},
  journal={The Review of Economic Studies},
  volume={84},
  number={1},
  pages={82--105},
  year={2017},
  publisher={Oxford University Press}
}

@article{kakade2004economic,
  title={Economic properties of social networks},
  author={Kakade, Sham M and Kearns, Michael and Ortiz, Luis E and Pemantle, Robin and Suri, Siddharth},
  journal={Advances in Neural Information Processing Systems},
  volume={17},
  year={2004}
}

@inproceedings{even2007network,
  title={A network formation game for bipartite exchange economies},
  author={Even-Dar, Eyal and Kearns, Michael J and Suri, Siddharth},
  booktitle={SODA},
  pages={697--706},
  year={2007}
}

@article{mckinsey2021,
	title={Ordering in: The rapid evolution of food delivery}, 
	author={Kabir Ahuja and Vishwa Chandra and Victoria Lord and Curtis Peens},
	year={2021},
	journal = {McKinsey \& Company},
}

@inproceedings{ZhangC20,
  author       = {Hanrui Zhang and
                  Vincent Conitzer},
  title        = {Learning the Valuations of a k-demand Agent},
  booktitle    = {Proceedings of the 37th International Conference on Machine Learning,
                  {ICML} 2020, 13-18 July 2020, Virtual Event},
  series       = {Proceedings of Machine Learning Research},
  volume       = {119},
  pages        = {11066--11075},
  publisher    = {{PMLR}},
  year         = {2020},
  url          = {http://proceedings.mlr.press/v119/zhang20f.html},
  bibsource    = {dblp computer science bibliography, https://dblp.org}
}

@inproceedings{BergerEF20,
  author       = {Ben Berger and
                  Alon Eden and
                  Michal Feldman},
  editor       = {Xujin Chen and
                  Nikolai Gravin and
                  Martin Hoefer and
                  Ruta Mehta},
  title        = {On the Power and Limits of Dynamic Pricing in Combinatorial Markets},
  booktitle    = {Web and Internet Economics - 16th International Conference, {WINE}
                  2020, Beijing, China, December 7-11, 2020, Proceedings},
  series       = {Lecture Notes in Computer Science},
  volume       = {12495},
  pages        = {206--219},
  publisher    = {Springer},
  year         = {2020},
  url          = {https://doi.org/10.1007/978-3-030-64946-3\_15},
  doi          = {10.1007/978-3-030-64946-3\_15},
  bibsource    = {dblp computer science bibliography, https://dblp.org}
}

@article{kranton2001theory,
  title={A theory of buyer-seller networks},
  author={Kranton, Rachel E and Minehart, Deborah F},
  journal={American economic review},
  volume={91},
  number={3},
  pages={485--508},
  year={2001}
}

@article{kranton2000competition,
  title={Competition for goods in buyer-seller networks},
  author={Kranton, Rachel E and Minehart, Deborah F},
  journal={Review of Economic Design},
  volume={5},
  number={3},
  pages={301--331},
  year={2000},
  publisher={Springer}
}

@misc{maxweightsubmod,
  title = {Maximum Weight Matching and submodular functions},
  year={2016},
  author={George Octavian Rabanca},
  howpublished = {\url{https://cstheory.stackexchange.com/questions/34654/maximum-weight-matching-and-submodular-functions}},
  note = {Accessed: 2023-08-01}
}

@article{FeeCapNews,
 author  = {Rich Calder},
 date    = {2021-06-24},
 year    = {2021},
 title   = {Permanent Cap on Delivery-App Fees Proposed for New York City},
 journal = {The Wall Street Journal},
 url     = {https://www.wsj.com/articles/permanent-cap-on-delivery-app-fees-proposed-for-new-york-city-11624549165},
 urldate = {2023-09-07}
}

@article{felix2020us,
  title={US food supply chain: Disruptions and implications from COVID-19},
  author={Felix, Ignacio and Martin, Adrian and Mehta, Vivek and Mueller, Curt},
  journal={McKinsey \& Company, July},
  year={2020}
}

@inproceedings{platformEq,
author = {Eden, Alon and Ma, Gary Qiurui and Parkes, David C.},
title = {Platform Equilibrium: Analyzing Social Welfare in Online Market Places},
year = {2024},
isbn = {9798400707049},
publisher = {Association for Computing Machinery},
address = {New York, NY, USA},
url = {https://doi.org/10.1145/3670865.3673496},
doi = {10.1145/3670865.3673496},
booktitle = {Proceedings of the 25th ACM Conference on Economics and Computation},
pages = {542},
numpages = {1},
location = {New Haven, CT, USA},
series = {EC '24}
}

@inproceedings{platformDisruption,
author = {D'Amico-Wong, Luca and Gonczarowski, Yannai A. and Ma, Gary Qiurui and Parkes, David C.},
title = {Disrupting Bipartite Trading Networks: Matching for Revenue Maximization},
year = {2024},
isbn = {9798400707049},
publisher = {Association for Computing Machinery},
address = {New York, NY, USA},
url = {https://doi.org/10.1145/3670865.3673567},
doi = {10.1145/3670865.3673567},
booktitle = {Proceedings of the 25th ACM Conference on Economics and Computation},
pages = {545–546},
numpages = {2},
location = {New Haven, CT, USA},
series = {EC '24}
}

@inproceedings{platformSim,
author = {Wang, Xintong and Ma, Gary Qiurui and Eden, Alon and Li, Clara and Trott, Alexander and Zheng, Stephan and Parkes, David},
title = {Platform Behavior under Market Shocks: A Simulation Framework and Reinforcement-Learning Based Study},
year = {2023},
isbn = {9781450394161},
publisher = {Association for Computing Machinery},
address = {New York, NY, USA},
url = {https://doi.org/10.1145/3543507.3583523},
doi = {10.1145/3543507.3583523},
booktitle = {Proceedings of the ACM Web Conference 2023},
pages = {3592–3602},
numpages = {11},
location = {Austin, TX, USA},
series = {WWW '23}
}

@article{gonczarowski2025pricing,
  title={Pricing with Tips in Three-Sided Delivery Platforms},
  author={Gonczarowski, Yannai A and Ma, Gary Qiurui and Parkes, David C},
  journal={arXiv preprint arXiv:2507.10872},
  year={2025}
}

@article{OptimalSourcing2026,
  title={Buyer to Bundle: Optimal Sourcing from Monopolistic Sellers},
  author={Immorlica, Nicole and Lucier, Branden and Ma, Gary Qiurui},
  journal={},
  year={2026}
}

@book{ben1976tip,
  title={" Tip" payments and the quality of service},
  author={Ben-Zion, Uri and Karni, Edi},
  year={1976},
  publisher={Foerder Institute of Economic Research, Tel-Aviv University}
}

@article{azar2005social,
  title={The social norm of tipping: does it improve social welfare?},
  author={Azar, Ofer H},
  journal={Journal of Economics},
  volume={85},
  pages={141--173},
  year={2005},
  publisher={Springer}
}

@inproceedings{randomizedFifo,
author = {Castro, Francisco and Ma, Hongyao and Nazerzadeh, Hamid and Yan, Chiwei},
title = {Randomized FIFO Mechanisms},
year = {2022},
isbn = {9781450391504},
publisher = {Association for Computing Machinery},
address = {New York, NY, USA},
url = {https://doi.org/10.1145/3490486.3538353},
doi = {10.1145/3490486.3538353},
booktitle = {Proceedings of the 23rd ACM Conference on Economics and Computation},
pages = {60},
numpages = {1},
location = {Boulder, CO, USA},
series = {EC '22}
}

@online{NoTipNoTripNewsweek,
  author       = {Nick Mordowanec},
  title        = {{``No Tip, No Trip'': DoorDash Orders Go Undelivered as Drivers Refuse Low Pay}},
  year         = {2022},
  url = {https://www.newsweek.com/no-tip-no-trip-undelivered-doordash-orders-spark-debate-viral-video-1697322},
  note         = {Newsweek}
}

@article{stuart1997supplier,
  title={The supplier--firm--buyer game and its m-sided generalization},
  author={Stuart Jr, Harborne W},
  journal={Mathematical Social Sciences},
  volume={34},
  number={1},
  pages={21--27},
  year={1997},
  publisher={Elsevier}
}

@article{quint1991core,
  title={The core of an m-sided assignment game},
  author={Quint, Thomas},
  journal={Games and Economic Behavior},
  volume={3},
  number={4},
  pages={487--503},
  year={1991},
  publisher={Elsevier}
}

@article{atay2023matching,
  title={Matching markets with middlemen under transferable utility},
  author={Atay, Ata and Bahel, Eric and Solymosi, Tam{\'a}s},
  journal={Annals of Operations Research},
  volume={322},
  number={2},
  pages={539--563},
  year={2023},
  publisher={Springer}
}

@article{oishi2014middlemen,
  title={Middlemen in the Shapley--Shubik competitive markets for indivisible goods},
  author={Oishi, Takayuki and Sakaue, Shin},
  journal={Mathematical Economics Letters},
  volume={2},
  number={1-2},
  pages={19--26},
  year={2014},
  publisher={De Gruyter}
}

@article{atay2016generalized,
  title={Generalized three-sided assignment markets: core consistency and competitive prices},
  author={Atay, Ata and Llerena, Francesc and N{\'u}{\~n}ez, Marina},
  journal={Top},
  volume={24},
  number={3},
  pages={572--593},
  year={2016},
  publisher={Springer}
}

@article{snitkovsky2021modeling,
  title={A modeling framework for tipping in the presence of a social norm},
  author={Snitkovsky, Ran I and Debo, Laurens},
  journal={Available at SSRN 3932570},
  year={2021}
}

@article{debo2018tipping,
  title={Tipping in service systems: The role of a social norm},
  author={Debo, Laurens and Snitkovsky, Ran I},
  journal={Tuck School of Business Working Paper},
  number={3287862},
  year={2018}
}

@online{notipnotrip,
  author = {Braden Bjella},
  title = {doordasher-shows-piles-of-non-tipping-orders},
  year = 2023,
  url = {https://www.dailydot.com/irl/doordasher-shows-piles-of-non-tipping-orders/},
  urldate = {2025-01-25}
}

@online{notipwaitlonger,
  author = {Christine Hauser},
  title = {No Tip for your Delivery Driver? Then Be Prepared to Wait},
  year = 2023,
  url = {https://www.nytimes.com/2023/11/02/business/doordash-tip-warning.html},
  note = {The New York Times},
  urldate = {2025-03-29}
}

@article{CherryPickingExample,
  author = {Adriana Diaz},
  title = {I deliver food--- I only accept orders in wealthy areas to ensure good tips},
  year = 2023,
  journal = {New York Post},
  url = {https://nypost.com/2023/05/05/i-deliver-food-i-only-accept-orders-in-wealthy-areas-to-ensure-good-tips/},
  note = {New York Post},
  urldate = {2025-03-29}
}

@online{EEOCEqualPayAct,
  author       = {{U.S. Equal Employment Opportunity Commission}},
  title        = {Equal Pay Act of 1963},
  year         = {2025},
  url = {https://www.eeoc.gov/statutes/equal-pay-act-1963},
  note         = {Accessed: 2025-12-24}
}

@online{EUPlatformWorkDirective2024,
  author       = {{European Union}},
  title        = {Directive on improving working conditions in platform work},
  year         = {2024},
  howpublished = {Official Journal of the European Union},
  note         = {Establishes fairness and transparency principles for digital labor platforms}
}

@article{NYCUber,
  author = {Matthew Haag},
  title = {Uber and DoorDash Try to Halt N.Y.C. Law
That Encourages Tipping},
  year = 2025,
  journal = {New York Times},
  url = {https://www.nytimes.com/2025/12/16/nyregion/uber-doordash-nyc-tipping.html},
  note = {The New York Times},
  urldate = {2025-12-24}
}

@online{UberEatTipPolicy,
  author = {UberEats},
  title = {How do I tip my delivery person?},
  url ={https://help.uber.com/ubereats/restaurants/article/how-do-i-tip-my-delivery-person?nodeId=d36fe0b4-2b19-4cbb-83d3-37dd72ed1bdf},
  note = {Accessed: 2025-12-24},
  year={2025},
  urldate = {2025-12-24}
}

@online{UberEatTipDriverView,
  author = {UberEats},
  title = {Delivery Fares Explained},
  url ={https://www.uber.com/blog/how-delivery-fares-works/},
  note = {Accessed: 2025-12-24},
  year={2025},
  urldate = {2025-12-24}
}

@online{UberEatSurgePricing,
  author = {UberEats},
  title = {Higher delivery fee},
  url ={https://help.uber.com/ubereats/restaurants/article/higher-delivery-fee?nodeId=4938ca61-e0d6-493a-9384-eb005c2eb6e5},
  note = {Accessed: 2025-12-24},
  year={2025},
  urldate = {2025-12-24}
}

@online{UberEatBuyerPricing,
  author = {UberEats},
  title = {What fees may apply to my orders?},
  url ={https://help.uber.com/ubereats/restaurants/article/what-fees-may-apply-to-my-order?nodeId=65d229e2-a2b4-4fa0-b10f-b36c9546cf55},
  note = {Accessed: 2025-12-24},
  year={2025},
  urldate = {2025-12-24}
}

@online{UberEatDeliveryRadius,
  author = {UberEats},
  title = {Flexible pricing and fees to help you reach your goals
},
  url ={https://merchants.ubereats.com/us/en/pricing/},
  note = {Accessed: 2025-12-24},
  year={2025},
  urldate = {2025-12-24}
}

@online{DoorDashTipPolicy,
  author       = {{DoorDash}},
  title        = {How Dasher pay works},
  year         = {2025},
  url = {https://help.doordash.com/dashers/s/article/How-is-Dasher-pay-calculated?language=en_US},
  note         = {Accessed: 2025-12-24},
  urldate = {2025-12-24}
}

@online{DoorDashOneMatch,
  author       = {{DoorDash}},
  title        = {How to accept an offer?},
  year         = {2025},
  url = {https://help.doordash.com/dashers/s/article/How-to-Accept-an-Order?language=en_US},
  note         = {Accessed: 2025-12-24},
  urldate = {2025-12-24}
}

@online{DoorDashSurgePricing,
  author = {DoorDash},
  title = {Peak Pay},
  url ={https://help.doordash.com/dashers/s/article/Peak-Pay?language=en_US},
  note = {Accessed: 2025-12-24},
  year={2025},
  urldate = {2025-12-24}
}

@online{DoorDashBuyerPricing,
  author = {DoorDash},
  title = {What fees do I pay?},
  url ={https://help.doordash.com/consumers/s/article/What-fees-do-I-pay?language=en_US},
  note = {Accessed: 2025-12-24},
  year={2025},
  urldate = {2025-12-24}
}

@online{DoorDashPositiveEarnings,
  author = {DoorDash},
  title = {DoorDash Releases Third Quarter 2024 Financial Results},
  url ={https://ir.doordash.com/news/news-details/2024/DoorDash-Releases-Third-Quarter-2024-Financial-Results/default.aspx},
  note = {Accessed: 2025-12-24},
  year={2024},
  urldate = {2025-12-24}
}

@online{DoorDashDeliveryRadius,
  author = {DoorDash},
  title = {What is a “Delivery Radius” or a “Delivery Area” on DoorDash?},
  url ={https://help.doordash.com/consumers/s/article/What-is-a-Delivery-Radius-or-a-Delivery-Area-on-DoorDash?language=en_US},
  note = {Accessed: 2025-12-24},
  year={2025},
  urldate = {2025-12-24}
}

@online{DoorDashDriverAPI,
  author = {DoorDash},
  title = {Driver API, Pricing and payment},
  url ={https://developer.doordash.com/en-US/docs/drive/overview/pricing_payment},
  note = {Accessed: 2025-12-24},
  year={2025},
  urldate = {2025-12-24}
}

@online{GrubhubTipPolicy,
  author       = {{Grubhub}},
  title        = {Tip policy},
  year         = {2025},
  url = {https://driver-support.grubhub.com/hc/en-us/articles/20977077466516-Seattle-Tip-Policy},
  note         = {Accessed: 2025-12-24}
}

@online{GrubhubOneMatch,
  author       = {{Grubhub}},
  title        = {How do I use the Grubhub for Drivers app to complete deliveries?},
  year         = {2025},
  url = {https://driver-support.grubhub.com/hc/en-us/articles/360030830171-How-do-I-use-the-Grubhub-for-Drivers-app-to-complete-deliveries},
  note         = {Accessed: 2025-12-24}
}

@online{GrubhubDriverView,
  author       = {{Grubhub}},
  title        = {Do Grubhub drivers know if you tip?},
  year         = {2025},
  url = {https://www.grubhub.com/answers/do-grubhub-drivers-know-if-you-tip},
  note         = {Accessed: 2025-12-24}
}

@online{GrubhubSurgePricing,
  author       = {{Grubhub}},
  title        = {Courier Pay},
  year         = {2025},
  url = {https://driver-support.grubhub.com/hc/en-us/articles/4407240376980-What-is-a-Mission},
  note         = {Accessed: 2025-12-24}
}

@online{GrubhubSurgePricingSecond,
  author       = {{Grubhub}},
  title        = {Tips to Maximize your Time on the Road},
  year         = {2025},
  url = {https://driver.grubhub.com/blog/tips-to-maximize-your-time/.},
  note         = {Accessed: 2025-12-24}
}

@online{GrubhubDeliveryRadius,
  author       = {{Grubhub}},
  title        = {Grubhub pricing and fees},
  year         = {2025},
  url = {https://get.grubhub.com/grubhub-pricing-and-fees/},
  note         = {Accessed: 2025-12-24}
}

@online{GrubhubNegativeEarnings,
  author       = {{Just Eat Takeaway}},
  title        = {Full Year 2024 Results
},
  year         = {2025},
  url = {https://newsroom.justeattakeaway.com/en-WW/247233-full-year-2024-results/},
  note         = {Accessed: 2025-12-24}
}

@online{GrubhubNegativeEarnings2025,
  author       = {{Just Eat Takeaway}},
  title        = {Half Year 2025 Results
},
  year         = {2025},
  url = {https://newsroom.justeattakeaway.com/en-WW/252645-half-year-2025-results/},
  note         = {Accessed: 2025-12-24}
}

@online{InstacartDriverView,
  author       = {{Instacart}},
  title        = {How earning with Instacart works?},
  year         = {2025},
  url = {https://www.instacart.com/company/shoppers/shopper-earnings},
  note         = {Accessed: 2025-12-24}
}

@online{UberRideTipPolicy,
  author = {{Uber}},
  title = {How tips work},
  url ={https://www.uber.com/nz/en/drive/driver-app/how-tips-work/},
  year={2025},
  note = {Accessed: 2025-12-24},
  urldate = {2025-12-24}
}

@online{UberTripRadar,
  author = {{Uber}},
  title = {Introducing Trip Radar},
  url ={https://www.uber.com/en-AU/blog/introducing-trip-radar/},
  year={2025},
  note = {Accessed: 2025-12-24},
  urldate = {2025-12-24}
}

@online{UberPositiveEarnings,
  author = {{Uber}},
  title = {Uber Announces Results for Second Quarter 2023},
  url ={https://investor.uber.com/news-events/news/press-release-details/2023/Uber-Announces-Results-for-Second-Quarter-2023/default.aspx},
  year={2023},
  note = {Accessed: 2025-12-24},
  urldate = {2025-12-24}
}

@article{UberEatRobotDelivery,
  author  = {Dan Raby},
  title   = {{Uber Eats launches new robot delivery service in Atlanta}},
  journal = {FOX5 Atlanta},
  year    = {2025},
  url     = {https://www.fox5atlanta.com/news/uber-eats-robot-delivery-service-atlanta},
  note    = {Accessed: 2025-12-24}
}

@article{UberEatRobotDeliveryNoTip,
  author  = {Kurt Knutsson},
  title   = {{Your next takeout burger could arrive at your doorstep via robot delivery}},
  journal = {FOX News},
  year    = {2024},
  url     = {https://www.foxnews.com/tech/your-next-takeout-burger-could-arrive-your-doorstep-via-robot-delivery},
  note    = {Accessed: 2025-12-24}
}

@online{LyftTipPolicy,
  author = {{Lyft}},
  title = {How to tip your driver},
  url ={https://help.lyft.com/hc/en-us/all/articles/115013081368-How-to-tip-your-driver},
  note = {Accessed: 2025-12-24},
year={2025},
  urldate = {2025-12-24}
}

@online{LyftBusinessTipPolicy,
  author = {{Lyft}},
  title = {Introducing tipping for concierge rides},
  url ={https://www.lyft.com/business/resources/product-updates/introducing-tipping-for-concierge-rides},
  note = {Accessed: 2025-12-24},
year={2025},
  urldate = {2025-12-24}
}

@online{InstacartAccessBatches,
  author = {Instacart},
  title = {How accessing batches works at Instacart},
  url ={https://www.instacart.com/company/shoppers/access-batches},
  note = {Accessed: 2025-12-24},
  urldate = {2025-12-24},
  year={2025}
}

@online{InstacartPositiveEarnings,
  author = {Instacart},
  title = {Lettor to shareholders},
  url ={https://investors.instacart.com/static-files/45c59490-b81c-4401-ba22-be639847baa7},
  note = {Accessed: 2025-12-24},
  urldate = {2025-12-24},
  year={2024}
}

@online{InstacartDistanceBasePrice,
  author = {Instacart},
  title = {Instacart fees and taxes},
  url ={https://www.instacart.com/help/section/360007902791/360039164252},
  note = {Accessed: 2025-12-24},
  urldate = {2025-12-24},
  year={2025}
}

@article{InstacartPeakBoost,
  author = {Brett Helling},
  title = {How Instacart Peak Boost Can Help Shoppers Earn More Money},
  url ={https://teachmedelivery.com/courier/instacart-peak-boost/},
  journal = {TeachMeDelivery},
  note = {Accessed: 2025-12-24},
  urldate = {2025-12-24},
  year={2024}
}

@article{InstacartBasePay,
  author  = {Alex Bitter and Nancy Luna},
  title   = {{Instacart just slashed pay for delivery drivers—from a minimum of \$7 per order to \$4. Now they're more reliant on your tips}},
  journal = {Business Insider},
  year    = {2023},
  url     = {https://www.businessinsider.com/instacart-cuts-minimum-base-pay-for-drivers-2023-7},
  note    = {Accessed December 24, 2025}
}

@article{InstacartCourierParkingLot,
  author  = {Alex Bitter},
  title   = {{Instacart shoppers are camping out in parking lots to compete for orders
}},
  journal = {Business Insider},
  year    = {2023},
  url     = {https://www.businessinsider.com/instacart-shoppers-wait-in-parking-lots-to-claim-orders-2023-7},
  note    = {Accessed December 25, 2025}
}

@article{BuyerCompete,
  author  = {Nora Redmond},
  title   = {Food deliveries are getting more expensive — but we can't stop ordering
},
  journal = {Business Insider},
  year    = {2024},
  url     = {https://www.businessinsider.com/us-diners-splurge-food-delivery-doordash-grubhub-ubereats-convenience-rules-2024-11},
  note    = {Accessed December 24, 2025}
}

@article{jacobs2024gig,
  title={Gig Passenger and Delivery Driver Pay in Five Metro Areas},
  author={Jacobs, Ken and Reich, Michael and Challenor, Tynan and Farmand, Aida},
  year={2024},
  publisher={< bound method Organization. get\_name\_with\_acronym of< Organization: Center~…}
}

@article{karp1975computational,
  title={On the computational complexity of combinatorial problems},
  author={Karp, Richard M},
  journal={Networks},
  volume={5},
  number={1},
  pages={45--68},
  year={1975},
  publisher={Wiley Online Library}
}

@inproceedings{guruswami2005profit,
  title={On profit-maximizing envy-free pricing.},
  author={Guruswami, Venkatesan and Hartline, Jason D and Karlin, Anna R and Kempe, David and Kenyon, Claire and McSherry, Frank},
  booktitle={SODA},
  volume={5},
  pages={1164--1173},
  year={2005}
}

@inproceedings{cheung2008approximation,
  title={Approximation algorithms for single-minded envy-free profit-maximization problems with limited supply},
  author={Cheung, Maurice and Swamy, Chaitanya},
  booktitle={2008 49th Annual IEEE Symposium on Foundations of Computer Science},
  pages={35--44},
  year={2008},
  organization={IEEE}
}

@article{im2012envy,
  title={Envy-free pricing with general supply constraints for unit demand consumers},
  author={Im, Sungjin and Lu, Pin-Yan and Wang, Ya-Jun},
  journal={Journal of Computer Science and Technology},
  volume={27},
  number={4},
  pages={702--709},
  year={2012},
  publisher={Springer}
}

@article{ma2020spatio,
  title={Spatio-temporal pricing for ridesharing platforms},
  author={Ma, Hongyao and Fang, Fei and Parkes, David C},
  journal={Operations Research},
  volume={70},
  number={2},
  pages={1025--1041},
  year={2022},
  publisher={INFORMS}
}

@article{azar2007pay,
  title={Why pay extra? Tipping and the importance of social norms and feelings in economic theory},
  author={Azar, Ofer H},
  journal={The Journal of Socio-Economics},
  volume={36},
  number={2},
  pages={250--265},
  year={2007},
  publisher={Elsevier}
}

@article{lynn2015service,
  title={Service gratuities and tipping: A motivational framework},
  author={Lynn, Michael},
  journal={Journal of Economic Psychology},
  volume={46},
  pages={74--88},
  year={2015},
  publisher={Elsevier}
}

@article{lei2023two,
  title={Two-sided platform competition in the presence of tip baiting},
  author={Lei, Yongqin and Pun, Hubert},
  journal={Available at SSRN 4528715},
  year={2023}
}

@article{alexander2021effects,
  title={The effects of tip recommendations on customer tipping, satisfaction, repatronage, and spending},
  author={Alexander, Damon and Boone, Christopher and Lynn, Michael},
  journal={Management Science},
  volume={67},
  number={1},
  pages={146--165},
  year={2021},
  publisher={INFORMS}
}

@article{shy2015tips,
  title={Do tips increase workers' income?},
  author={Shy, Oz},
  journal={Management Science},
  volume={61},
  number={9},
  pages={2041--2051},
  year={2015},
  publisher={INFORMS}
}

@article{chen2024courier,
  title={Courier dispatch in on-demand delivery},
  author={Chen, Mingliu and Hu, Ming},
  journal={Management Science},
  volume={70},
  number={6},
  pages={3789--3807},
  year={2024},
  publisher={INFORMS}
}

@article{chen2022food,
  title={Food delivery service and restaurant: Friend or foe?},
  author={Chen, Manlu and Hu, Ming and Wang, Jianfu},
  journal={Management Science},
  volume={68},
  number={9},
  pages={6539--6551},
  year={2022},
  publisher={INFORMS}
}

@article{feldman2023managing,
  title={Managing relationships between restaurants and food delivery platforms: Conflict, contracts, and coordination},
  author={Feldman, Pnina and Frazelle, Andrew E and Swinney, Robert},
  journal={Management Science},
  volume={69},
  number={2},
  pages={812--823},
  year={2023},
  publisher={INFORMS}
}

@article{bahrami2023three,
  title={The three-sided market of on-demand delivery},
  author={Bahrami, Sina and Nourinejad, Mehdi and Yin, Yafeng and Wang, Hai},
  journal={Transportation Research Part E: Logistics and Transportation Review},
  volume={179},
  pages={103313},
  year={2023},
  publisher={Elsevier}
}

@article{liu2023operating,
  title={Operating three-sided marketplace: Pricing and spatial staffing in food delivery systems},
  author={Liu, Zhe and Shen, Yiwen and Sun, Yanwei},
  journal={Available at SSRN 4668867},
  year={2023}
}

@article{wang2025recommending,
  title={Recommending for a multi-sided marketplace: A multi-objective hierarchical approach},
  author={Wang, Yuyan and Tao, Long and Zhang, Xian Xing},
  journal={Marketing Science},
  volume={44},
  number={1},
  pages={1--29},
  year={2025},
  publisher={INFORMS}
}

@article{castillo2022designing,
  title={Designing technology for on-demand delivery: The effect of customer tipping on crowdsourced driver behavior and last mile performance},
  author={Castillo, Vincent E and Mollenkopf, Diane A and Bell, John E and Esper, Terry L},
  journal={Journal of Operations Management},
  volume={68},
  number={5},
  pages={424--453},
  year={2022},
  publisher={Wiley Online Library}
}

@article{haggag2014default,
  title={Default tips},
  author={Haggag, Kareem and Paci, Giovanni},
  journal={American Economic Journal: Applied Economics},
  volume={6},
  number={3},
  pages={1--19},
  year={2014},
  publisher={American Economic Association 2014 Broadway, Suite 305, Nashville, TN 37203-2425}
}

@article{arkin1998local,
  title={On local search for weighted k-set packing},
  author={Arkin, Esther M and Hassin, Refael},
  journal={Mathematics of Operations Research},
  volume={23},
  number={3},
  pages={640--648},
  year={1998},
  publisher={INFORMS}
}

@inproceedings{halldorsson1995approximating,
  title={Approximating discrete collections via local improvements},
  author={Halld{\'o}rsson, Magn{\'u}s M},
  booktitle={Proceedings of the sixth annual ACM-SIAM symposium on Discrete algorithms},
  pages={160--169},
  year={1995}
}

@phdthesis{chan2009linear,
  title={On Linear Programming Relaxations of Hypergraph Matching},
  author={Chan, Yuk Hei},
  year={2009},
  school={Chinese University of Hong Kong}
}

@inproceedings{hatfield2011multilateral,
  title={Multilateral matching},
  author={Hatfield, John and Kominers, Scott Duke},
  booktitle={Proceedings of the 12th ACM conference on Electronic commerce},
  pages={337--338},
  year={2011}
}

@article{hatfield2013stability,
  title={Stability and competitive equilibrium in trading networks},
  author={Hatfield, John William and Kominers, Scott Duke and Nichifor, Alexandru and Ostrovsky, Michael and Westkamp, Alexander},
  journal={Journal of Political Economy},
  volume={121},
  number={5},
  pages={966--1005},
  year={2013},
  publisher={University of Chicago Press Chicago, IL}
}

@article{ostrovsky2008stability,
  title={Stability in supply chain networks},
  author={Ostrovsky, Michael},
  journal={American Economic Review},
  volume={98},
  number={3},
  pages={897--923},
  year={2008},
  publisher={American Economic Association}
}

@article{alkan1988nonexistence,
  title={Nonexistence of stable threesome matchings},
  author={Alkan, Ahmet},
  journal={Mathematical social sciences},
  volume={16},
  number={2},
  pages={207--209},
  year={1988},
  publisher={Elsevier}
}

@article{shapley1971assignment,
  title={The assignment game I: The core},
  author={Shapley, Lloyd S and Shubik, Martin},
  journal={International Journal of game theory},
  volume={1},
  number={1},
  pages={111--130},
  year={1971},
  publisher={Springer}
}

@misc{raj2021,
	title={COVID-19 and Digital Resilience: Evidence from Uber Eats}, 
	author={Manav Raj and Arun Sundararajan and Calum You},
	year={2021},
	URL = {https://ssrn.com/abstract=3625638},
	journal = {Available at SSRN 3625638},
}

@book{lovasz2009matching,
  title={Matching theory},
  author={Lov{\'a}sz, L{\'a}szl{\'o} and Plummer, Michael D},
  volume={367},
  year={2009},
  publisher={American Mathematical Soc.}
}

@article{icsik2021impact,
  title={The impact of the COVID-19 pandemic on Amazon's business},
  author={I{\c{s}}{\i}k, S{\i}la and {\.I}bi{\c{s}}, Hazal and Gulseven, Osman},
  journal={Available at SSRN 3766333},
  year={2021}
}

@inproceedings{tiananalysis,
  title={Analysis of the Impact of COVID-19 on the Share Investment of Top Five US Tech Companies},
  author={Tian, Yanqiang and Talukder, Dayal},
  booktitle={ICL},
  pages={67}
}

@article{chen2014envy,
  title={Envy-free pricing in multi-item markets},
  author={Chen, Ning and Deng, Xiaotie},
  journal={ACM Transactions on Algorithms (TALG)},
  volume={10},
  number={2},
  pages={1--15},
  year={2014},
  publisher={ACM New York, NY, USA}
}

@article{veljanovski2022algorithmic,
  title={Algorithmic antitrust: a critical overview},
  author={Veljanovski, Cento},
  journal={Algorithmic Antitrust},
  pages={39--64},
  year={2022},
  publisher={Springer}
}

@inproceedings{mladenov2020optimizing,
  title={Optimizing long-term social welfare in recommender systems: A constrained matching approach},
  author={Mladenov, Martin and Creager, Elliot and Ben-Porat, Omer and Swersky, Kevin and Zemel, Richard and Boutilier, Craig},
  booktitle={International Conference on Machine Learning},
  pages={6987--6998},
  year={2020},
  organization={PMLR}
}

@article{manea2018intermediation,
  title={Intermediation and resale in networks},
  author={Manea, Mihai},
  journal={Journal of Political Economy},
  volume={126},
  number={3},
  pages={1250--1301},
  year={2018},
  publisher={University of Chicago Press Chicago, IL}
}

@article{halaburda2018competing,
  title={Competing by restricting choice: The case of matching platforms},
  author={Halaburda, Hanna and Jan Piskorski, Miko{\l}aj and Y{\i}ld{\i}r{\i}m, P{\i}nar},
  journal={Management Science},
  volume={64},
  number={8},
  pages={3574--3594},
  year={2018},
  publisher={INFORMS}
}

@article{huttenlocher2023matching,
  title={Matching of users and creators in two-sided markets with departures},
  author={Huttenlocher, Daniel and Li, Hannah and Lyu, Liang and Ozdaglar, Asuman and Siderius, James},
  journal={arXiv preprint arXiv:2401.00313},
  year={2023}
}

@inproceedings{blume2007trading,
  title={Trading networks with price-setting agents},
  author={Blume, Larry and Easley, David and Kleinberg, Jon and Tardos, Eva},
  booktitle={Proceedings of the 8th ACM Conference on Electronic Commerce},
  pages={143--151},
  year={2007}
}

@inproceedings{balcan2008item,
  title={Item pricing for revenue maximization},
  author={Balcan, Maria-Florina and Blum, Avrim and Mansour, Yishay},
  booktitle={Proceedings of the 9th ACM Conference on Electronic Commerce},
  pages={50--59},
  year={2008}
}

@article{ke2022information,
  title={Information design of online platforms},
  author={Ke, T Tony and Lin, Song and Lu, Michelle Y},
  journal={Available at SSRN},
  year={2022}
}

@article{hu2022dynamic,
  title={Dynamic type matching},
  author={Hu, Ming and Zhou, Yun},
  journal={Manufacturing \& Service Operations Management},
  volume={24},
  number={1},
  pages={125--142},
  year={2022},
  publisher={INFORMS}
}

@inproceedings{immorlica2021designing,
  title={Designing approximately optimal search on matching platforms},
  author={Immorlica, Nicole and Lucier, Brendan and Manshadi, Vahideh and Wei, Alexander},
  booktitle={Proceedings of the 22nd ACM Conference on Economics and Computation},
  pages={632--633},
  year={2021}
}

@misc{li2021,
	title={Regulating Powerful Platforms: Evidence from Commission Fee Caps in On-Demand Services}, 
	author={Zhuoxin Li and Gang Wang},
	year={2021},
	URL = {https://ssrn.com/abstract=3871514},
	journal = {Available at 3871514},
}

@misc{oblander2022,
	title={Persistence of Consumer Lifestyle Choices: Evidence from Restaurant Delivery During COVID-19}, 
	author={E. Shin Oblander and Daniel McCarthy},
	year={2022},
	URL = {https://ssrn.com/abstract=3836262},
	journal = {Columbia Business School Research Paper Forthcoming},
}

@article{NRA2020,
	title={Restaurant Industry in Free Fall; 10,000 Close in Three Months}, 
	author={{National Restaurant Association}},
	year={2020},
}

@inproceedings{Zhan2021,
	author = {Zhan, Ruohan and Christakopoulou, Konstantina and Le, Ya and Ooi, Jayden and Mladenov, Martin and Beutel, Alex and Boutilier, Craig and Chi, Ed and Chen, Minmin},
	title = {Towards Content Provider Aware Recommender Systems: A Simulation Study on the Interplay between User and Provider Utilities},
	year = {2021},
	booktitle = {Proceedings of the Web Conference 2021},
	pages = {3872–3883}
}

@inproceedings{Tang17abc,
  author    = {Pingzhong Tang},
  title     = {Reinforcement mechanism design},
  booktitle = {Proceedings of the Twenty-Sixth International Joint Conference on
               Artificial Intelligence},
  pages     = {5146--5150},
  year      = {2017}
}

@inproceedings{BreroEGPR21,
  author    = {Gianluca Brero and
               Alon Eden and
               Matthias Gerstgrasser and
               David C. Parkes and
               Duncan Rheingans{-}Yoo},
  title     = {Reinforcement Learning of Sequential Price Mechanisms},
  booktitle = {Thirty-Fifth {AAAI} Conference on Artificial Intelligence},
  pages     = {5219--5227},
  year      = {2021}
}

@inproceedings{ShenPLZQHGDLT20,
  author    = {Weiran Shen and
               Binghui Peng and
               Hanpeng Liu and
               Michael Zhang and
               Ruohan Qian and
               Yan Hong and
               Zhi Guo and
               Zongyao Ding and
               Pengjun Lu and
               Pingzhong Tang},
  title     = {Reinforcement Mechanism Design: With Applications to Dynamic Pricing
               in Sponsored Search Auctions},
  booktitle = {The Thirty-Fourth {AAAI} Conference on Artificial Intelligence},
  pages     = {2236--2243},
  publisher = {{AAAI} Press},
  year      = {2020}
}

@inproceedings{Chen2019,
	author = {Minmin Chen and Alex Beutel and Paul Covington and Sagar Jain and Francois Belletti and Ed Chi},
	title = {Top-K Off-Policy Correction for a REINFORCE Recommender System},
	year = {2019},
	booktitle = {Proceedings of the 12th ACM International Conference on Web Search and Data Mining.},
	pages = {456--464}
}

@inproceedings{Salakhutdinov2007,
	author = {Salakhutdinov, Ruslan and Mnih, Andriy},
	title = {Probabilistic Matrix Factorization},
	year = {2007},
	booktitle = {Proceedings of the 20th International Conference on Neural Information Processing Systems},
	pages = {1257--1264}
}

@article{Zheng2021,
	author    = {Stephan Zheng and
	Alexander Trott and
	Sunil Srinivasa and
	David C. Parkes and
	Richard Socher},
	title     = {The {AI} Economist: Taxation policy design via two-level deep multiagent reinforcement learning},
year=2022,
journal={Science Advances},
volume=8,
number = {18},
pages = {eabk2607},}

@article{Ie2019,
	author    = {Eugene Ie and
	Chih{-}Wei Hsu and
	Martin Mladenov and
	Vihan Jain and
	Sanmit Narvekar and
	Jing Wang and
	Rui Wu and
	Craig Boutilier},
	title     = {RecSim: {A} Configurable Simulation Platform for Recommender Systems},
	journal   = {CoRR},
	volume    = {abs/1909.04847},
	year      = {2019},
}

@misc{MacKay2021,
	title={Consumer Inertia and Market Power}, 
	author={MacKay, Alexander and Remer, Marc},
	year={2021},
	URL = {https://ssrn.com/abstract=3380390},
	journal = {Available at SSRN 3380390},
}

@misc{Sullivan2022,
	title={Price Controls in a Multi-Sided Market}, 
	author={Michael Sullivan},
	year={2022},
	URL = {https://m-r-sullivan.github.io/assets/papers/food_delivery_cap.pdf},
}

@article{dube2010state,
  title={State dependence and alternative explanations for consumer inertia},
  author={Dub{\'e}, Jean-Pierre and Hitsch, G{\"u}nter J and Rossi, Peter E},
  journal={The RAND Journal of Economics},
  volume={41},
  number={3},
  pages={417--445},
  year={2010},
  publisher={Wiley Online Library}
}

@article{handel2013adverse,
  title={Adverse selection and inertia in health insurance markets: When nudging hurts},
  author={Handel, Benjamin R},
  journal={American Economic Review},
  volume={103},
  number={7},
  pages={2643--82},
  year={2013}
}

@inproceedings{PapadimitriouPP16,
  author    = {Christos H. Papadimitriou and
               George Pierrakos and
               Christos{-}Alexandros Psomas and
               Aviad Rubinstein},
  editor    = {Robert Krauthgamer},
  title     = {On the Complexity of Dynamic Mechanism Design},
  booktitle = {Proceedings of the Twenty-Seventh Annual {ACM-SIAM} Symposium on Discrete
               Algorithms},
  pages     = {1458--1475},
  publisher = {{SIAM}},
  year      = {2016},
 }

@article{honka2014quantifying,
  title={Quantifying search and switching costs in the US auto insurance industry},
  author={Honka, Elisabeth},
  journal={The RAND Journal of Economics},
  volume={45},
  number={4},
  pages={847--884},
  year={2014},
  publisher={Wiley Online Library}
}

@article{shum2004does,
  title={Does advertising overcome brand loyalty? Evidence from the breakfast-cereals market},
  author={Shum, Matthew},
  journal={Journal of Economics \& Management Strategy},
  volume={13},
  number={2},
  pages={241--272},
  year={2004},
  publisher={Wiley Online Library}
}

@article{farrell1988dynamic,
  title={Dynamic competition with switching costs},
  author={Farrell, Joseph and Shapiro, Carl},
  journal={The RAND Journal of Economics},
  pages={123--137},
  year={1988},
  publisher={JSTOR}
}

@article{dube2009switching,
  title={Do switching costs make markets less competitive?},
  author={Dub{\'e}, Jean-Pierre and Hitsch, G{\"u}nter J and Rossi, Peter E},
  journal={Journal of Marketing research},
  volume={46},
  number={4},
  pages={435--445},
  year={2009},
  publisher={SAGE Publications Sage CA: Los Angeles, CA}
}

@Misc{bo,
	author = {Fernando Nogueira},
	title = {{Bayesian Optimization}: Open source constrained global optimization tool for {Python}},
	year = {2014},
	url = " https://github.com/fmfn/BayesianOptimization"
}

@article{KAELBLING1998,
	title = {Planning and acting in partially observable stochastic domains},
	journal = {Artificial Intelligence},
	volume = {101},
	number = {1},
	pages = {99-134},
	year = {1998},
	issn = {0004-3702},
	author = {Leslie Pack Kaelbling and Michael L. Littman and Anthony R. Cassandra},
}

@article{Heess2015,
	author    = {Nicolas Heess and
	Jonathan J. Hunt and
	Timothy P. Lillicrap and
	David Silver},
	title     = {Memory-based control with recurrent neural networks},
	journal   = {CoRR},
	volume    = {abs/1512.04455},
	year      = {2015},
}

@InProceedings{Wierstra2007,
	author="Wierstra, Daan
	and Foerster, Alexander
	and Peters, Jan
	and Schmidhuber, J{\"u}rgen",
	editor="de S{\'a}, Joaquim Marques
	and Alexandre, Lu{\'i}s A.
	and Duch, W{\l}odzis{\l}aw
	and Mandic, Danilo",
	title="Solving Deep Memory POMDPs with Recurrent Policy Gradients",
	booktitle="Artificial Neural Networks -- ICANN 2007",
	year="2007",
	pages="697--706",
	}

@article{Hausknecht2015,
	author    = {Matthew J. Hausknecht and
	Peter Stone},
	title     = {Deep Recurrent Q-Learning for Partially Observable MDPs},
	journal   = {CoRR},
	volume    = {abs/1507.06527},
	year      = {2015},
	eprinttype = {arXiv},
}

@article{Armstrong2006,
 ISSN = {07416261},
 URL = {http://www.jstor.org/stable/25046266},
 author = {Mark Armstrong},
 journal = {The RAND Journal of Economics},
 number = {3},
 pages = {668--691},
 publisher = {[RAND Corporation, Wiley]},
 title = {Competition in Two-Sided Markets},
 volume = {37},
 year = {2006}
}

@BOOK{rlbook,
	TITLE = {Reinforcement Learning: an Introduction},
	SUBTITLE = {Second Edition},
	AUTHOR = { Richard S. Sutton and Andrew G. Barto},
	YEAR = {1998},
	PUBLISHER = {MIT Press},
}

@INPROCEEDINGS{Degris2012,
	author={Degris, Thomas and Pilarski, Patrick M. and Sutton, Richard S.},
	booktitle={2012 American Control Conference (ACC)}, 
	title={Model-Free reinforcement learning with continuous action in practice}, 
	year={2012},
	pages={2177-2182},
}

@article{stigler1963united,
  title={United States v. Loew's Inc.: A note on block-booking},
  author={Stigler, George J},
  journal={The Supreme Court Review},
  volume={1963},
  pages={152--157},
  year={1963},
  publisher={The University of Chicago Press}
}

@article{adams1976commodity,
  title={Commodity bundling and the burden of monopoly},
  author={Adams, William James and Yellen, Janet L},
  journal={The quarterly journal of economics},
  volume={90},
  number={3},
  pages={475--498},
  year={1976},
  publisher={MIT Press}
}

@article{mcafee1989multiproduct,
  title={Multiproduct monopoly, commodity bundling, and correlation of values},
  author={McAfee, R Preston and McMillan, John and Whinston, Michael D},
  journal={The Quarterly Journal of Economics},
  volume={104},
  number={2},
  pages={371--383},
  year={1989},
  publisher={MIT Press}
}

@article{salinger1995graphical,
  title={A graphical analysis of bundling},
  author={Salinger, Michael A},
  journal={Journal of Business},
  pages={85--98},
  year={1995},
  publisher={JSTOR}
}

@article{schmalensee1984gaussian,
  title={Gaussian demand and commodity bundling},
  author={Schmalensee, Richard},
  journal={Journal of business},
  pages={S211--S230},
  year={1984},
  publisher={JSTOR}
}

@techreport{pavlov2010optimal,
  title={Optimal mechanism for selling two goods},
  author={Pavlov, Gregory},
  year={2010},
  institution={Research Report}
}

@article{menicucci2015optimality,
  title={On the optimality of pure bundling for a monopolist},
  author={Menicucci, Domenico and Hurkens, Sjaak and Jeon, Doh-Shin},
  journal={Journal of Mathematical Economics},
  volume={60},
  pages={33--42},
  year={2015},
  publisher={Elsevier}
}

@article{bankos1997aggregation,
  title={Aggregation and Disaggregation of Information Goods: Implications for Bundling},
  author={Bankos, Y and Brynjolffson, E},
  journal={Site Licensing and Micropayment Systems},
  year={1997}
}

@article{geng2005bundling,
  title={Bundling information goods of decreasing value},
  author={Geng, Xianjun and Stinchcombe, Maxwell B and Whinston, Andrew B},
  journal={Management science},
  volume={51},
  number={4},
  pages={662--667},
  year={2005},
  publisher={INFORMS}
}

@article{fang2006bundle,
  title={To bundle or not to bundle},
  author={Fang, Hanming and Norman, Peter},
  journal={The RAND Journal of Economics},
  volume={37},
  number={4},
  pages={946--963},
  year={2006},
  publisher={Wiley Online Library}
}

@article{ibragimov2010optimal,
  title={Optimal bundling strategies under heavy-tailed valuations},
  author={Ibragimov, Rustam and Walden, Johan},
  journal={Management Science},
  volume={56},
  number={11},
  pages={1963--1976},
  year={2010},
  publisher={INFORMS}
}

@article{ghili2023characterization,
  title={A characterization for optimal bundling of products with nonadditive values},
  author={Ghili, Soheil},
  journal={American Economic Review: Insights},
  volume={5},
  number={3},
  pages={311--326},
  year={2023},
  publisher={American Economic Association 2014 Broadway, Suite 305, Nashville, TN 37203}
}

@article{haghpanah2021pure,
  title={When is pure bundling optimal?},
  author={Haghpanah, Nima and Hartline, Jason},
  journal={The Review of Economic Studies},
  volume={88},
  number={3},
  pages={1127--1156},
  year={2021},
  publisher={Oxford University Press}
}

@article{sun2025partition,
  title={Partition and prosper: design and pricing of single bundle},
  author={Sun, Hailong and Li, Xiaobo and Teo, Chung-Piaw},
  journal={Operations Research},
  year={2025},
  publisher={INFORMS}
}

@article{babaioff2020simple,
  title={A simple and approximately optimal mechanism for an additive buyer},
  author={Babaioff, Moshe and Immorlica, Nicole and Lucier, Brendan and Weinberg, S Matthew},
  journal={Journal of the ACM (JACM)},
  volume={67},
  number={4},
  pages={1--40},
  year={2020},
  publisher={ACM New York, NY, USA}
}

@article{deng2024procurement,
  title={Procurement Auctions via Approximately Optimal Submodular Optimization},
  author={Deng, Yuan and Karbasi, Amin and Mirrokni, Vahab and Leme, Renato Paes and Velegkas, Grigoris and Zuo, Song},
  journal={arXiv preprint arXiv:2411.13513},
  year={2024}
}

@inproceedings{bei2012budget,
  title={Budget feasible mechanism design: from prior-free to bayesian},
  author={Bei, Xiaohui and Chen, Ning and Gravin, Nick and Lu, Pinyan},
  booktitle={Proceedings of the forty-fourth annual ACM symposium on Theory of computing},
  pages={449--458},
  year={2012}
}

@article{light2024quality,
  title={Quality selection in two-sided markets: A constrained price discrimination approach},
  author={Light, Bar and Johari, Ramesh and Weintraub, Gabriel},
  journal={Operations Research},
  volume={72},
  number={5},
  pages={1928--1957},
  year={2024},
  publisher={INFORMS}
}

@article{hagiu2015marketplace,
  title={Marketplace or reseller?},
  author={Hagiu, Andrei and Wright, Julian},
  journal={Management Science},
  volume={61},
  number={1},
  pages={184--203},
  year={2015},
  publisher={INFORMS}
}

@article{alaei2022revenue,
  title={Revenue-sharing allocation strategies for two-sided media platforms: Pro-rata vs. user-centric},
  author={Alaei, Saeed and Makhdoumi, Ali and Malekian, Azarakhsh and Peke{\v{c}}, Sa{\v{s}}a},
  journal={Management Science},
  volume={68},
  number={12},
  pages={8699--8721},
  year={2022},
  publisher={INFORMS}
}

@article{guan2025platform,
  title={Platform Certification and Seller Disclosure in Online Selling},
  author={Guan, Xu and Jiang, Yuan and Wang, Yulan},
  journal={Manufacturing \& Service Operations Management},
  year={2025},
  publisher={INFORMS}
}

@article{du2025originality,
  title={Originality vs. licensing: Optimal strategies of streaming platforms},
  author={Du, Siyu and Dong, Jiru and Li, Mingjun and Li, Shichang},
  journal={Transportation Research Part E: Logistics and Transportation Review},
  volume={201},
  pages={104260},
  year={2025},
  publisher={Elsevier}
}

@article{wu2024optimal,
  title={The optimal content provision strategy for a streaming platform: Pure agency, self-production, or collaboration},
  author={Wu, Jie and Du, Siyu and Ji, Xiang and Li, Mingjun},
  journal={IEEE Transactions on Engineering Management},
  volume={71},
  pages={3712--3726},
  year={2024},
  publisher={IEEE}
}

@misc{adgate2024content,
  author       = {Adgate, Brad},
  title        = {Report: Big Media Companies Are Increasing Their Spending On Content},
  year         = {2024},
  month        = nov,
  howpublished = {\emph{Forbes}},
  url          = {https://www.forbes.com/sites/bradadgate/2024/11/01/report-big-media-companies-are-increasing-their-spending-on-content/},
  note         = {Accessed: 2026-02-07},
}

@techreport{leung2025dissecting,
  title       = {Dissecting Netflix's Self-Preferencing: Evidence from Viewer-Level Data},
  author      = {Leung, Tin Cheuk and Qi, Shi and Strumpf, Koleman},
  year        = {2025},
  type        = {Working Papers},
  number      = {25-08},
  institution = {NET Institute},
}

@article{Anderson2024hybrid,
  title={Hybrid platform model: monopolistic competition and a dominant firm
},
  author={Simon P. Anderson and {\"O}zlem Bedre-Defolie},
  journal={The RAND Journal of Economics},
  volume={55},
  number={4},
  pages={684--718},
  year={2024}
}

@article{milgrom2002envelope,
  title={Envelope theorems for arbitrary choice sets},
  author={Milgrom, Paul and Segal, Ilya},
  journal={Econometrica},
  volume={70},
  number={2},
  pages={583--601},
  year={2002},
  publisher={Wiley Online Library}
}

@article{myerson1981optimal,
  title={Optimal auction design},
  author={Myerson, Roger B},
  journal={Mathematics of operations research},
  volume={6},
  number={1},
  pages={58--73},
  year={1981},
  publisher={INFORMS}
}

@article{hartline2013bayesian,
  title={Bayesian mechanism design},
  author={Hartline, Jason D},
  journal={Foundations and Trends{\textregistered} in Theoretical Computer Science},
  volume={8},
  number={3},
  pages={143--263},
  year={2013},
  publisher={Emerald Publishing Limited}
}

@article{baricz2010mills,
  title={Mills' ratio: Reciprocal convexity and functional inequalities},
  author={Baricz, Arp{\'a}d},
  journal={arXiv preprint arXiv:1010.3267},
  year={2010}
}

@article{sampford1953some,
  title={Some inequalities on Mill's ratio and related functions},
  author={Sampford, Michael R},
  journal={The Annals of Mathematical Statistics},
  volume={24},
  number={1},
  pages={130--132},
  year={1953},
  publisher={JSTOR}
}

@article{fu2017notes,
  title={Notes on Myerson’s Revenue Optimal Mechanisms},
  author={Fu, Hu},
  journal={Lecture Note},
  year={2017}
}

@article{rothschild2026agentic,
  title={The Agentic Economy},
  author={Rothschild, David M and Mobius, Markus M and Hofman, Jake M and Dillon, Eleanor and Goldstein, Daniel G and Immorlica, Nicole and Jaffe, Sonia and Lucier, Brendan and Slivkins, Aleksandrs and Vogel, Matthew},
  journal={Communications of the ACM},
  volume={69},
  number={2},
  pages={39--42},
  year={2026},
  publisher={ACM New York, NY, USA}
}

@article{lucier2026agentic,
  title={Agentic Markets: Equilibrium Effects of Improving Consumer Search},
  author={Lucier, Brendan and Immorlica, Nicole and Mobius, Markus and Slivkins, Aleksandrs and Goldstein, Daniel G and Hofman, Jake M and Jaffe, Sonia and Rothschild, David M},
  journal={arXiv preprint arXiv:2603.25893},
  year={2026}
}

@article{hagiu2006merchant,
  title={Merchant or two-sided platform?},
  author={Hagiu, Andrei},
  year={2006},
  publisher={Harvard NOM Working Paper}
}

@book{marshall1961principles,
  title={Principles of economics},
  author={Marshall, Alfred and Guillebaud, Claude William and others},
  volume={1},
  year={1961},
  publisher={Springer}
}

@article{roth2018marketplaces,
  title={Marketplaces, markets, and market design},
  author={Roth, Alvin E},
  journal={American Economic Review},
  volume={108},
  number={7},
  pages={1609--1658},
  year={2018},
  publisher={American Economic Association 2014 Broadway, Suite 305, Nashville, TN 37203}
}

@article{diamond1982aggregate,
  title={Aggregate demand management in search equilibrium},
  author={Diamond, Peter A},
  journal={Journal of political Economy},
  volume={90},
  number={5},
  pages={881--894},
  year={1982},
  publisher={The University of Chicago Press}
}

@article{bakos1997reducing,
  title={Reducing buyer search costs: Implications for electronic marketplaces},
  author={Bakos, J Yannis},
  journal={Management science},
  volume={43},
  number={12},
  pages={1676--1692},
  year={1997},
  publisher={INFORMS}
}

@article{ellison2009search,
  title={Search, obfuscation, and price elasticities on the internet},
  author={Ellison, Glenn and Ellison, Sara Fisher},
  journal={Econometrica},
  volume={77},
  number={2},
  pages={427--452},
  year={2009},
  publisher={Wiley Online Library}
}

@article{johnson2017agency,
  title={The agency model and MFN clauses},
  author={Johnson, Justin P},
  journal={The Review of Economic Studies},
  volume={84},
  number={3},
  pages={1151--1185},
  year={2017},
  publisher={Oxford University Press}
}

@article{abhishek2016agency,
  title={Agency selling or reselling? Channel structures in electronic retailing},
  author={Abhishek, Vibhanshu and Jerath, Kinshuk and Zhang, Z John},
  journal={Management Science},
  volume={62},
  number={8},
  pages={2259--2280},
  year={2016},
  publisher={INFORMS}
}

@article{derakhshan2022product,
  title={Product ranking on online platforms},
  author={Derakhshan, Mahsa and Golrezaei, Negin and Manshadi, Vahideh and Mirrokni, Vahab},
  journal={Management Science},
  volume={68},
  number={6},
  pages={4024--4041},
  year={2022},
  publisher={INFORMS}
}

@article{chu2020position,
  title={Position ranking and auctions for online marketplaces},
  author={Chu, Leon Yang and Nazerzadeh, Hamid and Zhang, Heng},
  journal={Management Science},
  volume={66},
  number={8},
  pages={3617--3634},
  year={2020},
  publisher={INFORMS}
}

\end{document}